**UNIVERSIDAD DE MÁLAGA**

**ESCUELA TÉCNICA SUPERIOR DE INGENIERÍA DE TELECOMUNICACIÓN**

**PROYECTO FIN DE CARRERA:**

# Diseño de un Enlace de Comunicaciones Ópticas con Marte

**INGENIERÍA TÉCNICA DE TELECOMUNICACIÓN**
**SISTEMAS DE TELECOMUNICACIÓN**

Málaga, 2005 Alberto Carrasco Casado

UNIVERSIDAD DE MÁLAGA

ESCUELA TÉCNICA SUPERIOR DE INGENIERÍA DE TELECOMUNICACIÓN

**Titulación: Ingeniería Técnica de Telecomunicación**
**Sistemas de Telecomunicación**

Reunido el tribunal examinador en el día de la fecha, constituido por:

D. ________________________________________________

D. ________________________________________________

D. ________________________________________________

para juzgar el Proyecto Fin de Carrera titulado:

# DISEÑO DE UN ENLACE DE COMUNICACIONES ÓPTICAS CON MARTE

del alumno D. Alberto Carrasco Casado

dirigido por D. Pedro Lázaro Legaz

ACORDÓ POR: ______________ OTORGAR LA CALIFICACIÓN DE

______________________________________________

y, para que conste, se extiende firmada por los componentes del Tribunal, la presente diligencia

Málaga, a ____de______________de ______

El Presidente | El Vocal | El Secretario

Fdo.: | Fdo.: | Fdo.:

_______________ _______________ _______________

UNIVERSIDAD DE MÁLAGA

ESCUELA TÉCNICA SUPERIOR DE INGENIERÍA DE TELECOMUNICACIÓN

# DISEÑO DE UN ENLACE DE COMUNICACIONES ÓPTICAS CON MARTE

REALIZADO POR:
Alberto Carrasco Casado

DIRIGIDO POR:
Pedro Lázaro Legaz

DEPARTAMENTO DE: Ingeniería de Comunicaciones

RESUMEN:

*En este proyecto se analiza la posibilidad del empleo de comunicaciones ópticas en espacio profundo como mejora de los actuales sistemas de comunicaciones de microondas. Se ha aplicado al caso particular de un enlace Marte-Tierra y se ha diseñado un enlace basado en el futuro proyecto MLCD de la NASA actualmente en fase de desarrollo. Para ello se ha realizado un simulador orbital, se ha evaluado la órbita de transferencia del satélite, se han analizado las pérdidas producidas en el canal de transmisión incluyendo varios modelos de atmósfera, se han seleccionado los elementos más idóneos para el transmisor y receptor, se ha calculado el efecto Doppler durante el transcurso de la misión y se ha realizado un balance de potencia en función de las posiciones orbitales. De los resultados obtenidos se ha evaluado la máxima tasa binaria de transmisión a lo largo del tiempo de misión, en función de los observatorios astronómicos escogidos como antenas receptoras.*

*Llegará un tiempo en que los hombres*
*serán capaces de ampliar su mirada...*
*y podrán ver los planetas como a nuestra propia Tierra.*

**Christopher Wren, *Discurso inaugural*,**
Gresham College, 1657.

# Prefacio

La capacidad de los enlaces actuales de telecomunicaciones en espacio profundo es comparable a los accesos de banda estrecha habituales para conexión a Internet. En un futuro próximo el creciente número de misiones de exploración espacial, junto a un aumento en los servicios y prestaciones requeridas en las mismas, precisarán de un significativo aumento en la capacidad de este tipo de enlaces.

El presente proyecto trata de demostrar la posibilidad de realizar una mejora cualitativa y cuantitativa en los enlaces de espacio profundo mediante el empleo de comunicaciones ópticas como principal canal tanto descendente como ascendente.

Se ha realizado el diseño de un enlace de comunicaciones ópticas entre la Tierra y Marte basado en el proyecto *MLCD* de la NASA, actualmente en fase de diseño, cuyo objetivo es alcanzar un régimen binario mínimo de 1 Mbit/s en las peores situaciones de distancia y ruido de fondo entre un satélite orbitando Marte y una estación terrestre ubicada en un observatorio astronómico. Se ha demostrado que un enlace de comunicaciones ópticas puede lograr los requerimientos exigidos por dicho proyecto.

Para llevar a cabo dicho diseño ha sido necesaria la implementación de un simulador de dinámica orbital que proporcionara las posiciones instantáneas de los terminales de comunicación durante el transcurso de los dos años considerados para la duración del experimento así como el viaje desde que la nave abandona su órbita terrestre hasta que se inserta en la de Marte. Este simulador es capaz de realizar animaciones con el objetivo de almacenar los datos que genera.

En base a los datos obtenidos del simulador mencionado y de los requerimientos del proyecto *MLCD*, se ha realizado un balance de potencia, para lo cual también ha sido necesario contemplar los efectos que la atmósfera terrestre induce a las señales que la atraviesan. Por último, se ha diseñado el receptor terrestre empleando fotodetectores comerciales y observatorios astronómicos reales como estaciones receptoras y se han evaluado las prestaciones del enlace en vista de los requerimientos mencionados. El enlace diseñado ha demostrado cumplir con las especificaciones durante los dos años considerados en la simulación a excepción de los periodos de conjunción en los que el Sol se interpone entre los dos terminales.

# Acrónimos

| Acrónimo | Español | Inglés |
|---|---|---|
| **APD** | Fotodiodo de avalancha | Avalanche Photo-Diode |
| **BER** | Tasa de error de bit | Bit Error Rate |
| **CC.OO.** | Comunicaciones ópticas | |
| **CCD** | Dispositivo de carga acoplada | Charged Couple Device |
| **DFB** | Láser de realimentación distribuida | Distributed Feedback |
| **DSN** | Red de espacio profundo | Deep Space Network |
| **EDFA** | Amplificador de fibra dopada con Erbio | Erbium-Doped Fiber Amplifier |
| **ESA** | Agencia espacial europea | European Space Agency |
| **FI** | Frecuencia Intermedia | |
| **FOV** | Campo de visión | Field Of View |
| **GEO** | Órbita geoestacionaria | Geostationary Earth Orbit |
| **GPS** | Sistema de posicionamiento global | Global Positioning System |
| **GTC** | Gran Telescopio de Canarias | |
| **IM/DD** | Modulación en intensidad y detección directa | Intensity Modulation/Direct Detection |
| **IR** | Infrarrojo | |
| **JPL** | | Jet Propulsion Laboratory |
| **LEO** | Órbita de baja altura | Low Earth Orbit |
| **MEO** | Órbita de media altura | Medium Earth Orbit |
| **MLCD** | | Mars Laser Communications Demonstration |
| **MOPA** | | Master Oscilator Power Amplifier |
| **MTO** | | Mars Telecommunications Orbiter |
| **NASA** | | National Aeronautics and Space Administration |
| **Nd:YAG** | Láser de Neodimio Yag | |
| **OGS** | Estación terrena óptica | Optical Ground Station |
| **OOK** | Modulación de apagado-encendido | On-Off Keying |
| **OPLL** | Bucle enganchado en fase óptico | Optical Phase-Locked Loop |
| **OSA** | Analizador de espectros óptico | Optical Spectrum Analyzer |
| **PIN** | Fotodiodo de unión p-i-n | |
| **PLL** | Bucle enganchado en fase | Phase-Locked Loop |
| **PMT** | Fotomultiplicador | Photo-Multiplier Tube |
| **PPM** | Modulación por posición de pulsos | Pulse Position Modulation |
| **PSF** | Función de dispersión puntual | Point Spread Function |
| **RF** | Radiofrecuencia | |
| **SNR** | Relación señal-ruido | Signal to Noise Ratio |
| **YDFA** | Amplificador de fibra dopada con Yterbio | Ytterbium-Doped Fiber Amplifier |

# Símbolos

| Símbolo | Significado |
|---|---|
| **a** | Semieje mayor de una elipse |
| **a** | Tamaño de partícula |
| $A_d$ | Área del fotodetector |
| $A_r$ | Área del espejo primario del telescopio receptor |
| $A_s$ | Área del objeto observado |
| **B** | Ancho de banda |
| **b** | Semieje menor de una elipse |
| **c** | Velocidad de la luz en el vacío ≈ $3 \cdot 10^8$ m/s |
| **D** | Diámetro del espejo primario de un telescopio |
| **E** | Anomalía excéntrica de una órbita elíptica |
| **e** | Carga del electrón = $1{,}601 \bullet 10^{-19}$ As |
| **e** | Excentricidad de una elipse |
| **erfc(x)** | Función de error complementaria de x |
| **F** | Factor de exceso de ruido de un fotodiodo |
| $f_c$ | Distancia focal de una lente |
| $f_R$ | Frecuencia recibida |
| $f_T$ | Frecuencia transmitida |
| **G** | Constante de Cavendish = $6{,}672 \cdot 10^{-11}$ $m^3 kg^{-1} s^{-1}$ |
| **g** | Eficiencia de la ganancia de una antena |
| $G_r$ | Ganancia de recepción de una antena |
| $G_t$ | Ganancia de transmisión de una antena |
| **h** | Constante de Planck = $6{,}624 \bullet 10^{-34}$ $Ws^2$ |
| **H(f)** | Irradiancia espectral de una fuente de ruido |
| **i** | Inclinación del plano orbital de una elipse |
| $I_d$ | Corriente de oscuridad de un fotodiodo |
| $I_{ph}$ | Fotocorriente generada por un fotodiodo |
| $J_1(x)$ | Función de Bessel de primer orden de x |
| **k** | Constante de Boltzman = $1{,}379 \bullet 10^{-23}$ Ws/K |
| $L_{apunt}$ | Pérdidas por apuntamiento de una antena |
| $L_{atm}$ | Pérdidas atmosféricas |
| $L_{fs}$ | Pérdidas por espacio libre |
| $L_M$ | Pérdidas por scattering Mie |
| $L_R$ | Pérdidas por scattering Rayleigh |
| $L_s$ | Pérdidas por scattering |
| **M** | Anomalía media de una órbita elíptica |
| **M** | Factor de multiplicación de un APD |
| **M** | Margen de seguridad |
| **m** | Número de símbolos en una modulación m-PPM |
| **M(z)** | Concentración másica a una altura z |
| $M_m$ | Peso molecular medio del aire = 28,94 gramos/mol |
| **n** | Número de bits de una modulación m-PPM |
| **N(f)** | Radiancia espectral de una fuente de ruido |
| $N_A$ | Número de Avogadro = $6{,}022 \bullet 10^{23}$ moléculas/mol |
| $N_g$ | Concentración numérica de gases o moléculas de gas por unidad de volumen |
| $N_p$ | Concentración numérica de partículas o partículas por unidad de volumen |
| $N_t$ | Densidad espectral de potencia de ruido térmico |
| $P_{em}$ | Probabilidad de error de símbolo |
| $P_l$ | Potencia del láser |
| $P_{opt}$ | Potencia óptica |
| $P_r$ | Potencia recibida |
| $P_t$ | Potencia transmitida |
| **Q** | Eficiencia de scattering |
| $r_0$ | Parámetro de Fried |
| $R_0$ | Responsividad de un fotodetector |
| $\mathbf{R_L}$ | Resistencia de carga |
| $\mathbf{S_n}$ | Densidad espectral de potencia de ruido |
| **T** | Periodo órbital |
| **T** | Temperatura |
| **v** | Anomalía verdadera de una órbita elíptica |
| $\mathbf{v_R}$ | Velocidad de rotación |
| $\mathbf{v_r}$ | Velocidad radial |
| $\mathbf{v_{rR}}$ | Velocidad radial por rotación |
| $\mathbf{v_{rT}}$ | Velocidad radial por traslación |
| $\mathbf{v_T}$ | Velocidad de traslación |
| $\boldsymbol{\alpha}$ | Factor de truncamiento de un telescopio |
| $\boldsymbol{\alpha_M}$ | Coeficiente de atenuación de scattering Mie |
| $\boldsymbol{\alpha_R}$ | Coeficiente de atenuación de scattering Rayleigh |
| $\boldsymbol{\gamma}$ | Factor de bloqueo de un telescopio |
| $\boldsymbol{\delta}$ | Ganancia unitaria por dínodo en un fotomultiplicador |
| $\boldsymbol{\Delta f}$ | Variación de frecuencia |
| $\boldsymbol{\eta}$ | Eficiencia cuántica de un fotodetector |
| $\boldsymbol{\theta}$ | Ángulo cenital |
| $\boldsymbol{\lambda}$ | Longitud de onda |
| $\boldsymbol{\nu}$ | Frecuencia óptica |
| $\boldsymbol{\rho}$ | Densidad |
| $\boldsymbol{\sigma_{geom}}$ | Sección geométrica |
| $\boldsymbol{\sigma_M}$ | Sección eficaz de scattering Mie |
| $\boldsymbol{\sigma_R}$ | Sección eficaz de scattering Rayleigh |
| $\boldsymbol{\omega}$ | Argumento del perihelio de una órbita elíptica |
| $\boldsymbol{\omega}$ | Velocidad angular |
| $\boldsymbol{\Omega}$ | Ascensión recta o longitud del nodo ascendente de una órbita elíptica |
| $\boldsymbol{\Omega_r}$ | Campo de visión del receptor (FOV) |
| $\boldsymbol{\Omega_s}$ | Ángulo sólido del objeto observado desde el detector |

# 1. Introducción

La Sociedad de la Información del siglo XXI, en un mundo cada vez más globalizado, se apoya en una compleja red de comunicaciones que abarcan zonas cada vez más amplias del espectro electromagnético y emplean los más diversos medios de transmisión. Entre ellas, las comunicaciones por satélite han conocido un despliegue espectacular, procurando al gran público un buen número de servicios que han modificado sustancialmente la vida cotidiana.

Las comunicaciones vía satélite se apoyan fundamentalmente en enlaces de microondas. Esta solución, que resulta satisfactoria para satélites en órbita terrestre, se ve cada vez más limitada a medida que se aleja de la Tierra el emisor remoto. La exploración de otros planetas del sistema solar, a pesar de sus impresionantes logros, se ve limitada por la reducida capacidad de los canales de comunicación. Dicha capacidad podría incrementarse sustancialmente reduciendo la longitud de onda de transmisión.

En este proyecto se aborda la posibilidad de utilizar comunicaciones ópticas por láser para enlaces en el espacio profundo. Se analizan los distintos problemas que presentan tales enlaces, y se aplican los resultados a un proyecto concreto de la NASA (*MLCD*), actualmente en fase de diseño, que pretende demostrar la viabilidad de un enlace óptico en una misión a Marte a finales de esta década.

Este capítulo se dedica a hacer una breve introducción a los temas principales que integran el presente proyecto (las comunicaciones por satélite, las comunicaciones en espacio profundo y las comunicaciones ópticas no guiadas), así como un resumen del trabajo llevado a cabo en el mismo. En sucesivos capítulos se estudiarán los distintos elementos del enlace, y finalmente se integrarán tales elementos en un ejemplo concreto, el proyecto *MLCD*.

## 1.1. COMUNICACIONES POR SATÉLITE

Se utiliza el término ***comunicaciones por satélite*** cuando, para establecer un enlace de comunicación, se emplean una o más naves espaciales. De esta manera se logran cubrir muy grandes distancias mediante el empleo de satélites como repetidores de las señales de comunicación. Por su parte, se usa el término ***comunicaciones en espacio profundo*** cuando se trata de realizar un enlace entre una nave a gran distancia de la Tierra (Luna, planetas, etc.) y una estación terrestre o en órbita terrestre. Estos dos tipos de comunicaciones comparten los fundamentos básicos, y los conceptos que se manejan suelen ser aplicables en los dos campos indistintamente. Sin embargo, los ámbitos de aplicación son radicalmente distintos: frente a las aplicaciones de uso masivo de las comunicaciones por satélite, las comunicaciones en espacio profundo están íntimamente ligadas *–al menos por ahora–* a la **exploración espacial**.

Los primeros satélites de comunicaciones proporcionaban poca capacidad a un alto coste (consecuencia de la necesidad de emplear lanzaderas para situarlos en órbita) con una duración muy corta (alrededor de un año y medio). El creciente desarrollo de las lanzaderas y los avances en la tecnología de microondas, han proporcionado una importante reducción de costes, y en la actualidad este tipo de comunicaciones está ampliamente extendida hasta el punto que cualquier persona puede poseer su propio terminal de comunicaciones por satélite, ya sea para recibir televisión, conectarse a Internet, hablar por teléfono u orientarse mediante *GPS*. Estos sistemas de comunicación complementan a los clásicos sistemas en ámbitos como la difusión de televisión y también proveen la única solución en otros ámbitos como la comunicación con determinadas zonas tales como el mar o regiones deshabitadas, donde otros sistemas de comunicación son inviables.

### 1.1.1. Reseña histórica

El nacimiento de las comunicaciones por satélite fue la consecuencia del gran desarrollo que la segunda guerra mundial propició a dos tipos de tecnologías diferentes: los **misiles** y las **microondas**. Los conocimientos combinados de ambos campos abrieron el camino a la posibilidad de usar satélites para establecer ***enlaces de comunicación.***

En el número de octubre de 1945 de la revista *Wireless World* se publicó un artículo titulado *"Relés extraterrestres"* en el que se expuso por primera vez la posibilidad de las comunicaciones vía satélite (aunque el concepto de órbita geoestacionaria ya había sido descrito en 1929 por Hermann Noordnung [1]). En el artículo citado, el autor Arthur C. Clarke (más conocido por sus novelas de ciencia ficción) proponía la puesta en órbita de tres satélites a 36.000 kilómetros y separados entre sí 120 grados en un plano paralelo al ecuador terrestre. La idea no se pudo llevar a cabo ya que en esa época no existía la tecnología para situar satélites en órbita y mucho menos a 36.000 kilómetros.

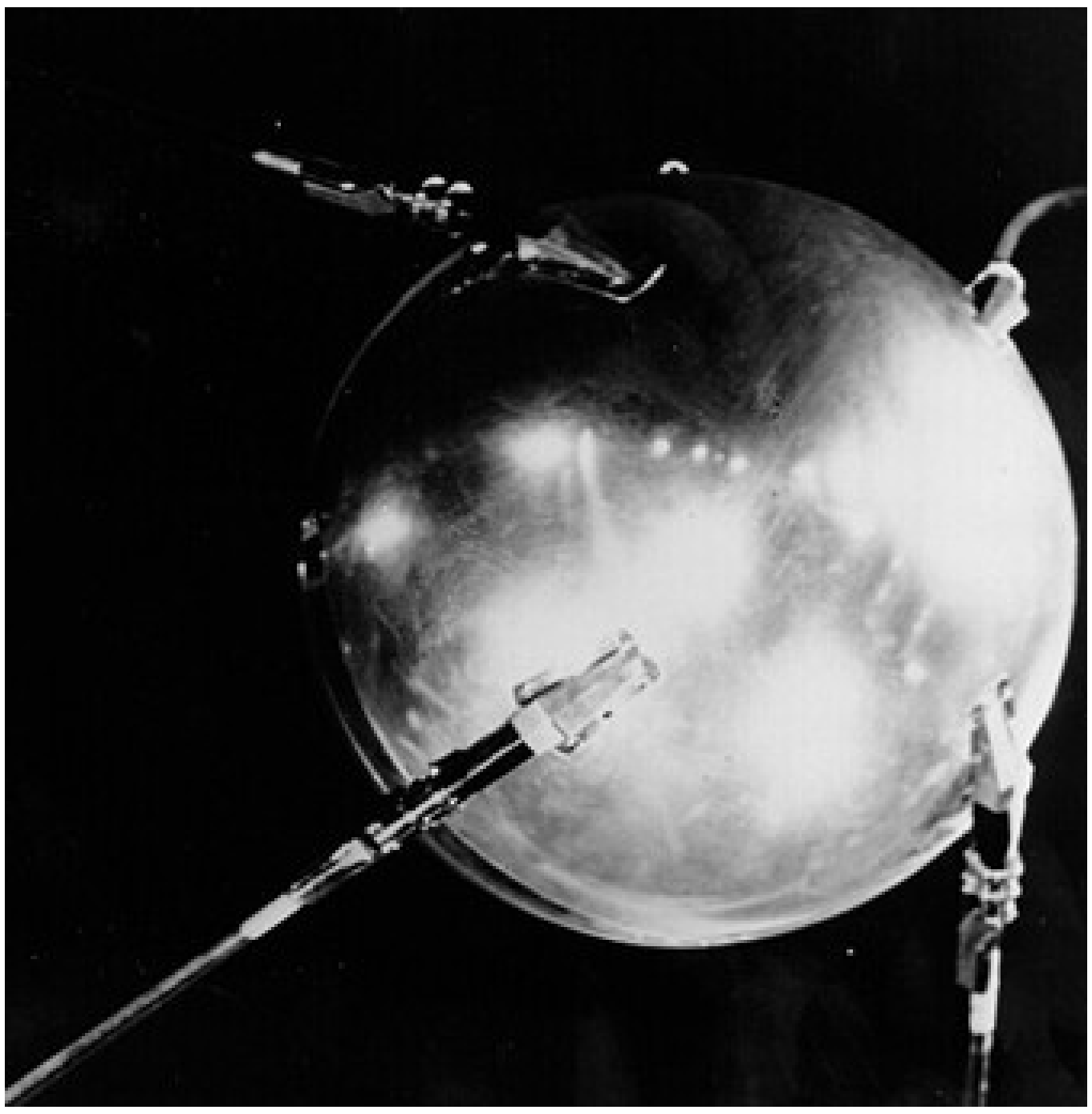

***Figura 1.1. Primer satélite artificial Sputnik 1 [2].***

La era espacial comenzó en 1957 con el lanzamiento del primer satélite artificial por parte de la Unión Soviética, el *Sputnik 1*. Era una esfera de aluminio de 58 cm con un peso de 83 kg que sólo emitía un tono intermitente. En su órbita elíptica baja tardaba 96,2 minutos en dar una vuelta a la Tierra [3].

Al *Sputnik 1* le sucedieron varios satélites mejorados pero siempre pasivos, se limitaban a reflejar las ondas electromagnéticas emitidas desde las estaciones terrestres. El primer satélite que actuó de repetidor activo fue el *Courier 1B*, lanzado en 1960 por el Departamento de Defensa de Estados Unidos. Además fue el primero que usó paneles solares, en lugar de baterías, para obtener la energía.

El *Telstar 1* (1962) fue el primer satélite en transmitir señales de televisión; el primer satélite geoestacionario fue el *Syncom 2* (1963), y el *Intelsat 1* (conocido como *early bird*) fue el primer satélite geoestacionario comercial de comunicaciones. Desde entonces la sucesión de satélites de comunicaciones cada vez más sofisticados fue constante. Cabe señalar que sólo transcurrieron once años entre el lanzamiento del primer satélite artificial (*Sputnik 1*) y la realización efectiva de un sistema global de comunicaciones por satélite plenamente operacional (*Intelsat 3*) en 1968 [1]. Actualmente hay cientos de satélites en órbita, incluyendo satélites de comunicaciones, militares, de ayuda a la navegación, científicos, etc.

## 1.1.2. Conceptos básicos

Las señales en un enlace de comunicaciones por satélite se propagan en espacio libre, esto es, el camino no cuenta con obstáculos (salvo la atmósfera terrestre) ni se trata de guiar a la señal una vez ha sido radiada. El trayecto que recorren las señales está formado siempre por una línea de visión directa, es decir, una línea imaginaria que une ambos terminales.

La arquitectura básica de un sistema de comunicaciones por satélite se puede dividir en dos partes diferenciadas: el ***segmento espacial*** y el ***segmento de Tierra***. El primero lo constituye el satélite y todo lo que transporta; el segmento de Tierra está formado por las estaciones terrestres que se comunican directamente con el terminal espacial, así como por toda la infraestructura de control, que consiste en las estaciones encargadas de monitorizar y controlar la posición y orientación del satélite y de gestionar las comunicaciones y demás recursos a bordo del satélite.

El satélite está formado por la ***plataforma*** y la ***carga útil***, conociéndose como carga útil al conjunto de antenas de recepción y transmisión y todo el equipo de comunicaciones. La plataforma la constituyen tanto la estructura básica del satélite como los equipos que hacen que la carga útil pueda operar.

La comunicación se establece a través de dos canales: uno de ***subida*** o ascendente en el que se establece la comunicación en dirección Tierra-satélite y otro de ***bajada*** o descendente con sentido contrario. En el canal de subida suelen emplearse señales de frecuencias mayores a las del canal de bajada debido a que los efectos perjudiciales que introduce la atmósfera en las señales, y que se traducen en pérdida de potencia, aumentan con la frecuencia. De esta manera se aprovecha mejor la potencia del satélite, que es un recurso limitado en el espacio.

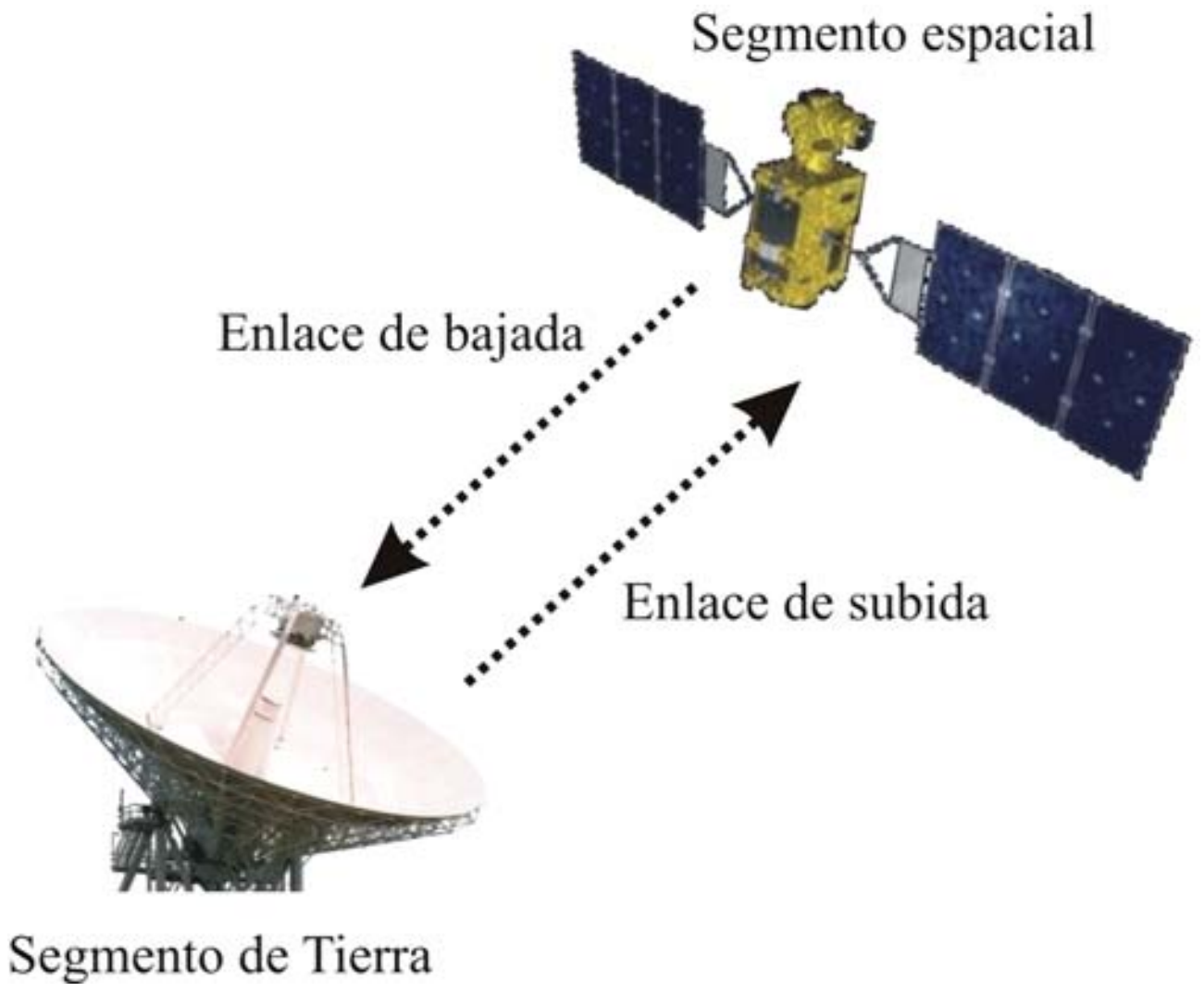

***Figura 1.2. Arquitectura básica de un sistema de comunicaciones por satélite.***

Los satélites de comunicaciones se suelen situar en varios tipos distintos de órbitas nombradas en referencia a su posición con respecto a la Tierra. Las principales, como se pueden observar en la Figura 1.3, son

- ***Geoestacionaria*** (**GEO** del inglés *Geoestationary Earth Orbit*). Órbita circular paralela al ecuador a una altura de 35.786,557 kilómetros con una duración de 24 horas. Se llama así porque el satélite, visto desde la superficie terrestre, aparenta estar siempre en la misma posición, lo que hace a la órbita muy atractiva para la difusión continuada de algún servicio a una misma área.

- ***Geosíncrona***. A la misma altura que la órbita geoestacionaria, aunque el plano orbital puede no coincidir con el plano ecuatorial.

- ***De media altura*** (**MEO** del inglés *Medium Earth Orbit*). A una altura de unos 10.000 kilómetros, los satélites tardan varias horas en dar una vuelta completa a la Tierra.

- ***De baja altura*** (**LEO** del inglés *Low Earth Orbit*). Estas órbitas se encuentran entre unos 500 y 2.000 kilómetros entre la región de densidad atmosférica constante y el cinturón de Van Allen. Los periodos orbitales, debido a la baja altura, son muy pequeños, entre los 90 minutos y las dos horas.

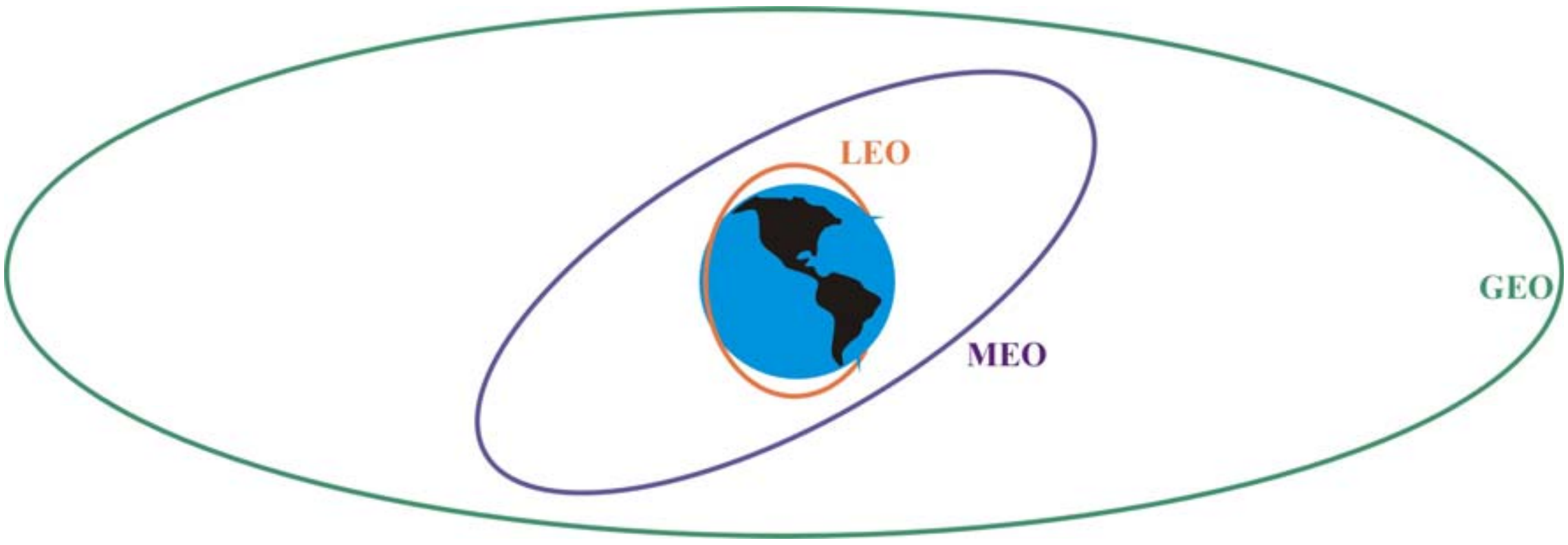

***Figura 1.3. Tipos de órbitas más comunes para satélites de comunicaciones [4].***

La vida de un satélite termina cuando el combustible destinado a mantener la posición orbital se agota. Desde ese momento se convierte en chatarra espacial.

# 1.2. COMUNICACIONES EN ESPACIO PROFUNDO

El rápido desarrollo que tuvo la tecnología de satélites proporcionó muy pronto la posibilidad de realizar misiones de exploración en espacio profundo. Con la sucesión de este tipo de misiones, se fue experimentando con tecnología de comunicaciones cada vez más avanzada. A continuación se ofrece una breve reseña histórica de los hitos más importantes en el ámbito de las comunicaciones en el espacio, así como una serie de conceptos básicos en los que se han venido basando este tipo de comunicaciones.

## 1.2.1. Reseña histórica

Los primeros programas espaciales se concentraron en naves tripuladas a poca distancia de la Tierra y en misiones a la Luna. Si se considera, como es usual, al espacio profundo como las distancias más allá de la Luna, se puede decir que el primer enlace de comunicación en espacio profundo se estableció en 1960 entre la Tierra y la primera nave espacial que logró **abandonar la atracción gravitatoria** terrestre y realizar un viaje interplanetario, la *Pioneer 5*. Esta misión sirvió para probar las comunicaciones a grandes distancias y se pudo establecer contacto con la Tierra hasta una distancia de 36,2 millones de kilómetros utilizando un transmisor de radio de 5 W.

En la Figura 1.4 se muestra la evolución de los enlaces de comunicaciones de las principales misiones en espacio profundo de la historia. Puede observarse el progreso de las prestaciones mediante el aumento del régimen binario y los avances tecnológicos más destacables.

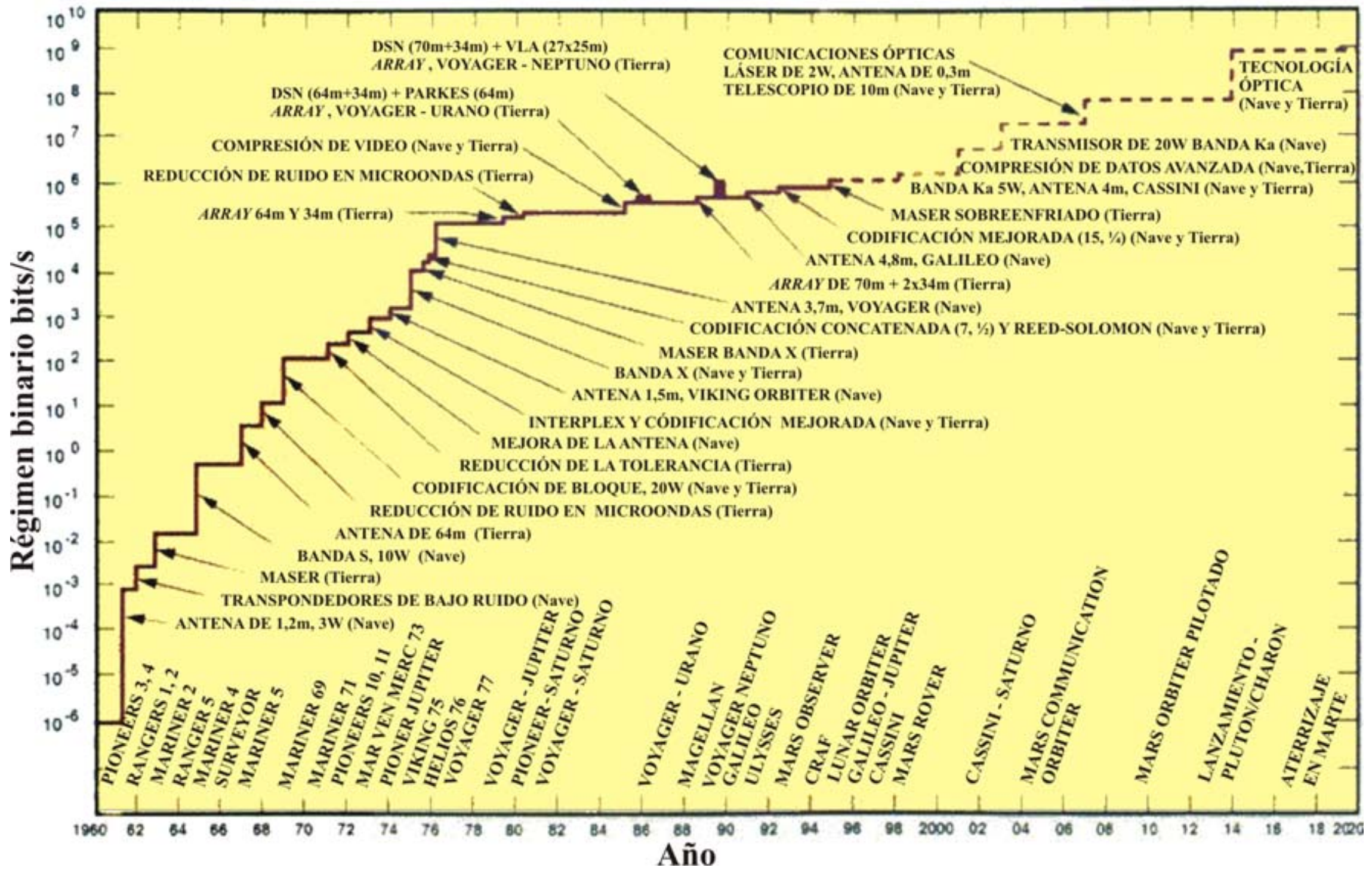

*Figura 1.4. Evolución histórica de las comunicaciones en espacio profundo [5].*

El principal aumento en la capacidad de las comunicaciones en espacio profundo ha venido marcado por la sucesiva *migración a frecuencias cada vez superiores* (las ventajas que supone el aumento de la frecuencia de las señales se puede ver en el apartado siguiente). En la década de 1960 se usaban frecuencias de 890 MHz dentro de la banda L y 2,3 GHz en la banda S. En 1977, las naves *Voyager* emplearon por vez primera la banda X a 8,4 GHz como principal enlace descendente tras la exitosa demostración probada en las misiones *Mariner* y *Viking* a principios de los 70.

En la década de 1990, la *NASA* desarrolló la idea de utilizar naves más pequeñas con una reducción de la carga útil. Con la migración a la banda Ka (32 GHz) pudo conseguirse una significativa disminución en la potencia consumida y la masa a bordo de la nave manteniendo los mismos volúmenes de datos que se conseguían en banda X. Los equipos experimentales de comunicaciones utilizando banda Ka de las misiones *Cassini* y *Mars Global Surveyor* ayudaron a establecer la viabilidad de esta banda en los siguientes enlaces de comunicaciones de espacio profundo.

En 2002 se celebraron los **25 años de funcionamiento** del enlace de comunicación entre la *DSN* (del inglés *Deep Space Network*, la red de espacio profundo de la *NASA*) y las naves *Voyager 1* y *Voyager 2*, que actualmente se encuentran en el exterior del sistema solar. Ambas naves están tan lejos que únicamente las mayores antenas de la *DSN* pueden detectar su señal.

Las futuras misiones tripuladas, así como los establecimientos de larga duración a distancias como Marte o mayores exigirán enlaces de comunicaciones fiables y de gran capacidad. Será necesaria una mejora cualitativa en las técnicas y tecnologías de telecomunicación en espacio profundo, tanto en los **terminales espaciales** –que deberán ser cada vez más pequeños, ligeros, eficientes e intercomunicados (surgiendo el concepto de red en el espacio)– como en los **terminales terrestres**, que tendrán que soportar una gran cantidad de tráfico y dar cobertura a todas las naves en activo (Figura 1.5). Todo ello, junto al permanente aumento en la frecuencia de las señales, revela a las comunicaciones ópticas como probables sucesoras de las actuales comunicaciones por microondas.

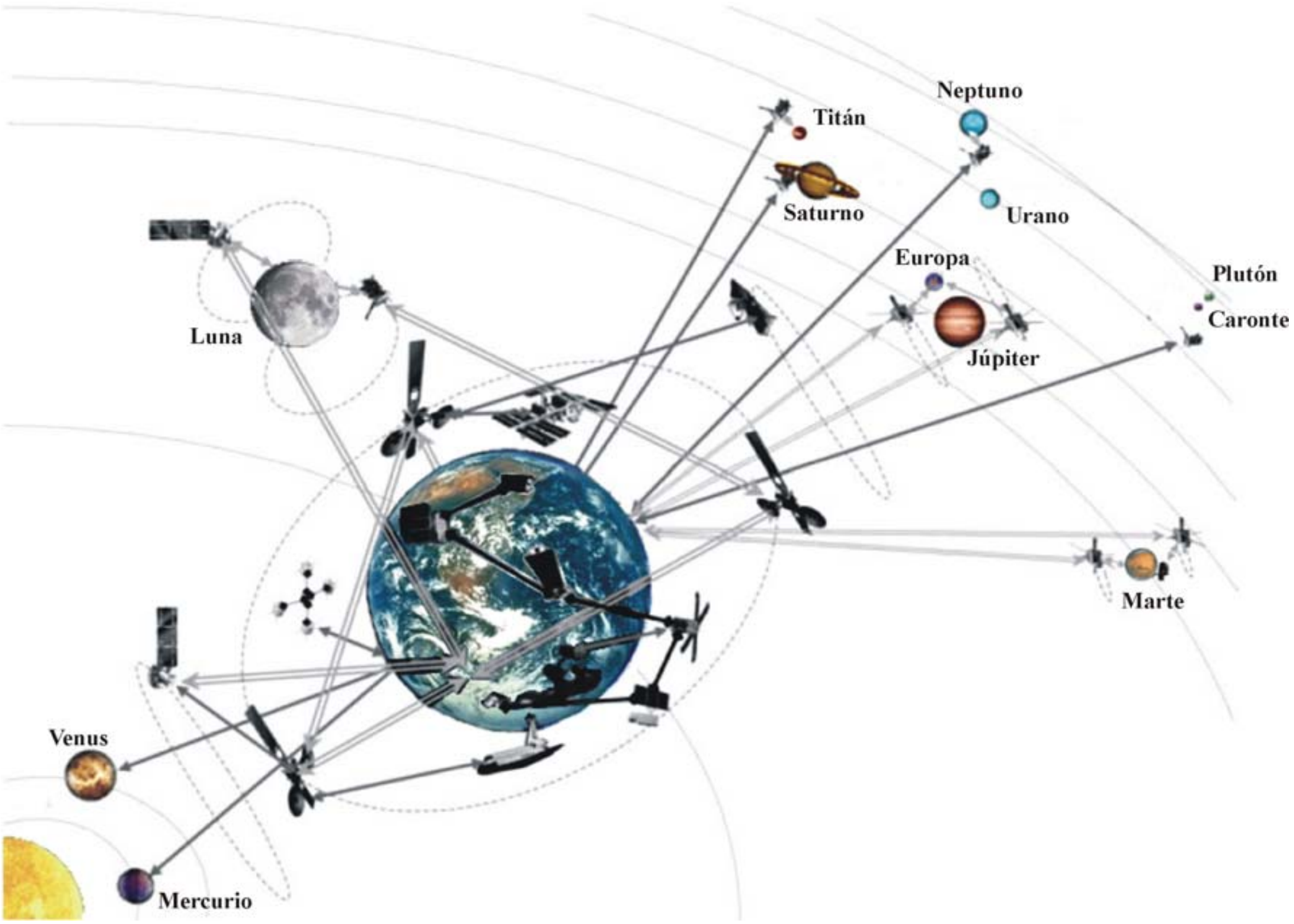

***Figura 1.5. Representación de una posible red de comunicaciones en espacio profundo [6].***

## 1.2.2. Conceptos básicos

En el espacio la potencia es un *bien escaso*. Sin embargo, para detectar una señal a millones de kilómetros se necesita recibir la suficiente potencia como para poder distinguirla del ruido. Una de las claves para poder establecer un enlace de comunicaciones en espacio profundo es la capacidad de ***confinar toda la potencia*** que se genere en un haz muy estrecho para intentar transportarla, a través de las grandes distancias involucradas, con las mínimas pérdidas posibles. Esto se consigue utilizando ondas electromagnéticas de alta frecuencia y reflectores que concentran dicha radiación en una única dirección.

Las enormes distancias que tiene que recorrer la onda electromagnética hacen que el haz en el que se concentra la potencia vaya abriéndose siguiendo la conocida **ley del cuadrado de la distancia**, según la cual la densidad de señal existente a una determinada distancia desde la fuente es inversamente proporcional al cuadrado de la misma. Por ello, en la Tierra la estrategia es utilizar antenas lo más grandes posibles para recoger la máxima cantidad de radiación.

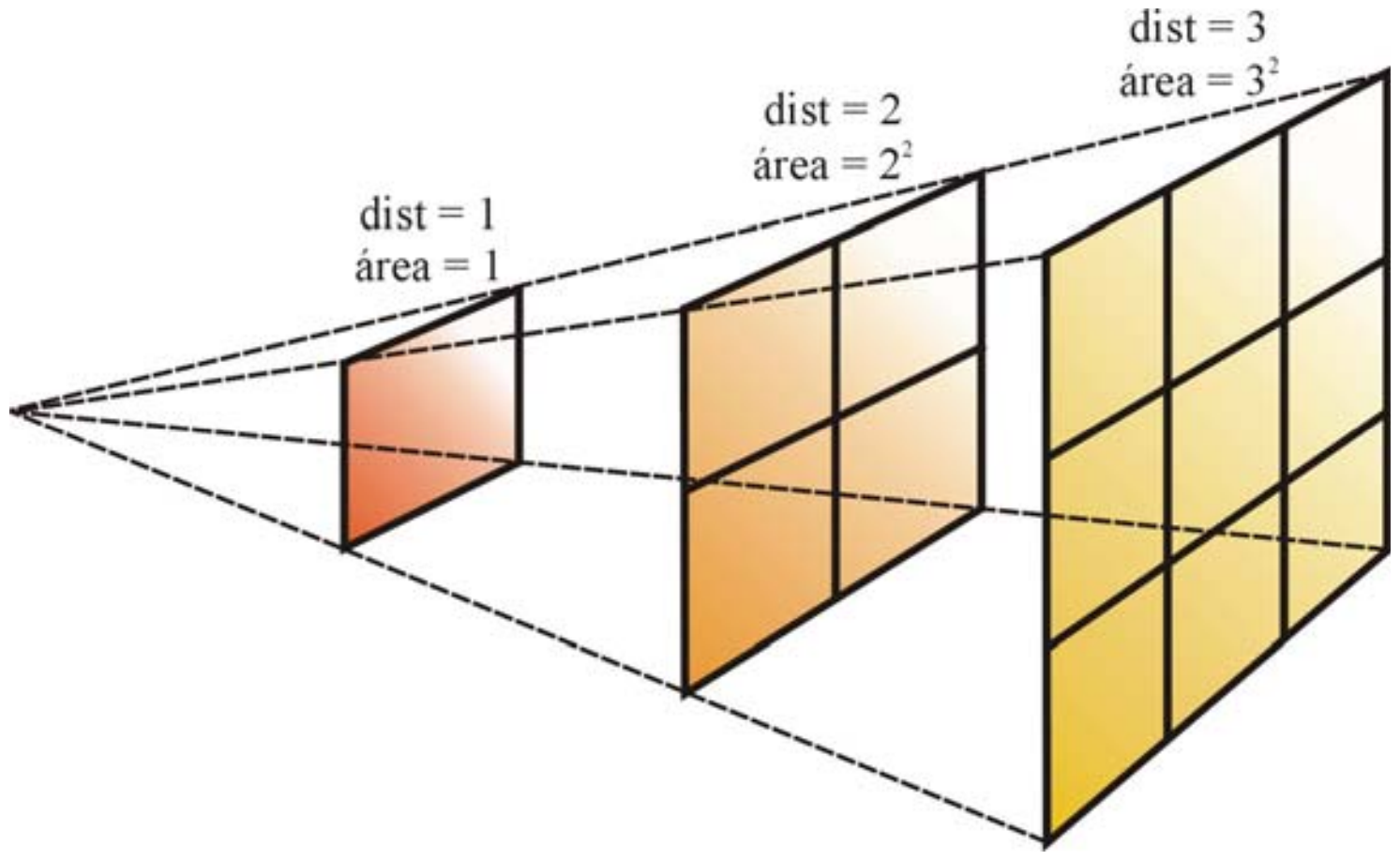

***Figura 1.6. Ley del cuadrado de la distancia.***

En el espacio todo está siempre moviéndose con respecto a la Tierra. Los planetas se mueven alrededor del Sol y generalmente las naves se moverán alrededor de los planetas siguiendo órbitas. Estos dos movimientos relativos introducen unos cambios en la **velocidad relativa** entre receptor y transmisor en rangos muy diferentes que habrá que tener bajo control. Para recuperar la frecuencia de la portadora, que debido al ***efecto Doppler*** varía continuamente, se emplean *PLL* (*Phase-Lock Loops* o bucles enganchados en fase), que son dispositivos que combinan un oscilador controlado por voltaje y un comparador de fase, diseñados para que el oscilador genere exactamente la misma frecuencia de la señal recibida y que es distinta a la señal emitida por la variación Doppler inducida. Estos *PLL* se usan tanto en los receptores terrestres como en las naves espaciales.

Para enviar una señal con información se necesita una señal electromagnética, llamada ***tono***, que oscila a una única frecuencia. Esta señal es modulada por la señal de información para enviarla al receptor, pero también se usa con otros fines tales como experimentos científicos o para los propios datos de telemetría que asisten la navegación de la nave. Por ello se necesita una frecuencia lo más estable posible. Sin embargo, sería imposible que las naves transportaran los equipos necesarios para obtener la precisión requerida. La solución está en que la señal se genere en la Tierra utilizando un máser de hidrógeno en ambientes de temperatura controlada. Esta señal **extremadamente estable** (equivalente a ganar o perder un segundo en 30 millones de años) es enviada a la nave en el enlace de subida. Una vez que se recibe esta señal en el espacio, para evitar problemas de interferencia entre el enlace de bajada y el de subida, se multiplica por una *constante predeterminada* (por ejemplo, la constante que usa la nave *Cassini* en su transpondedor de banda X es 1,1748999). De esta forma se consigue en la nave un enlace de bajada excepcionalmente estable.

La nave también incorpora un oscilador propio de menor precisión para generar el enlace de bajada en los momentos en los que no está disponible la señal del máser. Sin embargo, debido a los grandes cambios de temperatura que soportan las naves, la frecuencia generada nunca es perfectamente estable.

Las modulaciones que se usan en comunicaciones en espacio profundo son del tipo de las que se usan en enlaces no guiados de radiofrecuencia en la Tierra, es decir, modulaciones de fase. La mejora importante que añade este tipo de modulación respecto a las de amplitud es que presentan **mayor inmunidad al ruido** que se añade a la señal en el canal. Este ruido, al ser aditivo, tiene mayor efecto en las señales moduladas en amplitud, ya que las señales moduladas en fase no llevan información en su amplitud. Por otra parte, la codificación que se usa para garantizar transmisiones libres de errores suelen ser **códigos convolucionales**. Recientemente se han empezado a emplear los ***turbo códigos*** [7], un tipo de codificación convolucional adecuada para situaciones de baja relación señal-ruido, como lo es el espacio libre, que reduce la complejidad y se aproxima al límite teórico de Shannon.

## 1.3. COMUNICACIONES ÓPTICAS NO GUIADAS

Los sistemas actuales de comunicación utilizan una onda electromagnética sobre la que se superpone la información que se desea transmitir. A esta señal que no contiene información útil se le conoce como ***portadora*** y la técnica habitual consiste en la **modulación de la portadora** mediante una señal de información. Una vez que la señal se ha propagado hasta su destino, se extrae la información de ésta a través de un proceso de **demodulación**. La elección de la frecuencia de la señal portadora original de entre todo el espectro electromagnético determina las técnicas y las tecnologías que se pueden usar en el sistema de comunicación.

Cuando la frecuencia de la señal portadora cae dentro del rango de frecuencias visibles por el ojo humano o cercano a ellas, al tipo de comunicaciones que resulta del empleo de estas frecuencias se le conoce como ***comunicaciones ópticas***. La banda propia de las comunicaciones ópticas (incluyendo sistemas guiados y no guiados) se extiende desde el infrarrojo próximo ($\nu = 3 \cdot 10^{13}$ Hz) hasta el ultravioleta próximo ($\nu = 1{,}5 \cdot 10^{15}$ Hz).

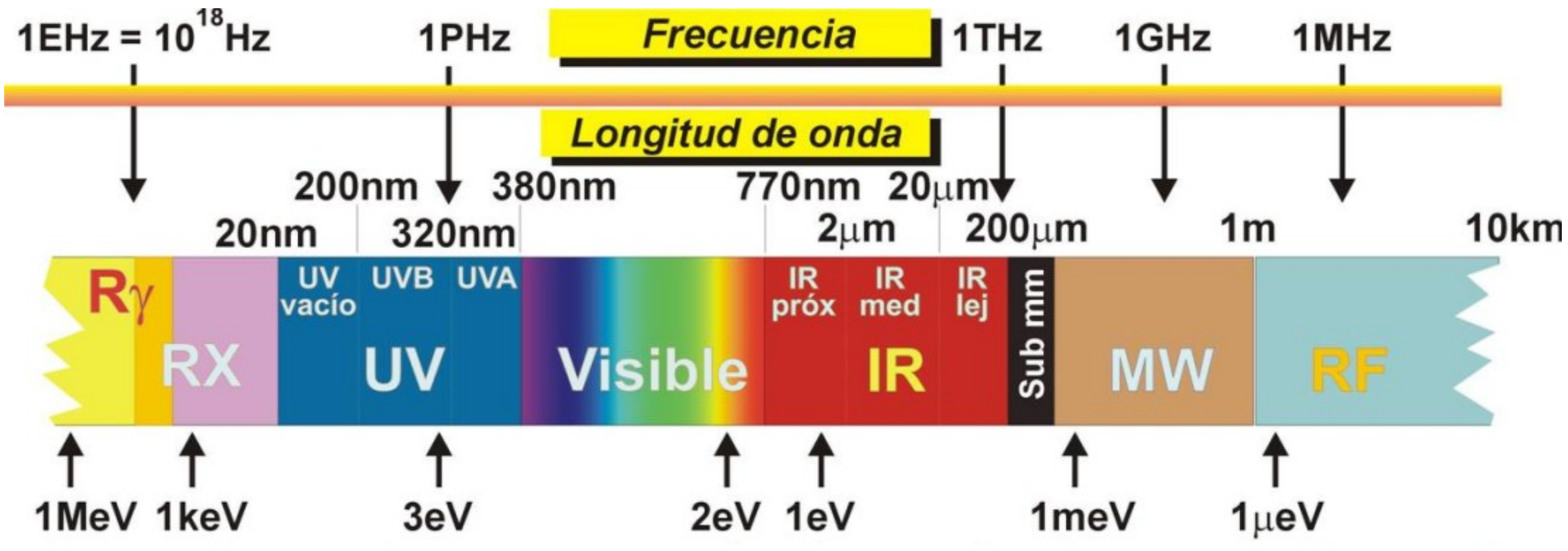

*Figura 1.7. Espectro electromagnético (no a escala).*

A diferencia de las comunicaciones ópticas ***guiadas*** en las que las señales ópticas son conducidas a través de un camino establecido en el interior de **fibras ópticas**, las comunicaciones ópticas no guiadas constituyen el tipo de comunicación en el que se usan portadoras a frecuencias ópticas desplazándose en un **medio libre**, como puede ser la atmósfera, el mar o el vacío interplanetario.

### 1.3.1. Reseña histórica

Los orígenes de las comunicaciones ópticas no guiadas coinciden con el origen de las comunicaciones ópticas mismas. Desde épocas prehistóricas el hombre ha usado señales ópticas para comunicarse, si bien esta comunicación consistía en muy poca información y cubría muy cortos alcances. Las comunicaciones ópticas no guiadas al menos cuentan con cerca de tres milenios de historia ya que ha sido verificado [9] el empleo sobre el 800 A.C. de señales ópticas creadas con fuego para transmitir información formada por un limitado número de mensajes conocidos previamente por ambas partes.

A caballo entre la realidad histórica y la mitología, varios siglos antes pudo llevarse a cabo el enlace de comunicaciones que el poeta griego Esquilo describe en su obra *Agamenón* (primera de la trilogía conocida como la *Orestiada*). En esta obra el autor, por boca de la reina Clitemnestra, cuenta la forma por la que ésta ha tenido noticia en Atenas de la caída de Troya. A lo largo de más de sesenta versos va describiendo minuciosamente el paso de una señal luminosa que se había ido propagando por medio de ocho hogueras sucesivas. Todos los nombres que cita se corresponden con la cima de montes consecutivos o valles desde donde existe línea de visión directa con los montes adyacentes, situados a distancias de entre veinte y cien kilómetros unos de otros (Figura 1.8). Lo destacable de este enlace es la gran distancia cubierta (más de 500 kilómetros) y el primitivo empleo de algunas técnicas modernas tales como la *regeneración de señales*, el *uso de repetidores* y la *transmisión digital* (un bit de información).

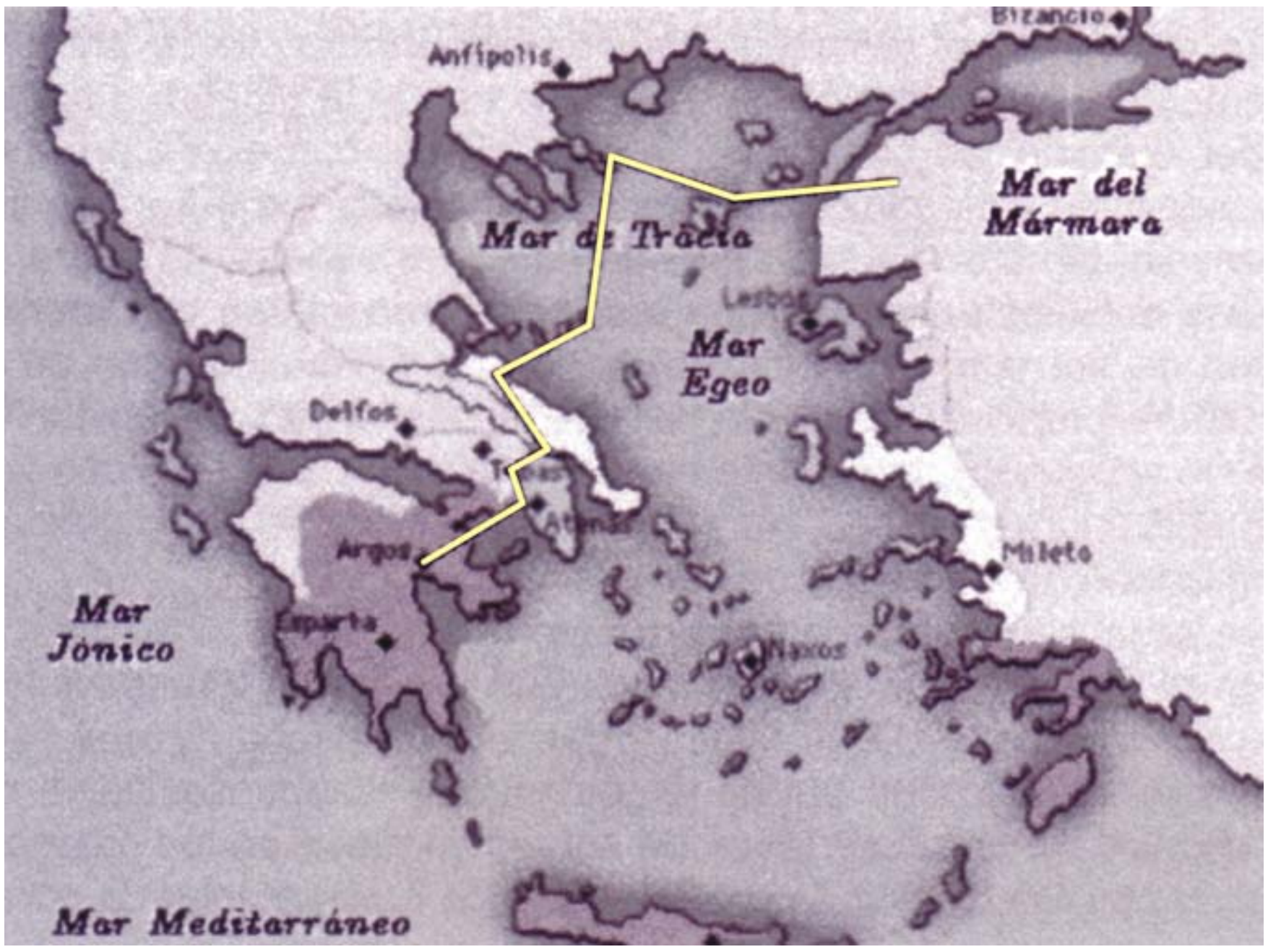

***Figura 1.8. Trazado de la transmisión de señales luminosas entre Troya y Atenas, de acuerdo con lo escrito por Esquilo [8].***

Seguramente muchos enlaces ópticos de este tipo fueron llevados a cabo en épocas antiguas. Sin embargo, todos mostraban el mismo inconveniente. Sólo se podía comunicar cada vez uno de entre el limitado número de mensajes que estaban disponibles por previo acuerdo.

Sobre el 200 A.C. el historiador griego Polibio desarrolló un sistema que era capaz de transmitir letras en lugar de mensajes fijos. El funcionamiento se basaba en una tabla de códigos como la de la Figura 1.9. En función del número de antorchas que hubiera encendidas a la derecha y a la izquierda, se seleccionaba una letra determinada. Con operadores entrenados se lograba una comunicación de unas ocho letras por minuto.

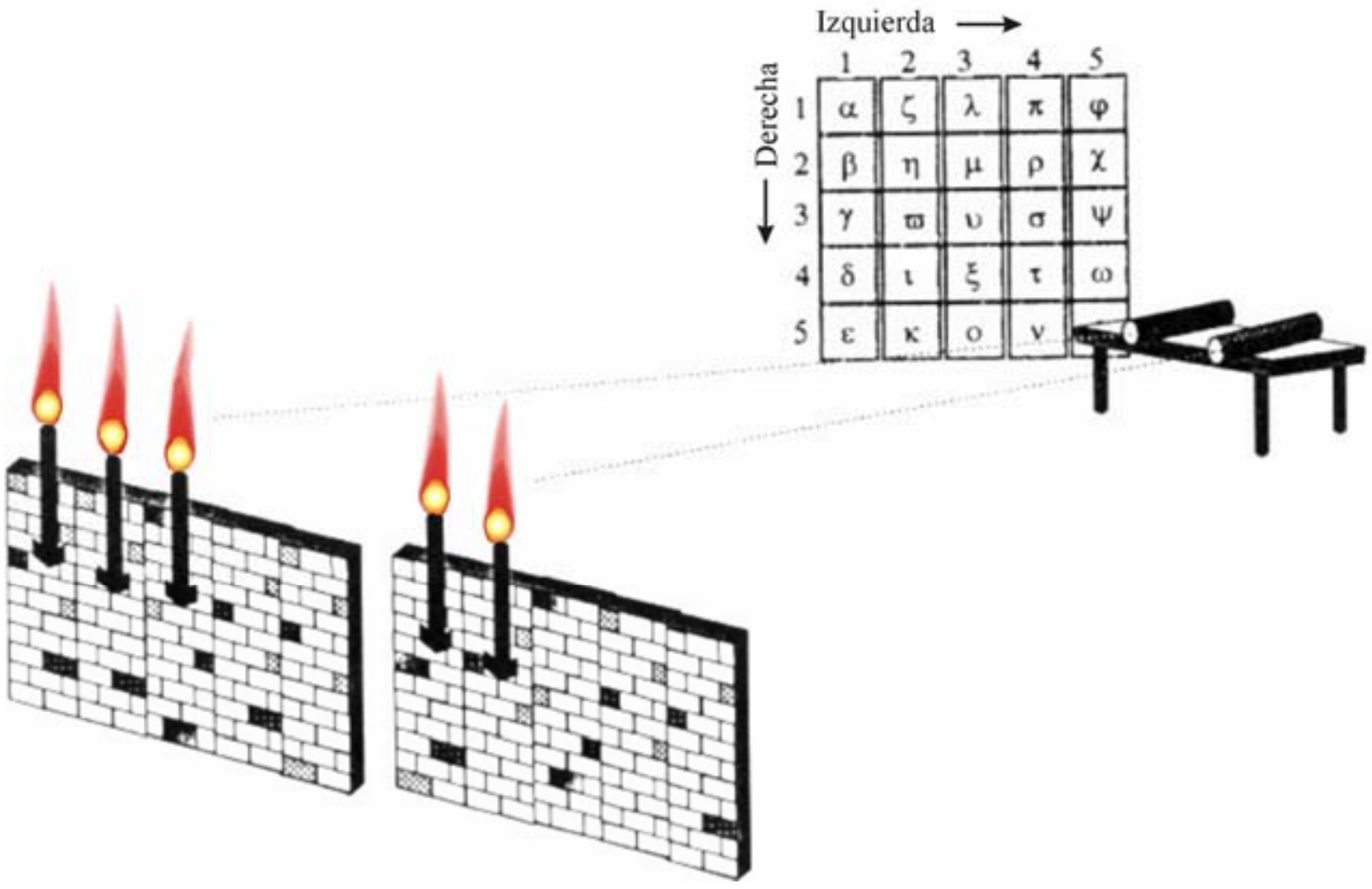

*Figura 1.9. Sistema de comunicaciones ópticas de Polibio [9].*

El avance de las comunicaciones ópticas prácticamente se detuvo hasta el siglo XVII con el desarrollo del telescopio. En 1.791 se dio un gran paso con la invención del primer telégrafo (que fue óptico, no eléctrico) por parte de Claude Chappe.

Consistía en un mástil fijo que soportaba dos piezas móviles con las que se podían adaptar una gran cantidad de configuraciones diferentes. La disposición era observada a través de un telescopio por un operador situado en otra estación, actuando de repetidora a la siguiente estación. Este sistema propició la creación de una red de comunicaciones ópticas en Francia ( Figura 1.10) que en 1.844 cubría unos 5.000 kilómetros con más de 500 estaciones [10]. Aunque este tipo de comunicación tuvo gran éxito en su época, el descubrimiento de la electricidad y el telégrafo eléctrico propició su final.

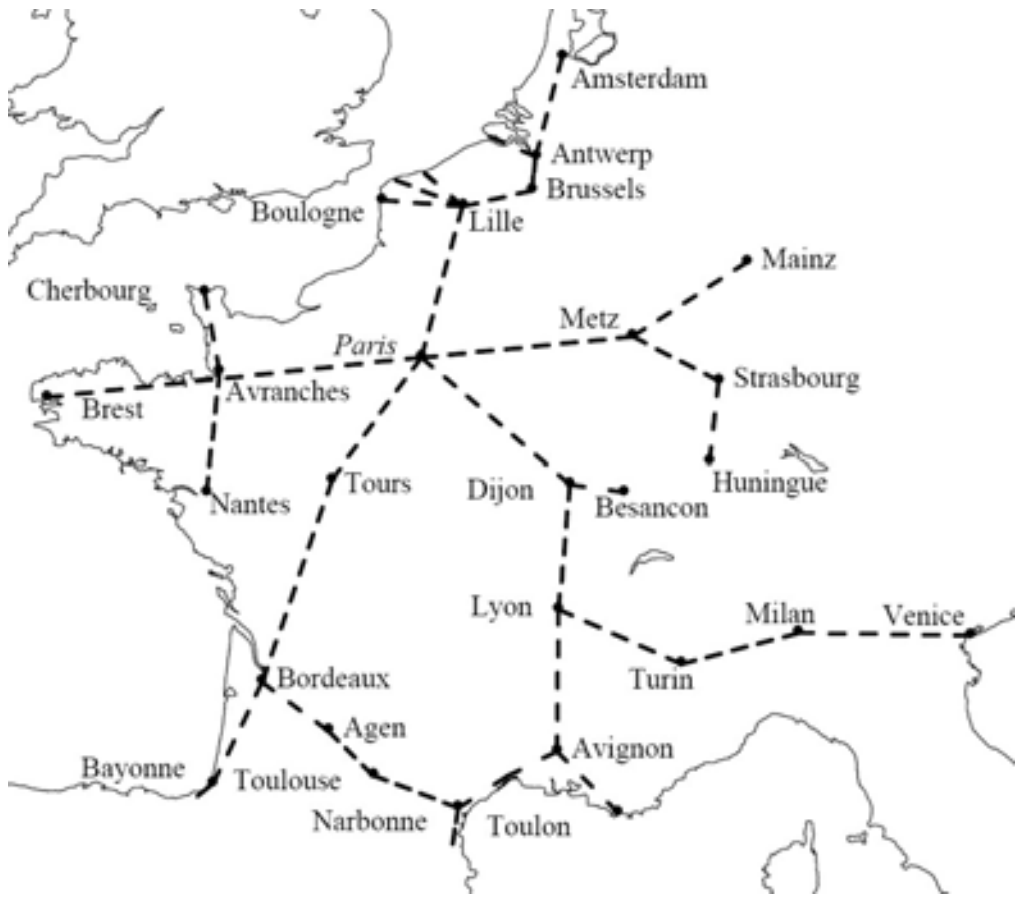

*Figura 1.10. Red francesa de telégrafos ópticos en 1846.*

El telégrafo óptico de Claude Chappe se extendió más adelante a otros países (Suecia, Rusia). En España se inauguró en octubre de 1846 la primera línea de telégrafo óptico Madrid-Irún [11]. Diez años después la red se extendía hasta Valencia, Barcelona, Cádiz y Badajoz (Figura 1.11). El director del proyecto fue el entonces coronel José María Mathé y Aragua, el mismo que se encargó posteriormente de desplegar el telégrafo eléctrico, y que fue el primer director de la Escuela de Telégrafos, embrión de las actuales Escuelas de Telecomunicación. El laboratorio de Comunicaciones Ópticas de la ETSI de Telecomunicación de la U. Politécnica de Madrid se denomina *Laboratorio Brigadier Mathé* en su honor.

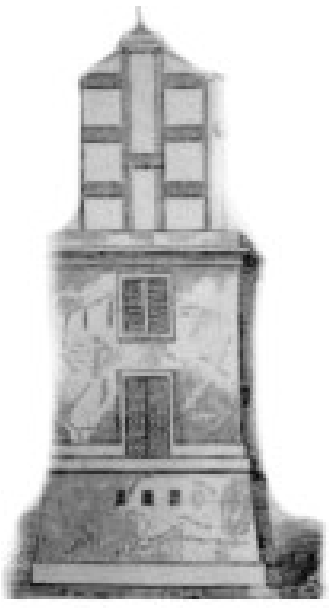

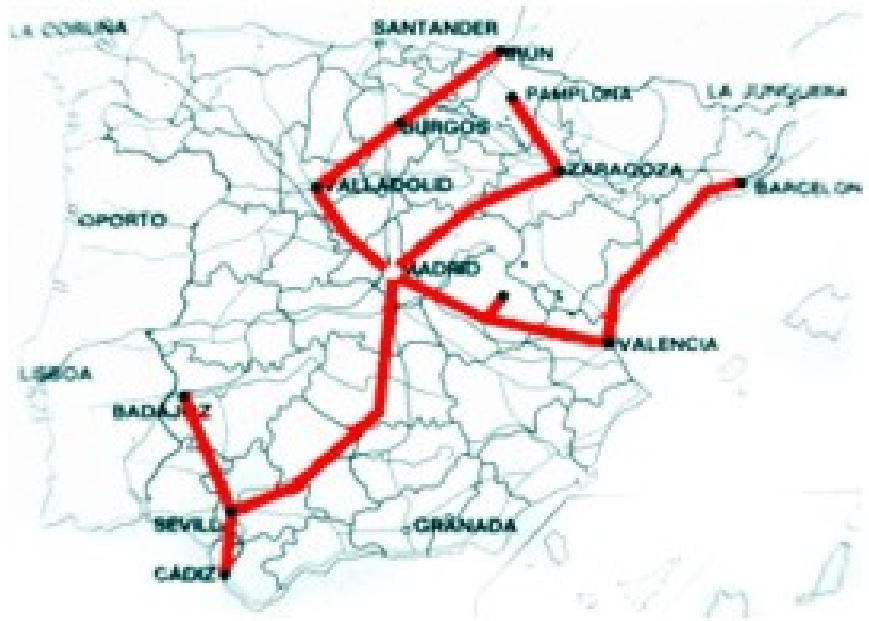

***Figura 1.11. Torre de telégrafo óptico reconstruida en la línea Madrid-Irún y máximo despliegue del telégrafo óptico en España.***

El primer sistema de comunicaciones ópticas no guiadas en el sentido actual fue el ***fotófono***, construido por Alexander Graham Bell en 1880 [12], cuatro años después del teléfono. Consistía en un espejo reflector de la luz del Sol que podía ser modulado acústicamente con una membrana. La señal modulada, después de avanzar unos 200 m como máximo, incidía sobre una placa cuya resistencia eléctrica (de selenio) dependía de la intensidad luminosa que incidiera sobre ella. De esta forma se obtenía una corriente eléctrica que podía ser convertida, mediante un altavoz, en una señal acústica.

En 1916 Albert Einstein presentó su trabajo teórico sobre emisión estimulada de la radiación electromagnética. Hasta entonces se creía que un fotón sólo podía interaccionar con un átomo de dos formas: podía ser **absorbido**, elevando el átomo a un nivel de energía superior, o podía ser **emitido**, de forma que el átomo pasase a un nivel inferior. Einstein propuso una tercera posibilidad: que un fotón con energía correspondiente al valor de una transición entre niveles podía ***estimular*** que un átomo pasase a un nivel energético inferior, emitiendo de esta manera otro fotón con idéntica energía que el primero. Este trabajo teórico supuso la base para el posterior desarrollo del ***láser*** (del inglés *Light Amplification by Stimulated Emission of Radiation*). Sin embargo, tuvieron que pasar más de 40 años hasta que el primer láser de gas fuese construido. Sólo dos años más tarde, se construyó el primer láser de semiconductor.

Un primer experimento (llevado a cabo por la NASA) de comunicaciones ópticas no guiadas se realizó en 1967 [9] usando un **láser de gas $CO_2$** y técnicas coherentes, y un segundo (por parte de la Fuerza Aérea de los EEUU) empleando un **láser Nd:YAG** y técnicas de detección directa. Sin embargo los efectos de la atmósfera sobre las señales hicieron decaer el interés en esa época ya que los enlaces quedaban seriamente perjudicados en presencia de lluvia o niebla. En su lugar, en los años posteriores a 1970 se produjo un enorme desarrollo en las fibras ópticas, por lo que las comunicaciones ópticas no guiadas quedaron relegadas a unas pocas aplicaciones muy específicas. De esta manera quedó casi totalmente detenido su desarrollo hasta el resurgimiento en la actualidad de una serie de proyectos que sitúan a las comunicaciones ópticas como una ventajosa alternativa en determinadas aplicaciones a las radiocomunicaciones no guiadas.

## 1.3.2. Ventajas e inconvenientes de las CC.OO. no guiadas

Las comunicaciones ópticas no guiadas presentan ventajas significativas en relación a los sistemas de microondas en aplicaciones tales como enlaces entre satélites o en espacio profundo. La mejora conseguida con este tipo de enlaces es tanto mayor cuanto mayores sean la capacidad requerida del enlace y la distancia a superar.

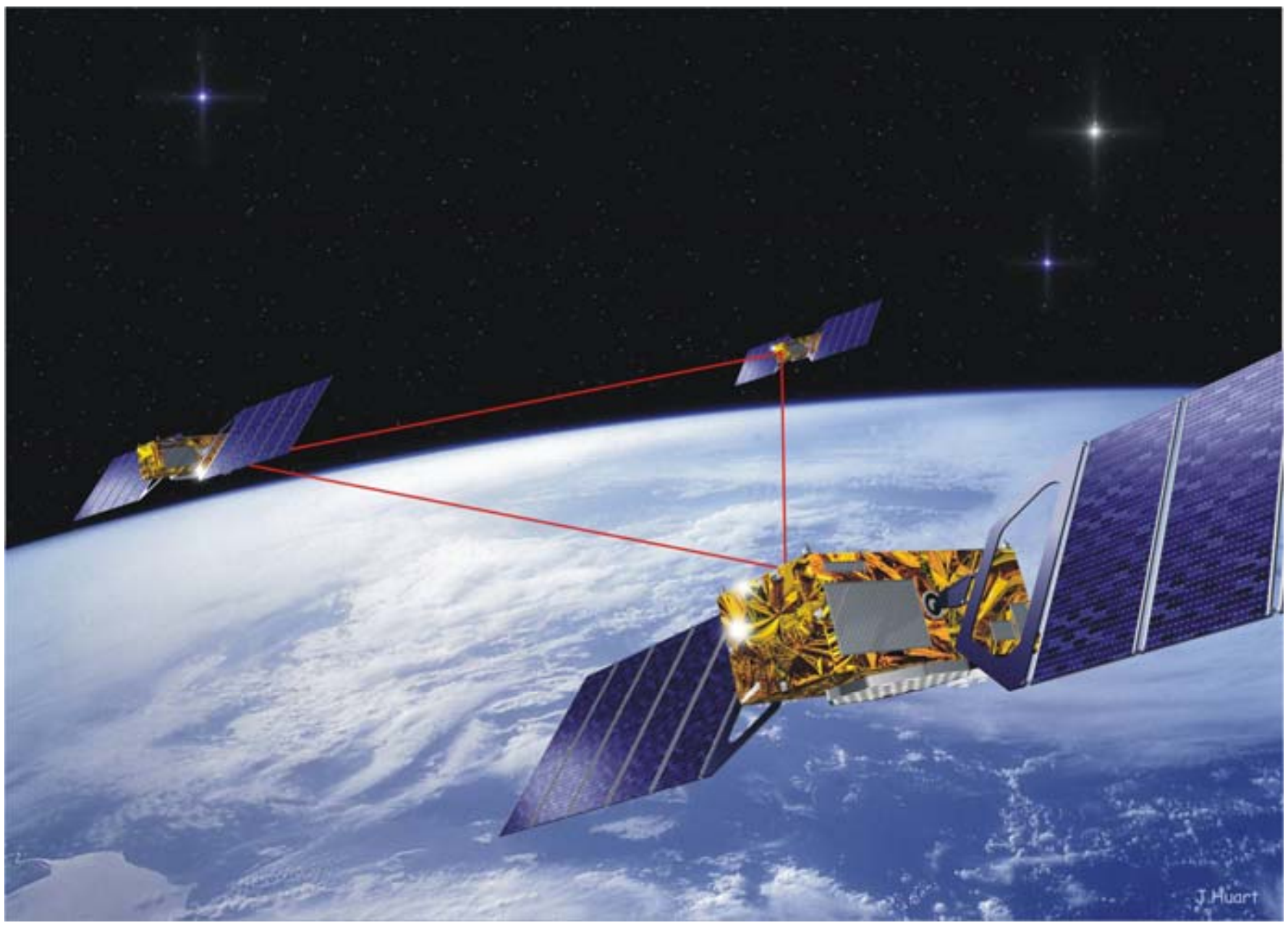

***Figura 1.12. Ejemplo de comunicaciones ópticas entre satélites [17].***

La principal ventaja proviene del hecho de que a frecuencias ópticas, como se analizará en adelante, la divergencia del haz es *proporcionalmente menor* que a frecuencias menores, y por lo tanto también es menor el tamaño del *spot* detectado en el receptor. El pequeño tamaño de *spot* se traduce en un gran aumento de la densidad de potencia recibida, consiguiéndose así una importante mejora en las prestaciones del sistema. Esta gran ventaja también se traduce en la necesidad de disponer de un sistema de apuntamiento extremadamente preciso.

La cantidad de información que se puede transmitir en un sistema de comunicación está directamente relacionada con el ancho de banda de la señal modulada y este ancho de banda viene limitado por la frecuencia de la señal portadora. Al incrementarse ésta, se puede **incrementar el ancho de banda** transmitido. Las frecuencias ópticas están varios órdenes de magnitud por encima de las frecuencias de microondas, por lo que la cantidad de información potencial que pueden transportar es considerablemente mayor. En cualquier caso, las tasas binarias empleadas en espacio profundo están muy por debajo de las frecuencias de microondas, por lo que esta ventaja potencial no es relevante actualmente.

Otra consecuencia de la menor longitud de onda de la señal portadora es que son necesarias antenas (telescopios) de menor tamaño que a frecuencias de microondas para obtener las mismas ganancias. A esto se le suma el hecho de que los equipos son más ligeros y pequeños que los empleados en microondas (alrededor de un 60% más pequeños [13]). Esta reducción de carga supone una gran ventaja económica en el caso de situarlos a bordo de satélites o naves espaciales donde cada gramo de más eleva considerablemente el presupuesto necesario.

En comunicaciones por fibra óptica, el empleo de las fuentes y detectores ópticos viene impuesto por las limitaciones de la propia fibra. En comunicaciones ópticas no guiadas se dispone de un rango más amplio de elección, aunque hay que señalar que si el medio de propagación incluye la atmósfera, ésta también impondrá limitaciones en la longitud de onda, como se verá en adelante.

Frente a las virtudes de los sistemas ópticos no guiados comentadas, hay que mencionar también algún inconveniente. El principal lo constituye la atmósfera terrestre, que provoca determinados efectos indeseados en las señales ópticas, deteriorando las prestaciones del enlace, y en ciertas condiciones bloqueándolo por completo. Dejando aparte el efecto de la absorción (que, como se verá, se puede minimizar haciéndose despreciable) las **turbulencias** atmosféricas consiguen que, al distorsionar el frente de ondas por variaciones aleatorias en el índice de refracción, *la intensidad y fase de la señal recibida en el telescopio fluctúen*. Por supuesto todos estos efectos no existen si el medio es exterior a la atmósfera.

En la Tabla 1.1 se presenta un resumen de las ventajas e inconvenientes de las comunicaciones ópticas no guiadas frente a los clásicos sistemas no guiados de microondas.

| Ventajas | Inconvenientes |
|---|---|
| Ancho de banda mucho mayor. | Los efectos de la atmósfera degradan y hasta pueden llegar a bloquear el enlace óptico. |
| Elevada potencia recibida respecto a la transmitida debido a la pequeña divergencia del haz óptico. | Necesidad de un sistema de apuntamiento muy preciso debido al estrecho haz óptico. |
| Elevadas ganancias de transmisión y recepción con tamaños de antenas (telescopios) muy pequeños. | Elevado ruido recibido junto a la señal óptica. |
| Terminales transmisores y receptores más ligeros y pequeños. | Poco apropiadas para difusión debido al estrecho haz óptico. |
| Explota una parte del espectro electromagnético no usada. | Posible riesgo de radiación. |

***Tabla 1.1. Ventajas e inconvenientes de las comunicaciones ópticas no guiadas frente a los sistemas no guiados de microondas.***

## 1.4. COMUNICACIONES ÓPTICAS EN EL ESPACIO

El gran auge que han tenido las comunicaciones ópticas guiadas (por medio de fibra óptica) ha provocado un enorme avance en la tecnología de las fuentes láser y los detectores ópticos, lo que ha marcado el resurgimiento de numerosos proyectos de investigación en el campo de las comunicaciones ópticas no guiadas. Este renovado interés se hace patente en los actuales proyectos llevados a cabo por las principales agencias espaciales, tales como la americana *NASA*, la europea *ESA* o la japonesa *NASDA*.

La primera demostración de comunicaciones ópticas no guiadas en el espacio profundo tuvo lugar en diciembre de 1992 dentro del proyecto ***GOPEX*** (del inglés *Galileo Optical Experiment*). Cuando la nave *Galileo* iniciaba su viaje a Júpiter, durante un periodo de ocho días, en los que se la distancia varió desde los 600.000 kilómetros hasta los seis millones de kilómetros, dos estaciones terrestres transmitieron pulsos ópticos en el rango de los MW que fueron detectados satisfactoriamente por una cámara de imágenes instalada como receptor en la nave. Este experimento fue realizado siempre de noche con longitudes de onda de 532 nm (un láser Nd:YAG de frecuencia doblada) emitiendo con una frecuencia de repetición de 15 a 30 Hz [14].

Entre noviembre de 1995 y enero de 1996 se llevó a cabo una segunda demostración con el proyecto ***GOLD*** (*Ground-Orbiter Lasercomm Demonstration*) que consistió en un enlace de un megabit por segundo entre una estación terrestre y el satélite japonés *ETS 6*. Problemas en la puesta en órbita geoestacionaria, dejaron al satélite en una órbita elíptica de transferencia, por lo que las pruebas eran difíciles de realizar y mantener, y el proyecto acabó antes de lo esperado [15].

Los primeras concepciones de proyectos de comunicaciones ópticas entre satélites datan de los años sesenta [16]. Sin embargo no ha sido hasta noviembre de 2001 (aunque el desarrollo del proyecto comenzó a mediados de los ochenta) en que la ESA ha demostrado el **primer sistema civil de comunicaciones por láser** entre satélites: el proyecto ***Silex*** (*Semiconductor Laser Intersatellite Link Experiment*). Consiste en dos terminales láser; uno a bordo del satélite cuasi geoestacionario (situado a unos 31.000 kilómetros) *Artemis* y otro a bordo del satélite francés de órbita baja *SPOT 4* (a unos 800 kilómetros). Ambos terminales llevan telescopios de 25 cm y emplean diodos láser de GaAs de 60 mW de potencia media que emiten luz en torno a 800 nm con modulación en intensidad y detección directa (IM/DD). El enlace logra superar distancias de hasta 45.000 kilómetros con regímenes binarios de hasta 50 megabits por segundo y una probabilidad de error de $10^{-9}$.

El proyecto *Silex* también ha demostrado la comunicación láser entre el satélite *Artemis* y un telescopio de 1 metro situado en la *OGS* (*Optical Ground Station*) de la ESA en el observatorio del Teide, Tenerife, situado a casi 2.400 metros y seleccionado por contar con unas condiciones meteorológicas excepcionales.

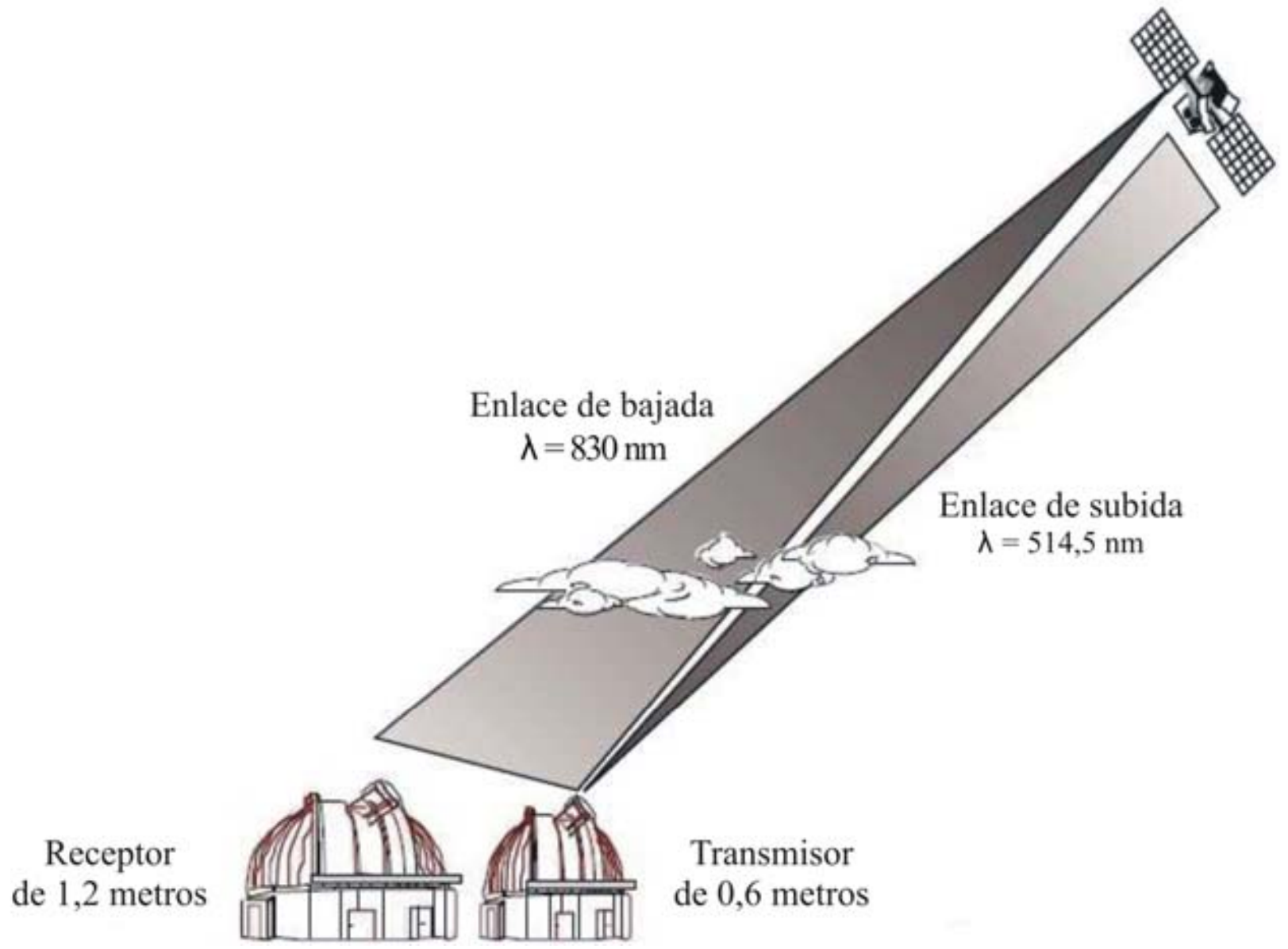

***Figura 1.13. Esquema básico del proyecto GOLD [15].***

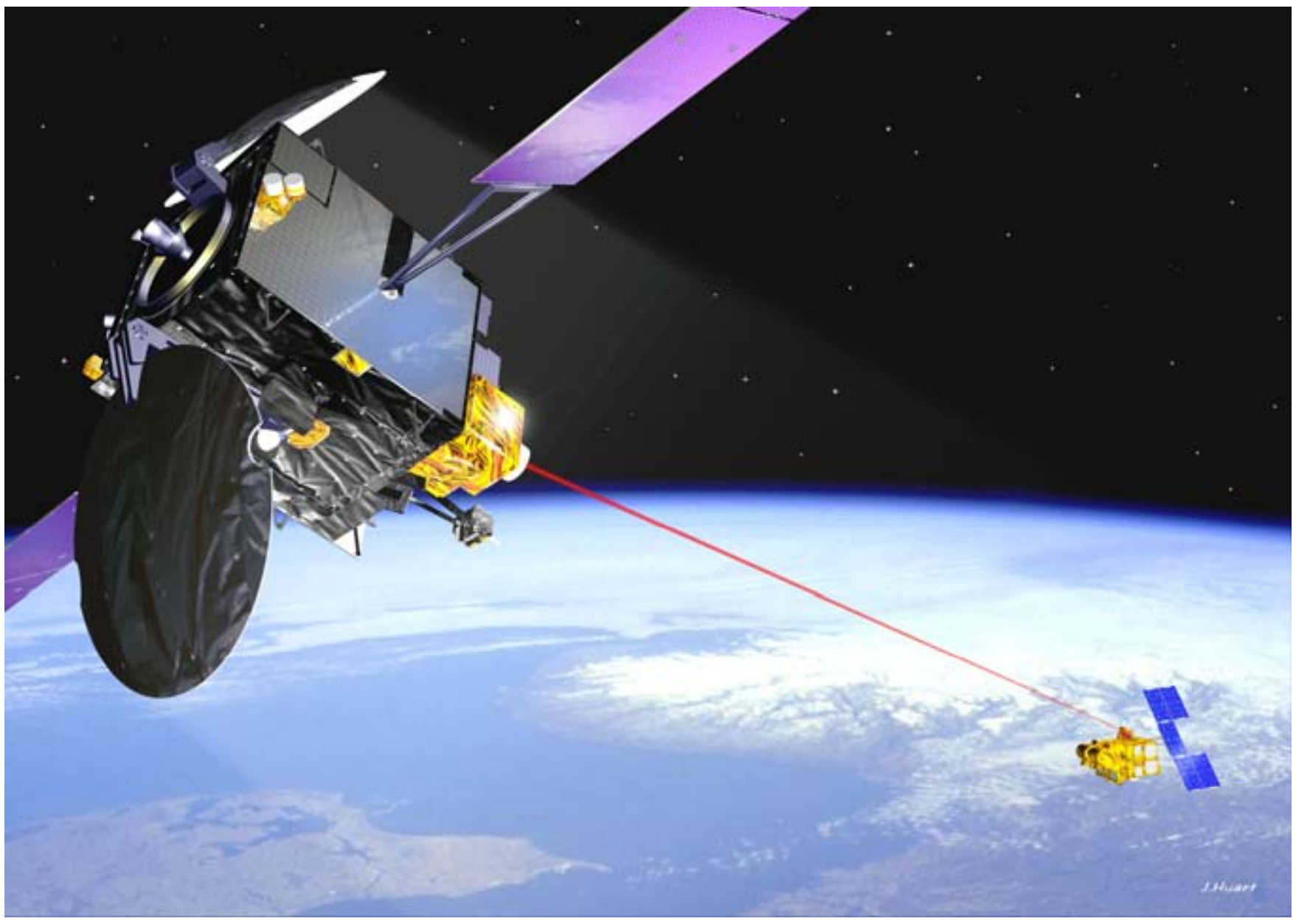

***Figura 1.14. Ilustración artística del satélite Artemis comunicándose mediante un láser con el satélite SPOT 4 (proyecto Silex de la ESA) [17].***

Japón también ha participado en el proyecto *Silex* con otro terminal láser a bordo del satélite de órbita baja (unos 600 kilómetros) *OICETS* (*Optical Inter-orbit Communications Engineering Test Satellite*), formado por un telescopio de 26 cm, un diodo láser de 200 mW y un sistema IM/DD compatible con los otros terminales del proyecto. Este terminal estableció comunicación con el satélite *Artemis* y con la estación terrena *OGS*.

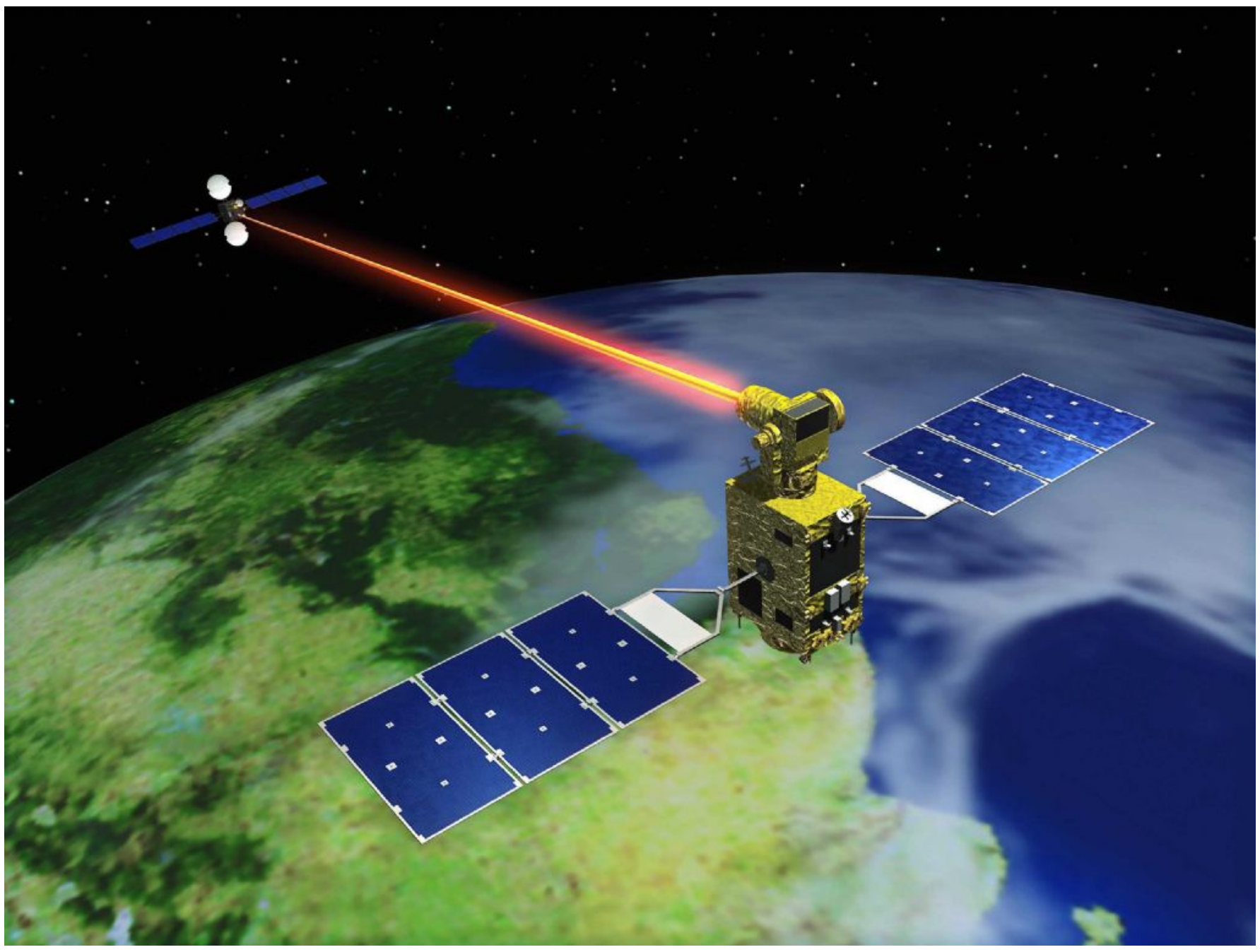

***Figura 1.15. Ilustración artística del satélite Oicets comunicándose mediante un láser con el satélite Artemis [18].***

Unos meses antes del éxito logrado por el experimento *Silex*, un satélite en órbita geosíncrona llamado *Geolite* (*Geosynchronous Lightweight Technology Experiment*) y patrocinado por la *NRO* (*National Reconnaissance Office*) supuestamente ha llevado a cabo una demostración de comunicaciones por láser entre el satélite y una estación receptora óptica en la superficie terrestre. Debido a la naturaleza militar de este proyecto, no se han hecho públicos detalles del mismo.

Actualmente la *NASA* tiene en fase de desarrollo un proyecto llamado *MLCD* (*Mars Laser Communications Demonstration*) para demostrar la viabilidad de un enlace de comunicaciones ópticas de gran capacidad en espacio profundo. Este proyecto se llevará a cabo mediante un terminal láser instalado en el satélite *MTO* (*Mars Telecommunications Orbiter*) que entrará en la órbita de Marte entre agosto y septiembre de 2010. Los detalles de este proyecto pueden consultarse en el capítulo 6, en el que se analiza la posibilidad de un enlace de comunicaciones ópticas Marte-Tierra basado en el proyecto *MLCD*.

## 1.5. RESUMEN DEL TRABAJO REALIZADO

El proyecto intenta demostrar la ***viabilidad de un enlace*** de CC.OO. Marte-Tierra operando 24 horas al día durante un periodo superior a un año, con un régimen binario superior a 1 Mbit/s en las peores condiciones, y especificando la duración de los periodos en que el enlace se verá cortado por la interposición del Sol entre ambos planetas. Se tomará como referencia el proyecto *MLCD*, actualmente en fase de diseño, que intentará demostrar precisamente este enlace a finales de la presente década. Dado que los mayores requerimientos (escasez de potencia, alto régimen binario) se producen en el canal de bajada Marte-Tierra, el proyecto se concentrará en este canal.

Para la realización del proyecto será necesaria la preparación de un ***simulador orbital***, capaz de situar tanto los planetas como la nave en el espacio tridimensional, en función del tiempo, con el fin de evaluar las distancias y velocidades relativas entre los mismos.

Seguidamente se realizará un estudio de los distintos elementos del enlace de comunicaciones, a saber, el canal, el equipo emisor y el equipo receptor. De este estudio se derivará la ***selección*** de los componentes más idóneos, teniendo en cuenta las limitaciones impuestas en la potencia del emisor y en la antena remota, lo que hará necesario aumentar las prestaciones de los equipos de tierra. En todo caso, se intentará trabajar en todo momento con componentes, equipos e instalaciones existentes en la actualidad, empleando datos reales y optando por las soluciones más idóneas para cada situación.

Tanto la selección de componentes como su posterior aplicación al enlace requerirán un notable ***acopio*** de material bibliográfico, que incluya datos de carácter general así como detalles técnicos de las especificaciones del proyecto *MLCD*.

Se procurará, a lo largo del proyecto, evolucionar desde la generalidad a casos concretos, con el fin de profundizar al máximo en los detalles técnicos, sin intentar abarcar el sinnúmero de diferentes posibilidades que podrían tomarse en consideración. Será necesario seleccionar una, o dos como máximo, ***longitudes de onda*** de trabajo, puesto que tal decisión condiciona en gran medida el resto del diseño. Por otra parte, deberán establecerse las ubicaciones de la o las ***estaciones terrestres***, ya sean satélites en órbita u observatorios astronómicos. Según sea la opción elegida, se deberá incluir o no un estudio de la influencia de la atmósfera en la transmisión.

Finalmente, se aplicarán los resultados obtenidos al ***enlace Tierra-Marte***, intentando alcanzar una solución compatible con las prestaciones exigidas a partir de componentes comerciales. Para ello se desarrollarán los ***modelos de cálculo*** necesarios para evaluar los distintos parámetros de diseño: efecto Doppler, filtros ópticos, diámetros de antena, condiciones meteorológicas, etc. Una vez conocidos estos datos, se realizará un ***balance de potencia*** proyectado temporalmente a los meses de duración del proyecto *MLCD*, y se calculará el régimen binario alcanzable en función del tiempo y de los factores ambientales que intervengan.

# 2. Simulación de dinámica orbital

Para la realización de los cálculos de un enlace de comunicaciones en el espacio profundo se necesita conocer con exactitud la dinámica planetaria, de modo que puedan situarse los diferentes cuerpos celestes en las posiciones que ocupan en cada momento durante la duración de la misión.

Se ha realizado un simulador de la dinámica orbital en el entorno *Visual Basic for Aplications-Microsoft Office Excel*. Su objetivo último ha sido la obtención de un conjunto de datos necesarios para llevar a cabo el cálculo del enlace de comunicación por láser del proyecto *MLCD* descrito en el capítulo 6. Aún así, se ha desarrollado el simulador con carácter general, puesto que el conjunto de ecuaciones de movimiento empleadas son válidas para cualquier cuerpo celeste situado en órbita alrededor del Sol. En este capítulo se describen las ecuaciones empleadas, se muestra el desarrollo del simulador, y se aplica a varios ejemplos comparando los datos obtenidos con otros extraídos de la bibliografía o de programas de cálculo orbital comerciales, con el fin de validar sus predicciones y evaluar los errores de cálculo cometidos.

# 2.1. NECESIDAD DE UN SIMULADOR

La implementación de un simulador de dinámica orbital ha venido determinada por la necesidad de conocer las posiciones **reales** e **instantáneas** de cada planeta en su órbita alrededor del Sol. Este dato es necesario para la realización de dos cálculos fundamentales en el enlace de comunicación

- La ***distancia relativa*** entre el planeta y la Tierra. Este dato determina la variación de las pérdidas por espacio libre en cada momento de la duración de la misión que se trate (el proyecto *MLCD* en nuestro caso) a partir del instante que el satélite entra en la órbitas del planeta. Se trata de un elemento muy variable y fundamental en el cálculo del balance de potencia del enlace.

- Las ***velocidades relativas*** entre el planeta, el satélite de comunicaciones y la Tierra. Son necesarias para el cálculo de la variación que producen las velocidades variables del terminal espacial y terrestre en la longitud de onda (por efecto **Doppler**) de las señales que se transmiten.

Dado que ambos datos son variables en el tiempo y se necesitan durante el periodo de tiempo específico ha sido necesaria la implementación de un simulador que generara estos datos de forma que se pudiera conocer la posición instantánea de los planetas y así el valor instantáneo de los valores anteriormente descritos. Los resultados que se presentan en el capítulo 6 aplicados al proyecto *MLCD* se apoyan de forma constante en los datos de este simulador.

## 2.1.1. Elección del entorno de programación

Tras un análisis preliminar del objetivo y futura implementación de la simulación y sus necesidades en cuanto a entorno de programación se concluyó que la mayor parte de la simulación precisaría cálculo aritmético simple además de la posibilidad de sistematizar procedimientos con algún lenguaje de programación. Además el objetivo de la simulación era la obtención de una gran cantidad de datos, de los que sería necesario, además de conocer su valor, generar representaciones gráficas.

Teniendo en cuenta estas razones se optó por el programa *Microsoft Office Excel 2003*, parte de *Microsoft Office Professional Edition 2003*. Esta potente aplicación proporciona a los objetivos del simulador las siguientes características

- Capacidad de trabajo con **hojas de cálculo**, cumpliendo así con los requisitos de cálculo necesarios para la implementación del simulador.

- Acompañamiento de un **entorno de programación** llamado *Visual Basic for Applications* que se integra con *Microsoft Office Excel 2003* permitiendo así la comunicación entre ambos programas.

- Posibilidad de crear **gráficos a partir de los datos** de la hoja de cálculo, permitiendo de esta manera generar las representaciones gráficas de los resultados de la simulación, que posteriormente se presentarán en el capítulo 6.

# 2.2. DINÁMICA ORBITAL

A continuación se presenta un repaso de los conceptos básicos de dinámica orbital necesarios para comprender los parámetros orbitales que se utilizan en la simulación.

## 2.2.1. Leyes de Kepler

Los fundamentos de la dinámica orbital fueron establecidos por Johannes Kepler en el siglo XVII. Kepler, mediante datos observacionales de los movimientos planetarios obtenidos por el astrónomo Tycho Brahe, comprendió que los planetas orbitaban alrededor del Sol siguiendo elipses y no círculos (como se creía desde que Copérnico descubrió que los planetas se mueven alrededor del Sol). Este movimiento elíptico lo explicó mediante sus tres leyes

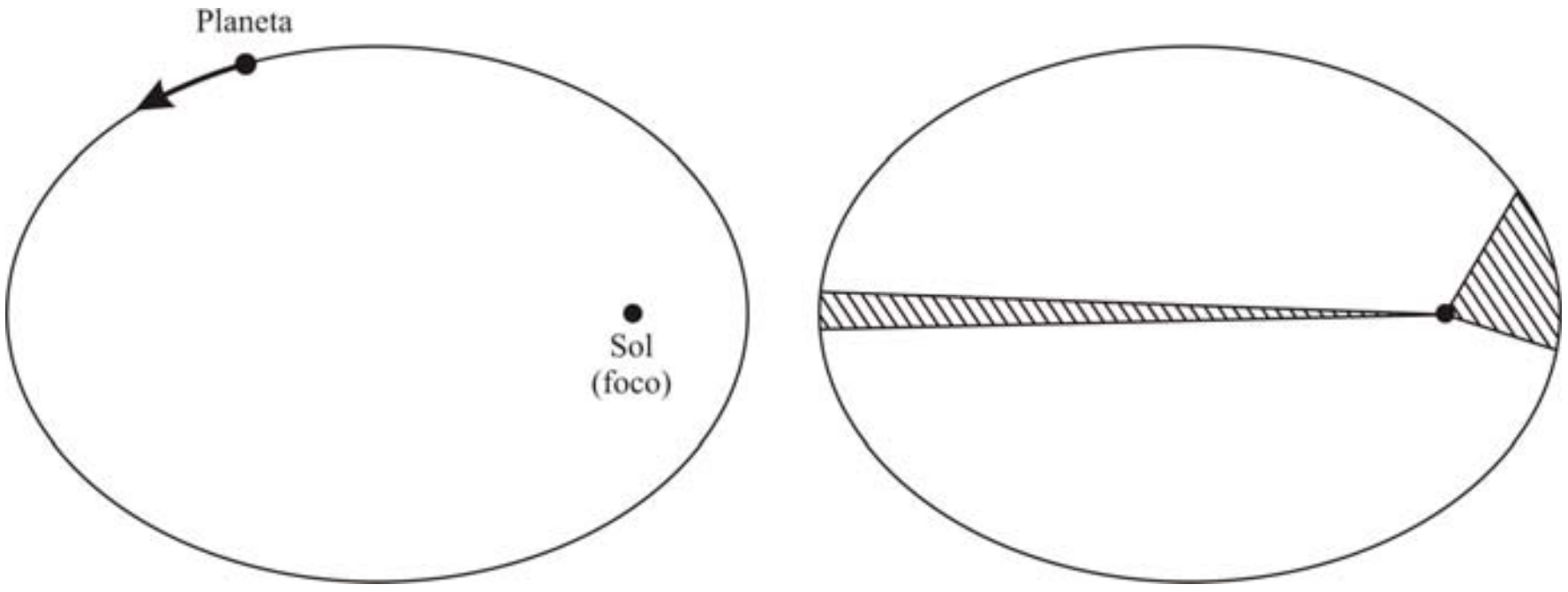

***Figura 2.1. Primera y segunda ley de Kepler.***

1. La órbita de un planeta sigue una **trayectoria elíptica** con el Sol en uno de sus focos (Figura 2.1).

2. La línea que une el planeta y el Sol barre **áreas iguales** en tiempos iguales (Figura 2.1).

3. El **cuadrado del periodo** de un planeta es proporcional al **cubo del semieje** mayor de la elipse. (Ecuación (1)).

$$T^2 \propto a^3 \qquad (1)$$

## 2.2.2. Ley de gravitación de Newton

Algunos años más tarde del descubrimiento de Kepler, Newton enunció la ley de gravitación universal que establece que dos cuerpos de masa m y M se atraen mutuamente con una fuerza proporcional a sus masas e inversamente proporcional al cuadrado de sus distancias. Esta ley viene dada por la siguiente expresión

$$F = -G\frac{Mm}{r^2} \quad (N) \qquad (2)$$

donde G es la ***constante de gravitación universal*** o de Cavendish, cuyo valor es $6{,}672 \cdot 10^{-11}$ $m^3/(kg \cdot s^2)$.

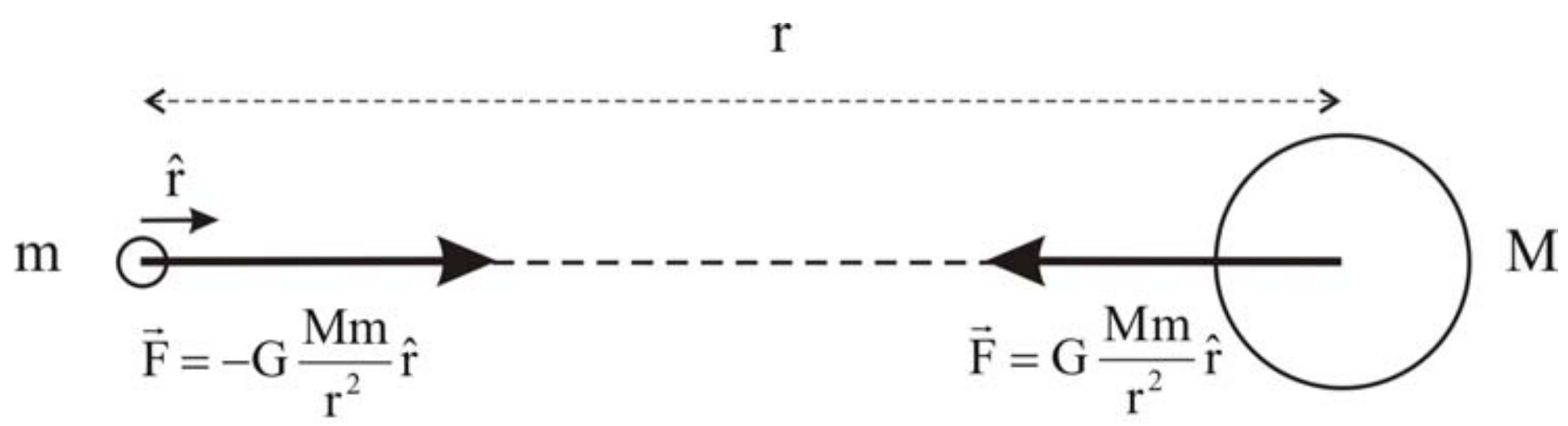

***Figura 2.2. Atracción gravitatoria entre dos cuerpos de masa m y M a una distancia r.***

## 2.2.3. Movimiento relativo entre dos cuerpos

Para abordar este estudio, conviene realizar una serie de suposiciones que simplifican el cálculo en cierta medida

- La ***única fuerza*** que actúa entre los dos cuerpos es la gravitatoria, es decir, se considera un movimiento en espacio libre con la única presencia de estos dos cuerpos.
- El cuerpo de mayor masa M de la ecuación (2) se supondrá ***fijo en el espacio*** y el único movimiento lo llevará a cabo el satélite (o planeta) que orbita al cuerpo de masa mayor. En realidad ambos orbitan alrededor del centro de gravedad común, pero puede suponerse que dicho centro coincide con el centro de gravedad del cuerpo mayor cuando M>>m.
- En la ecuación (2) el cuerpo de masa M se considera ***perfectamente esférico*** y homogéneo, de forma que la fuerza gravitatoria en cualquier punto únicamente varía con la distancia entre el punto y el centro de masas del dicho cuerpo.

Se considera un sistema de coordenadas ortogonal como el de la Figura 2.3 cuyo origen es el centro de gravedad del cuerpo de masa M. La fuerza gravitatoria que actúa sobre el satélite será

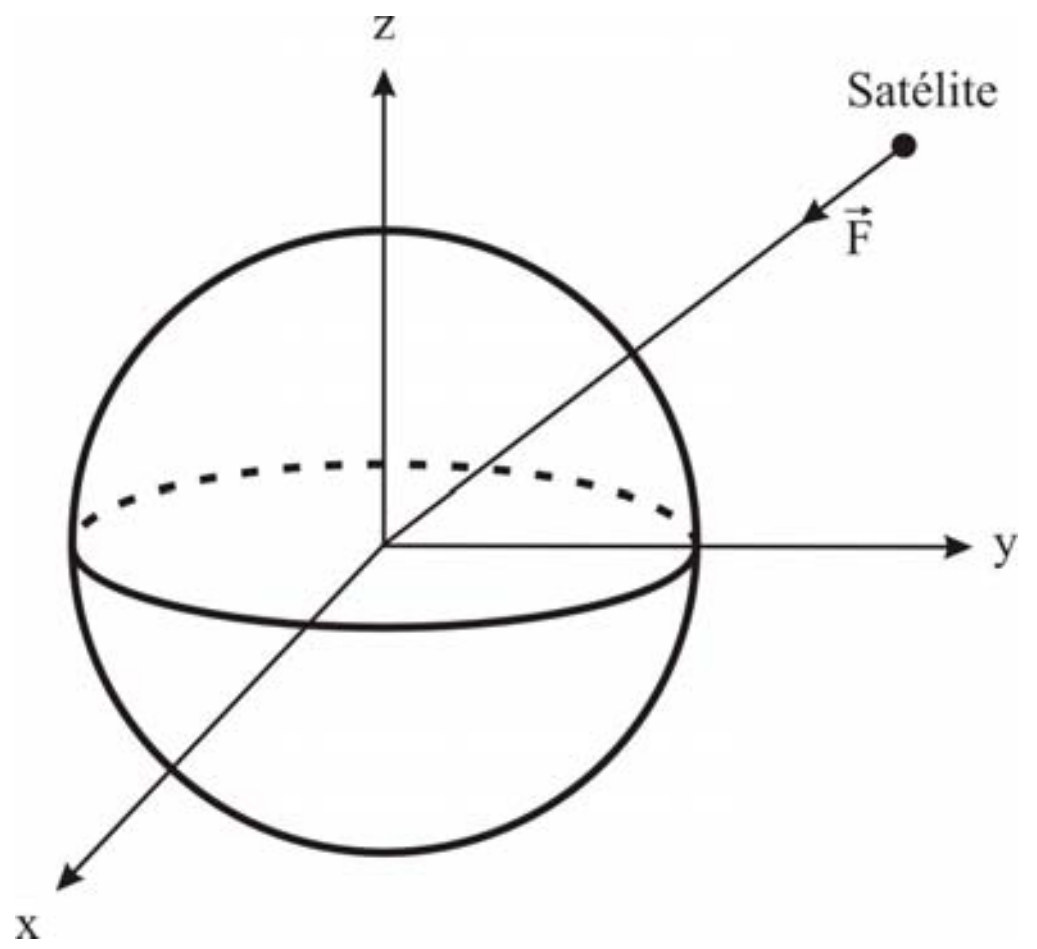

***Figura 2.3. Sistema de coordenadas ortogonales para la modelización del movimiento relativo entre dos cuerpos.***

$$\vec{F} = -G\frac{Mm}{r^2}\hat{r} \qquad \text{(N)} \qquad (3)$$

siendo $\hat{r}$ el vector unitario en la dirección de la aplicación de la fuerza.

Para que el satélite[1] no se precipite hacia el cuerpo de mayor masa y se mantenga en una órbita estable debe anular la fuerza gravitatoria que lo atrae. La fuerza que la anula deriva de la velocidad que lleva el satélite en su órbita y viene dada por [19]

$$\vec{F} = m\frac{d^2r}{dt^2}\hat{r} - mr\left(\frac{d\theta}{dt}\right)^2\hat{r} \qquad \text{(N)} \qquad (4)$$

donde el primer término representa la variación de la velocidad radial y el segundo representa la fuerza centrípeta a la que se ve sometido el satélite por su movimiento. Para que se anulen ambas fuerzas, se igualan las ecuaciones (3) y (4) quedando

$$-G\frac{M}{r^2} = \frac{d^2r}{dt^2} - r\left(\frac{d\theta}{dt}\right)^2 \qquad \text{(m/s}^2\text{)} \qquad (5)$$

Resolviendo esta ecuación diferencial se obtiene la ecuación buscada, para lo cual es necesario eliminar de la ecuación el tiempo con los siguientes cambios

$$\begin{cases} \dfrac{dr}{dt} = \dfrac{dr}{d\theta}\dfrac{d\theta}{dt} \\ \rho = \dfrac{1}{r} \Rightarrow \dfrac{dr}{d\theta} = -\dfrac{1}{\rho^2}\dfrac{d\rho}{d\theta} \end{cases}$$

Para sustituir el factor dθ/dt se puede usar la expresión del momento angular del sistema satélite-masa M que viene dado por [19]

$$H = \vec{r} \times m\vec{V} = \vec{r} \times m\left(\vec{V}_R + \vec{V}_T\right) = \vec{r} \times m\vec{V}_R + \vec{r} \times m\vec{V}_T \qquad \text{(N·m·s)} \qquad (6)$$

donde $\vec{V}$ es el vector velocidad del satélite (Figura 2.4). El producto vectorial $r \times m\vec{V}_R$ es cero porque $\vec{V}_R$ tiene la misma dirección que r y el producto vectorial $r \times m\vec{V}_T$ es igual al producto escalar $r \cdot m\vec{V}_T$ porque $\vec{V}_T$ es perpendicular a $\vec{r}$.

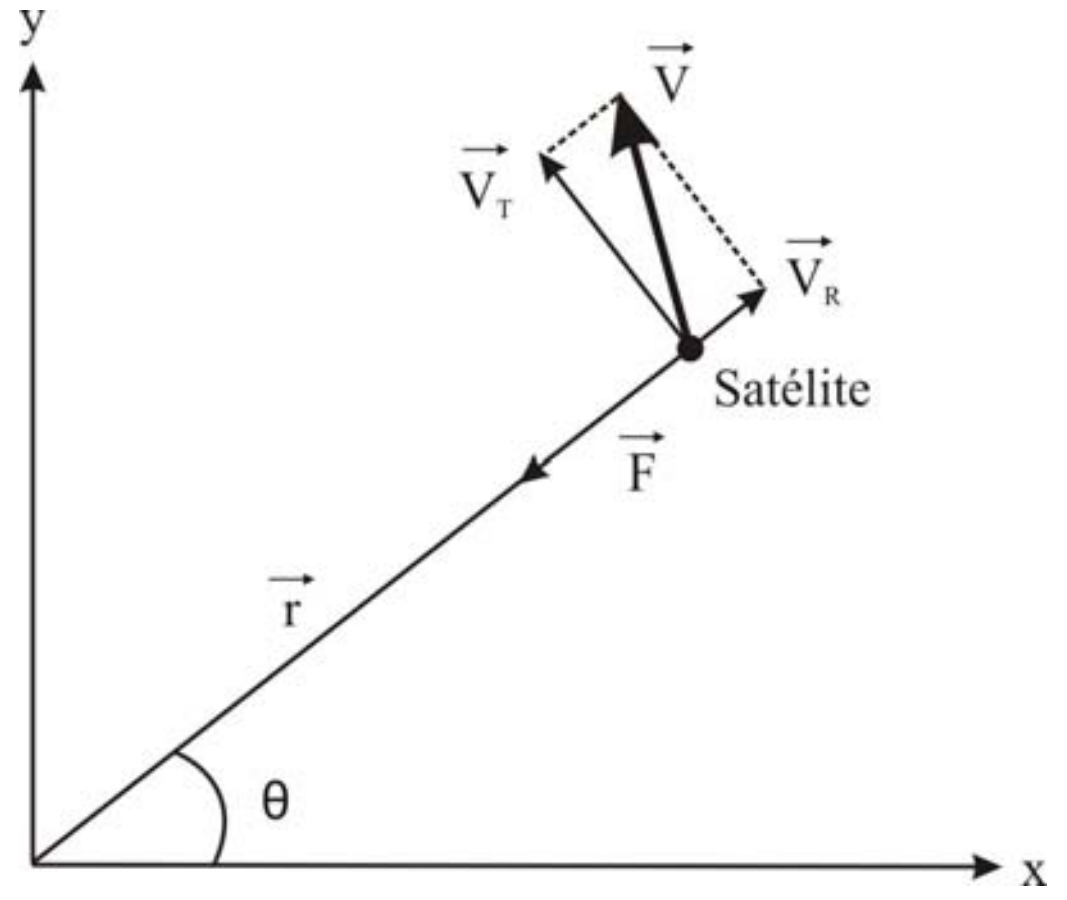

***Figura 2.4. Localización del satélite en coordenadas polares.***

Por lo tanto el momento angular queda

[1] Las ecuaciones son obviamente idénticas si se trata de un planeta, un satélite o un satélite artificial. Se emplea en estos apartados el término *"satélite"* con un sentido genérico.

$$H = \vec{r} \cdot m \cdot \vec{V}_T = r^2 \cdot m \frac{d\theta}{dt} \qquad (N{\cdot}m{\cdot}s) \qquad (7)$$

de donde puede despejarse dθ/dt e introducirse en la ecuación (5) resultando

$$\begin{cases} \dfrac{d\theta}{dt} = \dfrac{H}{r^2 m} \\ \dfrac{dr}{dt} = -\dfrac{H}{m}\dfrac{d\rho}{d\theta} \Rightarrow \dfrac{d^2 r}{dt^2} = -\dfrac{H^2}{m^2}\rho^2 \dfrac{d^2\rho}{d\theta^2} \end{cases} \Rightarrow \frac{G\,M\,m^2}{H^2} = \frac{d^2\rho}{d\theta^2} + \rho \qquad (8)$$

La resolución de esta ecuación puede obtenerse mediante el método de cálculo de solución homogénea y particular [20] quedando

$$\rho = \rho_0 \cos(\theta - \theta_0) + \frac{G\,M\,m^2}{H^2} \qquad (9)$$

Por último, deshaciendo el cambio ρ=1/r se obtiene

$$r = \frac{\dfrac{H^2}{G\,M\,m^2}}{1 + \rho_0 \dfrac{H^2}{G\,M\,m^2}\cos(\theta - \theta_0)} = \frac{p}{1 + e\cos(\theta - \theta_0)} \qquad (m) \qquad (10)$$

donde $p = \dfrac{H^2}{G\,M\,m^2}$ y $e = \dfrac{\rho_0 H^2}{G\,M\,m^2}$

## 2.2.4. Trayectoria de un satélite en una órbita elíptica

La ecuación (10) se corresponde con la ecuación general en coordenadas polares de una sección cónica cuyo foco se encuentra en el origen del sistema de coordenadas, es decir, el centro de masas del cuerpo de masa M se encuentra en uno de los focos. El parámetro **e** determina la ecuación como sigue

- **e = 0** ***círculo***
- **0 < e < 1** ***elipse***
- **e = 1** ***parábola***
- **e > 1** ***hipérbola***

Los planetas siguen una trayectoria elíptica alrededor del Sol, por lo tanto la ecuación de la trayectoria buscada es la de una elipse. El valor de e que hace que la ecuación (10) defina una elipse está comprendido entre 0 y 1. Los parámetros de la elipse pueden apreciarse en la Figura 2.5.

La forma de una elipse la determinan la **excentricidad** y el **semieje mayor** (**e** y **a** respectivamente). La ecuación (10) puede escribirse en función de estos parámetros tal como se expone a continuación.

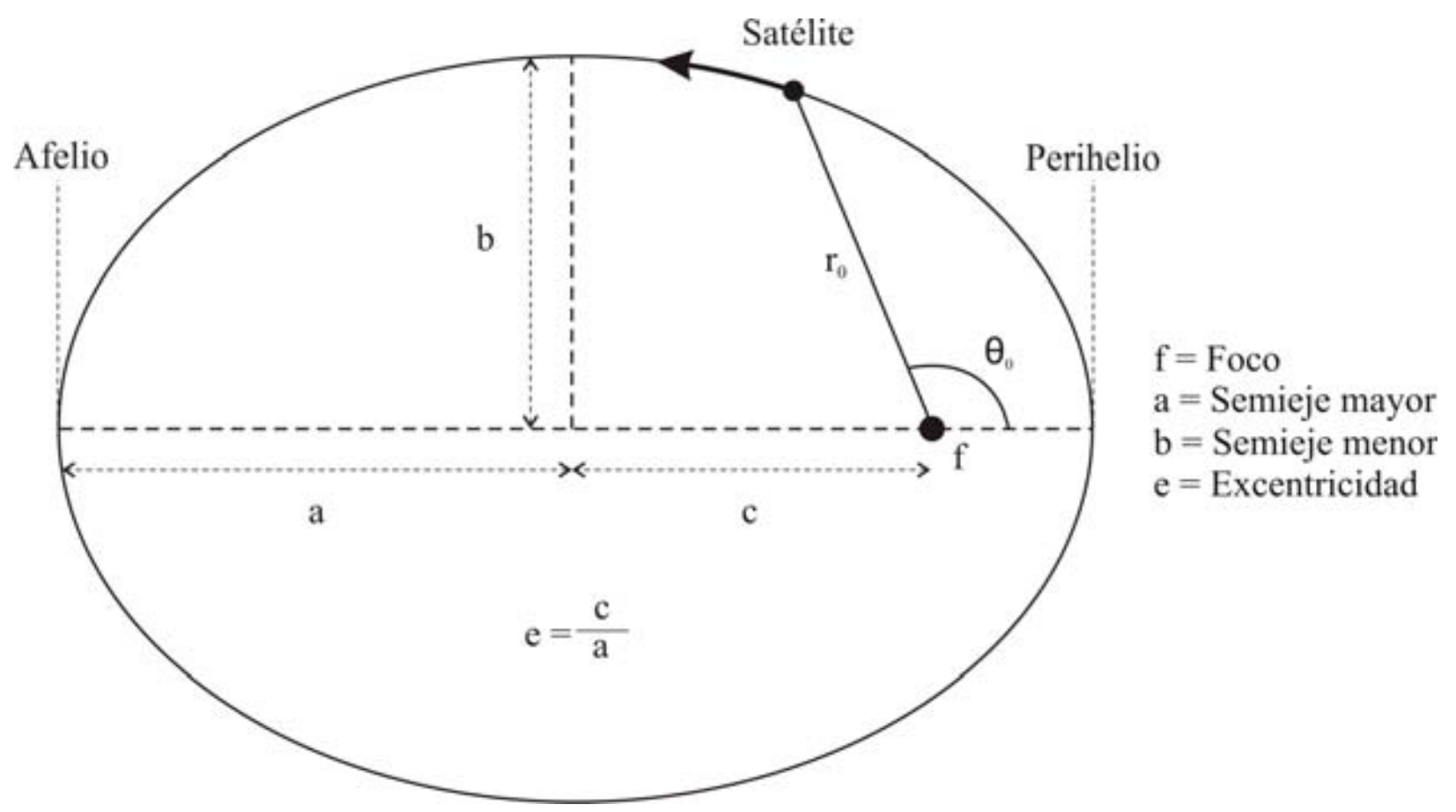

***Figura 2.5. Órbita elíptica y sus parámetros.***

Cuando el satélite está en el punto más cercano al foco (si el foco es el Sol este punto es conocido como ***perihelio***) el radio vector r es mínimo y la ecuación (10) se transforma de la siguiente manera

$$\theta - \theta_0 = 0 \Rightarrow \cos(\theta - \theta_0) = 1 \Rightarrow r = \frac{p}{1+e} \qquad (m) \qquad (11)$$

y cuando el satélite está en el más lejano (conocido como ***afelio*** si el foco es el Sol) queda

$$\theta - \theta_0 = \pi \Rightarrow \cos(\theta - \theta_0) = -1 \Rightarrow r = \frac{p}{1-e} \qquad (m) \qquad (12)$$

Como se observa en la Figura 2.5 la suma de la distancia entre el foco y el afelio y la distancia entre el foco y el perihelio es igual a 2a. Por lo tanto se puede despejar p y obtener así su expresión en función de e y a

$$\frac{p}{1-e} + \frac{p}{1+e} = 2a \Rightarrow p = a(1-e)(1+e) = a(1-e^2) \qquad (13)$$

Sustituyendo el parámetro p en la ecuación (10) se obtiene la ecuación general de la trayectoria del satélite en una órbita elíptica

$$r = \frac{a(1-e^2)}{1+e\cos(\theta - \theta_0)} \qquad (m) \qquad (14)$$

## 2.2.5. Velocidad orbital

La energía que posee un satélite en un punto x determinado de su órbita está formada por dos componentes: la ***energía cinética*** $E_c$, debida únicamente al movimiento del satélite (por lo que para un sistema determinado varía con su velocidad), y la ***energía gravitatoria*** $E_g$, debida a la atracción gravitatoria del cuerpo alrededor del cual orbita, por lo que varía según cual sea su posición en la órbita. La suma de ambas componentes determina la energía del satélite en un punto determinado de la órbita.

$$E_x = E_c + E_g = \frac{1}{2} m v_x^2 - G\frac{Mm}{r_x} \qquad (J) \qquad (15)$$

Cuando el satélite está en el afelio y en el perihelio posee una energía que viene dada por

$$E_a = \frac{1}{2} m v_a^2 - G\frac{Mm}{r_a} \qquad (J) \qquad (16)$$

$$E_p = \frac{1}{2} m v_p^2 - G\frac{Mm}{r_p} \qquad (J) \qquad (17)$$

Como la energía gravitatoria y cinética son conservativas, según la ley de la conservación de la energía

$$E_p = E_a \Rightarrow \frac{1}{2} v_a^2 - G\frac{M}{r_a} = \frac{1}{2} v_p^2 - G\frac{M}{r_p} \qquad (J) \qquad (18)$$

Que ambas energías sean iguales significa que, al ser $r_a$ mayor que $r_p$, entonces $v_p$ tiene que ser mayor que $v_a$. Con ello se deduce que el satélite se mueve más rápidamente cuando pasa cerca del perihelio y más lentamente cuando está próximo al afelio. La velocidad máxima y mínima se corresponde exactamente con la posición del perihelio y afelio respectivamente (Figura 2.6).

Por otro lado, según la ley de conservación del momento angular

$$m\, v_a r_a \mathrm{sen}(\theta_a) = m\, v_p r_p \mathrm{sen}(\theta_p) \Rightarrow v_a r_a = v_p r_p \qquad (J) \qquad (19)$$

Si se introducen en las ecuaciones (16) y (17) la ecuación (19) se puede obtener la energía de la órbita

$$E = -\frac{GMm}{2a} \qquad (J) \qquad (20)$$

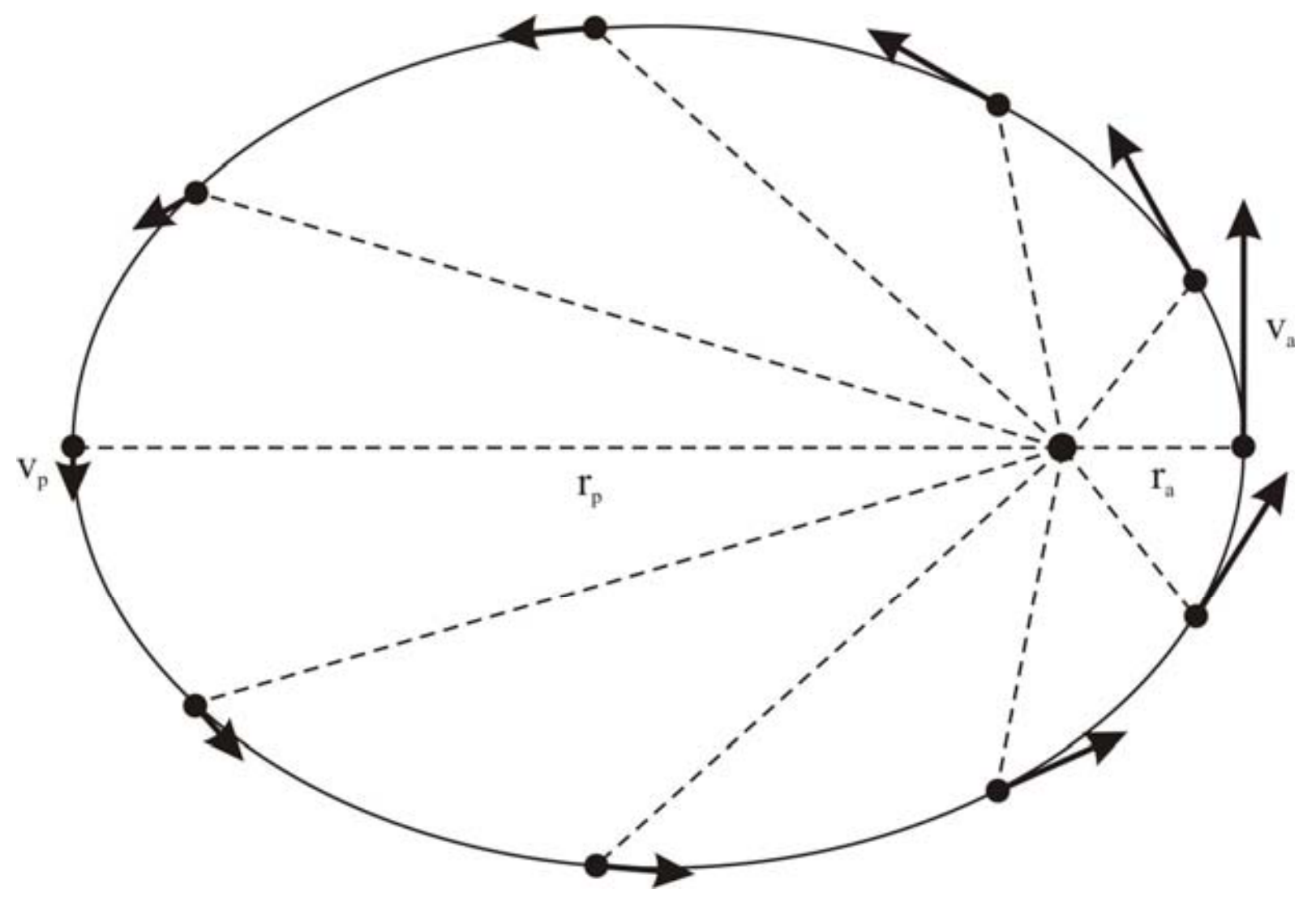

***Figura 2.6. Velocidad orbital del satélite.***

Con lo que puede calcularse la velocidad del satélite en cualquier posición de la órbita de la siguiente manera

$$E = \frac{1}{2}mv^2 - G\frac{Mm}{r} = -\frac{GMm}{2a} \Rightarrow v = \sqrt{GM\left(\frac{2}{r} - \frac{1}{a}\right)} \qquad \text{(m/s)} \qquad (21)$$

## 2.2.6. Periodo de la órbita

El periodo de la órbita es igual al tiempo que tarda el satélite en pasar dos veces por el mismo punto. Utilizando la tercera ley de Kepler puede calcularse este tiempo de la siguiente manera.

Como se observa en la ecuación (1), la tercera ley de Kepler dice que el cuadrado del periodo de una órbita y el cubo del semieje mayor de todas las órbitas posibles guardan una relación constante. Si se considera una órbita circular de radio a y velocidad angular ω, el periodo viene determinado por

$$T = \frac{2\pi}{\omega} \qquad \text{(s)} \qquad (22)$$

Para un movimiento circular, la velocidad angular ω se define como

$$\omega = \frac{v}{a} \qquad \text{(rad/s)} \qquad (23)$$

siendo a el radio del círculo. Por lo tanto (dado que en una órbita circular el radio vector r es igual al radio del círculo) sustituyendo la ecuación (21) y (23) en la ecuación (22) se obtiene la expresión del periodo

$$T = \frac{2\pi}{\omega} = \frac{2\pi a}{v} = \frac{2\pi a}{\sqrt{\frac{GM}{a}}} = 2\pi\sqrt{\frac{a^3}{GM}} \qquad \text{(s)} \qquad (24)$$

También se puede observar que la constante que relaciona en la tercera ley de Kepler el periodo y el semieje mayor de la órbita será

$$k = \frac{T^2}{a^3} = \frac{(2\pi)^2}{GM} \qquad (25)$$

## 2.2.7. Posición de un satélite en una órbita elíptica

Para realizar el cálculo de la posición de un satélite en una órbita elíptica general se hace uso de un círculo circunscrito a la órbita como se puede apreciar en la Figura 2.7.

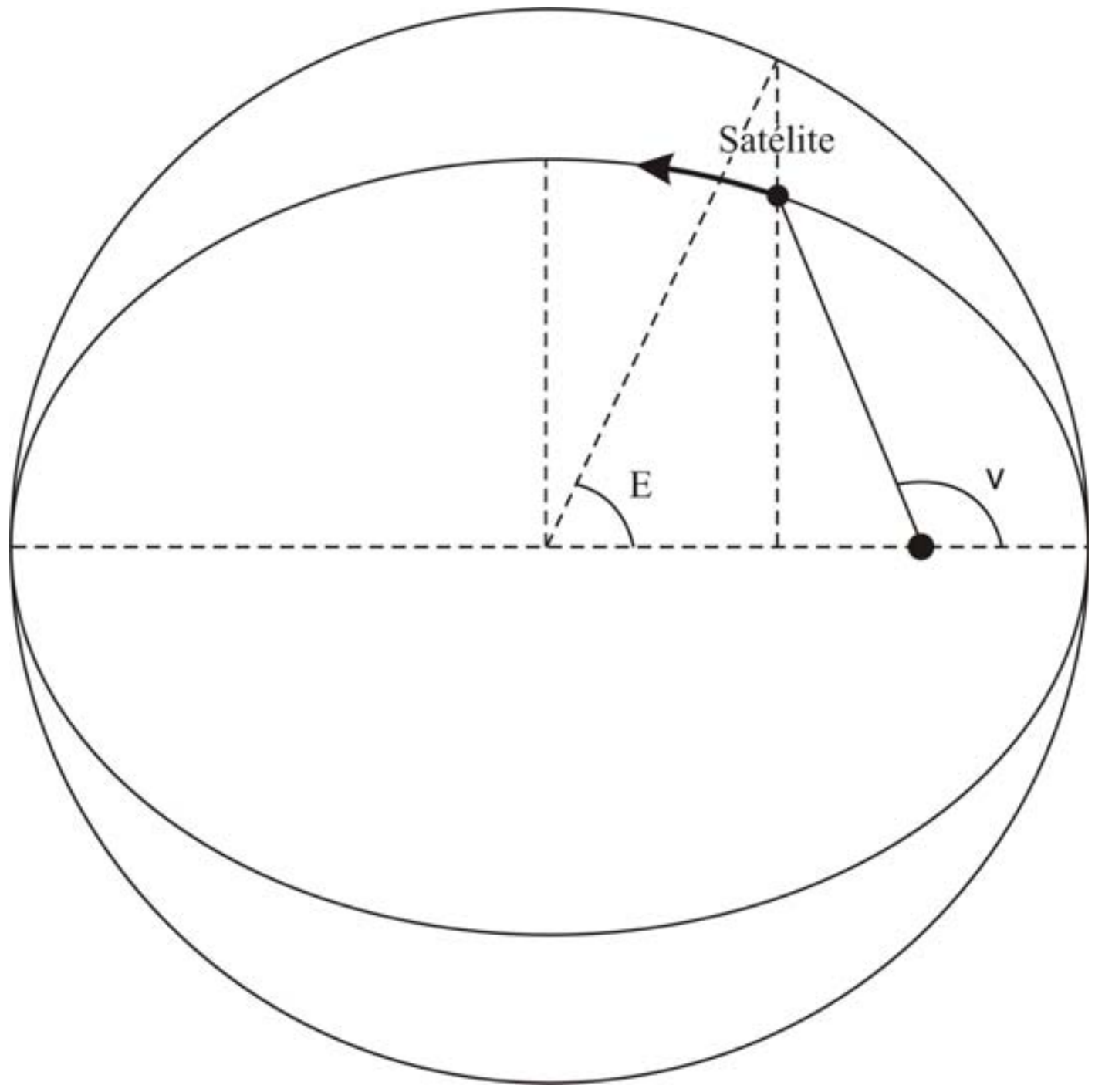

***Figura 2.7. Anomalía excéntrica y anomalía verdadera de una órbita elíptica.***

Haciendo uso de la notación de la Figura 2.7 se puede apreciar que el ángulo v corresponde al ángulo declarado anteriormente como θ. Este ángulo se denomina ***anomalía verdadera*** y está definido como el ángulo entre la dirección del perihelio y la dirección del satélite en la órbita. Su valor es cero en el perihelio y aumenta hasta 2π por cada órbita. Tal como predice la segunda ley de Kepler, este ángulo se incrementa más rápidamente cerca del perihelio y más lentamente cerca del afelio.

Son necesarios dos ángulos auxiliares para el cálculo de la posición del satélite en la órbita

- ***Anomalía media M***. Definida como la anomalía verdadera que tendría el satélite en una órbita circular del mismo periodo que la elíptica y cuya expresión, por tanto, es

$$M = \frac{2\pi}{T} t \qquad \text{(rad)} \qquad (26)$$

  donde T es el periodo de la órbita y t se incrementa desde 0 hasta T. Esto proporciona una variación uniforme del ángulo M entre 0 y 2π.

- ***Anomalía excéntrica E***. Se define, tal y como se observa en la Figura 2.7, como el ángulo entre la dirección del perihelio y la línea que une el centro de la elipse con el punto que resulta de la intersección de la circunferencia circunscrita y una línea vertical trazada desde la posición del satélite hasta el círculo.

Por simple geometría se demuestra que la anomalía excéntrica está relacionada con la anomalía verdadera de la siguiente forma

$$\cos(v) = \frac{\cos(E) - e}{1 - e \cdot \cos(E)} \tag{27}$$

y con la anomalía media a través de la ecuación de Kepler, cuya expresión es

$$M = E - e \cdot \text{sen}(E) \qquad \text{(rad)} \tag{28}$$

**CÁLCULO DE LA ANOMALÍA EXCÉNTRICA**

Debido a la forma de la ecuación (28), el valor de E tendrá que ser forzosamente una aproximación. Este valor se calculará mediante el **método de aproximaciones sucesivas** (que, para los valores de las excentricidades de los planetas del sistema solar, garantiza que la solución converge sin un número elevado de iteraciones), realizando un número determinado de iteraciones con la ecuación de Kepler hasta que se reduzca el error por debajo de un cierto límite. El procedimiento es el siguiente.

El objetivo es hallar el valor de E, por lo que despejándolo de la ecuación de Kepler se obtiene

$$E = M + e \cdot \text{sen}(E)$$

Como una primera aproximación se puede suponer que, si la excentricidad e no es muy grande, la elipse no es muy diferente a un círculo, por lo que los valores de M y E son muy parecidos. En la primera iteración $E_0$ será igual a M.

$$E_0 = M$$

Un segundo valor de E mejorado será

$$E_1 = M + e \cdot \text{sen}(E_0)$$

y otro valor mejorado

$$E_2 = M + e \cdot \text{sen}(E_1)$$

y así sucesivamente. El valor de E va convergiendo hacia un valor más correcto cada vez. Se realizan tantas iteraciones como sean necesarias hasta garantizar que el error no supera un cierto límite máximo.

Nótese que todas las anomalías valen cero en el perihelio y que en el caso de una órbita circular todas tienen el mismo valor.

La posición del satélite en el plano orbital, es decir, sus coordenadas bidimensionales x e y, pueden calcularse directamente, mediante la anomalía verdadera, de la siguiente manera

$$\begin{cases} x = r \cdot \cos(v) \\ y = r \cdot \text{sen}(v) \end{cases} \tag{29}$$

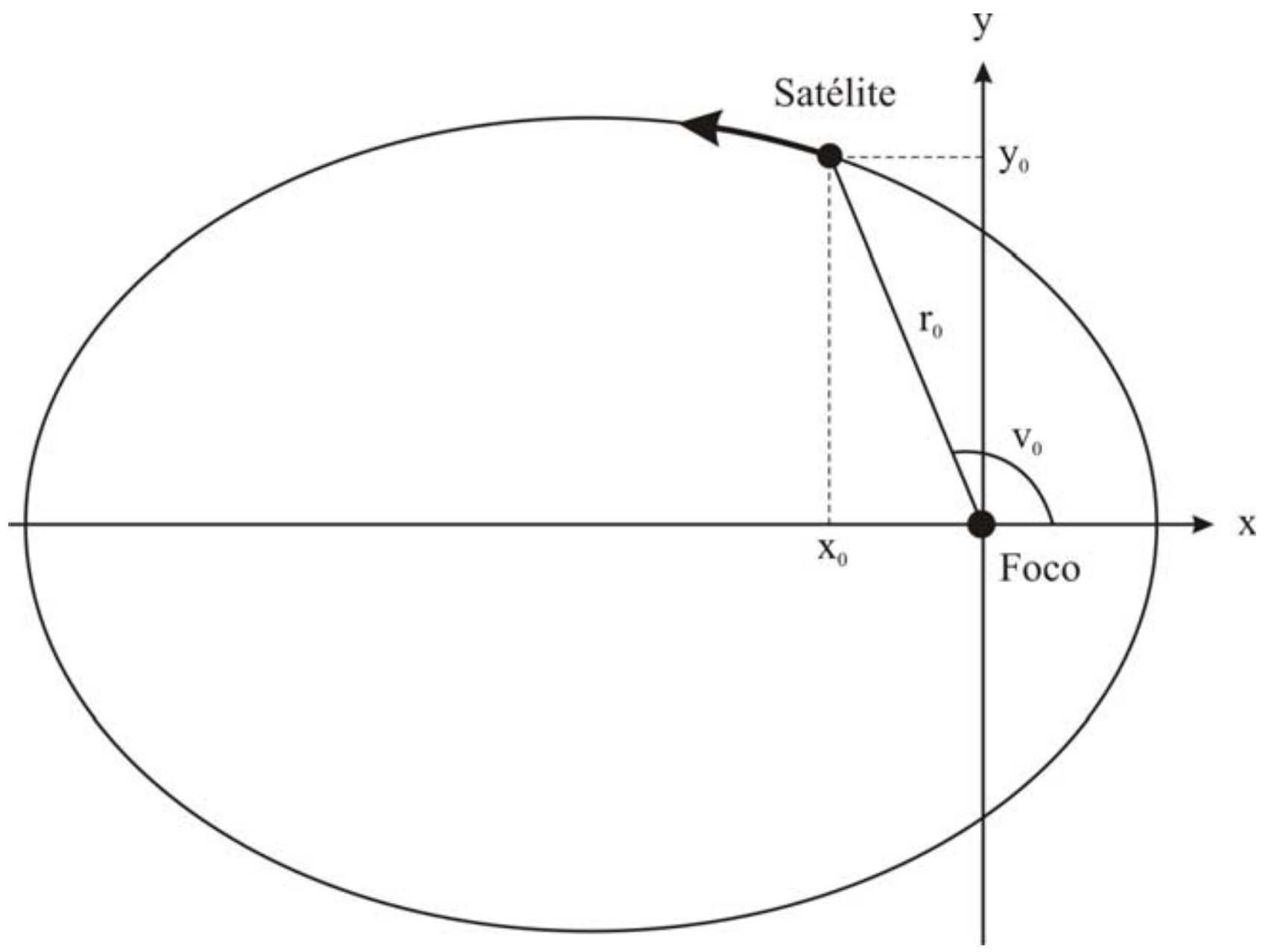

***Figura 2.8. Posición del satélite en el plano orbital de una órbita elíptica en coordenadas cartesianas bidimensionales.***

Estas coordenadas cartesianas, como se observa en la Figura 2.8, corresponden a la posición del satélite en el plano orbital con un sistema de referencia cuyo centro se encuentra en el foco de la elipse.

## 2.2.8. Localización de la órbita en el espacio

Orientar la órbita en un espacio de tres dimensiones requiere un **plano** y una **dirección de referencia**. En el simulador implementado para este proyecto los satélites son planetas y el foco de sus órbitas es el Sol. Por ello se emplea un ***sistema de coordenadas eclípticas***, en el que el plano de referencia es el plano de la eclíptica (el plano en el que la Tierra gira alrededor del Sol o el plano sobre el que se sitúa la órbita de la Tierra). La dirección de referencia viene marcada por el equinoccio vernal o de primavera en el hemisferio norte. Esta dirección la determina la línea que une el Sol y el centro de la Tierra en el equinoccio de Marzo. Esta dirección marca la dirección del eje x en el eje de coordenadas cartesianas.

La línea formada por la intersección del plano orbital con el plano de la eclíptica se conoce como ***línea de nodos***. El origen de coordenadas de nuevo se sitúa en el centro del Sol, que a su vez es el foco de la elipse que sigue la órbita. Este punto pertenece a ambos planos y por lo tanto siempre pertenece a la línea de nodos. El **nodo ascendente** se define como el punto de la línea de nodos por el que pasa el satélite cuando asciende del semiplano de la eclíptica inferior al superior. Este punto sirve como referencia desde donde medir los ángulos.

Por lo tanto, tal y como se observa en la Figura 2.9, los parámetros que sitúan a la órbita en el espacio tridimensional son los siguientes

- ***Inclinación (i).*** Es el ángulo que forman el plano de referencia (el de la eclíptica) y el plano de la órbita del satélite (del planeta). En el caso de la Tierra es claro que este ángulo es cero.

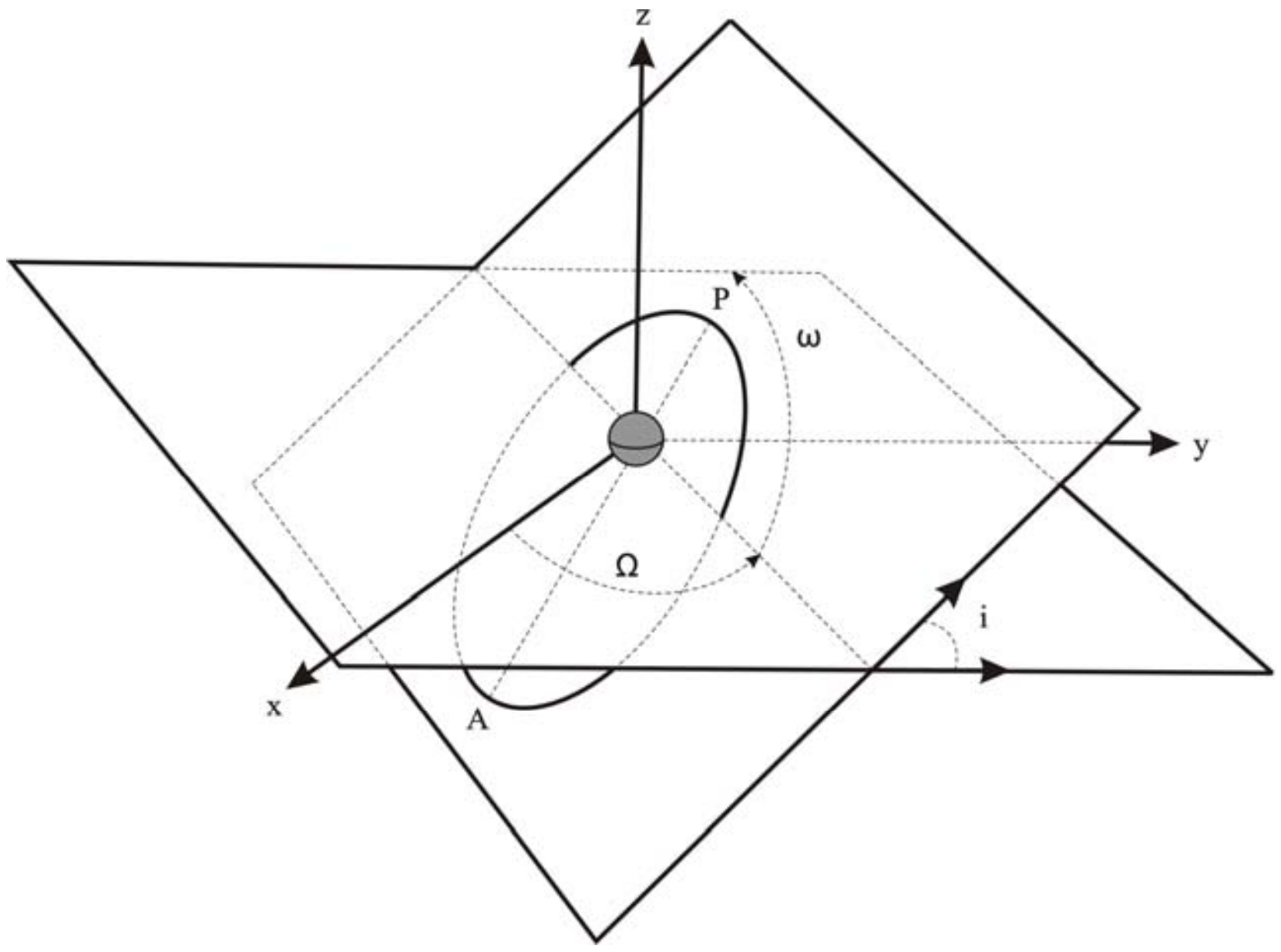

*Figura 2.9. Localización de una órbita en un sistema de coordenadas heliocéntrico.*

- ***Ascensión recta o longitud del nodo ascendente (Ω).*** Este ángulo se mide desde el punto vernal (la dirección de referencia) hasta el nodo ascendente. De nuevo en el caso de la Tierra este ángulo es cero.

- ***Argumento del perihelio (ω).*** Es el ángulo que indica cómo se orienta la órbita en su propio plano orbital. Se calcula entre el nodo ascendente y el punto de la órbita más cercano al Sol (perihelio).

## CALENDARIOS DE REFERENCIA

La posición del punto vernal varía con el tiempo por lo que es necesario establecer para los parámetros orbitales una referencia temporal. Existen varios sistemas diferentes, los más conocidos son el B1950, el J2000 y el TOD.

- **B1950.** Era el sistema más usado en el pasado antes de ser sustituido por el sistema J2000. La B denota época Besseliana.

- **J2000.** Actualmente el más extendido. Casi todos los datos astronómicos han sido actualizados a este sistema, por lo que es de uso común. Como referencia toma el día juliano 2451545 (el día juliano es el número de días transcurridos desde el día cero juliano que es el día 1 de enero del 4713 AC a las 12). El día juliano 2451545 es el 1 de enero de 2.000 a las 12:00 UT (*Universal Time*).

- **TOD ("*True Of Date*").** La referencia es una fecha determinada para la cual se dan los datos.

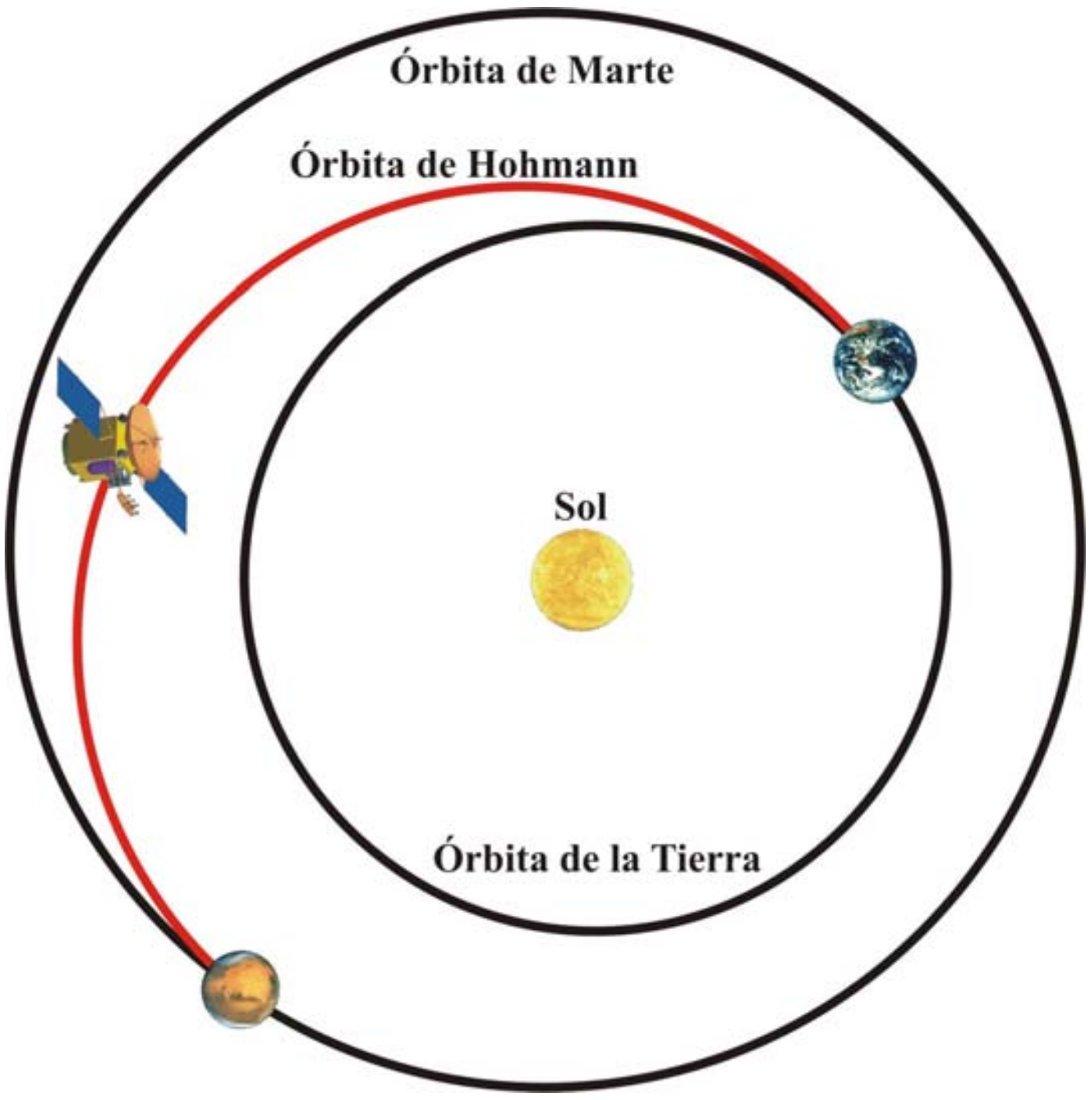

***Figura 2.10. Viaje Tierra-Marte mediante una órbita de transferencia de Hohmann.***

## 2.2.9. Órbita de transferencia

Como se menciona en el Capítulo 6, la nave espacial *MTO* estará disponible para realizar pruebas del sistema de comunicaciones por láser durante el tiempo que dura el viaje de la Tierra a Marte. Por este motivo también se ha simulado dicho viaje, que como se explica a continuación, está basado en la dinámica orbital vista hasta ahora.

Dado que el sistema solar está dominado por la gravedad del Sol, cualquier objeto dentro de él se desplaza siguiendo trayectorias que son parte de *secciones cónicas*. Esto significa que el viaje que realiza la nave espacial desde la Tierra hasta Marte seguirá un camino *elíptico*, *hiperbólico* o *parabólico*. El tipo de trayectoria que se suele utilizar para un viaje típico Tierra-Marte sigue la llamada ***órbita de transferencia de Hohmann***. Esta trayectoria elíptica es óptima ya que minimiza la cantidad de energía necesaria para realizar el viaje. La mayor parte de la energía necesaria para completar el recorrido se obtiene aprovechando la atracción gravitatoria del Sol. El resto se genera de los motores de la nave y resulta de la necesidad de cambiar de la órbita inicial alrededor de la Tierra a la órbita de Hohmann y de ésta a la órbita en Marte.

Como se observa en la Figura 2.10, el *perihelio* de la órbita de Hohmann coincide con el punto en el que la nave abandona la órbita alrededor de la Tierra para introducirse en la de transferencia. De igual modo, el *afelio* de la órbita de Hohmann tiene lugar en el punto en el que la nave abandona la órbita de transferencia para su inserción en la órbita alrededor de Marte. Por lo tanto, existe una conjunción imaginaria entre la Tierra en el momento de iniciar el viaje y Marte en el momento en el que llega la nave. Por ello también se denomina **trayectoria de conjunción**.

La duración del viaje siguiendo esta trayectoria es mínima cada 26 meses (ya que las posiciones relativas entre la Tierra y Marte se repiten aproximadamente cada 26 meses), por lo que se aprovecha esta circunstancia para realizar los viajes Tierra-Marte con estos intervalos.

# 2.3. REALIZACIÓN DEL SIMULADOR

Se ha llevado a cabo la implementación de un simulador de órbitas planetarias cuyo objetivo es la obtención de la posición instantánea de cada planeta en su órbita en un entorno con tres dimensiones espaciales (con el Sol en el origen de coordenadas) y una temporal con el objetivo de precisar el momento para el cual se desea conocer la posición y tener además la posibilidad de animar la simulación.

## 2.3.1. Elementos orbitales

Como se ha visto en la parte de dinámica orbital, una órbita queda completamente definida en el espacio mediante los siguientes seis parámetros o elementos orbitales

- ***Semieje mayor (a)***
- ***Excentricidad (e)***
- ***Anomalía media (M)***
- ***Inclinación (i)***
- ***Longitud del nodo ascendente (Ω)***
- ***Argumento del perihelio (ω)***

Los tres primeros definen la forma de la órbita y el movimiento del satélite en ella en un sistema de referencia de dos dimensiones espaciales y una temporal, y los tres últimos sitúan a la órbita en el espacio de tres dimensiones espaciales más la temporal.

Los elementos orbitales que se han empleado en la simulación, y que se muestran en la Tabla 2.1, han sido obtenidos de la base de datos planetaria *"National Space Science Data Center"* [21] de la NASA. Estos datos son producto de un ajuste por mínimos cuadrados en un periodo de 250 años referidos al sistema J2000, esto es, el día de referencia es el 1 de Enero de 2.000 a las 12:00 UT.

| Planeta | a (UA) | e | M (grados) | i (grados) | Ω (grados) | ω (grados) |
|---|---|---|---|---|---|---|
| *Mercurio* | 0,38709893 | 0,20563069 | 174,79439 | 7,00487 | 48,33167 | 29,12478 |
| *Venus* | 0,72333199 | 0,00677323 | 50,44675 | 3,39471 | 76,68069 | 54,85229 |
| *Tierra* | 1,00000011 | 0,01671022 | 357,51716 | 0,00005 | -11,26064 | 114,20783 |
| *Marte* | 1,52366231 | 0,09341233 | 19,41248 | 1,85061 | 49,57854 | 286,4623 |
| *Júpiter* | 5,20336301 | 0,04839266 | 19,65053 | 1,30530 | 100,55615 | -85,8023 |
| *Saturno* | 9,53707032 | 0,05415060 | 317,51238 | 2,48446 | 113,71504 | -21,2831 |
| *Urano* | 19,19126393 | 0,04716771 | 142,26794 | 0,76986 | 74,22988 | 96,73436 |
| *Neptuno* | 30,06896348 | 0,00858587 | 259,90868 | 1,76917 | 131,72169 | -86,75034 |
| *Plutón* | 39,48168677 | 0,24880766 | 14,86205 | 17,14175 | 110,30347 | 113,76329 |

***Tabla 2.1. Elementos orbitales de los planetas del sistema solar [21].***

La base de datos de la que se han obtenido los datos proporciona la longitud media en lugar de la anomalía media y la longitud del perihelio en lugar del argumento del perihelio. En la Tabla 2.1 se presentan los datos según los parámetros que se han visto en la exposición teórica de dinámica orbital en este proyecto, por lo que se han hecho las siguientes conversiones

$$M = L - \varpi \qquad \text{(rad)} \qquad (30)$$

$$\omega = \varpi - \Omega \qquad \text{(rad)} \qquad (31)$$

donde $\varpi$ es la longitud del perihelio y L la longitud media.

### 2.3.2. Escala temporal

El sistema temporal utilizado es el **J2000** mencionado anteriormente. Así pues, los elementos o parámetros orbitales usados en la simulación toman como referencia el día *1 de enero de 2.000 a las 12:00 UT*, o lo que es lo mismo, el ***día juliano 2451545***. Este instante determinará el día 0,0 en la escala temporal usada en esta simulación. El día, denotado como d, por tanto, será el número de días desde el día cero. Esto significa que puede ser calculado a partir del día juliano de la siguiente forma

$$d = \text{día juliano} - 2451545 \qquad (32)$$

El día juliano se puede convertir a partir de una fecha de calendario gregoriano[1] mediante la siguiente expresión [22]

$$\text{día juliano} = \frac{1461}{4} \cdot \left( \text{año} + 4800 + \frac{\text{mes - 14}}{12} \right)$$
$$+ \frac{367}{12} \cdot \left( \text{mes - 2 - 12} \cdot \left( \frac{\text{mes - 14}}{12} \right) \right) - \frac{3}{4} \cdot \left( \frac{y + 4900 + \frac{\text{mes - 14}}{12}}{100} \right) + \text{día - 32075} \qquad (33)$$

donde día, mes y año representan la fecha del calendario gregoriano.

### 2.3.3. Variación temporal

Los parámetros que definen las órbitas de los planetas no son constantes. Existen fenómenos que los alteran con el paso del tiempo tales como los efectos gravitatorios de los demás elementos del sistema solar. En esta simulación no se ha considerado ningún tipo de rectificación temporal a los parámetros orbitales. La razón es que los elementos orbitales utilizados en la simulación toman como referencia al año 2.000 y el periodo para el que se ha implementado la simulación está suficientemente próximo (2.009-2.010) para la precisión que se necesita en este proyecto, por lo que el efecto puede ser considerado despreciable.

Por lo tanto, el único parámetro orbital que varía con el tiempo es la **anomalía media**, que determina la posición instantánea de cada planeta en su órbita. Debido a que la unidad mínima de la escala temporal, tal como se ha explicado en el apartado anterior, es el día, la anomalía media varía cada día. El movimiento diario de un planeta viene dado por

$$n = \frac{2\pi}{T} \qquad \text{(rad/día)} \qquad (34)$$

[1] El calendario gregoriano es el usado actualmente en la mayor parte del mundo occidental y parte de Asia. Establecido en el siglo XVI, corrigió al calendario juliano por no tener su año la misma duración que el solar. Esta corrección se basa en que los años centenarios divisibles por 400 sean bisiestos y que todos los demás años centenarios sean normales. Por ejemplo, 2.000 fue bisiesto, pero 1.900 y 1.800 no.

siendo T el periodo orbital en días que puede ser calculado mediante la fórmula (24), anteriormente deducida, que ha de convertirse de segundos a días dividiendo entre 3.600·24. Este valor de n se le suma a M cada vez que pase un día, quedando así establecida la variación temporal de la simulación. La implementación de esta variación temporal consiste en la adición de un día al día d anterior dentro de un bucle.

## 2.3.4. Posición del planeta en el espacio bidimensional

En el apartado 2.2.7 se explicó la forma de localizar a un satélite en una órbita elíptica situada en un espacio bidimensional. Las coordenadas (x, y) eran obtenidas con la ecuación (29) en función de la anomalía verdadera. Se pueden expresar estas coordenadas en función de la anomalía excéntrica utilizando la ecuación (14) de la trayectoria del satélite (planeta) en función de la distancia al foco (Sol) y de la anomalía verdadera y la ecuación (27) que relaciona la anomalía verdadera y la excéntrica, tal como se indica a continuación. Si

$$\begin{cases} x = r \cdot \cos(v) \\ y = r \cdot \mathrm{sen}(v) = r \cdot \sqrt{1-\cos^2(v)} \end{cases} \tag{35}$$

y

$$r = \frac{a(1-e^2)}{1+e\cos(\theta-\theta_0)} \tag{36}$$

$$\cos(v) = \frac{\cos(E)-e}{1-e\cdot\cos(E)} \tag{37}$$

entonces sustituyendo las ecuaciones (36) y (37) en la ecuación (35) se llega a

$$\begin{cases} x = a \cdot (\cos(E) - e) \\ y = a \cdot \mathrm{sen}(E) \cdot \sqrt{1-e^2} \end{cases} \tag{38}$$

Recuérdese que la anomalía excéntrica era calculada (en el apartado 2.2.7) mediante el método de aproximaciones sucesivas en un bucle de este tipo

```
En-1 = M
salir = NO
REPETIR
        En = M + e · sen(En-1)
        SI |En - En-1| < ε
                salir = SI
        FIN SI
        En-1 = En
MIENTRAS salir = NO
```

donde E es la anomalía excéntrica, M la anomalía media, e la excentricidad y ε el error permitido en la aproximación de E. En la simulación implementada para este proyecto se ha empleado un error de $10^{-3}$. A la salida del bucle, la variable $E_n$ tiene almacenado el valor aproximado de la anomalía excéntrica.

### 2.3.5. Posición del planeta en el espacio tridimensional

Debido a que las órbitas del sistema solar están en diferentes planos y por lo tanto tienen diferentes orientaciones en el espacio, las coordenadas que definen la posición deben situarse en un espacio tridimensional común y estar referidas al mismo punto. El origen del sistema es el Sol, que tiene las coordenadas (0, 0, 0). Esto quiere decir que las coordenadas tridimensionales serán heliocéntricas.

Las coordenadas x e y instantáneas sirven para calcular directamente el valor de la distancia y la anomalía verdadera en cada momento

$$\begin{aligned} r &= \sqrt{x^2 + y^2} \\ v &= \operatorname{arctg}\left(\frac{y}{x}\right) \end{aligned} \tag{39}$$

Conocidos r y v, las coordenadas rectangulares, que contemplan todos los parámetros orbitales, se pueden expresar de la siguiente forma [23]

$$\begin{cases} x_h = r \cdot [\cos(\Omega) \cdot \cos(v + \omega) - \operatorname{sen}(\Omega) \cdot \operatorname{sen}(v + \omega) \cdot \cos(i)] \\ y_h = r \cdot [\operatorname{sen}(\Omega) \cdot \cos(v + \omega) + \cos(\Omega) \cdot \operatorname{sen}(v + \omega) \cdot \cos(i)] \\ z_h = r \cdot [\operatorname{sen}(v + \omega) \cdot \operatorname{sen}(i)] \end{cases} \tag{40}$$

donde el conjunto ($x_h$, $y_h$, $z_h$) representa las coordenadas heliocéntricas que definen la posición de un punto en un sistema tridimensional heliocéntrico. Recuérdese que el eje x es la dirección de referencia que apunta al equinoccio vernal y que el eje z es perpendicular al plano de referencia de la eclíptica. Como comprobación se puede verificar que se cumple

$$r = \sqrt{{x_h}^2 + {y_h}^2 + {z_h}^2} \qquad \text{(m)} \tag{41}$$

De esta manera se obtienen las coordenadas en tres dimensiones que determinan en la simulación la variación temporal del movimiento de los planetas en sus órbitas. Para conocer la distancia entre dos puntos ($x_{h0}$, $y_{h0}$, $z_{h0}$) y ($x_{h1}$, $y_{h1}$, $z_{h1}$) definidos por sus coordenadas heliocéntricas basta con aplicar la siguiente fórmula

$$\text{dist} = \sqrt{(x_{h1} - x_{h0})^2 + (y_{h1} - y_{h0})^2 + (z_{h1} - z_{h0})^2} \qquad \text{(m)} \tag{42}$$

### 2.3.6. Velocidad radial Marte-Tierra por traslación

Como se demostró en la ecuación (21), la velocidad de un satélite alrededor de una órbita elíptica no es constante; la distancia al foco hace que varíe con el tiempo. Además, debido al movimiento relativo entre la Tierra y Marte, la velocidad en la dirección de la línea imaginaria que une el centro de ambos planetas también varía al desplazarse éstos.

Teniendo en cuenta los dos efectos mencionados sobre la velocidad, es posible definir para cada planeta un vector **velocidad radial** $\vec{v}_{rT}$ debido a la traslación de los planetas cuya magnitud sea el valor absoluto de la velocidad en la dirección hacia el otro planeta y cuyo ángulo quede determinado por la dirección mencionada y por el sentido de la velocidad en dicha dirección.

El cálculo de la velocidad radial para cada planeta surge de la necesidad de conocer la **velocidad instantánea** de emisor y receptor en la dirección que los une para determinar la variación entre la longitud de onda transmitida y la recibida a consecuencia del ***efecto Doppler***.

## 2.3.6.a. Módulo

La velocidad orbital calculada en el apartado 2.2.5 se corresponde con la velocidad de traslación del planeta. El vector de esta velocidad tiene la dirección de la recta *tangente a la órbita* en ese punto, cuya expresión puede ser deducida en función de la posición del planeta y los semiejes mayor (a) y menor (b) de la elipse. La ecuación de esta recta tiene la siguiente forma

$$\frac{x_0 \cdot x}{a^2} + \frac{y_0 \cdot y}{b^2} = 1 \qquad (43)$$

donde ($x_0$, $y_0$) son las coordenadas rectangulares –bidimensionales, ya que para el cálculo de la velocidad es indiferente la orientación de la órbita en el espacio– que definen la posición del planeta, a es el semieje mayor de la elipse y b el semieje menor, que viene dado por

$$b = a\sqrt{1 - e^2} \qquad \text{(m)} \qquad (44)$$

Por lo tanto la velocidad radial será la proyección vectorial de la velocidad de traslación del planeta sobre la recta que une ambos planetas. Dado que las coordenadas de ambos planetas son conocidas, la ecuación de esta recta puede calcularse directamente de esta manera

$$\frac{x - x_0}{x_1 - x_0} = \frac{y - y_0}{y_1 - y_0} \qquad (45)$$

siendo ($x_1$, $y_1$) las coordenadas del otro planeta.

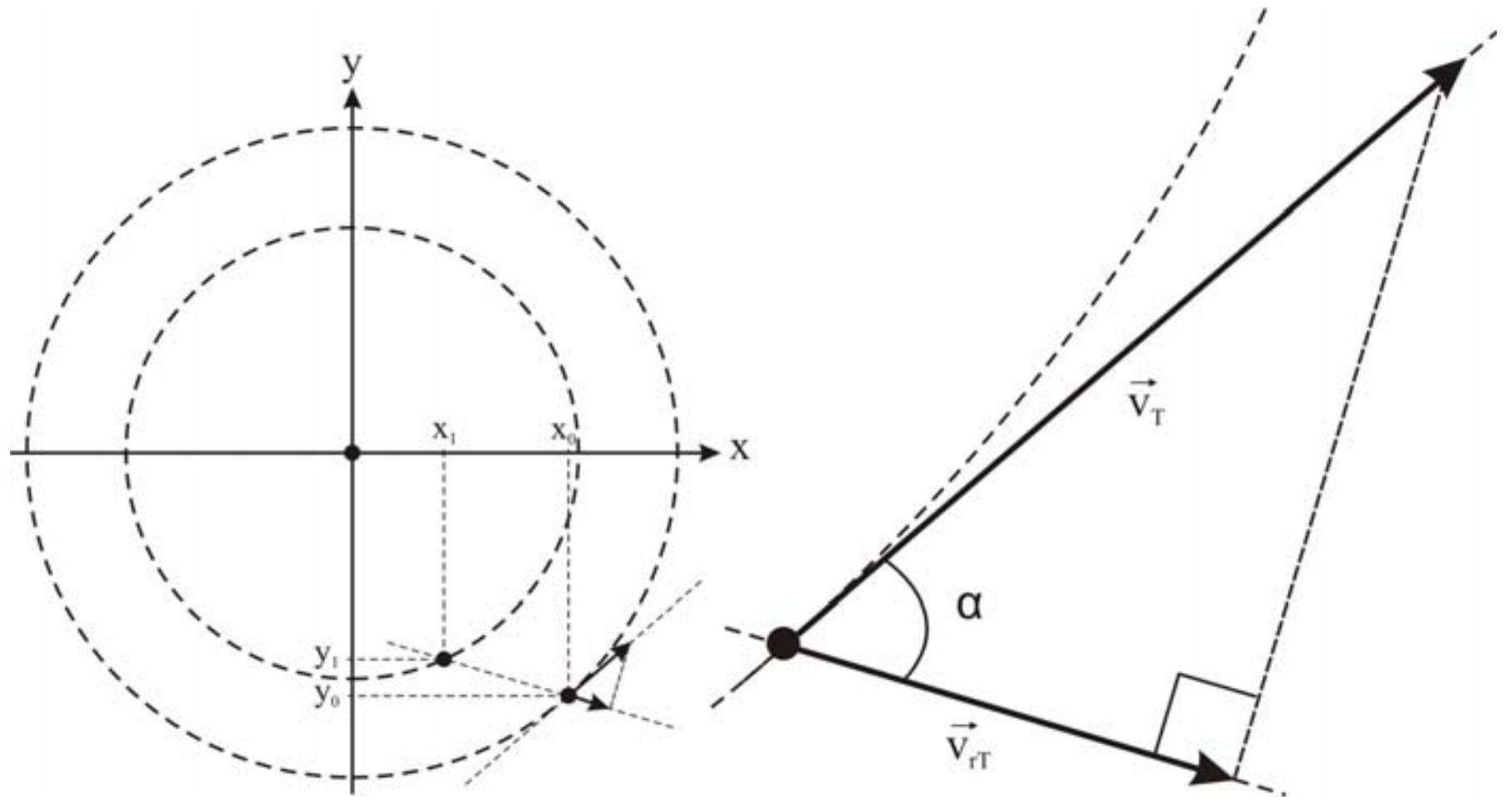

***Figura 2.11. Velocidad radial de Marte con respecto a la Tierra.***

Como se observa en la Figura 2.11, el vector $\vec{v}_{rT}$ es la proyección de la velocidad de traslación $\vec{v}_T$, por lo que el módulo se puede calcular como

$$\left|\vec{v}_{rT}\right| = \left|\vec{v}_T\right| \cdot \cos(\alpha) \qquad \text{(m/s)} \tag{46}$$

Sólo queda conocer el valor del ángulo α formado entre ambas rectas. Este ángulo viene dado por

$$\alpha = \arctan\left|\frac{m_r - m_t}{1 - m_r m_t}\right| \qquad \text{(rad)} \tag{47}$$

siendo $m_r$ la pendiente de la recta que une ambos planetas y $m_t$ la pendiente de la recta tangente a la elipse en el punto ($x_0$, $y_0$). Nótese que, dado que el argumento de la función arco tangente está en valor absoluto, podrían intercambiarse las pendientes $m_r$ y $m_t$ dando el mismo resultado. Estas pendientes se deducen directamente de la ecuación de cada recta expresándolas en la forma $y = m \cdot x + b$

$$m_r = \frac{y_1 - y_0}{x_1 - x_0} \tag{48}$$

$$m_t = \frac{x_0}{y_0}\left(e^2 - 1\right) \tag{49}$$

Por lo tanto la expresión final para el módulo de la velocidad radial queda

$$\left|v_{rT}\right| = \left|v_T\right| \cdot \cos\left(\arctan\left|\frac{\left(\frac{y_1 - y_0}{x_1 - x_0}\right) - \left(\frac{x_0}{y_0}\left(e^2 - 1\right)\right)}{1 - \left(\frac{y_1 - y_0}{x_1 - x_0}\right)\left(\frac{x_0}{y_0}\left(e^2 - 1\right)\right)}\right|\right) \tag{50}$$

Esta expresión sirve para el cálculo del módulo de la velocidad radial $\vec{v}_{rT}$ tanto para el caso de Marte que se acaba de ver, como para el caso de la Tierra, que se puede observar en la Figura 2.12.

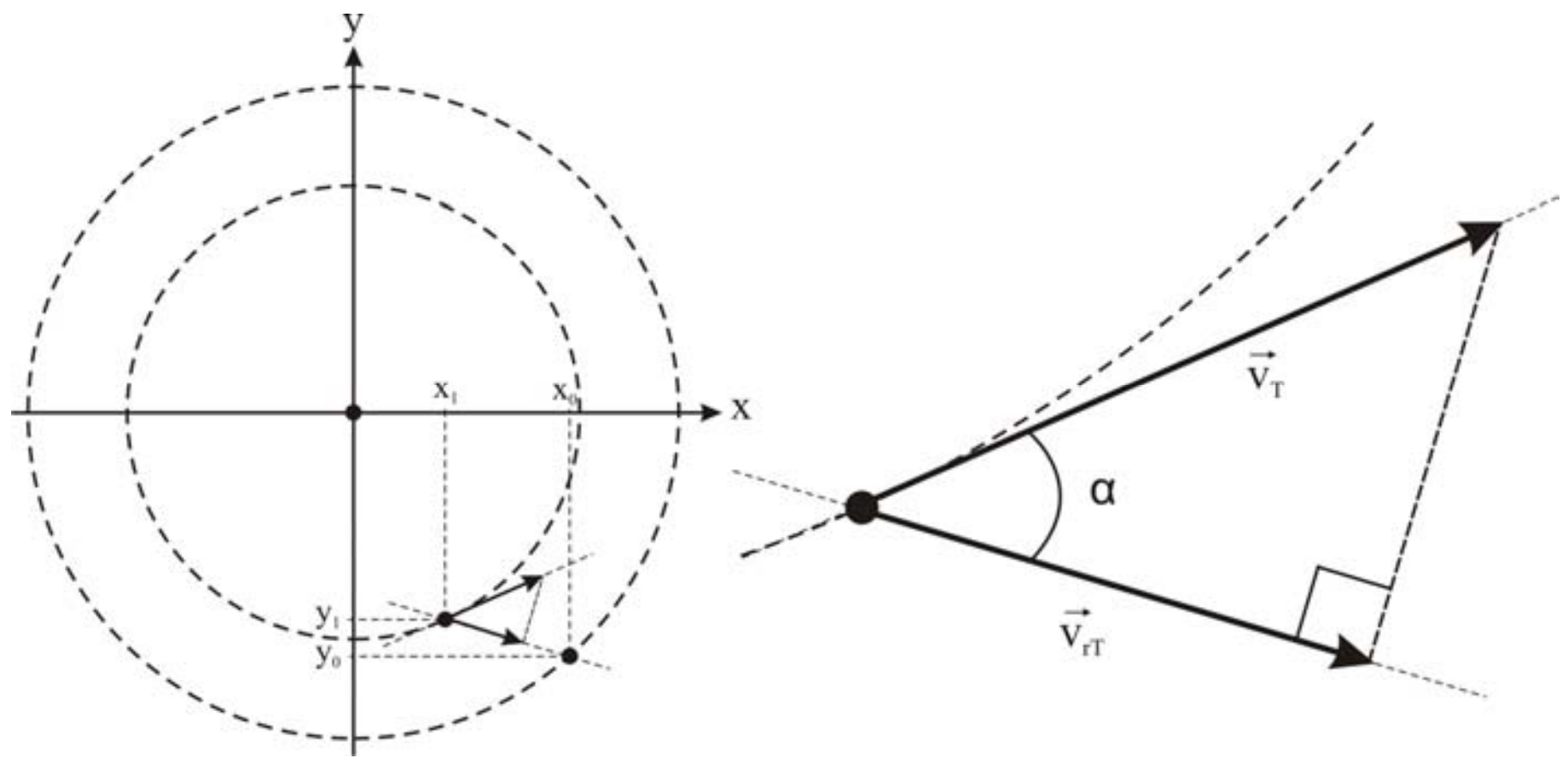

*Figura 2.12. Velocidad radial de la Tierra con respecto a Marte.*

## 2.3.6.b. Dirección

Conocido el módulo del vector velocidad radial $\vec{v}_{rT}$ se requiere saber si el sentido de la velocidad es hacia el otro planeta o en sentido contrario ya que para el cálculo del efecto Doppler sólo es necesario distinguir si el transmisor y el receptor se alejan o se acercan para, según sea el caso, sumar o restar la velocidad.

Para saber si uno de los dos planetas se aleja o se acerca con respecto al otro se puede emplear el algoritmo que se muestra a continuación. Está basado en comprobar si en cada instante de tiempo la distancia entre ambos planetas crece o disminuye respecto al instante anterior manteniendo fijo a uno de los planetas.

SI $dist(P_{t+1}, Q_t) < dist(P_t, Q_t)$ ENTONCES
    P se acerca
SI NO
    P se aleja

donde dist(P, Q) es una función que devuelve la distancia entre ambos puntos y puede calcularse mediante la ecuación (42).

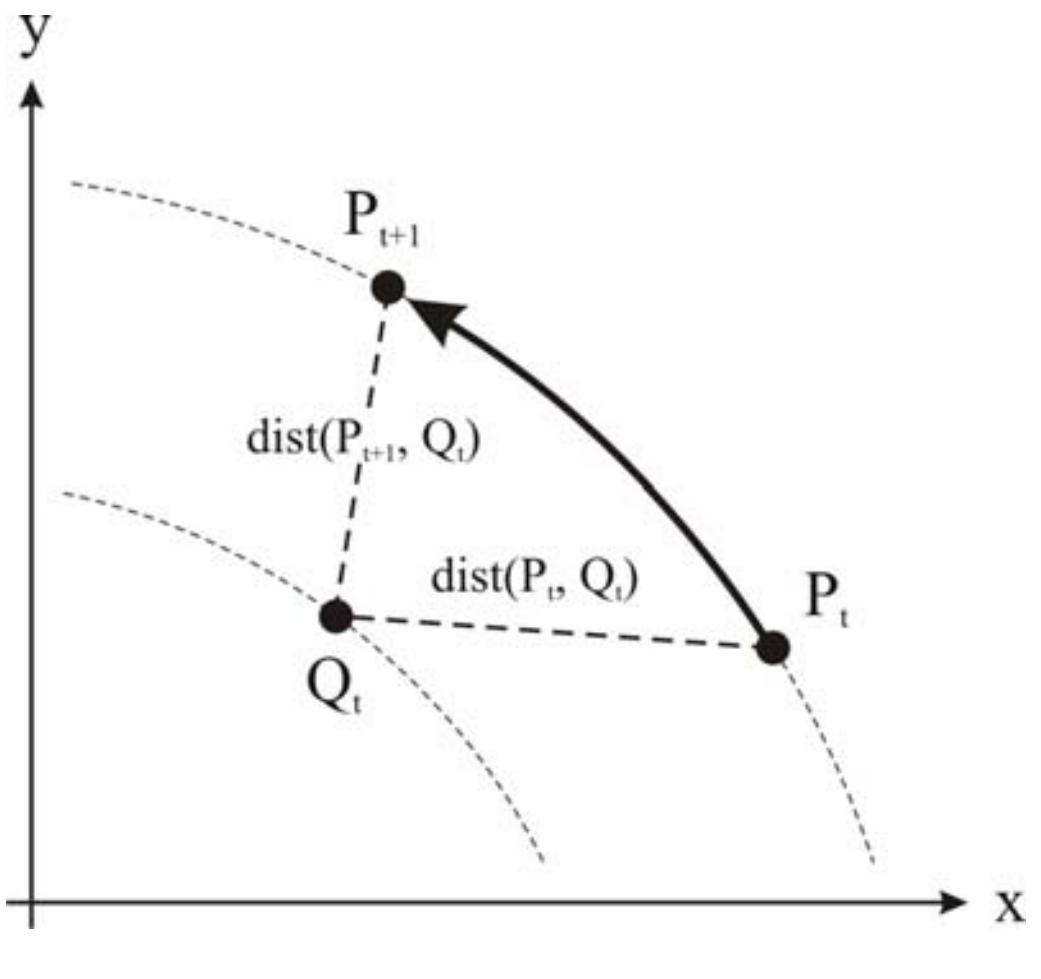

*Figura 2.13. Acercamiento radial de P hacia Q.*

## 2.3.7. Velocidad radial Marte-Tierra por rotación de la Tierra

En el apartado anterior se calculó el vector velocidad radial $\vec{v}_{rT}$ debido a la traslación de los planetas. Al vector $\vec{v}_{rT}$ de la Tierra es necesario sumarle otro vector velocidad radial $\vec{v}_{rR}$ debido a la rotación de la Tierra sobre su eje.

La velocidad angular de la Tierra debido a su rotación puede ser calculada, si se aproximan las secciones paralelas al ecuador a un círculo, como sigue

$$\omega_R = \frac{2\pi}{T} \qquad \text{(rad/s)} \qquad (51)$$

donde T representa el periodo de la Tierra en segundos. Por lo tanto, la velocidad de rotación de cualquier punto de la superficie viene dada por

$$v_R = \omega \cdot R \qquad \text{(m/s)} \qquad (52)$$

siendo R la distancia en metros desde el eje de rotación terrestre hasta el punto de la superficie considerado.

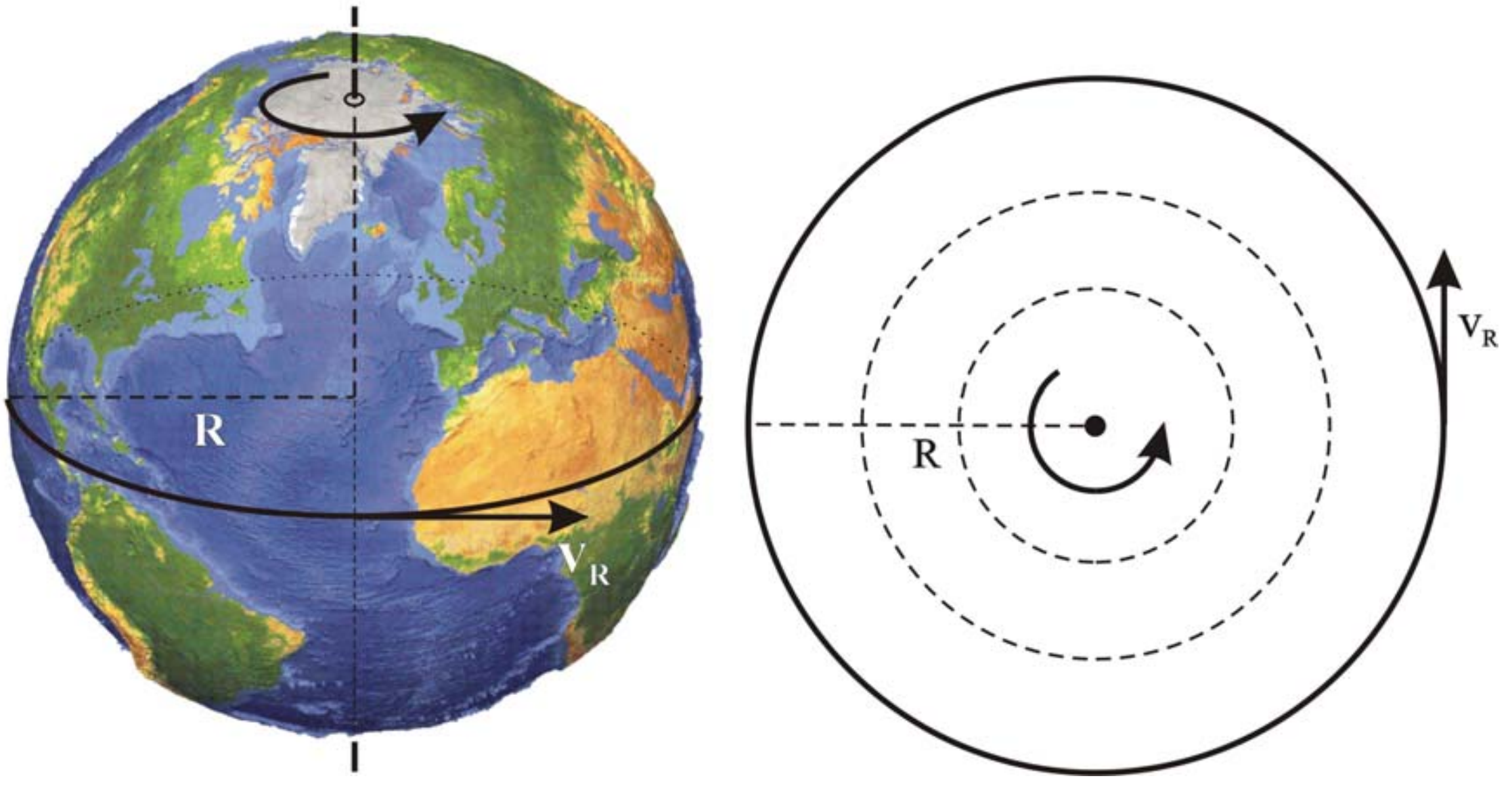

***Figura 2.14. Velocidad de rotación de la Tierra.***

### 2.3.7.a. Variación temporal

La variación temporal de la velocidad radial debida a la rotación es muy rápida en comparación con la de la velocidad radial debida a la traslación de los planetas ya que oscila entre valores máximos y mínimos cada 24 horas. Como puede verse en la Figura 2.15, la máxima velocidad radial por rotación diaria (*en valor absoluto*), se corresponde con el momento en que el vector $v_R$ (perpendicular a la línea que une cada punto con el centro de la Tierra) **se alinea exactamente** con la recta que marca la dirección hacia Marte. En cualquier otro momento la velocidad radial es menor, ya que el no estar alineadas, la velocidad efectiva sería una proyección vectorial de la misma, lo que hace disminuir el valor absoluto de la velocidad radial. La velocidad mínima coincide con el momento en que el vector $v_R$ es perpendicular a dicha dirección.

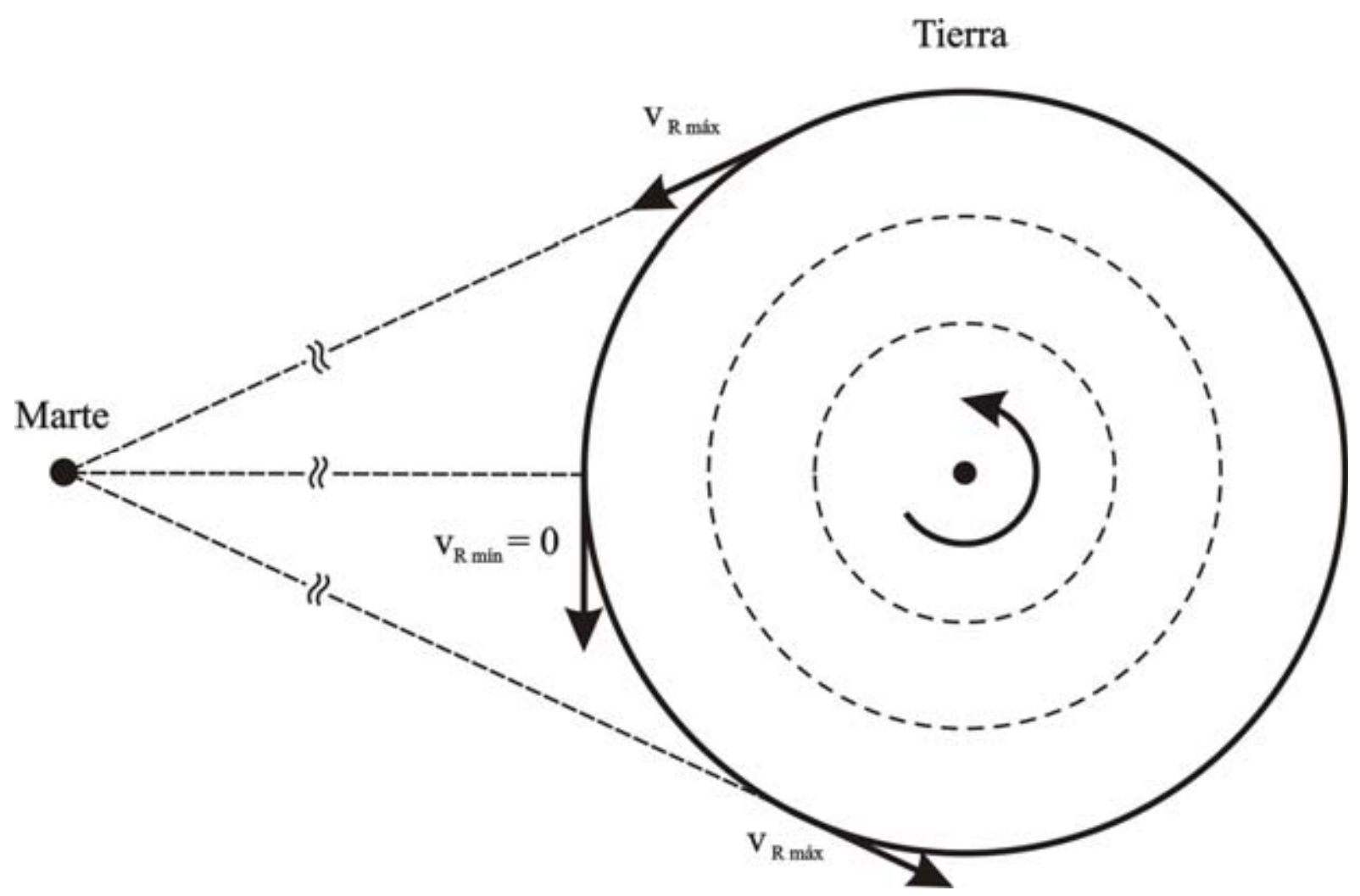

***Figura 2.15. Velocidad radial máxima y mínima por rotación de la Tierra.***

Para los propósitos de este proyecto sólo tiene utilidad conocer la ***velocidad radial máxima*** que coincide con la velocidad de rotación de la Tierra. Esto es así debido al hecho de que una vez al día se alcanza esta velocidad máxima en el sentido hacia el planeta y también en el contrario, por lo que, si la velocidad mínima (en valor absoluto) es cero, se asume que cada día se dan todos los demás valores intermedios.

## 2.3.7.b. Inclinación del eje de rotación

El eje sobre el que rota la Tierra no es paralelo al plano de la eclíptica. Posee una inclinación de 23,45 grados [21], que produce una influencia en la velocidad radial de la Tierra respecto a Marte. En este cálculo se obvian los 1,85 grados [21] de inclinación entre los planos de Marte y la Tierra por introducir un efecto mucho más pequeño que el calculado en este apartado.

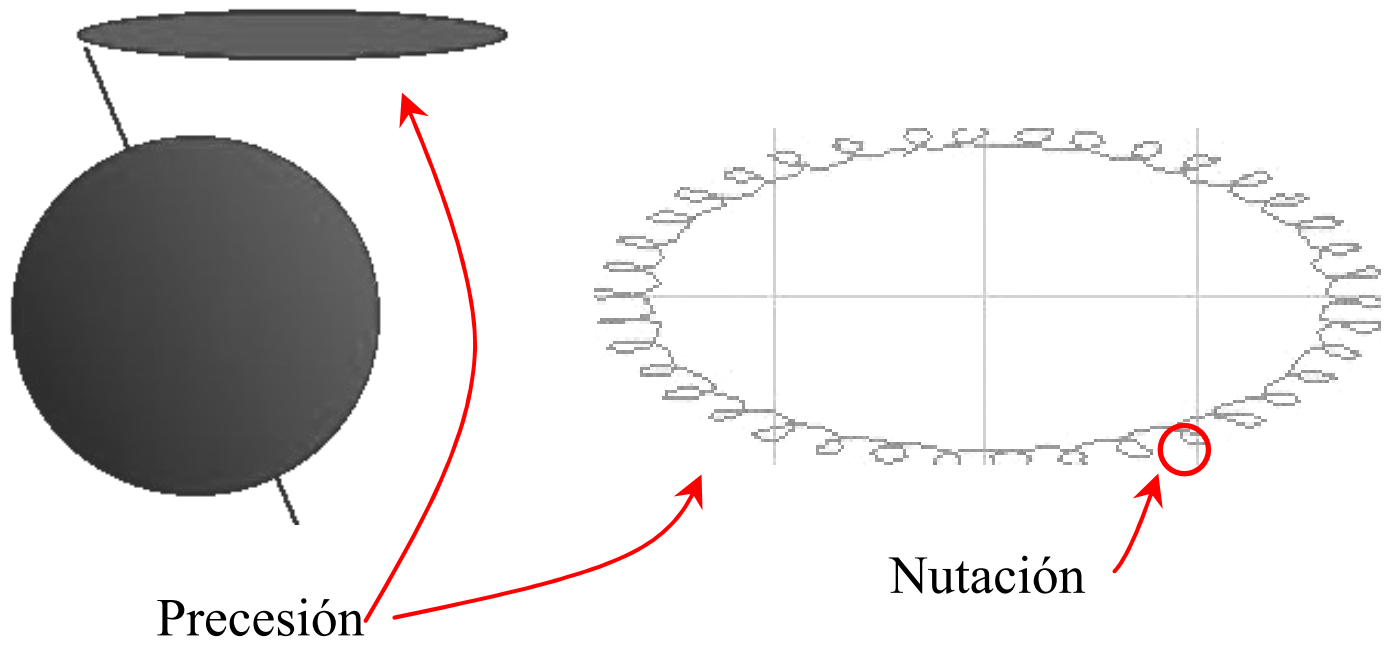

***Figura 2.16. Precesión y nutación del eje de la Tierra.***

Los efectos de la ***nutación*** y ***precesión*** del eje terrestre también se han despreciado. La nutación tiene un periodo de unos 19 años y la precesión de unos 26.000, por lo que en los menos de dos años considerados en la simulación no introducen alteraciones apreciables.

Existen dos situaciones extremas que sirven para evaluar el efecto de la inclinación del eje terrestre. Una forma muy útil para visualizarlas es imaginar que Marte es el Sol y pensar en términos de *equinoccios* y *solsticios*.

La situación en que el eje de rotación es tal que se alinea exactamente con la recta que une el centro de la Tierra y el centro de Marte se puede tratar como un ***solsticio*** de la Tierra con respecto a Marte. En esta situación la dirección de la velocidad de rotación de la Tierra, al ser paralela al ecuador, siempre tiene una inclinación con respecto a la eclíptica igual a la inclinación del eje de rotación. Por ello la velocidad radial es siempre la proyección de la velocidad de rotación tal como se observa en la Figura 2.17. La máxima velocidad radial, de la misma forma que en el apartado anterior, coincide con el momento en que el vector $v_R$ apunta en la dirección de Marte. Por lo tanto, teniendo en cuenta la proyección antes mencionada sobre esta velocidad de rotación, se puede calcular la velocidad radial máxima en las situaciones de solsticio respecto a Marte de la siguiente manera

$$v_{rR} = v_R \cos(23{,}45) = 0{,}92 \cdot v_R \tag{53}$$

En relación a la velocidad radial es indiferente si el polo norte apunta hacia Marte o lo hace el polo Sur, ya que únicamente tiene relevancia el valor absoluto de la velocidad máxima de rotación y éste será idéntico en ambos casos.

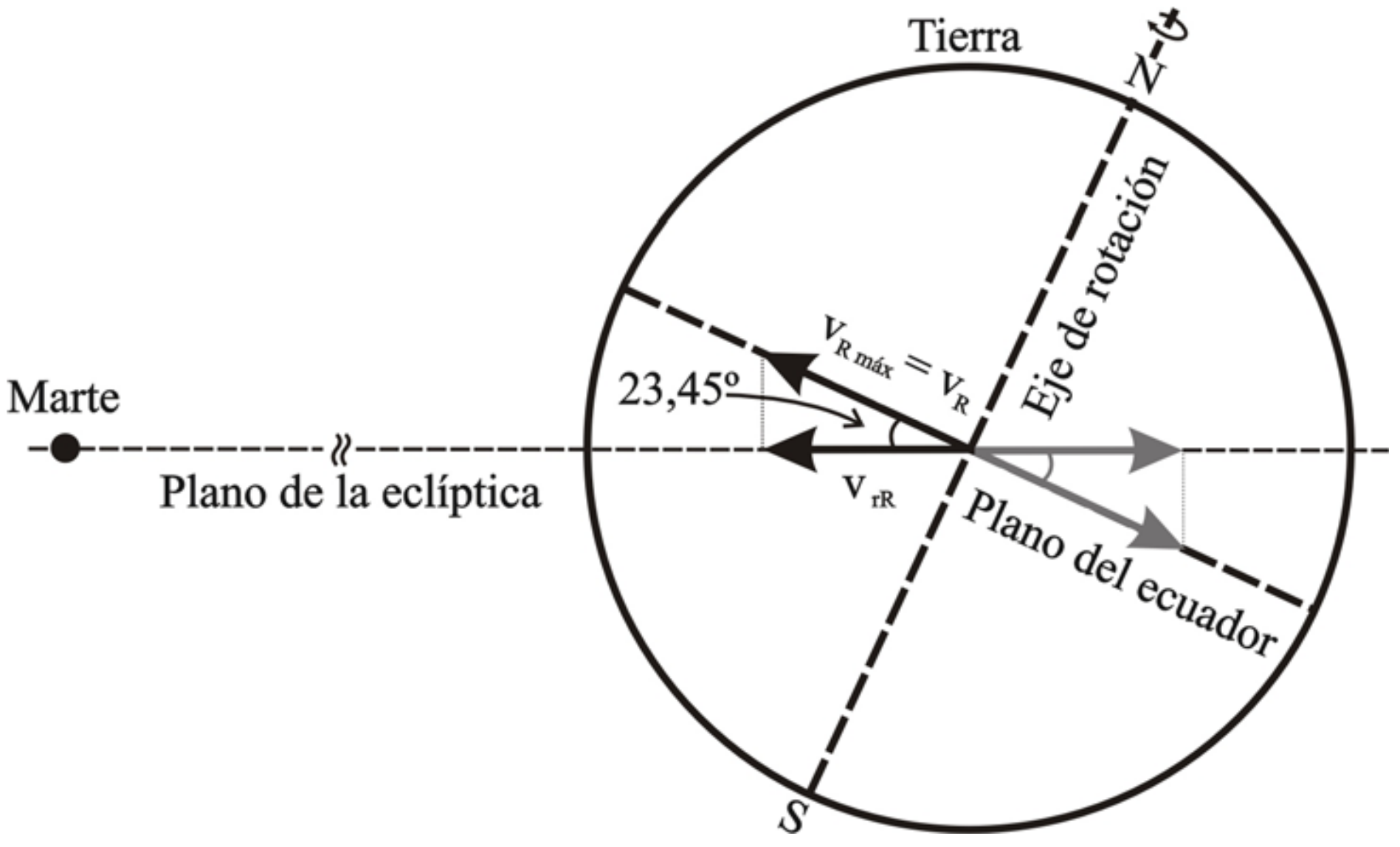

***Figura 2.17. Influencia del eje de rotación en la velocidad radial en los solsticios con Marte.***

Por su parte, la situación en que el eje terrestre hace que la distancia entre Marte y el polo norte y entre Marte y el polo sur sea la misma, se puede tratar como un ***equinoccio*** de la Tierra respecto a Marte. En este caso se comprueba que existen dos instantes durante cada día en que la velocidad radial máxima debida a la rotación de la Tierra coincide con la velocidad de rotación, sin necesidad de hacer ninguna proyección ya que la dirección es la misma.

Se deduce que, aunque en la primera situación tratada la velocidad radial es menor que la velocidad de rotación (menos de un uno por ciento), en la segunda situación ambas coinciden. Como ya se ha dicho, en este proyecto únicamente se considera la velocidad máxima de rotación terrestre en valor absoluto, por lo que la velocidad radial por rotación terrestre incluyendo el efecto de la inclinación del eje, será *idéntica a la velocidad de rotación* de la Tierra.

## 2.3.7.c. Influencia de la latitud

Como se observa en la ecuación (52), la velocidad de rotación $v_R$ depende de R. De ello se deduce que ubicaciones a diferentes latitudes tendrán diferentes rangos de velocidades ya que la distancia al eje de rotación varía con la latitud. Por lo tanto, los puntos sobre los polos tendrán la mínima velocidad en valor absoluto (nula, de hecho) y los puntos sobre el ecuador la máxima (en valor absoluto). Por supuesto, una vez considerado un emplazamiento determinado, este rango de velocidades será *constante*.

La Tierra está ligeramente achatada por los polos, por lo que el radio polar (6356,784 km) es menor que el radio ecuatorial (6378,188 km). Teniendo esto en cuenta, puede aproximarse la superficie de la Tierra a una elipse, cuya ecuación viene dada por

$$\frac{x^2}{a^2} + \frac{y^2}{b^2} = 1 \Rightarrow y = R_p \sqrt{1 - \frac{x^2}{R_e}} \qquad (54)$$

donde a es el semieje mayor de la elipse que coincide con el radio ecuatorial $R_e$ por ser mayor que el radio polar $R_p$ que es representado por b.

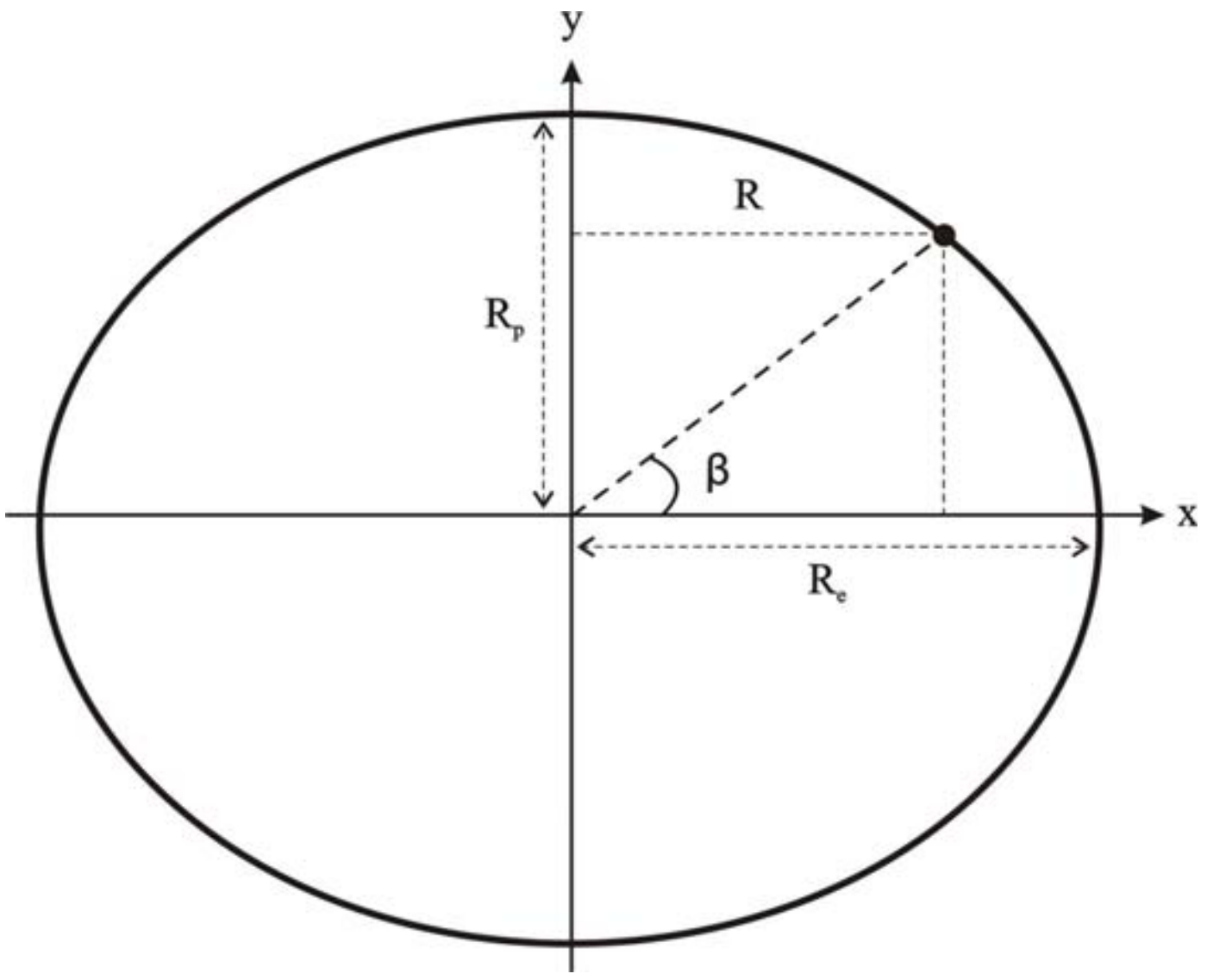

***Figura 2.18. Influencia de la latitud en la distancia al eje.***

Por otro lado, tal como se observa en la Figura 2.18, la ecuación de la recta que une el centro de la elipse con la superficie a una latitud β determinada es

$$y = m \cdot x + b = \tan(\beta) \cdot x \qquad (55)$$

ya que la pendiente m de la recta viene dada por la tangente del ángulo β y al no estar desplazada con respecto al centro, b es cero.

Si se igualan las ecuaciones (54) y (55) se obtiene despejando la coordenada x

$$x = R = \frac{R_p}{\sqrt{\tan^2(\beta) + \frac{R_p^2}{R_e^2}}} \quad (m) \qquad (56)$$

que coincide con la distancia R desde la superficie a la latitud β hasta el eje de rotación terrestre. Por lo tanto el efecto de la latitud en la velocidad radial por rotación terrestre queda contemplado empleando R, tal como se acaba de calcular, en la ecuación (52) para el cálculo de la velocidad de rotación.

Téngase en cuenta que la latitud β suele venir expresada en el sistema sexagesimal (grados, minutos y segundos), por lo que, para ser utilizada en la fórmula que se acaba de deducir, ha de ser convertida a centesimal de la siguiente forma

$$\text{latitud} = \text{grados} + \frac{\text{minutos}}{60} + \frac{\text{segundos}}{3600} \qquad \text{(grados centesimales)} \qquad (57)$$

## 2.3.8. Velocidad radial por rotación del satélite en Marte

La velocidad de rotación del planeta (Marte en este caso) no influye en el cálculo, ya que se supone que el emisor no está sobre su superficie, sino en órbita. Sin embargo, de la misma forma que se hizo en el apartado anterior con la Tierra, al vector velocidad radial $\vec{v}_{rT}$ de Marte debido a la traslación de los planetas es necesario sumarle otro vector velocidad radial $\vec{v}_{rR}$ debido a la **rotación del satélite** en su órbita alrededor de Marte.

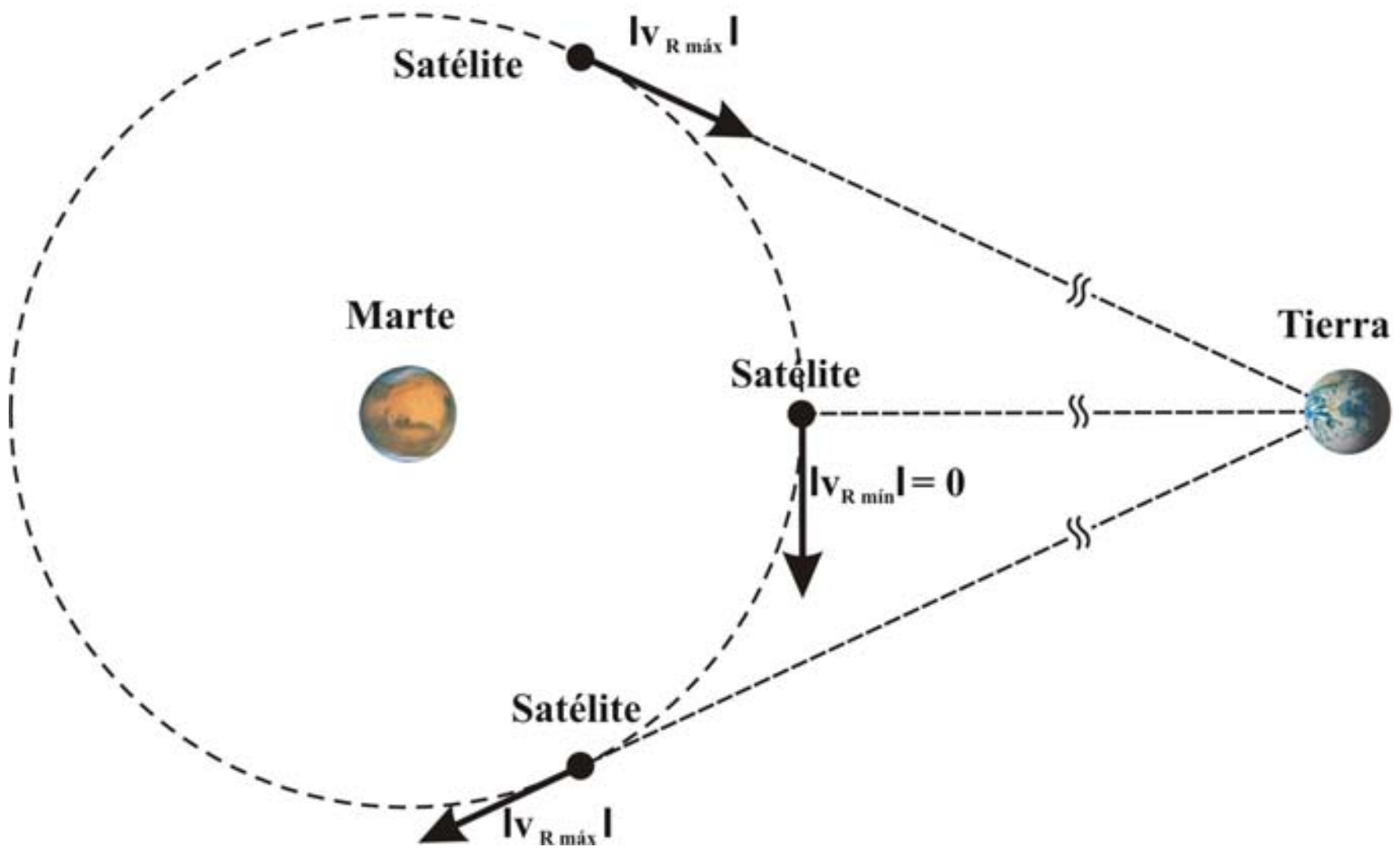

***Figura 2.19. Velocidad radial máxima y mínima por rotación del satélite de Marte.***

También de igual modo que en el caso de la Tierra, sólo va a considerarse la velocidad radial $\vec{v}_{rR}$ máxima, que va a ser la que determine (en ambos sentidos) la variación de velocidad en el tiempo. Esta velocidad máxima coincide con la velocidad tangencial de rotación del satélite que se calcula de la misma forma que en el apartado anterior con las ecuaciones (51) y (52). En este caso T será el periodo orbital del satélite y R la distancia de éste al centro de Marte.

## 2.3.9. Velocidad radial nave-Tierra durante el viaje a Marte

Como se explicó en el apartado 2.2.9, la trayectoria que sigue la nave durante el viaje Tierra-Marte viene descrita por una ***órbita de transferencia de Hohmann***, esto es, una elipse. Por lo tanto, el cálculo de la velocidad radial nave-Tierra será análogo al de la velocidad radial por traslación entre dos planetas, ya calculada en el apartado 2.3.6 y que, como se dedujo, puede calcularse con la ecuación (50). Para aplicar esta ecuación a la nave, falta por conocer su posición y su velocidad de traslación. Para ello se simula la órbita de Hohmann tal como sigue.

Inicialmente se calculan los parámetros orbitales que describen la forma de la órbita y el movimiento de la nave dentro de ella, como se muestra en la Figura 2.20.

El semieje mayor a puede calcularse como la mitad de la distancia entre la posición en que está la Tierra en el momento en que la nave inicia la órbita de Hohmann (esta posición se denominará $T_0$) y la posición en que se halla Marte en el momento de salir de dicha órbita (posición $M_0$).

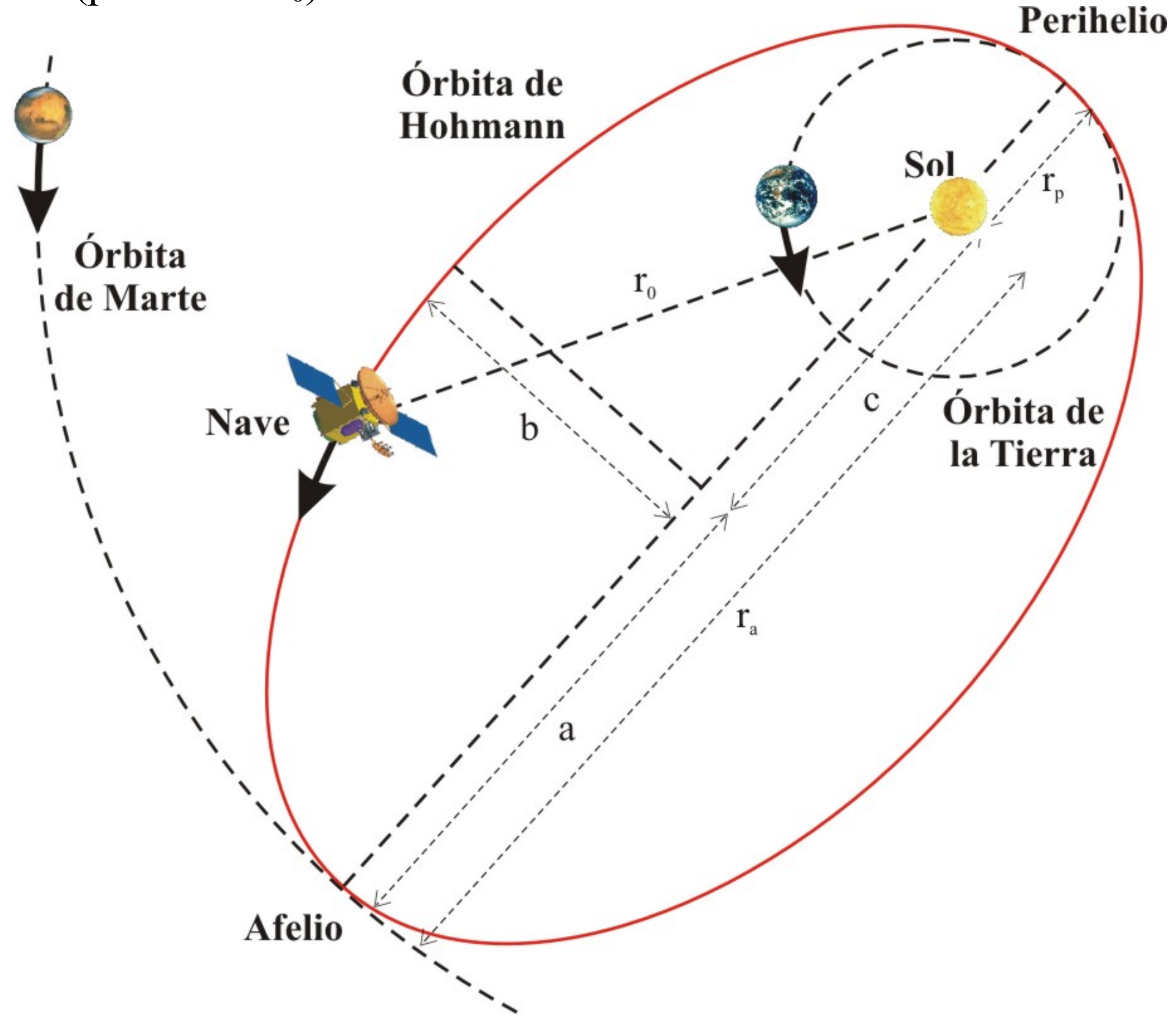

***Figura 2.20. Parámetros orbitales de la trayectoria de transferencia de Hohmann.***

La posición $M_0$ es conocida si se establece la posición $T_0$ de inserción en la órbita de Hohmann, ya que en ese momento *se produce una conjunción solar* entre T0 y M0. Por lo tanto el valor de a vendrá dado por

$$a = \frac{\text{dist}(T_0, M_0)}{2} = \frac{\sqrt{(T_{x0} - M_{x0})^2 + (T_{y0} - M_{y0})^2}}{2} \quad \text{(m)} \qquad (58)$$

Si la distancia c entre el centro de la elipse y el foco puede expresarse como

$$c = \text{dist}(M_0, \text{Sol}) - a = \sqrt{M_x^{\ 2} + M_y^{\ 2}} - \frac{\sqrt{(T_{x0} - M_{x0})^2 + (T_{y0} - M_{y0})^2}}{2} \quad \text{(m)} \qquad (59)$$

entonces la excentricidad e se puede calcular en función de a y c tal como se muestra a continuación

$$e = \frac{c}{a} = 2\sqrt{\frac{M_x{}^2 + M_y{}^2}{(T_{x0} - M_{x0})^2 + (T_{y0} - M_{y0})^2}} - 1 \qquad (60)$$

Conocido a, el periodo T se puede calcular mediante la fórmula (24) del periodo de una órbita elíptica. La duración del viaje Tierra-Marte será la mitad de ese periodo.

La anomalía media M inicialmente será cero por coincidir el punto de partida de la nave con el perihelio de la órbita. Al valor de la anomalía media cada día se le sumará el valor del movimiento angular diario, calculado mediante la ecuación (34) empleando el valor de periodo T anteriormente calculado. De esta forma, el valor de la anomalía media, al igual que el de la anomalía excéntrica y verdadera, variarán durante todo el viaje desde 0 radianes al comienzo del mismo hasta $\pi$ radianes al final.

La inclinación de la órbita de Hohmann se considera cero y los parámetros de longitud del nodo ascendente $\Omega$ y argumento del perihelio $\omega$ son ajustados combinadamente de forma que se consiga la conexión, mediante la órbita de Hohmann descrita, entre las órbitas de la Tierra y de Marte.

Una vez simulada la órbita de Hohmann, ya se tienen todos los datos necesarios para calcular la velocidad radial nave-Tierra. Estos datos, como se vio al principio de este apartado, eran la **posición de la nave** en cualquier momento del viaje y la **velocidad de traslación instantánea** (que puede calcularse mediante la fórmula (21), en función de la distancia r al Sol).

## 2.4. RESULTADOS DEL SIMULADOR

A continuación se presentan varias muestras de los resultados obtenidos con el simulador de dinámica orbital desarrollado para este proyecto, así como una verificación del mismo mediante su comparación con un simulador comercial similar.

### 2.4.1. Órbitas de la Tierra y Marte

Uno de los dos objetivos para los que se implementó el simulador de dinámica orbital era la obtención de la ***distancia instantánea*** entre la Tierra y Marte. Para ello, dado que los planos orbitales en los que se mueve cada planeta no son coincidentes, sino que están inclinados, ha sido necesario, tal como se ha visto en el apartado 2.3.5, que la simulación contara con las tres dimensiones espaciales. A continuación se presenta el resultado obtenido con el simulador para las órbitas de la Tierra y Marte en distintos planos.

En la Figura 2.21 se muestran las órbitas de Marte y de la Tierra representadas en el plano x-y. Las unidades de ambos ejes vienen expresadas en UA. Este es el plano que **mayor información** aporta debido a que el plano de la órbita de Marte únicamente tiene una inclinación de 1,82 grados respecto al de la Tierra, que coincide con el de la eclíptica. En la Figura 2.22 se muestran otra vez las órbitas de Marte y de la Tierra representadas esta vez en el plano x-z. Nótese que el rango de ambos ejes (también en UA) es diferente. Se ha representado de esta manera con el objetivo de mostrar la pequeña inclinación entre los planos de ambas órbitas.

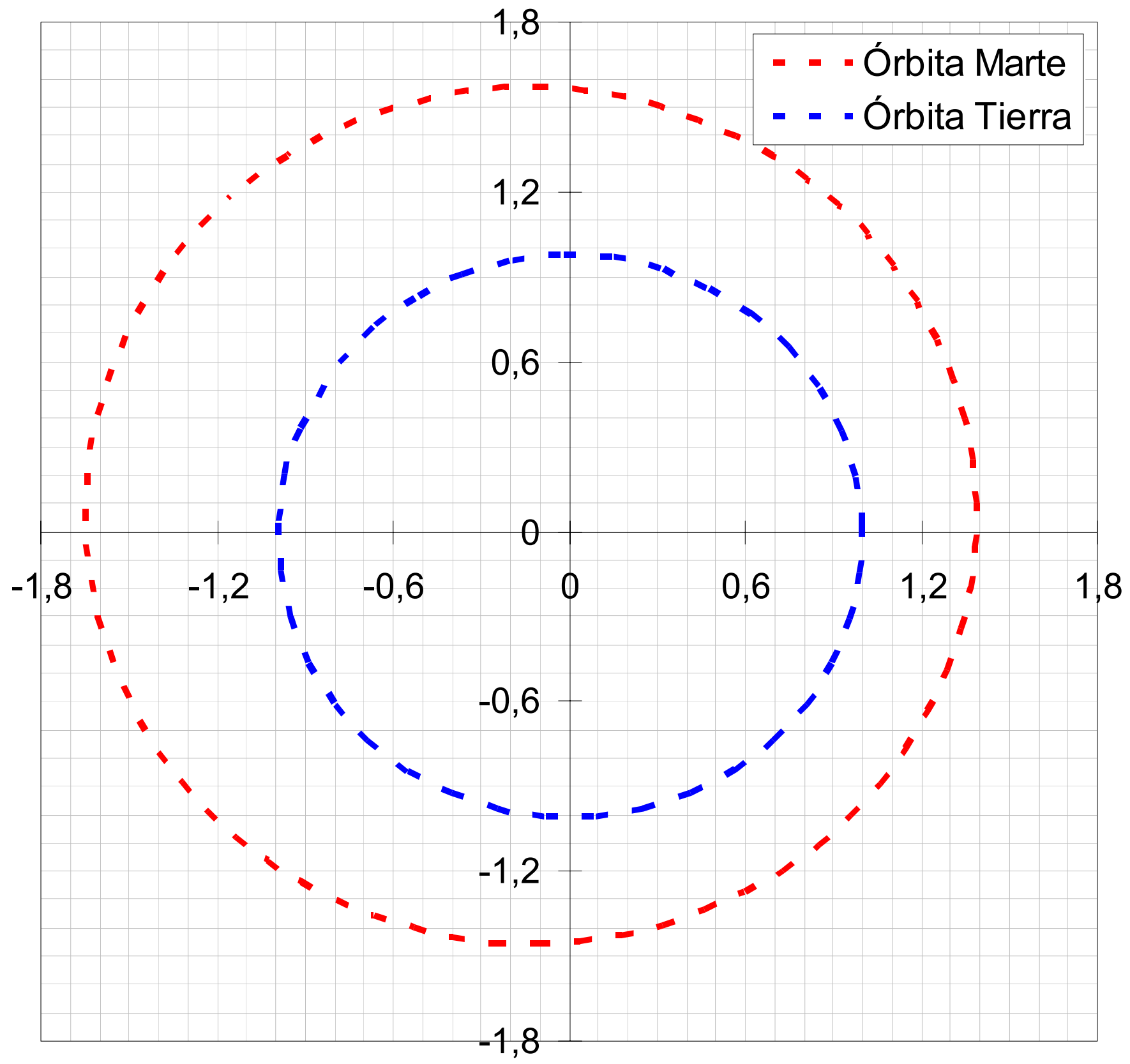

*Figura 2.21. Órbitas de Marte y la Tierra en el plano x-y.*

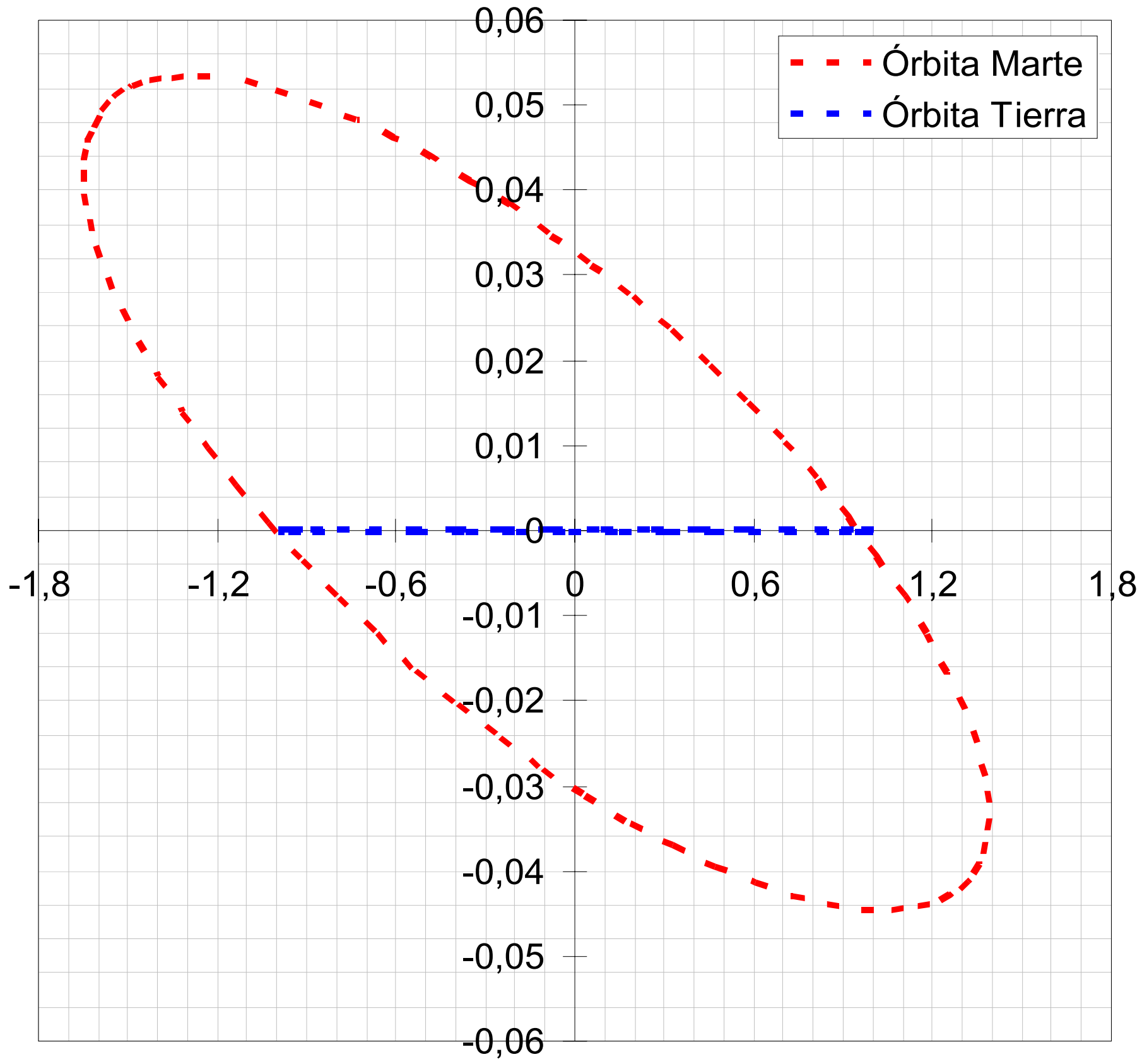

*Figura 2.22. Órbitas de Marte y la Tierra en el plano x-z.*

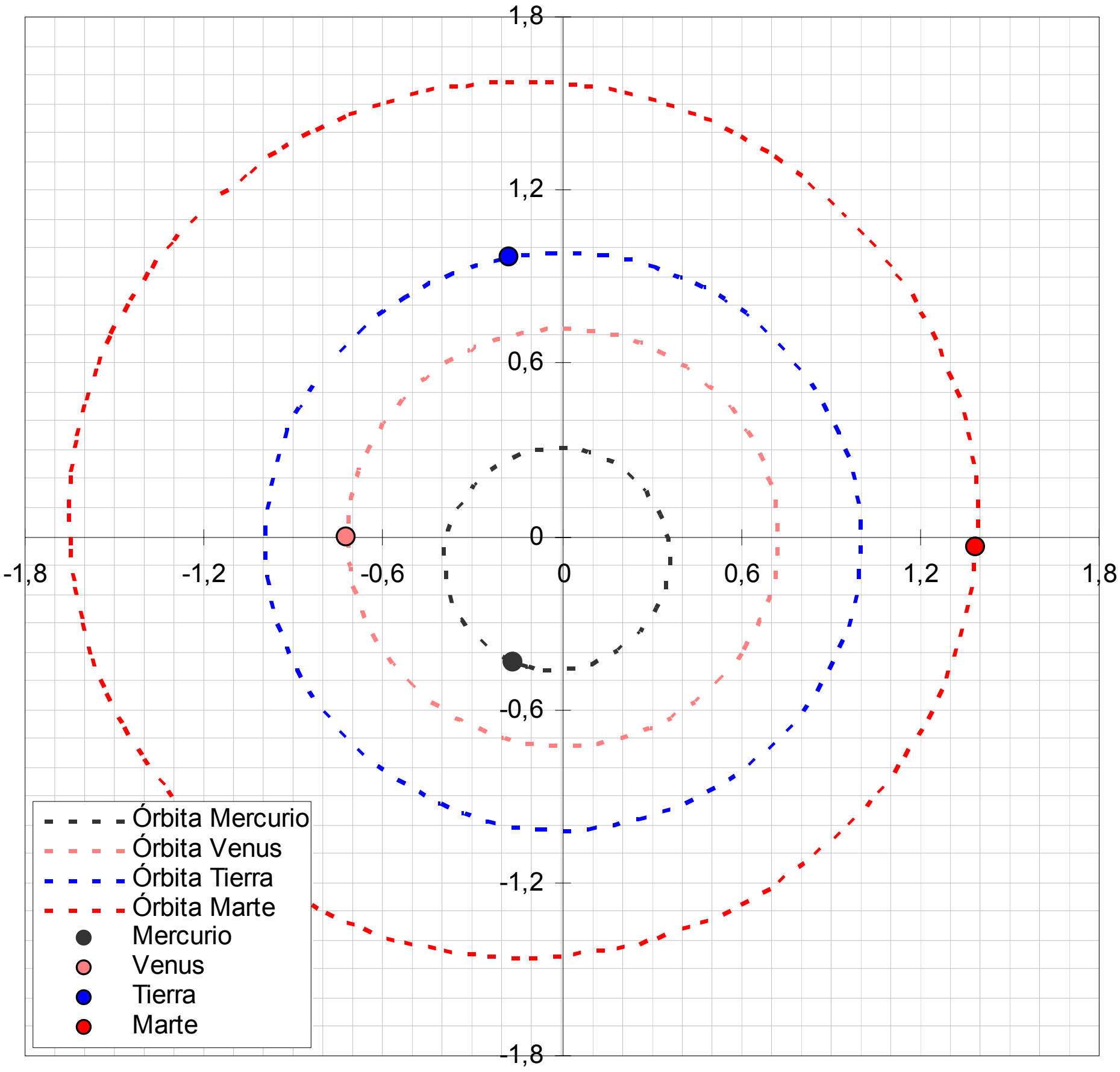

*Figura 2.23. Órbitas de los planetas interiores del sistema solar.*

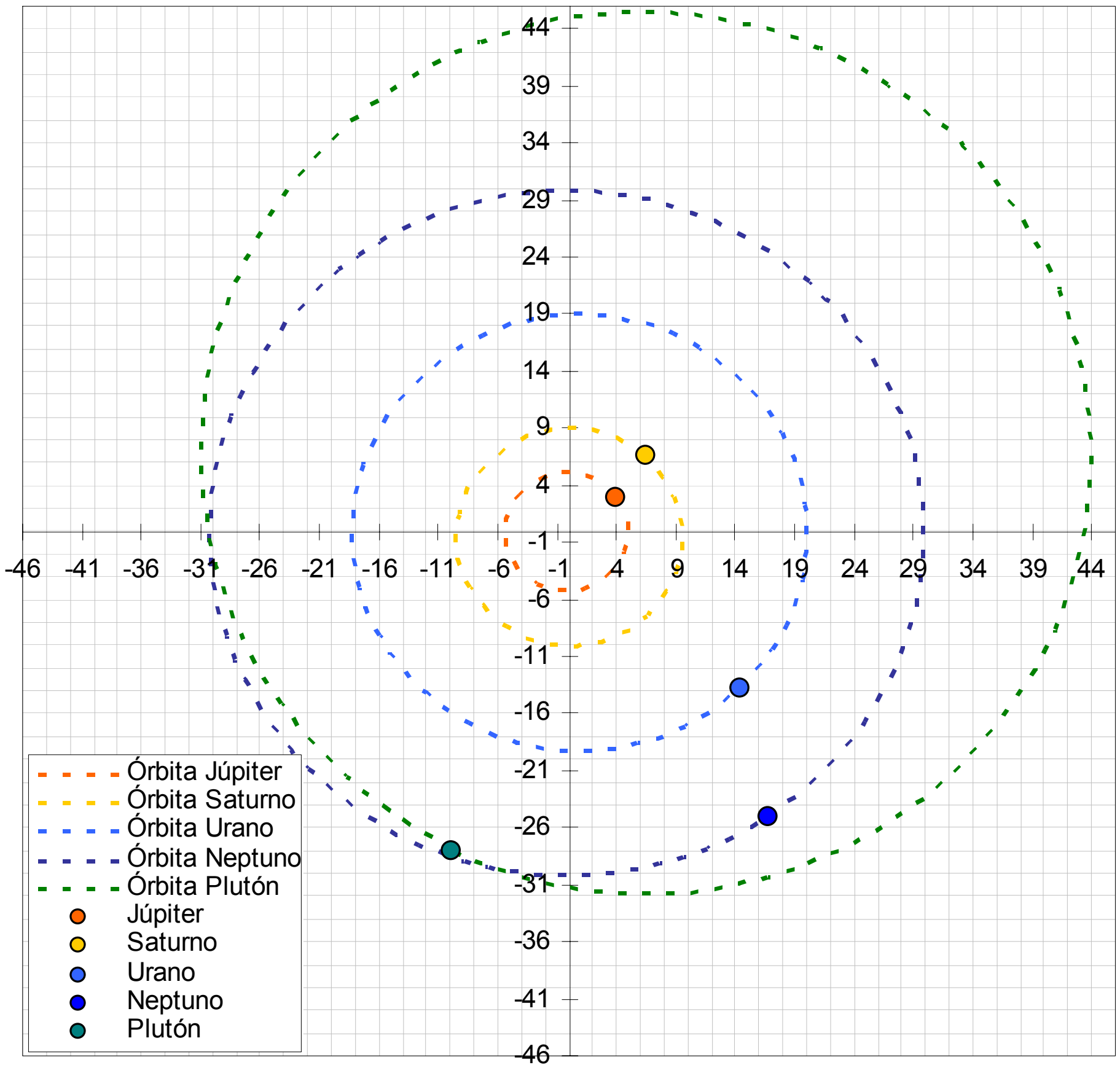

*Figura 2.24. Órbitas de los planetas exteriores del sistema solar.*

## 2.4.2. Órbitas de los planetas del sistema solar

Mediante el simulador de dinámica orbital no sólo se puede calcular la posición instantánea de Marte y la Tierra, también se han simulado las órbitas de todos los planetas del sistema solar, con el objetivo de calcular las distancias máximas y mínimas que se presentan en el apartado 3.1.3. A continuación se presenta una muestra gráfica de dicha simulación. En la Figura 2.23 se pueden observar las órbitas de los planetas interiores del sistema solar: **Mercurio**, **Venus**, **Tierra** y **Marte**. También se representan los planetas en las posiciones que corresponden al 1 de enero de 2.000.

En la Figura 2.24 se presentan las órbitas de los planetas exteriores del sistema solar: **Júpiter**, **Saturno**, **Urano**, **Neptuno** y **Plutón**, junto a la posición de los planetas correspondientes también el 1 de enero de 2.000.

## 2.4.3. Órbita de Hohmann de la nave MTO

La ***órbita de transferencia*** de un satélite puede también calcularse con el simulador, ofreciendo resultados de posiciones y distancias con el transcurso del tiempo. En la se presenta una gráfica de la distancia alcanzada por nave *MTO* del proyecto *MLCD* desde su lanzamiento, previsto para el 13 de Octubre de 2.009, hasta su inserción en órbita marciana, a finales de Agosto de 2.010. La trayectoria en **órbita de Hohmann** comienza a finales del mes de Octubre; durante el tiempo anterior, el satélite pasa por sucesivas órbitas terrestres elípticas.

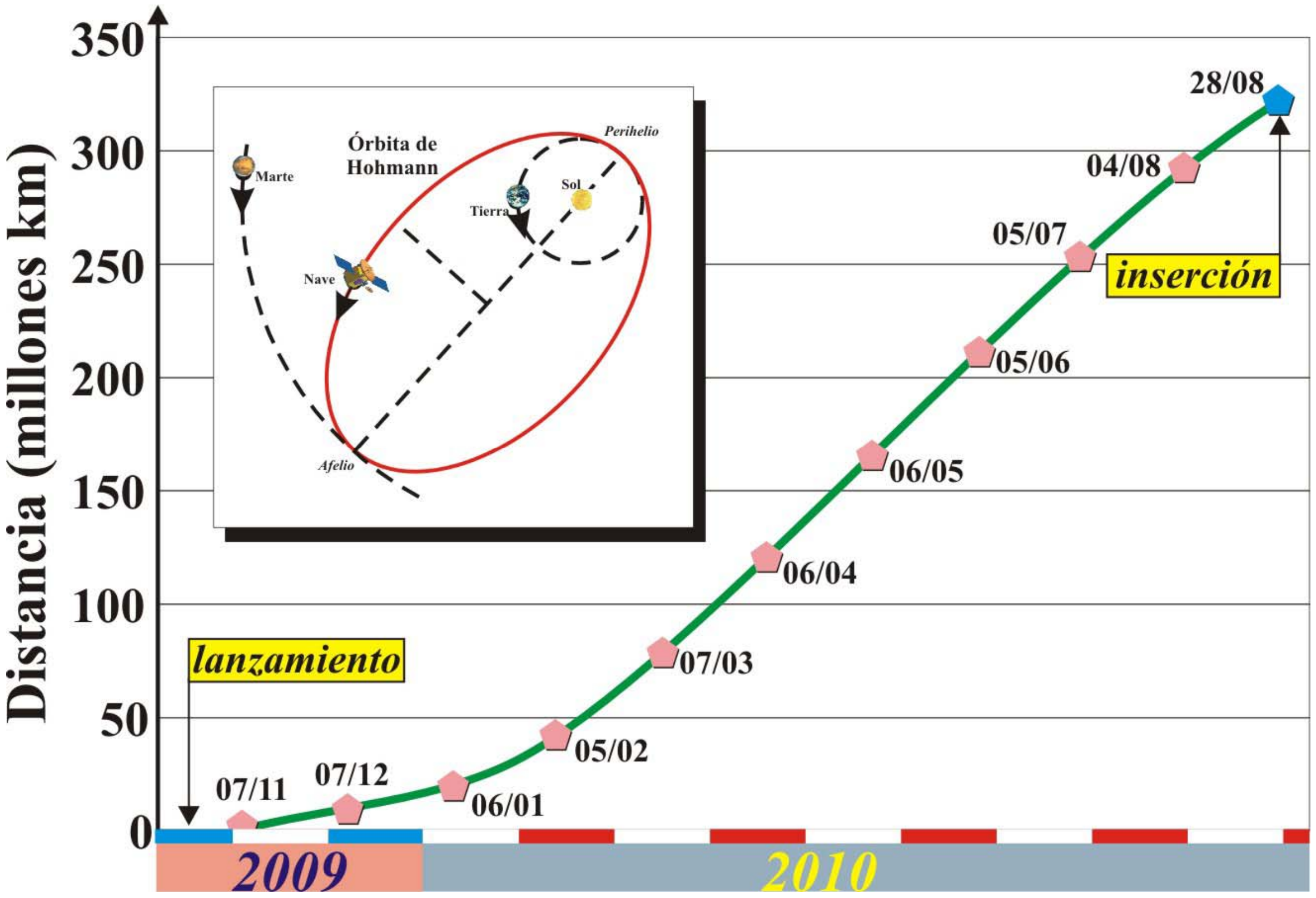

***Figura 2.25. Cálculo de distancias Tierra-satélite cada 30 días durante el viaje Tierra-Marte por órbita de Hohmann.***

### 2.4.4. Variación de las posiciones planetarias con el tiempo

Como se explicó en el apartado 2.3.3, el simulador de dinámica orbital también es capaz de realizar ***animaciones*** de los planetas en su movimiento orbital. El aumento del tiempo hace variar con él las anomalías medias de forma que la posición de los planetas cambia con la fecha de cada instante. También es posible introducir la fecha para la cual se quiere conocer la posición o a partir de la cual se desea realizar la animación. En la Tabla 2.2 se presentan los valores obtenidos (cada 30 días) al simular un periodo de dos años desde el 28 de agosto de 2.010, fecha prevista para el comienzo de la maniobra de inserción en órbita de Marte.

| | Marte | | | | Tierra | | | |
|---|---|---|---|---|---|---|---|---|
| Fecha | x (UA) | y (UA) | z (UA) | Velocidad (m/s) | x (UA) | y (UA) | z (UA) | Velocidad (m/s) |
| 28/08/2010 | -1,10 | -1,11 | 0,00 | 23514,92 | 0,92 | -0,42 | 0,00 | 29482,56 |
| 27/09/2010 | -0,77 | -1,32 | -0,01 | 24088,83 | 1,00 | 0,06 | 0,00 | 29706,46 |
| 27/10/2010 | -0,37 | -1,44 | -0,02 | 24714,71 | 0,83 | 0,54 | 0,00 | 29959,73 |
| 26/11/2010 | 0,07 | -1,45 | -0,03 | 25325,50 | 0,44 | 0,88 | 0,00 | 30168,89 |
| 26/12/2010 | 0,49 | -1,33 | -0,04 | 25862,43 | -0,07 | 0,98 | 0,00 | 30275,18 |
| 25/01/2011 | 0,87 | -1,09 | -0,04 | 26262,81 | -0,56 | 0,81 | 0,00 | 30247,46 |
| 24/02/2011 | 1,16 | -0,75 | -0,04 | 26468,79 | -0,89 | 0,44 | 0,00 | 30100,69 |
| 26/03/2011 | 1,35 | -0,29 | -0,04 | 26450,71 | -0,99 | -0,12 | 0,00 | 29841,96 |
| 25/04/2011 | 1,39 | 0,15 | -0,03 | 26215,49 | -0,82 | -0,58 | 0,00 | 29599,14 |
| 25/05/2011 | 1,30 | 0,59 | -0,02 | 25791,58 | -0,44 | -0,91 | 0,00 | 29397,44 |
| 24/06/2011 | 1,08 | 0,97 | -0,01 | 25240,03 | 0,05 | -1,02 | 0,00 | 29293,83 |
| 24/07/2011 | 0,78 | 1,28 | 0,01 | 24623,61 | 0,53 | -0,87 | 0,00 | 29312,13 |
| 23/08/2011 | 0,41 | 1,48 | 0,02 | 23999,83 | 0,88 | -0,50 | 0,00 | 29448,20 |
| 22/09/2011 | 0,02 | 1,57 | 0,03 | 23433,67 | 1,00 | -0,03 | 0,00 | 29662,14 |
| 22/10/2011 | -0,39 | 1,55 | 0,04 | 22904,30 | 0,87 | 0,48 | 0,00 | 29924,47 |
| 21/11/2011 | -0,75 | 1,44 | 0,05 | 22503,22 | 0,52 | 0,84 | 0,00 | 30137,94 |
| 21/12/2011 | -1,09 | 1,24 | 0,05 | 22191,88 | 0,01 | 0,98 | 0,00 | 30267,83 |
| 20/01/2012 | -1,35 | 0,97 | 0,05 | 22019,91 | -0,48 | 0,86 | 0,00 | 30262,30 |
| 19/02/2012 | -1,53 | 0,65 | 0,05 | 21968,11 | -0,84 | 0,52 | 0,00 | 30134,92 |
| 20/03/2012 | -1,64 | 0,26 | 0,05 | 22051,94 | -1,00 | -0,01 | 0,00 | 29895,39 |
| 19/04/2012 | -1,64 | -0,11 | 0,04 | 22253,52 | -0,88 | -0,49 | 0,00 | 29649,89 |
| 19/05/2012 | -1,54 | -0,50 | 0,03 | 22594,52 | -0,52 | -0,87 | 0,00 | 29427,10 |
| 18/06/2012 | -1,36 | -0,83 | 0,02 | 23020,19 | -0,06 | -1,01 | 0,00 | 29305,89 |
| 18/07/2012 | -1,07 | -1,13 | 0,00 | 23570,62 | 0,45 | -0,91 | 0,00 | 29299,83 |

***Tabla 2.2. Posición tridimensional heliocéntrica y velocidad de traslación de la Tierra y de Marte durante los dos años de proyecto MLCD.***

Realizando la misma simulación con intervalos de días en lugar de meses se pueden comparar los valores de velocidad de traslación obtenidos por el simulador de dinámica orbital con los valores de la base de datos planetaria *National Space Science Data Center* [21] de la *NASA*. En la Tabla 2.3 se muestra el resultado de esta comparación.

| | National Space Science Data Center | | Simulador | |
|---|---|---|---|---|
| | Marte | Tierra | Marte | Tierra |
| *Velocidad orbital máx. (km/s)* | 26.50 | 30.29 | 26,4948 | 30,2874 |
| *Velocidad orbital mín. (km/s)* | 21.97 | 29.29 | 21,9678 | 29,286 |

***Tabla 2.3. Comparación entre la velocidad orbital obtenida por el simulador y los datos proporcionados por el National Space Science Data Center de la NASA.***

Ambos resultados son idénticos si los obtenidos por el simulador se expresan con el número de decimales empleado en los datos de la *Nacional Space Science Data Center*. Una comparación más precisa puede verse en el siguiente apartado.

## 2.4.5. Comparación con otro simulador comercial

Para la verificación final de los resultados obtenidos por el simulador se ha efectuado una comparación con otro programa comercial dedicado a la simulación de órbitas de planetas, cometas, etc. en el sistema solar. Este programa se llama *Planet's Orbits* [24] de *Alcycone Astronomical Software* y la versión utilizada para la prueba ha sido la 1.6.1.

A continuación se presenta una comparación entre los resultados obtenidos por ambos programas para la misma simulación. Se ha introducido como fecha la correspondiente al inicio del sistema **J2000**, esto es, el día juliano 2451545 o el 1 de enero de 2.000 a las 12:00 UT. En la Figura 2.26 se observa la representación gráfica bidimensional (ejes x-y) obtenida por el programa *Planet's orbits* y en la Figura 2.27 la equivalente obtenida con el simulador.

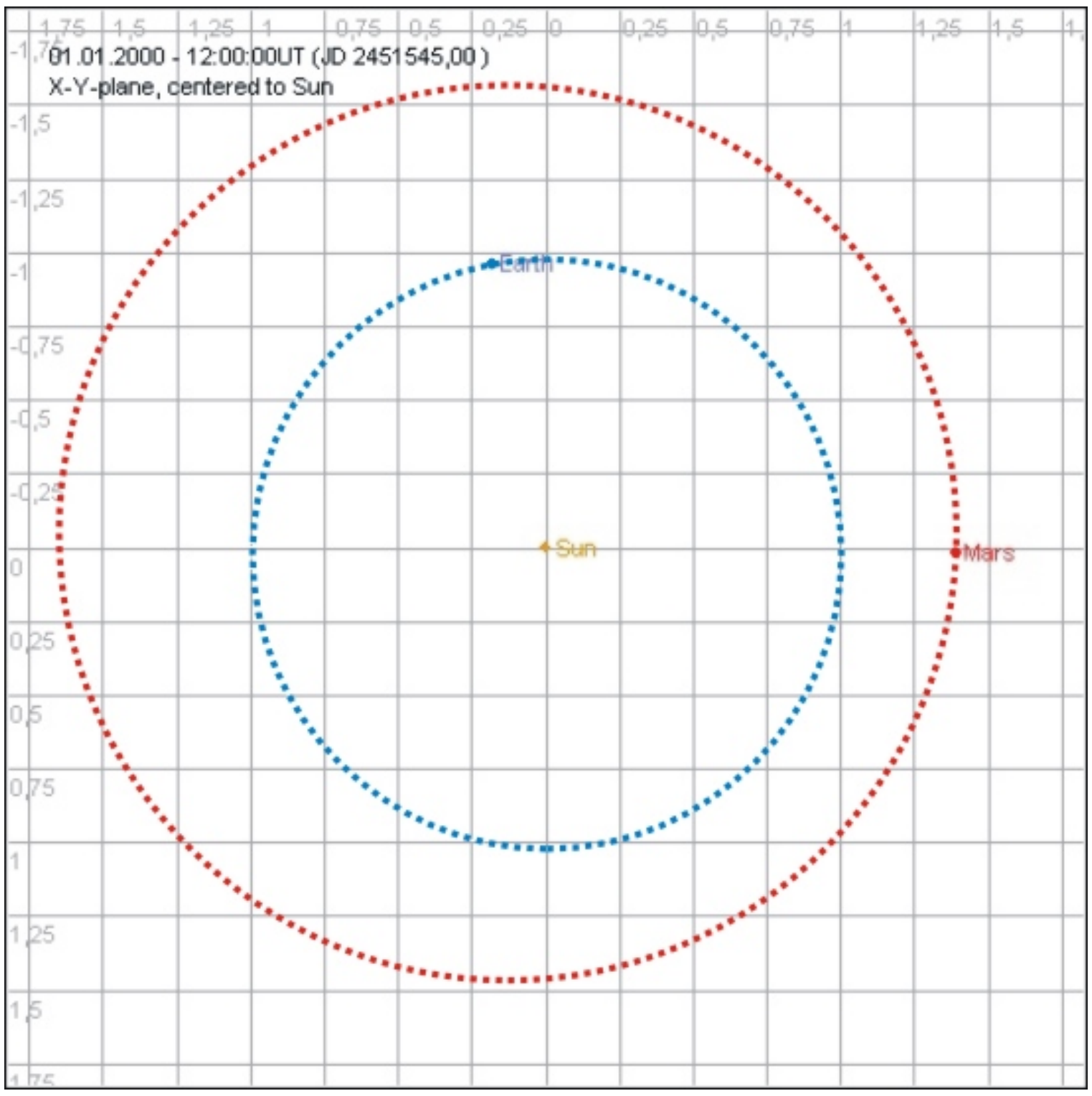

***Figura 2.26. Órbita y posición de la Tierra y de Marte el 1-1-2.000 12:00 (día juliano 2451545) en el plano x-y obtenida por el programa comercial Planet's Orbits V1.6.1.***

| | Planet's Orbits V1.6.1. | | Simulador | |
|---|---|---|---|---|
| | Tierra | Marte | Tierra | Marte |
| *Coordenada x (UA)* | -0,177147 | 1,390716 | -0,177162 | 1,390623 |
| *Coordenada y (UA)* | 0,967239 | -0,013405 | 0,967215 | -0,013100 |
| *Coordenada z (UA)* | 0,000004 | -0,034467 | 0,000001 | -0,034481 |
| *Distancia Tierra-Marte (UA)* | 1,849607 | | 1,849366 | |

***Tabla 2.4. Comparación del resultado de la simulación correspondiente al 1-1-2000 12:00 (día juliano 2451545) en el plano x-y en ambos programas.***

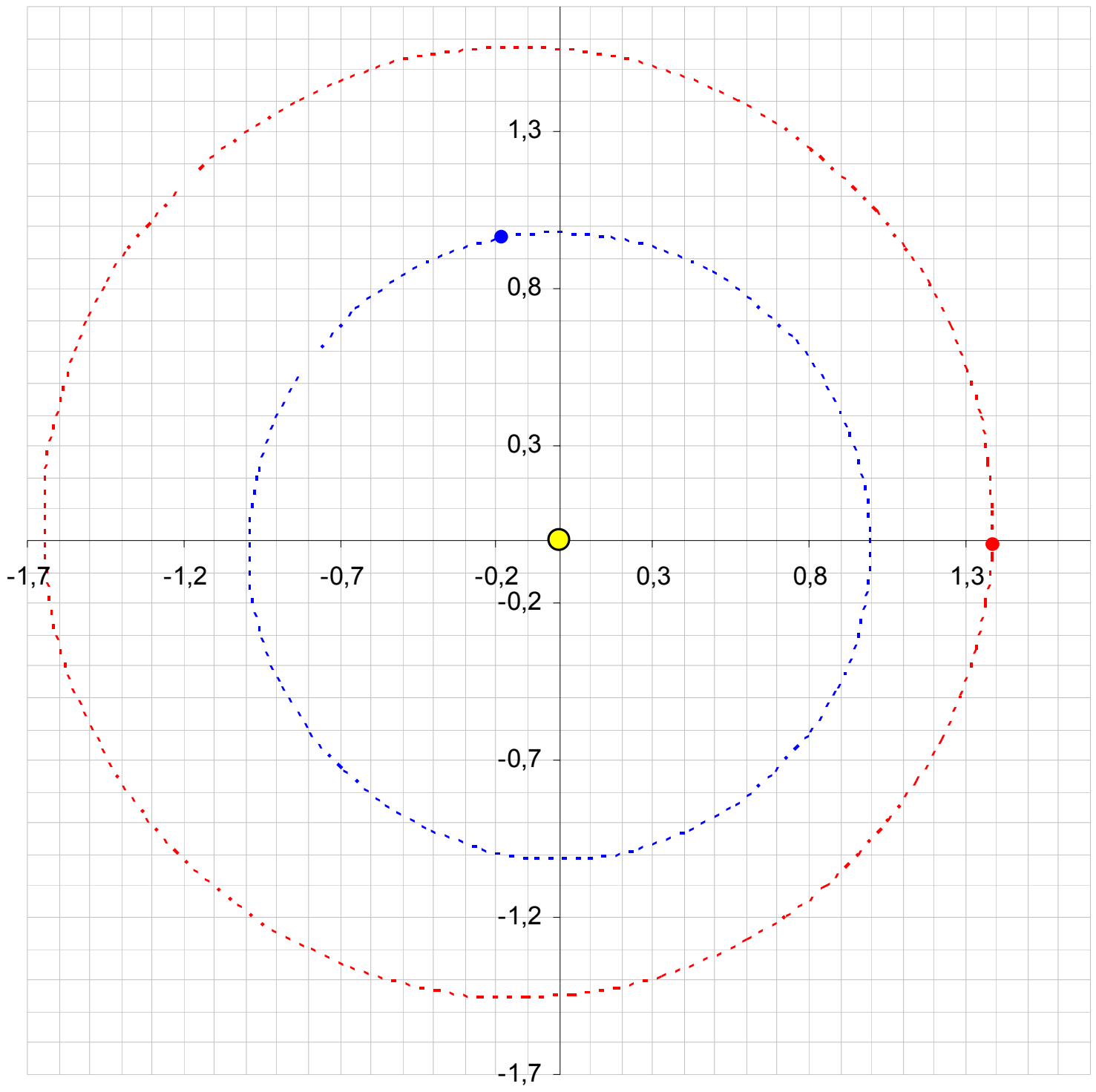

***Figura 2.27. Órbita y posición de la Tierra y de Marte el 1-1-2.000 12:00 (día juliano 2451545) en el plano x-y obtenida por el simulador realizado para este proyecto.***

En la Tabla 2.4 se presenta la comparación numérica entre los resultados de ambos programas para la simulación antes mencionada.

Como se observa, la diferencia entre las distancias Tierra-Marte obtenidas por cada uno de los simuladores es de 1,849607 - 1,849366 = 0,000241 unidades astronómicas (es decir, un 0,013 %). Para evaluar el efecto que esta discrepancia puede tener en el enlace de comunicación conviene conocer esta diferencia en términos de *pérdidas en el balance de potencia*. En este caso, empleando la fórmula (65) de las pérdidas por espacio libre calculada en el apartado 3.1.2 puede obtenerse fácilmente la diferencia de la siguiente manera

$$\Delta_L = 10\log_{10}\left(\frac{\lambda}{4\pi d_i}\right)^2 - 10\log_{10}\left(\frac{\lambda}{4\pi d_j}\right)^2 = 10\log_{10}\left(\frac{d_j}{d_i}\right)^2 \quad \text{(dB)} \qquad (61)$$

donde $d_i$ y $d_j$ son las diferentes distancias Tierra-Marte en unidades naturales y $\Delta_L$ es la diferencia de pérdidas por espacio libre usando una u otra distancia. Si se introducen los datos de distancia Tierra-Marte de la Tabla 2.4 en la ecuación (61) se obtiene una diferencia de 0,001 dB de pérdidas debidas a espacio libre entre usar una distancia u otra. Frente a pérdidas por espacio libre del orden de 360-370 dB, puede perfectamente despreciarse este error. Se concluye pues que la precisión del simulador cumple sobradamente los requisitos necesarios para el empleo adecuado de sus resultados en este proyecto.

# 3. El canal de transmisión

Las comunicaciones ópticas en espacio profundo representan una interesante alternativa a las comunicaciones por radiofrecuencia al uso. La principal ventaja estriba en las pérdidas inducidas por la ***divergencia del haz***, que dependen directamente de la **longitud de onda**. Las longitudes de onda empleadas en comunicaciones ópticas (dentro del infrarrojo próximo), son varios órdenes de magnitud inferiores a las empleadas habitualmente en comunicaciones por satélite. La menor atenuación derivada de esta diferencia puede traducirse, potencialmente, en **regimenes binarios más elevados** (y en consecuencia más adecuados) para la transmisión de los grandes volúmenes de información que se requieren en las misiones de exploración en espacio profundo. Se dispone actualmente de fuentes láser de gran potencia y alto rendimiento, fiables y resistentes, que pueden emplearse en el emisor remoto. Además, su longitud de onda puede seleccionarse entre un amplio rango de posibilidades. Finalmente, las técnicas de **apuntamiento** son asimismo precisas y seguras, extremo de gran importancia al tratarse de radiación altamente colimada.

En un enlace de comunicaciones en el espacio profundo, el canal que reviste mayor importancia es el **canal de bajada** (*downlink*), basado en un emisor remoto situado en el planeta o satélite que se está explorando, y un receptor terrestre. La utilidad del enlace está determinada por la capacidad y fiabilidad de este enlace. El **canal de subida** (*uplink*) operará con una tasa binaria sustancialmente menor, por lo que usualmente no será el factor que limite las prestaciones del enlace.

En cualquier caso, el planteamiento de un enlace de comunicaciones ópticas con un emisor en el espacio profundo presenta dificultades formidables, que explican la razón última por la que no se han ensayado hasta ahora como opción preferida en este tipo de enlaces. El problema del apuntamiento ya mencionado se agrava con las **distorsiones** introducidas por la atmósfera terrestre. Por otra parte, si la misión se prolonga en el tiempo, será necesario enfrentarse a situaciones muy cambiantes, como transmisiones diurnas y nocturnas. En el primer caso surgirán distintas orientaciones con respecto al Sol, con ángulos de recepción que deben minimizarse para mantener activo el enlace el mayor tiempo posible. Esta cuestión es especialmente problemática cuando se consideran enlaces con los planetas interiores, cuya posición angular es siempre próxima a la del propio Sol.

En el presente capítulo se revisan los distintos aspectos que se han de afrontar en el canal de un enlace de comunicaciones ópticas en el espacio. En los capítulos sucesivos se estudian las peculiaridades del transmisor y receptor.

## 3.1. DISTANCIA ENTRE TERMINALES

El mayor inconveniente que presentan las comunicaciones en espacio profundo es la **gran distancia** existente entre el transmisor y el receptor –millones de km, si se trata de otros planetas. Un haz de radiación electromagnética propagándose por el espacio exterior está prácticamente exento de alteraciones de dirección, y tampoco sufre atenuaciones debidas a absorción de la radiación ni a *scattering*. Sin embargo, su propagación produce un ensanchamiento paulatino, ya que se trata de un haz de tamaño finito, que no puede asimilarse a un frente de onda plano. Esta ***divergencia*** presenta un valor que, como mínimo, está determinado por el **límite de difracción** del haz, el cual depende, entre otras cosas, de la *longitud de onda* de la radiación empleada. Ésa es precisamente la mayor ventaja potencial que presenta un canal de comunicaciones ópticas en comparación con un canal de radiofrecuencia. En este apartado se estudia el efecto que tiene la distancia en el rendimiento del enlace de comunicación.

### 3.1.1. Efecto de la distancia en la divergencia del haz

Cuando un haz de luz pasa por una apertura siempre se produce un proceso de **difracción** que hace que el haz no salga perfectamente colimado. Esta difracción es la causante de que el haz diverja al aumentar la distancia. Al tratarse de tan grandes distancias la divergencia del haz debida a la difracción producida en la fuente se convierte en un factor crítico en el sistema ya que, cuanto mayor sea el área en el que está confinada la potencia, menor será la *densidad de potencia* por unidad de superficie, es decir, menor señal alcanzará la superficie de la antena receptora.

Si la apertura del transmisor es circular, la superficie transversal a la propagación en la que se concentra la mayor parte de la señal mantiene la misma forma. Como el área de un círculo es $\pi r^2$ y el radio r del área se puede calcular como $d \cdot tg(\theta)$, la variación del área con la distancia d para un ángulo θ fijo de divergencia del haz vendrá dada por

$$A = \pi d^2 tg^2(\theta) \qquad (m^2) \qquad (62)$$

con lo que se comprueba que el área del *spot* aumenta proporcionalmente con el cuadrado de la distancia. Esto es, el área aumenta cuadráticamente al aumentar la distancia. Si la densidad de potencia viene dada por

$$S = \frac{P}{A} \qquad (W/m^2) \qquad (63)$$

siendo P la potencia y A el área sobre la que se distribuye, se deduce que con el aumento de la distancia, la potencia (que permanece constante) debe repartirse sobre superficies cada vez mayores, por lo que la densidad de potencia disminuye linealmente con el área y cuadráticamente con la distancia.

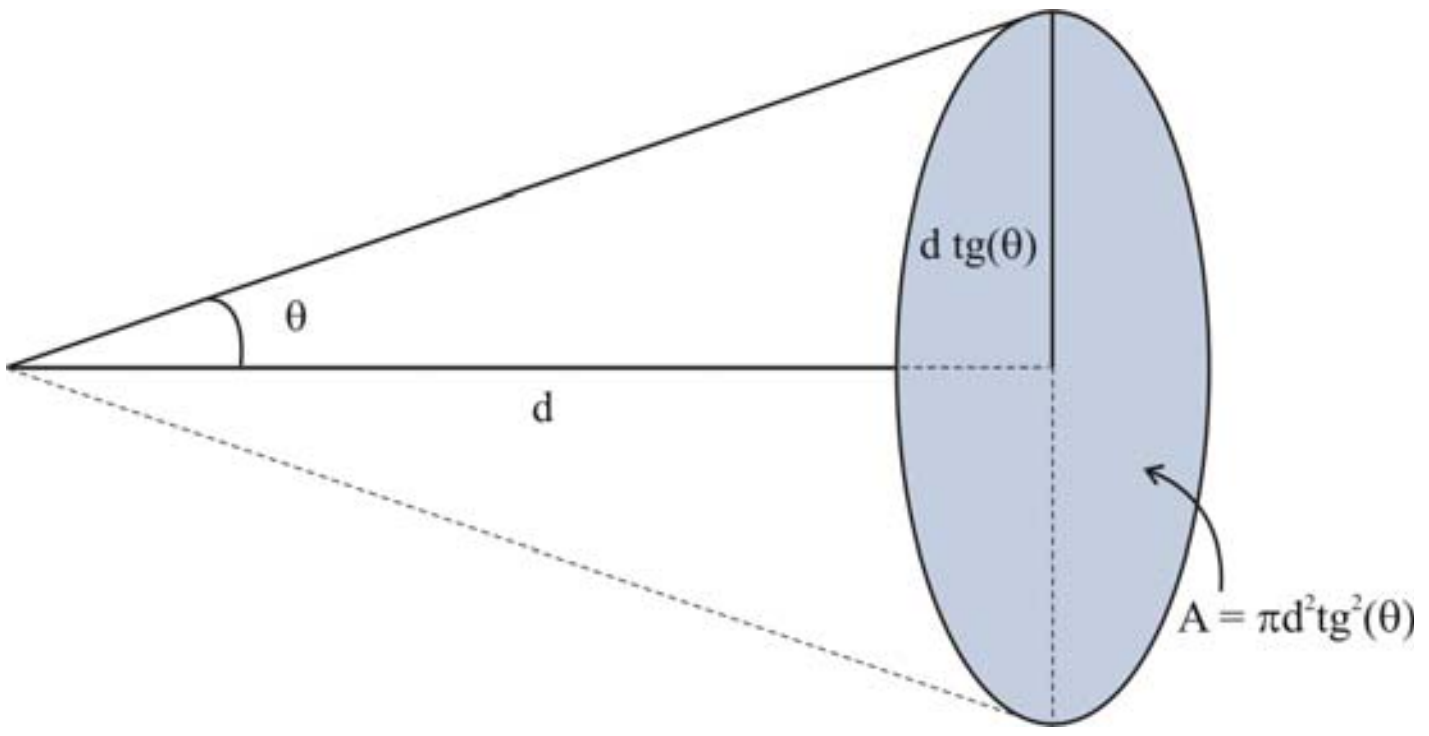

***Figura 3.1. Área del spot.***

### 3.1.2. Pérdidas por espacio libre

Conocido el fundamento físico de la influencia que tiene la distancia en la potencia recibida en un área colectora, para poder evaluar numéricamente este efecto se va a utilizar la fórmula que se ha venido empleando tradicionalmente en comunicaciones por radiofrecuencia. Ésta es la fórmula de ***pérdidas por espacio libre***.

La expresión clásica para el cálculo de las pérdidas por espacio libre se deduce como sigue a continuación[1] [25]. Se parte del supuesto de que –si el diámetro de antena no es notablemente mayor que la longitud de onda– la potencia de la fuente se radiaría uniformemente en todas las direcciones. Esto es, se supone que la antena transmisora es **isótropa**, con lo que la potencia radiada a una distancia d de la fuente se distribuiría uniformemente en un área $4\pi d^2$ (correspondiente al área de la esfera cuyo centro es la antena). Asumiendo que la antena receptora también es isótropa, y que su tamaño está ajustado a la longitud de onda que recibe, el área efectiva que presentará a la radiación que le llega será $\lambda^2/(4\pi)$. Por lo tanto el cálculo de la fracción de potencia que le llega a la antena receptora queda como sigue

$$\frac{P_{rec}}{P_{emi}} = \frac{1}{4\pi d^2} \cdot \frac{\lambda^2}{4\pi} = \left(\frac{\lambda}{4\pi d}\right)^2 \qquad (64)$$

[1] Se emplea esta fórmula por su carácter general, puesto que es válida para cualquier longitud de onda, ya sea óptica o de radiofrecuencia. No obstante, el propio autor de la referencia advierte que este modo de calcular pérdidas en comunicaciones ópticas es poco habitual.

La ecuación (64) es la fracción de potencia que le llega al receptor, de la que se infieren directamente las pérdidas expresándolas en unidades logarítmicas (***Nota****: en otras referencias se utiliza la expresión inversa, con lo que el valor de pérdidas tiene signo positivo*). Esta fórmula, conocida como **pérdidas por espacio libre** (suele expresarse en unidades logarítmicas) es la siguiente

$$L_{fs} = 10\log_{10} P_{rec} - 10\log_{10} P_{emi} = 10\log_{10}\left(\frac{\lambda}{4\pi d}\right)^2 \quad \text{(dB)} \qquad (65)$$

Se ha empleado la notación clásica $L_{fs}$, que significa en inglés *Loss by free space.*

## EJEMPLO DE CÁLCULO DE PÉRDIDAS POR ESPACIO LIBRE

Como se ha visto en el apartado 3.1.1, dedicado al efecto de la distancia en la divergencia del haz, el área que ocupa la señal *aumenta cuadráticamente* con la distancia y la densidad de potencia *disminuye linealmente* con el área. Esto quiere decir que la pérdida de potencia es **proporcional al cuadrado de la distancia**, lo que también se acaba de comprobar en el apartado 3.1.2, dedicado a la fórmula de las pérdidas por espacio libre.

Para evaluar numéricamente este efecto es muy útil apreciar que si duplicamos la distancia, las pérdidas aumentan en 6 dB. Esta regla es la misma ya se trate de pasar de 30 a 60 km o de 30 millones de km a 60 millones de km.

Las pérdidas por espacio libre, calculadas de este modo, son *catastróficas* para un enlace de comunicaciones ópticas en espacio profundo. Así, con una longitud de onda de 1 μm y una distancia de 100 millones de km (una distancia del rango Marte-Tierra), las pérdidas superan los 360 dB. Un láser de 1 W funcionando en continua emite a esa λ aproximadamente $5{\cdot}10^{18}$ fotones por segundo. A la antena receptora llegaría, en estas condiciones, *¡menos de un fotón cada 10.000 millones de años!* Naturalmente, el resultado queda drásticamente modificado cuando se considera la colimación del haz del emisor y el tamaño de la antena receptora –un telescopio astronómico– cuyo tamaño es muchos órdenes de magnitud mayor que la longitud de onda. Es explicable, pues, que en comunicaciones ópticas se prefiera *tratar directamente con los haces colimados*, sin considerar la emisión isótropa. En todo caso, los resultados obtenidos por uno y otro método son idénticos salvo ligeras desviaciones derivadas de las aproximaciones asumidas en los cálculos.

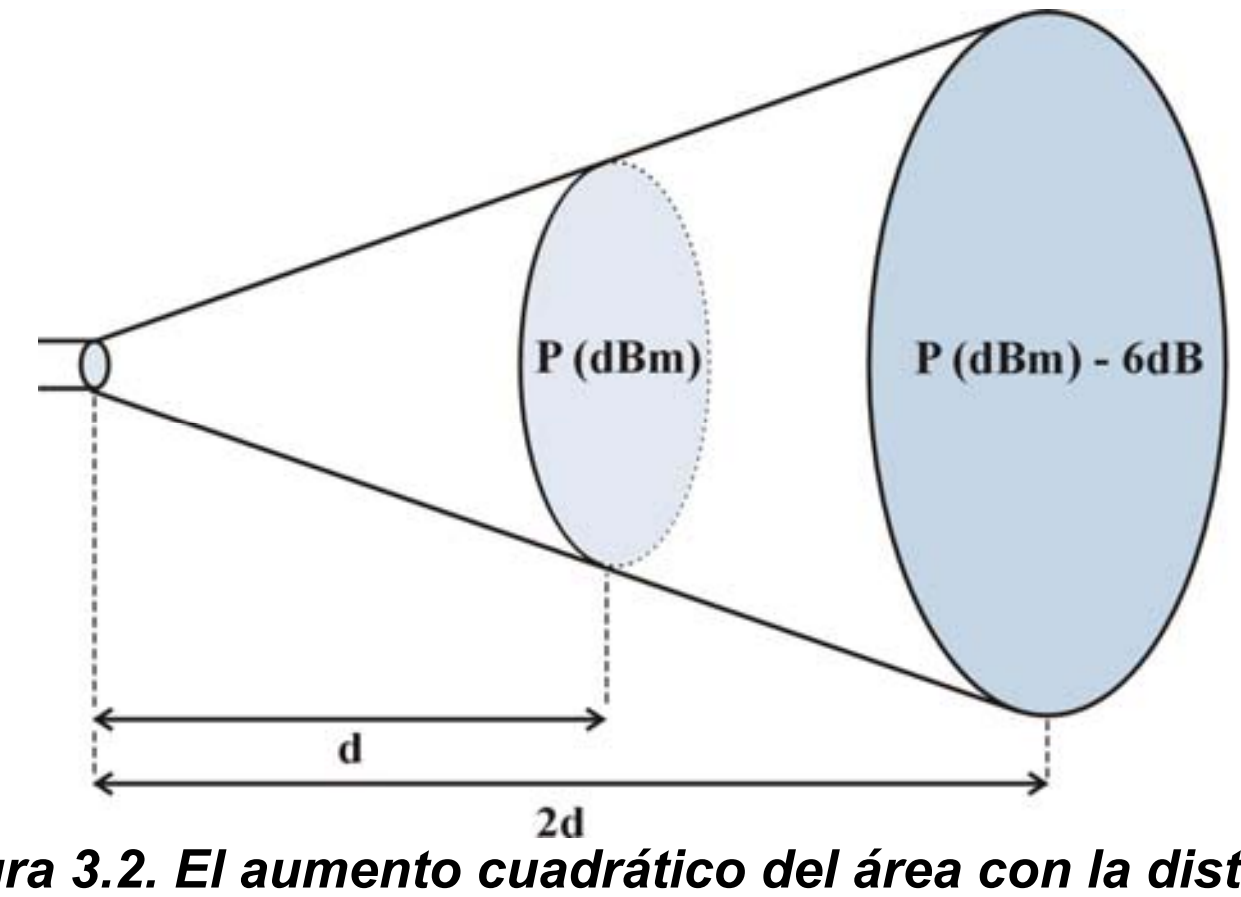

***Figura 3.2. El aumento cuadrático del área con la distancia determina que al duplicarla se pierdan 6 dB de potencia.***

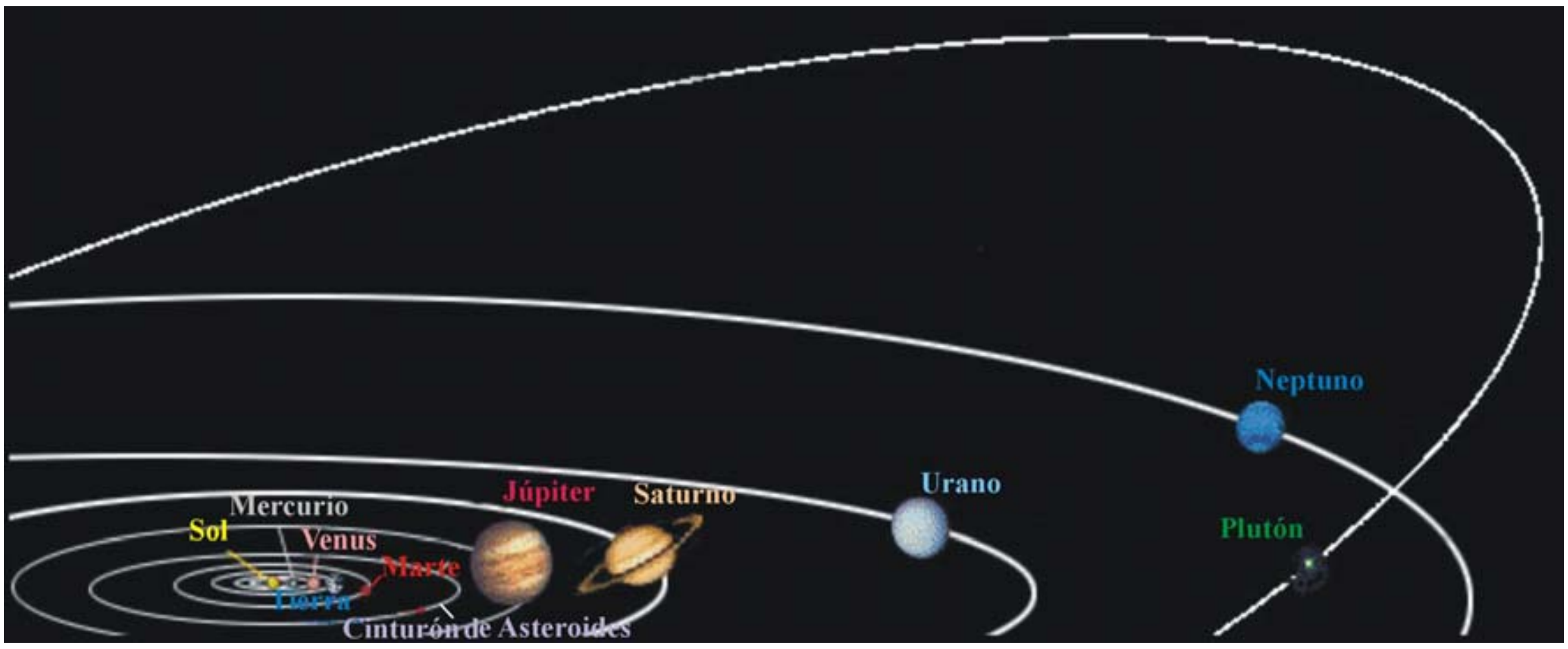

*Figura 3.3. Vista general del sistema solar [26].*

## 3.1.3. Sistema solar

Una vez vista la manera de cuantificar el efecto que tiene la distancia en términos de atenuación en el balance de potencia, en la Tabla 3.1 se presentan las pérdidas (para caso mejor y peor) que habría que afrontar a la hora de diseñar un enlace de comunicaciones ópticas entre la Tierra y algún lugar del sistema solar (Figura 3.3).

Los valores de la columna de distancias mínimas y máximas se han calculado empleando el simulador descrito en el capítulo 2. Para obtener valores extremos de distancias máximas y mínimas se han simulado **4 siglos**, desde el año 2.000 hasta el 2.400.

| | | | $\lambda = 1\ \mu m$ | | $\lambda = 10\ \mu m$ | |
|---|---|---|---|---|---|---|
| **Satélite o planeta** | **Distancia mín. (millones de km)** | **Distancia máx. (millones de km)** | **Pérdidas mín/máx. (dB)** | | **Pérdidas mín/máx. (dB)** | |
| ***Luna*** | 0,36 | 0,41 | 313,1 | 314,2 | 293,1 | 294,2 |
| ***Mercurio*** | 82,15 | 217,15 | 360,3 | 368,7 | 340,3 | 348,7 |
| ***Venus*** | 39,56 | 259,67 | 353,9 | 370,3 | 333,9 | 350,3 |
| ***Marte*** | 55,71 | 400,36 | 356,9 | 374,0 | 336,9 | 354,0 |
| ***Júpiter*** | 590,91 | 965,97 | 377,4 | 381,7 | 357,4 | 361,7 |
| ***Saturno*** | 1203,37 | 1655,60 | 383,6 | 386,4 | 363,6 | 366,4 |
| ***Urano*** | 2581,81 | 3156,03 | 390,2 | 392,0 | 370,2 | 372,0 |
| ***Neptuno*** | 4309,05 | 4685,08 | 394,7 | 395,4 | 374,7 | 375,4 |
| ***Plutón*** | 4304,73 | 7467,46 | 394,7 | 399,4 | 374,7 | 379,4 |

***Tabla 3.1. Pérdidas mínimas y máximas por espacio libre a 1 y 10 µm en distancias mínimas y máximas desde la Tierra a varios emplazamientos del sistema solar.***

Los valores de la columna de pérdidas mínimas y máximas se han calculado empleando la fórmula de pérdidas por espacio libre descrita en el apartado inmediatamente anterior. Se utilizan como ejemplo las longitudes de onda de 1 y 10 µm por existir tecnología láser adecuada al sistema de comunicación y coincidir con ventanas atmosféricas de transmisión. En todo caso, conocidas las distancias, podrían obtenerse las pérdidas por espacio libre para otras longitudes de onda sin más que cambiar este término en la ecuación (65). Consecuencia lógica de esta ecuación, como se observa en las columnas de pérdidas, es que *dividir la longitud de onda por 10* significa *sumar 20 dB* a las pérdidas. Un enlace de microondas, cuya longitud de onda es unas 10.000 veces mayor, tendría por idéntica razón *unos 80 dB menos* de pérdidas. Obsérvese además que, dada la escala logarítmica empleada, las pérdidas de un emplazamiento relativamente próximo como la Luna, en torno a 310 dB, *no difieren demasiado* de las pérdidas del lejanísimo Plutón (unos 390 dB).

### 3.1.4. Comparación con microondas

Como se desarrollará en el capítulo 4, dedicado al emisor espacial, el mínimo ángulo de divergencia θ con el que el *spot* es emitido en un colimador es *proporcional a la longitud de onda* e *inversamente proporcional al diámetro del espejo* primario. Esto es válido tanto para frecuencias de microondas como para frecuencias ópticas (Figura 3.4). Se puede concluir por tanto que para una misma apertura del telescopio y transmitiendo con el mínimo ángulo de divergencia, el haz (que será el más colimado posible) ***diverge menos con la distancia*** usando frecuencias ópticas que usando frecuencias de microondas. Ésa es precisamente la ventaja que compensa las enormes pérdidas por espacio libre mostradas en el apartado anterior.

El límite de difracción impone el área mínima del *spot* que forma el haz a la distancia a la que se encuentra el receptor. Como se aprecia en el ejemplo de la Figura 3.5, para una misma apertura de transmisión y una misma distancia Marte-Tierra, el haz óptico que se recibe en la Tierra es un 10 por ciento del diámetro de la misma, mientras que en el caso del haz de microondas es unas 1000 veces mayor que el diámetro terrestre. De esta forma, la potencia *se distribuye en un área mucho menor* en el caso de las frecuencias ópticas, quedando una densidad de potencia mucho mayor y por tanto una mayor potencia recibida en la superficie receptora.

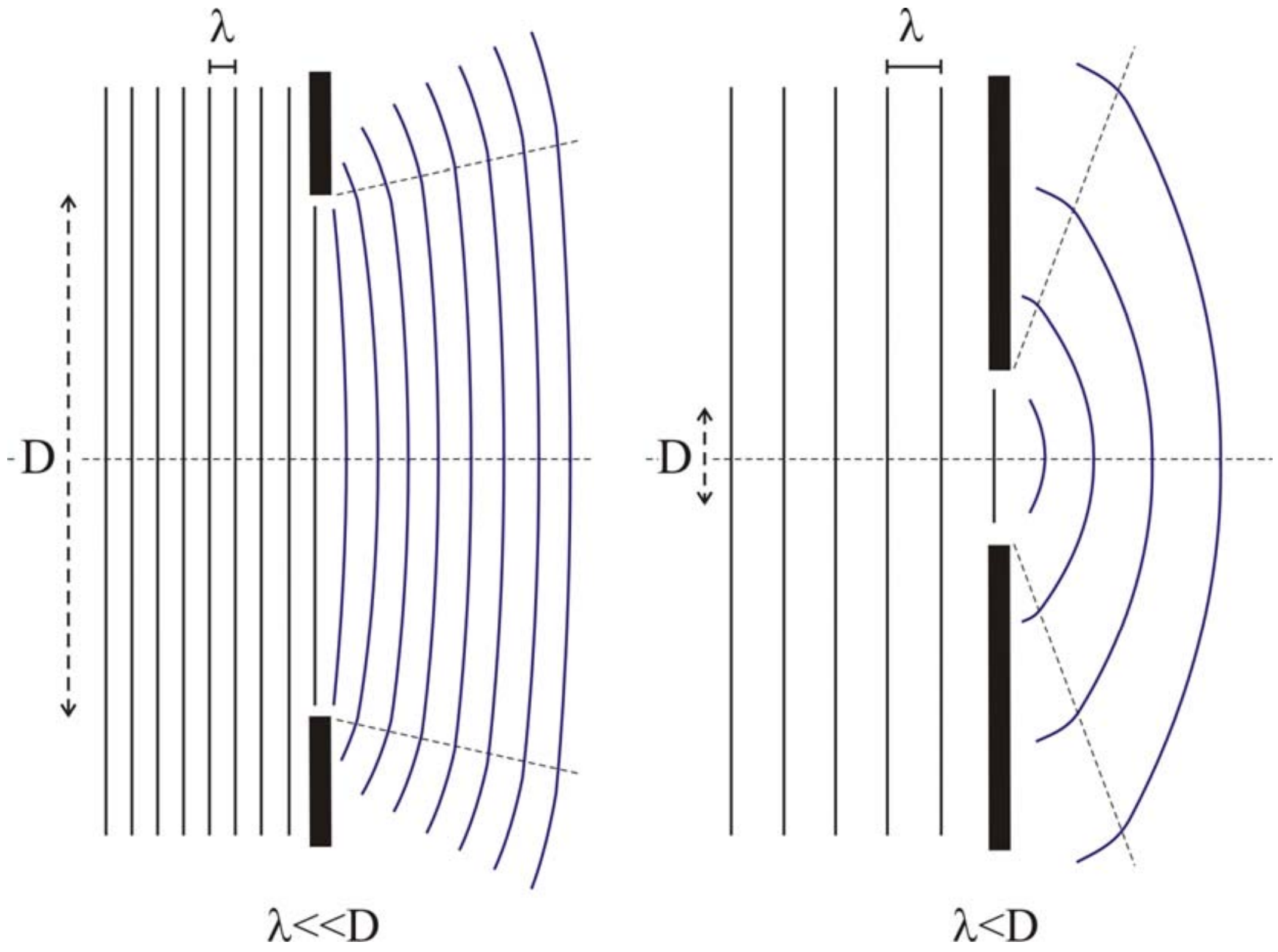

***Figura 3.4. Difracción de un frente de ondas planas producida por una apertura para longitudes de onda ópticas y microondas.***

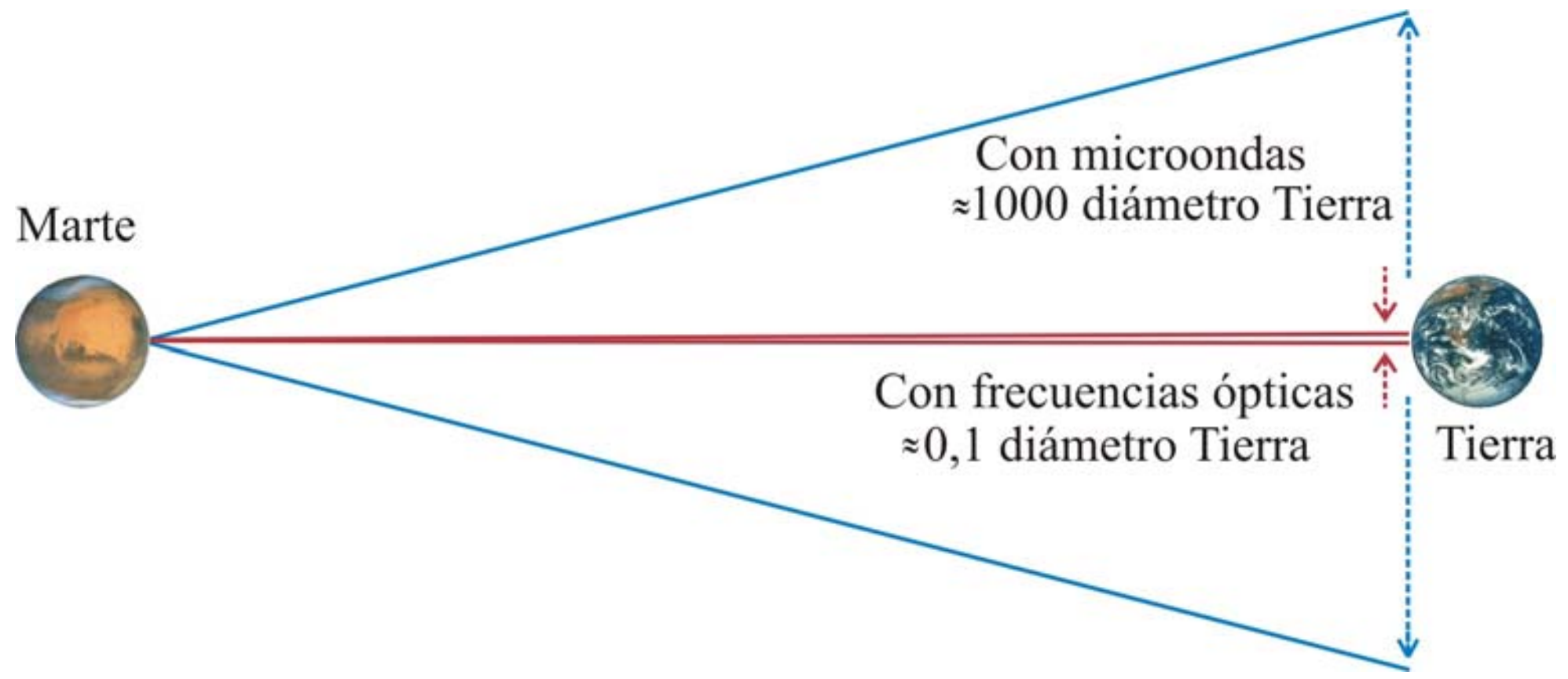

***Figura 3.5. Divergencia de un haz de frecuencias ópticas y de microondas en un enlace Marte-Tierra.***

## 3.2. BLOQUEO DEL SOL

La única razón que inevitablemente ocasiona un bloqueo temporal del enlace cortando toda comunicación se deberá al hecho de que el Sol se interpone entre el terminal terrestre y el terminal espacial. Este hecho se produce *para todos los planetas* del sistema solar ya que orbitan alrededor del Sol aproximadamente en el mismo plano (**eclíptica**).

El bloqueo directo del sol ocurre cuando éste se halla en algún punto intermedio de la línea imaginaria que uniría a ambos terminales (Figura 3.6). Para los planetas exteriores a la Tierra, el bloqueo se da cada vez que hay una ***conjunción solar***, y coincide con la situación en la que los terminales se hallan a distancias próximas a la máxima. Para los planetas interiores sucede cuando se da una ***conjunción superior*** coincidiendo también con la distancia máxima. En estas ocasiones la comunicación es imposible, aunque se puede minimizar el periodo de tiempo que dura esta situación. El mayor problema estará en el receptor que tendrá que apuntar con *ángulos muy pequeños* entre el Sol y el transmisor, con lo que se deberá filtrar cuidadosamente la radiación solar recibida. En cualquier caso, llegará un momento en el que tenga que detener la observación por estar apuntando demasiado cerca del Sol.

La situación opuesta para los planetas exteriores tiene lugar cuando se hallan en ***oposición*** y para los interiores cuando se hallan en ***conjunción inferior***. En estos casos el receptor terrestre recibe la señal de noche y distancias próximas a la mínima. Es el caso más favorable.

**ALINEAMIENTOS**

Los términos ***conjunción*** y ***oposición*** siempre van referidos a la Tierra por lo que se hablará de **conjunción solar** cuando, visto desde la Tierra, un planeta se halle alineado con el Sol, y de **oposición** cuando desde la Tierra se tenga que dar la espalda al Sol para orientarse hacia el otro planeta que, aún con la Tierra interponiéndose, estará también alineado con el Sol.

Según esta definición los planetas interiores, al estar dentro de la órbita de la Tierra, nunca se hallan en oposición, por lo que se habla de **conjunción inferior** y **conjunción superior**. La primera se da cuando el planeta está, visto desde la Tierra, delante del Sol y la segunda cuando está detrás. En la Figura 3.6 se pueden apreciar todos los posibles alineamientos entre la Tierra, el Sol y otro planeta.

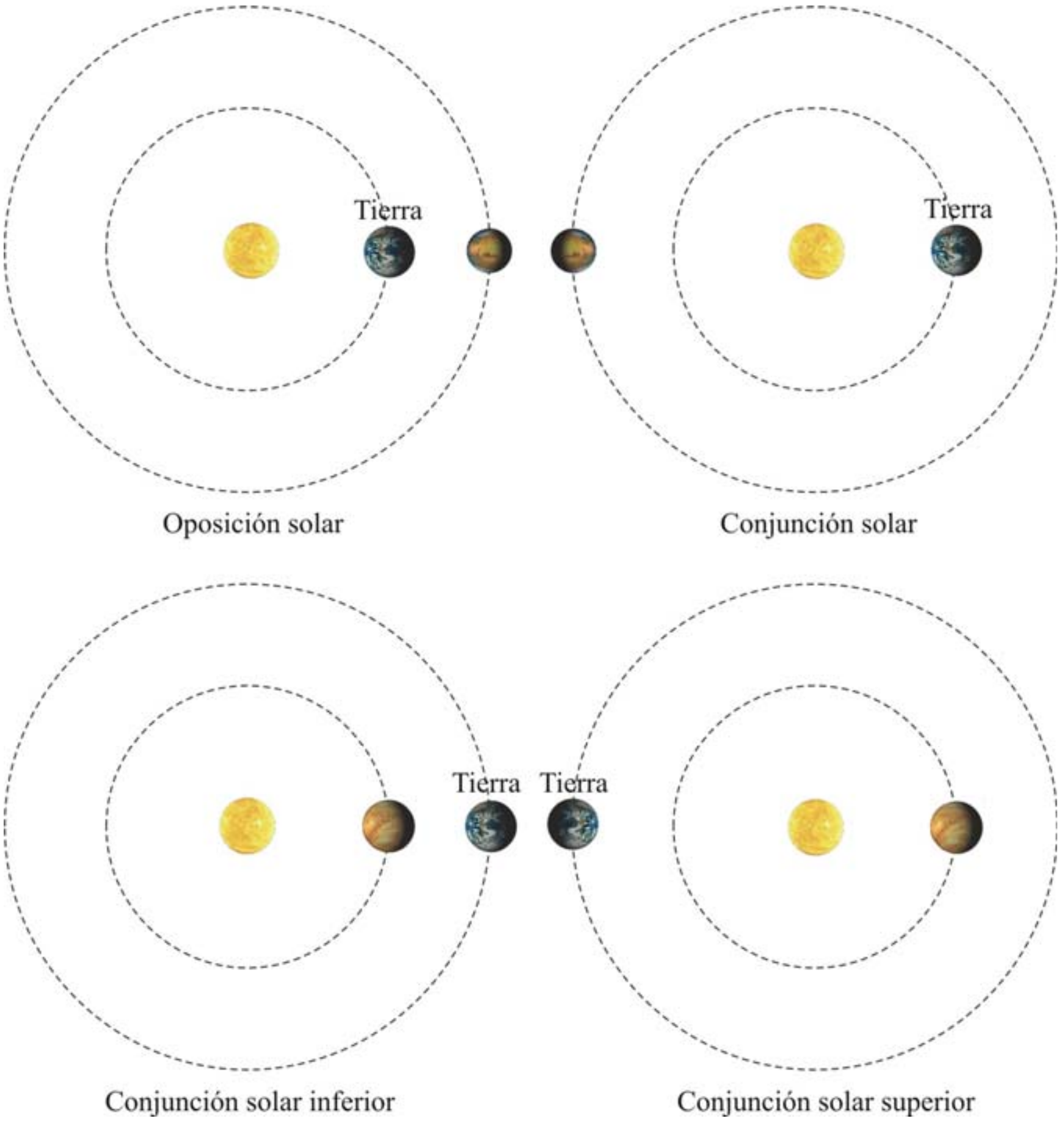

***Figura 3.6. Posibles alineamientos de la Tierra, el Sol y un planeta exterior o interior.***

## 3.3. LA ATMÓSFERA TERRESTRE

Las comunicaciones con longitudes de onda *dentro del espectro óptico* se ven fuertemente afectadas por la presencia de la atmósfera terrestre. En este apartado se asume que el terminal terrestre se encuentre en la superficie de la Tierra y se da cuenta de los efectos que tiene la atmósfera en la comunicación debidos al hecho de que la señal tiene que atravesarla.

### 3.3.1. Nubes

La gran variabilidad de presencia de nubes en la atmósfera terrestre y su aparente aleatoriedad pueden hacer que el enlace quede completamente cortado debido a la pérdida de la línea de visión directa. Como se estudiará en el apartado 5.7.3, dedicado a las posibles alternativas de situación del terminal terrestre en la superficie, la única posibilidad de mitigar este problema con terminales en la superficie de la Tierra –en caso de nubosidad sobre el emplazamiento receptor– es recibir la señal simultáneamente en otra ubicación en la que en ese momento no haya presencia de nubes que dificulten la comunicación. Esta estrategia consiste en añadir redundancia en el número de terminales que en un momento determinado pueden recibir la señal.

En la Figura 3.7 se observan dos alternativas de diseño para la creación de una red de receptores terrestres cuyo fin es proporcionar una recepción con mínima probabilidad de interrupción del enlace.

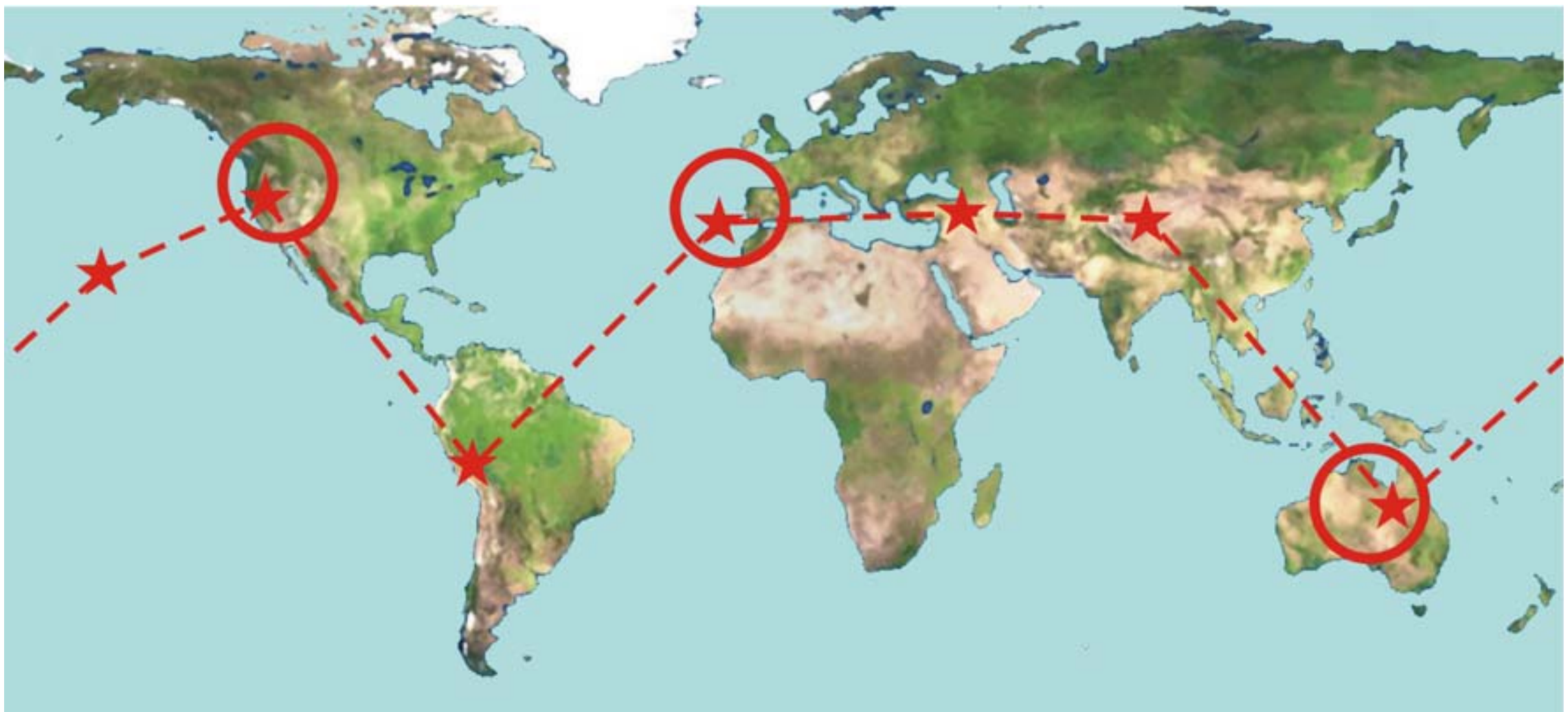

*Figura 3.7. Dos alternativas de diseño para una red de receptores terrestres [27].*

- ***Red linealmente distribuida.*** Formada por terminales lo más equidistantes posible unos de otros. Añadiendo un número suficiente se consigue que el terminal remoto sea visible en todo momento por dos o más terminales terrestres
- ***Agrupaciones de receptores.*** Formadas por series de receptores situados relativamente próximos unos a otros dentro de cada agrupamiento.

En la figura se representan con estrellas los terminales que formarían la red linealmente distribuida y con círculos los que integran la red formada por agrupaciones de receptores. La ubicación de cada terminal tanto en la red linealmente distribuida como en la agrupación de terminales obedece a la misma exigencia, esto es, hacer la distribución *lo más equidistante posible* para afrontar el problema de la rotación de la Tierra sobre su eje. Por otra parte, en la red formada por agrupaciones de receptores, la disposición de terminales dentro de cada una de las agrupaciones debe hacerse procurando que *no haya correlación* en la variabilidad atmosférica existente en la ubicación de uno de los terminales en relación a los otros que se encuentran en su misma agrupación. Debido a la distancia existente entre los terminales de la red linealmente distribuida, la variabilidad atmosférica en dos ubicaciones diferentes se supone independiente.

Para evaluar el efecto que tiene la redundancia en el aumento de probabilidad de una recepción continuada se presenta el siguiente resultado [28]: Las ubicaciones simples con menos nubosidad presentan una **disponibilidad** del 70%, esto es, el 30% del tiempo hay nubes. En la Figura 3.8 se aprecia cómo aumenta la disponibilidad con el incremento del número de terminales para distintas altitudes. La menor disponibilidad a grandes alturas se explica por el hecho de que el número de ubicaciones posibles es menor en zonas muy altas y la correcta elección de unos sitios con respecto a otros es fundamental para asegurar que cuando en uno hay nubes en otro haya la máxima probabilidad de que no las haya.

## 3.3.2. Absorción

La atmósfera no es un medio transparente a la radiación electromagnética. Dentro del espectro, existen longitudes de onda que ven su paso notablemente más obstaculizado que otras. El Sol emite radiación en todo el espectro electromagnético (su espectro es muy semejante al de un cuerpo negro), pero la radiación que alcanza la superficie terrestre difiere del espectro solar por la **absorción** de los gases atmosféricos (y también por *scattering*, como se verá). Aunque la absorción de determinadas regiones del espectro por cada tipo de molécula de gas sólo puede explicarse mediante cálculos de química cuántica, el espectro de absorción de cada molécula de gas es conocido. En la Figura 3.9 puede observarse una estimación del espectro de absorción de los gases más importantes en este proceso y el espectro de absorción total de la atmósfera.

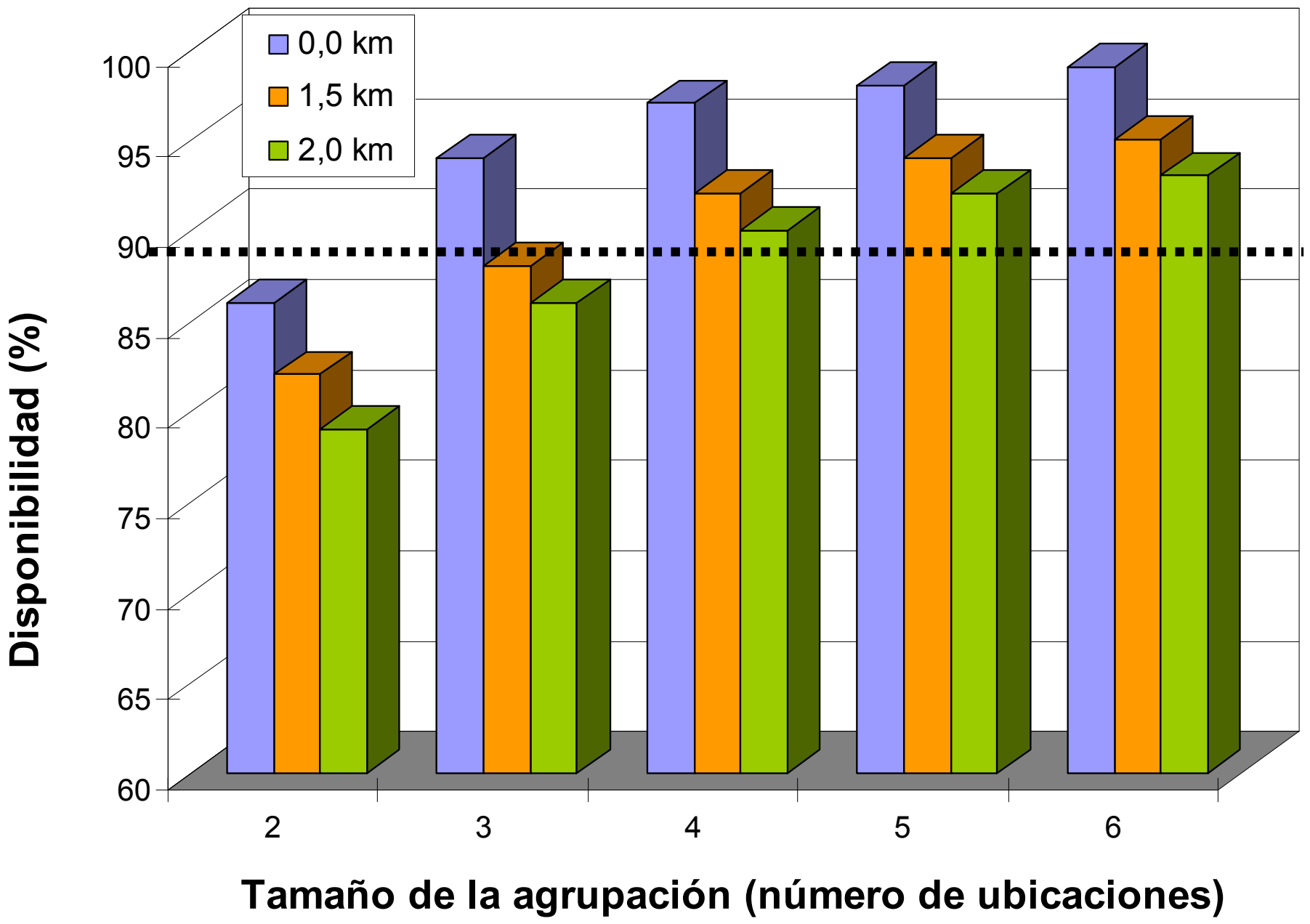

***Figura 3.8. Disponibilidad del enlace en presencia de nubes para distintos tamaños de agrupaciones de terminales a distintas alturas [28].***

Es de destacar el hecho de que los gases que más radiación electromagnética absorben en la región visible y en sus regiones laterales –UV próximo e IR próximo– *no son los gases mayoritarios en la atmósfera* –oxígeno ($O_2$) y nitrógeno ($N_2$)–, sino principalmente **dióxido de carbono** ($CO_2$), **vapor de agua** ($H_2O$) y **ozono** ($O_3$) como se aprecia en la Figura 3.9.

En el espectro total de absorción atmosférica de la Figura 3.9 se pueden apreciar zonas en las que la absorción es menor, o "***ventanas***" de mínima absorción. Por supuesto una de esas zonas corresponde al intervalo que va de unos 350 a unos 750 nm correspondiente al espectro de luz visible. En comunicaciones ópticas *es imprescindible que la longitud de onda de trabajo esté situada en una de las ventanas* de mínima absorción, de lo contrario los gases de la atmósfera absorberían la señal resultando en una atenuación que dificultaría en gran medida la detección en superficie.

Existen ventanas adecuadas para las comunicaciones ópticas (aparte del rango de luz visible ya mencionado) en el infrarrojo próximo y medio:

**0,85 µm—1,06 µm—1,22 µm—1,6 µm—2,2 µm—3,7 µm.**

En la Figura 3.10 se pueden apreciar en detalle estas ventanas de transmisión. Estas gráficas han sido obtenidas mediante el programa de simulación *6S* para un modelo de atmósfera estándar *US-1962* [29].

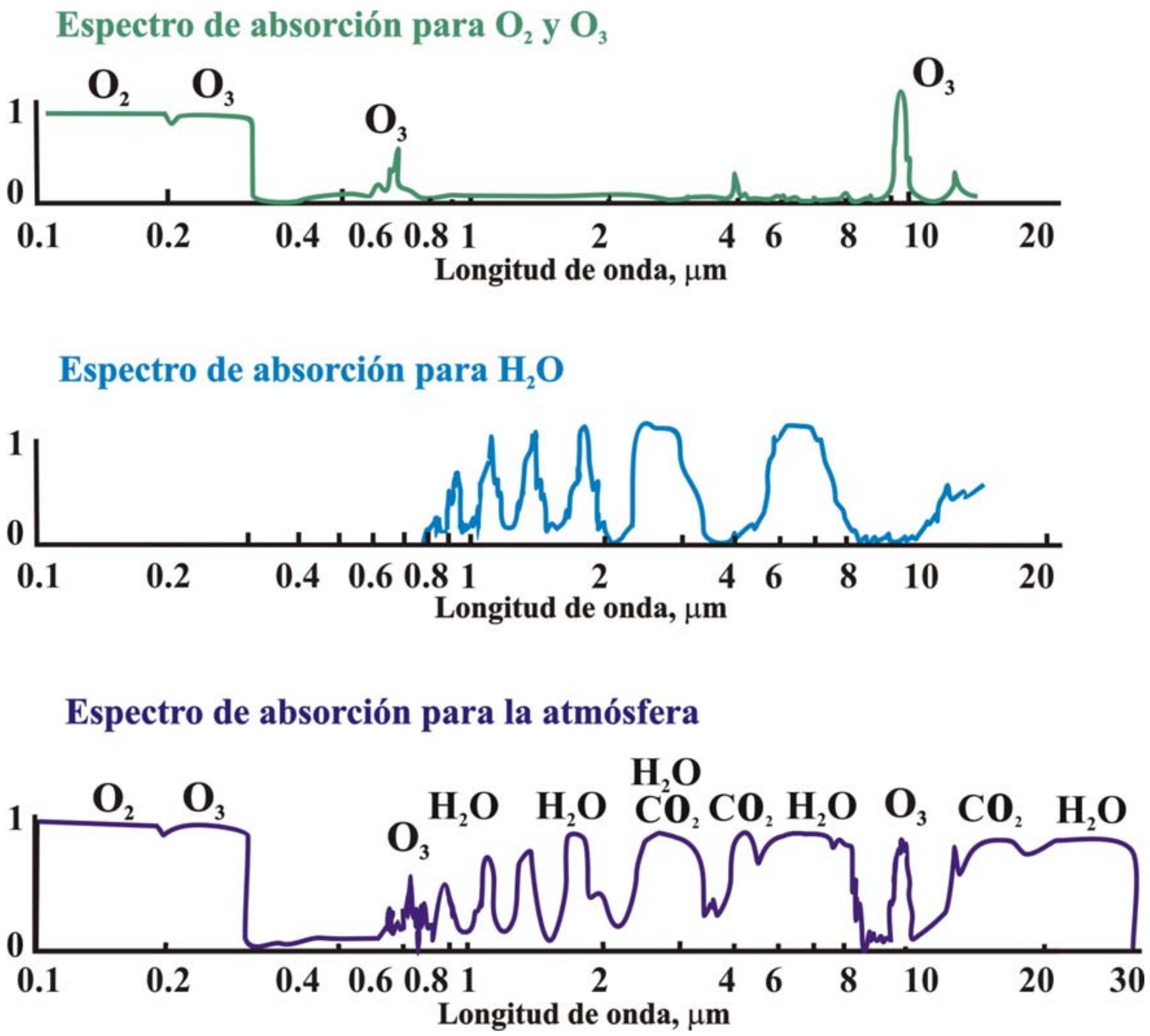

***Figura 3.9. Espectro de absorción por gases de la atmósfera [30].***

La manera de evaluar la influencia que tiene la absorción atmosférica depende de la ventana en la que se produzca la comunicación y de la precisión requerida para dicha evaluación. En general el efecto de la absorción en términos de pérdida de potencia en la señal se puede expresar de la siguiente forma

$$L_a = 10 \cdot \log\left(\frac{1}{T}\right) \quad \text{(dB)} \qquad (66)$$

donde la T representa el valor de la transmitancia normalizada entre 0 y 1 tal como se muestra en la Figura 3.10.

En la práctica se puede aceptar que las pérdidas debidas a absorción ***pueden ser despreciadas*** sin problemas –siempre que se emplee una de las ventanas de transmisión atmosférica, obviamente–. Por ejemplo, si se toma la peor de las ventanas que se muestran en la Figura 3.10, se observa que la transmitancia vale 0,95 a la longitud de onda de 3,7 μm (Figura 3.11). Aplicando la fórmula (66) se obtiene un valor de pérdidas por absorción de 0,2 dB. Cabe añadir que este valor quedaría aún más reducido en una ubicación a **mayor altitud** sobre el nivel del mar, para el cual están obtenidas las gráficas de la Figura 3.10.

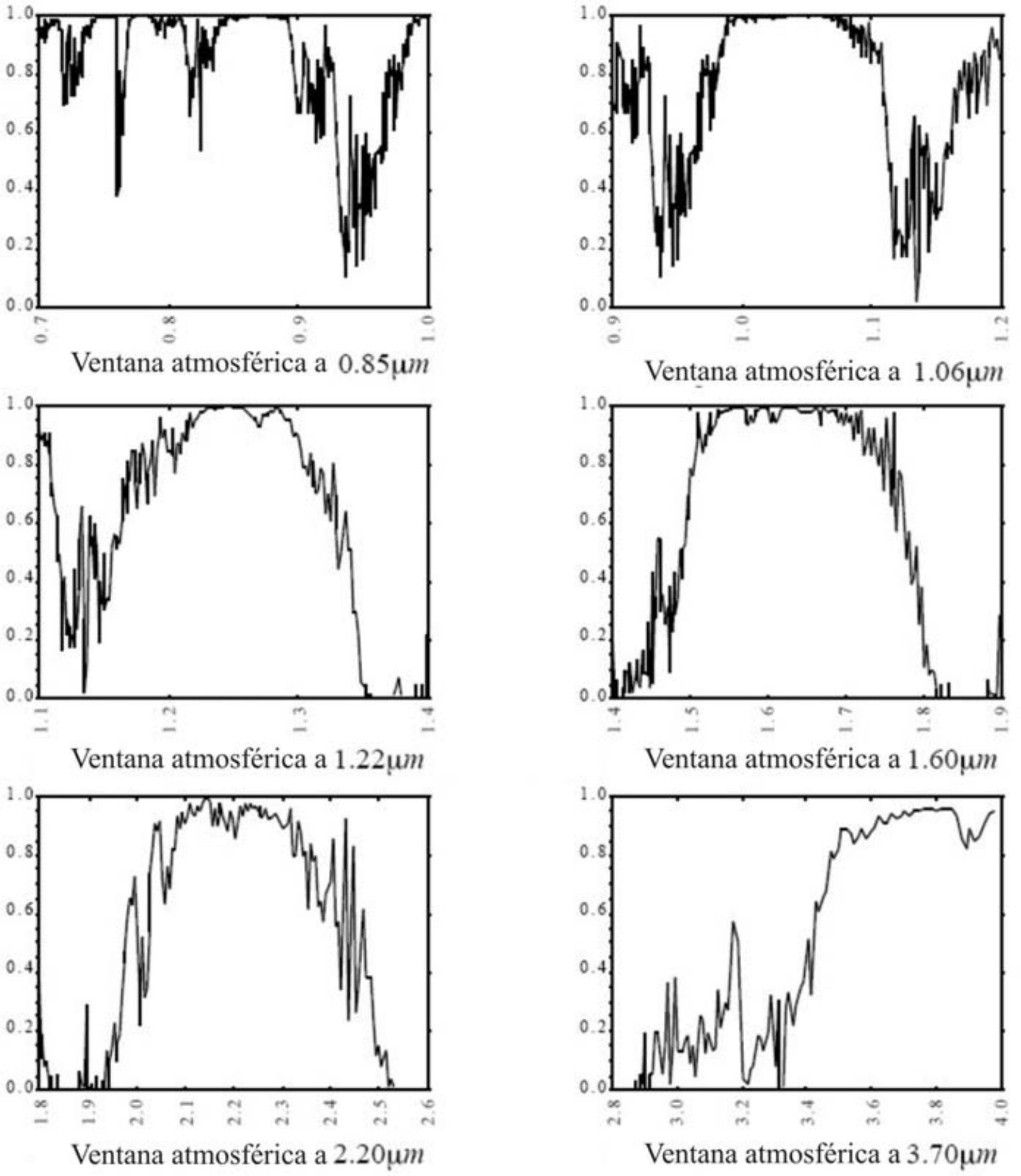

*Figura 3.10. Transmitancia debida a absorción atmosférica para las ventanas de transmisión en infrarrojo cercano y medio [29].*

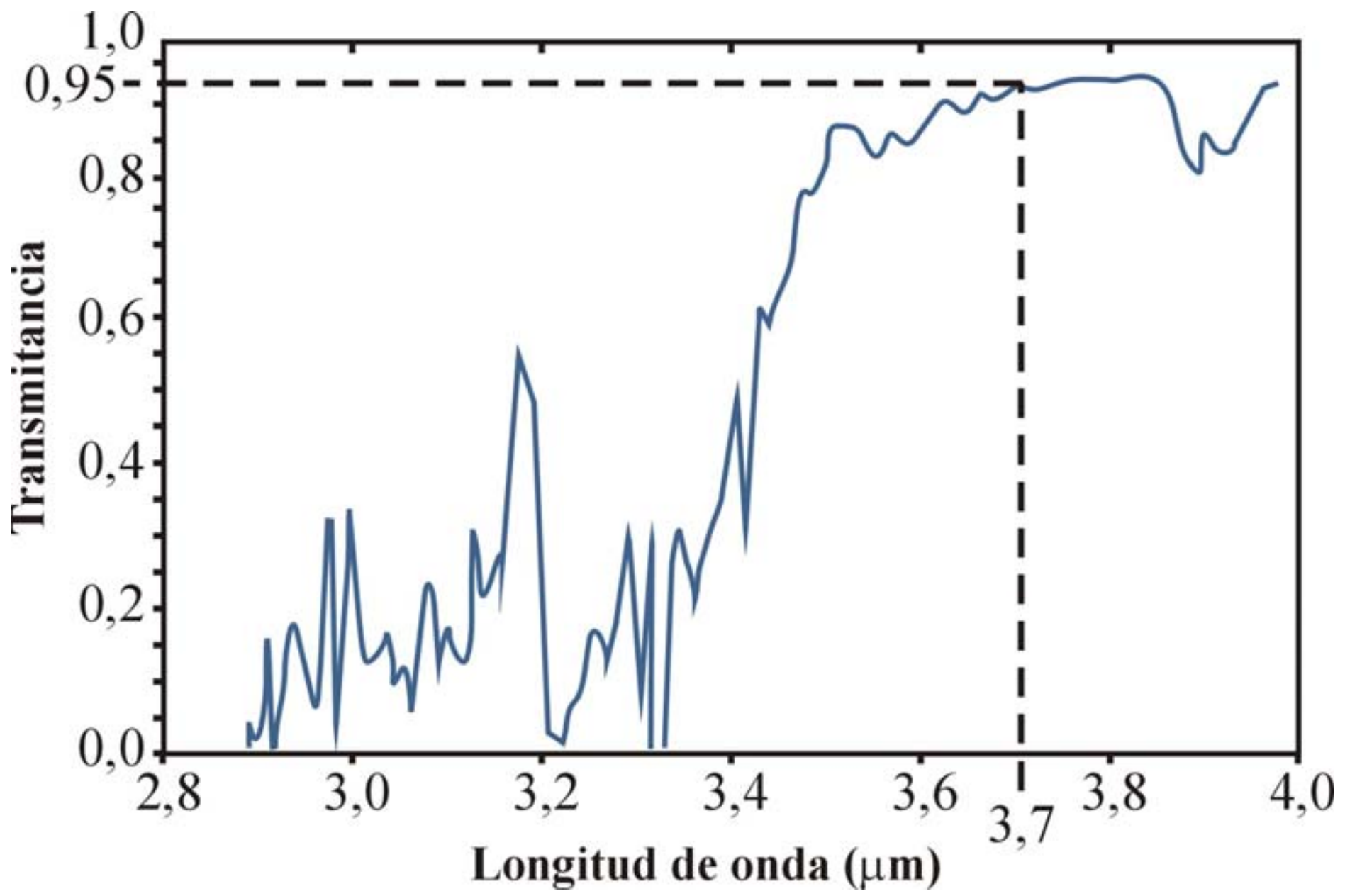

*Figura 3.11. Pérdidas debidas a absorción atmosférica en la ventana de transmisión de 3,7 µm [29].*

### 3.3.3. *Scattering*

Otro de los efectos atmosféricos que afecta a cualquier señal óptica que se propaga a través de la atmósfera terrestre es la ***dispersión de la luz*** por tratarse de un medio en el que se hallan partículas con un índice de refracción distinto. La palabra *dispersión*, en comunicaciones ópticas, se reserva tradicionalmente a los *ensanchamientos temporales* que experimenta la señal al propagarse por el medio (por lo general, una fibra óptica). Obviamente, la **dispersión temporal** puede despreciarse en un enlace de espacio profundo, cuya propagación se realiza casi exclusivamente a través del vacío interplanetario. Aún así, se seguirá la nomenclatura habitual en comunicaciones ópticas, empleando el término en inglés ***scattering*** para referirnos a **dispersión espacial**[1].

La razón de la luz dispersada respecto a la incidente depende esencialmente de la **relación entre el tamaño de la partícula** causante del *scattering* y la **longitud de onda**, y de la **relación entre los índices** de refracción de la partícula y el medio (Figura 3.12). Cuando el radio a de la partícula es pequeño comparado con la longitud de onda de la luz en el medio ($a \ll \lambda$), la luz dispersada depende **inversamente de la cuarta potencia** de λ. En estas condiciones el *scattering* recibe el nombre de ***Rayleigh*** en honor a Lord Rayleigh, que fue el primero en desarrollar la teoría matemática correspondiente. En el caso de tamaños de partícula mayores que la longitud de onda ($a \gg \lambda$), la potencia luminosa dispersada se hace paulatinamente independiente de λ con oscilaciones amortiguadas. Esta zona se conoce como ***scattering Mie***.

Cuando se utilizan radiaciones próximas al visible, la atmósfera presenta **los dos tipos de *scattering***, Rayleigh y Mie. El *scattering* Rayleigh se debe principalmente a **moléculas de aire** y el Mie a partículas de mayor tamaño como **aerosoles**, **polvo** y **nubes**. La diferente dependencia con λ de uno y otro tipo de *scattering* explica fenómenos muy habituales en la vida cotidiana, como el color azul del cielo, el color blanco de las nubes, o los tonos rojizos del cielo al atardecer.

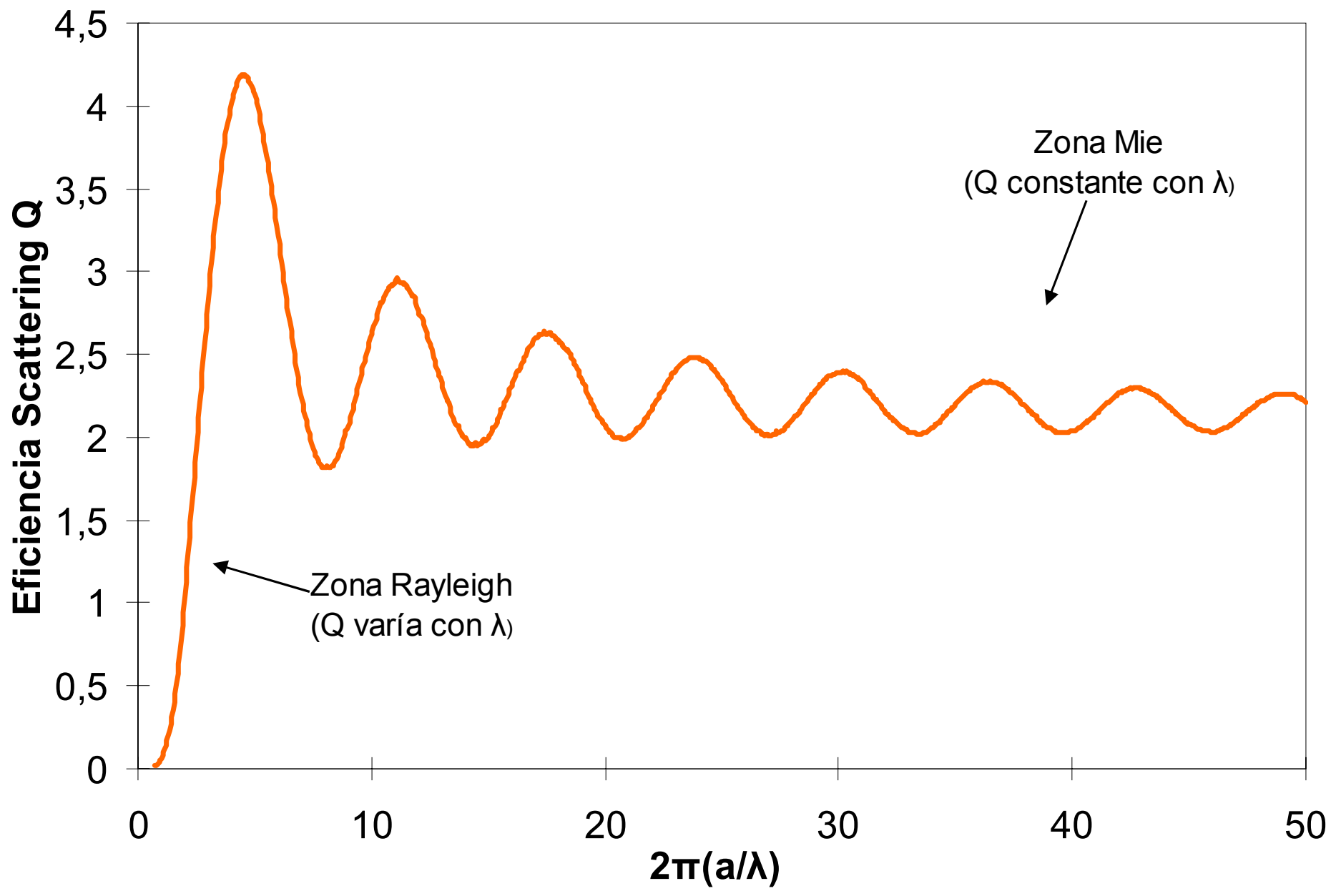

***Figura 3.12. Resultado del cálculo de la eficiencia normalizada de* scattering *en función de la relación entre tamaño de partícula y longitud de onda.***

[1] Algunos autores emplean los términos *reflexión difusa*, *difusión* o incluso *esparcimiento*. Se trata en todos los casos del mismo fenómeno.

En este apartado se estudiará la forma de cuantificar las pérdidas de potencia debidas a la atenuación por *scattering* asociada a la atmósfera. El procedimiento descrito a continuación se apoya en el cálculo del ***coeficiente de atenuación*** atmosférica α, para ambos tipos de *scattering*, en unidades de inverso de metros, y en la ***altura de atmósfera*** z que la señal atraviesa hasta finalmente alcanzar el receptor terrestre. De esta forma, la expresión que calcula las pérdidas para cada tipo de *scattering* es la siguiente

$$L_s = \exp\left[\int_{x_0}^{z} \alpha \cdot dr\right] \tag{67}$$

donde $L_s$ (expresada en unidades naturales) debe calcularse para cada tipo de *scattering* de forma que las pérdidas totales por ambos tipos será la suma de cada aportación por separado. De esta manera cada tipo de *scattering* tendrá asociado un coeficiente de atenuación propio que será necesario integrar en el rango de altura que corresponda. En el caso de que el receptor *esté situado a una cierta altitud sobre el nivel del mar*, el límite $x_0$ inferior de la integral pasaría a ser dicha altura.

### 3.3.3.a. *Scattering* Rayleigh

El componente más abundante de la atmósfera son moléculas de gas, de tamaño muy inferior a las longitudes de onda del espectro visible e infrarrojo próximo. El coeficiente de atenuación $\alpha_R$ debido a dichas moléculas se puede calcular por medio de la ***sección eficaz de scattering*** $\sigma_R$ ($m^2$)

$$\alpha_R = N_g \cdot \sigma_R \qquad (m^{-1}) \tag{68}$$

donde $N_g$ es la concentración numérica de moléculas de gas (moléculas/$m^3$) y la expresión de $\sigma_R$, para alturas inferiores a 100 km, viene dada por [31]

$$\sigma_R = \left(4{,}59 \cdot 10^{-27}\right) \cdot \left[\frac{\lambda(\mu m)}{0{,}55}\right]^{-4} \qquad (cm^2) \tag{69}$$

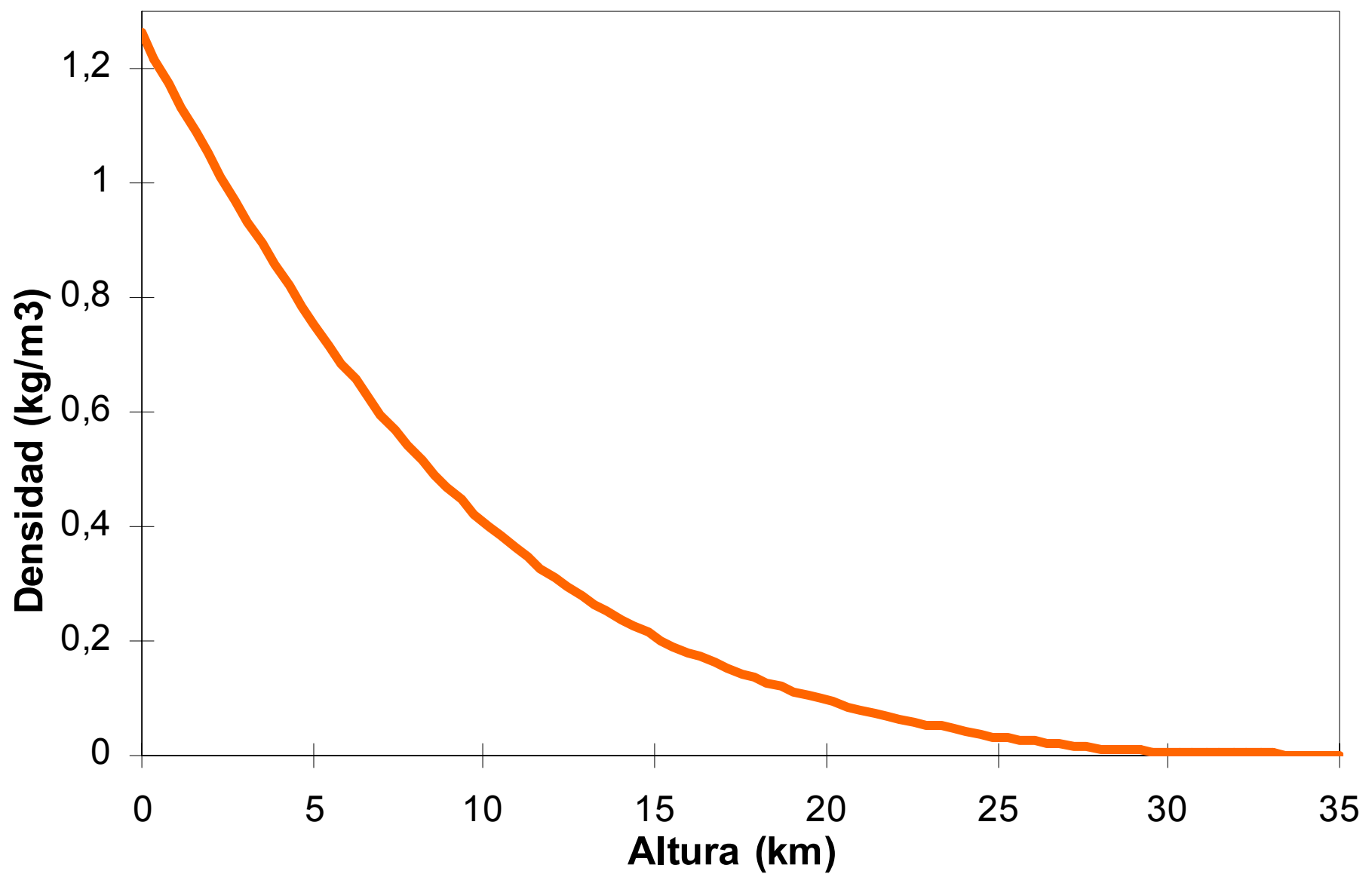

***Figura 3.13. Densidad de la atmósfera en función de la altura.***

En la Figura 3.13 se muestra una representación gráfica de la densidad D de moléculas de gas en la atmósfera en función de la altura.

Conocida la densidad D, es posible determinar la concentración numérica a través del **número de Avogadro** ($N_A = 6{,}022 \cdot 10^{23}$ moléculas/mol) y el **peso molecular medio** del aire ($M_m = 28{,}94$ gramos/mol) de la siguiente manera

$$N_g = \frac{D\,N_A}{M_m} \qquad (\text{moléculas/m}^3) \qquad (70)$$

Puesto que la densidad de moléculas de gas es distinta según la altura de la atmósfera que se considere (Figura 3.13), el coeficiente de atenuación será función también de la altura.

### 3.3.3.b. *Scattering* Mie

A diferencia de los gases atmosféricos, en el estudio de la atenuación de partículas de mayor tamaño que se encuentran en la atmósfera, como son los ***aerosoles***, su concentración de masa depende no sólo de la altura de la atmósfera, sino también del **modelo** al que pertenece la ubicación receptora (Figura 3.14). Así la concentración de masa de aerosoles M(z) a una cierta altitud z, se puede calcular por una *exponencial decreciente* con la altura y dependiente de un parámetro fijo $H_p$ que depende del modelo en cuestión [30]

$$M(z) = M(0)\exp\left(-\frac{z}{H_p}\right) \qquad (\text{g/m}^3) \qquad (71)$$

donde M(0) es la concentración másica en la superficie g/m$^3$ y $H_p$ toma los siguientes valores: 900 metros para modelo **marítimo**, 730 metros para modelo **continental**, 2.000 metros para un modelo de **desierto** y 30 metros para un modelo **polar**. La variación la introduce la altura z expresada también en metros.

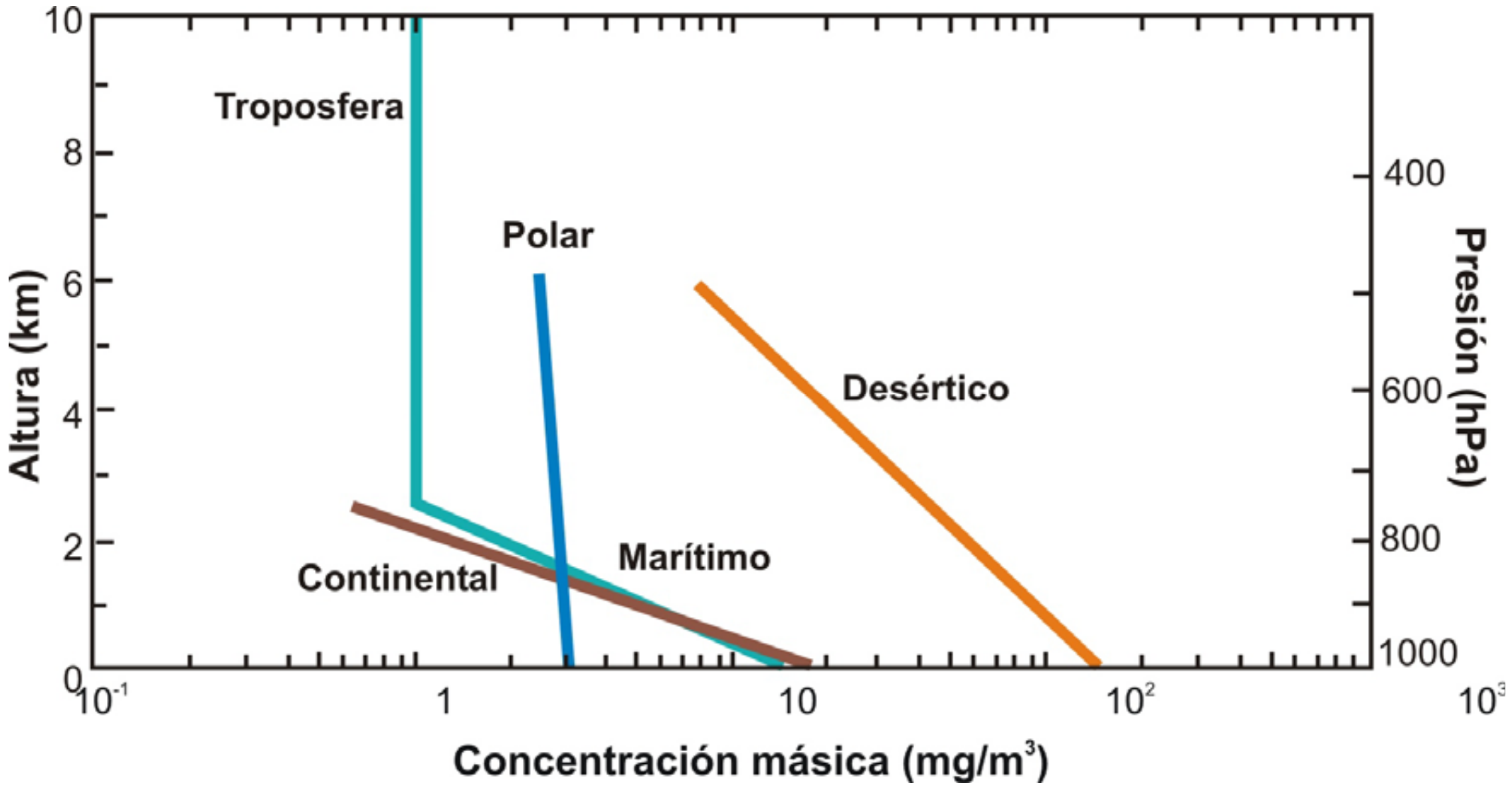

***Figura 3.14. Distribución de aerosoles en función de la altura para varios modelos de atmósfera [30].***

Conocido el volumen V de las partículas, o lo que es lo mismo la distribución de tamaños de los aerosoles que se encuentran en la atmósfera, y la densidad D de dichas partículas, es posible calcular la ***concentración numérica*** (en unidades de partículas/$m^3$), a través de la concentración másica de aerosoles del modelo bajo estudio, de la siguiente manera

$$N_P = \frac{M(z)}{V\,D} \qquad (\text{moléculas}/m^3) \qquad (72)$$

De forma similar al apartado anterior, el coeficiente de atenuación $\alpha_M$ de dichas partículas puede determinarse por el producto de la concentración numérica $N_p$ de aerosoles y la sección eficaz de *scattering* correspondiente $\sigma_M$ como sigue

$$\alpha_M = N_p \cdot \sigma_M \qquad (m^{-1}) \qquad (73)$$

Otro parámetro de interés es la ya citada ***eficiencia de scattering Q*** normalizada, que adquirirá distintos valores según la relación entre el radio a de la partícula y la longitud de onda, y el cociente m entre los índices de refracción de la partícula y del medio. Esta dependencia se muestra en la Figura 3.12, para un valor m de 1,5. En dicha figura se distinguen claramente los dos tipos de *scattering*: la **zona lineal** con la cuarta potencia de λ correspondiente al *scattering* Rayleigh y la **zona sinusoidal** estable para un valor de Q en torno a 2 representativa del *scattering* Mie.

Existe una relación entre la **sección eficaz** de *scattering* y la **eficiencia** a través de la sección geométrica de la partícula ($\pi \cdot a^2$ para el caso de una partícula esférica):

$$\sigma_R = Q \cdot \sigma_{\text{geométrica}} \qquad (m^2) \qquad (74)$$

Fijado un modelo, puesto que la concentración de masa de aerosoles depende de la altura de la atmósfera, el coeficiente de atenuación debido a los aerosoles no será el mismo a una cierta altitud atmosférica que a otra.

### 3.3.4. Turbulencia

El origen de este fenómeno se debe al calentamiento del aire de la atmósfera por el Sol. Durante el día la radiación solar *calienta la superficie* de la Tierra y durante la noche *la superficie se enfría* cediendo calor a la atmósfera. Estos procesos producen turbulentos movimientos de masas de aire, creándose así regiones en la atmósfera a diferentes temperaturas. El movimiento de estas regiones y su ubicación son **aleatorios**. Como el índice de refracción del aire depende de la densidad y las variaciones aleatorias de temperatura mencionadas hacen variar la densidad, el índice de refracción también **varía de forma aleatoria** con las variaciones de temperatura de las masas de aire. Esto significa que el frente de ondas que atraviesa la atmósfera se encuentra con un ***medio no homogéneo***, cuyo índice de refracción varía aleatoriamente en tiempos característicos de varios milisegundos, por lo que se producirá cierta refracción que hará que la potencia recibida en la superficie del telescopio receptor fluctúe. Estas fluctuaciones de potencia harán que también fluctúe la tasa de bits erróneos (*BER,* del inglés *Bit Error Rate*). Si se producen grandes fluctuaciones que resulten en tasas de error de bit relativamente altas se generaría un deterioro considerable de las prestaciones del enlace de comunicaciones.

Debido a la enorme complejidad del proceso que determina la dinámica de las turbulencias atmosféricas, no existe una forma válida para evaluar con precisión el efecto que provocan éstas en el enlace de comunicaciones. Generalmente se emplean modelos empíricos o aproximaciones simples asentadas en la teoría clásica de turbulencia atmosférica de Kolmogorov [32]. A continuación se presenta una manera de estimar el efecto de la turbulencia traducido a pérdidas en el enlace.

En la atmósfera existen capas con diferentes grados de turbulencia para diferentes alturas. $C^2_N$ es un parámetro muy importante, conocido como la ***constante de estructura del índice de refracción***, que describe el perfil vertical de la fuerza o intensidad de la turbulencia. Este perfil se obtiene experimentalmente y varía con el tiempo y la ubicación, por lo que se suelen emplear modelos o directamente mediciones empíricas. En la Figura 3.15 se presenta el modelo de Hufnagel-Valley, que es el más ampliamente difundido.

Para cuantificar la intensidad de la turbulencia sobre el frente de ondas suele utilizarse el **parámetro $r_0$**, denominado ***parámetro de Fried***. Entre las interpretaciones que pueden darse a este parámetro hay una muy intuitiva: representa el *diámetro* que tendría que poseer un telescopio para recibir en su superficie, en presencia de atmósfera turbulenta, *un frente de ondas sin alteraciones debidas a turbulencias*[1]. Esto es, el parámetro de Fried determina la cantidad del frente de ondas que se propaga sin verse distorsionado por la turbulencia. Un valor elevado de $r_0$ se traducirá en un efecto moderado de la turbulencia sobre el frente de ondas. La expresión para hallar $r_0$ suponiendo un frente de ondas plano (esta suposición es perfectamente correcta en comunicaciones en espacio profundo debido a que la gran distancia a la fuente hace que el frente de ondas incidente se vea como un frente de ondas planas) viene dada por [47]

$$r_0 = 0{,}185\,\lambda^{6/5}\left[\frac{\cos(\theta)}{\int_0^h C_N^2(z)dz}\right]^{\frac{3}{5}} \qquad (m^{6/5}) \tag{75}$$

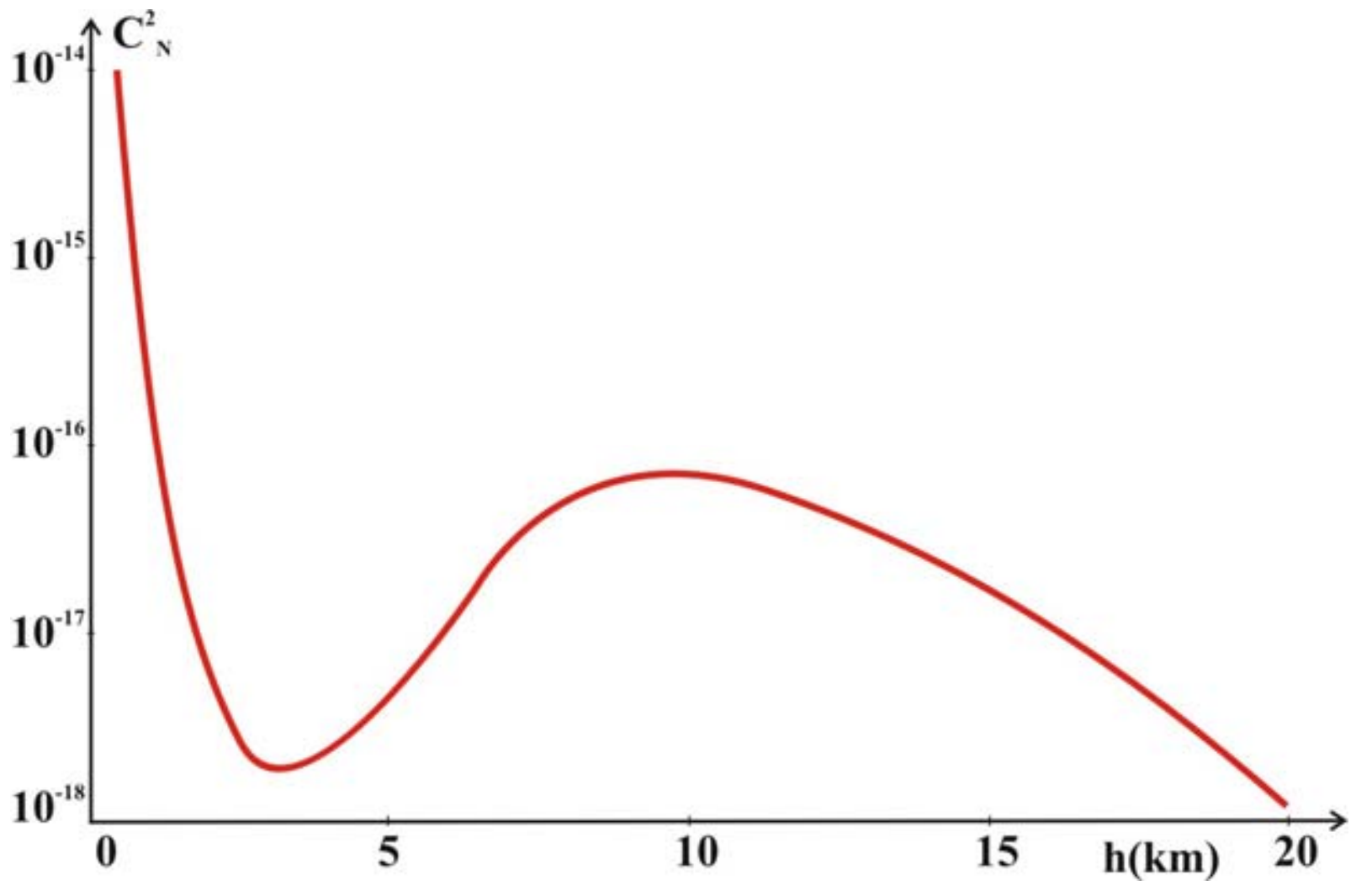

*Figura 3.15. Constante de estructura del índice de refracción [33].*

[1] Por ello el parámetro de Fried determina el tamaño de los elementos ópticos que forman el corrector de haz en un sistema de óptica adaptativa.

donde θ representa el ángulo cenital (cuyo valor crece de 0º a 90º hasta llegar al horizonte), h la altura de la parte de la atmósfera turbulenta y $C^2_N$ la ya mencionada constante de estructura del índice de refracción.

Se puede comprobar en la ecuación (75) que cuanto más fuerte es la turbulencia (un valor de $C^2_N$ elevado) más pequeño es el parámetro $r_0$. También se observa que $r_0$ disminuye a medida que el ángulo cenital aumenta, o lo que es lo mismo, a medida que la dirección desde la que se recibe el frente de ondas se acerca al horizonte (el espesor de atmósfera que tiene que recorrer es mayor). Por último, la dependencia de λ muestra que la misma turbulencia afecta en mayor medida a longitudes de onda menores, lo cual supone un inconveniente en comunicaciones ópticas, donde se trabaja con longitudes de onda muy pequeñas.

En función de cual sea el campo de visión del telescopio se recibirá una potencia u otra en el plano focal. Por lo tanto las pérdidas en el plano focal para distintos ángulos de visión se obtendrán al evaluar la energía concentrada de la ***función de dispersión puntual*** (PSF del inglés *Point Spread Function*) calculada como [34]

$$PSF = 2\pi\int_0^1 v\frac{2}{\pi}\left[\cos^{-1}(v) - v\sqrt{1-v}\right]\exp\left[-3{,}44\left(\frac{D}{r_0}v\right)^{5/3}\right]J_0(2\pi vr)dv \qquad (76)$$

donde r es el desplazamiento angular desde el centro del *spot* (normalizado por λ/D), D el diámetro del telescopio y $r_0$ el parámetro de Fried.

Se ha presentado una de las formas que existen para estimar el efecto que la turbulencia tendría sobre el enlace de comunicación. Este efecto se traduce en unas pérdidas adicionales que habrá de afrontar el sistema para que no se vea limitado por la turbulencia atmosférica. Por lo tanto habrá que llegar a un compromiso entre el **ángulo de visión** del telescopio y las **pérdidas** asociadas a éste. Cabe añadir que el efecto de la turbulencia atmosférica en las fluctuaciones aleatorias de potencia en el telescopio puede mitigarse recurriendo a mayores aperturas de telescopio y/o a mayor cantidad de receptores, pero la técnica más efectiva (obviando la posibilidad de sacar al receptor de la atmósfera situándolo en un satélite) es la óptica adaptativa ya mencionada anteriormente.

## 3.4. ENLACE DE SUBIDA

La función principal del enlace de subida es la de asistir al transmisor del terminal espacial para efectuar el **apuntamiento** más preciso posible y también para controlar el mismo durante la comunicación. También debe tener la posibilidad de establecer comunicación con el terminal espacial (para su control), aunque siempre de menor capacidad que el canal descendente. Este proyecto está fundamentalmente dedicado al canal descendente, que es el que supone un mayor reto tecnológico, y el que determina fundamentalmente la calidad del enlace.

El enlace ascendente es de ***muy baja velocidad*** en comparación con el de bajada. El motivo principal es que no hay razón para que sea rápido. No se precisa gran cantidad de información en el terminal espacial ya que el flujo de datos es principalmente descendente.

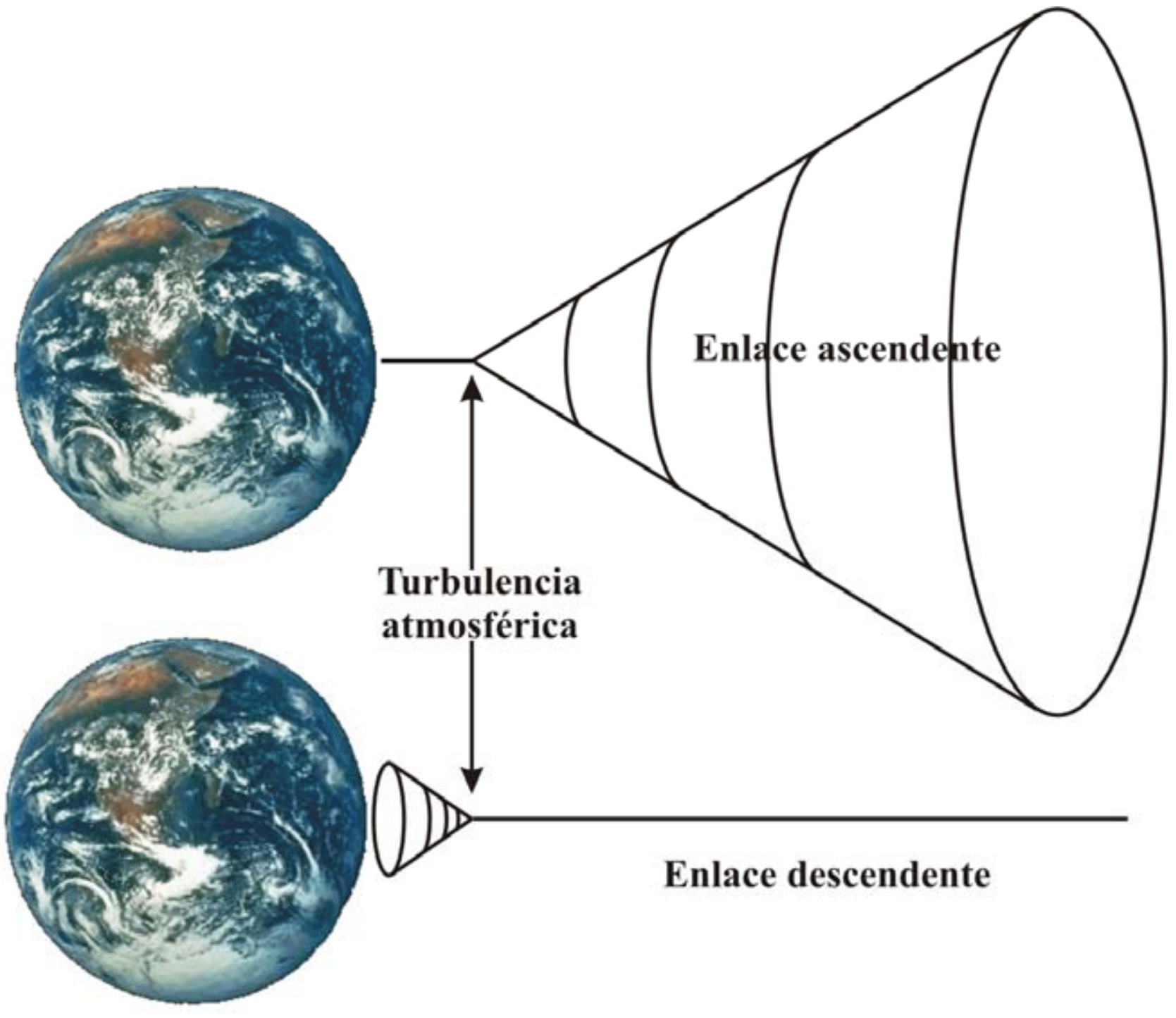

***Figura 3.16. Refracción del haz en el enlace ascendente y descendente debida a la turbulencia atmosférica.***

Por otro lado, en el caso de que el transmisor esté en la superficie de la Tierra, la turbulencia atmosférica juega un papel fundamental. Como se ha comentado en el apartado 3.3.4, la turbulencia atmosférica induce una refracción aleatoria en el haz óptico que la atraviesa. Por supuesto esta refracción es la misma para el enlace ascendente como para el descendente, pero en el caso del segundo el efecto sobre la comunicación es mucho menor. La razón, como se puede observar en la Figura 3.16, es que en el enlace de bajada la refracción se produce en el *último tramo del trayecto*. En cambio, en el caso del enlace de subida, la refracción del haz tiene lugar en los primeros kilómetros, lo que se traduce en una **gran divergencia** a lo largo de todo el camino ya que cualquier pequeña desviación inducida al principio provoca una gran desviación a lo largo de la gran distancia existente entre los terminales. En consecuencia, la densidad de potencia que llega al terminal espacial se puede llegar a reducir considerablemente.

Esta refracción del enlace ascendente puede llegar a comprometer la recepción de un canal con la suficiente capacidad para permitir un control continuo del apuntamiento, por lo que el terminal espacial deberá disponer de un buen sistema autónomo de estabilización frente a los movimientos y vibraciones de la nave. En todo caso, al disponer de un emisor en tierra, no existen las grandes limitaciones de potencia del canal de bajada. Si es preciso, el canal de subida puede operarse con potencias de MW o más altas, lo que simplifica decisivamente su diseño. Por otra parte, la antena emisora puede ser también mucho mayor si se desea. A diferencia del emisor remoto, el canal de subida puede disponer de un telescopio astronómico como antena si resulta necesario. Alternativamente, se puede evitar el problema de la turbulencia atmosférica si el transmisor estuviera situado en un satélite alrededor de la Tierra.

Como ejemplo de canal de subida (aunque no de comunicaciones) cabe citar que aún en la actualidad se realizan sistemáticamente medidas de distancia a la Luna empleando láseres de potencia que inciden sobre los ***retrorreflectores*** dejados en la superficie lunar por los astronautas del Apollo XI (Figura 3.17). Tales medidas, de gran precisión (2 cm), han permitido estudiar fenómenos sísmicos lunares y sutiles alteraciones de su órbita debidas a varias causas, así como realizar medidas precisas del parámetro de Fried [35].

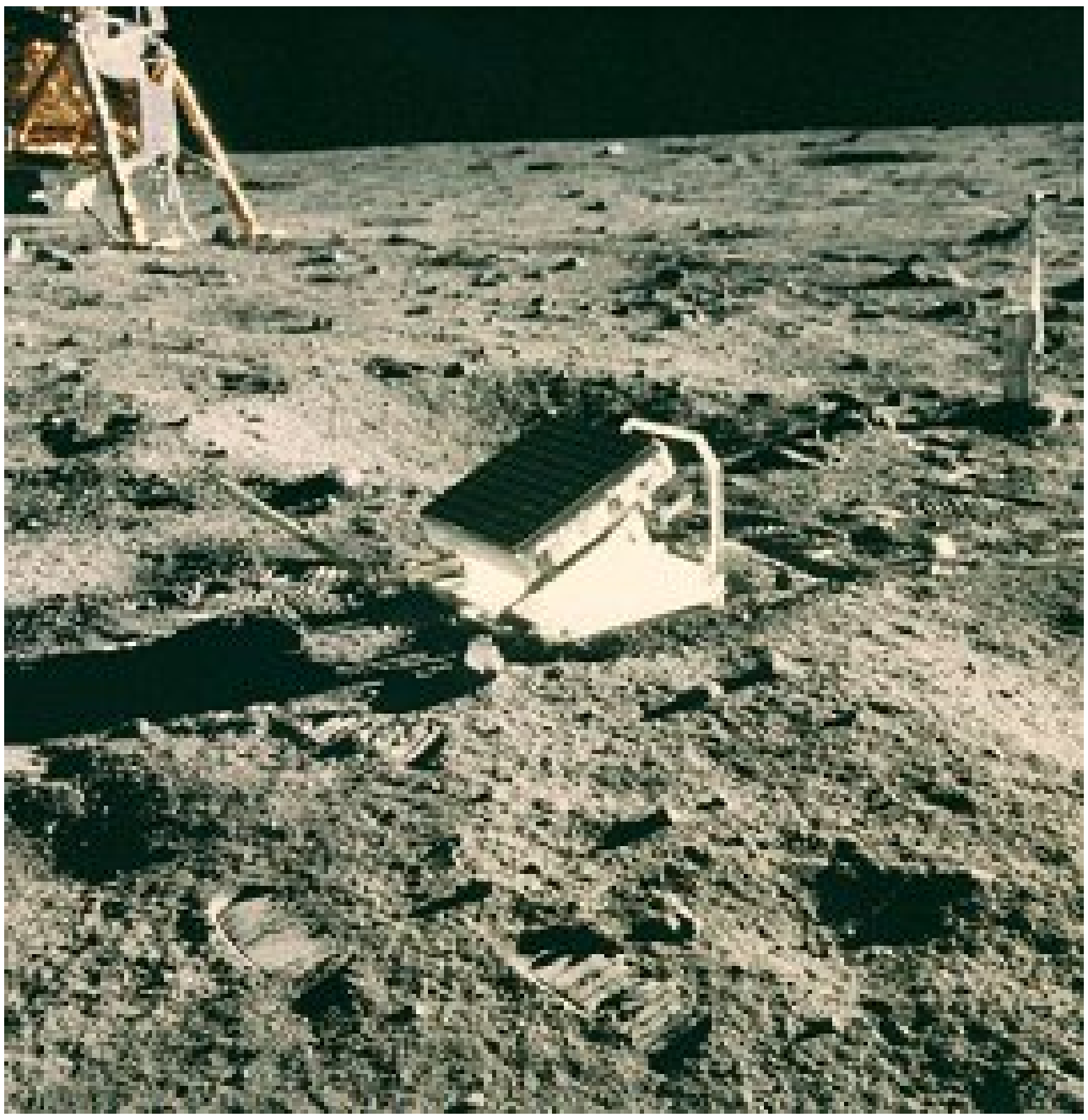

***Figura 3.17. Agrupamiento de retrorreflectores dejados en la Luna por Neil Armstrong durante la misión Apollo XI [36]***

# 4. El transmisor remoto

En este capítulo se analizan las distintas posibilidades por las que puede optarse para seleccionar un transmisor para el canal de bajada. Como ya se ha indicado en el capítulo anterior, el proyecto se concentra en este canal, que determina la factibilidad y calidad del enlace.

Un transmisor óptico de un enlace a través del espacio profundo, se compone básicamente de

- ***Emisor óptico***
- ***Modulador***
- ***Codificador***
- ***Antena de transmisión.***

La función de la antena de transmisión, compuesta por un conjunto de lentes o espejos, es enfocar el haz óptico para llevar a cabo la transmisión por el canal espacial. Para su diseño será necesario considerar los límites de difracción en haces que se propagan en el vacío. Estos límites están condicionados por la **longitud de onda de trabajo**. Como se explica en el apartado correspondiente, la frecuencia óptica de transmisión es uno de los parámetros más críticos que determinan las prestaciones del enlace. El uso de longitudes de onda menores, en comunicaciones ópticas espaciales, hace que la anchura del haz sea menor y en consecuencia también el tamaño del *spot*. Como contrapartida, la menor anchura dificulta el apuntamiento.

## 4.1. FUENTES ÓPTICAS

Las tres tecnologías láser existentes para comunicaciones espaciales son

- ***Láser de $CO_2$***. Es un láser de gas (una mezcla de $CO_2$, $N_2$ y metano $CH_4$) con bombeo eléctrico, que trabaja en el infrarrojo medio, en torno a 10,6 µm o a 9,6 µm.

- ***Láser de Nd:YAG***. Es un láser (bombeado por otro diodo láser) de estado sólido cuyo material activo está formado por un granate de aluminio e ytrio dopado con neodimio. Funciona con bombeo óptico (una lámpara o un diodo láser), y trabaja a 1,064 µm (también puede trabajar al doble de frecuencia). Existen otros tipos semejantes de láseres de estado sólido dopados con tierras raras, que operan en distintas longitudes de onda del IR próximo.

- ***Láser de semiconductor***. Existen diferentes tipos y configuraciones, desarrollados fundamentalmente como fuentes para comunicaciones ópticas guiadas por fibra, que trabajan en la región del IR próximo, entre 0,78 µm y 1,55 µm.

Inicialmente, los láseres de $CO_2$ resultaban bastante atractivos para su uso en comunicaciones debido a su gran eficiencia de conversión electroóptica y a su largo tiempo de vida. Los problemas con estos láseres eran su elevada longitud de onda y baja fiabilidad. El rápido desarrollo que experimentó la tecnología de diodos láser a partir de la década de los 80, ha hecho que ahora las principales fuentes de luz sean dispositivos basados en láseres de semiconductor y láseres de Nd:YAG.

La alta fiabilidad, eficiencia y el pequeño tamaño y peso, hacen de los láseres de semiconductor una de las fuentes de luz por excelencia en comunicaciones ópticas espaciales. Se consiguen actualmente potencias del orden de decenas de W, aunque la fiabilidad y tiempo de vida de estos dispositivos es bastante mayor en potencias en torno a cientos de mW a 1 W. Si se necesitan potencias instantáneas de centenares de W, como puede ser el caso en comunicaciones en el espacio profundo, la alternativa actual más plausible sería un *láser de estado sólido bombeado por diodos láser*. Como alternativa, podría estudiarse combinar la potencia de varios láseres de semiconductor formando un *array*.

Por otra parte, se debe establecer el **modo de funcionamiento** del láser, que puede ser ***continuo*** o ***pulsado***. En el primer caso, el flujo de datos se introduce por medio de un **modulador externo**, mientras que en el segundo, la señal **modula directamente** el propio láser. La principal ventaja de un modulador externo es que se estabiliza notablemente la emisión del láser. Sin embargo, el funcionamiento en continua supone un *aprovechamiento poco eficiente* de la potencia disponible. Tradicionalmente, los láseres de semiconductor se utilizan en modulación directa, a menos que se empleen tasas binarias muy elevadas (por encima de 10 Gb/s, fuera del rango esperable, por el momento, en misiones en espacio profundo). Los láseres de estado sólido, como el Nd:YAG, pueden operar tanto de forma continua como pulsada, con el auxilio para esta última de un dispositivo intracavidad para generación de pulsos, como el conmutador de Q (*Q-switching*) o el anclaje de modos (*mode-locking*).

## 4.2. EL EMISOR EN EL ESPACIO

Un importante factor a considerar a la hora de seleccionar una fuente como emisor remoto es el comprensible *conservadurismo* que se aplica estrictamente a todos los elementos enviados al espacio. Como regla general, no se utilizarán tecnologías muy recientes, aunque resulten ventajosas, frente a otras más "clásicas" o probadas. En el caso que nos ocupa, el emisor remoto tendrá que basarse en una tecnología sobradamente conocida, y que preferentemente haya sido ya experimentada en el espacio (que "*haya volado*", se dice en el argot). Estas estrictas normas se alivian en el caso del terminal receptor terrestre, el cual puede ser reparado o sustituido en caso de fallo.

### 4.2.1. Longitud de onda de trabajo

La elección de la longitud de onda de la señal dentro del rango óptico la determinarán fundamentalmente dos criterios

- La absorción de la atmósfera terrestre muestra para distintas longitudes de onda ***ventanas*** propicias para la comunicación ya que minimizan las pérdidas asociadas a este fenómeno. Por ello se hace necesaria la elección de una longitud de onda que esté dentro de una de dichas ventanas de mínima absorción en el caso de que la señal tenga que atravesar la atmósfera. En el apartado 3.3.2 dedicado a la absorción atmosférica se pueden apreciar estas ventanas. Este criterio no es aplicable en el caso de que el terminal terrestre sea un satélite orbitando alrededor de la Tierra.

- Debe existir una fuente óptica capaz de generar una señal a la ***longitud de onda elegida***, así como todos los demás elementos necesarios para trabajar con ella en el sistema completo. Todas las fuentes citadas en el apartado anterior cumplen este criterio. Existen ventanas atmosféricas en las regiones de 1,064 μm y 10,6 μm. Los láseres de semiconductor, por su parte, ofrecen la posibilidad de transmitir en un *gran número de longitudes de onda* del IR próximo, por lo que pueden identificarse sin dificultad fuentes que tengan su pico de emisión en alguna de las ventanas atmosféricas.

### 4.2.2. Selección del emisor

Entre todas las posibles alternativas, el ***láser de Nd:YAG*** podría ser el más adecuado. Se trata de una tecnología sobradamente probada (actualmente se encuentran en aplicaciones que van desde punteros láser de bolsillo a gigantescas instalaciones para fusión nuclear que generan TW de potencia). Su emisión se sitúa en el IR próximo (lo cual mejora sustancialmente la divergencia frente al $CO_2$). La longitud de onda de emisión, 1,064 μm, coincide con una de las ventanas atmosféricas de menor atenuación. Por otra parte, se posee gran experiencia en dispositivos de generación de pulsos adecuados para este láser, específicamente conmutadores de Q (la técnica de bloqueo de modos produce trenes de pulsos con una frecuencia fija de centenares de MHz, que resulta menos apropiada para el caso de las comunicaciones en el espacio).

Las bandas de bombeo más eficientes (Figura 4.1) para un láser de Nd:YAG están en 808, 869 y 885 nm [37] Estas longitudes de onda de bombeo se pueden conseguir fácilmente mediante diferentes tipos de láseres de semiconductor. Cabe destacar que los requisitos que se exige a un láser de semiconductor son menos estrictos cuando se usan como fuentes de bombeo que cuando se emplean como fuentes de luz. Así, este esquema funciona como un conversor de continua a pulsos muy eficiente.

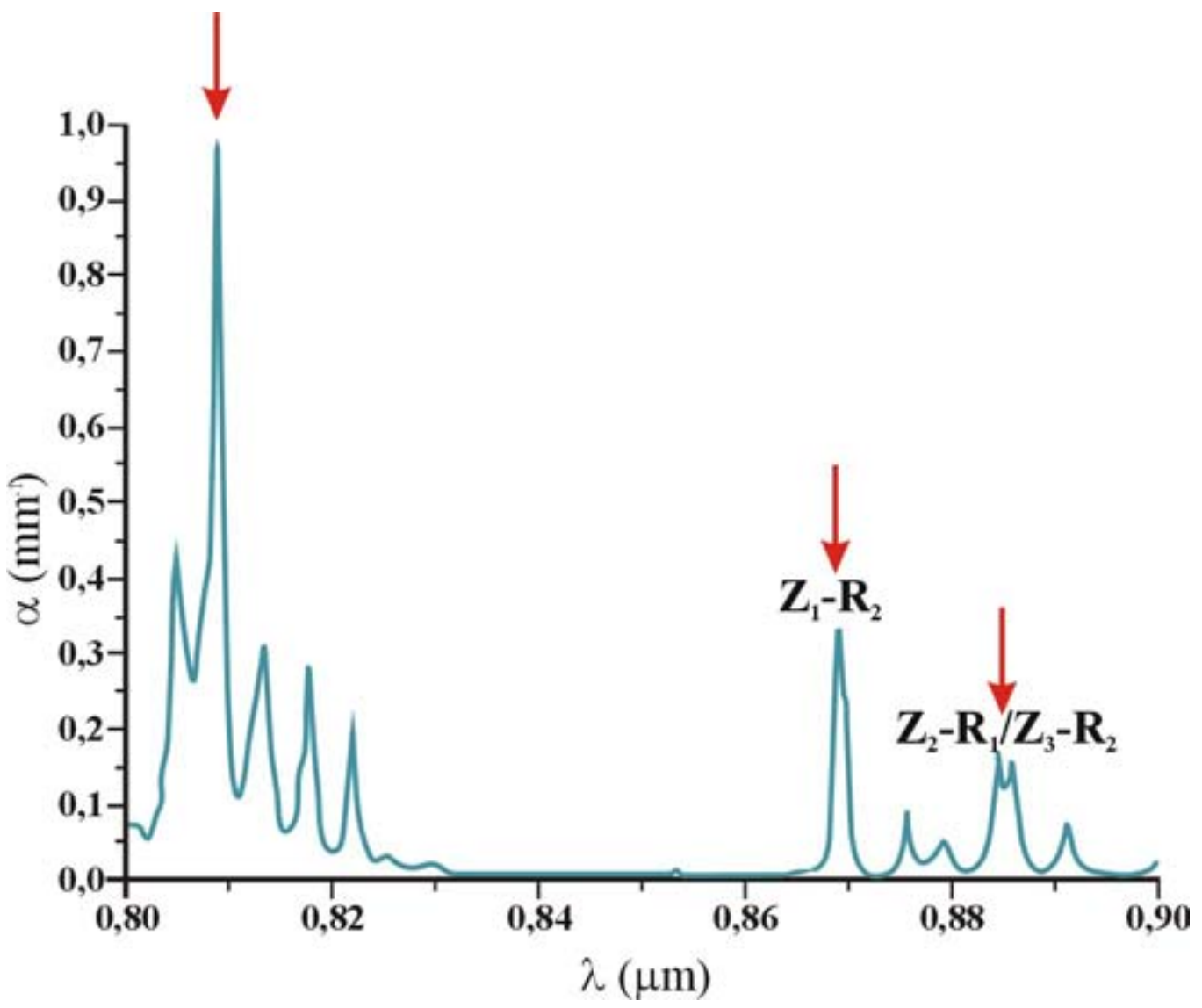

***Figura 4.1. Espectro de absorción del Nd:YAG.***

Las principales ventajas de un láser Nd:YAG bombeado por un diodo semiconductor son su reducida anchura espectral, del orden de MHz frente a los GHz de un diodo láser de semiconductor típico, y su casi despreciable ruido de exceso. Sin embargo, presentan un inconveniente, su baja eficiencia de conversión eléctrico-óptica. Actualmente, este problema está siendo superado, pues ya se han desarrollado y probado láseres con eficiencias del 8.5% al 17% [9].

Como se ha comentado más arriba, existe toda una familia de láseres de estado sólido basados en una matriz inerte (usualmente una piedra semipreciosa como el granate o el zafiro o un vidrio amorfo) y un dopante activo. El ejemplo más clásico es el láser de **rubí** (cuyo dopante es el Cr), que fue el primer láser que se demostró experimentalmente. Otro ejemplo más reciente, de amplio uso, es el láser vibrónico sintonizable de **Ti-zafiro**. El **Nd-vidrio** es un ejemplo de láser de matriz amorfa que se emplea en aplicaciones de potencia. Por lo tanto, la selección del láser puede extenderse, si se diera el caso, a otros láseres de estado sólido, sobre puesto que todos ellos comparten parecidas propiedades de estabilidad, fiabilidad y experiencia de uso. Así, en el proyecto *MLCD* que se desarrolla en el capítulo 6, el emisor seleccionado es un láser de fibra (matriz amorfa) cuyo elemento activo es el **Yterbio** (Yb), que emite a 1,064 μm, la misma longitud de onda que el Nd.

## 4.3. FORMATO DE MODULACIÓN

En un enlace de comunicaciones ópticas típico, la señal modula directamente el emisor láser, y el formato de modulación suele ser muy simple, transmitiéndose los bits por encendido y apagado del láser (como se verá a continuación, esta técnica se conoce como OOK). Cuando se trata de un enlace óptico en espacio profundo, sin embargo, los condicionantes del enlace pueden hacer que convenga elegir algún otro tipo de modulación.

El planteamiento del enlace de bajada se basa en *descargar* al máximo los requerimientos del emisor remoto, aunque sea a costa de *cargar* las prestaciones exigidas al receptor. Se requiere además que la **potencia de pico** de los pulsos sea elevada, con el fin de salvar con éxito las enormes distancias del canal. No obstante la **potencia media** de transmisión debe ser moderada (unos pocos W) para reducir el consumo eléctrico, que debe ser compatible con la limitada potencia que proveen las baterías solares.

En este contexto, surgen como interesantes alternativas otros tipos de modulación cuyo denominador común es la posibilidad de ***codificar más de un bit*** por pulso. A continuación se analizan varios formatos posibles para la elección de la modulación de las señales.

### 4.3.1. Modulación de apagado-encendido (OOK)

La técnica de modulación OOK (del inglés *On-Off Keying*) es la más simple que se puede utilizar. Conocida vulgarmente como modulación por "señales de humo" consiste únicamente, como se acaba de mencionar, en encender o apagar el láser, de manera que un bit "1" corresponde al encendido del láser con la consiguiente transmisión de un pulso y un bit "0" corresponde a no transmitir nada durante el periodo de bit.

El esquema de modulación OOK puede ser **NRZ** (sin retorno a cero), o **RZ** (con retorno a cero). En el esquema RZ la duración del pulso encendido es menor que el tiempo de bit, mejorando la eficiencia luminosa y la recuperación de reloj, a cambio de empeorar el ancho de banda del receptor. Estos dos esquemas son los más usados en comunicaciones ópticas guiadas.

Si se llama *slot* al mínimo intervalo de tiempo considerado, en OOK se transmite un bit por *slot*. El concepto de *slot* sólo tiene utilidad en OOK para compararla con otras técnicas, como se verá a continuación. En cierto modo las técnicas de modulación que se tratan a continuación podrían verse como versiones de OOK precodificadas más que como modulaciones en sí mismas. Dado que las demás modulaciones implican imposiciones que no afectan a OOK, esta técnica puede ser considerada como la que mayor tasa de bits por *slot* alcanza (como se ve en la Figura 4.2 en OOK-NRZ coincide la duración de slot con la duración de bit T). Si se supone la misma probabilidad para bits "0" que para bits "1", para OOK-NRZ se concluye que

$$\left(\frac{P_{pico}}{P_{media}}\right)_{OOK} = 2 \tag{77}$$

Por último, si la duración de bit es T, la expresión para el ancho de banda queda

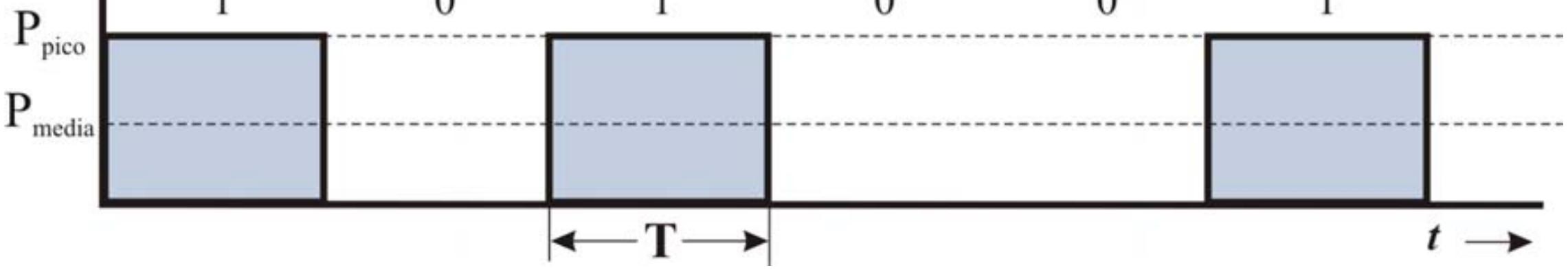

*Figura 4.2. Modulación de la secuencia 101001 en OOK-NRZ.*

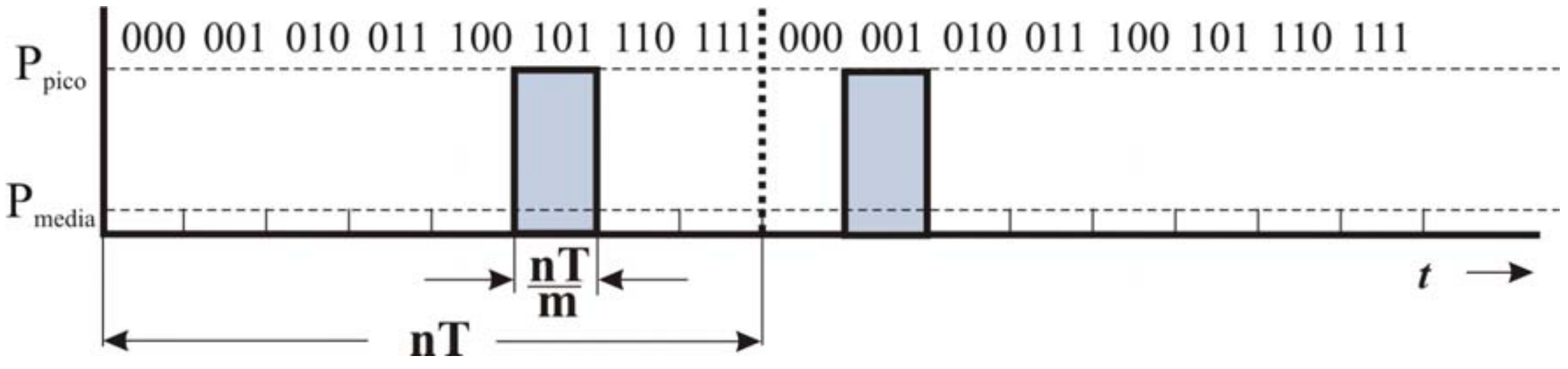

*Figura 4.3. Modulación de la secuencia 101001 en 8-PPM.*

$$B_{OOK} = \frac{1}{T} \tag{78}$$

## 4.3.2. Modulación por posición de pulsos (PPM)

La modulación PPM m-aria o m-PPM (del inglés *Pulse Position Modulation*) consiste en dividir (si se compara con una OOK de igual régimen binario) la duración de cada secuencia de n bits de OOK en $m = 2^n$ *slots*, correspondientes a los m símbolos que se pueden codificar. Cada vez que se envía un pulso, éste va colocado en uno de estos *slots*, definiendo así su valor por su posición dentro del intervalo. La posición del pulso dentro del intervalo de tiempo formado por los m *slots*, codifica simultáneamente n bits. Por lo tanto, dividiendo el periodo, por ejemplo, en $m = 8$ *slots* se pueden codificar simultáneamente 3 bits con cada pulso (Figura 4.3). Los bits que se codifican con m slots son $n = \log_2(m)$, por lo tanto la tasa de bits por *slot* de m-PPM será de $\log_2(m)/m$.

Dado que la modulación PPM emite un pulso cada m *slots*, se concluye que

$$\left(\frac{P_{pico}}{P_{media}}\right)_{m-PPM} = m \tag{79}$$

Cada periodo de símbolo tiene una duración de n·T (siendo T la duración de un bit en una OOK de igual régimen binario), por lo que cada *slot* dura (n·T)/m. Esto quiere decir que el ancho de banda respecto a una OOK del mismo régimen binario es

$$B_{m-PPM} = \frac{m}{n} B_{OOK} \tag{80}$$

## 4.3.3. Modulación por posición de multipulsos (MPPM)

La modulación MPPM puede verse como una generalización de m-PPM. Un grupo de $\log_2\binom{m}{k}$ bits se codifican mediante k pulsos en m *slots*. La modulación m-PPM es una modulación MPPM en que k = 1 o una MPPM de un pulso.

Cada grupo de $\log_2\binom{m}{k}$ bits determina la localización de k pulsos en los m *slots* (Figura 4.4). Por lo tanto se deduce que [38]

$$\left(\frac{P_{pico}}{P_{media}}\right)_{MPPM} = \frac{m}{n} \quad (81)$$

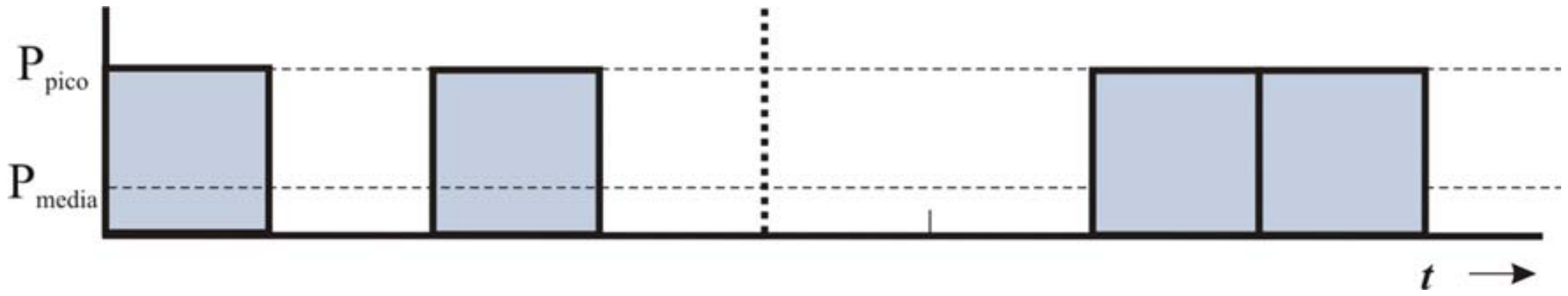

*Figura 4.4. Modulación de la secuencia 101001 en MPPM.*

## 4.3.4. Modulación diferencial por posición de pulsos (DPPM)

Es similar a m-PPM excepto que después del *slot* que contiene el pulso, los siguientes *slots* que no contienen pulso no son transmitidos y en su lugar comienza un nuevo símbolo (Figura 4.5).

La relación entre potencia media y pico queda

$$\left(\frac{P_{pico}}{P_{media}}\right)_{DPPM} = \frac{m}{2} \quad (82)$$

De esta manera, la duración de los símbolos es la mitad que en m-PPM (suponiendo equiprobabilidad para "1" y "0"), permitiendo un mayor régimen binario. El inconveniente de DPPM es que la potencia media también es el doble y puede ser difícil detectar la duración variable de los símbolos.

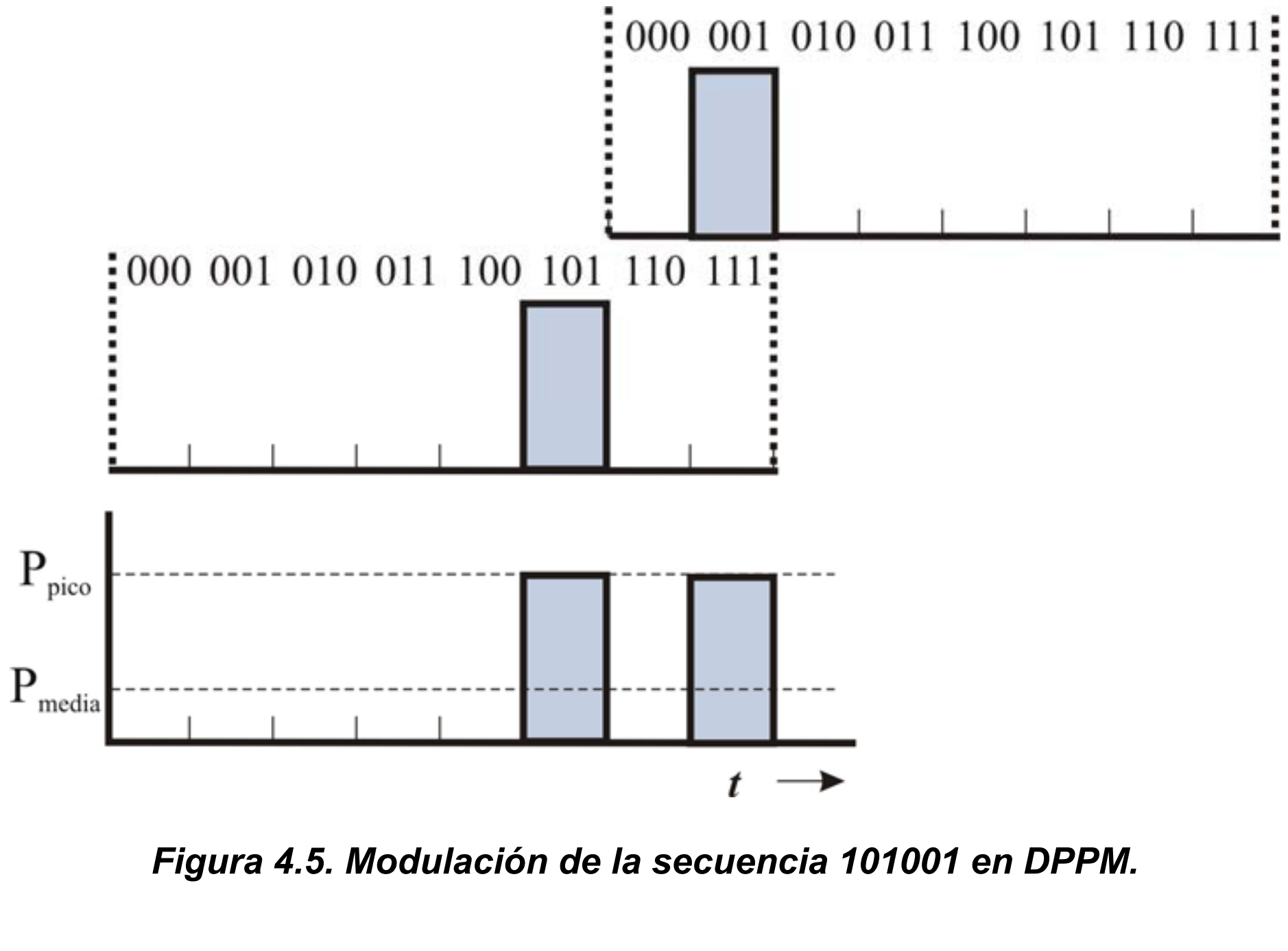

*Figura 4.5. Modulación de la secuencia 101001 en DPPM.*

**COMPARACIÓN ENTRE OOK Y PPM**

La modulación **PPM** es una posible alternativa a **OOK** que puede reducir el ciclo de trabajo del láser o mejorar la relación señal-ruido. Para evaluarla se puede comparar, por ejemplo, un sistema OOK-NRZ con un sistema PPM de 64 símbolos (64-PPM). En OOK se necesitarían 6 bits para codificar uno de los 64 símbolos. Si se supone que se codifica el símbolo en PPM en el tiempo de duración de los 6 bits de OOK, el incremento de ancho de banda en PPM sería, como ya se ha visto, de 64/6. A cambio, en PPM el gasto en potencia sería 1/32 (suponiendo la misma probabilidad de "1" y "0" para OOK y PPM). Es decir, se podría emitir (para la misma potencia media del emisor) con una **potencia de pico 32 veces mayor** sacrificando el ancho de banda en un factor 64/6.

Esta mejora también se traduce en una ***mejora de la relación señal/ruido***. Si se supone que el ruido predominante es el ruido *shot*, la señal aumenta en un factor de $32^2 = 1024$ (la potencia eléctrica es proporcional al cuadrado de la corriente) frente al ruido que aumenta en un factor de $32 \cdot (64/6) = 341,3$ (el factor 2eBI del ruido *shot* se transforma en $2e(B \cdot 64/6)(I \cdot 32)$. Si el ruido predominante fuera térmico, la mejora sería de 1024 frente a $64/6 = 10,7$ (el ruido térmico es proporcional a B, pero no a I).

Obviamente el aumento de ancho de banda en PPM no es asumible de forma indefinida. Así, en comunicaciones ópticas guiadas, la ***"estrechez"*** de los pulsos viene limitada por la **dispersión temporal** de los mismos y por el ancho de banda del receptor. En sistemas de comunicaciones ópticas no guiadas, en cambio, la dispersión es despreciable en todos los casos, y la única limitación viene dada por el ancho de banda del receptor.

## 4.3.5. Conclusión

La obvia ventaja de un sistema de codificación de bits múltiples como PPM y sus variantes es el **ahorro de potencia** del emisor. Alternativamente, en comunicaciones en espacio profundo, es más interesante poder incrementar la potencia de pico manteniendo la potencia media, al dedicar la energía necesaria para la transmisión de varios bits a un solo pulso de corta duración y elevada potencia de pico. El inconveniente principal de la técnica es que se incrementa considerablemente el ancho de banda necesario para el detector.

Como se indicaba más arriba, esta situación puede ser deseable en un emisor remoto en espacio profundo, ya que resulta mucho más simple (y económico) mejorar las prestaciones del receptor local que las del emisor. Además, por lo que respecta a la relación señal-ruido, la modulación PPM ***es preferible*** a la OOK, a pesar del incremento de nivel de ruido que produce el aumento de ancho de banda del detector. En el cuadro resumen se realiza un análisis comparativo de ambas técnicas.

El formato de modulación sugerido para el proyecto *MLCD* (véase capítulo 6) es 256-PPM debido a que optimiza de forma eficiente la escasa potencia media (5 W) del terminal espacial. Teniendo en cuenta las especificaciones exigidas al enlace, tal modulación requiere que el ancho de banda del receptor sea **superior a 1 GHz**. Como se muestra en ese capítulo, el diseño de un receptor con esas características ha resultado muy complejo, debido a la escasez de componentes comerciales con tales características que pudieran adaptarse al enlace propuesto.

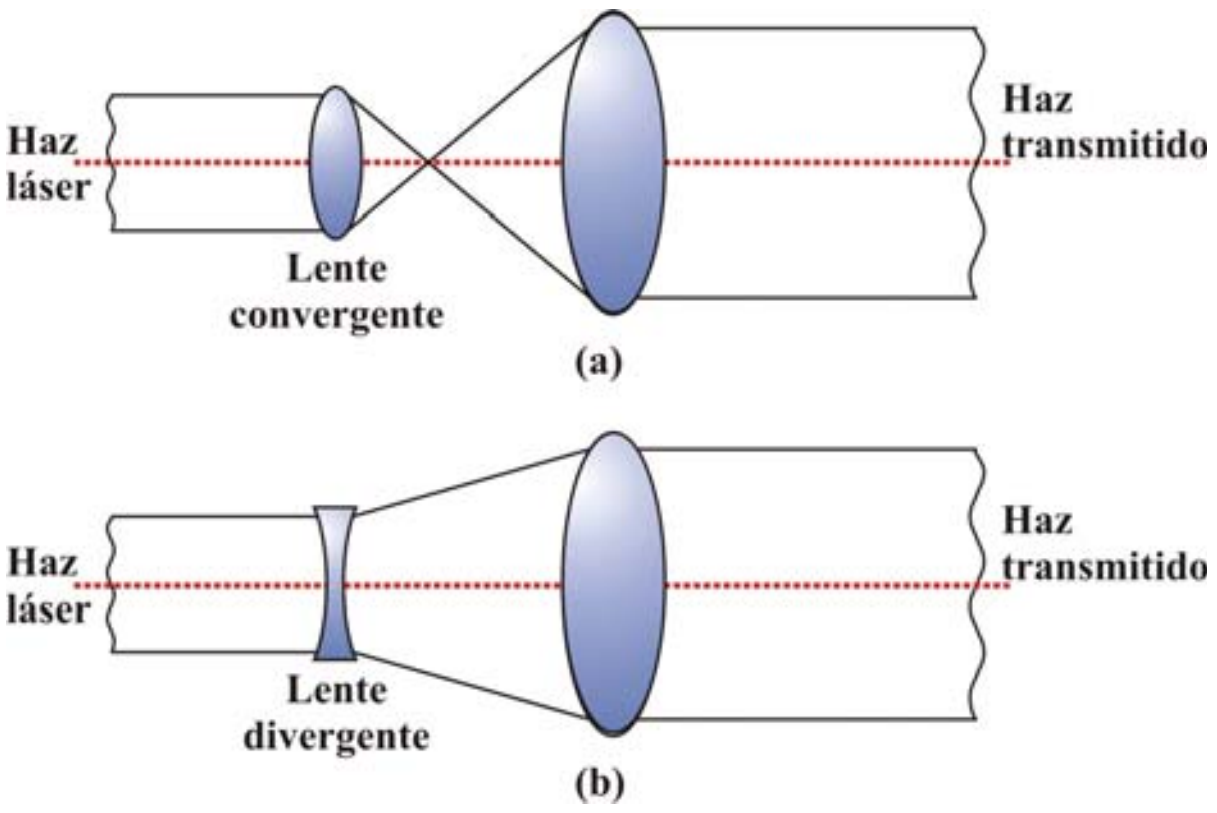

***Figura 4.6. Telescopio transmisor con óptica refractiva (a) Newton y (b) Cassegrain.***

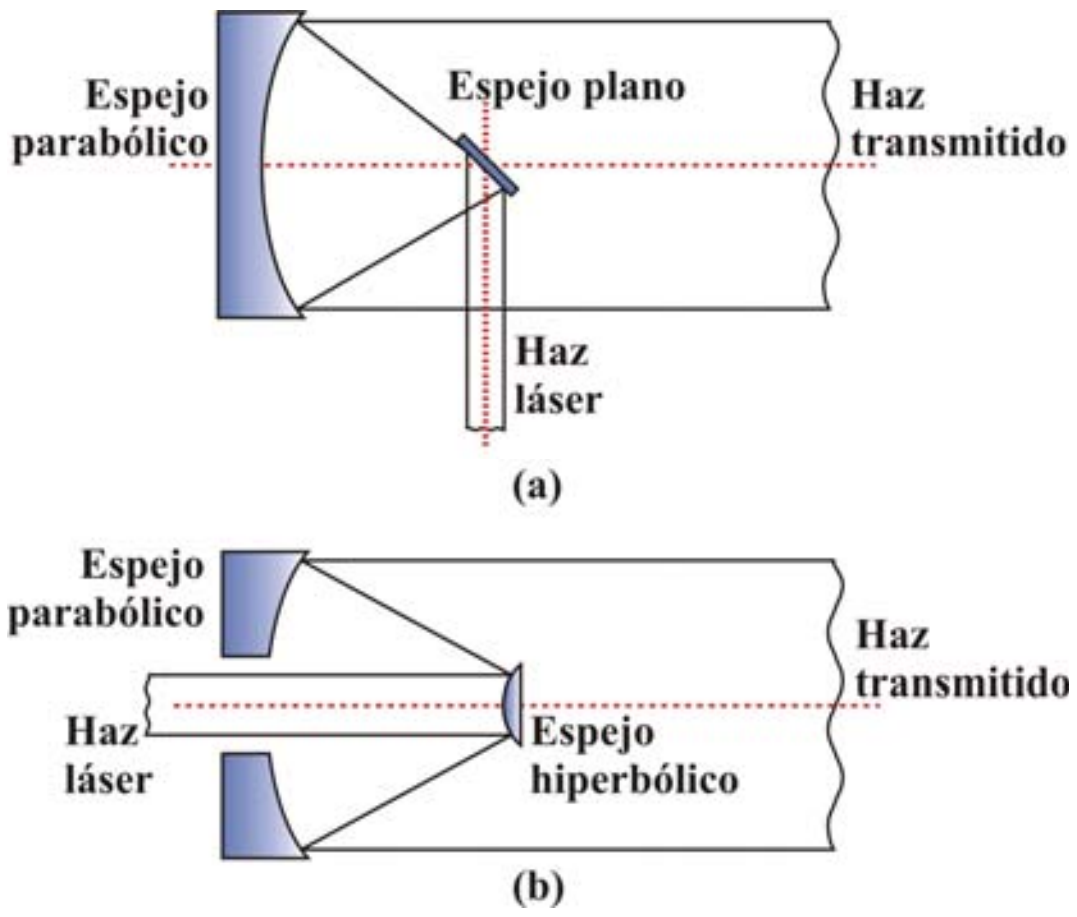

***Figura 4.7. Telescopio transmisor con óptica reflexiva (a) convergente y (b) divergente.***

## 4.4. LA ANTENA

Las antenas del emisor y el receptor de un sistema de comunicaciones ópticas no guiadas son básicamente ***colimadores***, es decir, instrumentos ópticos cuya función es ampliar el **diámetro** del haz para reducir proporcionalmente su **divergencia**.

Cuando se trata de un enlace en el espacio profundo, la actuación de la antena se hace esencial. Ya se han calculado en el capítulo anterior las enormes pérdidas de espacio libre que presenta un enlace de comunicaciones ópticas a tan larga distancia. La colimación del haz del emisor (que determina en último término la **ganancia** de la antena) es fundamental para reducir esas pérdidas a niveles que permitan la consecución del enlace dentro de las especificaciones que se solicitan.

Si, como en este caso, el emisor es un láser, el haz emergente está ya apreciablemente colimado. La antena del emisor se configura entonces como un ***telescopio astronómico***. Es éste un instrumento óptico no formador de imagen que se puede hacer equivaler a un conjunto de dos **lentes** –o dos **espejos**, o un espejo primario y una lente ocular– confocales (es decir, con un foco común). Introduciendo el haz por la lente de menor focal –el ***ocular***–, el haz de salida del ***objetivo*** tiene un **diámetro** que se ve incrementado por un factor

$$\Delta D = \frac{f_1}{f_2} \tag{83}$$

siendo $f_1$ y $f_2$ las distancias focales del objetivo y el ocular respectivamente. Por su parte, la apertura angular (la ***divergencia***) del haz *se ve reducida precisamente por el mismo factor*, que es el efecto que se desea.

El telescopio transmisor puede basarse en óptica reflexiva o refractiva. Estos dos tipos de telescopios se muestran en la Figura 4.6 y la Figura 4.7 respectivamente. Para tamaños de apertura pequeños, son más prácticas las lentes (óptica refractiva), que pese a introducir aberración cromática en la óptica de formación de imágenes (debido a que el haz debe atravesar una zona con cambios en el índice de refracción (las lentes) y éste depende de la longitud de onda). En comunicaciones cuasimonocromáticas este efecto carece de importancia. Para tamaños de apertura por encima de unos decímetros de diámetro, la óptica reflexiva es preferible debido a su menor coste y peso.

Como se verá en el siguiente apartado, conviene hacer que la lente o espejo de salida del telescopio sea lo más grande posible, con el fin de reducir la divergencia. Sin embargo, el tamaño de este elemento está limitado cuando se trata de un emisor en el espacio, por consideraciones de peso y volumen del satélite. Por otra parte, la divergencia no puede reducirse arbitrariamente, ya que se complica sobremanera el problema del ***apuntamiento*** del haz hacia la estación terrestre. Por ejemplo, el diámetro del espejo de salida propuesto para el proyecto *MLCD* es de 12” (30,5 cm).

En todo caso, el único haz que puede propagarse sin divergencia es una **onda plana**. Para cualquier otro frente de ondas existe un ***límite de difracción*** ligado al diámetro del objetivo del telescopio, que se estudia en el siguiente apartado.

### 4.4.1. Límite de difracción de un telescopio

A todos los efectos la lente o espejo de un telescopio puede ser considerado una abertura circular, ya que produce luz dentro del círculo que describe su espejo primario. El patrón de difracción de menor apertura angular se produce cuando dicha abertura se ilumina homogéneamente con una onda plana, y consiste en un conjunto de anillos concéntricos conocido como disco de Airy (Figura 4.8).

Si el diámetro de la abertura es D y $\lambda$ la longitud de onda, la variación angular de la intensidad de radiación viene dada por la fórmula [39]

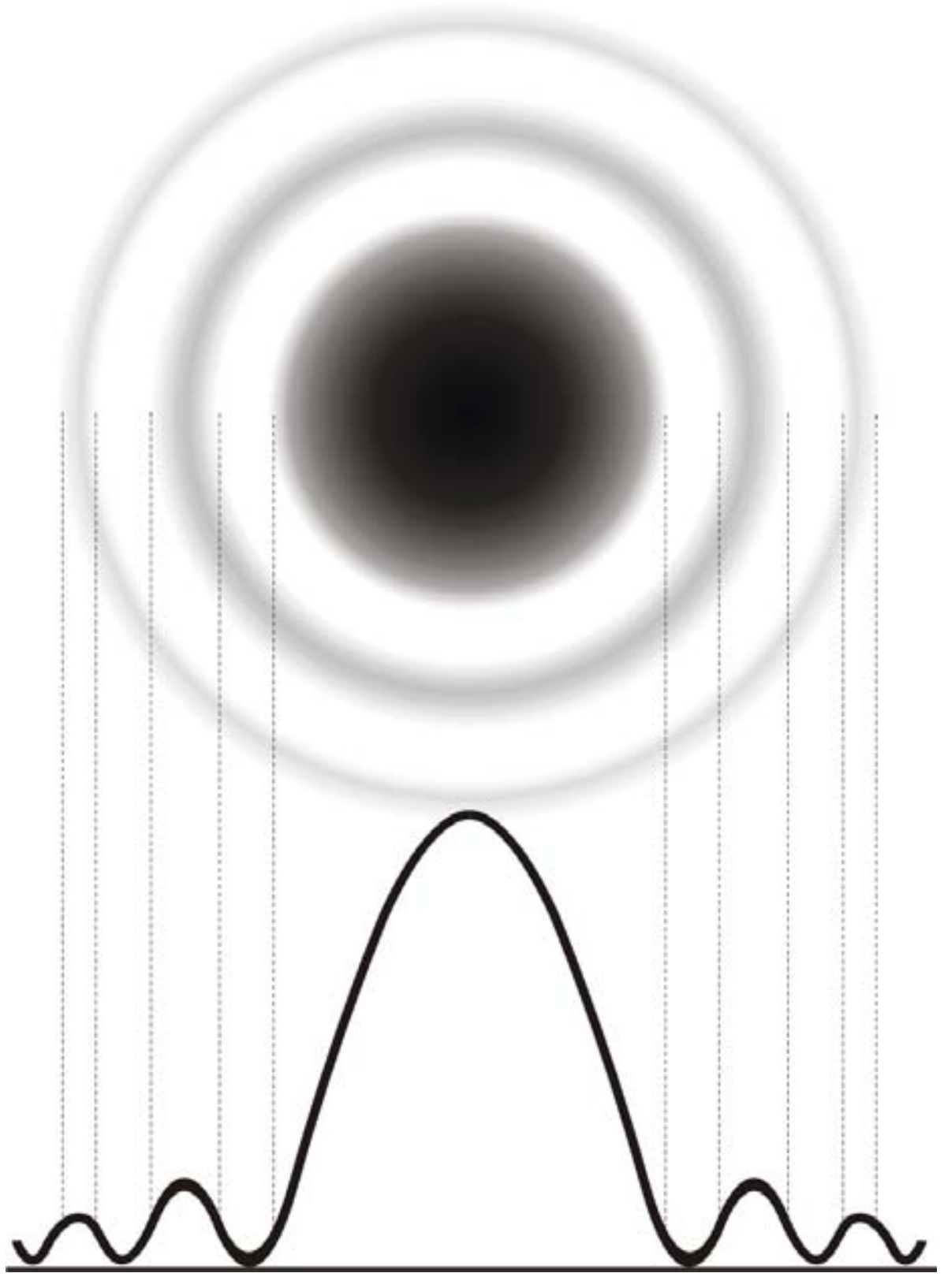

***Figura 4.8. Patrón de difracción producido por una apertura circular (disco de Airy).***

$$\frac{I(\theta)}{I(0)}=\left[2\frac{J_1\left(\frac{\pi D}{\lambda}\operatorname{sen}(\theta)\right)}{\frac{\pi D}{\lambda}\operatorname{sen}(\theta)}\right]^2 \qquad (84)$$

donde $J_1\left(\frac{\pi D}{\lambda}\operatorname{sen}(\theta)\right)$ es la función de Bessel de primer orden de $\frac{\pi D}{\lambda}\operatorname{sen}(\theta)$.

El primer cero corresponde a $\frac{\pi D}{\lambda}\operatorname{sen}(\theta)=3{,}832$. Utilizando la aproximación $\operatorname{sen}(\theta)\approx\theta$ se obtiene el ***límite de difracción del telescopio***, que viene dado por

$$\theta = 1{,}22\left(\frac{\lambda}{D}\right) \qquad (\text{rad}) \qquad (85)$$

Este límite determina el ***mínimo ángulo de difracción*** y por lo tanto la mínima divergencia del haz con el aumento de la distancia.

En la Figura 4.10 puede apreciarse una representación del límite de difracción. Se observa cómo la anchura (sea cual sea el criterio para su medición) del lóbulo principal de la función de Bessel aumenta al disminuir la abertura (D) y/o aumentar la longitud de onda (λ).

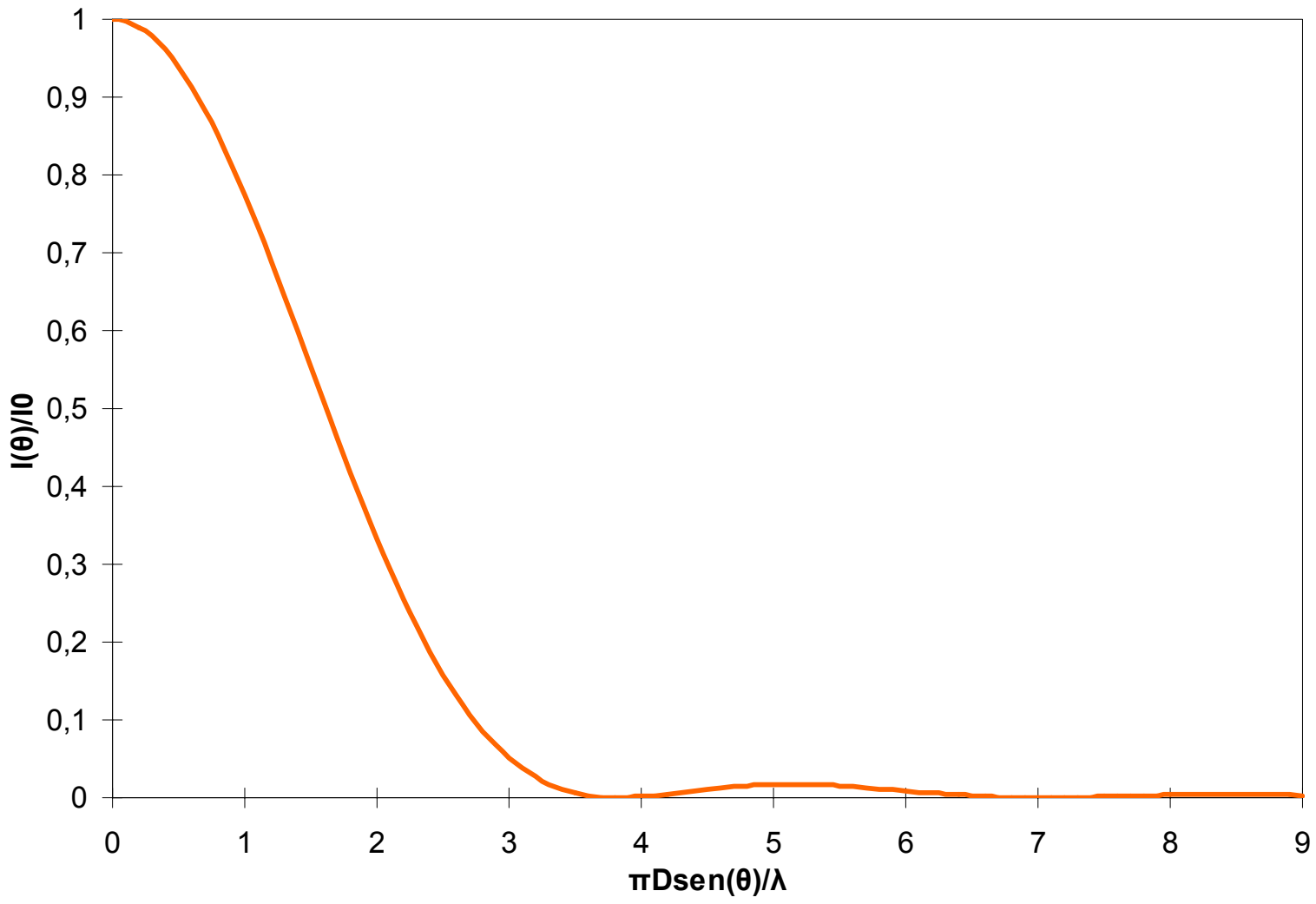

***Figura 4.9. Representación gráfica de la intensidad de radiación calculada a partir de la ecuación (84).***

<u>CRITERIOS PARA EL CÁLCULO DEL LÍMITE DE DIFRACCIÓN</u>

La fórmula del límite de difracción calculada en este apartado se corresponde con el **criterio del primer cero** de la función de Bessel. Si se usara otro criterio el factor multiplicador de λ/D sería diferente. Por ejemplo [9], si en lugar de tomar el punto en el que se encuentra el primer cero, se tomase el punto en el que la potencia cae a la mitad, el factor multiplicador sería 1,03.

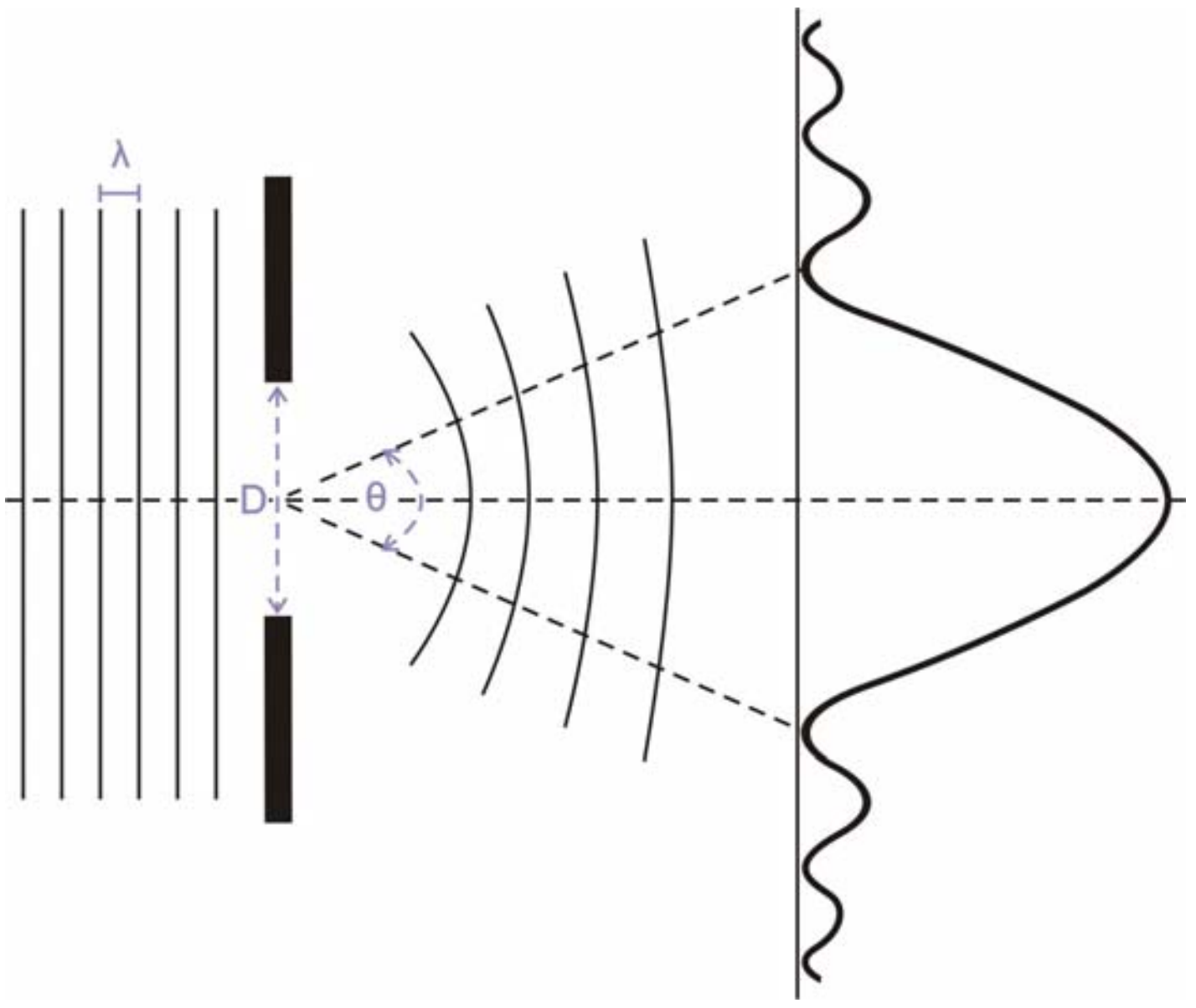

*Figura 4.10. Límite de difracción de un telescopio [40].*

## 4.4.2. Apuntamiento

La gran ventaja que proporciona el pequeño tamaño del *spot* a frecuencias ópticas (como ya se vio en el apartado 3.1.1 dedicado al efecto de la distancia en la divergencia del haz) representa también una desventaja en términos de apuntamiento.

El área del *spot* en la ubicación en la que se encuentra el receptor aumenta al aumentar la distancia o la longitud de onda y al disminuir el diámetro del telescopio emisor. Por lo tanto, dado un límite de difracción para el diámetro D del espejo primario del telescopio usado y la λ de trabajo, en función de la precisión en el apuntamiento que pueda alcanzar el transmisor y de la distancia a la que se encuentre el receptor puede interesar transmitir al límite de difracción o hacerlo por encima de éste.

- Si se transmite **cerca del límite de difracción**, se consigue hacer *máxima la potencia de pico* del *spot*, esto es, se hace máxima la potencia que hay en el centro del mismo. Sin embargo se obliga a una mayor precisión en el apuntamiento ya que la anchura del *spot* es la mínima posible. Como regla general se puede decir que, siempre que la tecnología de apuntamiento disponible permita que el haz caiga encima del receptor, *se emitirá el haz al límite de difracción* (Figura 4.11.a). Normalmente éste será el caso de naves espaciales a muy grandes distancias o de terminales espaciales a menor distancia pero asentados en alguna superficie firme como la de un planeta que les permita mantener un apuntamiento suficientemente preciso.

- Por otro lado si el **error de apuntamiento del emisor** es tal que la probabilidad de que el receptor quede iluminado por el haz es baja (Figura 4.11.b), conviene emitir la señal por encima del límite de difracción de forma que el área del *spot* aumente lo suficiente como para que los errores de apuntamiento no dejen al receptor fuera de la cobertura de la señal. Este será el caso de naves espaciales relativamente cercanas que debido a sus propias vibraciones no puedan mantener un correcto apuntamiento.

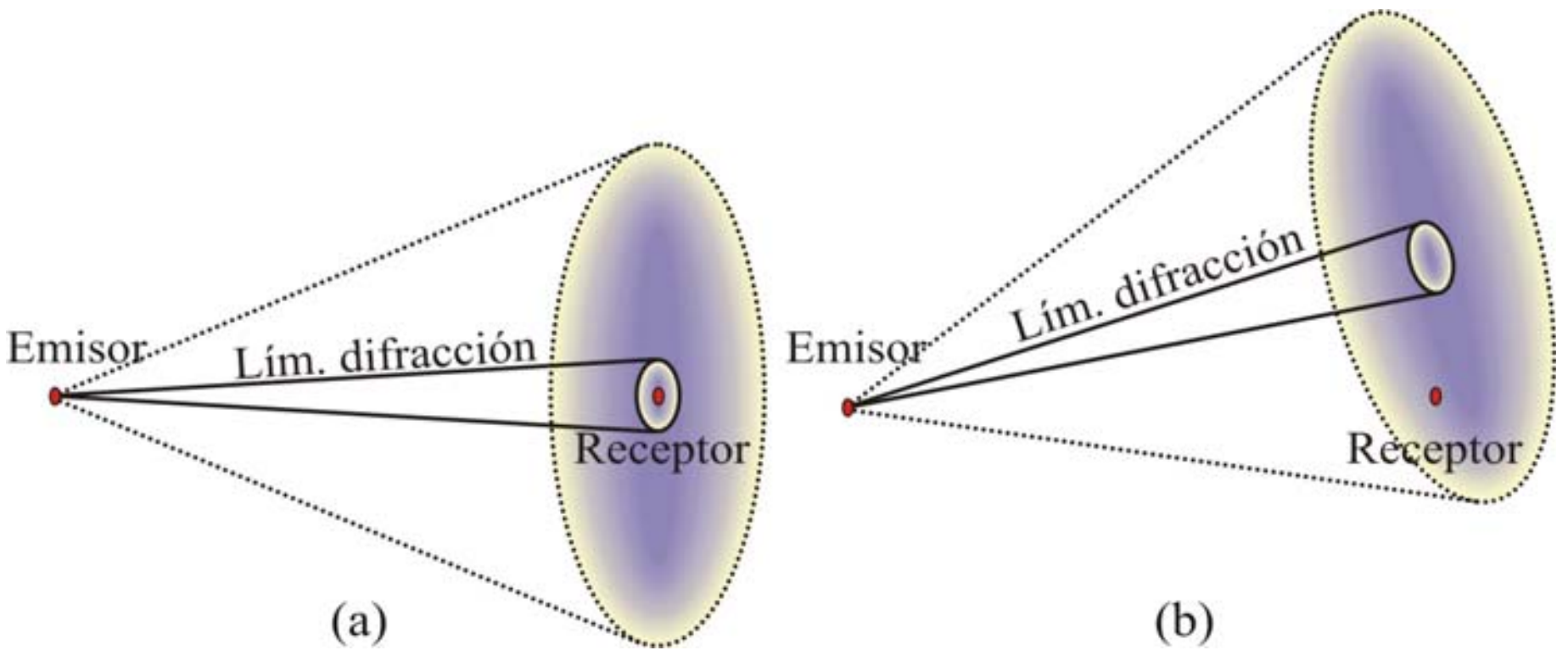

***Figura 4.11. Efecto del límite de difracción con (a) suficiente precisión de apuntamiento y (b) sin suficiente precisión de apuntamiento.***

Es necesario resaltar, sin embargo, que aún disponiendo de un sistema de apuntamiento ideal, las **turbulencias atmosféricas** producirán –en el caso de que el terminal terrestre esté sobre la superficie– fluctuaciones que alterarían aleatoriamente la posición de la señal. Este efecto y la manera de cuantificarlo se trata en el apartado dedicado a las turbulencias atmosféricas.

En la Figura 4.12 se muestra la evolución histórica en la máxima precisión de apuntamiento lograda en misiones de exploración en espacio profundo.

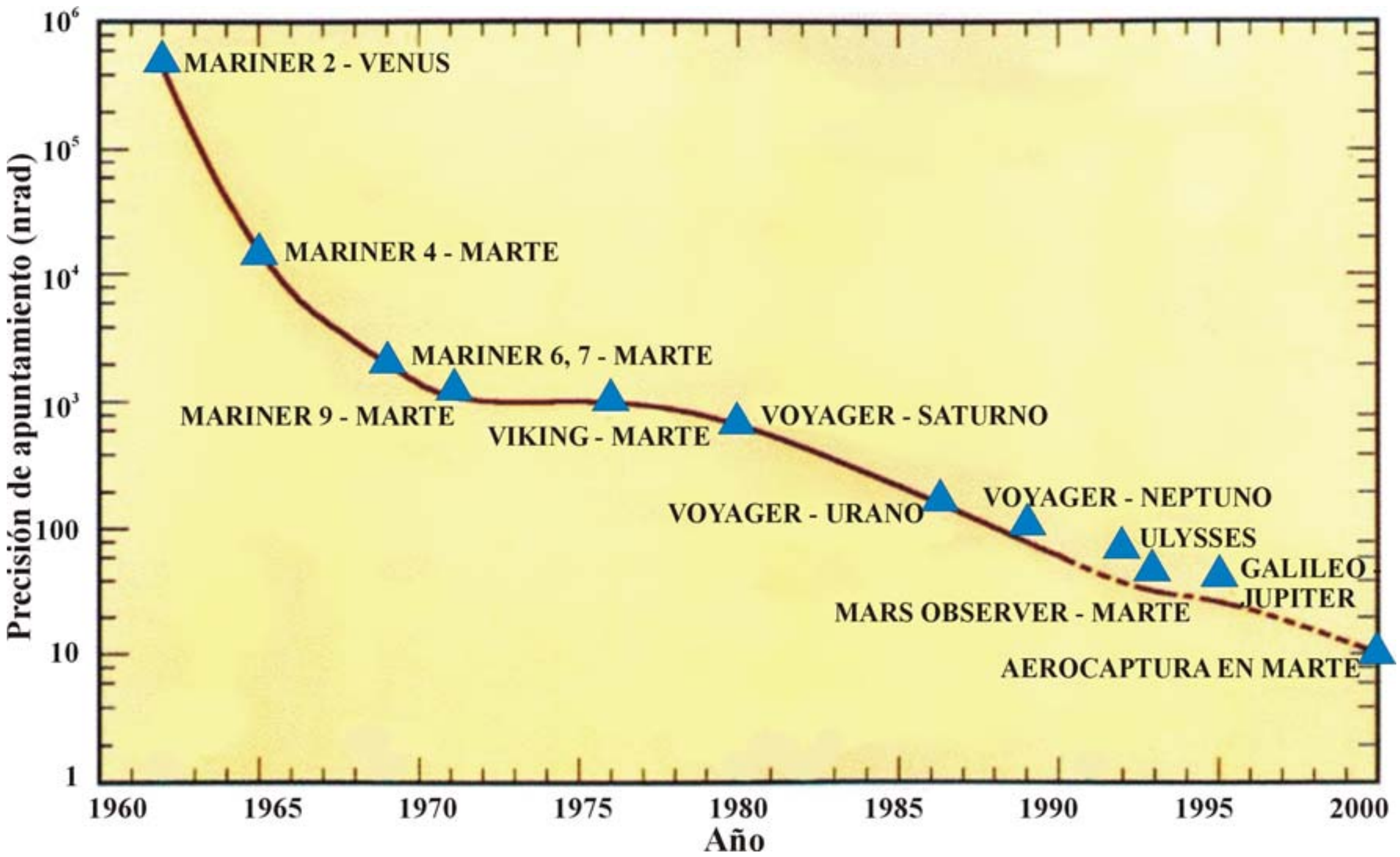

***Figura 4.12. Evolución histórica de la precisión de apuntamiento en espacio profundo [5].***

## 4.4.3. Ganancia

Empleando la forma clásica en la que se tratan las antenas en radiofrecuencia, un telescopio puede tratarse como una antena con una determinada ***ganancia***. Esta ganancia significaría una mejora que tendría el telescopio, por la capacidad de **concentrar** la radiación en un haz directivo, respecto a una que radiara la energía en todas las direcciones homogéneamente (antena isótropa). Según esta definición se puede expresar la ganancia como

$$G = \frac{I(r,\theta)}{I_0} \tag{86}$$

donde I(r, θ) es la intensidad de radiación en la dirección hacia el punto de observación y $I_0$ la intensidad de radiación de una antena isótropa, que es igual a

$$I_0 = \frac{1}{4\pi r^2} \tag{87}$$

La expresión general para la intensidad de radiación a través de una apertura circular viene dada por [41]

$$I(r,\theta) = \frac{k^2}{r^2}\left[\int_b^a \sqrt{\frac{2}{\pi}}\frac{1}{\omega}\exp\left(-\frac{r_0^2}{\omega^2}\right)\exp\left[j\frac{kr_0^2}{2}\left(\frac{1}{r}+\frac{1}{R}\right)\right]J_0\left(kr_0\,\mathrm{sen}(\theta)\right)r_0\,dr_0\right]^2 \tag{88}$$

donde k es el número de onda (2π/λ) para la longitud de onda λ; r y θ indican el punto de observación; ω es el radio $1/e^2$ del haz gaussiano acoplado a la óptica del transmisor; R es el radio de curvatura del frente del haz en el plano de la apertura del telescopio; a es el radio del espejo primario y b el radio del espejo secundario tal como se observa en la Figura 4.13.

Si A es el área de la apertura del telescopio, que en este caso es circular y por lo tanto $A = \pi a^2$, la expresión de la ganancia ideal de una antena es

$$G_{ideal} = \frac{4\pi A}{\lambda^2} = \left(\frac{2\pi a}{\lambda}\right)^2 \tag{89}$$

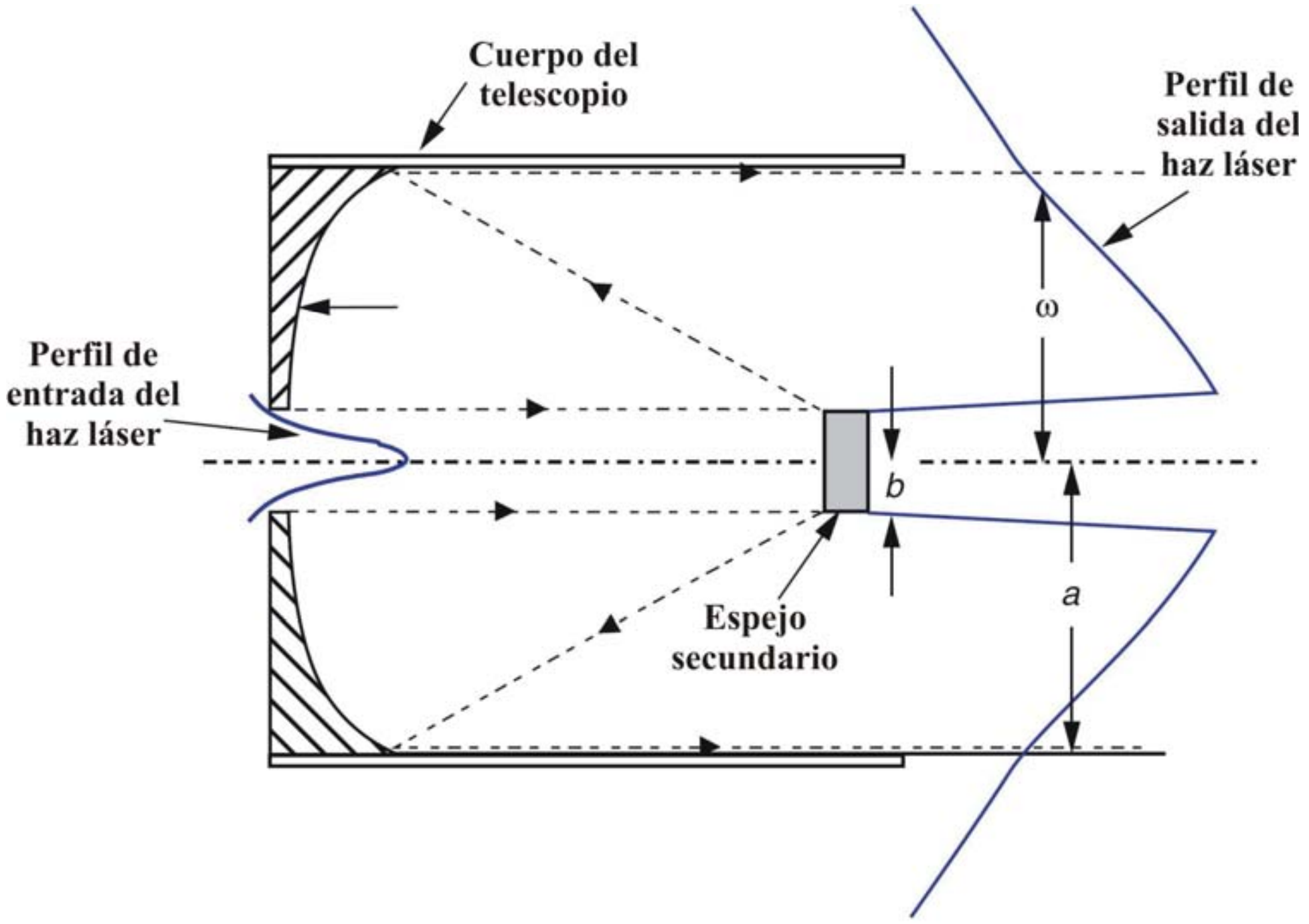

*Figura 4.13. Relación entre el haz gaussiano del láser y los elementos ópticos del telescopio transmisor.*

### 4.4.4. Eficiencia de la ganancia

Sea g un factor de eficiencia de la ganancia del telescopio, que varía entre 0 y 1; la **ganancia real** del telescopio se puede expresar de la siguiente manera

$$G = g \cdot G_{ideal} \quad (90)$$

El parámetro g va a tener principalmente dos aportaciones que van a hacer que la ganancia real no sea la ideal: el factor $\alpha = a/\omega$ cuantifica las pérdidas por ***truncamiento*** del haz gaussiano y el factor $\gamma = b/a$ cuantifica las pérdidas por ***bloqueo*** del espejo secundario. Sustituyendo las ecuaciones (87) y (88) en la ecuación (86) e igualando a la ecuación (90) tras sustituir en ésta la ecuación (89) se obtiene la expresión general para la ganancia real de un telescopio

$$G = \frac{\frac{k^2}{r^2}\left[\int_b^a \sqrt{\frac{2}{\pi}}\frac{1}{\omega}\exp\left(-\frac{r_0^2}{\omega^2}\right)\exp\left[j\frac{kr_0^2}{2}\left(\frac{1}{r}+\frac{1}{R}\right)\right]J_0\left(kr_0 \mathrm{sen}(\theta)\right)r_0 dr_0\right]^2}{\frac{1}{4\pi r^2}} = g\left(\frac{2\pi a}{\lambda}\right)^2 \quad (91)$$

de donde se puede despejar g. Si $k \cdot a \cdot \mathrm{sen}(\theta_1) = x$, se hace el cambio

$$\begin{cases} r_0 = \frac{b}{\gamma}\sqrt{u} \\ dr_0 = \frac{b^2}{2\gamma^2 r_0}du \end{cases} \quad y \quad \beta = \frac{ka^2}{2}\left(\frac{1}{r}+\frac{1}{R}\right) \quad (92)$$

y se simplifica, se obtiene la expresión para g

$$g = 2\alpha^2\left[\int_{\gamma^2}^{a} \exp\left(-\alpha^2 u\right)\exp\left(j\beta u\right)J_0\left[x\sqrt{u}\right]du\right]^2 \quad (93)$$

donde β puede aproximarse a cero para campo lejano y x se hace cero en el eje óptico (ángulo θ para el cual la ganancia es máxima) quedando la expresión de g que relaciona los dos factores correctores mencionados anteriormente

$$g = \frac{2}{\alpha^2}\left[\exp\left(-\alpha^2\right) - \exp\left(-\gamma^2\alpha^2\right)\right] \quad (94)$$

## 4.5. ARQUITECTURA DEL TERMINAL ESPACIAL

El tipo de terminal que se ha de situar en el espacio ofrece menos posibilidades que en el caso del terminal terrestre. Básicamente sólo existen dos alternativas: ***satélites en órbita*** o terminales situados sobre alguna ***superficie***. En muchos casos, estas dos alternativas se reducen sólo a la posibilidad del satélite. Por ejemplo, Venus –al margen de sus extremas temperaturas– cuenta con una **densa y turbulenta atmósfera** cuya masa nubosa oculta completa y permanentemente la superficie; similar es el caso de Titán, al que a su atmósfera de densidad unas cinco veces mayor que la terrestre, hay que sumar la gran distancia existente entre los terminales; o los **planetas gaseosos** como Júpiter o Saturno que carecen de superficie sólida.

Incluso en los casos en los que sea posible la comunicación desde superficie porque no exista atmósfera o la atmósfera no presente un problema (como es el caso de Marte con una atmósfera muy poco densa), hay una serie de inconvenientes de la opción en superficie a favor de la opción de satélite. Por un lado, la escasez de potencia de los terminales en superficie no suele justificar el empleo de éstos para comunicaciones (y en el caso en que se comuniquen directamente con el terminal terrestre, quizá la forma más eficiente sería por microondas). Además desde la superficie no se puede dar simultáneamente cobertura a las estaciones terrenas y los terminales de exploración terrestres y servir como enlace de comunicación con la Tierra. Uno o más satélites en órbitas adecuadas pueden prestar servicio a varias estaciones terrenas cumpliendo una función de repetidor.

Normalmente la mejor estrategia es realizar la comunicación *a través de un satélite que orbite alrededor del planeta*, ya sea el propio satélite el que obtenga directamente los datos a transmitir o los obtenga a través de un vehículo de exploración en superficie.

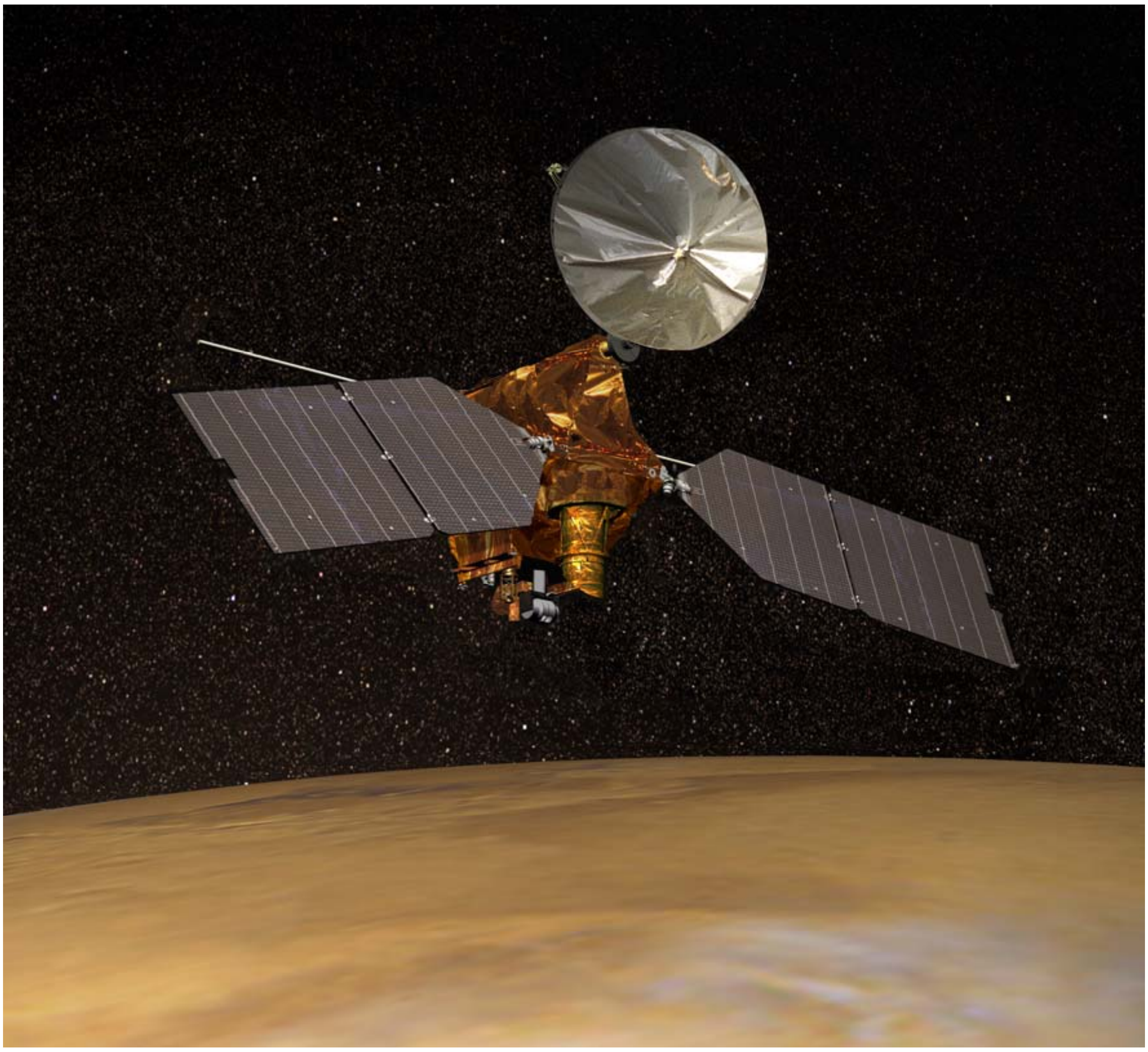

***Figura 4.14. Ilustración artística de Mars Reconnaissance Orbiter orbitando Marte [42].***

# 5. El receptor local

El receptor suele ser el ***elemento más crítico*** de diseñar en un enlace de comunicaciones ópticas, puesto que de su correcto comportamiento depende decisivamente la calidad del enlace. En el canal de bajada de un enlace en el espacio profundo, el diseño del receptor reviste aún más importancia. La razón es que la capacidad del emisor está fuertemente mediatizada por la **escasa potencia** disponible, el limitado **ciclo de trabajo** y el **tamaño** relativamente reducido de la antena. Tal como se comentaba en el capítulo anterior, se intentan *aliviar en lo posible* las exigencias en el transmisor remoto, haciendo recaer sobre el receptor terrestre la máxima carga de trabajo, a base de llevarlo al *límite de prestaciones* con la tecnología existente. En el fondo, se trata de aprovechar las máximas posibilidades de la unidad local, reparable o sustituible, evitando riesgos innecesarios al transmisor, que podrían dar al traste con la misión.

En este capítulo se estudian los posibles sistemas de recepción que podrían emplearse en un enlace óptico espacial, para particularizarlos posteriormente a la misión *MLCD*. Cabe señalar que, en el tiempo que falta hasta el lanzamiento, e incluso durante el tiempo de viaje de la nave, surgirán probablemente mejores alternativas que las aquí expuestas, que sin duda serán las que finalmente se seleccionen. Por el momento, se intentará demostrar la factibilidad de un enlace con las características solicitadas al *MLCD* empleando dispositivos comerciales ya existentes en la actualidad.

Los elementos básicos de un receptor óptico terrestre en un enlace a través del espacio profundo son

- ***Óptica de recepción*** (antena). Usualmente es un telescopio astronómico reflexivo

- ***Detector***, para transformar la señal recibida en el dominio óptico al eléctrico.

- ***Demodulador*** basado en técnica coherente o incoherente dependiendo del esquema de modulación usado.

- ***Decodificador*** correspondiente al codificador empleado en la transmisión.

- Y todos los ***filtros*** y ***etapas de amplificación*** necesarios para que la señal eléctrica tenga finalmente un nivel y ancho de banda suficiente como para que pueda ser identificada con la señal original transmitida.

## 5.1. ANTENA RECEPTORA

La antena receptora podría basarse, en principio, en un **telescopio** refractor o reflector. Se encarga de recoger en una apertura de diámetro $D_r$ la energía óptica fuertemente colimada del emisor y transformarla en un haz de diámetro *algo más pequeño que el área del fotodetector* de forma que se pueda corregir la inexactitud de apuntamiento del receptor. El fotodetector se situará a la distancia focal de la antena.

La opción principal que se contempla es un telescopio astronómico de gran área. Si se opta por esta opción, *las ópticas refractivas quedan descartadas*. Sin embargo, dado que el haz que se desea detectar es prácticamente paralelo, se puede sustituir opcionalmente el telescopio por una lente o espejo primario con el detector situado en el foco. Para aumentar el área, se puede emplear un conjunto de elementos en paralelo, cada uno equipado con su propio detector. Por ejemplo [43], una de las propuestas de antena de recepción del proyecto *MLCD* consiste en un conjunto de 80 tubos dotados cada uno de un espejo de 0,5 m y un fotodetector.

### 5.1.1. Configuración

En este proyecto se tomará como base de la antena receptora un telescopio astronómico, empleando datos reales de las instalaciones existentes. Las configuraciones de antena serán, pues, las mismas que se mostraban en la antena emisora (Figura 4.6). A la salida del ocular se colocará otra lente, cuya función sea enfocar la luz colimada (algo menos) de la salida sobre el detector.

Idealmente, la potencia óptica enfocada sobre el detector es igual a la potencia incidente sobre el espejo primario. Por tanto, la potencia fotodetectada puede ser igualmente calculada determinando la potencia recibida sobre el área del espejo en vez de la potencia enfocada sobre el área del detector.

De esta forma, suponiendo que la incidencia del láser está casi perfectamente colimada, el telescopio astronómico en configuración confocal (focos coincidentes), realizará básicamente una multiplicación de ángulos, la operación opuesta a la antena emisora. Este **poder multiplicador** del telescopio, como se vio en el apartado 4.4, está directamente relacionado con sus correspondientes distancias focales. Así, si la focal del secundario es 5 cm y la del primario 5 m, los ángulos se multiplican por 100.

Ahora bien, dicha amplificación de ángulos, tendrá una consecuencia inmediata. Ante una **turbulencia atmosférica**, que se puede modelizar como una interfase con cambio de índice, el haz incidente experimentará una desviación. Esta desviación aún en el caso de ser muy pequeña, aumentará considerablemente al llegar al secundario (con los datos del ejemplo anterior en un factor 100). Como resultado, el haz enfocado por la tercera lente sobre el plano focal podría salirse de la superficie activa del detector. Dicho problema se soluciona aumentando el ***campo de visión***.

## 5.1.2. Campo de visión

El campo de visión (FOV, *field of view*) se define como la ***apertura angular de la escena*** que se proyecta sobre la zona activa –una película fotográfica, un sensor CCD, o un fotodetector– del plano focal del sistema óptico. En nuestro caso no se pretende formar imagen, sino detectar potencia luminosa (Figura 5.1). Por ello el campo de visión podría ser, teóricamente, tan estrecho como se quisiera, ya que solo interesa la luz procedente del láser emisor, que es paralela a todos los efectos, y cualquier apertura superior del FOV serviría únicamente para introducir más ruido de fondo.

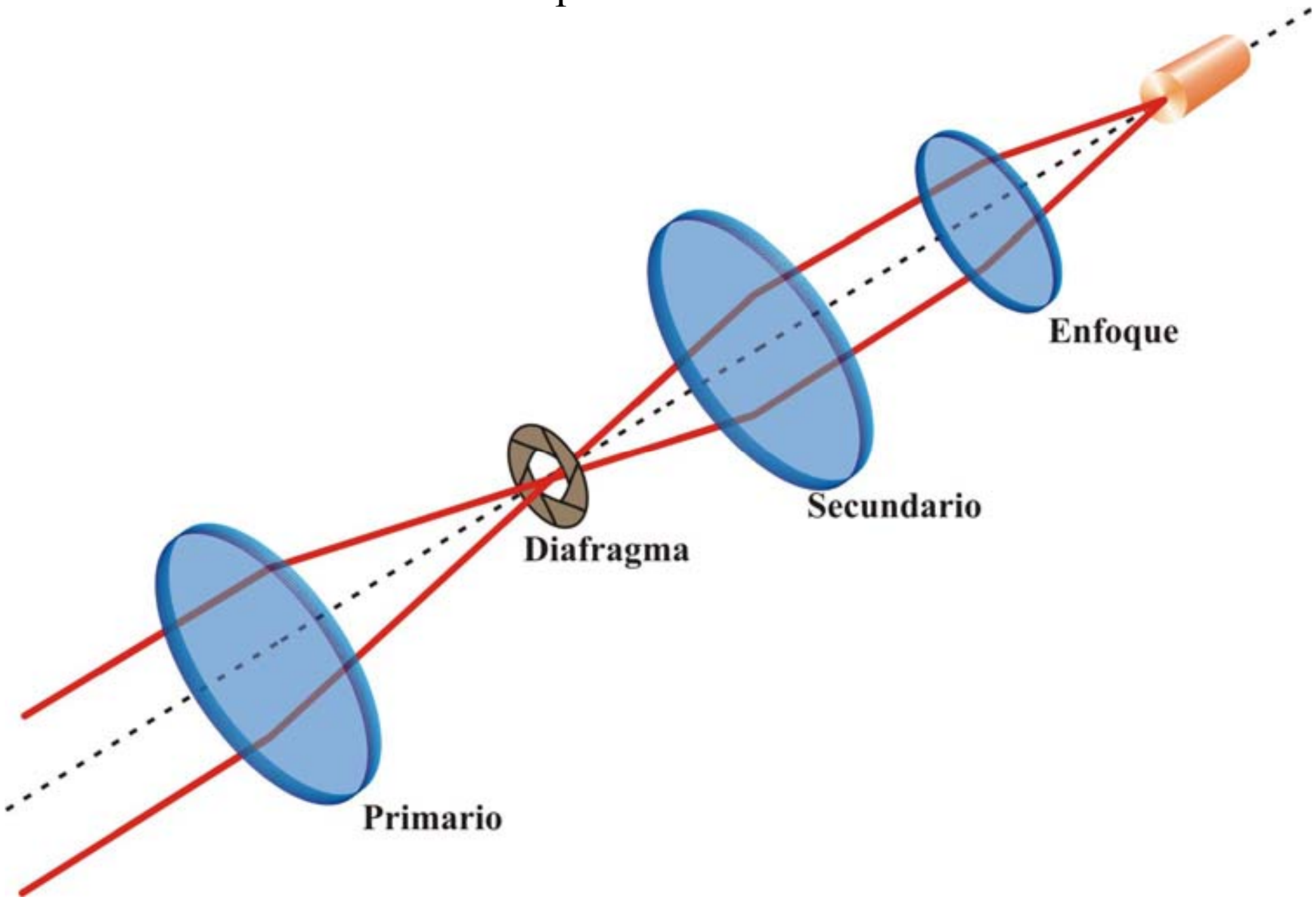

***Figura 5.1. Esquema simplificado de control del campo de visión con un diafragma entre el elemento primario y secundario.***

Sin embargo, el efecto de la **turbulencia atmosférica** apuntado antes hace que se deba utilizar un campo de visión con un cierto *margen de seguridad* para evitar que las fluctuaciones alteren el enfoque. Adicionalmente, se ha de considerar que el emisor puede estar ***en órbita*** sobre otro planeta, o bien ***sobre la superficie*** del mismo, sometido a su movimiento de rotación. En cualquiera de los casos, deberá optarse por una de las dos opciones siguientes

- Escoger un campo de visión que cubra **el planeta completo** o bien **el planeta y la órbita del satélite**, según esté localizado el emisor en uno o en otra. con lo que el telescopio tendría que realizar simplemente el desplazamiento habitual de seguimiento astronómico debido a la rotación terrestre.

- **Minimizar el campo de visión** al límite de resolución del telescopio. En ese caso, si el FOV del telescopio es inferior al indicado en el apartado anterior, *se deberá corregir el apuntamiento* del telescopio para que **siga al emisor** en todo momento. Esta opción implica un complicado algoritmo de orientación del telescopio, que debería evitarse, en principio, siempre que la relación señal-ruido del sistema tolere un campo de visión más amplio.

Un fotodetector responderá a toda la radiación que llegue a su superficie activa, (siempre que su $\lambda$ esté en el rango de detección). Por ello, el ***campo de visión del receptor***, $\Omega_r$, dependerá del **área del detector**, $A_d$, y de la **distancia focal** de la lente de enfoque, $f_c$, *y no del tamaño de la lente de recepción*, $A_r$.

$$\Omega_r \approx \frac{A_d}{f_c^2} \qquad \text{(sr)} \qquad (95)$$

Este planteamiento es diferente que el de una antena RF, cuyo FOV –cuando está limitada por difracción– depende de $\lambda^2 / A_r$. Como comparación, si suponemos que la distancia focal es del orden del diámetro de la lente –algo común en óptica– se puede considerar que aproximadamente $f_c = \sqrt{A_r}$. Con esa aproximación, el FOV del receptor queda

$$\Omega_r \approx \frac{A_d}{f_c^2} \approx \frac{A_d}{A_r} = \frac{A_d}{\lambda^2} \cdot \frac{\lambda^2}{A_r} \qquad \text{(sr)} \qquad (96)$$

A frecuencias ópticas, $A_d$ es del orden de mm, mientras que $\lambda$ es del orden de $\mu$m. Así pues, la primera fracción del segundo miembro de (96), $A_d/\lambda^2 >> 1$, y el campo de visión $\Omega_r$ es mucho ***mayor que el límite de difracción***. Recuérdese que, en todo caso, si se emplea un telescopio astronómico como antena receptora, el FOV del primario será una fracción del indicado aquí, ya que los ángulos considerados están multiplicados por la relación entre las focales del espejo primario y secundario.

### 5.1.3. Distancia focal y área del fotodetector

La dependencia del campo de visión con el área del detector y la distancia focal a la lente de enfoque, se puede ilustrar a través del estudio de la Figura 5.2. En las condiciones de medida se pueden utilizar sin errores aproximaciones de **lentes delgadas** y **óptica paraxial**. Por lo tanto, aplican las siguientes reglas simples de la óptica geométrica[1] (óptica de rayos)

- Rayos *paralelos al eje* se desvían pasando por el foco.
- Rayos que pasan por el *centro de la lente* no se desvían.
- Rayos paralelos entre sí *convergen en el mismo punto* del plano focal.
- El plano focal *carece de deformaciones*, y es perpendicular al eje del sistema

[1] Se utilizan lentes en el ejemplo por simplicidad de dibujo, pero el razonamiento es igualmente válido para espejos

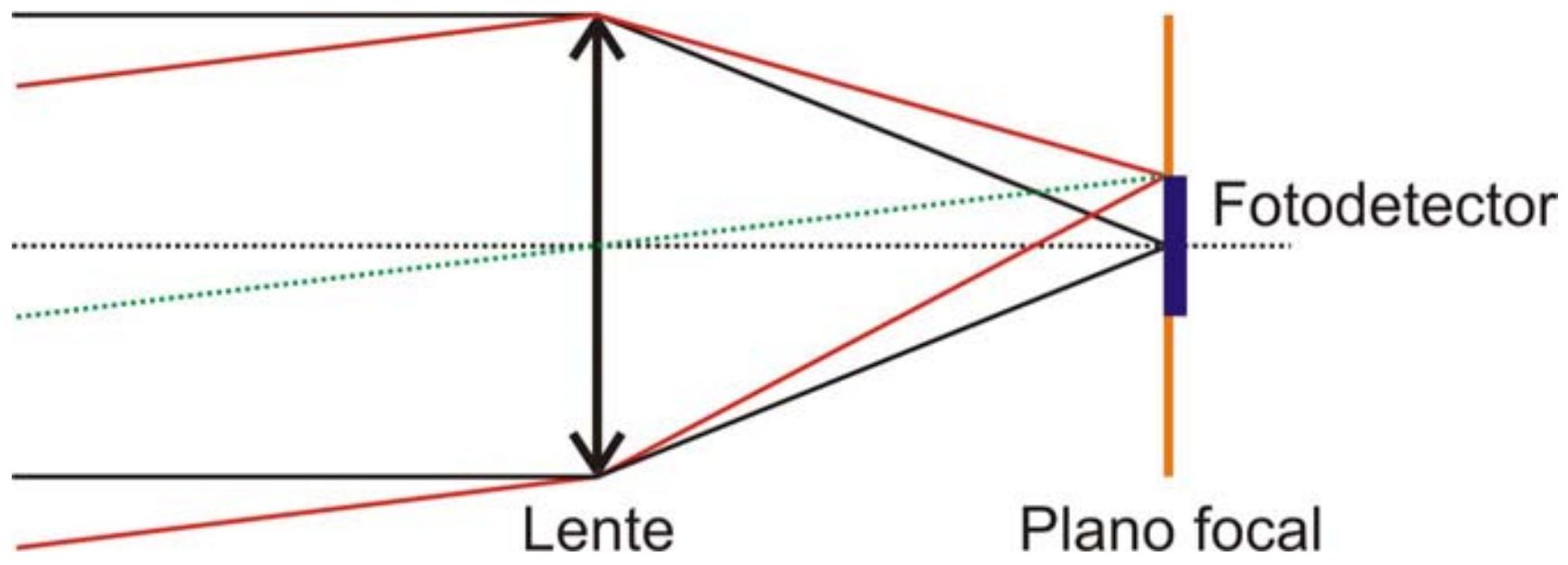

***Figura 5.2. Correspondencia entre el área del fotodetector y el campo de visión.***

En la Figura 5.2 aparece la lente de enfoque con un detector en su plano focal. Los rayos que entran paralelos al eje van a parar al **centro** del detector. Los rayos que inciden con determinada inclinación (rayos rojos de la figura), van a otro **punto del plano focal**, que se puede calcular trazando una paralela –línea verde– por el centro de la lente (que no se desvía).

Así, el campo de visión viene limitado por el ángulo con que inciden los rayos rojos de la Figura 5.2, porque llegan justo al borde del fotodetector. Si hubiesen entrado más inclinados (mayor FOV) habrían ido a parar a otro punto del plano focal situado fuera del fotodetector, por lo que no se detectarían. Por otra parte, si manteniendo la misma área del detector se aumenta la focal de la lente, se observa que los mismos rayos que antes alcanzaban el borde del detector, ahora se salen fuera (Figura 5.3).

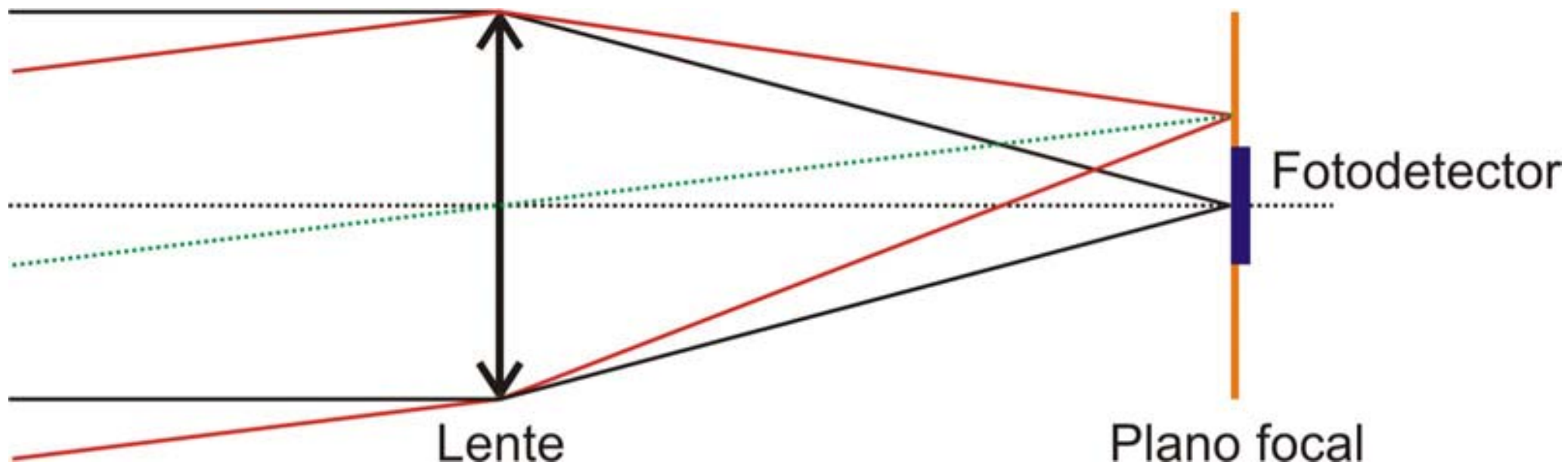

***Figura 5.3. Si la focal de la lente aumenta, la radiación se sale del área del fotodetector con el mismo campo de visión.***

Obsérvese que este hecho *no depende del tamaño de la lente* –los rayos paralelos de toda la lente se focalizan idealmente en el mismo punto del plano focal– sino de la relación entre el tamaño del detector y la distancia focal.

Si se quisiese aumentar el FOV para la misma área de detector, habría que disminuir la distancia focal a la lente de enfoque. Esta estrategia *permite solucionar el problema de la turbulencia atmosférica*. Aumentando adecuadamente el FOV, se evita que posibles desviaciones aleatorias de los rayos lleven a intervalos en los que se pierda o reduzca la incidencia en la superficie del detector.

Como contrapartida, este cambio de focal **empeora la relación señal-ruido**. La cantidad de señal que estaría recibiendo el fotodetector permanece invariable, puesto que depende de la superficie del primario, mientras que el nivel de ruido aumenta, porque al ser el campo de visión mayor se capta más cantidad de luz no deseada. Para compensarlo, habría que utilizar un primario más grande.

En conclusión, hay que llegar a un compromiso para el campo de visión: no se puede hacer muy grande porque se introduciría una gran cantidad de ruido, ni muy pequeño, porque cualquier turbulencia, podría desviar la señal fuera del detector. Consecuentemente, en un diseño real el campo de visión elegido será el mínimo que garantice la estabilidad del haz, en el área del detector, dentro del margen de seguridad que se establezca. Idealmente, el FOV debería ***ajustarse dinámicamente*** en cada sesión de detección, en función de las **condiciones meteorológicas**.

## 5.2. EL DETECTOR: FOTOMULTIPLICADORES

Un detector ideal debería reunir las siguientes características

- ***Alta sensibilidad*** en la región de trabajo para la que se diseña.
- ***Alta fidelidad*** de forma que se produzca una reproducción exacta de la señal óptica en un amplio margen.
- ***Alta respuesta eléctrica*** que conlleve un alto rendimiento cuántico.
- ***Bajo tiempo de respuesta***, lo que implica un gran ancho de banda.
- ***Bajo ruido***.
- ***Estabilidad*** frente a alteraciones de las condiciones ambientales.
- ***Fiabilidad***

Los detectores empleados en Comunicaciones Ópticas que más se aproximan a este ideal son los basados en ***semiconductores***, que se revisan en la sección 5.3.

Los ***tubos fotomultiplicadores*** presentan excelentes propiedades en relación a las características apuntadas, pero suelen adolecer de tiempos de respuesta demasiado largos, y resultan más difíciles de acoplar a fibras ópticas. Por otra parte, su ventaja más notable frente a los diodos semiconductores es la alta ganancia que presentan, que alcanza valores de $10^6$ o más. Esta propiedad, unida a las tasas binarias relativamente bajas que se buscan en un enlace en el espacio profundo, hacen que, en este caso, los tubos fotomultiplicadores sean un serio candidato a considerar.

En todo caso, la tasa binaria dependerá del formato de modulación elegido. En la aplicación práctica del proyecto *MLCD* se estudiará la pertinencia del uso de fotomultiplicadores, comparándolos con fotodiodos, en función de la modulación y los parámetros ambientales (distancia, posición planetaria y de Sol, etc.).

### 5.2.1. Estructura de un fotomultiplicador

Un fotomultiplicador (PMT, *photo-multiplier tube*) clásico [44] es un tubo de vacío compuesto de un **fotocátodo**, una serie de electrodos llamados **dínodos** y un **ánodo**. Los dínodos se mantienen a tensiones progresivamente más altas con respecto al cátodo, y el ánodo a una tensión más alta aún. El fotocátodo, en presencia de radiación luminosa, genera electrones por ***efecto fotoeléctrico***. Los electrones son acelerados por la tensión existente y se estrellan contra el primer dínodo, provocando la emisión de **electrones secundarios** que a su vez se aceleran hacia el siguiente dínodo repitiéndose el proceso. Así pues, si se generan como promedio δ electrones secundarios por cada electrón incidente, y hay un total de N dínodos, la ganancia es

$$G = \delta^N \tag{97}$$

Se observa que la ganancia puede ser muy alta. Así, si δ = 5 y N = 9, la ganancia total es $2 \cdot 10^6$.

En resumen, el PMT responde a la luz incidente entregando carga al ánodo. Esta corriente puede detectarse como tensión en una resistencia de carga $R_L$. El valor de la carga suele ser pequeño para reducir el tiempo de respuesta del sistema.

En la Figura 5.4 se muestran algunas de las configuraciones usuales en PMTs. El diseño intenta focalizar los electrones secundarios generados por cada dínodo sobre el siguiente, al tiempo que minimiza el número de **posibles trayectorias**. Este punto es importante para evitar la dispersión de los electrones, que aumentaría la anchura del pulso eléctrico de salida reduciendo el ancho de banda.

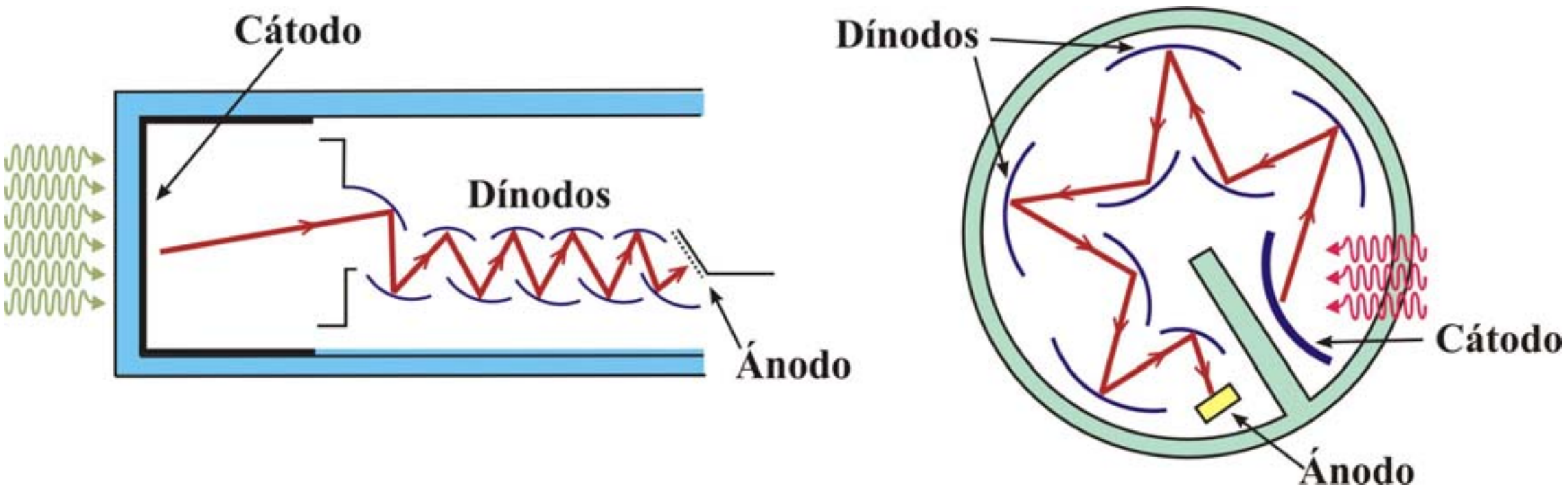

***Figura 5.4. Dos posibles configuraciones de fotomultiplicadores.***

## 5.2.2. Velocidad de respuesta

Los electrones consumen un tiempo finito, llamado ***tiempo de tránsito***, en atravesar la cadena de dínodos. El tiempo de tránsito en sí no influye en la respuesta (simplemente produce un retardo), pero el hecho es que este tiempo no es idéntico para todos los electrones. Existen dos razones para ello: la primera, ya apuntada, es que *las trayectorias de los electrones pueden ser diferentes*; la segunda, que la *velocidad de salida* de los electrones del fotocátodo y los dínodos es variable.

La dispersión de tiempos de tránsito se reduce reduciendo el número de dínodos, y diseñándolos con una forma adecuada para que produzcan **enfoque electrostático**. El enfoque, equivalente al de una lente en óptica, hace que las trayectorias se agrupen, disminuyendo la dispersión.

El problema de la reducción de dínodos –la limitación de la ganancia– se alivia actualmente incrementando la tensión entre dínodos y empleando materiales con altos coeficientes de emisión secundaria. Se observa asimismo una tendencia a la **miniaturización** de los tubos PMT, reduciendo notablemente el tiempo de tránsito. En la actualidad, como se verá en el estudio del proyecto *MLCD*, se ofrecen en el mercado fotomultiplicadores con ancho de banda superior a 1 GHz.

### 5.2.3. Placas microcanal

Las placas microcanal son un tipo especial de fotomultiplicadores que han experimentado un notable desarrollo en los últimos tiempos por su uso en **intensificadores de imagen**.

Estos dispositivos carecen de dínodos. En su lugar, el electrón primario se introduce en un tubo de pequeño diámetro (15 μm) –el ***microcanal***– sometido a una tensión entre sus extremos (Figura 5.5). El electrón tropieza con las paredes del tubo produciendo electrones secundarios que se multiplican a medida que avanzan por el mismo. La longitud de los microcanales es de unos 500 μm. Para aumentar la posibilidad de colisión, los microcanales se orientan oblicuamente a la dirección de acceso de los electrones primarios. Alternativamente, se pueden construir con forma de S con el mismo fin.

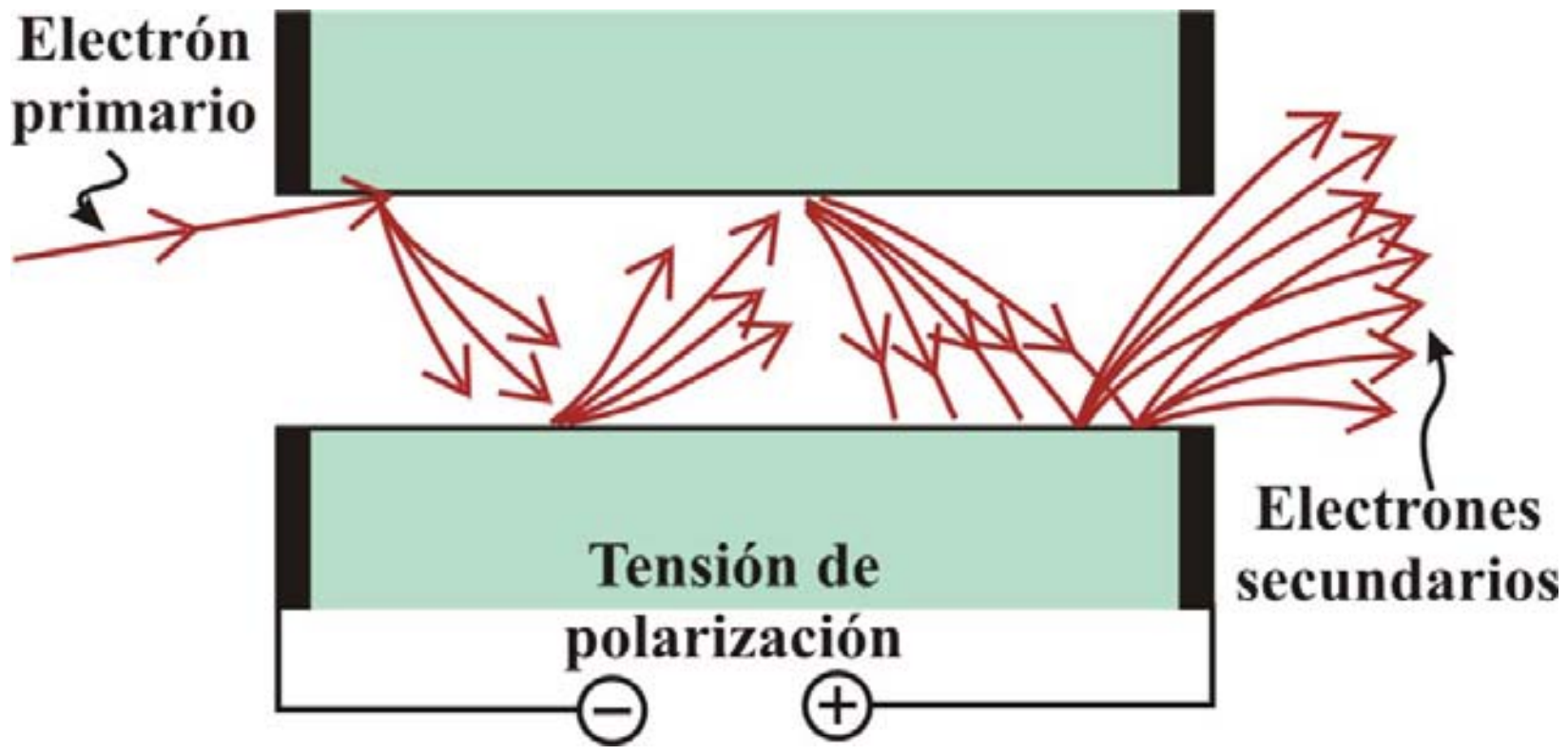

***Figura 5.5. Fotomultiplicador con estructura microcanal.***

Una ***placa microcanal*** se construye por agrupación de muchos miles de microcanales. La placa es originalmente dieléctrica, con una densidad alta de agujeros del diámetro y longitud anotados. Las superficies internas de los agujeros se metalizan ligeramente para construir los microcanales. La tensión de funcionamiento es de varios centenares de V.

Como ya se ha dicho, las placas microcanal se emplean habitualmente como intensificadores de imagen. Recientemente, sin embargo, han surgido *fotomultiplicadores comerciales*, de pequeño tamaño y muy rápidos, basados en estructuras microcanal. Su ganancia, entre $10^4$ y $2{,}5 \cdot 10^5$, es bastante apreciable, aunque sin llegar a los extraordinarios valores de los tubos fotomultiplicadores estándar.

## 5.3. DETECTORES DE SEMICONDUCTOR

Los detectores de semiconductor son los componentes usados tradicionalmente en los receptores de Comunicaciones Ópticas. En este apartado se revisan los tipos más comunes (PIN y APD), sus características, y su adecuación a un enlace no guiado en espacio profundo.

### 5.3.1. Parámetros de caracterización

Para determinar la bondad de un determinado diodo es necesario conocer cual es su capacidad para transformar la luz en corriente eléctrica y con qué rendimiento se hace. Los dos parámetros que determinan lo anterior son los conocidos como ***responsividad***, $R_0$, y ***eficiencia cuántica***, η.

La **eficiencia cuántica** es el número de pares de portadores electrón-hueco generados por cada fotón de energía hν incidente y viene dada por

$$\eta = \frac{\text{n}^{\text{o}}\ \text{electrones recogidos}}{\text{n}^{\text{o}}\ \text{fotones incidentes}} \qquad (98)$$

Igualmente, la **responsividad** se define como

$$R_0 = \frac{I_{ph}}{P_{opt}} \qquad \text{(A/W)} \qquad (99)$$

siendo $I_{ph}$ la fotocorriente, $P_{opt}$ la potencia óptica y la relación entre ambas magnitudes

$$\eta = R_0 \frac{h\nu}{e} = R_0 \frac{hc}{\lambda e} \qquad (100)$$

Idealmente, la responsividad debería ser una función lineal creciente de λ ( para una eficiencia cuántica constante) hasta caer a cero a la λ correspondiente a la energía del *gap* ($\lambda_c$ = hc/$E_g$ ), donde alcanzaría su valor máximo (Figura 5.6). En la realidad, dicha curva sí es aproximadamente lineal creciente pero experimenta un descenso brusco a la $\lambda_c$.

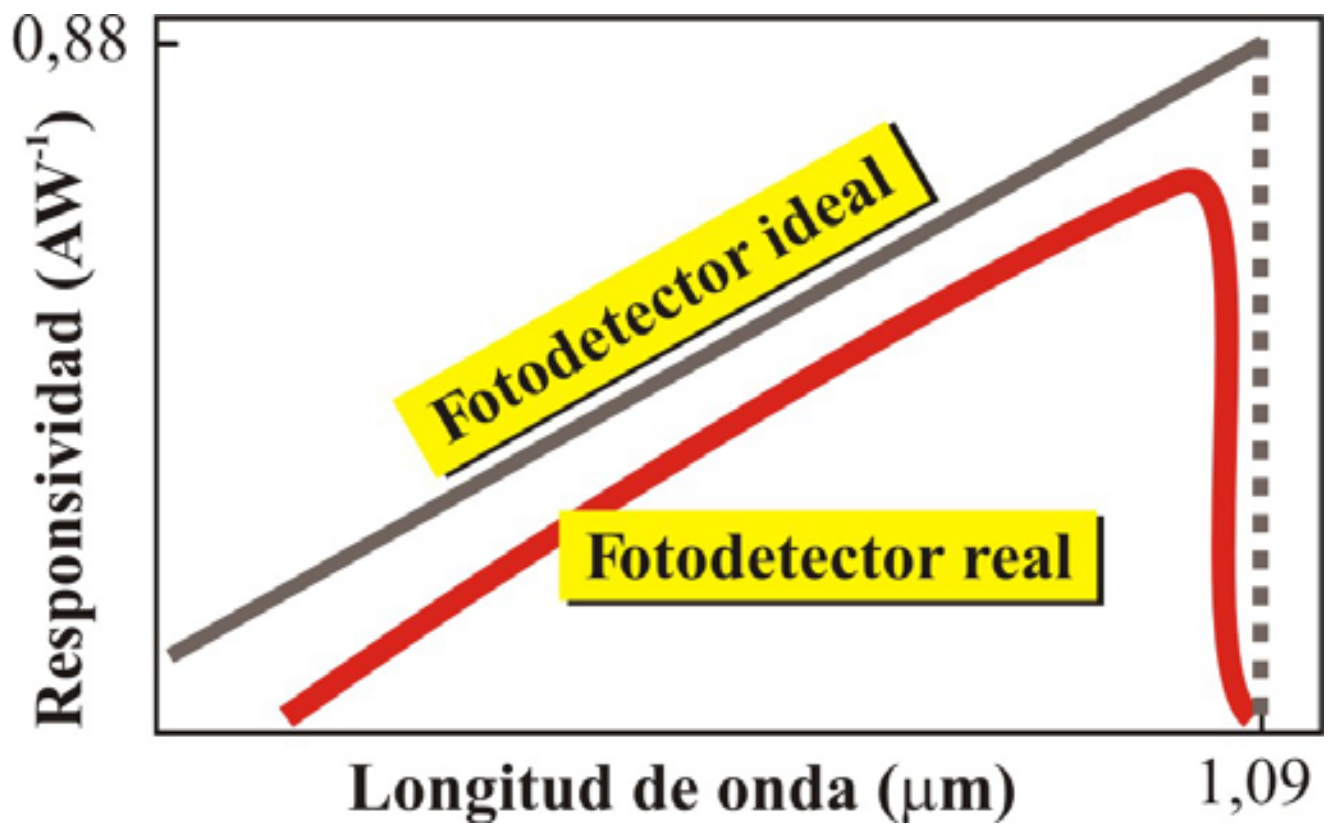

***Figura 5.6. Responsividad ideal y real de un fotodetector de Si.***

### 5.3.2. Respuesta optoelectrónica

Cuando el fotodiodo no está iluminado, su respuesta característica I-V es la típica de un diodo con una corriente residual, $I_d$, denominada **corriente de oscuridad**, de origen térmico.

Por el contrario, cuando incide una cierta radiación luminosa, aparece una fotocorriente inversa que desplaza toda la curva. La parte que queda en el tercer cuadrante muestra un comportamiento lineal, mientras que la zona del cuarto cuadrante es no lineal. Por esta razón, los fotodiodos se polarizan en inversa, de modo que la recta de carga discurra por la zona lineal. En todos los casos, la magnitud registrada es prácticamente constante, e independiente de la tensión inversa aplicada, hasta un cierto valor $V_a$, denominado de ***avalancha***. En ese momento, y debido al **efecto Zener**, el valor de la intensidad eléctrica crece de manera abrupta hasta valores muy elevados. Los fotodetectores PIN trabajan lejos de la zona Zener, mientras que los de avalancha (APD) se polarizan próximos a ella, con el fin de provocar ionización secundaria.

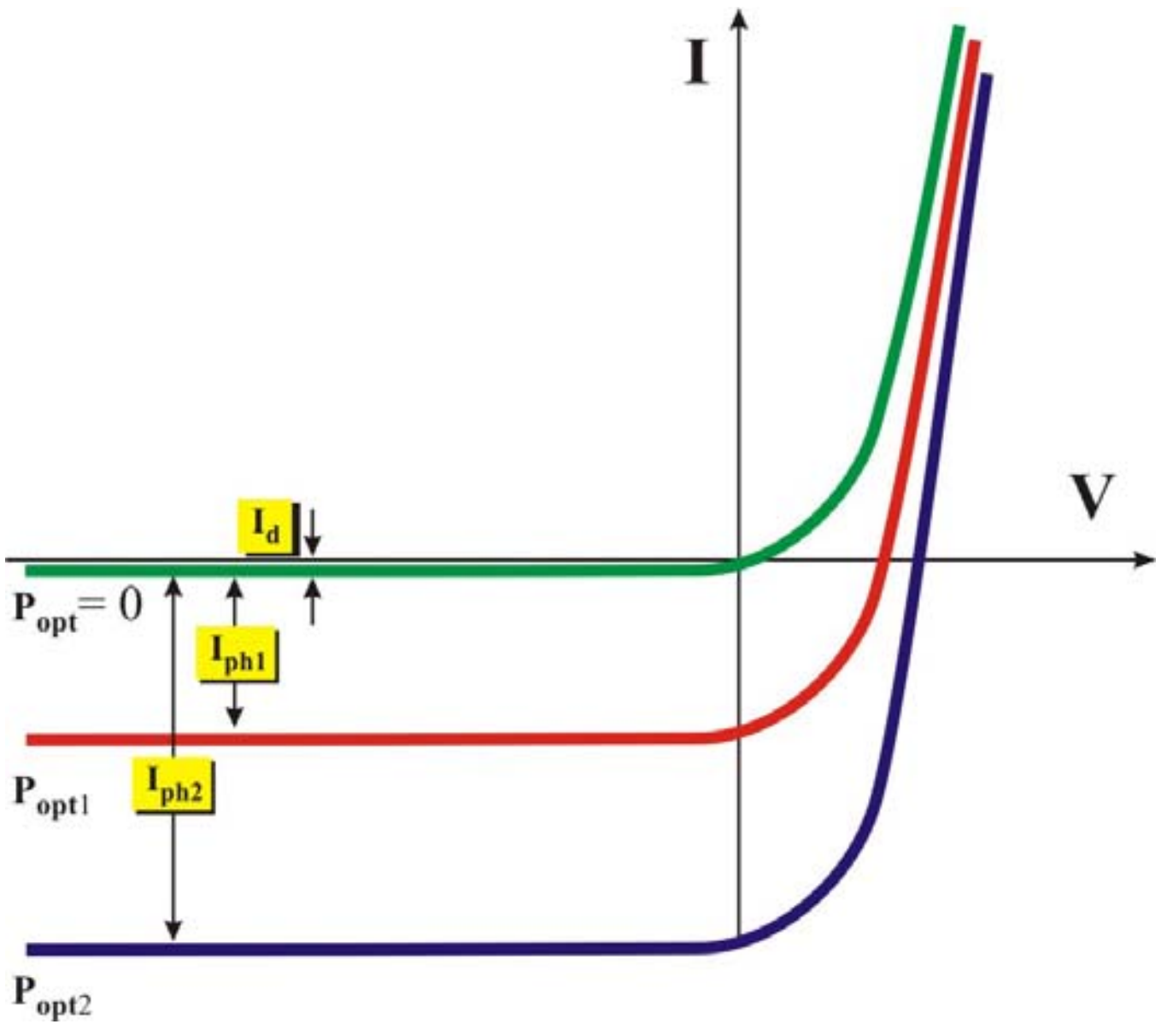

*Figura 5.7. Respuesta de un fotodiodo con y sin iluminación*

En definitiva, un fotodiodo en tercer cuadrante se comporta como un ***generador de corriente*** prácticamente ideal, cuya intensidad depende exclusivamente de la cantidad de luz recibida.

### 5.3.3. Materiales del detector

El material de fabricación de los fotodetectores Depende de la longitud de onda a detectar. Los detectores más comunes para el visible e IR próximo son de silicio. Este material presenta un **gap indirecto** de 1,14 eV, equivalente a una longitud de trabajo máxima de 1,09 µm. La ventana de trabajo del proyecto *MLCD* está en 1,064 µm, justamente por encima, por lo que teóricamente puede utilizar un detector de Si, si bien en la realidad su responsividad no será la máxima. Para longitudes de onda mayores se necesitan *gaps* más reducidos como los que ofrecen el Ge y los compuestos III-V ternarios y cuaternarios. Debido a que el Ge tiene un *gap* demasiado pequeño, y por lo tanto una mayor corriente de oscuridad, se prefieren los III-V (aunque puede optarse por un detector de Ge refrigerado).

### 5.3.4. Tipos de fotodiodos

Los dos tipos fundamentales de fotodiodos de semiconductor utilizados en Comunicaciones Ópticas son el fotodiodo *p-i-n* (conocido usualmente como PIN) y el fotodiodo de avalancha (APD, *avalanche photodiode*).

| Material | Rango longitud de onda (µm) | Longitud de onda de pico (µm) | Responsividad máxima (A/W) | Exponente x de exceso de ruido |
|---|---|---|---|---|
| ***Si*** | 0,3-1,1 | 0,8 | 0,5 | 0,3-0,5 |
| ***Ge*** | 0,5-1,8 | 1,5 | 0,6 | 1,0 |
| ***InGaAs*** | 1,0-1,8 | 1,7 | 0,75 | 0,5-0,8 |
| ***InGaAsP*** | 1,0-1,6 | 1,4 | 0,7 | 0,4-0,9 |

*Tabla 5.1. Algunas características de materiales de detectores*

El ***fotodiodo PIN*** está constituido por una unión p-n a la que se intercala una capa intrínseca con el fin de ensanchar la zona de deplexión (también llamada de **carga espacial**). De este modo se consigue hacer más ancha la zona activa, permitiendo que se incremente la radiación absorbida en la misma.

Este fotodiodo genera como máximo un solo par electrón-hueco por cada fotón absorbido (es decir, no tiene ganancia). Así pues, su responsividad máxima teórica (haciendo $\eta = 1$) será función de la longitud de onda. Para $\lambda = 1{,}064$ µm, la longitud de onda de trabajo en el proyecto *MLCD*, la responsividad máxima de un PIN es

$$R_0 = \eta \frac{\lambda e}{hc} = 0{,}858 \qquad (101)$$

ligeramente menor que el máximo valor teórico de un PIN de Si (Figura 5.6). Es importante observar que este valor es independiente del material utilizado, es decir, la responsividad de un PIN no puede sobrepasar ese valor a la $\lambda$ indicada, sea cual sea su composición o su tensión de polarización.

El ***fotodiodo de avalancha*** o ***APD*** es un fotodiodo PIN al que se ha agregado una capa adicional llamada **zona de ganancia**. En esa zona existe un campo eléctrico elevado, suficiente para que el par electrón-hueco generado como consecuencia del fotón absorbido puede adquirir la energía suficiente como para producir nuevos pares por ionización de impacto. Esta **ionización secundaria** suministra ganancia al APD, haciendo que por cada fotón se recoja un cierto número de electrones. La ganancia se denota con un parámetro característico de los APD, llamado ***factor de multiplicación***, **M**, que normalmente vale *algunos cientos* en APDs de Si y *algunas decenas* en APDs III-V.

Los APD pueden ser usados en dos modos de operación: ***modo de multiplicación*** y ***modo Geiger***. En el modo de multiplicación, la fotocorriente se amplifica dentro de la capa de deplexión del diodo por el proceso de multiplicación de avalancha, como ya se comentó anteriormente.

En el modo Geiger, el diodo opera por encima de la tensión de ruptura. La absorción de **un solo fotón** producirá la corriente de ruptura de avalancha, permitiendo la detección de fotones individuales. Cuando la polarización inversa excede la tensión de ruptura, los electrones y huecos se multiplican más rápido que la velocidad a la que pueden ser extraídos, en media. La cantidad de electrones y huecos en la región de campo intenso y la fotocorriente asociada crecen exponencialmente en el tiempo. Este crecimiento de la corriente continúa hasta que los campos eléctricos en el dispositivo se alteran por la presencia de la corriente creciente y la también creciente cantidad de electrones y huecos. Si se coloca una resistencia en serie con el diodo, se produce una **caída de tensión** cada vez mayor en la misma a medida que crece la corriente. Este efecto reduce la tensión de la región de multiplicación, extinguiendo la avalancha. Por lo tanto, una avalancha iniciada por la absorción de un solo fotón provoca que la corriente del diodo aumente hasta un cierto valor limitado por la resistencia. El tiempo de respuesta de esta corriente es bastante rápido, decenas de ps.

Este modo de funcionamiento, idéntico al de los contadores Geiger para rayos $\gamma$, puede resultar de interés para detección de señales muy débiles, como es el caso. Adviértase que un efecto similar puede conseguirse con un fotomultiplicador operando a la tensión adecuada. De hecho, existen numerosos equipos comerciales "contadores de fotones" (*photon-counting*) basados preferentemente en fotomultiplicadores.

### 5.3.5. Amplificador

El receptor de un sistema de Comunicaciones Ópticas dotado de un detector de fotodiodo se completa con un amplificador de baja impedancia, alta impedancia, o transimpedancia. Esta etapa es importante ya que interviene decisivamente en la evaluación de la relación señal-ruido.

En comunicaciones ópticas espaciales, la configuración de amplificación más adecuada es la de ***transimpedancia*** (Figura 5.8). Dicho amplificador presenta algunas de las ventajas de los amplificadores de baja impedancia de entrada –menor necesidad de ecualización– con una sensibilidad próxima a la de los amplificadores de alta impedancia. Como desventaja, la incorporación en este montaje de la nueva resistencia de realimentación $R_F$ introduce una fuente adicional de ruido.

## 5.4. RUIDOS DEL DETECTOR

La señal que se recibe en el fotodetector del sistema de comunicaciones ópticas, está formada por dos componentes. El primero lo constituye la verdadera **señal** en la que está contenida la información, que en su momento fue suministrada por el emisor. El segundo, por el conjunto de **señales no deseadas**, de origen óptico, que se hayan podido introducir bien a lo largo del camino de transmisión o bien en el momento de la propia detección.

Una vez que la luz ha incidido sobre el fotorreceptor, aparece un nuevo conjunto de señales indeseadas de naturaleza eléctrica, o lo que es lo mismo, de ruidos.

### 5.4.1. Sensibilidad

El parámetro más importante del receptor es su ***sensibilidad***, es decir, la potencia óptica que necesita para detectar que llega señal.

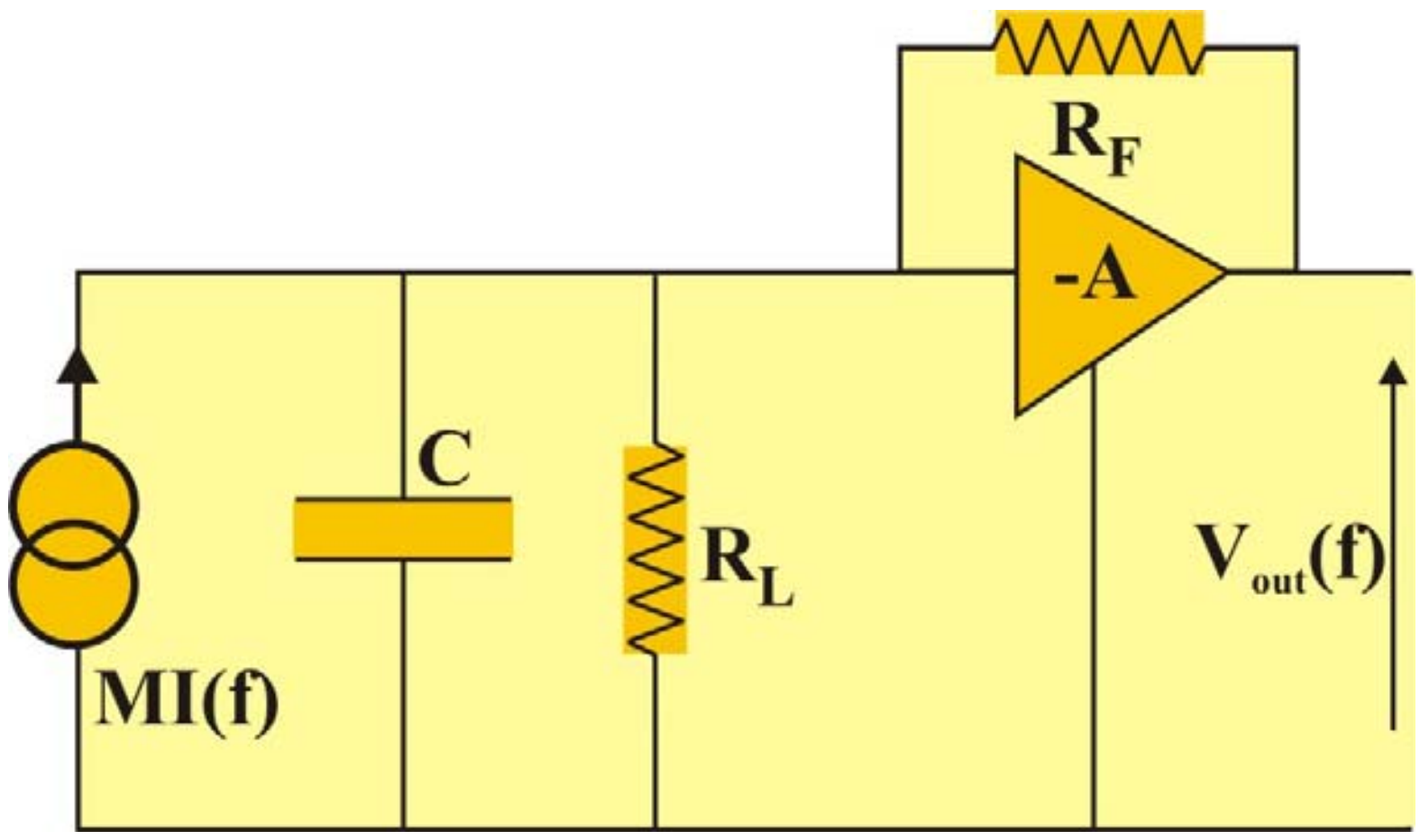

***Figura 5.8. Sistema receptor formado por un fotodetector APD y un amplificador en configuración de transimpedancia.***

La sensibilidad depende del tipo de detector empleado y de su modo de funcionamiento, así como de la velocidad de transmisión de los datos y de la tasa binaria de error (BER) que se requiera en las especificaciones del enlace.

El detector, con independencia del régimen binario, necesita un cierto número de fotones para detectar la señal. De esta forma y manteniendo la energía recibida por pulso constante, la potencia **crece linealmente** con el régimen binario. Esta dependencia podría aparentemente evitarse introduciendo, por ejemplo, un amplificador óptico a la entrada del detector. Sin embargo, las componentes de ruido que acompañan a la señal crecen también linealmente con el ancho de banda del receptor, por lo que la pretendida ventaja queda cancelada, y se mantiene una relación aproximadamente lineal con el régimen binario (Figura 5.9).

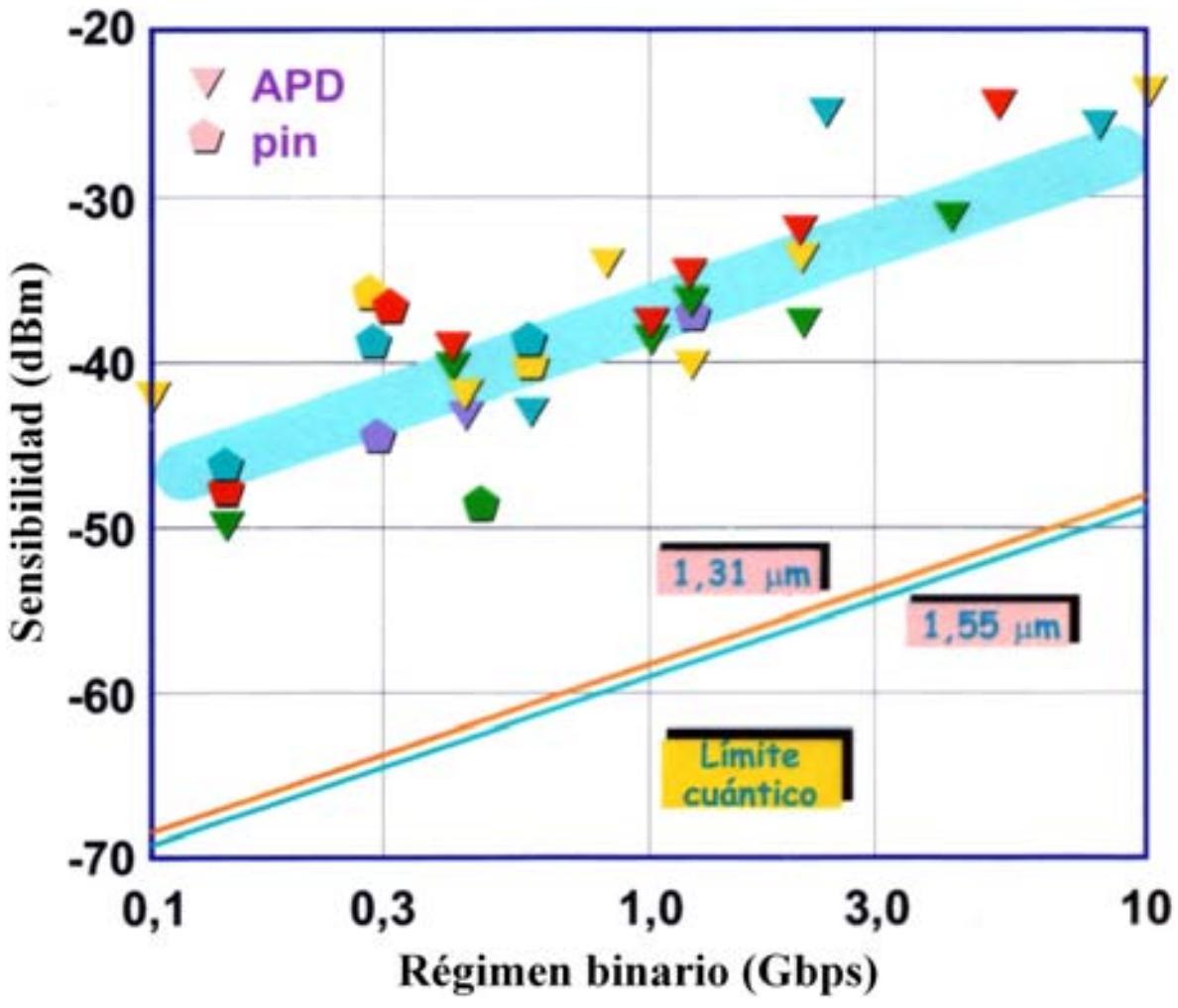

***Figura 5.9. Sensibilidad de algunos fotodetectores comerciales.***

## 5.4.2. El ruido *shot*

El origen del ruido *shot* o ***de granalla*** está en la **granularidad** de la materia (átomos) o radiación (fotones) que se reciben –aunque pueden citarse ejemplos más cotidianos, como la lluvia–. El ruido *shot* de los fotodetectores tiene un origen cuántico, que subyace en la propia incidencia de la luz sobre la superficie del detector. Una radiación monocromática de intensidad teóricamente constante, está en realidad compuesta por un conjunto de fotones cuya llegada se produce de manera aleatoria y según una cierta distribución probabilística de Poisson

$$P(z) = \frac{z_m^z \exp(-z_m)}{z!} \tag{102}$$

donde P(z) es la probabilidad de **detectar z fotones** durante el tiempo $\tau$ en que se están detectando $z_m$ como media.

Suponiendo un detector perfecto, es decir sin ningún ruido excepto esas fluctuaciones cuánticas, la única causa de error será que, durante el intervalo de tiempo ocupado por un bit, no se reciba **ni un solo fotón** cuando debería haberse recibido un pulso. Este evento *se produce inevitablemente con una cierta probabilidad*, que depende del número de fotones promedio por pulso, $z_m$. Se puede pues ajustar la probabilidad mínima de error modificando $z_m$ hasta alcanzar la BER deseada. Por ejemplo, si se quiere que la BER sea $10^{-9}$, y se comete un error únicamente cuando se reciben cero fotones

$$P(0) = \exp(-z_m) \;\Rightarrow\; \text{Para } P(0) = 10^{-9} : z_m = 20{,}7 \tag{103}$$

Es decir, el detector necesita 21 fotones en promedio para garantizar una BER<$10^{-9}$. Es importante percatarse de que esta limitación nada tiene que ver con imperfecciones del receptor o del sistema. Se trata de un límite absoluto, llamado ***límite cuántico*** (Figura 5.9), impuesto por la propia naturaleza corpuscular de la luz.

## 5.4.3. Ruido *shot* de la corriente de oscuridad

La ***corriente de oscuridad***, $I_d$, es la corriente eléctrica que aparece sin que haya incidido ningún impulso óptico. La corriente en sí es una continua que puede ser filtrada con facilidad. Sin embargo la corriente de oscuridad tiene también fluctuaciones *shot* que se traducen en ruido

$$< i_d^2 >= 2eBI_d \qquad (A^2) \qquad (104)$$

En un fotodiodo PIN, el ruido *shot* total se calcula sumando las contribuciones de las corrientes involucradas, es decir, la fotocorriente y la corriente de oscuridad

$$< i_{TS}^2 >= 2eB(I_{ph} + I_d) \qquad (A^2) \qquad (105)$$

siendo B el ancho de banda e $I_{ph}$ la corriente fotogenerada

$$I_{ph} = R_0 P_r \qquad (A) \qquad (106)$$

donde $P_r$ es la potencia óptica recibida en la superficie del fotodetector.

La corriente de oscuridad aparece también en los fotomultiplicadores, debida a **emisión termoiónica**. Este término constituye a menudo *la fuente de ruido más significativa* de estos dispositivos. La corriente de emisión termoiónica $i_T$ de un cátodo de área A a una temperatura T, con una función de trabajo φ viene dada por la ecuación de Richardson-Dushman

$$i_T = aAT^2 \cdot \exp\left( -\frac{e\phi}{kT} \right) \qquad (A) \qquad (107)$$

siendo **a** una constante que, para metales puros, toma el valor de $1{,}2 \cdot 10^{-6}$ A $m^{-2}$ $K^{-2}$. La expresión del ruido *shot* por corriente de fotoionización y de oscuridad de un fotomultiplicador es idéntica a la (105). Es importante señalar la dependencia de la corriente termoiónica de oscuridad con la temperatura: se puede reducir tal corriente (y su ruido asociado) **enfriando** el detector.

La presencia de ruido *shot* de la corriente de oscuridad limita la ***sensibilidad*** del fotomultiplicador. Es habitual considerar que el límite de detección de un PMT se alcanza cuando la corriente fotogenerada se iguala a la termoiónica. De este modo, se puede calcular la sensibilidad del PMT en función de su responsividad

$$I_{ph\ min} = R_0 S_{PMT} = \sqrt{2eBi_T} \Rightarrow S_{PMT} = \frac{\sqrt{2eBi_T}}{R_0} \qquad (W) \qquad (108)$$

## 5.4.4. Ruido *shot* en el factor de multiplicación

En los detectores de avalancha, las expresiones correspondientes aparecen multiplicadas por el factor de multiplicación M. El ruido *shot* de la corriente de oscuridad debería ser

$$< i_d^2 >= 2eBM^2 I_d \qquad (A^2) \tag{109}$$

En realidad, los APD multiplican la corriente por el factor M, pero el ruido no crece de forma lineal con la señal, ya que el propio proceso de multiplicación está sujeto a fluctuaciones shot. Este ruido adicional se denomina ***factor de exceso de ruido***, F, y es proporcional a una potencia de M

$$F = M^x \tag{110}$$

Los valores del exponente x de varios fotodiodos comunes se han indicado en la Tabla 5.1. El ruido total de un APD es

$$< i_{SA}^2 >= 2eB(I_{ph} + I_d)M^{2+x} = 2eBFM^2(I_{ph} + I_d) \qquad (A^2) \tag{111}$$

En condiciones normales, el término de ruido *shot* deberá ser el dominante en fotodiodos APD.

En fotomultiplicadores aparece también este **exceso de ruido**. Se demuestra que el ***factor de exceso*** es

$$F_{PMT} = \left( \frac{\delta}{\delta - 1} \right)^{1/2} \tag{112}$$

Se observa que el factor es relevante únicamente cuando $\delta \approx 1$. Por ejemplo, para un valor típico $\delta = 4$, el ruido se incrementa apenas en un 15%.

## 5.4.5. Ruido térmico o Johnson

Su origen son las fluctuaciones espontáneas debidas a las interacciones, entre por ejemplo los electrones libres y los iones de un medio conductor.

La tensión de ruido térmico producida por una resistencia $R_L$ puede expresarse según

$$< i_T^2 > = \frac{4kTB}{R_L} \qquad (A^2) \tag{113}$$

El ruido térmico es el predominante en los fotodiodos PIN.

Idéntica expresión se encuentra en los fotomultiplicadores, aunque en este caso (como en los fotodiodos APD), el ruido térmico suele ser despreciable frente al ruido *shot*.

## 5.4.6. Ruidos en el amplificador

Como se ha comentado anteriormente, el cálculo del ruido de un fotodiodo incluye el amplificador asociado. En la siguiente figura se muestra el circuito equivalente de ruido de un amplificador de transimpedancia

Aparecen *tres corrientes de ruido*, $i_{shot}$, $i_{term}$ e $i_A$, correspondientes al ruido *shot*, térmico y del amplificador, respectivamente, además de *dos términos de tensión de ruido* del amplificador con la resistencia y la capacidad parásita. Todos los términos de ruido dependen linealmente del ancho de banda, excepto el de la capacidad, que contiene un término $B^2$ adicional. La resistencia R, incluye el paralelo de la resistencia de carga $R_L$ con la resistencia de realimentación $R_F$ típica de una configuración de transimpedancia.

## 5.4.7. Ruido de fondo

Este tipo de ruido se debe a toda la radiación luminosa de fondo (sol, planetas, estrellas…) distinta de la correspondiente a la señal emitida y que puede ser relativamente significativa en Comunicaciones Ópticas a través del espacio profundo. Como siempre, son las **fluctuaciones**, y no la radiación luminosa en sí, lo que agrega nuevos términos al ruido *shot* ya presentado.

Durante el día, la fuente de ruido de fondo predominante es el Sol, mientras que por la noche, deben considerarse otras fuentes como la Luna, las estrellas, los planetas y la luz zodiacal [45]. Sin embargo, todas ellas resultan despreciables (excepto la Luna a muy bajo ángulo de observación) comparadas con la potencia luminosa del *planeta al que se está apuntando* (se supone que el satélite está en órbita alrededor de un planeta, que necesariamente habrá de estar incluido en el campo de visión).

La radiación de fondo puede entrar directamente en el telescopio receptor, cuando la fuente está dentro del campo de visión del receptor, o indirectamente, cuando la radiación sufre reflexión o *scattering* (la contribución diurna de la atmósfera terrestre se debe a este efecto). Si el campo de visión del receptor es lo suficientemente grande y el sátelite y el sol están suficientemente alineados, acabará produciéndose la interrupción de la comunicación debido a la gran cantidad de ruido de fondo procedente del Sol que incide directamente en el telescopio receptor. En el proyecto *MLCD* se establece un ángulo mínimo de 3º con respecto al Sol por encima del cual se exige que el enlace pueda establecerse.

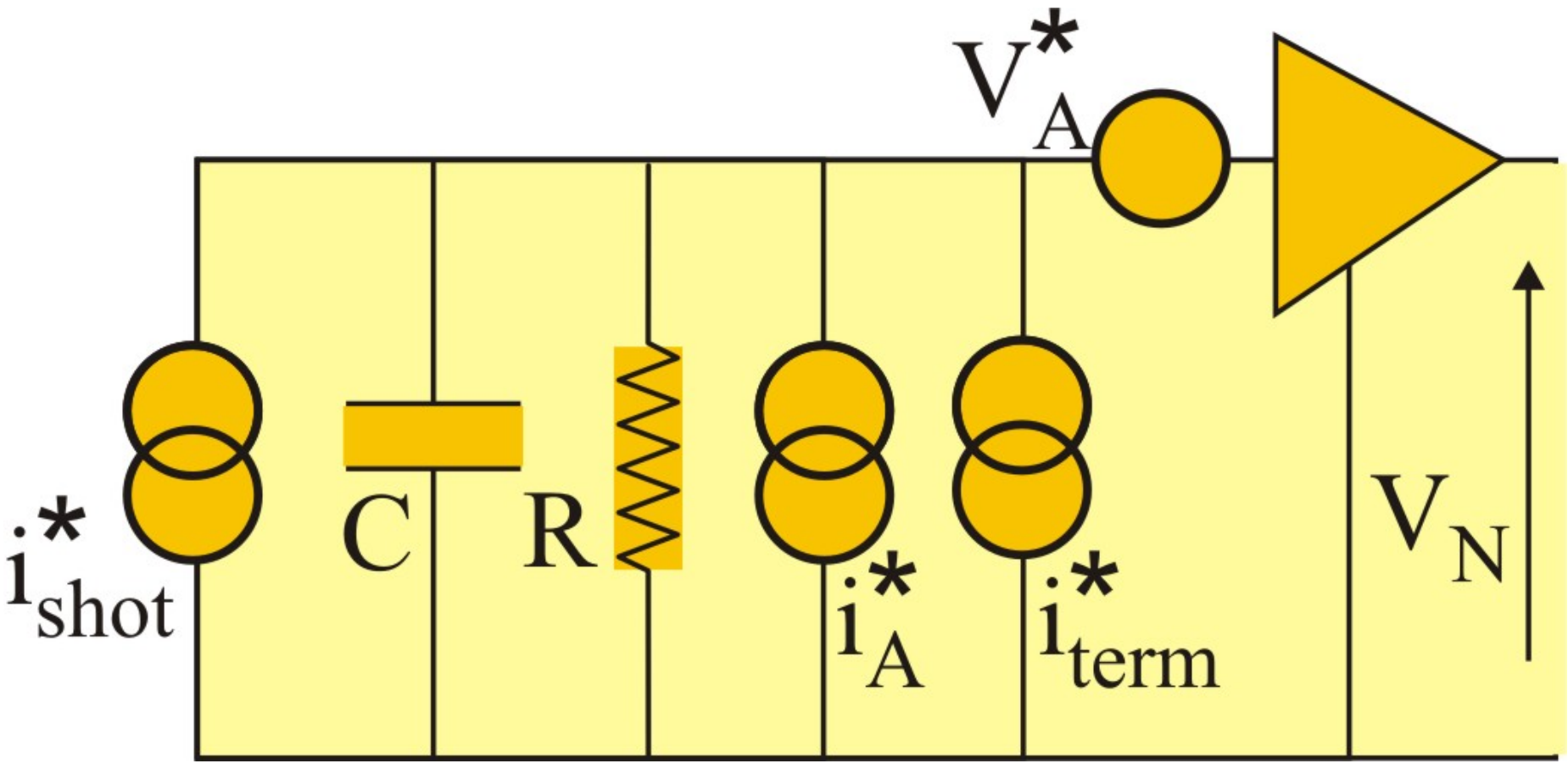

***Figura 5.10. Circuito equivalente de ruido de un receptor con amplificador de transimpedancia.***

El campo de visión del receptor deberá reducirse al máximo para minimizar el ruido de fondo. Sin embargo, las difíciles condiciones de apuntamiento y seguimiento con un campo de visión pequeño hace que se tenga que llegar a un compromiso.

La potencia de ruido de fondo, $P_b$, recogida en el receptor en un ancho de banda B viene dada por

$$P_b = \begin{cases} N(f)BA_r\Omega_r & \text{si } \Omega_r \in \Omega_s \\ N(f)BA_r\Omega_s & \text{si } \Omega_s \in \Omega_r \end{cases} \qquad (114)$$

donde N(f) es la ***función de radiancia espectral***, que representa la potencia radiada frente a la frecuencia, por unidad de ancho de banda, unidad de ángulo sólido y área de la fuente de ruido.

Igualmente la potencia de ruido puede ser expresada en función de la irradiancia espectral, H(f),es decir, la potencia radiada frente a la frecuencia, por unidad de ancho de banda y área de la fuente de ruido ($H(f) = N(f)\cdot\Omega_s$)

$$P_b = \begin{cases} H(f)BA_r\Omega_r / \Omega_s & \text{si } \Omega_r \in \Omega_s \\ H(f)BA_r & \text{si } \Omega_s \in \Omega_r \end{cases} \qquad (115)$$

El ángulo sólido $\Omega_s$, es el ángulo presentado por la fuente visto desde el detector, y su valor es

$$\Omega_s = \frac{A_s}{L^2} \qquad \text{(sr)} \qquad (116)$$

siendo L la distancia entre la fuente y el receptor y $A_s$ el área de la fuente.

El parámetro $\Omega_r$ representa el **campo de visión del receptor**, como ya se mencionó en el apartado dedicado a la antena receptora

$$\Omega_r \approx \frac{A_d}{f_c^{\,2}} \qquad \text{(sr)} \qquad (117)$$

donde $A_d$ es el área del detector y $f_c$ es la distancia focal de la lente en el receptor. El ruido de fondo del cielo es un caso en el que la extensión angular de la fuente es mayor que el campo de visión del receptor, $\Omega_r \in \Omega_s$. Otras fuentes como las estrellas o planetas pueden ser un ejemplo en el que la extensión angular de la fuente es menor que el campo de visión del receptor, $\Omega_s \in \Omega_r$.

Empleando la relación $S_n = P_b/B$, la densidad espectral de potencia de ruido de fondo para ambos tipos de fuentes (es decir, $\Omega_r \in \Omega_s$ y $\Omega_s \in \Omega_r$) viene dada por

$$S_n = \begin{cases} N(f)A_r\Omega_r & \text{si } \Omega_r \in \Omega_s \\ N(f)A_r\Omega_s & \text{si } \Omega_s \in \Omega_r \end{cases} \qquad (118)$$

o por

$$S_n = \begin{cases} H(f)A_r\Omega_r/\Omega_s & \text{si } \Omega_r \in \Omega_s \\ H(f)A_r & \text{si } \Omega_s \in \Omega_r \end{cases} \qquad (119)$$

Aplicando las relaciones

$$H(f) = \frac{H(\lambda)\lambda^2}{c} \qquad N(f) = \frac{N(\lambda)\lambda^2}{c} \qquad (120)$$

se obtienen las expresiones finales para la densidad espectral de potencia de ruido

$$S_n = \begin{cases} N(\lambda)A_r\Omega_r \dfrac{\lambda^2}{c} & \text{si } \Omega_r \in \Omega_s \\ N(\lambda)A_r\Omega_s \dfrac{\lambda^2}{c} & \text{si } \Omega_s \in \Omega_r \end{cases} \qquad (121)$$

$$S_n = \begin{cases} H(\lambda)A_r \dfrac{\Omega_r\lambda^2}{\Omega_s c} & \text{si } \Omega_r \in \Omega_s \\ H(\lambda)A_r \dfrac{\lambda^2}{c} & \text{si } \Omega_s \in \Omega_r \end{cases} \qquad (122)$$

## 5.5. SELECCIÓN DEL DETECTOR

El uso de un detector u otro en un enlace de comunicaciones a través del espacio profundo dependerá de las condiciones del sistema, pues cada tipo de detector tendrá sus propias ventajas e inconvenientes que le harán más o menos atractivo.

Así, para enlaces entre satélites con detección directa basados en un láser Nd:YAG ($\lambda$ = 1,064 µm) el detector más adecuado es en principio un ***fotodiodo PIN*** o un ***APD*** (aunque, como se discute en el capítulo 6, esa longitud de onda es conflictiva por estar cerca del límite de detección del silicio). Si el nivel de señal es bajo, es preferible el fotodiodo de avalancha debido a su ganancia interna. Aunque el factor de multiplicación M de un APD pueda llegar a 1000 o más, normalmente toma valores entre 50-400 en APDs de Si y de 10-30 en APDs III-V. Un aumento en dicha ganancia provoca un incremento en el factor de exceso de ruido en un rango de 2-5 para dicho margen de multiplicación. Conviene recordar, no obstante, que actualmente se encuentran ***fotomultiplicadores*** capaces de competir con ventaja (excepto en precio) con los APD dentro de este rango de $\lambda$.

Cuando la comunicación no precisa ganancia, sería más adecuado un fotodiodo PIN debido a su mejor comportamiento y simplicidad. Es decir, cuando no se requiere la gran sensibilidad que poseen los APD frente a los PIN, se evidencian sus principales inconvenientes: trabajan a tensiones mayores (decenas o centenas de voltios), son más ruidosos y también más lentos, a causa de la ionización secundaria que aumenta el tiempo de recolección de portadores. También son preferibles los PIN en sistemas de comunicaciones coherentes.

Si se emplea un láser Nd:YAG doblado en frecuencia ($\lambda$ = 532 nm) la ventaja del fotomultiplicador se hace decisiva. Sus principales características son su altísima ganancia ($\approx 10^4$–$10^6$) y su bajo factor de exceso de ruido (1-2). Los tradicionales problemas de los PMT clásicos –tamaño, fiabilidad de componentes de alta tensión, baja eficiencia cuántica por limitación de materiales, reducido ancho de banda– están actualmente superados en gran medida por las unidades miniaturizadas y microcanal existentes en el mercado.

# 5.6. DETECCIÓN DE LA SEÑAL

Existen dos tipos básicos de detección en comunicaciones ópticas, ***coherente*** e ***incoherente***. Se pretende en este apartado contrastar las cualidades de cada una; dado que la mejor forma de evaluar la calidad de una modulación es su **relación señal-ruido** (abreviada SNR o relación S/N), la descripción de las dos modulaciones se basará en el cálculo de este parámetro. Para este cálculo, en principio, se tendrá en consideración, tanto para modulación coherente como no coherente, un ***diodo fotodetector de avalancha***. De esta manera será fácil la comparación de las expresiones de relación señal-ruido de cada modulación para los dos tipos de fotodetectores más ampliamente difundidos, esto es, APD y PIN, ya que el segundo se puede tratar como un caso particular del primero.

Asimismo, una vez analizado por separado cada tipo de modulación, se realizará una comparación entre ambos para dar cuenta de las virtudes o defectos que conlleva el empleo de uno u otro dentro del ámbito de estudio de este proyecto, un sistema de comunicaciones ópticas en espacio profundo.

## 5.6.1. Detección no coherente

Las técnicas de modulación no coherentes (o **incoherentes**) se conocen en la terminología de comunicaciones ópticas como de ***detección directa*** debido al hecho de que en un receptor de este tipo la potencia óptica recibida es directamente transformada por un fotodetector en una corriente eléctrica. Esta corriente generada es proporcional a la potencia óptica a la entrada del mismo, o lo que es lo mismo, es proporcional al cuadrado del campo eléctrico incidente sobre el fotodetector.

Empleando el esquema básico de Figura 5.11, se realiza seguidamente un cálculo de la relación señal-ruido que servirá como referencia para evaluar la calidad de esta técnica de detección frente a la coherente.

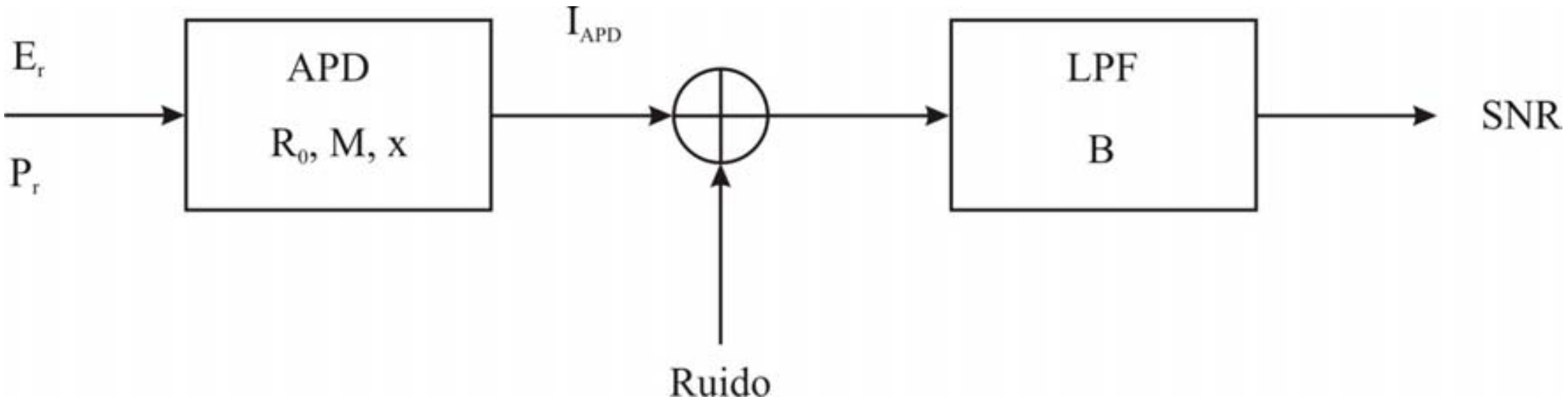

***Figura 5.11. Diagrama de bloques de un receptor básico de detección directa.***

## VALOR ÓPTIMO PARA EL FACTOR DE MULTIPLICACIÓN M DEL APD

Teniendo en consideración las dos fuentes de ruido ya mencionadas, esto es, térmico y *shot*, se demuestra a continuación que, en un fotodiodo APD, existe un ***valor óptimo*** para el factor multiplicador M. Un valor inferior o superior a este M óptimo supondrá la degradación de la relación señal-ruido, por lo que la máxima relación señal-ruido se obtendrá para un valor de M igual al óptimo.

Si se aprecia la componente de ruido *shot* en la ecuación (126) se comprueba que su valor es proporcional a $M^{2+x}$. Como x es siempre mayor que cero, $M^{2+x}$ será siempre mayor que $M^2$ (que es el factor que multiplica a la componente de señal), lo que significa que el término de ganancia M del APD multiplica en mayor medida a la componente de ruido *shot* que a la señal. Esto significa que, cuando el valor de M hace que la componente de ruido *shot* sobrepase a la de ruido térmico (que es constante), el aumento de M por encima de ese valor es favorable para el ruido *shot* y desfavorable para la señal. Por lo tanto el valor de M no debe permanecer por encima de ese M máximo, de otra manera la relación señal-ruido empeorará, acotando así el valor máximo de M.

Por otro lado, si el valor de M disminuye haciendo que la componente de ruido *shot* quede por debajo de la componente de ruido térmico, el valor de ruido total quedará determinado por la componente de ruido térmico, que al ser constante, hará que lo que disminuya M se traduzca únicamente en una disminución de la potencia de la señal, quedando el ruido invariable y dominado por el ruido térmico. Obviamente disminuir la potencia de la señal manteniendo una potencia de ruido constante se traduce, de nuevo, en un deterioro de la relación señal-ruido. Esto determina el valor mínimo de M.

En la Figura 5.12 puede comprobarse cómo por encima y por debajo del valor M óptimo, el efecto conseguido es la **disminución del valor de la relación señal-ruido** por lo que el valor de M óptimo maximiza la relación señal-ruido y viene dado por [9]

$$M_{opt} = \left( \frac{2\,N_t}{x\,e\left(R_0 P_r + I_d\right)} \right)^{\frac{1}{2+x}} \qquad (123)$$

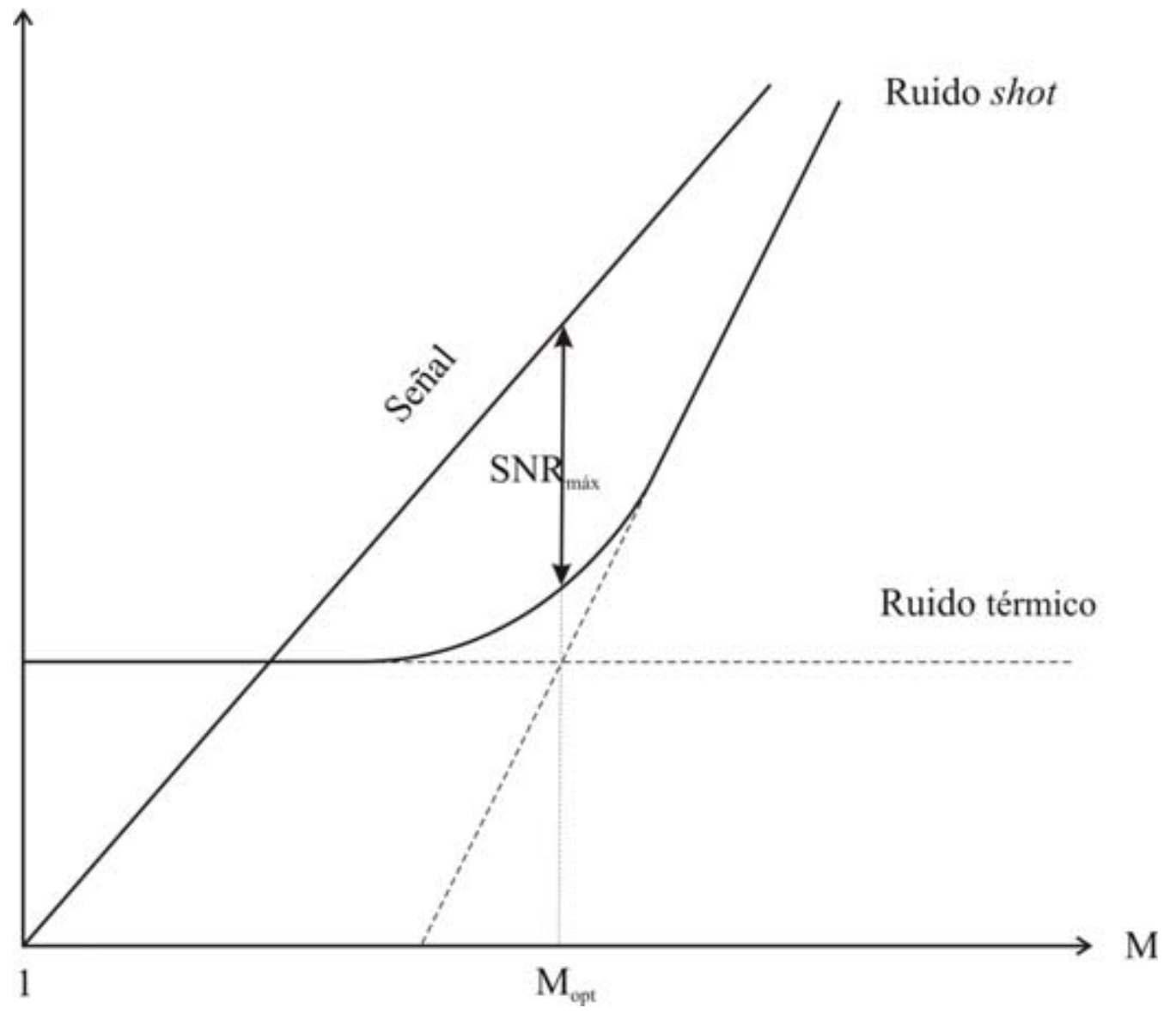

***Figura 5.12. Máxima relación señal-ruido en un APD para M = $M_{opt}$.***

Como se observa en la Figura 5.11, el campo eléctrico de la señal incide directamente sobre la superficie del fotodetector. Éste, de la forma que se explica en el apartado 5.3.2, transforma la potencia óptica en una fotocorriente, que viene dada por

$$I_{APD} = M\, R_0\, P_r \qquad (A) \qquad (124)$$

donde M representa el factor de multiplicación APD y $R_0$ la responsividad

Conocida la expresión de la fotocorriente generada, la potencia de la señal (ya en dominio eléctrico) será

$$P_S = \frac{I_{APD}{}^2}{R} = (M\, R_0\, P_r)^2 \qquad (W) \qquad (125)$$

refiriéndola a una resistencia de 1Ω.

Por otro lado, el ruido que se superpone a la señal tiene dos orígenes distintos, por una parte está el ruido *shot* del fotodiodo y por otra el ruido térmico del circuito eléctrico que sigue a la detección óptica. A su vez el ruido *shot* está compuesto por dos componentes, una que depende de la potencia óptica recibida y otra causada por la corriente de oscuridad del fotodiodo. Entonces la potencia de ruido total queda

$$\begin{aligned} P_N &= \left[e\, M^2\, F(M)\, R_0\, P_r + e\, M^2\, F(M)\, I_d + N_t\right] B = \\ &= e\, M^2\, F(M)\left[R_0\, P_r + I_d\right] B + N_t\, B \qquad (W) \end{aligned} \qquad (126)$$

donde $F(M)$ es el ya comentado ***factor de exceso de ruido***, que puede ser aproximado a $M^x$. El exponente x depende del material del fotodetector. El factor $I_d$ representa la corriente de oscuridad y $N_t$ la densidad espectral de potencia de ruido térmico. Esta potencia de ruido resulta de hacer el producto de la suma de la densidad espectral de potencia de cada ruido por el ancho de banda B del filtro paso bajo. La densidad espectral de potencia de todos los ruidos es aproximadamente independiente de la frecuencia, por lo tanto es una constante.

Conocidas la potencia de señal y de ruido, la relación señal-ruido no es más que el cociente de la primera y la segunda resultando en

$$SNR = \frac{(M\, R_0\, P_r)^2}{e\, M^{2+x}\left[R_0\, P_r + I_d\right] B + N_t\, B} \qquad (127)$$

## 5.6.2. Detección coherente

De la misma manera que se ha hecho en el apartado anterior, en este apartado se realiza un cálculo de la relación señal-ruido basado en el esquema básico de receptor coherente de la Figura 5.13. Como se observa en dicha figura, la diferencia entre el receptor coherente y el de detección directa radica en el hecho de que en el receptor coherente se realiza la superposición entre la señal recibida y una señal generada por un láser local, de manera que a la superficie del fotodiodo llega esta mezcla de señales en lugar de únicamente la señal recibida como era el caso en el receptor de detección directa.

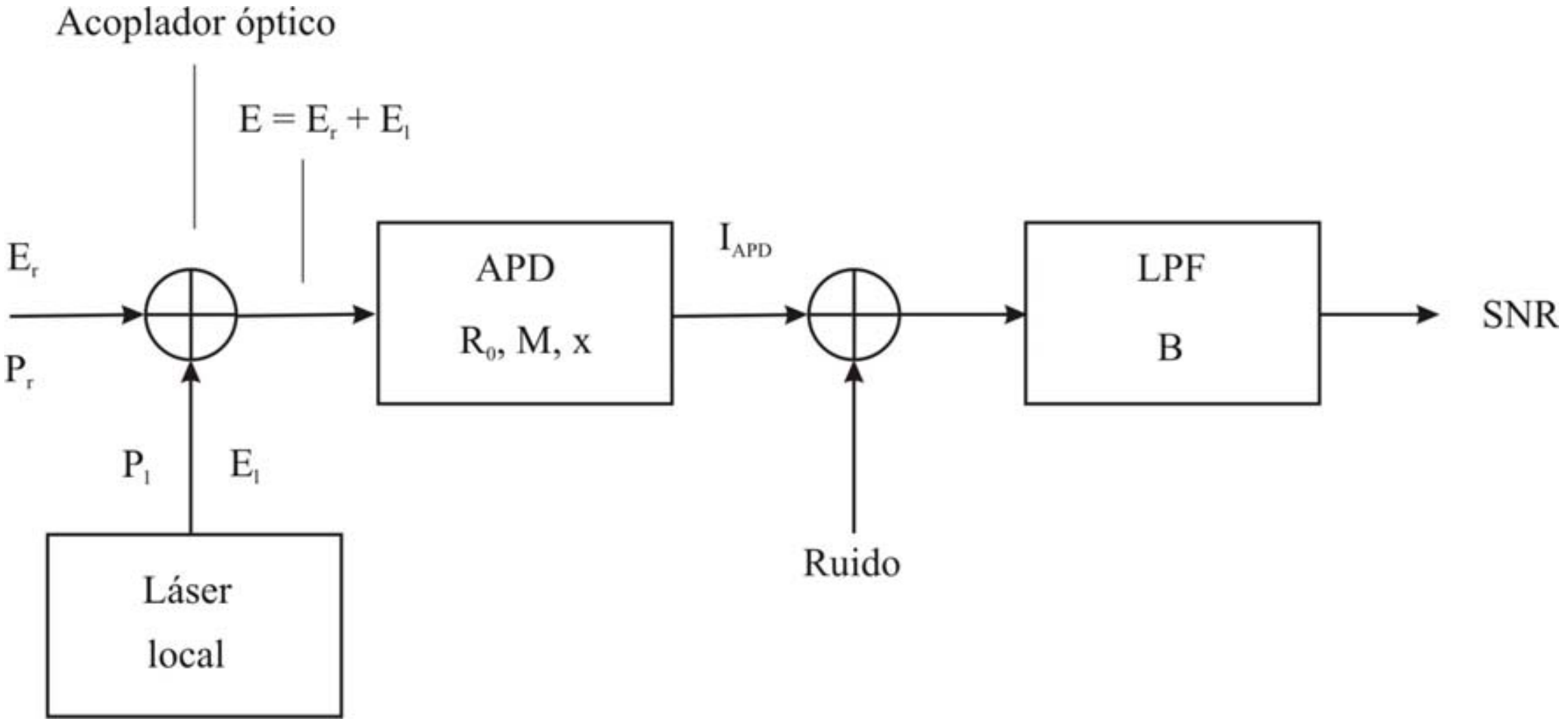

***Figura 5.13. Diagrama de bloques de un receptor básico de detección coherente.***

La señal generada como resultado de la superposición de la señal del láser y la recibida se expresa de la siguiente forma

$$E = E_r + E_l \qquad (V/m) \qquad (128)$$

siendo $E_r$ y $E_l$ los campos eléctricos de la señal recibida y la señal del láser local respectivamente y E el campo eléctrico resultante incidente en el fotodetector y cuyo cuadrado es proporcional a la potencia óptica y a la fotocorriente eléctrica generada, como se ve a continuación

$$I_{APD} \propto P_o \propto M\,R_0\,(E)^2 \qquad (A) \qquad (129)$$

donde $P_o$ representa la potencia óptica de la mezcla de señales y en términos de dicha potencia la fotocorriente a la salida del APD vendrá dada por

$$I_{APD} = M\,R_0\left(P_r + P_l + 2\sqrt{P_r\,P_l}\right) \qquad (A) \qquad (130)$$

Como se observa la fotocorriente generada tiene **tres componentes**, sin embargo sólo va a interesar la que incluye la potencia de las dos señales. Las otras dos pueden eliminarse fácilmente filtrando. Por lo tanto la expresión de la corriente de interés queda

$$I = 2\,M\,R_0\,\sqrt{P_r\,P_l} \qquad (A) \qquad (131)$$

De la misma manera que se hizo en el apartado de detección directa, si consideramos una resistencia de referencia de 1Ω la potencia de la señal resulta en

$$P_S = 4\,M^2\,R_0^{\,2}\,P_r\,P_l \qquad (W) \qquad (132)$$

La potencia de ruido se obtiene de la misma forma que en detección directa pero el factor que dependía de la potencia de la señal recibida ahora *también depende de la potencia de señal del láser*. El resto de los términos, que, como se vio, eran independientes de la señal recibida, permanecen igual que en la fórmula (126), correspondiente a la potencia de ruido en detección directa.

$$P_N = \left[e\, M^2\, F(M)\, R_0\, P_l + I_d\right] B + N_t\, B \qquad (W) \qquad (133)$$

Rigurosamente debería incluirse $P_r$ además de $P_l$ en la ecuación (133). Sin embargo, debido al hecho de que la potencia de la señal recibida es varios órdenes de magnitud menor que la de la señal del láser local, esto es, $P_r << P_l$, a efectos prácticos puede despreciarse el factor de la potencia recibida frente al de la potencia del láser.

Entonces la relación señal-ruido es

$$SNR = \frac{4\, M^2\, R_0^{\ 2}\, P_r\, P_l}{e\, M^{2+x}\left[R_0\, P_l + I_d\right] B + N_t\, B} \qquad (134)$$

Hay que señalar que, tanto en el caso del receptor coherente como el de detección directa, la expresión de la relación señal-ruido, en el caso de que se empleara un fotodiodo PIN en lugar del APD, quedaría de la misma manera con la excepción de que no llevarían el factor multiplicativo $M^2$ en el término de la señal ni el $M^{2+x}$ en el del ruido.

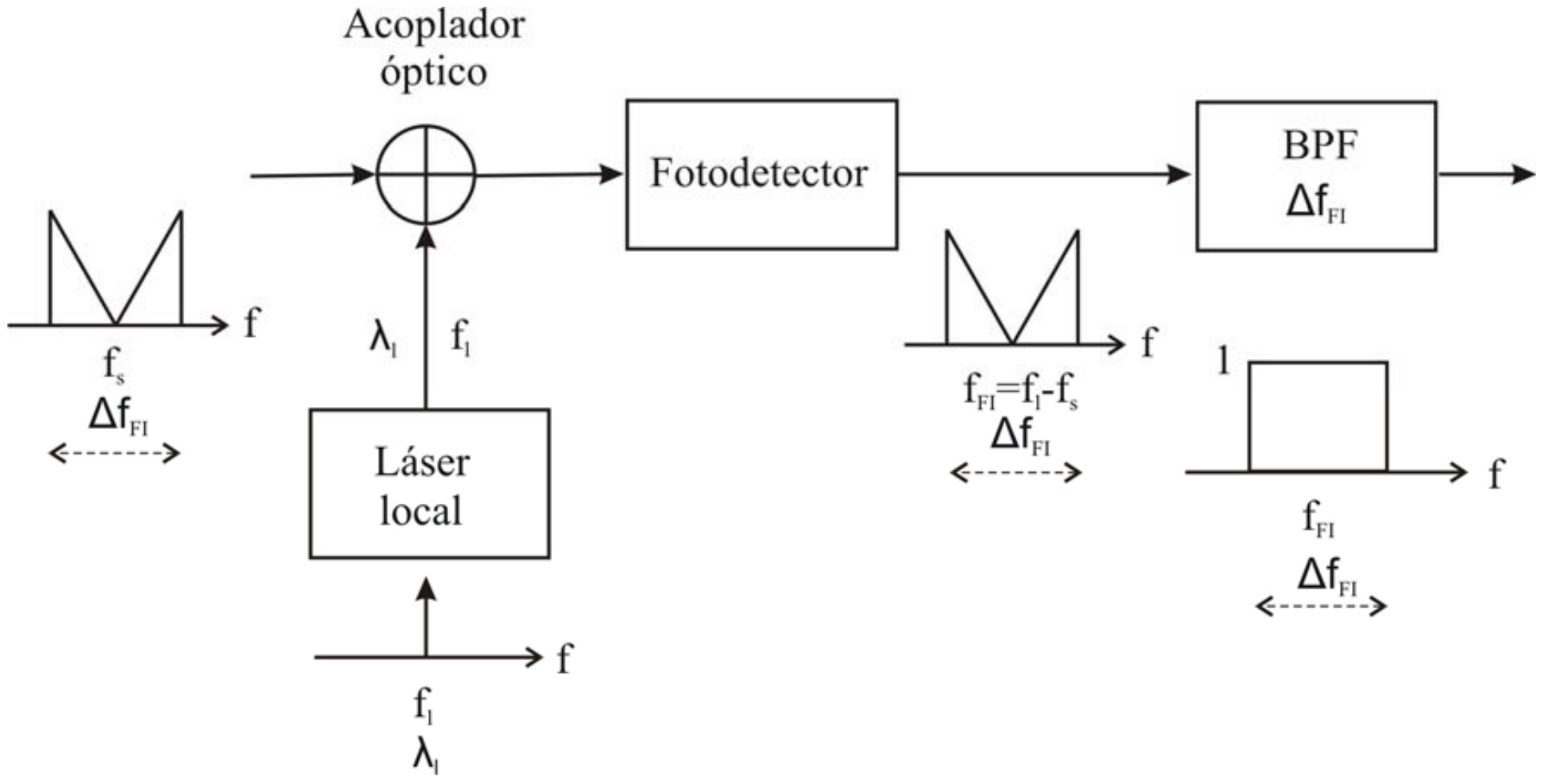

***Figura 5.14. Modulación coherente heterodina.***

## 5.6.3. Comparación

Como se aprecia en la expresión de la potencia de la señal detectada dada por la ecuación (132), en contraste con la expresión análoga en modulación no coherente (ecuación (126)), hay un término adicional de ganancia al depender la señal detectada de la potencia del láser, lo que significa que el láser actúa como amplificador de la señal recibida. Amplificación ésta que, en principio, puede ser incrementada tanto como se quiera mediante un incremento de la potencia del láser. Por otro lado, al aumentar la potencia del láser también se aumenta el ruido, sin embargo este aumento es favorable a la señal (aumenta más que el ruido). Con una potencia de láser suficientemente alta, el ruido dominante sería el ruido *shot*. Así la relación señal-ruido presenta *un máximo limitado por el ruido shot*. Sin embargo, por debajo de este límite, todo aumento en la potencia del láser se traducirá en un aumento de la relación señal-ruido, lo que se traduce en una ventaja respecto a la detección directa, en la que como se ha visto, el factor de ganancia M que introduce el APD tiene un valor óptimo que no se puede sobrepasar porque haría empeorar la relación señal-ruido.

**TÉCNICAS DE DETECCIÓN COHERENTE**

Existen dos diferentes tipos de detectores coherentes: ***homodino*** y ***heterodino***. La diferencia entre uno y otro es la misma que para los receptores de radiofrecuencia. Un **detector homodino** es aquel en el que la frecuencia del láser local es exactamente la misma que la frecuencia central del espectro de la señal recibida. Por el contrario, en el **detector heterodino**, la frecuencia del láser y la frecuencia central de la señal recibida son distintas.

Estas diferentes frecuencias del láser para cada tipo de modulación tienen el siguiente efecto en la señal que llega a la superficie del fotodetector: en el caso de la detección homodina, al tratarse de la misma frecuencia que la central de la señal recibida, la señal resultante de la mezcla es una **señal en banda base** mientras que en el caso de la detección heterodina se produce una **señal centrada en una frecuencia intermedia** (conocida como FI).

La razón de estas distintas frecuencias centrales de los espectros de las mezclas se debe a que idealmente el láser es una fuente coherente, es decir, su espectro es una delta de Dirac centrada a la frecuencia de trabajo del láser, por lo que la convolución en frecuencia entre este espectro y cualquier señal se traduce en un desplazamiento del espectro de la señal a la frecuencia de la delta. Por lo tanto, en el caso de la detección heterodina, la frecuencia intermedia será el resultado de la diferencia entre la frecuencia del láser y la de la señal recibida. Lo mismo puede decirse de la detección homodina, en cuyo caso la diferencia de frecuencias resultaría en una señal en banda base. En la Figura 5.14 se puede observar el diagrama de bloques de la modulación heterodina. El de la homodina sería idéntico salvo que $f_s$ sería igual a $f_l$, y por lo tanto, la diferencia de frecuencias sería cero (el filtro tendría que ser paso bajo en lugar de paso banda).

Como se acaba de ver, la modulación coherente homodina precisa que ambas señales, la del láser local y la recibida, tengan exactamente la misma frecuencia, por lo que se hace necesario el empleo de un OPLL (*Optical Phase-Locked Loop*), cuya función es idéntica a la de los tradicionales PLL de radio frecuencia, esto es, la de extraer a tiempo real la frecuencia central de la señal recibida para sintonizar a la misma frecuencia el oscilador local, que en este caso sería un láser sintonizable.

Cabe señalar que las modulaciones coherentes ofrecen varias técnicas para modular la señal. A diferencia de las técnicas de detección directa, en las que la única manera en que se puede modular es en intensidad (transmitiendo pulsos o ausencia de ellos,), las modulaciones coherentes admiten las técnicas de modulación conocidas de radiofrecuencia, tales como ASK (*Amplitude Shift Keying*), FSK (*Frequency Shift Keying*), PSK (*Phase Shift Keying*), DPSK (*Differential Phase Shift Keying*), etc.

Si se observan las ecuaciones (127) y (134) referentes a la relación señal-ruido de los receptores de detección directa y coherente respectivamente, cabe destacar que el ruido en el caso de la detección directa *depende de la señal recibida*, en contraste con el detector coherente cuyo ruido es *independiente de la señal*. Este hecho supone una ventaja a favor de la detección coherente. Sin embargo conviene advertir que podría convertirse en un inconveniente en presencia de ruido de láser (causado éste por la presencia inevitable y aleatoria de procesos de emisión espontánea en el láser).

Obviamente la técnica de detección directa es insensible a este ruido, pero en la detección coherente se convierte en un factor crítico si no es posible minimizarlo.

Una forma intuitiva de evaluar cómo de buena es una modulación respecto a la otra es mediante el **cociente** entre ambas relaciones señal-ruido. Operando con las ecuaciones (127) y (134) se llega a

$$\frac{SNR_c}{SNR_d} = \left(4\,\frac{P_l}{P_r}\right)\left(\frac{e\,M^{2+x}\left[R_0\,P_r + I_d\right] + N_t}{e\,M^{2+x}\left[R_0\,P_l + I_d\right] + N_t}\right) \tag{135}$$

donde $SNR_c$ representa la relación señal a ruido del detector coherente y $SNR_d$ la del detector no coherente. Se puede comprobar que si $P_l$ es suficientemente grande, la relación señal-ruido de la modulación coherente será siempre mayor que la del receptor no coherente debido a que el aumento de $P_l$ hace que el numerador aumente más rápido que el denominador.

Por ejemplo, para los valores $P_r$ = 10 nW, $N_t=10^{-23}$ $A^2/Hz$, $I_d=10^{-11}$ A, $R_0$=1 A/W, x=0,9, y $M=M_{opt}=27$ resulta que el valor de $P_l$ que hace que ambas relaciones señal-ruido sean iguales es 2,5 nW. Si se tiene en cuenta que la potencia real de un láser comercial pueden ser varios órdenes de magnitud mayor que eso, se comprueba que la relación señal-ruido sería siempre mucho mayor en detección coherente que en no coherente.

Para saber la máxima diferencia entre la relación señal-ruido del receptor coherente frente al no coherente, conviene recordar que el aumento de la potencia del láser supone un aumento de la relación señal-ruido pero hasta un cierto valor que viene limitado por el ruido *shot*. Si en la fórmula (134) de la relación señal-ruido del detector coherente se eliminan todos los ruidos a excepción del *shot* resultará que el valor máximo de relación señal-ruido viene dado por

$$SNR_{c\ máx} = \frac{4\,R_0\,P_r}{e\,M^x\,B} \tag{136}$$

Como se observa en esta ecuación el factor multiplicador del APD se convierte en un inconveniente al reducir la relación señal-ruido, por lo que será preferible usar un fotodiodo PIN, en el que M=1. Esto es debido a que la ventaja que proporcionaba el fotodiodo APD con su factor multiplicador M, en detección coherente la sustituye el factor de ganancia que introduce el láser local, que, a diferencia del factor M, puede aumentarse todo lo que se quiera, redundando en un aumento de la señal.

Si se vuelve a hacer el cociente de las relaciones señal-ruido de ambas modulaciones pero empleando ahora el valor máximo para la detección coherente, se obtendrá la máxima mejor de la modulación coherente en relación a la no coherente por aumento de la potencia del láser local. Esta expresión, por lo tanto, será

$$\frac{SNR_{c\ máx}}{SNR_d} = 4\left[\frac{R_0\,P_r + I_d}{R_0\,P_r} + \frac{N_t}{e\,M^{2+x} R_0\,P_r}\right] \tag{137}$$

Si se emplean los valores del ejemplo anterior (esta vez utilizando un fotodiodo PIN, por lo tanto M=1) para evaluar cuánto mejoraría la modulación coherente a la no coherente, se llega a una mejora de 112. Esto es, la relación señal-ruido para modulación coherente es 112 veces (o 20,5 dB) mejor que para detección directa.

Este valor es lo máximo que puede mejorar la relación señal-ruido una modulación coherente en relación a una no coherente (para los valores del ejemplo). Si en la fórmula (135) se despeja la potencia de láser que se necesitaría para alcanzar esta situación, se llega a una potencia de unos 0,3 μW, que es una potencia perfectamente alcanzable en un láser comercial.

### 5.6.4. Conclusión

Se ha demostrado que la relación señal-ruido de la modulación coherente siempre ***será mayor*** que en el caso de detección directa. Conviene advertir que, como se verá en el apartado dedicado a evaluar el efecto de la distancia, un aumento de la relación señal-ruido de 6 dB ofrece la posibilidad de duplicar la distancia entre los terminales, manteniendo la misma calidad del enlace, lo que supone una gran ventaja en espacio profundo.

Cabe destacar otra ventaja de la modulación coherente respecto a la no coherente. La capacidad de *aumentar la potencia del láser local* y así aumentar la relación señal-ruido hace que el ruido dominante sea el ruido *shot* por lo que la contribución de cualquier otra fuente de ruido quedaría por debajo de éste. En comunicaciones en el espacio hay momentos en que el receptor está apuntando al emisor con ángulos muy cercanos al Sol, el cual puede suponer una fuente considerable de ruido aún empleando filtros solares, por lo que la posibilidad de minimizar este ruido supone un gran beneficio. La modulación coherente permite minimizar este ruido haciendo que el ruido *shot* dependiente de la potencia del láser lo haga despreciable. Se ha estimado [46] que incluso apuntando directamente al Sol, la relación señal-ruido no se degradaría más de 3 dB. Por supuesto esto es válido para todas las fuentes de ruido de fondo que llegan al receptor, ya sea del Sol, de estrellas, de planetas, etc.

Por todas las virtudes expuestas, en términos de mejora de la relación señal-ruido, siempre será preferible utilizar una modulación coherente a una basada en la detección directa. Sin embargo, la implementación de sistemas de comunicaciones ópticas basados en modulaciones coherentes conlleva una serie de dificultades que les alejarían de su comportamiento ideal explicado anteriormente. A continuación se mencionan las más importantes.

Existe un inconveniente, para el caso de modulaciones coherentes, asociado a la posibilidad de que el terminal terrestre se halle sobre la superficie de la Tierra (lo que significaría que la señal recibida debe atravesar la atmósfera terrestre). El efecto pernicioso que introduce la atmósfera en la modulación coherente es la ***turbulencia***. La turbulencia hace que el frente de onda de la señal pierda coherencia espacial, esto es, debido a los cambios aleatorios de dirección del haz en cada punto, el frente de ondas dejará de ser plano. No obstante, existe una solución a este problema (aunque muy cara y compleja). El efecto de la turbulencia puede ser casi completamente eliminado mediante el uso de técnicas de corrección basadas en **óptica adaptativa** en los telescopios. En el caso de receptores de detección directa la coherencia espacial no tiene trascendencia, pues se trata únicamente de convertir la potencia óptica recibida en una corriente eléctrica de forma directa y sin ningún procesado de la señal.

## ÓPTICA ADAPTATIVA

La óptica adaptativa es una técnica relativamente reciente pero ampliamente extendida (pese a su elevado coste y complejidad técnica) en los mejores telescopios astronómicos debido a su gran eficiencia.

Como se verá en el apartado 3.3.4, dedicado a la turbulencia atmosférica, cuando un frente de ondas plano tiene que atravesar la atmósfera terrestre, uno de los efectos que ésta provoca es, debido a variaciones del índice de refracción, la alteración aleatoria de la dirección del haz en diferentes puntos de la sección perpendicular al desplazamiento de la onda. Esta refracción provoca que el frente de ondas deje de tener coherencia espacial como se puede apreciar en la parte superior de la Figura 5.15.

La óptica adaptativa es una técnica que permite medir y corregir en tiempo real la coherencia espacial del frente de onda que se recibe en el telescopio de manera que se puede decir que un telescopio equipado con un sistema de óptica adaptativa equivale a un telescopio recibiendo la señal desde el espacio, esto es, sin la perturbación de la turbulencia atmosférica.

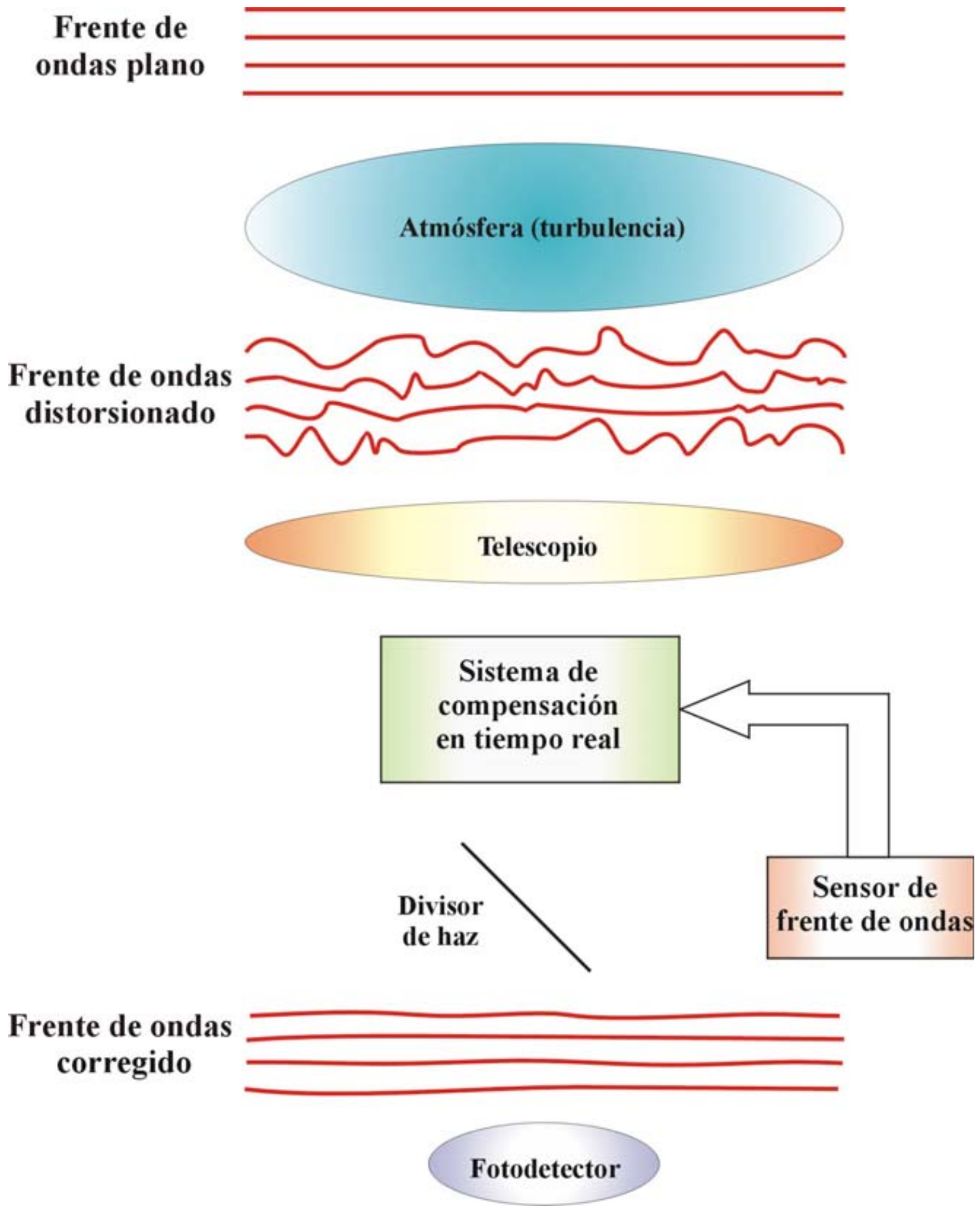

***Figura 5.15. Esquema básico de un telescopio con un sistema corrector de óptica adaptativa [47].***

Otro inconveniente de las modulaciones coherentes viene determinado por el **proceso de mezcla** de las dos señales a la entrada del fotodetector. Si las polarizaciones de ambas señales no coinciden o el alineamiento de los frentes de onda de los dos haces sobre el área del fotodetector es peor que un pequeño tanto por ciento de la longitud de onda, la sensibilidad del detector se verá *seriamente degradada*. Por lo tanto cuanto menor sea la longitud de onda de trabajo, tanto más difícil será llevar a cabo el alineamiento de las dos señales necesario para una modulación coherente. Como se ha comentado, básicamente hay dos rangos de longitud de onda con tecnología láser disponible, uno alrededor de 1 µm y otro alrededor de 10 µm. El alineamiento de las dos señales a 10 µm es ciertamente posible, sin embargo a 1 µm supone una mayor dificultad.

Relativo al proceso de mezcla de señales en modulación coherente, existe una dificultad añadida. Como se ha visto, el empleo de detectores coherentes implica el traslado del espectro de la señal recibida a otra frecuencia a la cual está centrado un filtro fijo. En consecuencia, es fundamental que el traslado en frecuencia se realice correctamente para que el espectro no salga del ancho de banda de dicho filtro. Sin embargo la frecuencia de la señal recibida cambia constantemente debido al **efecto Doppler** que provoca tanto la traslación de los planetas como la rotación de los terminales, ya sea según la órbita en la que se encuentre el satélite o sobre el eje del planeta si el terminal está en superficie. Si la frecuencia recibida varía y la frecuencia de la diferencia de señales debe ser constante, es preciso que la frecuencia del láser varíe de acuerdo con la señal recibida. Por ello al esquema del detector coherente es imprescindible añadirle otra exigencia más: que el láser local sea sintonizable mediante su correspondiente circuito de control de frecuencia de acuerdo a la frecuencia recibida alterada por efecto Doppler.

Conviene, por último, tener en consideración a la hora del diseño de un sistema de comunicaciones ópticas en espacio profundo, que los receptores basados en detección directa cuentan con la importante ventaja de su simplicidad y bajo coste en contraste con la gran complejidad técnica y elevado coste de los sistemas basados en modulaciones coherentes. Por tanto, siempre que sea posible, será preferible realizar el enlace utilizando detección directa. Sin embargo, en un enlace a muy largas distancias probablemente será la única alternativa para cumplir con los requisitos mínimos de comunicación.

## 5.7. ARQUITECTURA DEL TERMINAL TERRESTRE

Para que el sistema de comunicaciones ópticas sea eficiente se precisan receptores de gran tamaño. Además debería ser capaz de proporcionar un enlace de 24 horas al día para lo cual es necesario compensar los efectos de la rotación terrestre sobre su propio eje. Las alternativas consideradas para la arquitectura del terminal terrestre incluyen terminales espaciales (***satélites***), terminales situados en ***globos*** y ***terminales terrestres***.

### 5.7.1. Satélite

La mayor ventaja de colocar al terminal en un satélite en órbita alrededor de la Tierra consiste en eliminar todos los efectos que **la atmósfera terrestre** causa en el viaje de la señal óptica al atravesarla. Por una parte al no haber atmósfera, no hay nubes que puedan llegar a bloquear la recepción. Por otra, en el espacio la única fuente de luz indeseada es la de los planetas y estrellas, por lo que el ruido de fondo recibido en el detector sería el menor posible comparado con cualquier otra ubicación del terminal. Esto ofrece la posibilidad de reducir el tamaño del telescopio, que en el caso del satélite adquiere una gran importancia. Por ello, para un mismo enlace, la opción del satélite facilitará el empleo del menor telescopio posible.

Otra ventaja de que la señal no tenga que atravesar la atmósfera es la de contar con la posibilidad de ***disminuir la longitud de onda*** ya que la elección de ésta, viene determinada principalmente por las ventanas que minimizan la absorción atmosférica. Como se explica en el apartado 4.4.1 dedicado a la difracción producida en la apertura del emisor, una longitud de onda menor permitiría al transmisor en espacio profundo enviar un haz más estrecho con lo que la potencia recibida por el receptor sería mayor por estar más concentrada en un área más pequeña.

También la opción de satélite consigue el menor número de terminales para garantizar recepción continuada. Debido a que necesariamente habrá momentos en los que la Tierra eclipse la señal serían necesarios un mínimo de dos satélites para recibir la señal sin cortes.

Como mayores inconvenientes está el hecho de que sería la opción más costosa económicamente en cuanto a implementación y mantenimiento, además de su menor fiabilidad por la dificultad que entraña recuperar o reparar unidades defectuosas, y su menor precisión de apuntamiento. No obstante, en el caso de muy largas distancias (por ejemplo, un enlace entre la Tierra y los planetas más exteriores del sistema solar), es posible que fuese la única alternativa viable para mantener un enlace con un mínimo de calidad.

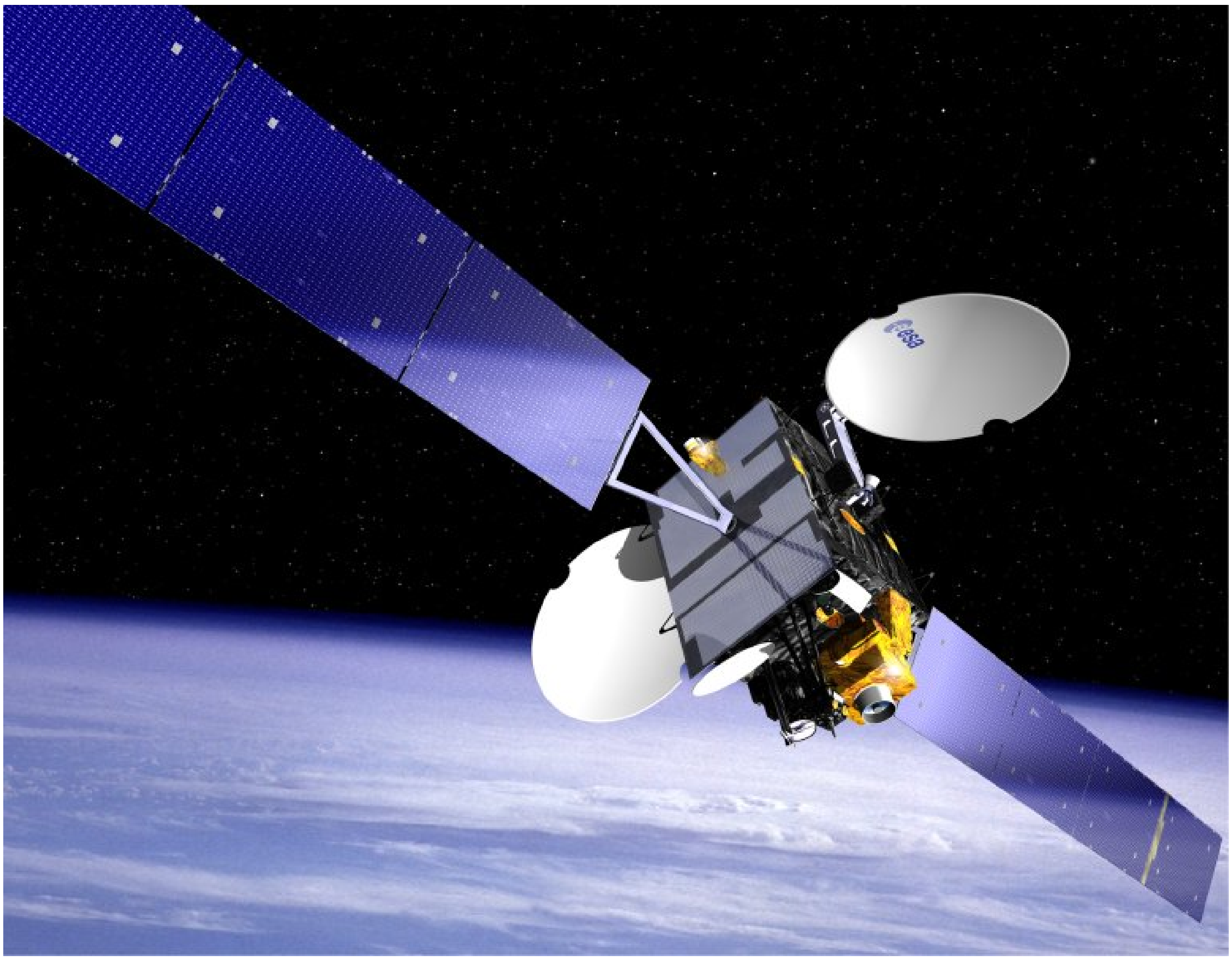

***Figura 5.16. Satélite de comunicaciones orbitando la Tierra [17].***

## 5.7.2. Globo

Los globos ofrecen una opción más económica que la del satélite aprovechando en gran parte su principal ventaja, esto es, la ausencia de atmósfera. La altitud a la que vuelan los globos en la estratosfera puede superar los 35 kilómetros, lo que se traduce en que el 99% de la atmósfera queda por debajo (la densidad del aire es del 1% de la que hay a nivel del mar). Además de eliminar por completo la posibilidad de interrupción de la comunicación debido a la presencia de nubes, podría eliminar en buena medida los demás efectos indeseados de la atmósfera como son la turbulencia y el ruido de fondo. Esto haría que el tamaño del telescopio fuera menor de lo que se necesitaría si hubiera atmósfera, aunque algo mayor del que se utilizaría en un satélite.

La duración máxima de vuelo puede prolongarse hasta unos 100 días si se usan los globos ULDB (del inglés *Ultra Long Duration Balloon*) de muy larga duración desarrollados por la NASA [48]. Sin embargo, en relación al mantenimiento permanente de un enlace de comunicación, esa duración tan corta puede resultar inconveniente, ya que los globos tendrían que ser sustituidos a menudo (aunque la carga es recuperable).

Además la trayectoria de un globo no puede gobernarse de forma precisa. En último término depende directamente del viento al que tenga que hacer frente, aunque actualmente existen modernos métodos para tener control sobre el desplazamiento del globo. En la Figura 5.18 puede observarse el resultado de una simulación [49] de la trayectoria de un globo sin control de trayectoria y con control. Sin embargo, aunque se pueda tener en cierta medida acotada la trayectoria, a diferencia del satélite, siempre existe un grado de imprecisión, debida a la aleatoriedad del viento, que dificultaría el apuntamiento del terminal espacial.

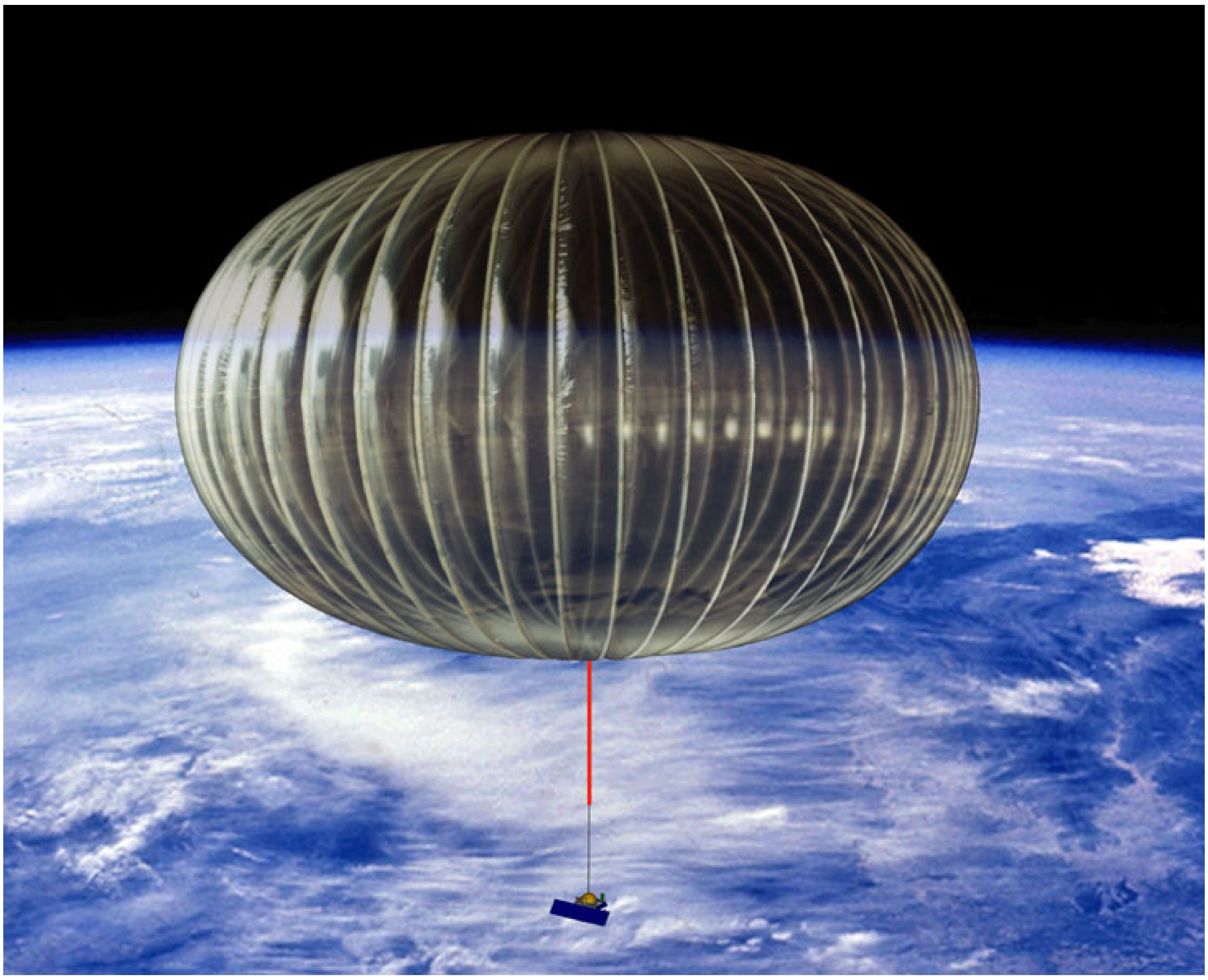

***Figura 5.17. Ilustración artística de un globo ULDB de muy larga duración [48].***

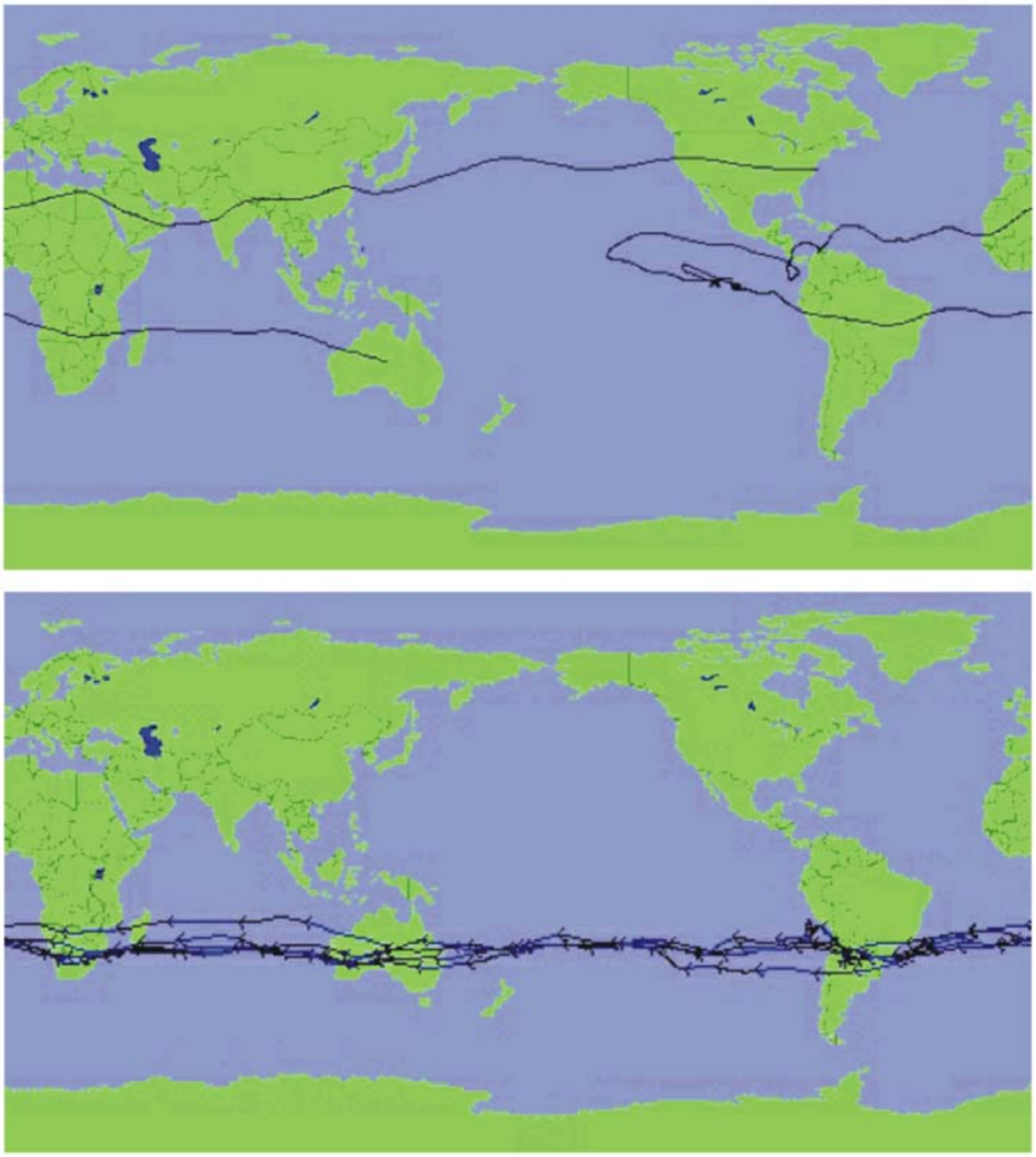

*Figura 5.18. Simulación del desplazamiento de un globo sin y con control de trayectoria [49].*

Otra desventaja del empleo de globos procede de su propia estructura al impedir la recepción cuando la señal llega de grandes ángulos de elevación, lo que hace necesaria la ubicación de este tipo de terminales lo más cercana posible a los polos.

## 5.7.3. Tierra

El mayor inconveniente de situar al terminal en la superficie de la Tierra es la presencia de la atmósfera terrestre que deberá atravesar la señal. Como ya se ha expuesto en las opciones de satélite y globo, para compensar los efectos de la atmósfera se hace necesario un mayor tamaño de telescopio.

Actualmente los telescopios más grandes que existen llegan a un diámetro del espejo primario de unos 10 metros. Además algunos disponen de óptica adaptativa, lo que reduce al mínimo el efecto de la turbulencia atmosférica. Los demás efectos perjudiciales de la atmósfera también quedan atenuados debido al hecho de que los observatorios se encuentran a gran altura (por encima de 2.000 metros en general) aunque sigue existiendo la posibilidad de bloqueo del enlace debido a la presencia de nubes, problema éste que no presentan las otras alternativas consideradas.

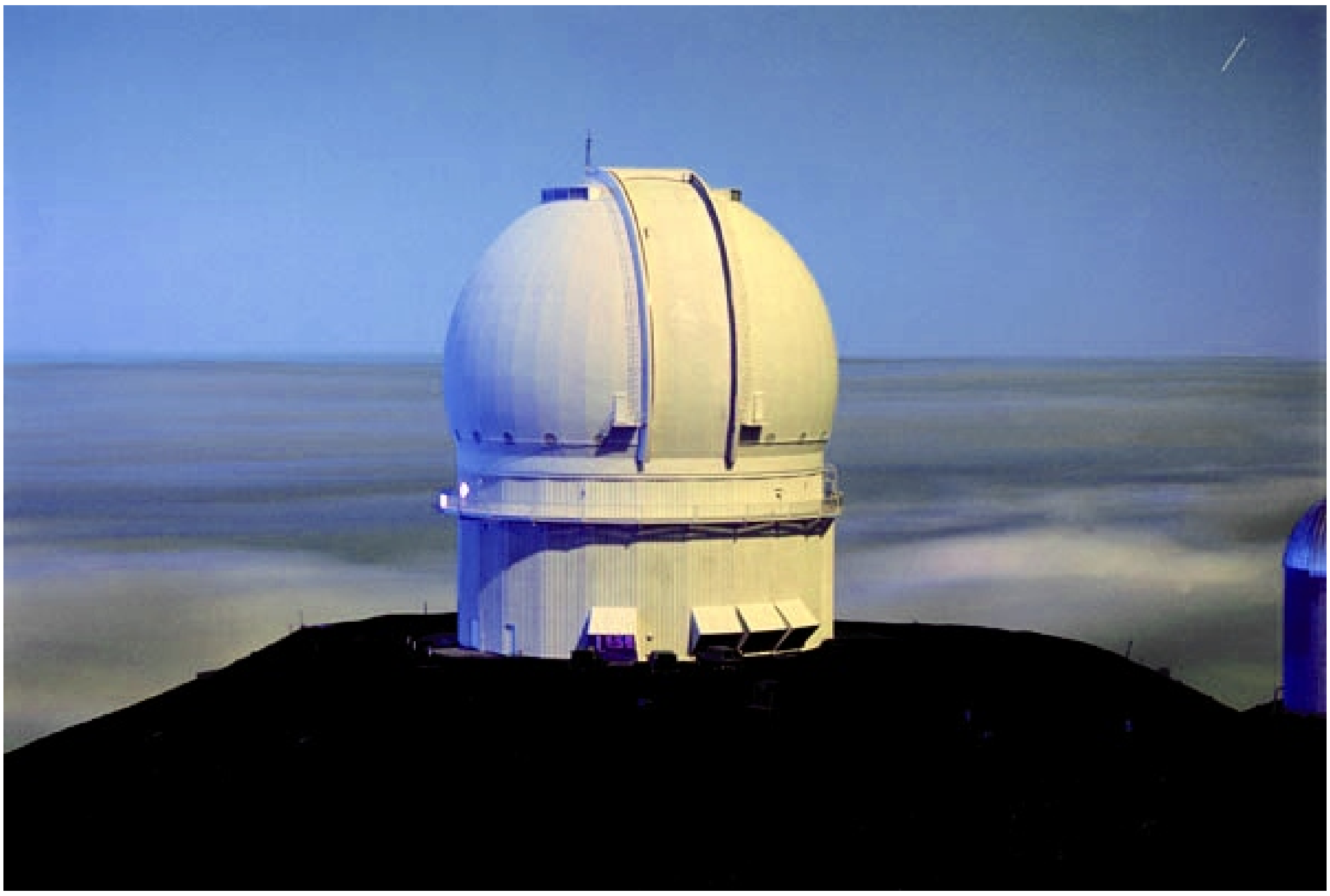

***Figura 5.19. Observatorio astronómico Keck (Hawaii) con las nubes a sus pies [50].***

Por otro lado, estos grandes telescopios, además de ser muy escasos, son excepcionalmente caros por su gran precisión y suelen estar muy solicitados para observaciones astronómicas que serían irrealizables en otros de menor tamaño. Además no son apropiados para operar a la luz del día, especialmente orientándolos con pequeñas separaciones del Sol. Por tanto, el uso rutinario de estos telescopios para comunicaciones no está justificado.

Los requisitos de un telescopio para comunicación son menos exigentes que para observación ya que es suficiente con el empleo de ***óptica anidólica*** (óptica que no forma imágenes). Esto posibilita el empleo de espejos esféricos[1] formados por segmentos idénticos y de baja calidad reduciendo así el coste de fabricación y mantenimiento. Por ello existe la posibilidad de construcción de grandes telescopios orientados en exclusiva a las comunicaciones ópticas en espacio profundo.

Otra posibilidad para el terminal terrestre es la arquitectura de receptor propuesta por el *Jet Propulsion Laboratory* (JPL) [43] consistente en un array de pequeños telescopios cada uno de ellos con un *array* de detectores "contadores de fotones" (de los que se da cuenta en el apartado 5.3.4 dedicado a los tipos de fotodiodos). De esta manera se suple la construcción de un gran espejo primario por la económica construcción de muchos pequeños espejos, con lo que se consigue que el coste aumente linealmente con el área efectiva del telescopio, la cual se podrá aumentar arbitrariamente con un número mayor de telescopios formando el *array*. Además esta arquitectura también se adapta perfectamente al caso de satélite o globo. Sería una forma eficiente de usar telescopios de grandes aperturas efectivas en el espacio sin suponer una pesada carga.

---

[1] Normalmente los espejos astronómicos, dado que su objetivo es la formación de imágenes, son asféricos o llevan correcciones asféricas. Este tipo de óptica minimiza la aberración esférica. La fabricación de un espejo asférico es mucho más costosa que la de uno esférico.

Además del principal inconveniente, que constituye la presencia de la atmósfera de la Tierra, un problema derivado directamente del anterior es que la alternativa terrestre es la que ***mayor número de terminales*** precisaría para recibir el haz óptico continuadamente. Por una parte serían los terminales que se encontrarían más próximos a la superficie terrestre lo que se traduce en periodos más prolongados de bloqueo de la Tierra al rotar sobre su eje. Sólo considerando este efecto serían suficientes 3 terminales receptores ubicados lo más equidistantes posible. Si bien la ubicación de los observatorios astronómicos constituye la mejor opción para situar un sistema de comunicaciones ópticas en espacio profundo (la nubosidad y turbulencia son las mínimas que se pueden encontrar en la superficie terrestre), la gran variabilidad en la meteorología de la atmósfera podría, en determinadas circunstancias, bloquear por completo el enlace. Esta inconveniencia se puede mitigar en buena medida añadiendo redundancia, esto es, aumentando el número de ubicaciones receptoras de forma que se reduzca al mínimo la probabilidad de que queden ciegos por la presencia de nubes en la atmósfera todos los receptores que potencialmente pueden recibir señal en un momento determinado. Esta estrategia se describe con más detalle en el apartado 3.3.1 dedicado al efecto de las nubes de la atmósfera terrestre en la comunicación.

### 5.7.4. Conclusión

Como se ha visto, el tipo de arquitectura debe ser un compromiso entre el coste y las necesidades particulares del enlace. A muy grandes distancias un receptor en la superficie terrestre podría llegar a ser inviable debido al efecto de la atmósfera en la muy débil señal recibida.

Como ejemplo estimativo [51], en una comunicación entre Marte y la Tierra un telescopio situado en un satélite alrededor de la Tierra necesitaría 2,6 metros para mantener un enlace de la misma capacidad que lo haría un telescopio de 3,3 metros en un globo o uno de 8,1 metros en la superficie terrestre.

En la Tabla 5.2 se presenta un resumen de las características de cada una de las arquitecturas descritas anteriormente.

| Arquitectura | Satélite | Globo | Superficie |
|---|---|---|---|
| **Nº de terminales (recepción continua) [58]** | 2-3 | 2-6 | 3-9 |
| **Tamaño relativo telescopio (mismo enlace) [51]** | 1 | 1,3 | 3,1 |
| **Coste** | Alto | Bajo | Medio |

***Tabla 5.2. Comparación entre las diferentes arquitecturas del terminal terrestre.***

# 6. El proyecto MLCD

El propósito del proyecto *MLCD* (*Mars Laser Communications Demonstration*) es demostrar la posibilidad del empleo de comunicaciones ópticas en espacio profundo. En este proyecto el enlace se establecerá entre un terminal de comunicaciones por láser a bordo del satélite *MTO* (*Mars Telecom Orbiter*) en órbita alrededor del planeta Marte y un terminal situado en la Tierra. El proyecto *MLCD* es fruto de la colaboración entre el *GSFC* (*Goddard Space Flight Center*) de la *NASA*, el *JPL* (*Jet Propulsion Laboratory*) del *CIT* (*California Institute of Technology)* y el *MIT/LL* (*Massachussets Institute of Technology/Lincoln Laboratory*).

En el futuro cercano, se prevé un importante aumento en la demanda de capacidad y servicios en las comunicaciones en espacio profundo. Un aumento de este tipo de misiones, junto a instrumentos más exigentes, necesidad de soporte a la exploración humana, mapeo de alta resolución de planetas y distancias más lejanas precisará un importante incremento en la capacidad de las telecomunicaciones, para lo cual las comunicaciones ópticas prometen ser la solución. El proyecto *MLCD* pretende convertirse en el primer sistema de comunicaciones por láser en espacio profundo, esto es, a distancias mayores que las de las órbitas alrededor de la Tierra a las cuales ya ha sido probado dicho tipo de comunicaciones.

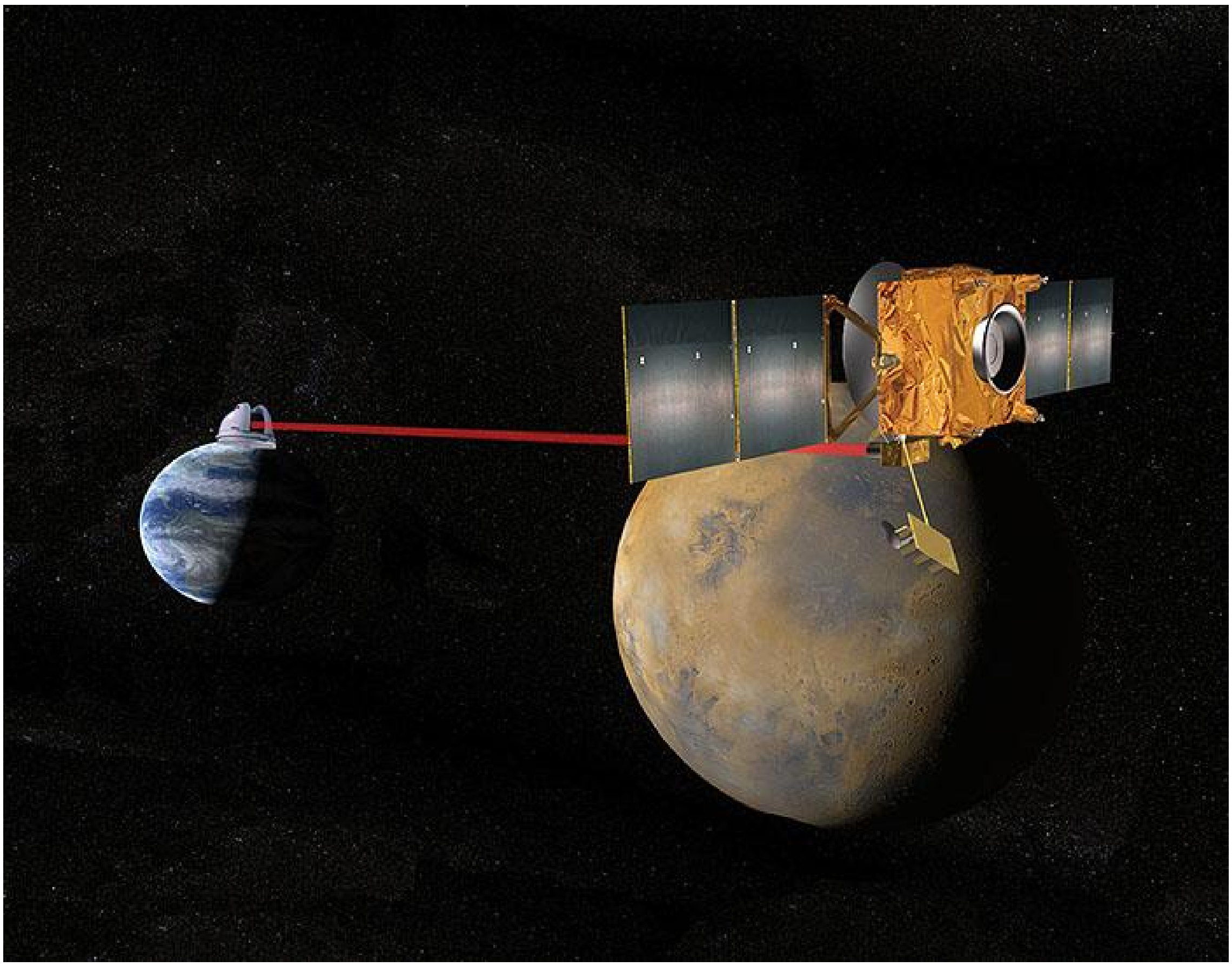

*Figura 6.1. Ilustración artística de la comunicación entre el terminal espacial MTO y un telescopio en la superficie de la Tierra [52].*

## 6.1. PERFIL DEL PROYECTO MLCD

El día 13 de Octubre de 2.009 está previsto el lanzamiento de una sonda espacial que transportará al *MTO* hasta Marte. En el periodo de tiempo desde el lanzamiento en la Tierra hasta la inserción en la órbita de Marte (durante la órbita de transferencia) se realizarán pruebas de funcionamiento del sistema de comunicación por láser.

La nave espacial *MTO* transporta el sistema de comunicaciones por láser y tiene prevista su inserción en la órbita marciana entre agosto y septiembre de 2.010, casi un año después de su lanzamiento en la Tierra. Desde entonces el proyecto *MLCD* estará operativo al menos un año (de la Tierra), tiempo en el que podrá someterse a todas las dificultades que presenta una misión a Marte. Se enfrentará a distancias máximas de casi 356 millones de km que además coincidirán con periodos de conjunción solar, que, como se ha descrito en el apartado 3.2, significa que la comunicación se tendrá que llevar a cabo con el Sol prácticamente en medio de los terminales.

El proyecto planea la utilización de terminales terrestres situados en la superficie, capaces de recibir el flujo de comunicaciones del *MTO* y de transmitir un enlace de subida dedicado al establecimiento y mantenimiento del apuntamiento del estrecho haz que emite el *MTO*. La función de los terminales terrestres la llevarán a cabo telescopios astronómicos ya existentes.

El terminal espacial será el satélite *MTO* que orbitará Marte con un periodo de seis horas. Las funciones que desempeñará para el proyecto *MLCD* serán las de generar eficientemente potencia óptica que pueda transportar información modulada, transmitir eficazmente esta potencia a través de los elementos ópticos apropiados y realizar y mantener (con la ayuda del canal de subida) el correcto enfoque del estrecho haz del láser a la Tierra evitando las vibraciones y movimientos de la estructura del satélite. La potencia media disponible para el transmisor será de cinco vatios y el telescopio tendrá un diámetro de espejo primario de 30,5 centímetros. Además del terminal de comunicaciones por láser, el *MTO* dispone de una antena de tres metros para comunicaciones en radio frecuencias (en banda X y en banda Ka) y otra para comunicaciones con la superficie de Marte. En la Figura 6.2 pueden apreciarse estos sistemas en una vista exterior del *MTO*.

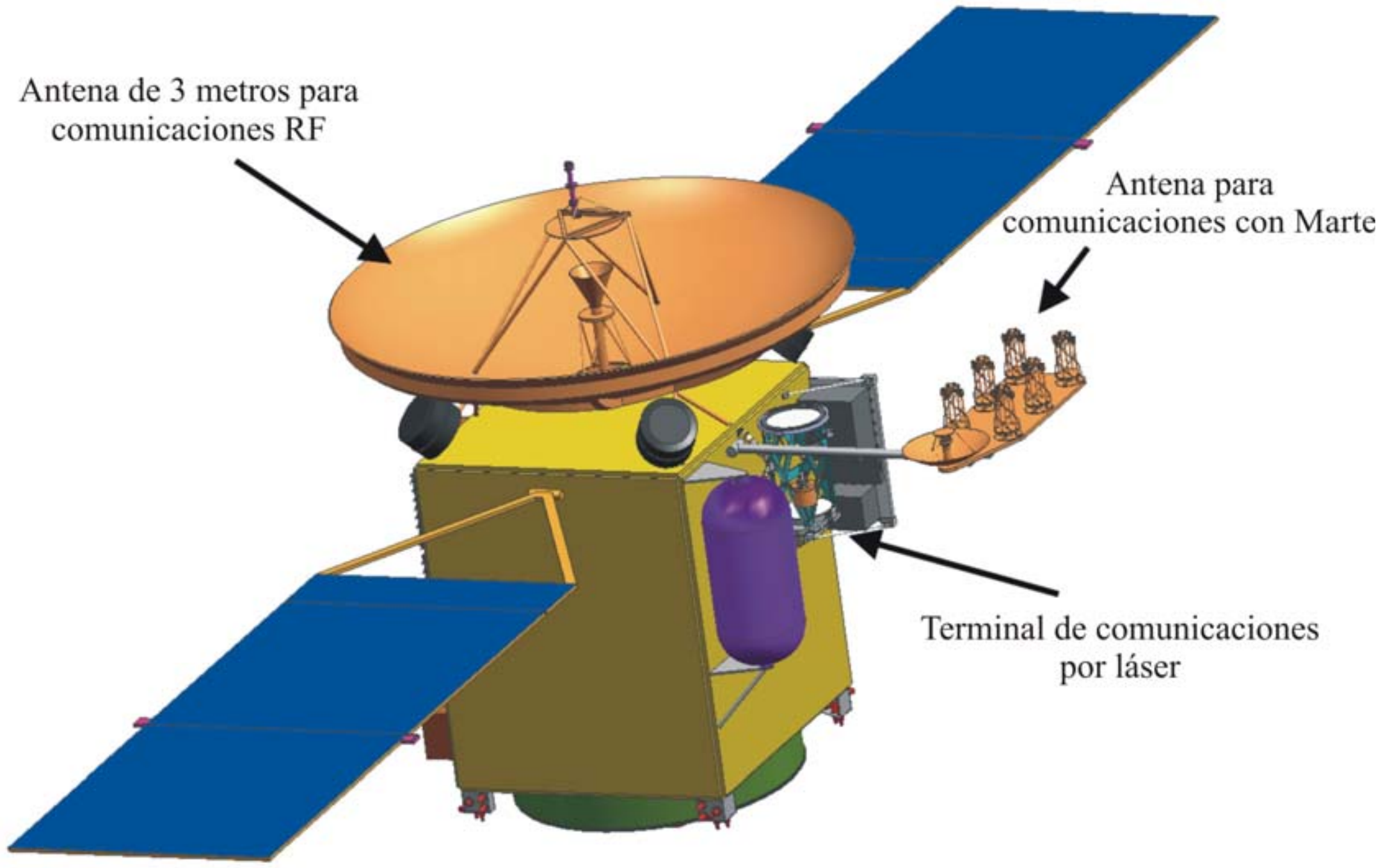

***Figura 6.2. Vista exterior de los sistemas de comunicaciones del MTO [53].***

Actualmente la velocidad de las comunicaciones por radiofrecuencia desde Marte es del orden de 128 kbits por segundo (*Mars Odyssey).* Durante el desarrollo del proyecto *MLCD* se requiere el mantenimiento de un régimen binario mínimo de 1 Mbit por segundo. Esto significa que incluso en las peores condiciones tendrá que asegurar esta velocidad (aunque existirán momentos en los que no haya comunicación en absoluto debido al bloqueo del Sol), lo que indica que en las mejores condiciones el régimen binario máximo podrá ser muy superior.

## 6.2. CONJUNCIÓN Y OPOSICIÓN SOLAR

Las situaciones de conjunción y oposición solar entre Marte y la Tierra, tal como se explicaron en el apartado 3.2, tienen dos consecuencias importantes. Por un lado, las conjunciones y oposiciones se repiten en un periodo de tiempo constante, por lo que un proyecto de demostración debería desarrollarse durante un periodo completo, formado por una conjunción y una oposición, para enfrentarse a todas las situaciones posibles.

Por otro lado, el alineamiento de los planetas con el Sol provoca que los terminales tengan que apuntarse mutuamente con ángulos muy pequeños de distancia al Sol, lo que hace que la óptica de los telescopios se caliente y deforme.

### 6.2.1. Distancia Marte-Tierra

El carácter periódico del movimiento orbital de Marte y la Tierra alrededor del Sol provoca que de forma también *periódica* se sucedan situaciones de oposición y conjunción solar. Debido a las excentricidades de las órbitas de Marte y la Tierra, la duración del ciclo es de unos 16,6 años. Esto quiere decir que aproximadamente la misma conjunción u oposición, y por ello la distancia máxima y mínima **absoluta** respectivamente entre Marte y la Tierra, tarda en repetirse entre 15 y 17 años. Sin embargo, cada 26 meses (2,2 años) aproximadamente se repiten las posiciones **relativas** de los dos planetas, por lo que las conjunciones y oposiciones con valores de distancia Marte-Tierra variables se repiten cada 26 meses.

En la Figura 6.3 se presenta el resultado, obtenido con el simulador de dinámica orbital implementado para este proyecto, de la distancia instantánea entre la Tierra y Marte durante un periodo de tiempo de 40 años (desde el 1 de enero del 2000 hasta el 1 de enero del 2040). Pueden apreciarse dos periodos distintos de oscilación: uno relativo a la distancia máxima y mínima absoluta (que corresponde a la envolvente de la oscilación) y otro de la repetición de la posición relativa de los planetas (correspondiente a la variación de periodo corto).

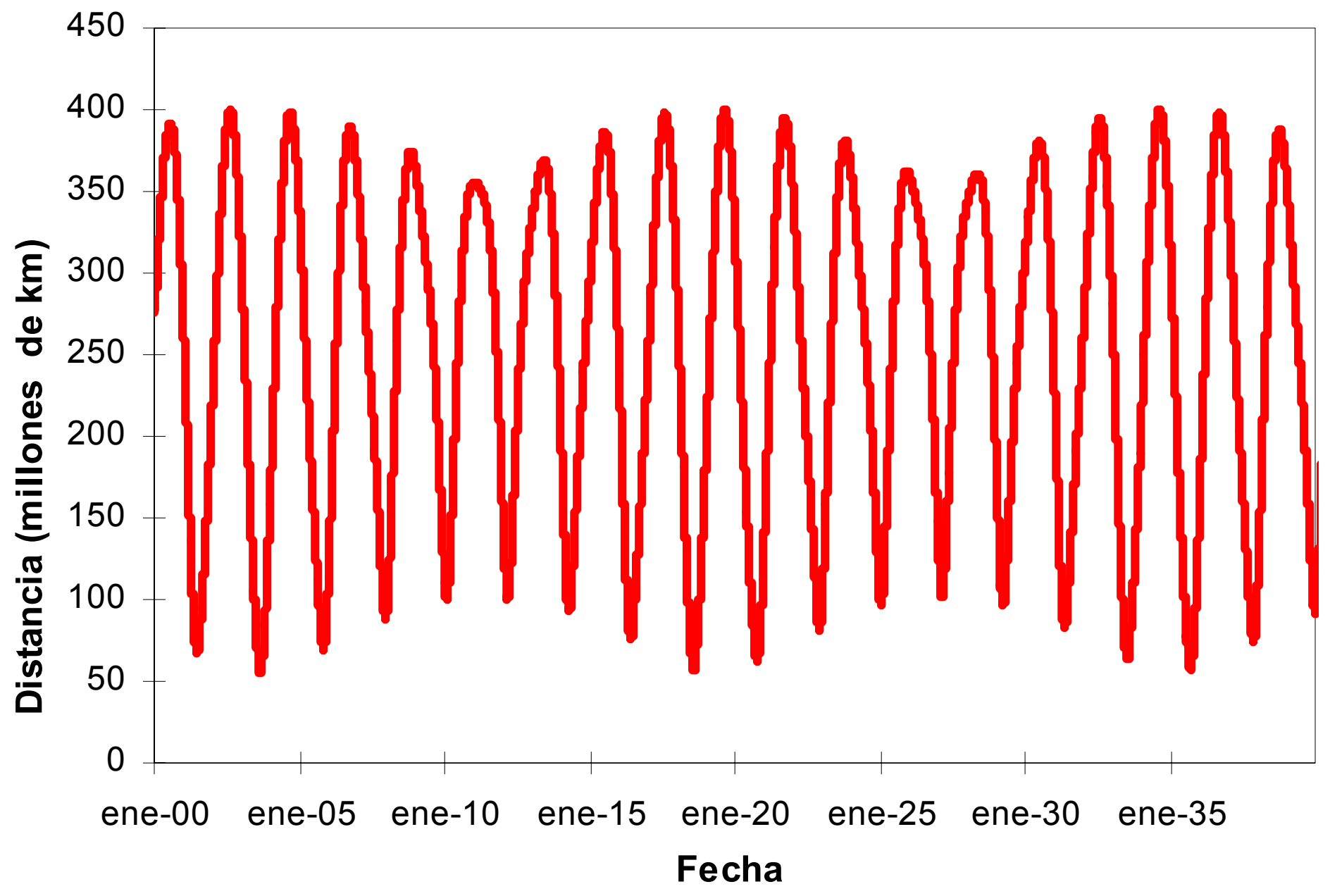

***Figura 6.3. Distancia Marte-Tierra en un periodo de tiempo de 40 años.***

Si de los resultados de la simulación mencionada, se aíslan la distancia máxima y mínima de cada oscilación de periodo corto, se puede obtener el periodo de variación de las conjunciones y oposiciones. Este resultado se presenta en la Figura 6.4.

De lo anterior se concluye que cada dos años aproximadamente se repiten cualitativamente los mismos acontecimientos críticos que influirán en el enlace de comunicación. Por lo tanto, pese a que la duración mínima del proyecto *MLCD* es de un año, en lo sucesivo se va a considerar un periodo de dos años desde que el satélite *MTO* se inserta en la órbita de Marte, con el objetivo de contemplar todas las situaciones a las que se verá sometido el proyecto.

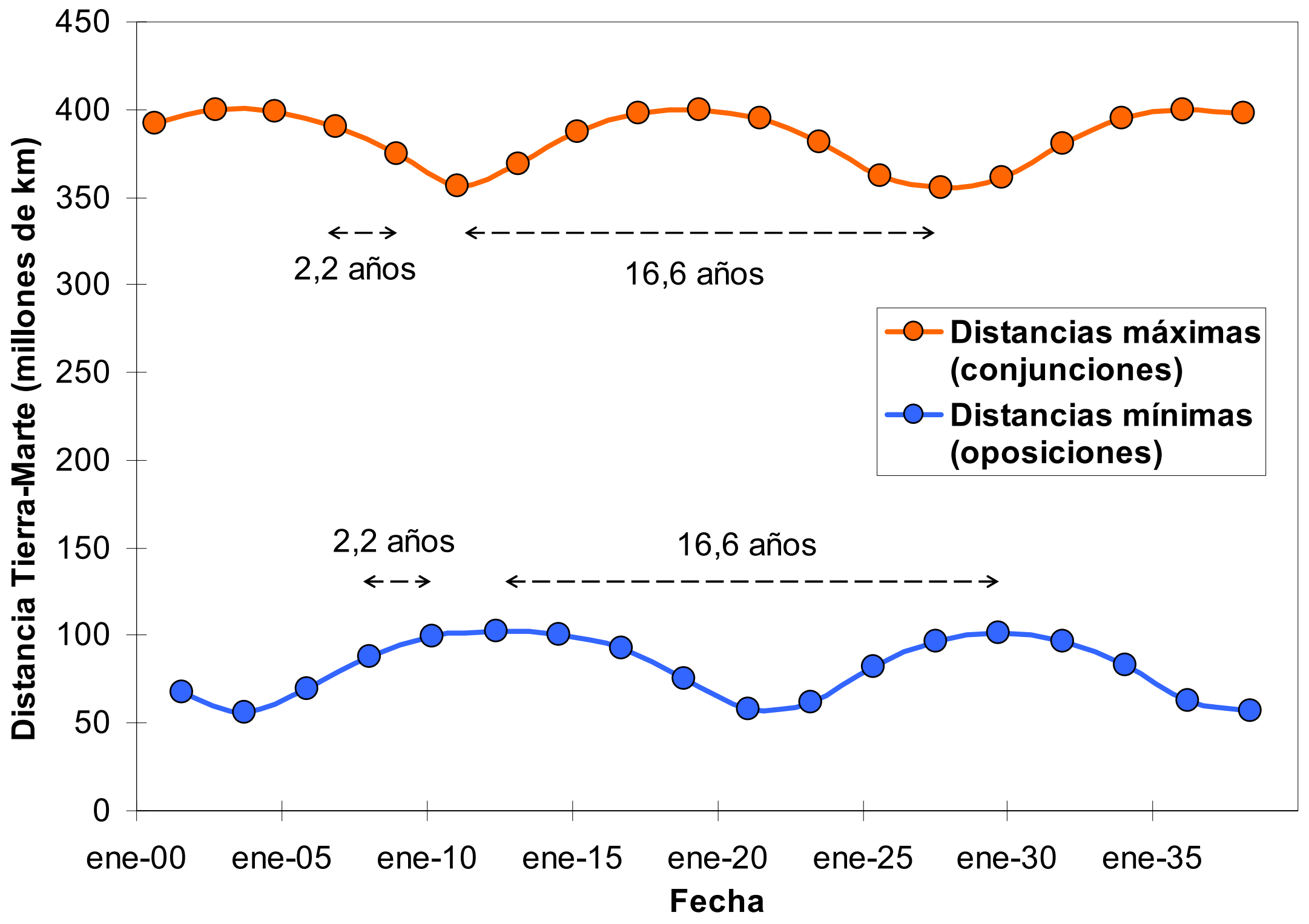

***Figura 6.4. Periodo de repetición de oposiciones y conjunciones solares entre Marte y la Tierra desde enero del 2000 hasta enero del 2040.***

## 6.2.2. Ángulos Sol-Tierra-MTO y Sol-MTO-Tierra

El hecho de que Marte y la Tierra se muevan en órbitas prácticamente *coplanares* (hay unos 1,8 grados de inclinación entre ambas) provoca que en las situaciones de conjunción solar, el Sol se interponga en la línea de visión directa que une ambos planetas, eclipsando a cada uno visto desde el opuesto. Por supuesto en estos periodos de tiempo es imposible la comunicación, sin embargo no es ésta la única causa que obliga a cortar el enlace en las conjunciones. El efecto más importante de apuntar con ángulos muy próximos al Sol es el calentamiento al que se somete la óptica de los telescopios. Un calentamiento que no se pudiera evitar empleando filtros solares, podría provocar la deformación de la óptica.

Se distingue los dos ángulos siguientes: ***Sol-Tierra-MTO*** es el ángulo medido desde la Tierra entre el Sol y el satélite *MTO*; ***Sol-MTO-Tierra*** es el ángulo medido desde el *MTO* entre el Sol y la Tierra. Estos ángulos se pueden observar en la Figura 6.5. Para evitar el calentamiento de la óptica del telescopio es necesario, además de emplear filtros solares (véase apartado 6.7.1), fijar una distancia de seguridad dentro de la cual nunca se apunta el telescopio. Según las especificaciones [53] del proyecto *MLCD*, esta distancia de seguridad se fija para los terminales terrestres a una anchura total de 3 grados que incluyen unos 0,5 grados del propio contorno del Sol y otros 2,5 que contemplan la corona solar más un margen de seguridad que cubre los variables límites de ésta. Es decir, cuando el ángulo Sol-Tierra-*MTO* sea inferior a **3 grados** el enlace de comunicación no estará disponible.

En la Figura 6.6 se presentan los valores para los ángulos Sol-MTO-Tierra y Sol-Tierra-MTO obtenidos con el simulador de dinámica orbital para un periodo de dos años de proyecto.

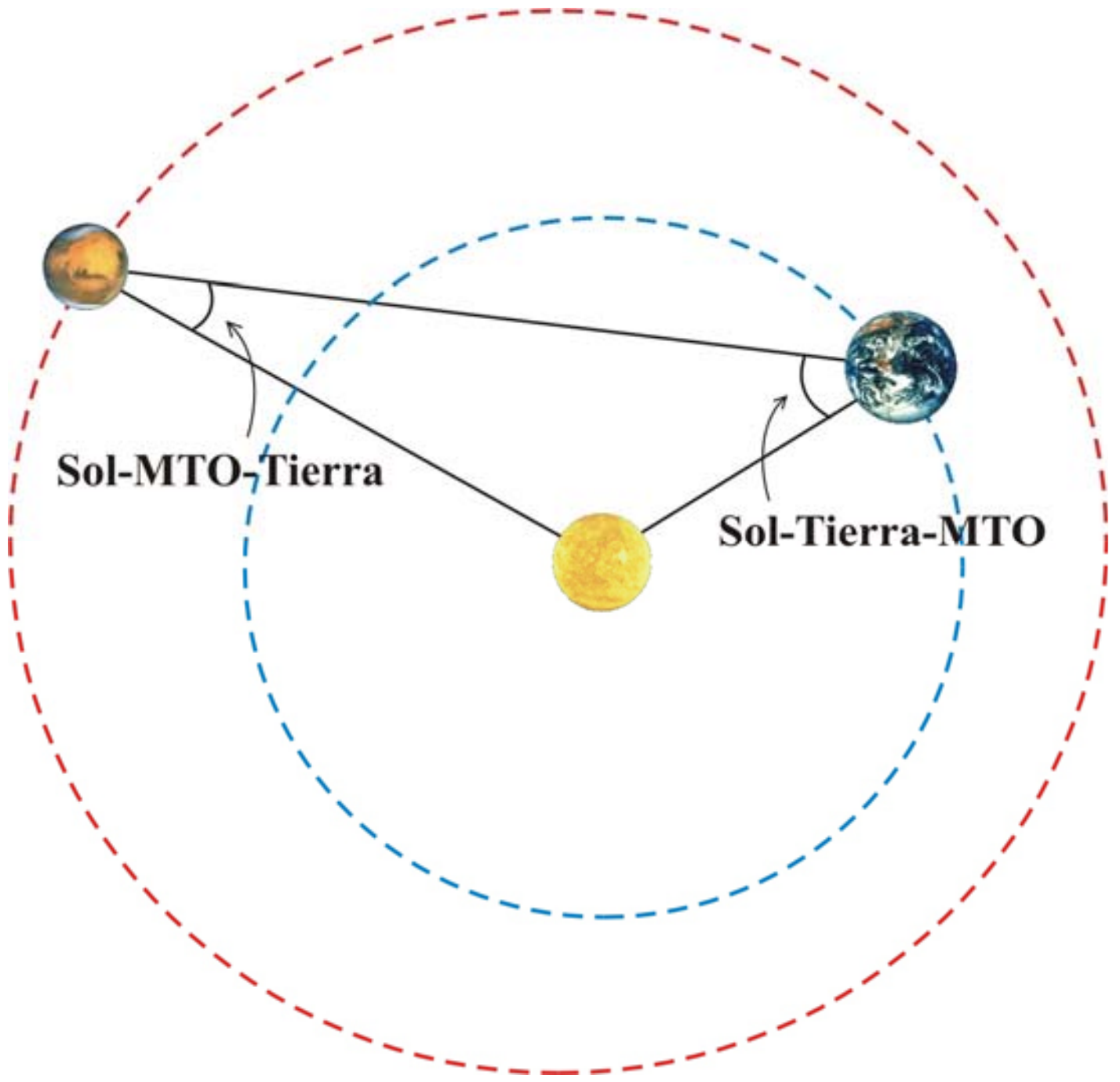

*Figura 6.5. Ángulos Sol- MTO-Tierra y Sol-Tierra-MTO.*

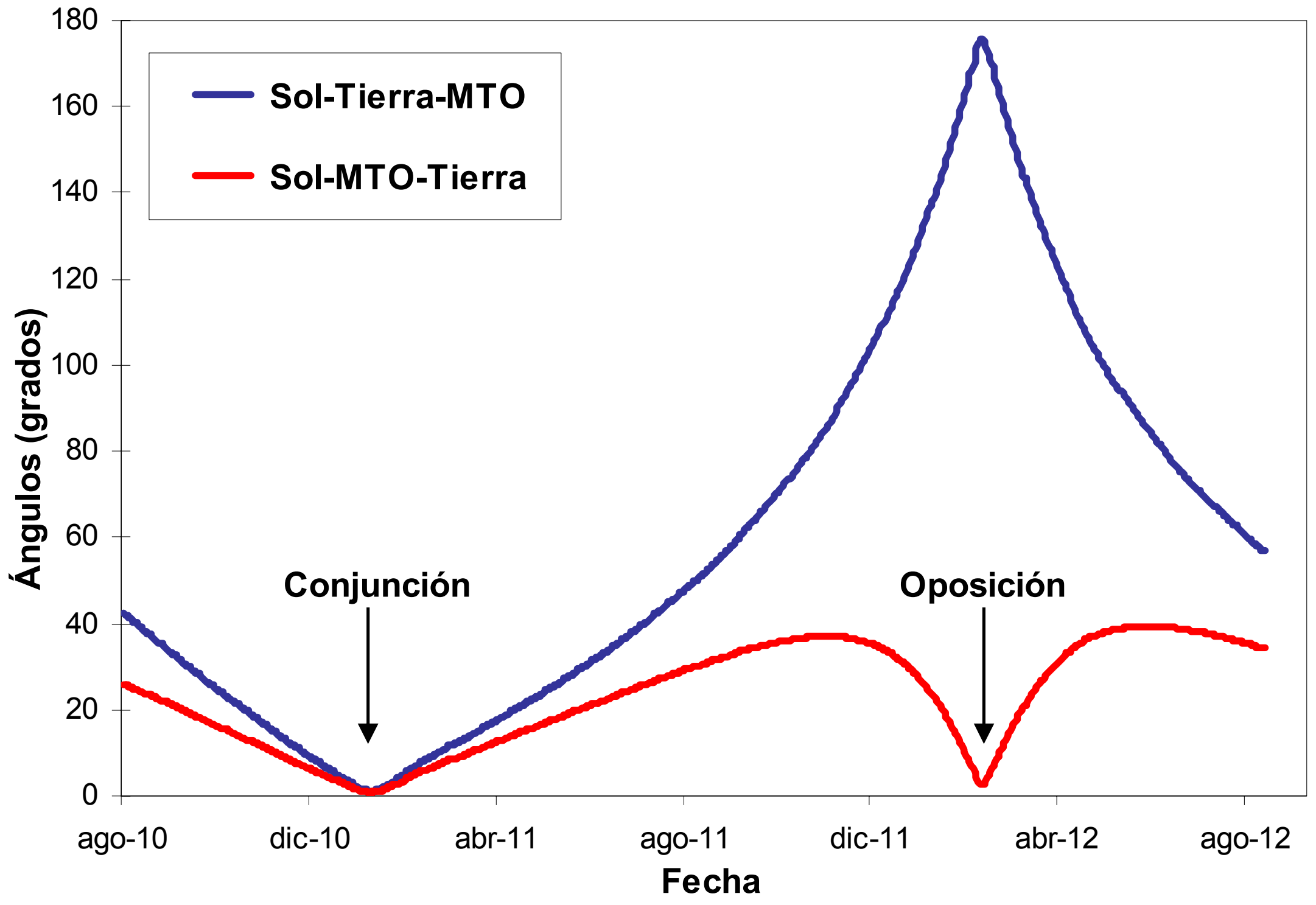

*Figura 6.6. Variación de los ángulos Sol-Tierra-MTO y Sol-MTO-Tierra durante dos años de proyecto.*

Según la simulación, el enlace no estará disponible los siguientes días

| | |
|---|---|
| Ángulo Sol-Tierra-MTO < 3º | 26 días (24/1/2011 -18/2/2011) |

Se puede comprobar en la Figura 6.6 que, debido a la simetría que se da en las conjunciones en la geometría Marte-Tierra en relación al Sol, ambos ángulos son igual a cero. En los periodos de tiempo alrededor de esta situación se dan los *momentos más críticos* del proyecto *MLCD*, ya que coinciden las máximas distancias (las pérdidas por espacio libre son las mayores posibles) y la necesidad de establecer el enlace durante el día (lo que provoca que la cantidad de luz de fondo del cielo que capta el telescopio suponga una importante fuente de ruido). Por otra parte, las oposiciones hacen que el ángulo Sol-Tierra-*MTO* sea máximo (180 grados) y el Sol-*MTO*-Tierra mínimo (0 grados). Alrededor de esta situación se dan las *mejores circunstancias* para el terminal terrestre ya que la distancia Marte-Tierra es la mínima (perdidas por espacio libre mínimas) y siempre se establece el enlace durante la noche, lo que hace que el ruido de fondo del cielo sea mínimo.

Durante los periodos intermedios, Marte será visible en tiempos variables durante el día y la noche. Este hecho, unido a la dificultad de disponer de tiempos prolongados en telescopios astronómicos, aconseja realizar los cálculos del enlace para ambas situaciones, día y noche, y para distintas horas del día –mediodía, crepúsculo, etc.– durante toda la duración del proyecto, a sabiendas que, en algunos de los periodos, el enlace será posible solo de día, y en otros solo de noche.

## 6.3. EFECTO DOPPLER

De acuerdo con la dinámica orbital que se vio en el capítulo 2, los terminales transmisor y receptor están permanentemente moviéndose uno respecto al otro. Recuérdese que este movimiento era debido a dos factores distintos. Por un lado, está el movimiento debido al desplazamiento por traslación de los dos planetas alrededor del Sol y por otro el movimiento debido a la rotación de los terminales alrededor de cada planeta (el terminal terrestre se mueve con la superficie de la Tierra al girar ésta sobre su propio eje y el terminal de Marte se mueve en su órbita alrededor del planeta). El movimiento relativo entre los dos terminales se tradujo en el apartado 2.3.6 a un vector velocidad radial $\vec{v}_r$ que servía para evaluar la fracción de movimiento de cada terminal en la dirección del otro. Este vector velocidad radial se calcula, para cada terminal, como

$$\vec{v}_r = \vec{v}_{rT} + \vec{v}_{rR} \quad (138)$$

donde $\vec{v}_{rT}$ representa la velocidad radial debida al movimiento de traslación mencionado anteriormente y $\vec{v}_{rR}$ la velocidad radial debida al movimiento de rotación. El rango de variación de la velocidad, así como el rango de tiempo en el que se produce esta variación, es muy diferente para cada uno de los vectores.

Por un lado, la velocidad $\vec{v}_{rT}$ varía muy lentamente en comparación a la velocidad $\vec{v}_{rR}$. La razón es que la variación de la velocidad $\vec{v}_{rT}$ depende únicamente del movimiento relativo entre Marte y la Tierra, que tarda **meses** en cubrir todo el rango de velocidades posibles, mientras que la variación $\vec{v}_{rR}$ depende del movimiento relativo entre el satélite y la Tierra o entre el receptor terrestre y Marte, según sea el terminal para el cual se calcula el vector $\vec{v}_r$. Ya sea el vector $\vec{v}_{rR}$ del terminal de Marte o de la

Tierra, ambos emplean tiempos del orden de **horas** en cubrir el rango completo de velocidades

Por otro lado, el rango de valores que toma la velocidad también es muy diferente para $\vec{v}_{rT}$ y para $\vec{v}_{rR}$. Según se vio en la ecuación (21), la velocidad de un cuerpo en una órbita elíptica general es proporcional a la masa del cuerpo sobre el que orbita e inversamente proporcional a la distancia a la que se encuentra de ese cuerpo. Por lo tanto se deduce que el rango de velocidades de $\vec{v}_{rR}$ es **mucho menor** que el de $\vec{v}_{rT}$ ya que la masa del Sol es más de seis órdenes de magnitud mayor que la de la Tierra o Marte y la distancia de cualquiera de los dos planetas al Sol también es mucho mayor que la de cualquiera de los terminales a su centro de gravedad.

La velocidad radial de cada uno de los terminales en la dirección que une el transmisor y el receptor provoca que la frecuencia que le llega al receptor sea diferente de la frecuencia que emitió el transmisor debido al conocido ***efecto Doppler.*** Dado que ninguna de las velocidades de los terminales es relativista (cercana a la de la luz), para el cálculo de la frecuencia recibida por efecto Doppler se emplea la fórmula clásica[1] que viene dada por [54]

$$f_R = f_T \frac{c \pm v_R}{c \mp v_T} \tag{139}$$

donde $f_T$ representa la frecuencia transmitida, $f_R$ la frecuencia recibida y $v_R$ y $v_T$ la velocidad del receptor y del transmisor respectivamente en la dirección que une ambos terminales. El convenio de signos para la suma de la velocidad de la luz y la velocidad de los terminales es el siguiente: si el receptor se aleja se usa signo negativo, si se acerca positivo; a la inversa para el terminal transmisor.

El cálculo de la frecuencia recibida por efecto Doppler interesa para conocer cual es la variación en la frecuencia recibida en la Tierra y en especial la variación $\Delta f$ en torno a la frecuencia original de emisión. Para todos los cálculos de efecto Doppler de este apartado se emplea la frecuencia correspondiente a la longitud de onda que transmite el terminal láser del *MTO*. Esta longitud de onda es 1,064 μm.

Dado que, como se acaba de decir, esta variación de frecuencia tiene su origen en diferentes variaciones de la velocidad de los terminales, va a tratarse cada una independientemente para evaluar su efecto por separado, de forma que cada vez se supondrá que el único movimiento es el debido a la contribución que se estudie en ese momento. Por último, el efecto total no será más que la superposición del efecto de cada contribución.

## 6.3.1. Rotación de la Tierra sobre su eje

Como ya se mencionó en el apartado 2.3.7, dedicado al cálculo de la velocidad radial debida a la Tierra, únicamente interesa la velocidad radial máxima y mínima, ya que en un periodo de 24 horas se cubre todo el rango de velocidades, y consecuentemente la frecuencia recibida máxima y mínima o la variación máxima y mínima de ésta respecto a la frecuencia emitida.

[1] Para las velocidades involucradas en este proyecto se ha calculado que la diferencia de resultados empleando la fórmula clásica y la relativista es menor que $10^{-8}$.

Dado que el movimiento de rotación de la Tierra es permanente en todo momento y además invariable, la componente de variación de frecuencia Δf introducida será un rango constante que se superpondrá a cualquier otra fuente de efecto Doppler. Esta variación de frecuencia se obtiene de la diferencia entre la frecuencia original de emisión y la frecuencia recibida. Una muestra de este rango de variación se puede observar en la Figura 6.7.

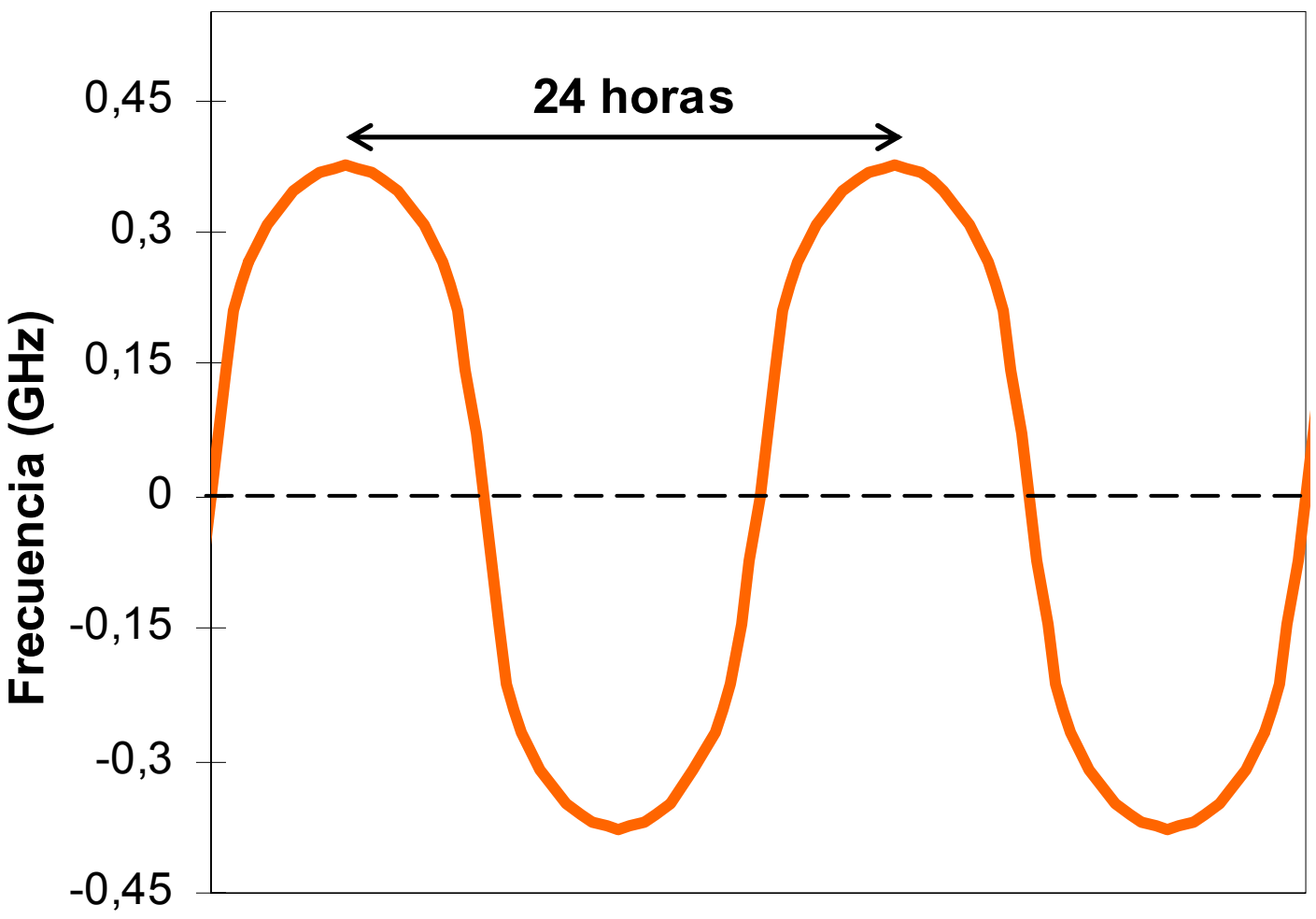

***Figura 6.7. Variación temporal de la frecuencia recibida en la Tierra debido a la rotación de la Tierra sobre su eje.***

Como también se vio en el apartado 2.3.7, el rango de velocidades depende de la ***latitud*** en que se encuentre el receptor. Para este proyecto se considera una distribución típica de tres ubicaciones receptoras como la que se muestra en la Figura 6.8. Como se puede observar, la latitud de las tres ubicaciones está en torno a los ± 30 grados, por lo que es este valor el empleado para los cálculos de este proyecto y el que corresponde a los resultados proporcionados en la Figura 6.7.

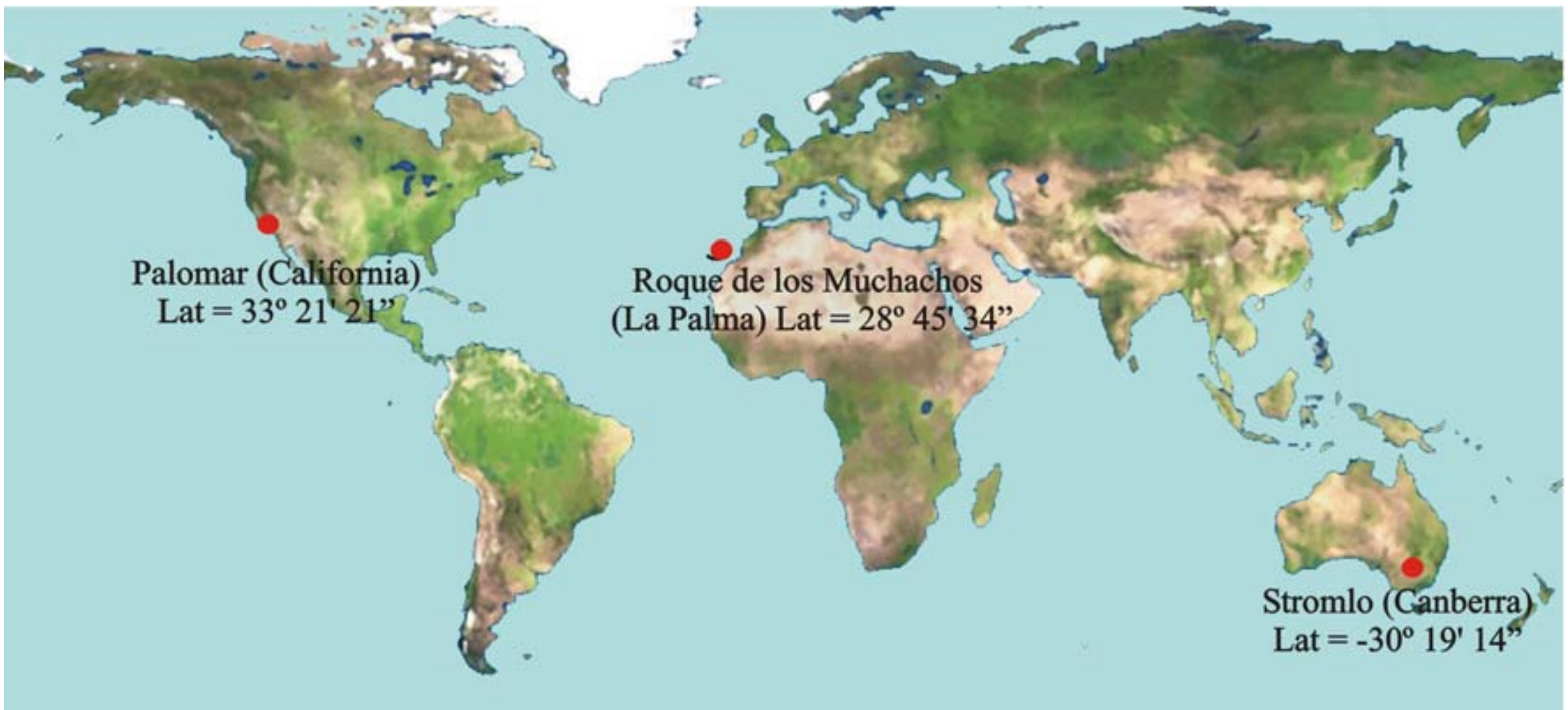

***Figura 6.8. Distribución para los terminales receptores terrestres.***

Para esta latitud de 30 grados la variación de la frecuencia debida a la rotación de la Tierra es

$$\Delta f = 0{,}378 - (-0{,}378) = 0{,}756 \text{ GHz} \tag{140}$$

Para otros terminales ubicados a distintas latitudes, el efecto Doppler inducido en la señal recibida sería cuantitativamente distinto. En la Figura 6.9 se puede observar cómo influye la latitud de los emplazamientos receptores en la variación (en valor absoluto) entre la frecuencia recibida y emitida.

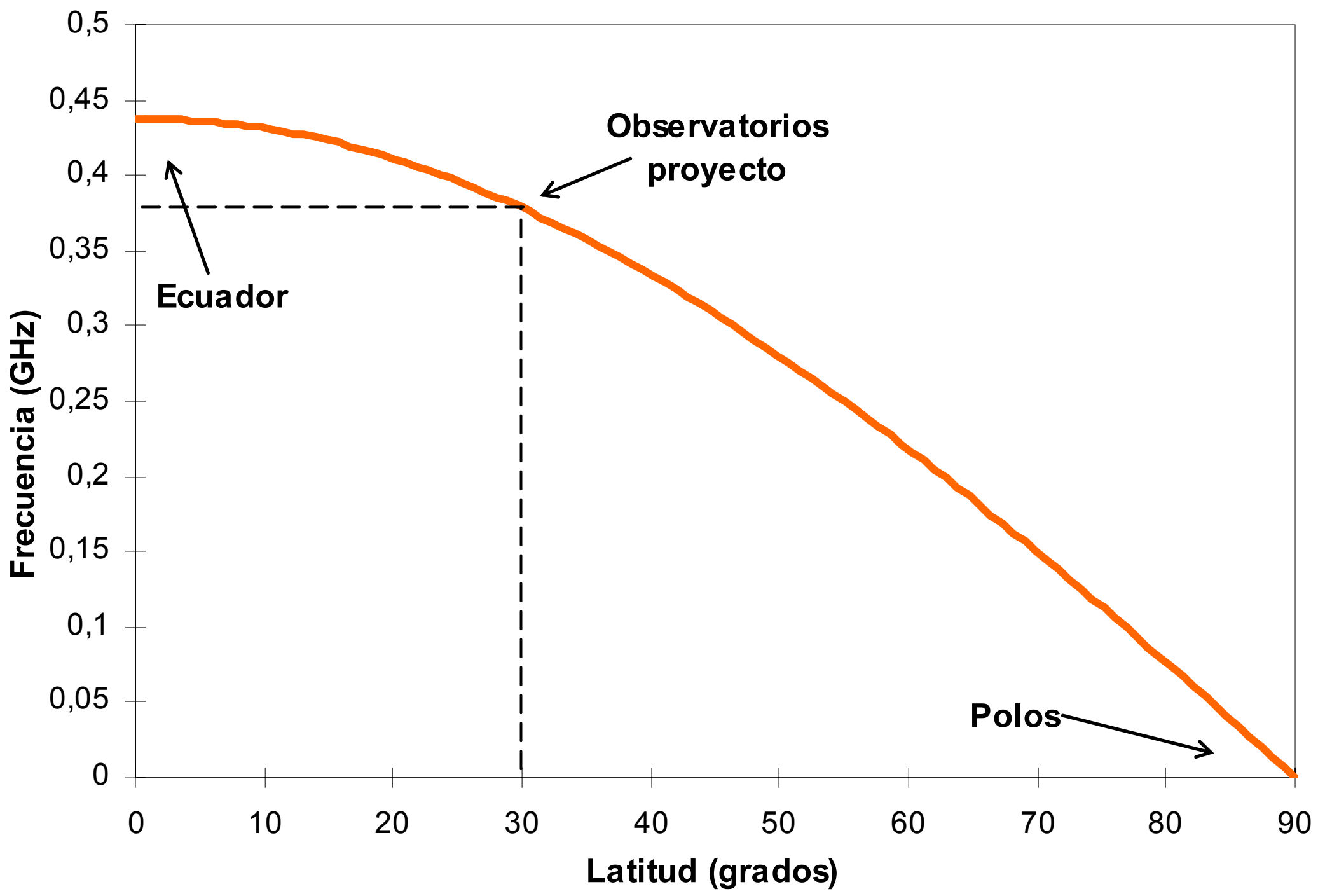

***Figura 6.9. Variación (en valor absoluto) con la latitud de la frecuencia recibida en la Tierra.***

## 6.3.2. Movimiento de traslación en el viaje Tierra-Marte

Desde la salida de la nave de la órbita terrestre, durante el viaje hacia Marte, se realizarán pruebas de funcionamiento del sistema de comunicación por láser. Por ello también se ha simulado, tal como se vio en el apartado 2.3.9, la trayectoria de la nave que describe una órbita de Hohmann.

El día previsto para la salida de la órbita terrestre es el 7 de noviembre del 2009. Según el simulador de dinámica orbital implementado para este proyecto, la posición en que se encuentran la Tierra ese día es $(Tx_0, Ty_0) = (0{,}7065, 0{,}6950)$. Trazando una línea entre esa posición y el punto (0, 0), correspondiente al Sol, se obtiene la posición en la que se encontrará Marte en el momento en que llegue la nave. Según el simulador, esta posición corresponde a $(Mx_1, My_1) = (-1{,}1043, -1{,}1057)$ que se da el 28 de agosto del 2010. Por lo tanto, el viaje Tierra-Marte tiene una duración de 294 días.

Haciendo que el simulador genere la trayectoria de la nave en la órbita de Hohmann desde el 7-11-2009 hasta el 28-8-2010, se obtiene todo el **rango de valores** de la velocidad orbital instantánea y la velocidad radial instantánea que se dan a lo largo del viaje Tierra-Marte. Estos resultados se presentan en la Figura 6.10.

Conocido el valor de la velocidad radial en un momento dado, puede calcularse, empleando la ecuación (139), la frecuencia recibida en el terminal terrestre. La diferencia $\Delta f$ entre esta frecuencia y la frecuencia emitida desde el terminal de Marte (correspondiente a la longitud de onda 1,064 $\mu$m) se presenta en la Figura 6.11 junto a la diferencia debida a la rotación terrestre para el rango de tiempo que dura el viaje Tierra-Marte.

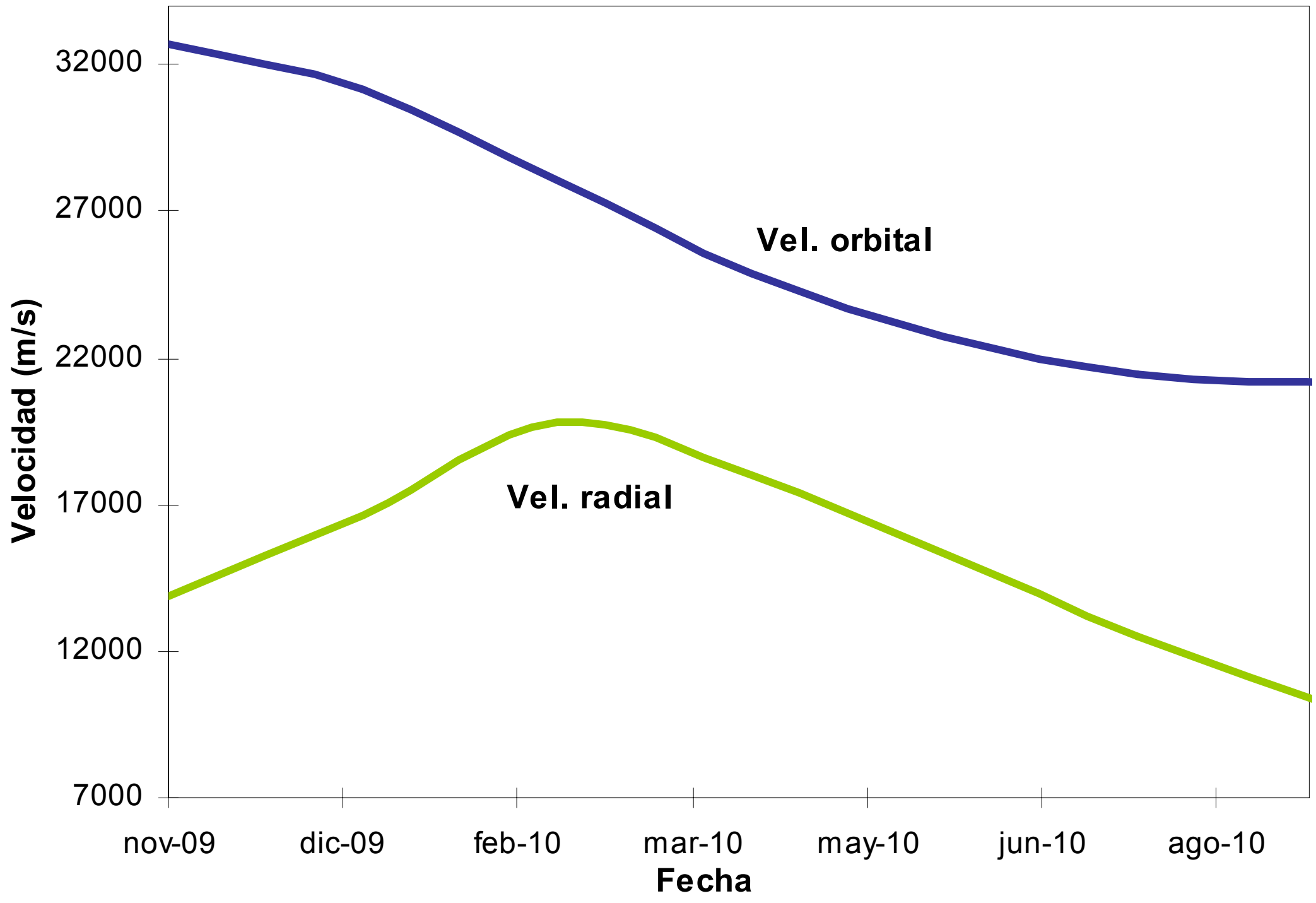

***Figura 6.10. Velocidad orbital y radial de la nave respecto a la Tierra durante el viaje Tierra-Marte.***

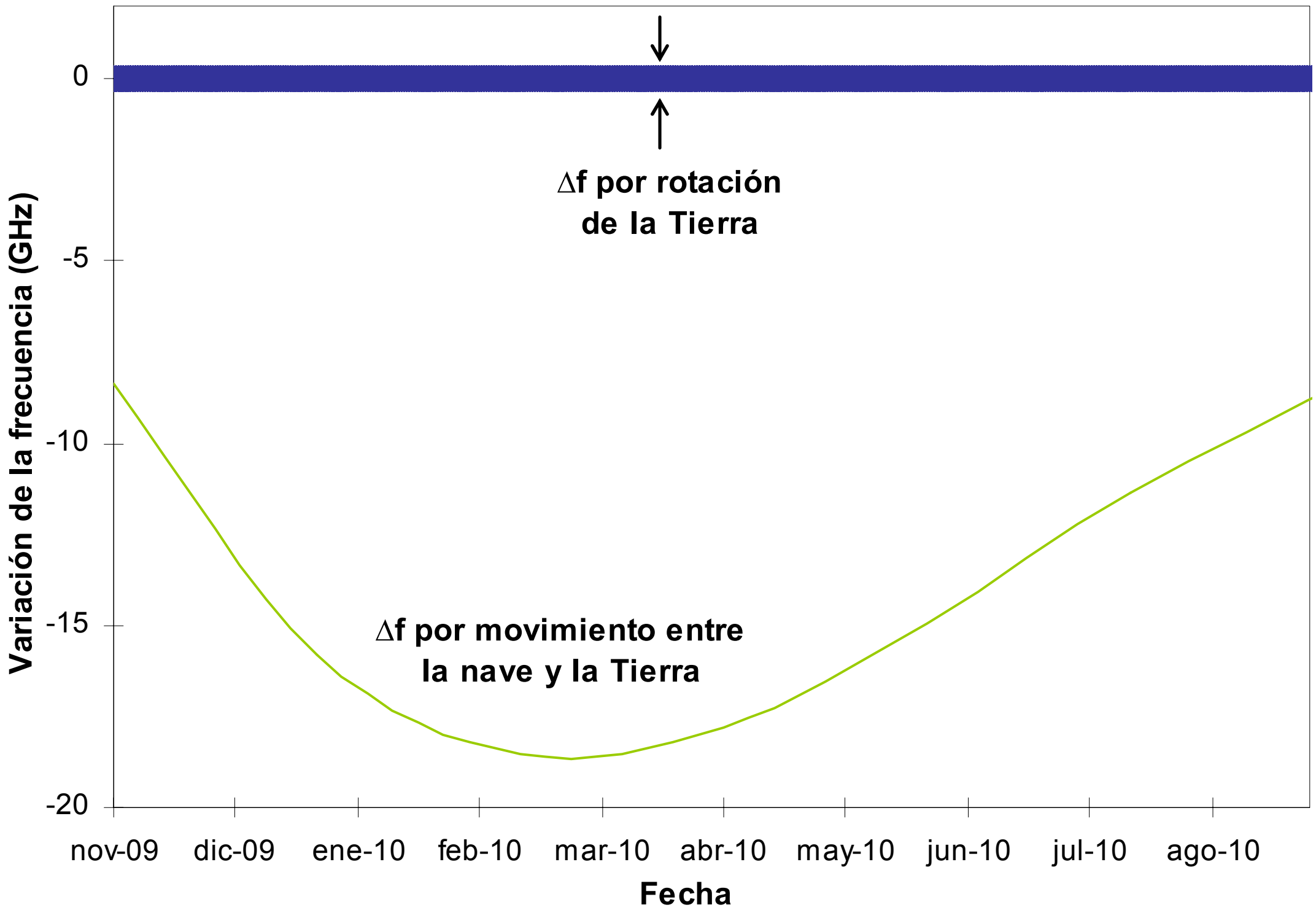

***Figura 6.11. Variación de la frecuencia recibida en la Tierra debida al movimiento relativo entre la nave y la Tierra y a la rotación de la Tierra.***

Realizando la superposición de ambas variaciones Δf, se obtiene la variación total de la frecuencia recibida en la Tierra presentada en la Figura 6.12.

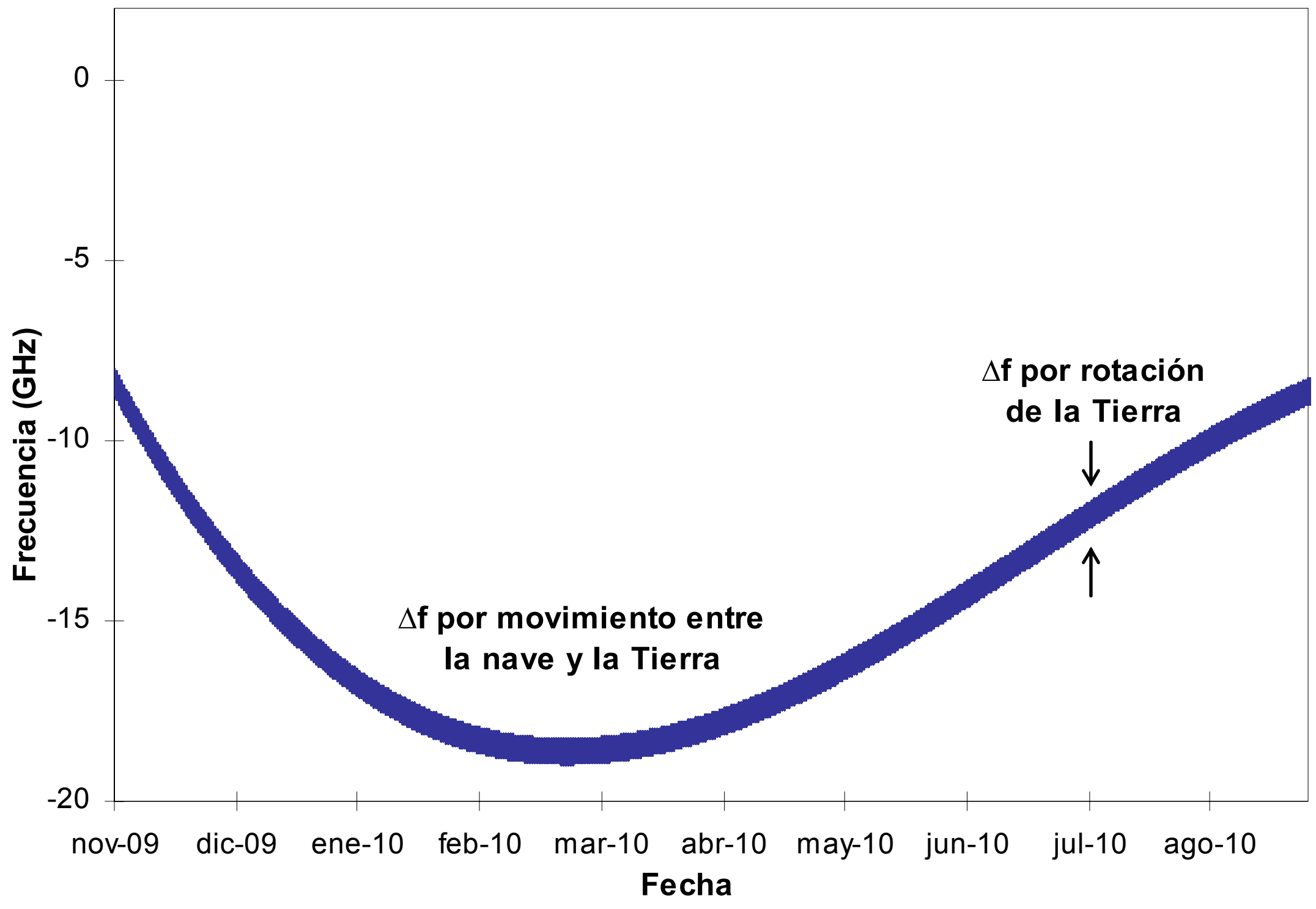

***Figura 6.12. Variación total de la frecuencia recibida en la Tierra durante el viaje Tierra-Marte.***

## 6.3.3. Rotación del satélite *MTO* alrededor de Marte

Según las especificaciones de la *NASA* [55], el satélite *MTO* se mueve según una órbita heliosíncrona, es decir, el plano orbital siempre está orientado formando el mismo ángulo respecto al Sol, tal como se puede observar en la Figura 6.13.

Esta órbita, casi **circumpolar**, resulta peculiar [56] porque *el plano de rotación varía a lo largo del año* (año marciano en este caso). Para conseguir alterar la órbita sin gasto de combustible se aprovechan las pequeñas aceleraciones y deceleraciones que induce en la nave el hecho de que el planeta –como la Tierra– no es exactamente esférico. El abultamiento ecuatorial provoca una ligera asimetría en la órbita, la cual, para una determinada orientación del plano orbital, induce una ***precesión*** en la órbita del satélite, haciendo que rote una determinada cantidad. En el caso de Marte, esa rotación deberá ser de algo menos de 1 grado por día terrestre, con el fin de completar los 360 grados en un año marciano. La ventaja más relevante de estas órbitas es que el satélite *pasa siempre por los mismos lugares a la misma hora* (marciana), lo que facilita en gran medida su comunicación con el módulo de superficie; también **minimiza los eclipses** marcianos de la nave respecto a la Tierra. Las órbitas heliosíncronas se emplean frecuentemente en satélites terrestres para meteorología y estudios medioambientales (y también para espionaje).

En el proyecto *MLCD*, esta órbita es circular y está situada a 4.450 km de altura sobre la superficie de Marte. Conocidos [21] la masa ($M_{Marte} = 0{,}64185 \cdot 10^{24}$ kg) y el radio medio (r = 3390 km) de Marte, puede calcularse el periodo del *MTO*, empleando la ecuación (24) del periodo de una órbita general elíptica

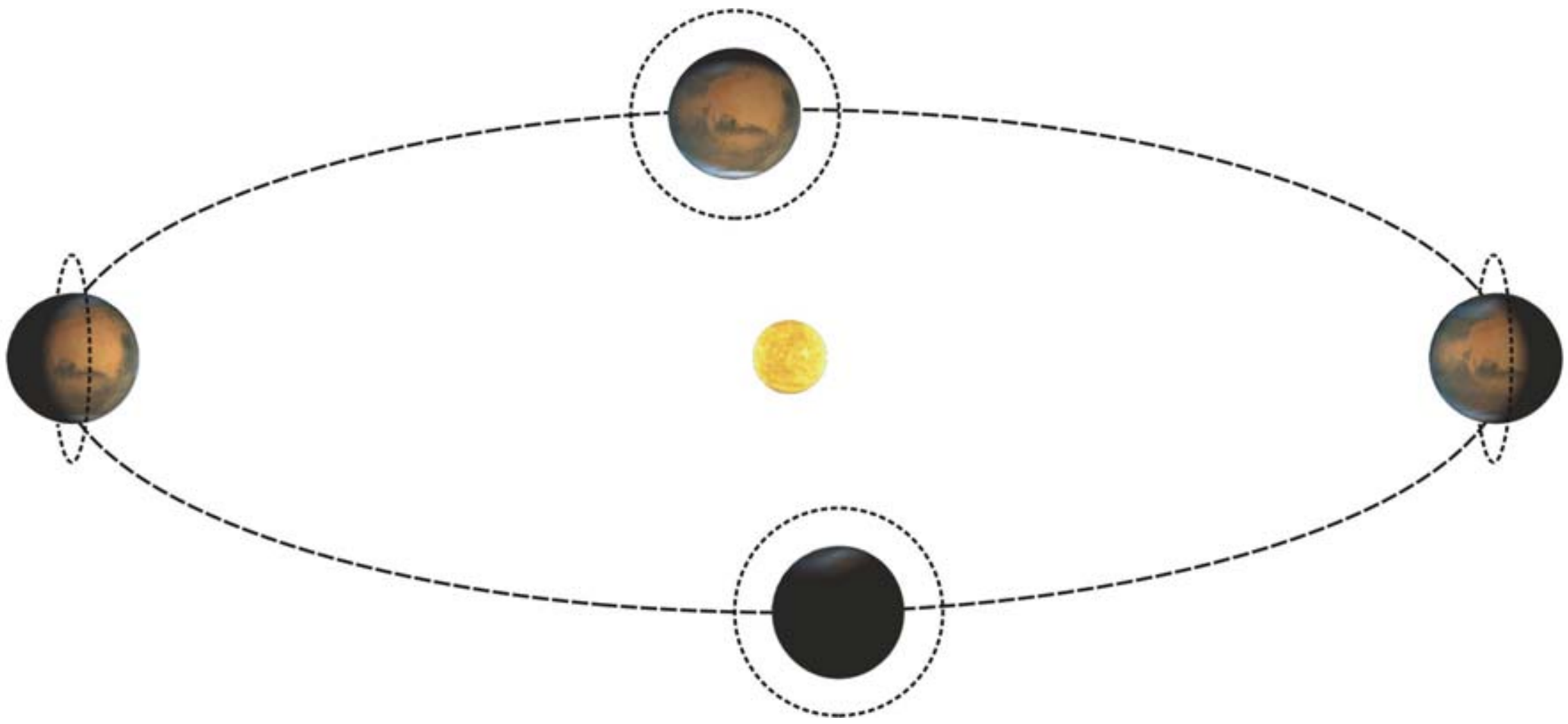

*Figura 6.13. Órbita heliosíncrona del MTO alrededor de Marte.*

$$T_{MTO} = 2\pi\sqrt{\frac{(r+h)^3}{GM_{Marte}}} = 21.077,02\ s = 5{,}85\ horas \qquad (141)$$

Se comprueba que la rotación del satélite *MTO* es la causa que genera la componente de efecto Doppler que más rápidamente cubre todo el rango de posibles frecuencias. En la Figura 6.14 se presenta una muestra calculada de la variación temporal de este rango.

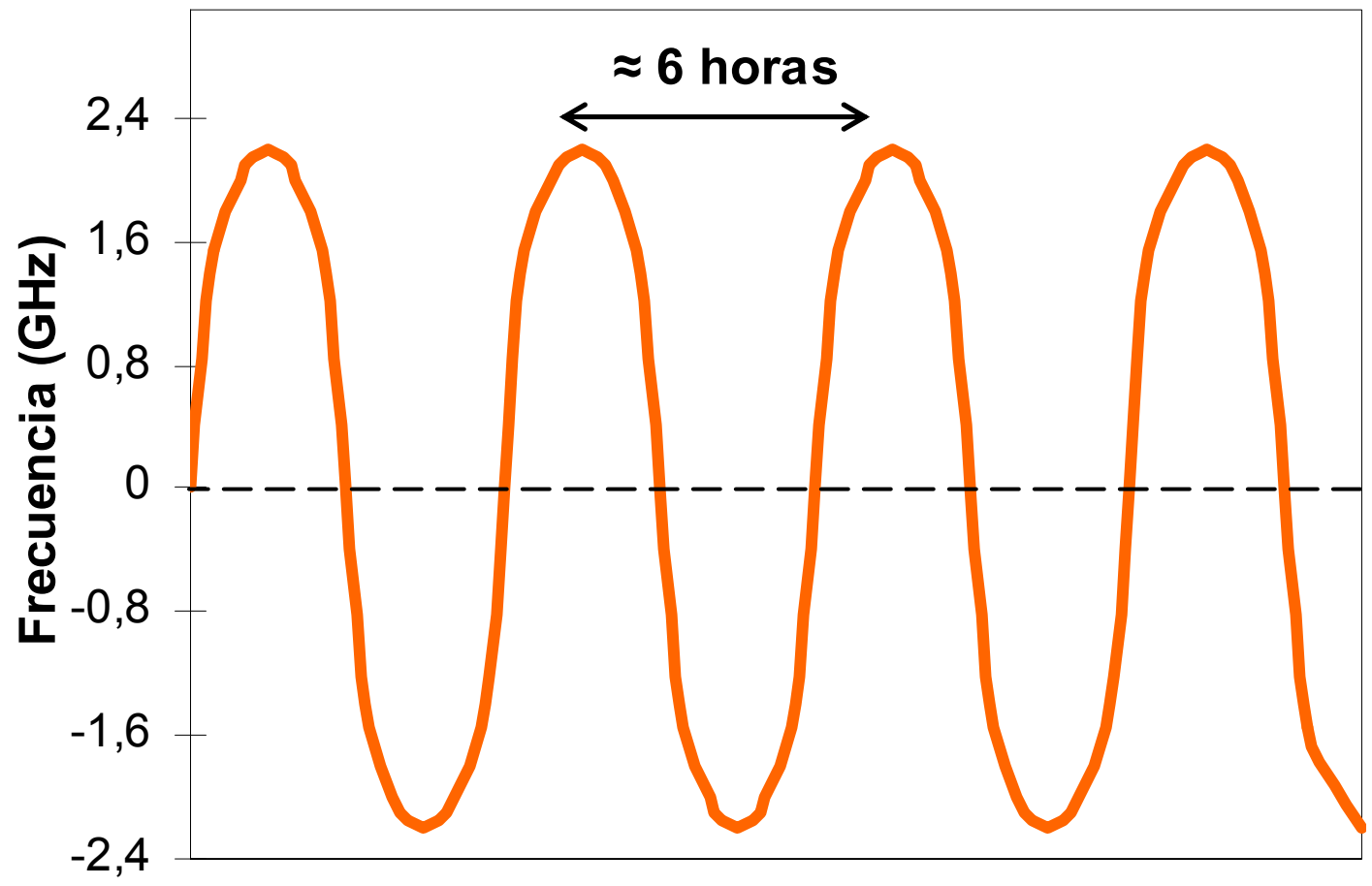

*Figura 6.14. Variación temporal de la frecuencia recibida en la Tierra debido a la rotación del satélite MTO.*

Conocido el periodo de rotación del satélite *MTO*, la velocidad radial debida a su movimiento orbital alrededor de Marte se calcula en el simulador, como se vio en el apartado 6.3.3, de idéntica forma que la análoga para la rotación de la Tierra, resultando en la siguiente diferencia de frecuencias recibidas

$$\Delta f = 2{,}197 - (-2{,}197) = 4{,}394\ GHz \qquad (142)$$

Esta contribución, al igual que la rotación terrestre, cubre todo su rango en periodos de tiempo muy pequeños, por lo que la combinación de estas dos componentes marcará el rango de *variaciones rápidas de frecuencias*, diferenciado del rango de variaciones lentas debidas al viaje Tierra-Marte y a la traslación de los planetas.

## 6.3.4. Movimiento de traslación de la Tierra y Marte

El mayor rango de valores para la velocidad radial de cualquiera de los terminales, y a su vez el que más lentamente varía, es el debido al movimiento de traslación de Marte y la Tierra en su órbita alrededor del Sol. En la Figura 6.15 y Figura 6.16 puede apreciarse cómo varían la velocidad orbital y la velocidad radial (de uno respecto al otro) de Marte y la Tierra. El rango de tiempo que se ha considerado es de dos años desde la inserción del satélite MTO en la órbita de Marte, tiempo en el que las posiciones relativas de Marte y la Tierra se repiten aproximadamente.

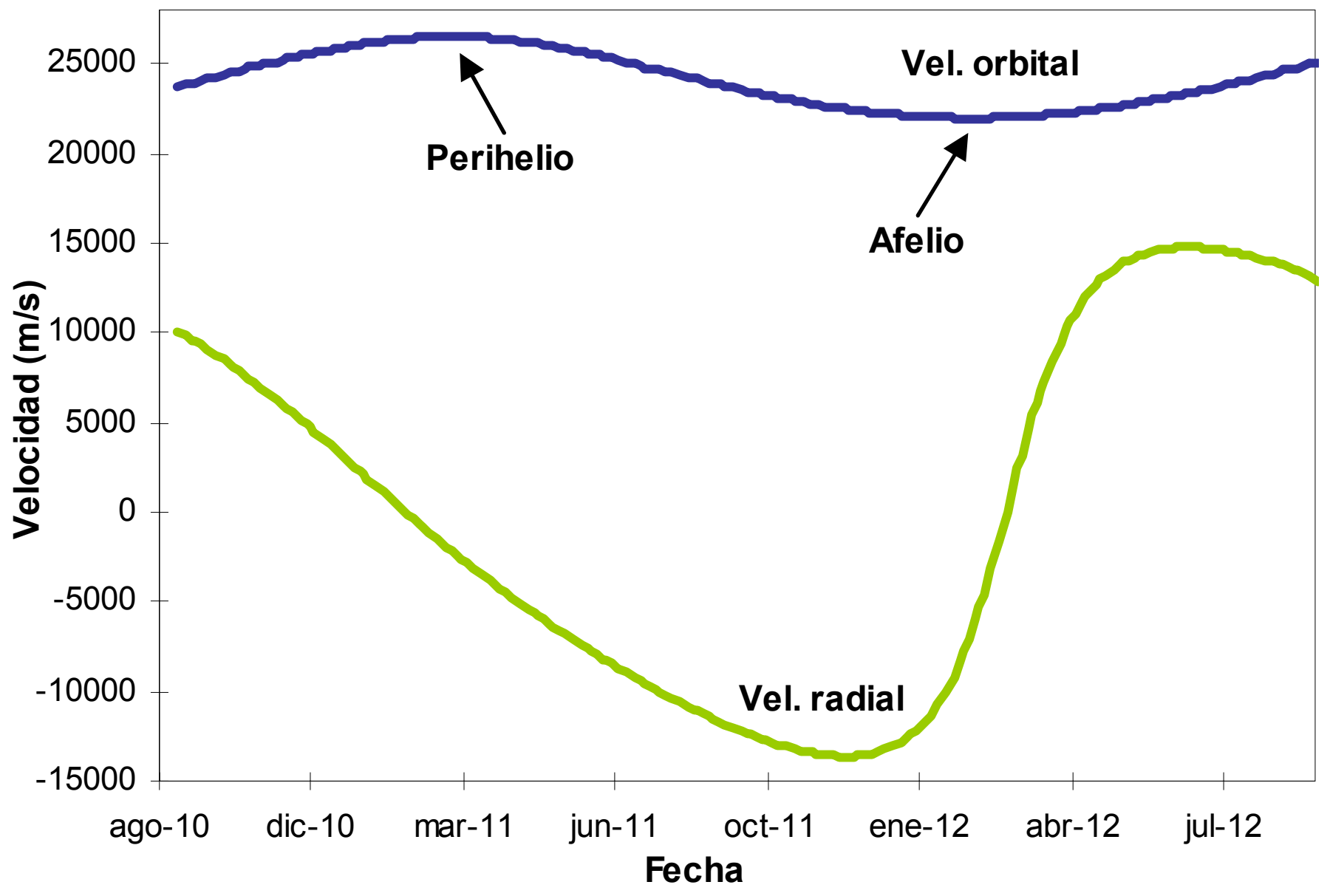

***Figura 6.15. Velocidad orbital y radial de Marte respecto a la Tierra.***

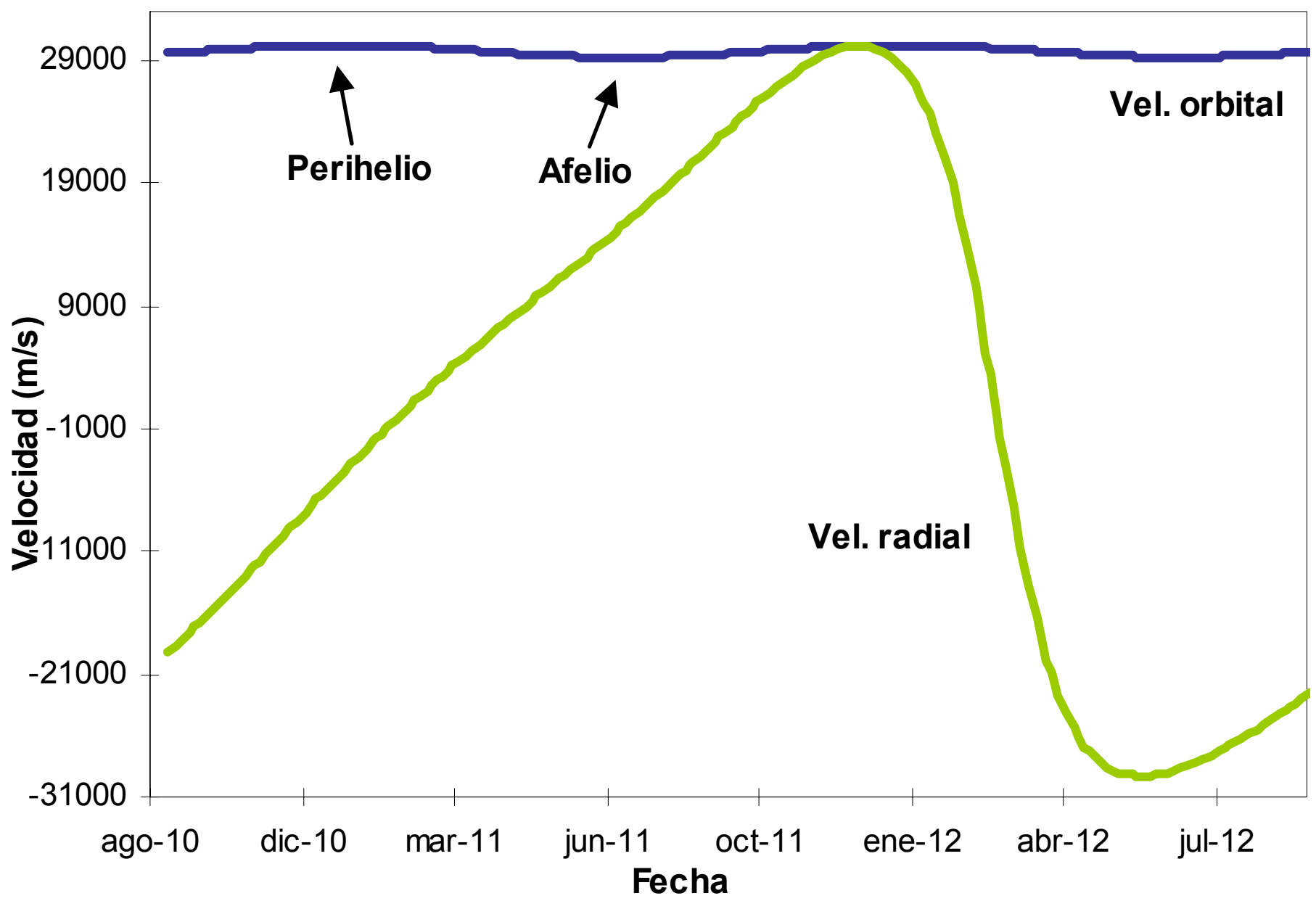

***Figura 6.16. Velocidad orbital y radial de la Tierra respecto a Marte.***

Empleando las velocidades radiales de Marte y la Tierra se obtiene la variación $\Delta f$ de la frecuencia que se recibe en la Tierra debida al movimiento de traslación entre los dos planetas. El resultado se presenta en la Figura 6.17. Si a esta variación se le superpone las correspondientes a la rotación de la Tierra y al movimiento orbital del satélite *MTO*, se obtiene la variación total de la frecuencia recibida en la Tierra que contempla todas las fuentes que introducen efecto Doppler. Esta variación se observa en la Figura 6.18.

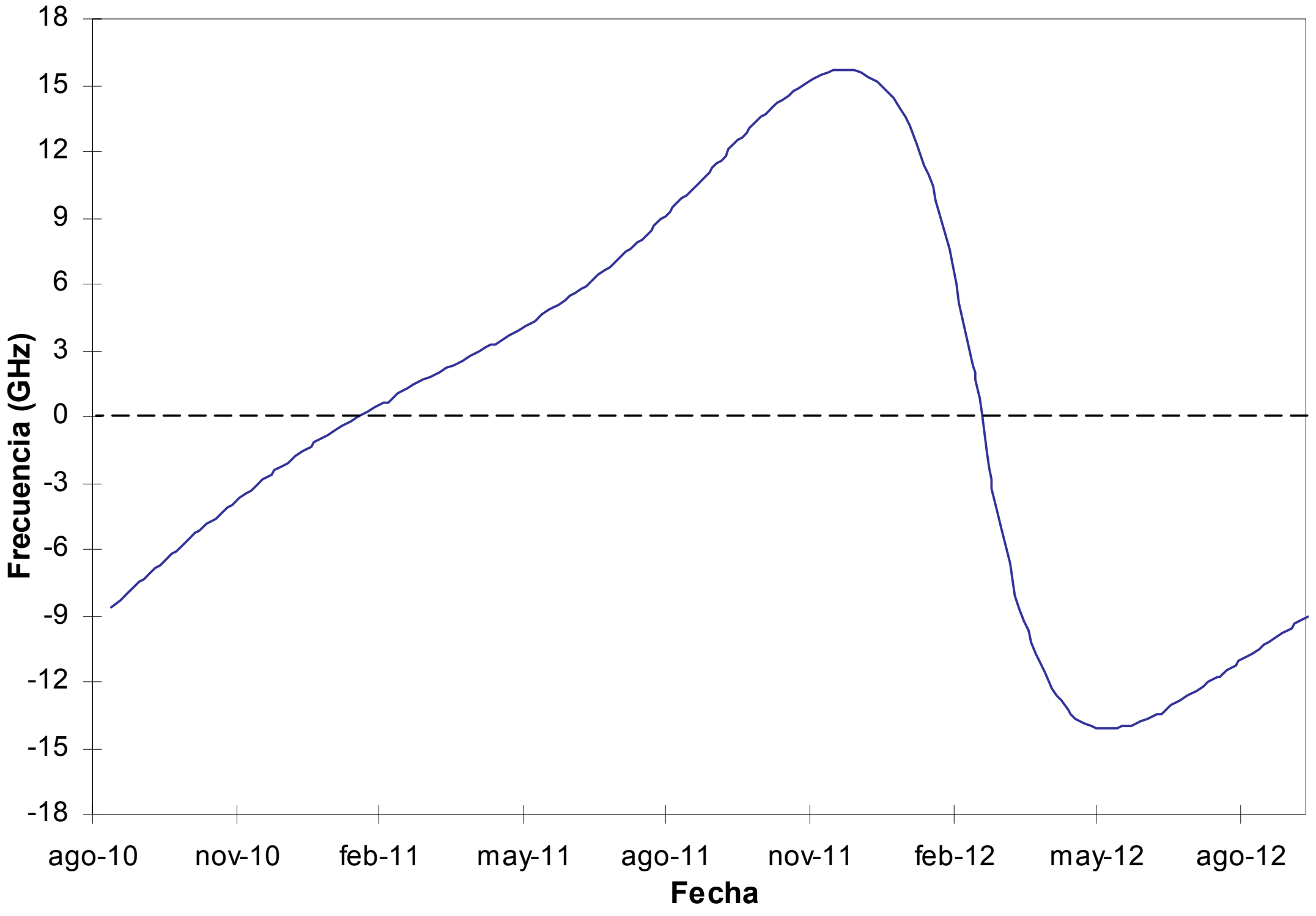

***Figura 6.17. Variación Δf de la frecuencia recibida en la Tierra debida a la traslación de Marte y la Tierra.***

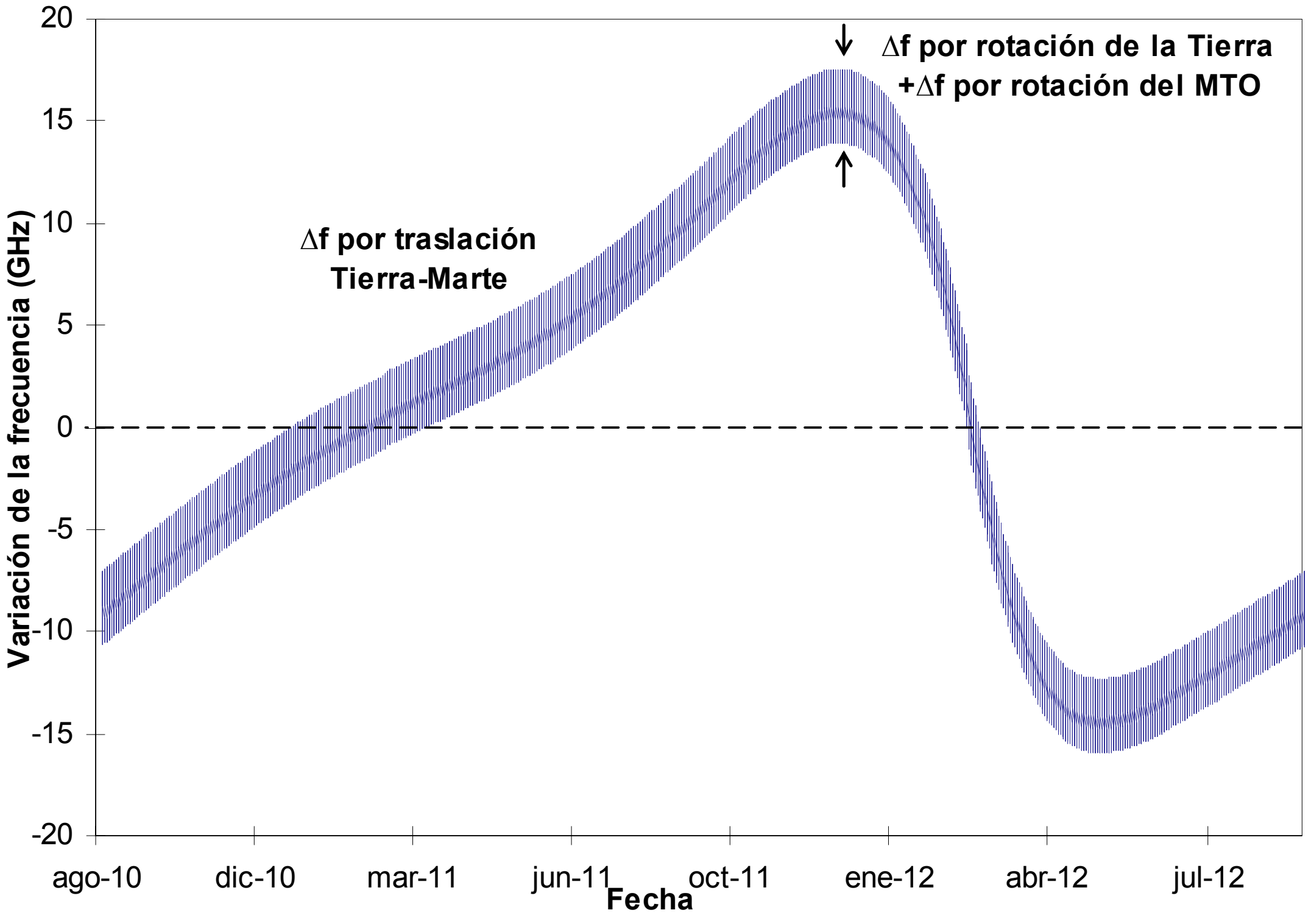

***Figura 6.18. Variación de la frecuencia recibida en la Tierra debida a la traslación de Marte y la Tierra, a la rotación de la Tierra y a la rotación del satélite MTO.***

## 6.3.5. Resultados finales del efecto Doppler

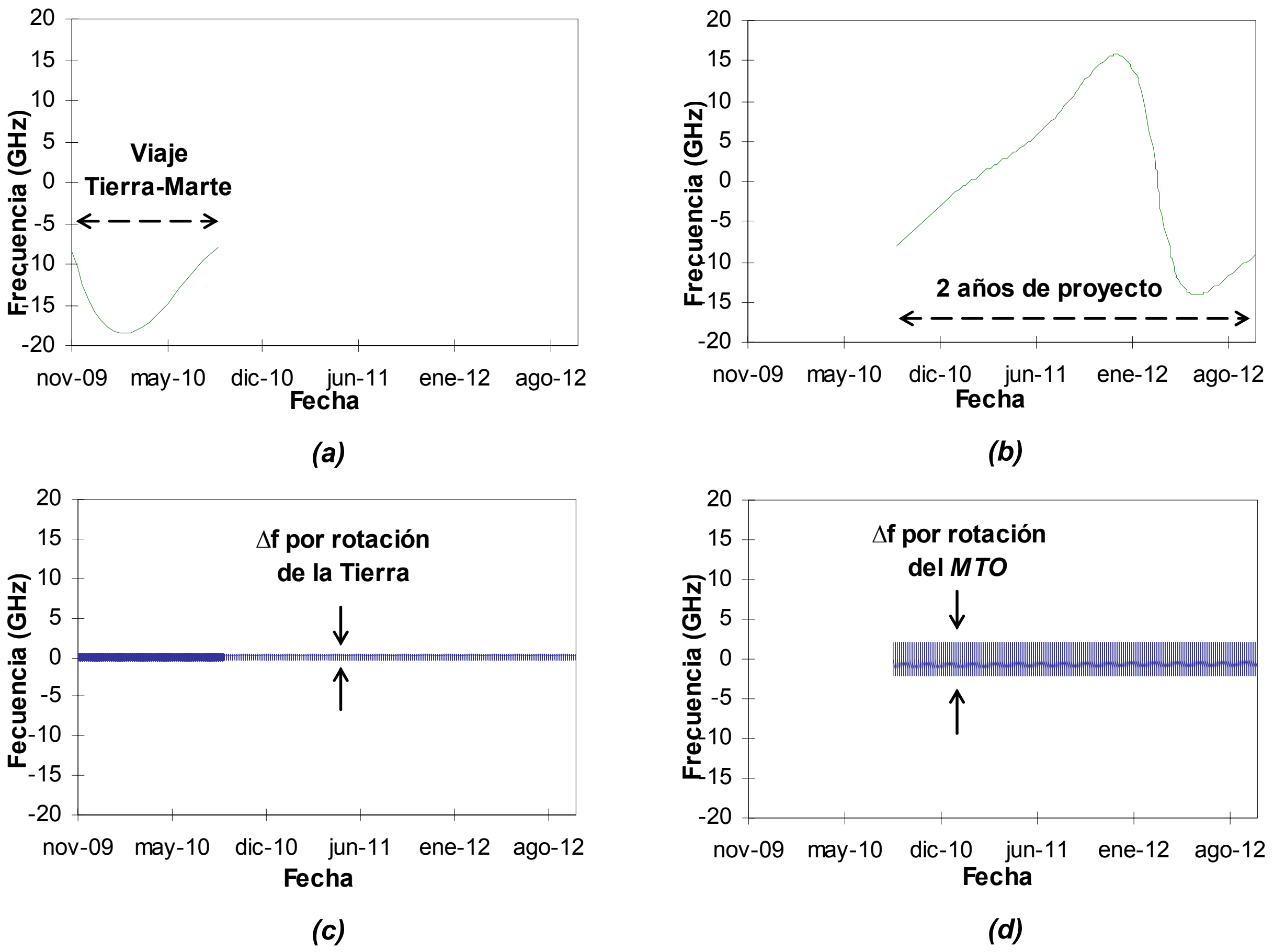

***Figura 6.19. Variación de la frecuencia recibida en la Tierra debida (a) al movimiento de la nave en el viaje Tierra-Marte, (b) a la traslación de Marte y Tierra durante dos años de proyecto, (c) a la rotación de la Tierra y (d) al movimiento orbital del satélite MTO.***

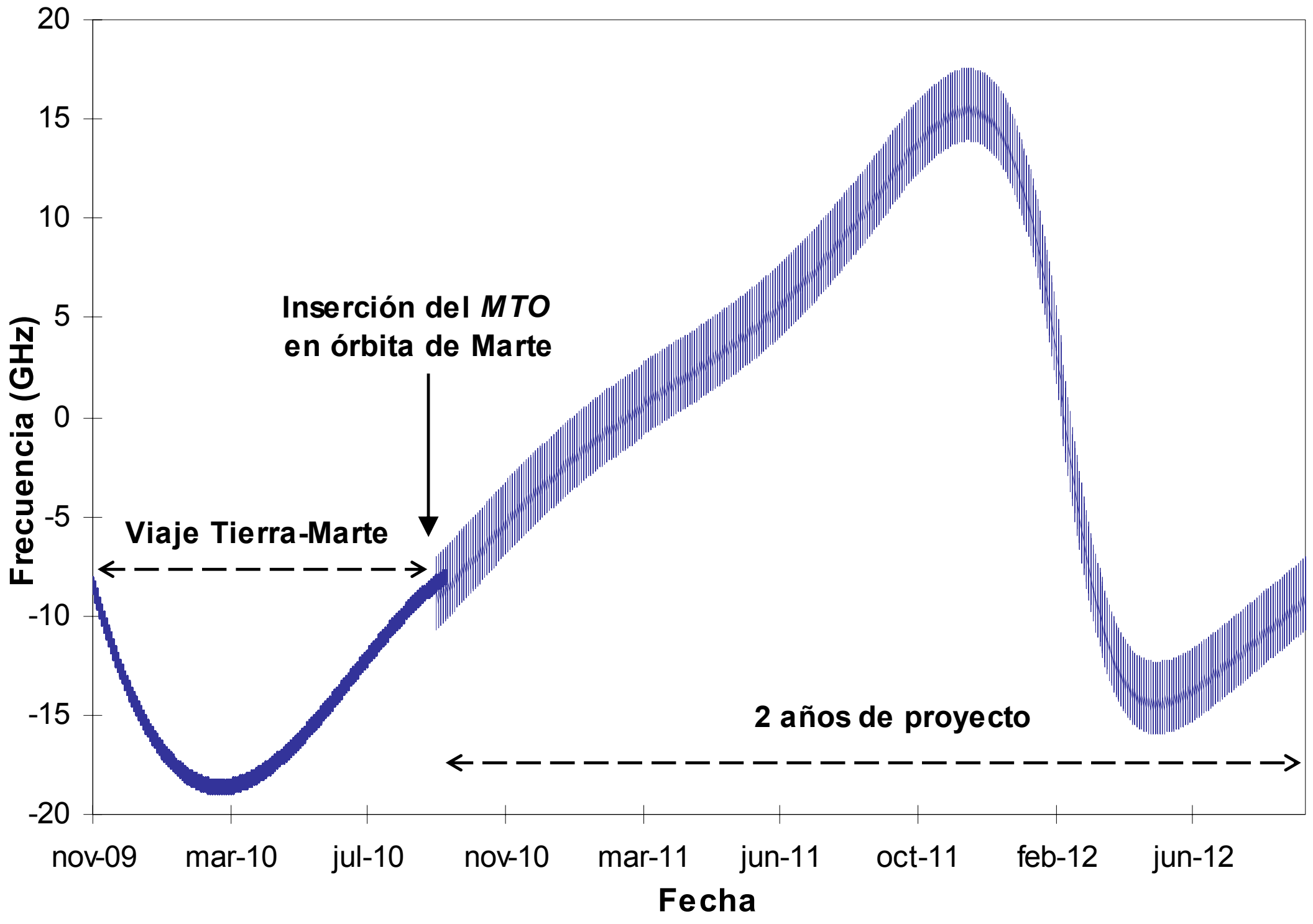

***Figura 6.20. Variación de la frecuencia recibida en la Tierra debida a la traslación de Marte y Tierra, al movimiento orbital del satélite MTO y a la rotación de la Tierra.***

En las figuras se puede observar la variación de la frecuencia recibida en la Tierra debida a la **traslación** de la Tierra y Marte, al **movimiento orbital** del satélite MTO y a la **rotación** de la Tierra sobre su eje. En la Figura 6.19 se presenta el efecto Doppler de cada fuente independientemente y en la Figura 6.20 la variación total de la frecuencia contemplando la superposición de todas las aportaciones anteriores. El rango de tiempo cubre desde el inicio del viaje Tierra-Marte cuando la nave abandona la órbita de la Tierra para tomar la trayectoria Hohmann hasta el momento que abandona esta trayectoria para introducirse en la órbita de Marte.

La frecuencia máxima y mínima, respecto a la correspondiente a la longitud de onda de 1,064 μm, recibida en la Tierra debida a cada fuente de efecto Doppler, así como el rango de variación para cada una de las fuentes, se muestra en la Tabla 6.1.

| | **Frec. máx. (GHz)** | **Frec. mín. (GHz)** | **Rango de variación (GHz)** |
|---|---|---|---|
| ***Traslación Tierra y Marte*** | 15,734 | -14,120 | 29,853 |
| ***Rotación de la Tierra*** | 0,378 | -0,378 | 0,756 |
| ***Movimiento orbital MTO*** | 2,197 | -2,197 | 4,394 |

***Tabla 6.1. Frecuencia máxima y mínima (respecto a 1,064 μm) recibida en la Tierra y debida a cada fuente de efecto Doppler y su rango de variación.***

## 6.4. PÉRDIDAS POR ESPACIO LIBRE

El factor que mayor variabilidad introduce en un enlace de comunicaciones con Marte es la distancia. Se explica esta influencia porque la distancia determina, como se dedujo en el apartado 3.1.2, el valor de las **pérdidas por espacio libre**. Es importante conocer cómo varían estas pérdidas ya que supone un factor fundamental en el balance de potencia.

Como también se explicó en el apartado 6.2.1, la distancia será máxima (y de la misma manera las pérdidas por espacio libre) en los periodos de conjunción solar y mínima en los de oposición.

En la Figura 6.21 se muestra el resultado, obtenido mediante la simulación de dinámica orbital realizada para este proyecto, del valor de la **distancia entre los dos terminales** de comunicaciones durante los casi diez meses de viaje Tierra-Marte hasta que se produce la inserción en la órbita marciana y posteriormente durante los dos años considerados de proyecto *MLCD*. La distancia en el caso del viaje Tierra-Marte se considera desde la posición de la nave hasta el centro de la Tierra (el radio terrestre puede despreciarse en comparación con las distancias involucradas) y en el caso de los dos años de proyecto entre el centro de la Tierra y el de Marte (el radio de Marte más la distancia desde la superficie hasta los 4.450 km de altura del satélite también son despreciables).

Conocida la distancia entre terminales correspondiente a cada fecha, y para la longitud de onda utilizada en el proyecto *MLCD* (1,064 μm), es posible calcular las pérdidas por espacio libre través de la ecuación (65). El resultado de aplicar esta ecuación, empleando el simulador de dinámica orbital, a cada uno de los días del proyecto, se muestra en la Figura 6.22.

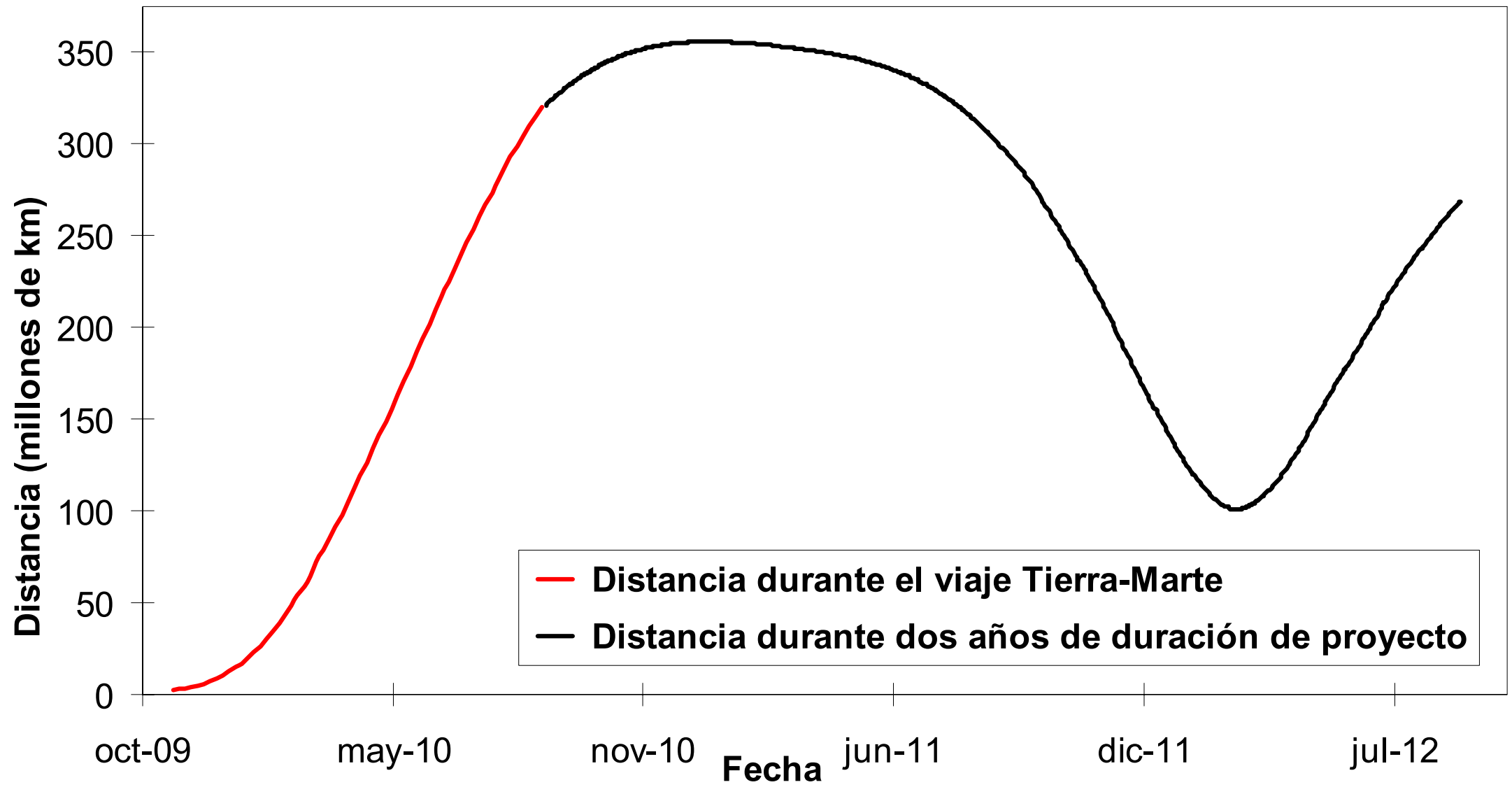

***Figura 6.21. Distancia entre ambos terminales de comunicaciones durante el transcurso del proyecto MLCD.***

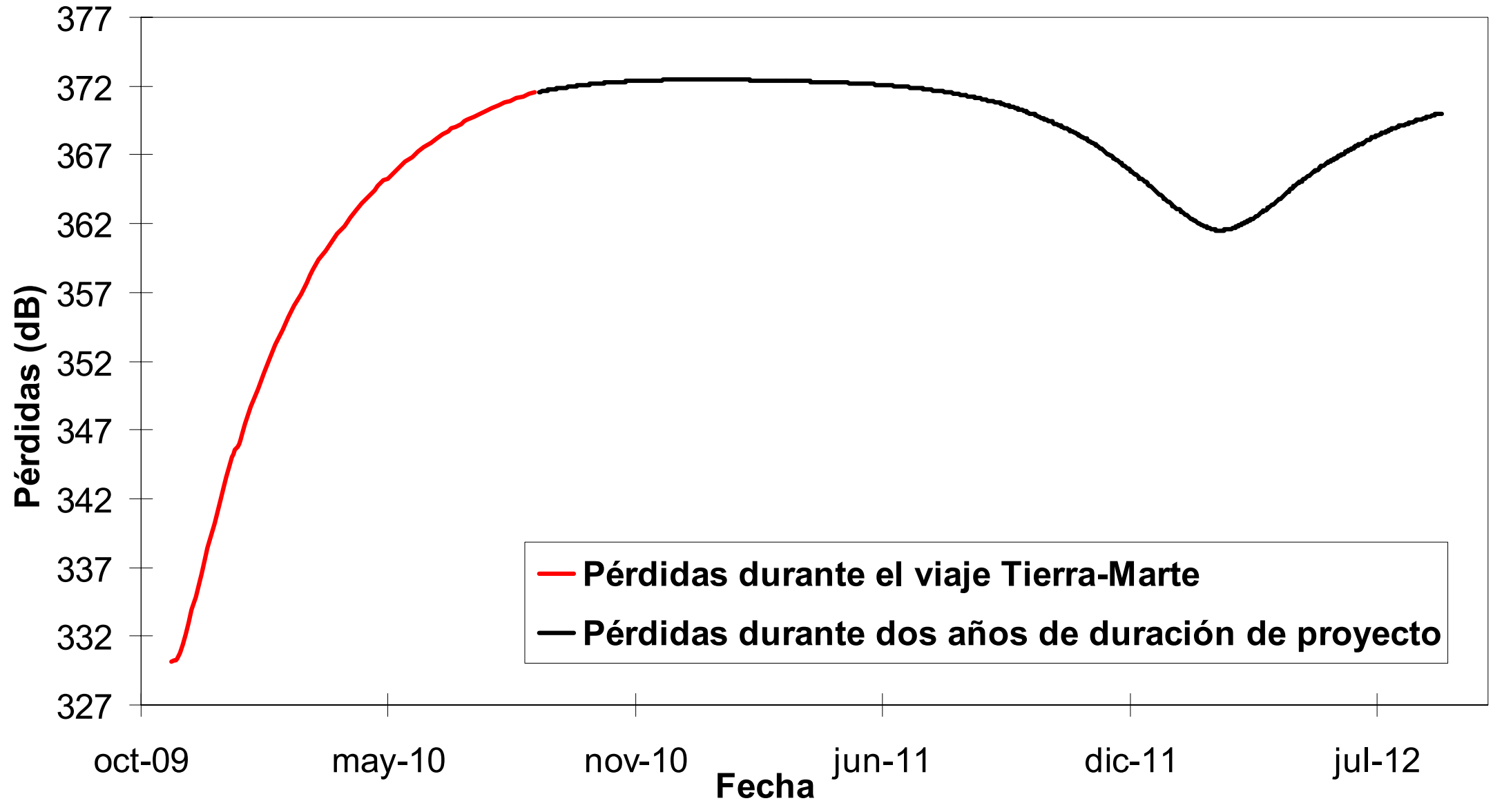

***Figura 6.22. Pérdidas por espacio libre entre ambos terminales de comunicaciones durante el transcurso del proyecto MLCD.***

Para el cálculo en el balance de potencia de caso peor y caso mejor interesa conocer, para los dos años considerados de duración de proyecto, la máxima distancia asociada a las máximas pérdidas por espacio libre y la mínima distancia y sus correspondientes mínimas pérdidas. Estos datos obtenidos del conjunto de valores proporcionados por el simulador y representados en las anteriores figuras, se presentan en la Tabla 6.2.

| | **Distancia (millones de km)** | **Pérdidas (dB)** | **Fecha** |
|---|---|---|---|
| ***Caso peor*** | 355,87 | 372,47 | 2/1/2011 |
| ***Caso mejor*** | 100,87 | 361,52 | 8/3/2012 |

***Tabla 6.2. Distancia y pérdidas máxima y mínima durante dos años de duración de proyecto tras la inserción del MTO en la órbita de Marte.***

# 6.5. INFLUENCIA DE LA ATMÓSFERA

A continuación se evalúa el efecto que tiene sobre el enlace de comunicación, en términos de pérdidas de la potencia transmitida, cada uno de los efectos de la atmósfera en la propagación de señales ópticas de comunicaciones. El desarrollo teórico de cada una de las aportaciones se realizó en el apartado 3.3.

## 6.5.1. Absorción

En la Figura 6.23 se puede observar la transmitancia para 1,064 µm. Como se observa está por encima de 0,99, y por lo tanto, aplicando la fórmula (66), las pérdidas debidas a atenuación atmosférica quedan por debajo de 0,044 dB. Una atenuación tan baja, frente al orden de magnitud de las demás atenuaciones, puede despreciarse sin problema.

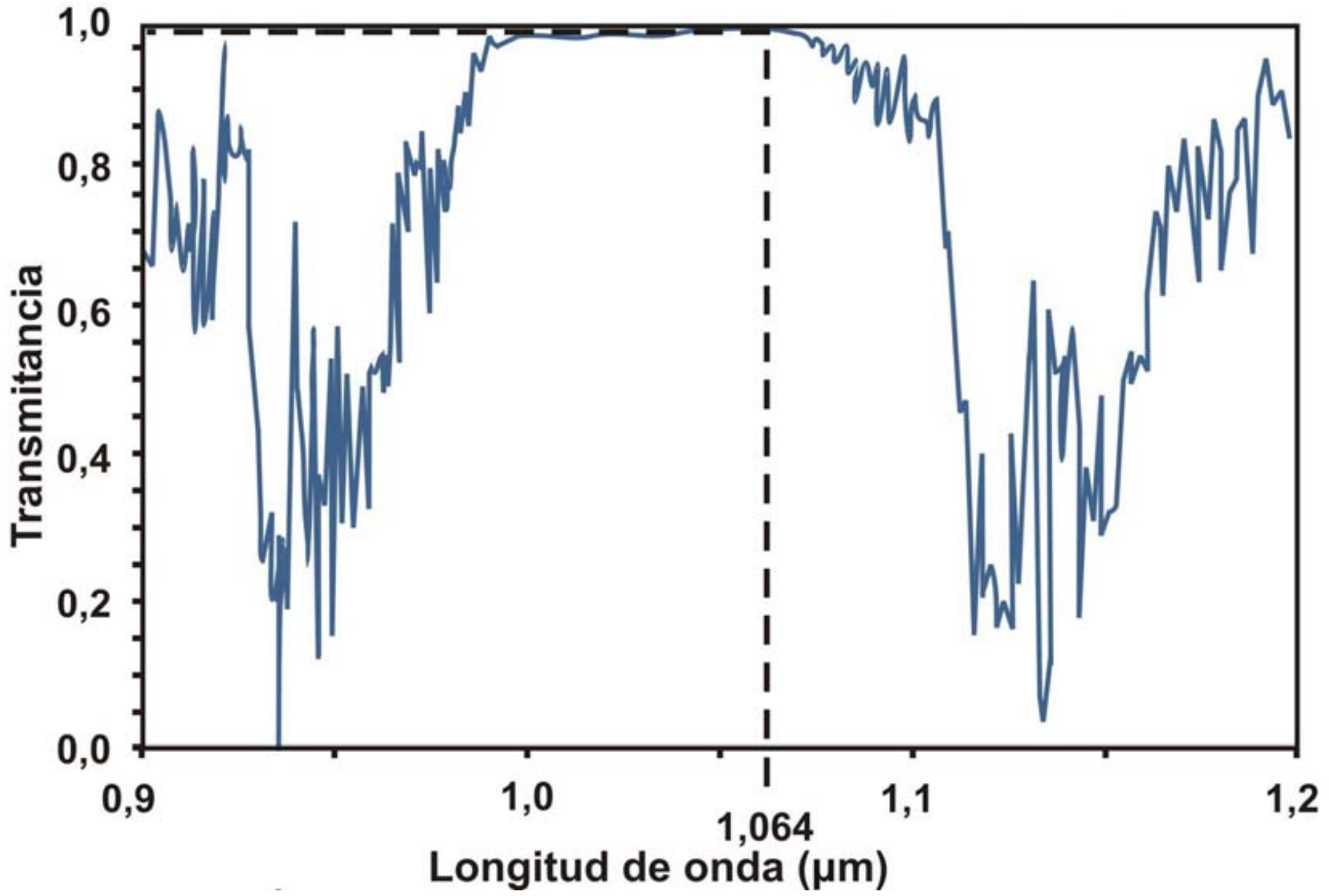

***Figura 6.23. Transmitancia debida a absorción atmosférica en la ventana de transmisión de 1,064 µm [29].***

## 6.5.2. Scattering Rayleigh

Como se explicó en el apartado dedicado al efecto de la atmósfera terrestre en el capítulo 3, la luz, al atravesar la atmósfera sufre una dispersión o *scattering* Rayleigh debida principalmente a moléculas de aire. Puesto que la densidad de estas moléculas no es constante a lo largo de toda la atmósfera, el coeficiente de atenuación por *scattering* Rayleigh variará en función de la altura de la atmósfera. En este apartado se considera la atmósfera *totalmente limpia*, sin nubes, aerosoles ni polvo, puesto que el cálculo Rayleigh aplica únicamente a partículas de tamaño muy inferior a λ (moléculas, en este caso). El resto de partículas de mayor tamaño se estudian en el apartado 6.5.3, dedicado a *scattering* Mie.

El punto de partida para cuantificar la atenuación Rayleigh es el valor de la densidad de moléculas de gas para distintas alturas atmosféricas. La representación gráfica de esta densidad se presentó en la Figura 3.13 dentro del apartado 3.3.3.a dedicado al estudio teórico del *scattering* Rayleigh. A continuación, a través del cálculo de la concentración numérica $N_g$ mediante la ecuación (70) y de la sección eficaz Rayleigh $\sigma_R$ a la longitud de onda de 1,064 μm mediante la ecuación (69), es posible determinar el coeficiente de atenuación $\alpha_R$ para distintas alturas mediante la ecuación (68).

| Altura (km) | D (kg/m$^3$) | $N_g$ (moléculas/m$^3$) | $\sigma_R$ (m$^2$) | $\alpha_R$ (Np/km) |
|---|---|---|---|---|
| 0 | 1,259 | 2,62E+25 | 3,28E-32 | 0,00086 |
| 2,00 | 1,051 | 2,19E+25 | | 0,00072 |
| 4,47 | 0,800 | 1,66E+25 | | 0,00055 |
| 7,90 | 0,529 | 1,10E+25 | | 0,00036 |
| 11,60 | 0,337 | 7,01E+24 | | 0,00023 |
| 15,64 | 0,188 | 3,91E+24 | | 0,00013 |
| 20,88 | 0,080 | 1,66E+24 | | 5,46E-05 |
| 24,74 | 0,035 | 7,28E+23 | | 2,39E-05 |
| 27,84 | 0,017 | 3,54E+23 | | 1,16E-05 |
| 29,90 | 0,007 | 1,46E+23 | | 4,77E-06 |
| 32,04 | 0,0001 | 2,08E+21 | | 6,82E-08 |
| 35 | 0 | 0 | | 0 |

***Tabla 6.3. Cálculo del coeficiente de atenuación Rayleigh para distintos valores de la altura de la atmósfera.***

De la Tabla 6.3 se puede representar la dependencia del coeficiente de atenuación $\alpha_R$ con la altura que se presenta en la Figura 6.24.

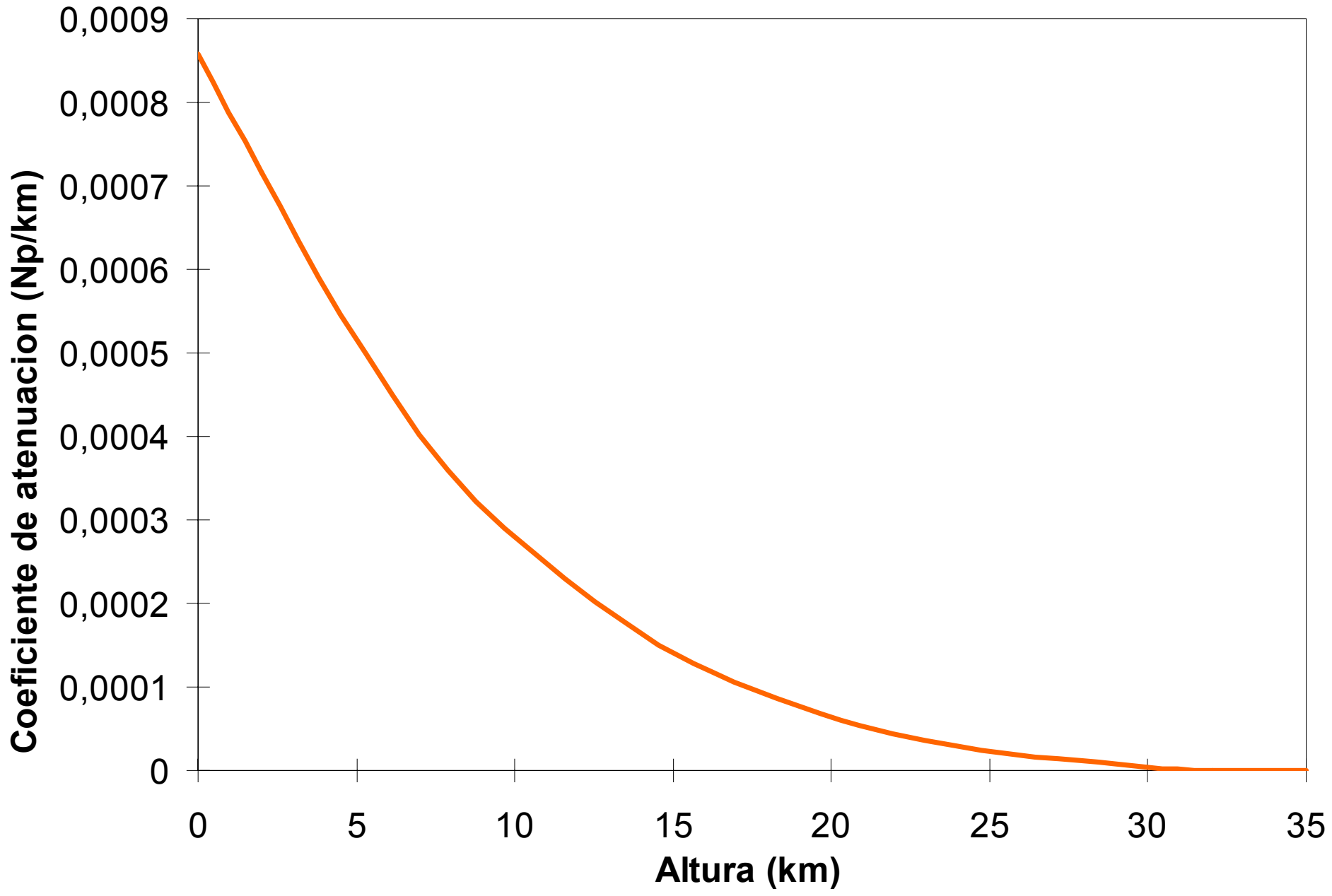

***Figura 6.24. Coeficiente de atenuación Rayleigh en función de la altura en la atmósfera.***

Para obtener el valor de la atenuación atmosférica se debe integrar el coeficiente de atenuación en el rango de alturas. Como se observa en la Figura 6.24, a partir de 30 km, el coeficiente de atenuación se hace despreciable, con lo que éste será el límite superior.

El límite inferior lo determinará la altura media de las ubicaciones receptoras consideradas, unos 2 km. Realizando un ajuste de la curva a un polinomio de sexto grado se puede realizar la integral del mismo entre los límites de integración mencionados, resultando

$$L_R = \int_2^{30} \alpha_R \, dr = 0{,}0071\,\text{Np} = 0{,}048\text{ dB} \tag{143}$$

A modo ilustrativo, se muestra a continuación la Tabla 6.4 en la que se puede comprobar el efecto que tiene el situar al receptor a nivel del mar o a dos kilómetros por encima. Las pérdidas para nivel del mar se han obtenido de idéntica forma que las calculadas a dos kilómetros pero como límite inferior se ha tomado un valor nulo. Se comprueba que los primeros kilómetros tienen mucha influencia en comparación con los restantes.

| Límite Inferior (km) | Límite Superior (km) | $L_R$ (dB) |
|---|---|---|
| 0 | 30 | 0,062 |
| 2 | 30 | 0,048 |

***Tabla 6.4. Pérdidas por scattering Rayleigh para nivel del mar y para 2 km por encima.***

Cabe destacar la influencia del ángulo cenital en el cálculo de la atenuación atmosférica. Para un ángulo cenital de cero grados, la longitud de atmósfera atravesada por la señal se corresponderá exactamente con los 28 km considerados, mientras que para valores de ángulo cenital superiores, la longitud de atmósfera a atravesar será mayor, lo que se traducirá en una atenuación mayor. A continuación se explica la manera de evaluar la distancia que atraviesa la señal en el rango de atmósfera considerado para distintos ángulos cenitales.

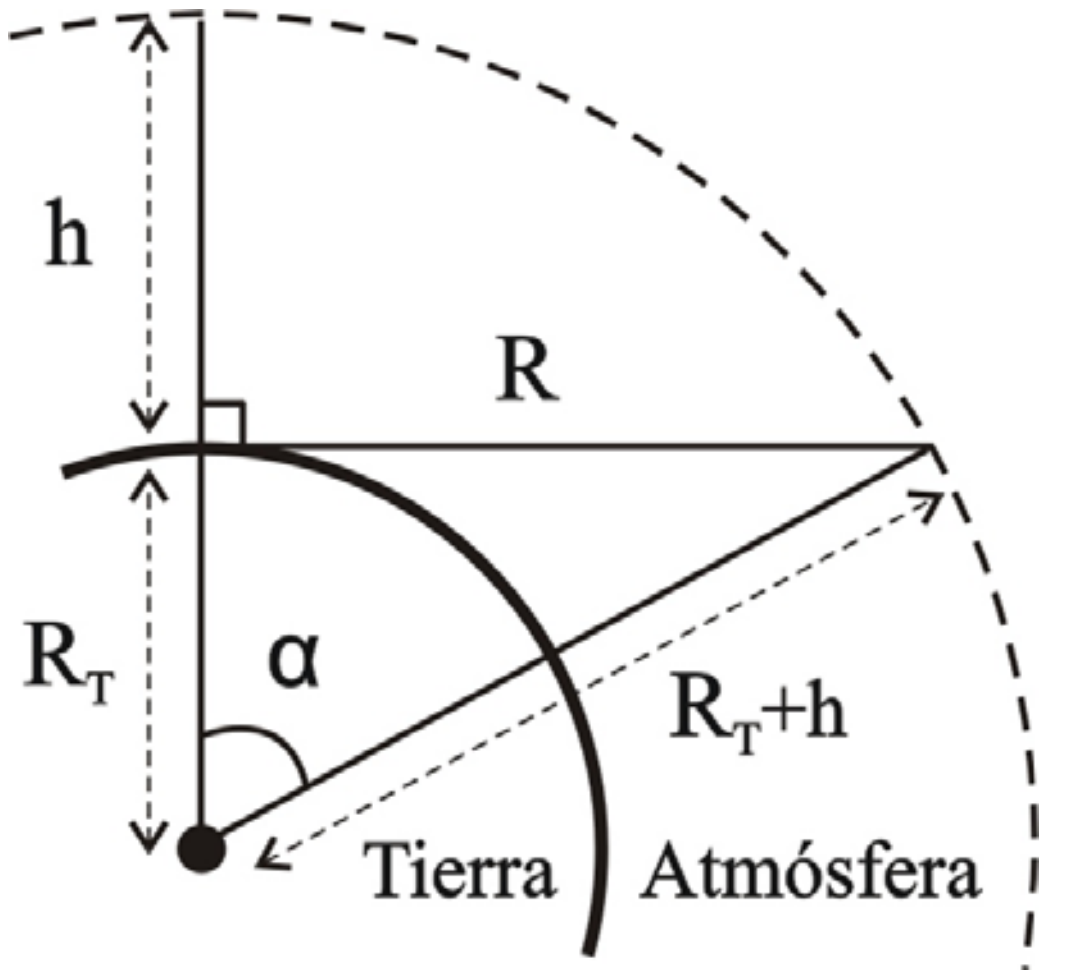

***Figura 6.25. Cálculo de la distancia auxiliar R.***

Para calcular la distancia real de atmósfera que atraviesa la señal se utiliza la distancia que atravesaría si el ángulo cenital fuera 90 grados (horizonte). Como se aprecia en la Figura 6.25, esta distancia R se puede calcular a partir del ángulo auxiliar α, del radio terrestre medio $R_T$ (6.371 km [21]) y de la altura vertical h considerada de atmósfera. El ángulo α se calcula de la siguiente manera

$$\cos(\alpha) = \frac{R_T}{R_T + h} \Rightarrow \alpha = a\cos\left(\frac{R_T}{R_T + h}\right) \tag{144}$$

Por lo que la expresión de la distancia auxiliar R queda

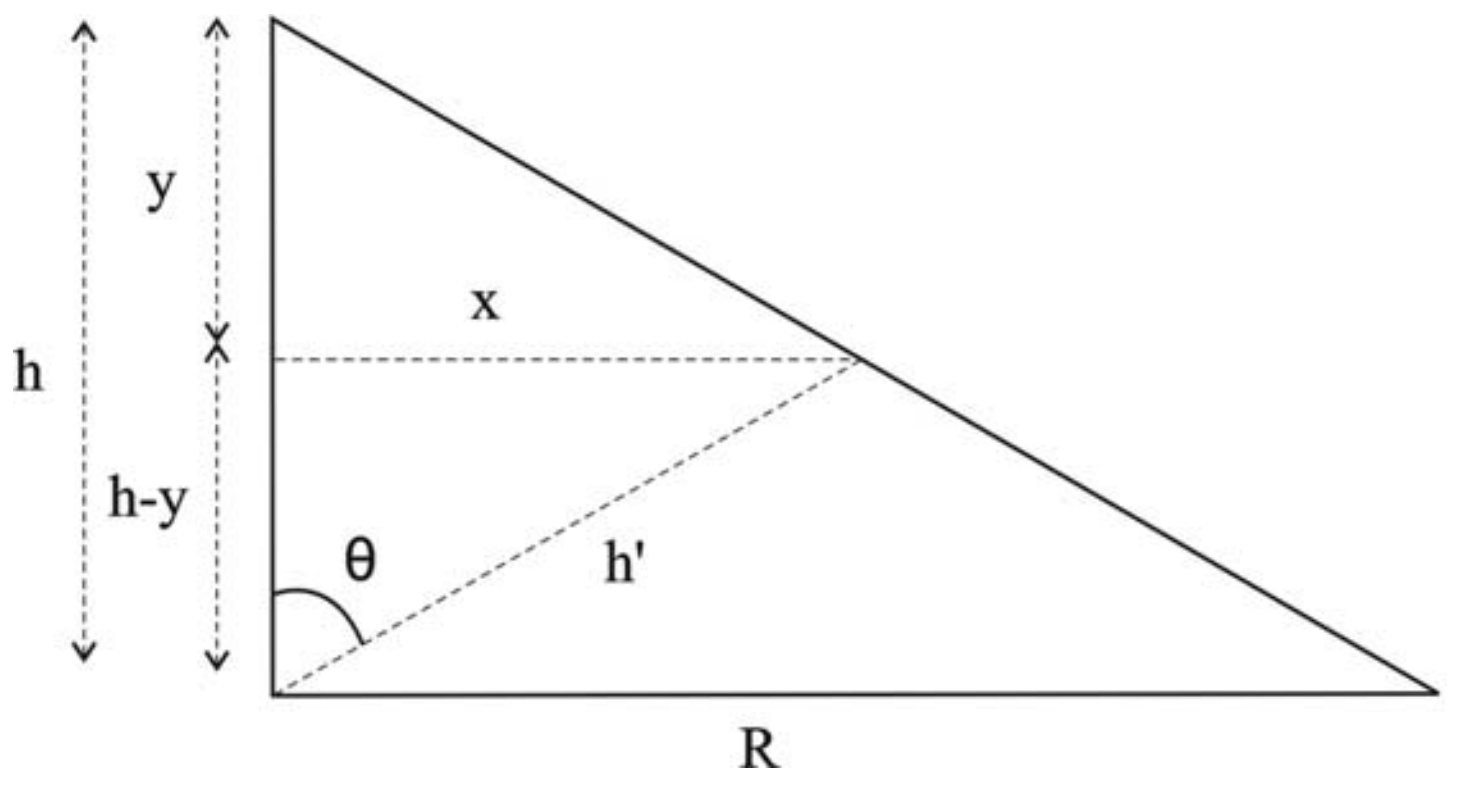

***Figura 6.26. Cálculo de la distancia real h' de atmósfera atravesada por la señal (la gráfica no está a escala, en realidad R>>h).***

$$\mathrm{sen}(\alpha) = \frac{R}{R_T + h} \Rightarrow R = (R_T + h)\mathrm{sen}\left[a\cos\left(\frac{R_T}{R_T + h}\right)\right] \tag{145}$$

El cálculo de la distancia real h' (Figura 6.26) se realiza tal como sigue

$$\begin{cases} \dfrac{h}{R} = \dfrac{y}{x} \\ \tan(\theta) = \dfrac{x}{h-y} \end{cases} \Rightarrow h' = \frac{h - \dfrac{h^2 \tan(\theta)}{h\tan(\theta) + R}}{\cos(\theta)} \tag{146}$$

donde θ es el ángulo cenital que varía entre 0 (en dirección al cenit) y 90 grados (hacia el horizonte). Una vez hallada la distancia real h', el coeficiente de atenuación $\alpha_R$ de la Tabla 6.3 hay que distribuirlo en un rango de distancias de 0 a h' kilómetros en lugar del rango original y proceder de manera similar: calcular el polinomio e integrarlo entre 2 y h'. Obsérvese que se ha aproximado la curvatura superior de la atmósfera a una recta (hipotenusa del triángulo de la Figura 6.26). Esta aproximación es válida teniendo en cuenta la gran diferencia entre la altura de la atmósfera y el radio terrestre.

Mediante el procedimiento descrito se llega a las pérdidas de la Figura 6.27.

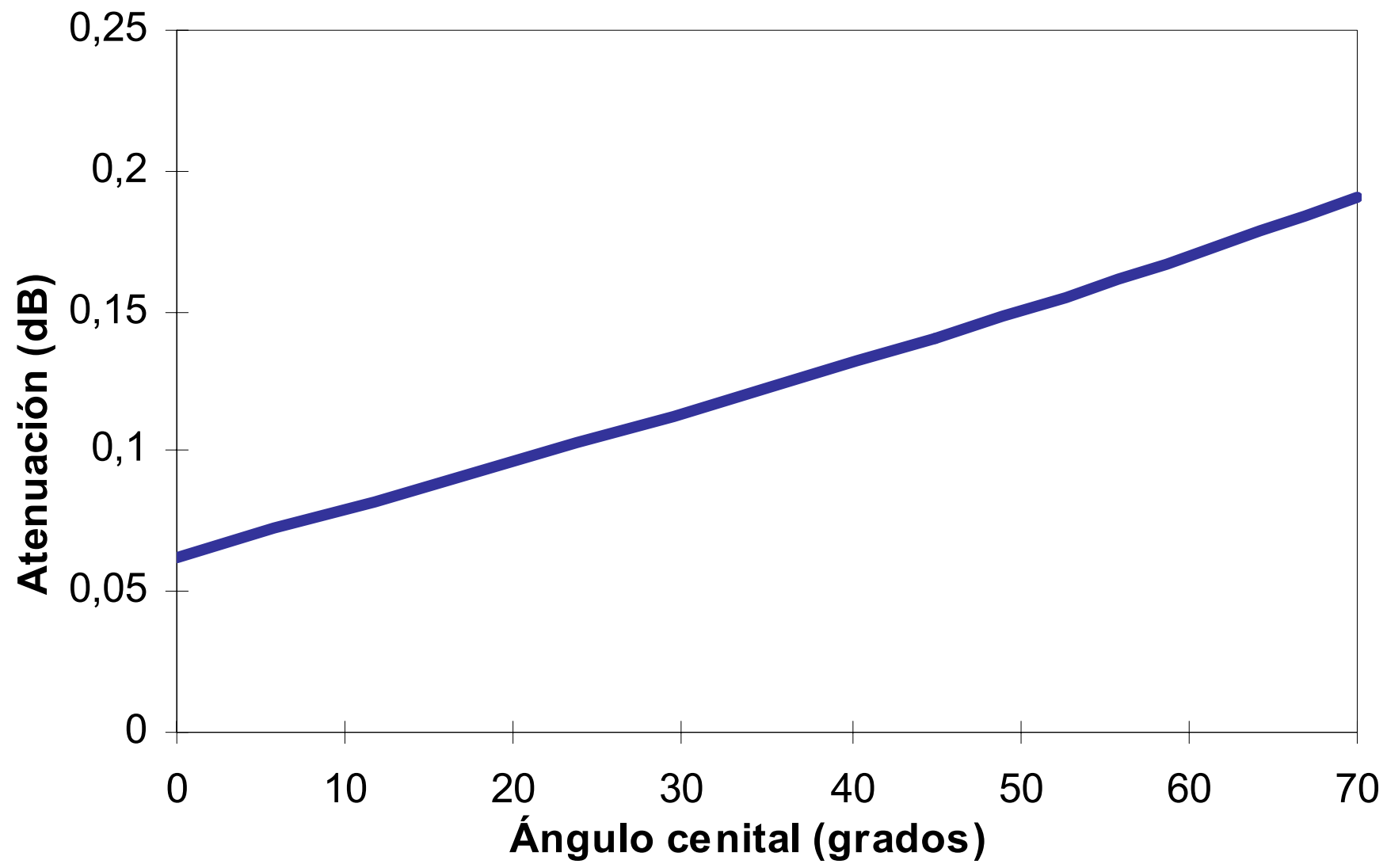

***Figura 6.27. Pérdidas por* scattering *Rayleigh a diferentes ángulos cenitales.***

Considerando que, para el peor de los casos, el ángulo cenital sea igual a 70 grados[1], las pérdidas debidas a *scattering* Rayleigh son de 0,191 dB. Este valor (caso peor) es el que se tomará para ser incluido en el balance de potencia, como ***atenuación*** atmosférica por moléculas de aire (*scattering* Rayleigh).

## 6.5.3. *Scattering* Mie

En la atmósfera no sólo están presentes moléculas de gas, sino también partículas en suspensión –polvo, gotículas de agua, hollín, cristales de hielo, etc.– a los que referiremos con el nombre genérico de ***aerosoles***. La concentración de masa de los aerosoles, en función de la altura, se puede determinar a través de la expresión (71), una vez fijado el modelo al que pertenece la ubicación receptora considerada.

El estudio de la atenuación correspondiente a los aerosoles se ha realizado para dos modelos extremos (Figura 6.28): ***modelo de desierto*** y ***modelo marítimo***. Estos dos modelos se corresponden, además, con las posibles situaciones que pueden darse en el observatorio considerado para la ubicación del telescopio receptor en el balance de potencia (para el cálculo de la ganancia). Este observatorio es el de **Roque de los Muchachos** en la isla de La Palma.

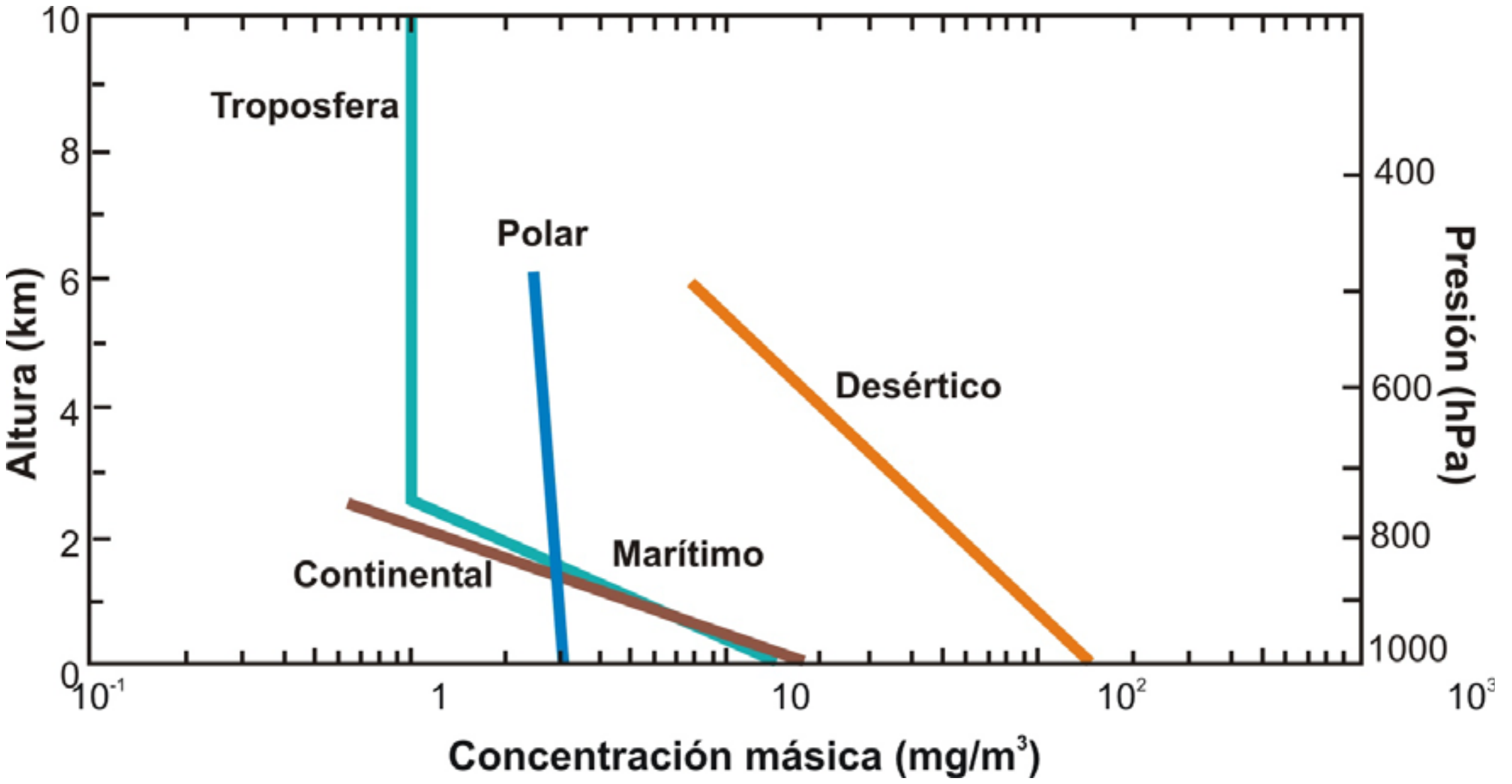

***Figura 6.28. Distribución de aerosoles en función de la altura para varios modelos de atmósfera (reproducción de la Figura 3.14).***

Los datos de partida para el cálculo incluyen la **concentración másica** del aerosol, variable según el modelo adoptado y la altura, y una **función de distribución** de tamaños, usualmente empírica y variable en función de las condiciones meteorológicas y de la altura. Esas dependencias hacen que el cálculo de la atenuación a través de la función de distribución sea extremadamente laborioso. En su lugar se ha optado por **acotar** la atenuación Mie de la forma siguiente: manteniendo constante la concentración másica, se ha calculado el *scattering* Mie para un conjunto de partículas del mismo tamaño. Se comienza con partículas de gran diámetro, y se van subdividiendo (el número de partículas se multiplica por $2^3 = 8$ cada vez que su diámetro se reduce a la mitad). Al principio la atenuación es muy baja, porque el número de partículas es reducido. A medida que se subdividen, el *scattering* va creciendo, alcanzando un máximo cuando el tamaño de partícula *es próximo a la longitud de onda*. Si se continúan subdividiendo las partículas, el *scattering* se reduce porque, aunque el número de partículas se hace muy grande, se abandona la zona Mie y se entra en zona Rayleigh, cuya eficiencia de *scattering* es sensiblemente menor. El **máximo de atenuación** se introduce en los cálculos del balance de potencia, garantizándose que *cualquier función de distribución con la misma concentración másica producirá una atenuación igual o inferior*.

---

[1] Considerando un mínimo de tres estaciones terrestres equiespaciadas, no se necesitan observaciones por encima de 60º de cenit. Se utiliza 70º como margen de seguridad.

En lo que respecta al procedimiento seguido para realizar dicho cálculo es idéntico para ambos modelos. A partir de la concentración de masa de aerosoles para el modelo en cuestión, se puede calcular el ***número de partículas*** por unidad de volumen $N_p$, para cada diámetro fijado de aerosol, a través de la ecuación (72). A continuación, nuevamente para cada radio a de partícula es posible determinar su ***sección geométrica*** ($\pi \cdot a^2$ suponiendo partículas esféricas) y la eficiencia de *scattering* Q mediante la curva representada en la Figura 6.29. Dicha figura relaciona la eficiencia Q con el parámetro $2\pi(a/\lambda)$ para una relación de índices de refracción de partícula y medio de 1,5 –es decir, para un índice de refracción de la partícula igual a 1,5, puesto que el medio tiene un índice muy próximo a 1. Ese valor es *razonable* para todos los modelos considerados (el agua tiene un índice 1,33, y ninguno de los materiales que intervienen en la composición del polvo supera el valor 1,6). En todo caso, la relación de índices actúa como **factor de escala** en la Figura 6.29, modificando el periodo de las oscilaciones, pero alterando mínimamente el valor del máximo.

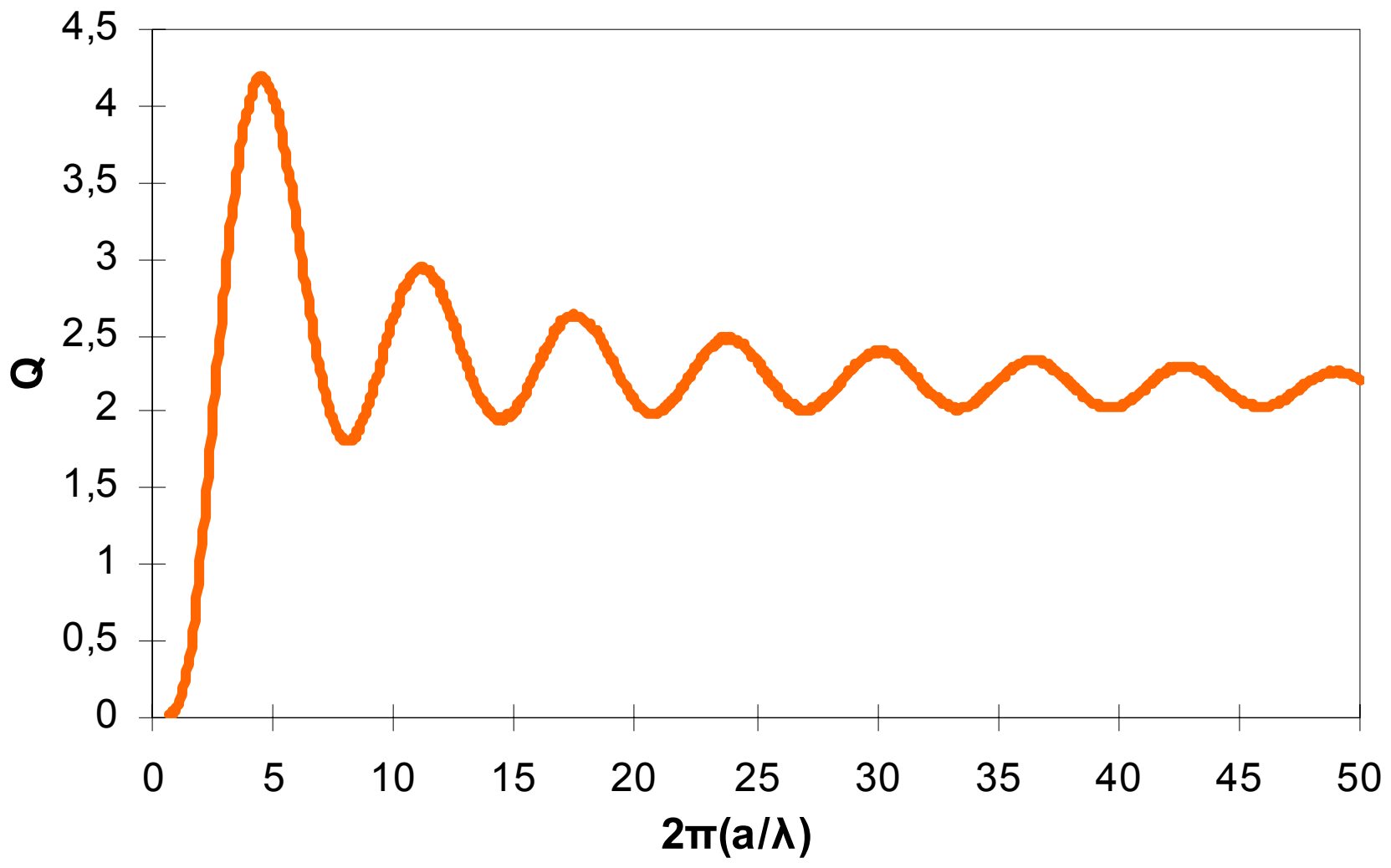

***Figura 6.29. Eficiencia de*** **scattering** ***para m = 1,5.***

Conocida la eficiencia de *scattering* es inmediato el cálculo de la sección eficaz $\sigma_M$, a través de la ecuación (74) y finalmente, conocida la concentración numérica, el coeficiente de atenuación se calcula mediante (73).

Por último, considerando el tramo de atmósfera que atraviesa la señal, se traduce el coeficiente de atenuación en un valor de pérdidas. Para acotar el tramo de atmósfera, se toma como límite inferior los dos kilómetros mencionados anteriormente y como límite superior de la atmósfera un valor promedio de la ***tropopausa*** (frontera de la estratosfera) de 14 km, pues éste se encuentra a unos 9 km en los polos y unos 18 km en el ecuador.

Recuérdese que los tres observatorios seleccionados tienen una latitud de unos 30º, por lo que es esperable que su tropopausa esté en el entorno del valor citado.

La influencia del ángulo cenital es la misma que en el apartado anterior y el cálculo de la distancia real h' que atraviesa la señal en la atmósfera es idéntico. Una vez conocido el coeficiente de atenuación $\alpha_M$, las pérdidas se calculan de la siguiente forma

$$L_M = \alpha_M \cdot h' \qquad (147)$$

### 6.5.3.a. Modelo marítimo

Siguiendo con el criterio de elección de considerar un rango de atmósfera atravesado por la señal transmitida de doce kilómetros (2 km-14 km), se puede apreciar en la Figura 6.28 que, por encima de dos kilómetros, la concentración de aerosoles es **constante con la altura** y adquiere un valor de 1 $\mu g/m^3$. Igualmente, se adopta una nueva suposición acerca de la ***densidad de partículas***. De esta forma, el cálculo de atenuación de este modelo se subdivide en otros dos estudios para acotar la posible densidad de las partículas. Se asumirán dos hipótesis: que todas las partículas son de ***sial*** ($\rho$ = 2,6 $kg/dm^3$) o que todas son de agua ($\rho$ = 1 $kg/dm^3$). El sial (término abreviado de **si**licatos **al**umínicos) [57] representa una composición media de la corteza terrestre, por lo que puede suponerse que el polvo atmosférico tiene una densidad parecida.

El procedimiento descrito anteriormente para el cálculo de la atenuación debida a aerosoles, en el caso de modelo marítimo, partículas de sial y de agua, ángulo cenital de 45 grados y concentración de partículas constante de 1 $\mu g/m^3$, queda resumido en la Tabla 6.5 para distintos diámetros de partícula.

| | | | | | *Sial* | | | *Agua* | | |
|---|---|---|---|---|---|---|---|---|---|---|
| **Ø (μm)** | **σ geom. ($m^2$)** | **2πa/λ** | **Efic. Q** | **σ ef. ($m^2$)** | **$N_p$ (part./$m^3$)** | **$L_M$ (Np)** | **$L_M$ (dB)** | **$N_p$ (part./$m^3$)** | **$L_M$ (Np)** | **$L_M$ (dB)** |
| 0,1 | 7,85E-15 | 0,3 | 0,002 | 1,49E-17 | 3,85E8 | 0,0001 | 0,0008 | 1,00E9 | 0,0002 | 0,0021 |
| 0,3 | 7,07E-14 | 0,89 | 0,13 | 9,19E-15 | 1,42E7 | 0,0022 | 0,0187 | 3,70E7 | 0,0056 | 0,0487 |
| 0,7 | 3,85E-13 | 2,07 | 1,88 | 7,25E-13 | 1,12E6 | 0,0134 | 0,1162 | 2,92E6 | 0,0348 | 0,3021 |
| 0,9 | 6,36E-13 | 2,66 | 2,96 | 1,86E-12 | 5,27E5 | 0,0162 | 0,1404 | 1,37E6 | 0,0420 | 0,3651 |
| 1,1 | 9,50E-13 | 3,25 | 3,59 | 3,41E-12 | 2,89E5 | 0,0162 | 0,1411 | 7,51E5 | 0,0422 | 0,3668 |
| 1,3 | 1,33E-12 | 3,84 | 4,10 | 5,44E-12 | 1,75E5 | 0,0157 | 0,1362 | 4,55E5 | 0,0408 | 0,3542 |
| 1,5 | 1,77E-12 | 4,43 | 4,30 | 7,60E-12 | 1,14E5 | 0,0143 | 0,1238 | 2,96E5 | 0,0371 | 0,3220 |
| 1,7 | 2,27E-12 | 5,02 | 3,93 | 8,92E-12 | 7,83E4 | 0,0115 | 0,0999 | 2,04E5 | 0,0299 | 0,2596 |
| 2 | 3,14E-12 | 5,91 | 3,11 | 9,78E-12 | 4,81E4 | 0,0077 | 0,0672 | 1,25E5 | 0,0201 | 0,1747 |
| 3 | 7,07E-12 | 8,86 | 2,36 | 1,67E-11 | 1,42E4 | 0,0039 | 0,0340 | 3,70E4 | 0,0102 | 0,0883 |
| 5 | 1,96E-11 | 14,8 | 2,10 | 4,12E-11 | 3,08E3 | 0,0021 | 0,0181 | 8,00E3 | 0,0054 | 0,0472 |
| 7 | 3,85E-11 | 20,7 | 2,08 | 8,00E-11 | 1,12E3 | 0,0015 | 0,0128 | 2,92E3 | 0,0038 | 0,0333 |
| 10 | 7,85E-11 | 29,5 | 2,48 | 1,95E-10 | 3,85E2 | 0,0012 | 0,0107 | 1,00E3 | 0,0032 | 0,0278 |

***Tabla 6.5. Modelo marítimo de* scattering**
***Mie para partículas de sial y agua.***

Repitiendo el cálculo recogido en las tablas anteriores para valores de ángulo cenital de 0 grados y 70 grados, es posible representar la atenuación atmosférica en función de distintos diámetros de partículas para diferentes ángulos. El resultado queda reflejado en la Figura 6.30 y Figura 6.31 correspondientes a partículas de sial y de $H_2O$ respectivamente.

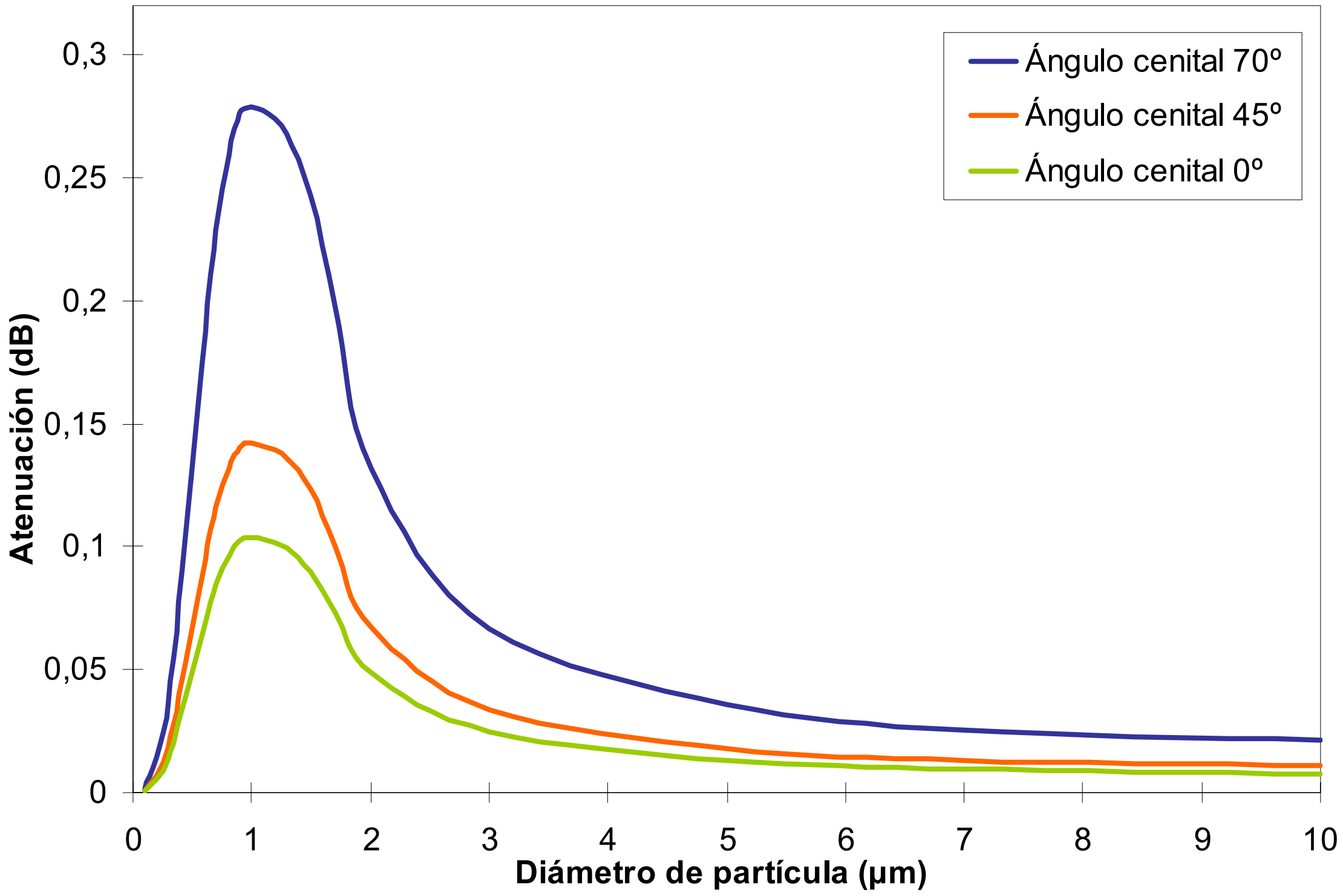

***Figura 6.30. Atenuación en modelo marítimo para distintos tamaños de partículas de sial y distintos ángulos cenitales.***

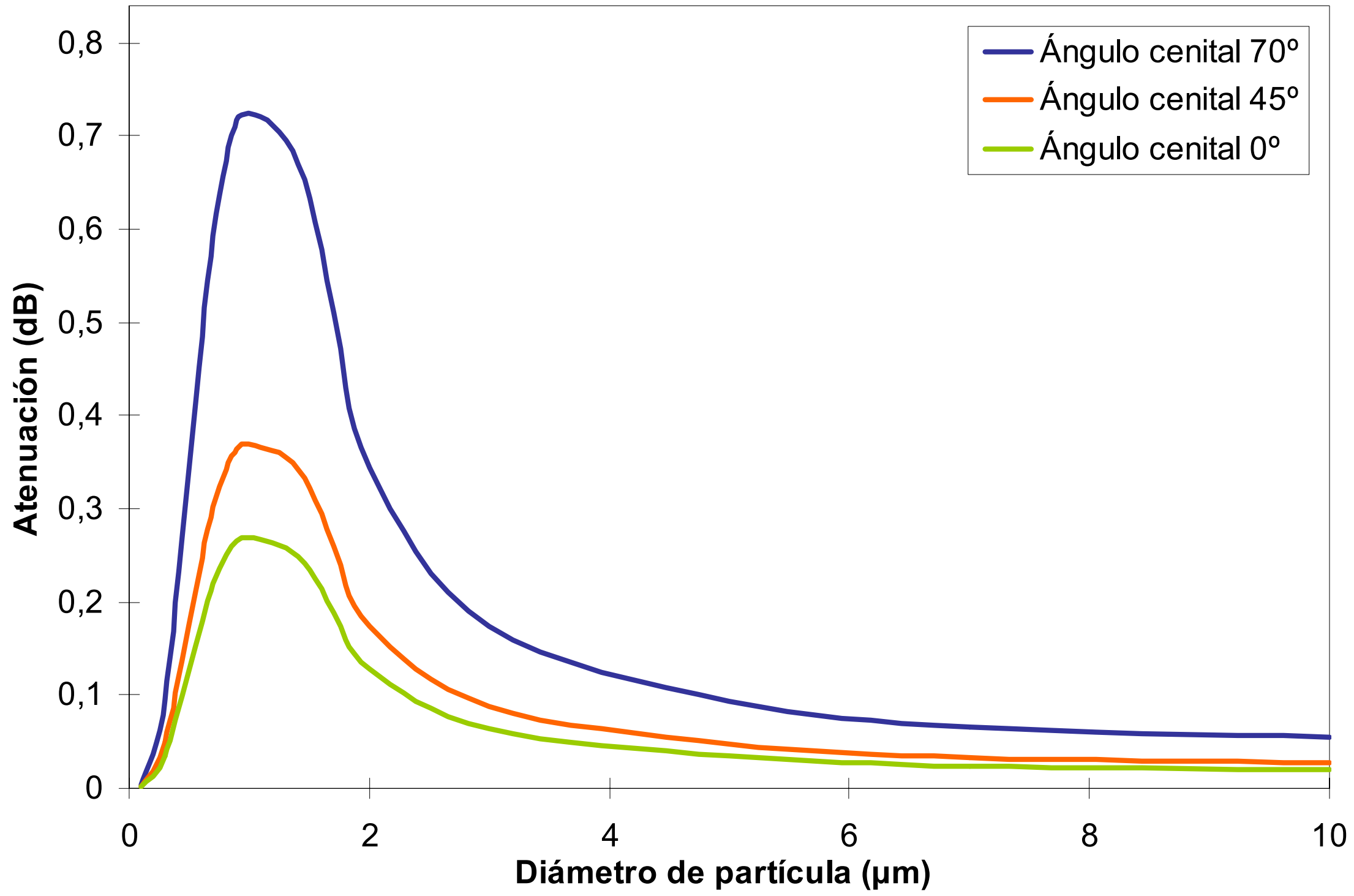

***Figura 6.31. Atenuación en modelo marítimo para distintos tamaños de partículas de $H_2O$ y distintos ángulos cenitales.***

De estas dos figuras se desprende como conclusión que se puede establecer una **cota superior** para la atenuación atmosférica de **0,30 dB** para partículas de sial de diámetro aproximadamente 1,1 µm y ángulo cenital de 70 grados, y una cota superior bajo las mismas condiciones, para partículas de agua, de **0,78 dB**.

### 6.5.3.b. Modelo de desierto

En el caso de que la ubicación receptora pertenezca a un modelo de desierto, en el rango de atmósfera considerado (2 km-14 km), la concentración másica de aerosoles no es constante en todo el tramo (Figura 6.32), como ocurría en el modelo anterior (véase Figura 6.28).

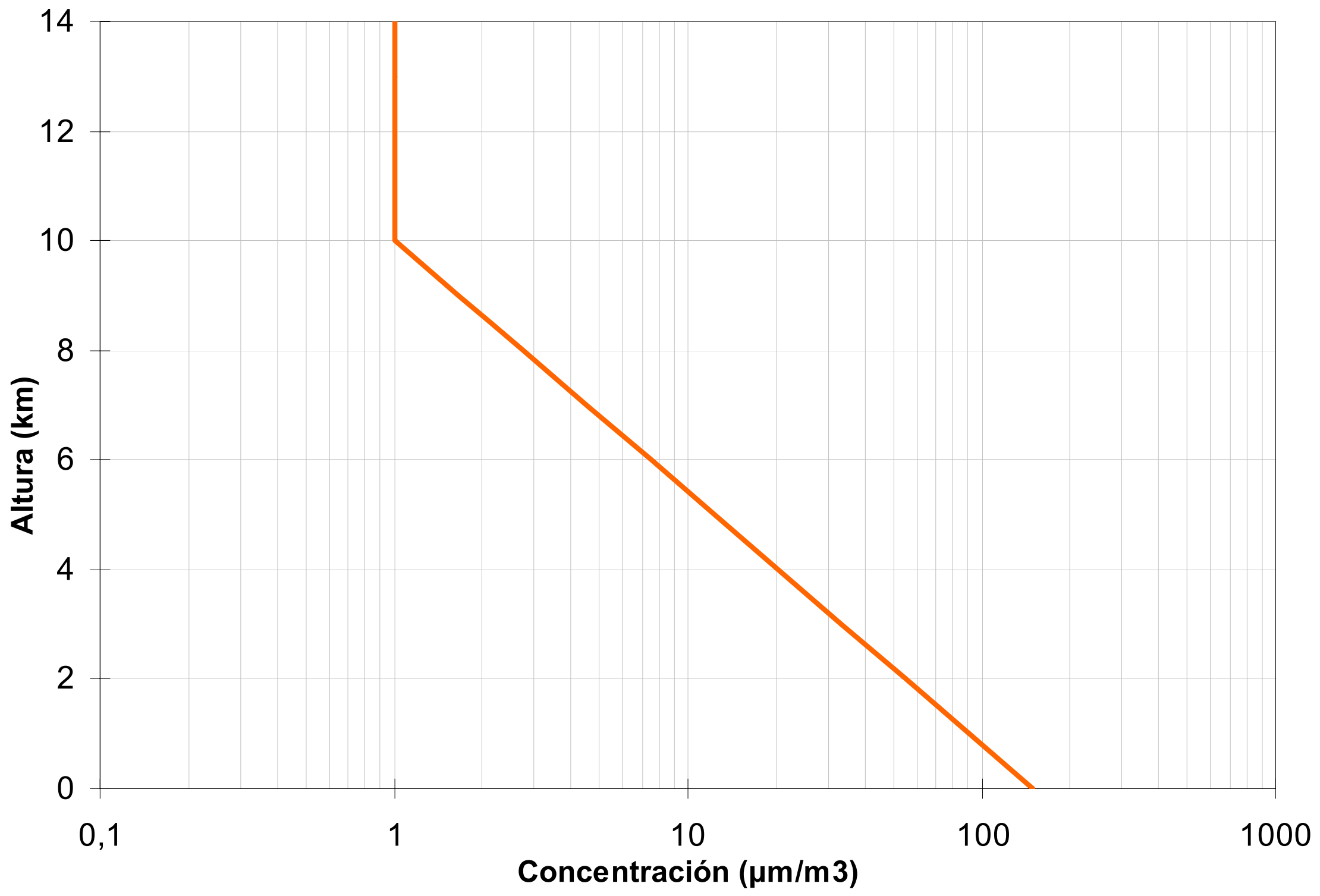

***Figura 6.32. Concentración másica de aerosoles en función de la altura para un modelo de desierto.***

Debido a la dependencia de la altura en el tramo de 2 a 10 km de la concentración de aerosoles, se consideró un conjunto de intervalos de 1 km cada uno en los que poder tomar la concentración como constante. Teniendo en cuenta estos intervalos, el procedimiento de cálculo de la atenuación atmosférica por aerosoles es análogo al modelo anterior, con la salvedad de que para determinar la atenuación total para cada tamaño de partícula fijado, habrá que sumar la atenuación correspondiente a cada escalón, es decir

$$L_M = \sum_{i=1}^{11} Q \cdot N_{p_i} \cdot \sigma_{geom} \cdot h' \qquad (148)$$

siendo i es el número de intervalo y h' el tamaño del intervalo, función del ángulo cenital según se vio en (146). Puesto que para cada tamaño de partícula la eficiencia de *scattering* Q y la sección geométrica $\sigma_{geom}$ es constante, la ecuación anterior se puede escribir como

$$L_M = Q \cdot \sigma_{geom} \sum_{i=1}^{11} N_{p_i} h' \qquad (149)$$

Bajo la suposición de que todas las partículas son de sial, el resultado del cálculo de la atenuación atmosférica debida a aerosoles en función del tamaño de la partícula y el ángulo cenital, se representa en la Figura 6.33.

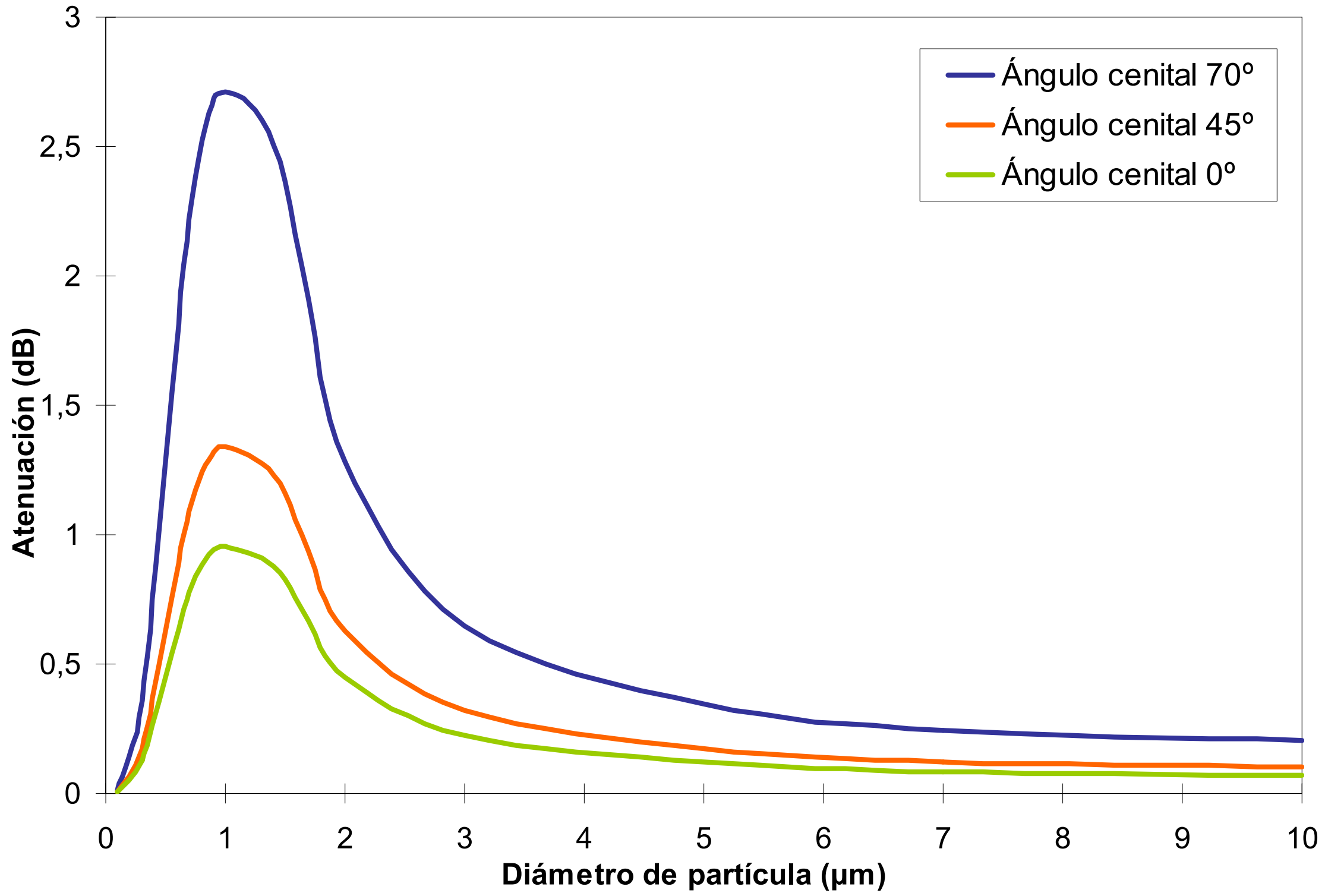

***Figura 6.33. Atenuación en modelo desértico para distintos tamaños de partículas de sial y distintos ángulos cenitales.***

Como se puede deducir de la Figura 6.33, el valor máximo de atenuación atmosférica se corresponde con un ángulo cenital de 70 grados y partículas de un diámetro de aproximadamente 1,1 µm. Se puede así establecer una cota superior para la atenuación de un modelo desértico de 2,76 dB.

## 6.5.4. Turbulencia

Tal como se comentó en el apartado 3.3.4, el cálculo del efecto de la turbulencia es el más complejo entre todas las contribuciones atmosféricas a la atenuación. Para poder incluir este término dentro del balance de potencia se ha aplicado una estrategia diferente.

Como se recordará, el efecto de la turbulencia es provocar fluctuaciones del haz incidente, haciendo que, en determinadas ocasiones, puede salirse parcial o totalmente del área activa del fotodetector, provocando variaciones de potencia.

Se conoce también [45] que tales fluctuaciones se reducen, en condiciones habituales, a 1” de arco (unos 5 µrad). El procedimiento seguido para garantizar que el haz queda siempre situado sobre el fotodetector consiste en seleccionar un campo de visión (FOV) que abarque el planeta, la órbita del satélite y un margen de seguridad, que se establece en 1.000 km por encima de la órbita en consonancia con el dato anterior. Dado que el diámetro de la órbita del *MTO* es de algo menos de 16.000 km, se ha ajustado el FOV de modo que, en cualquier posición relativa Marte-Tierra, la apertura angular abarque 18.000 km. Esta cifra corresponde a un FOV con un semiángulo de 30 μrad cuando el planeta está a $3 \cdot 10^8$ km, y 90 μrad en su situación más próxima durante la misión, a $10^8$ km. Estos FOV son relativamente pequeños en comparación con los habituales en astronomía (algunos minutos de arco), aunque perfectamente factibles, a condición de emplear un sistema de **autoguiado** (un bucle de realimentación) que fije el planeta en el centro del diafragma.

## 6.6. TERMINAL LÁSER DEL MTO

El terminal remoto del *MTO* estará formado por los siguientes elementos [58]:

- El ***transmisor*** propiamente dicho (MOPA, *master oscillator power amplifier*), formado por un oscilador láser, modulador y amplificador.
- El ***sistema de apuntamiento***, dotado de una unidad miniaturizada de referencia inercial y un conjunto de detectores y actuadores.
- La ***antena***, destinada a focalizar el haz sobre el terminal terrestre.

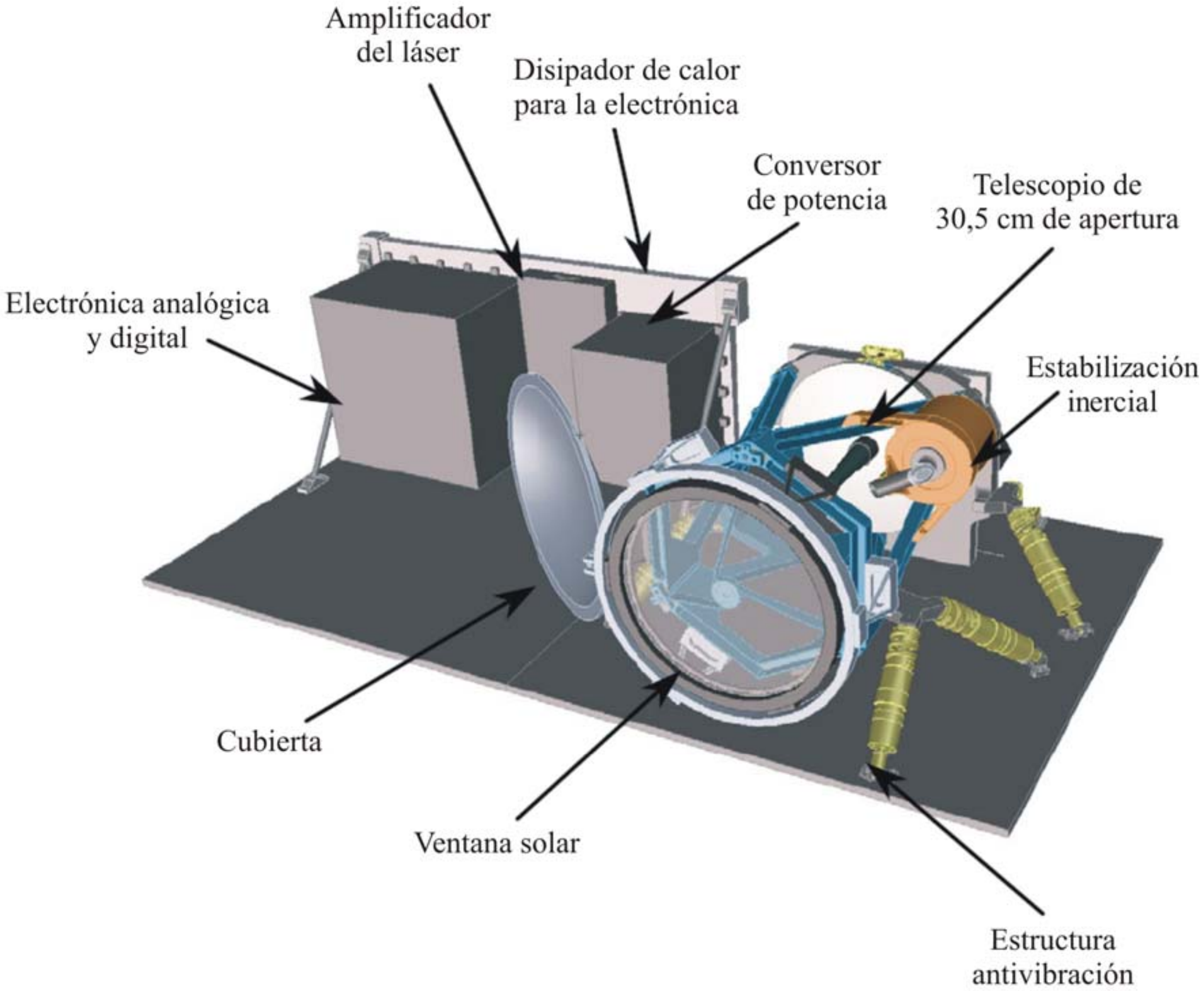

***Figura 6.34. Configuración del terminal de comunicaciones por láser del MTO [58].***

## 6.6.1. El transmisor

El equipo transmisor del *MTO*, llamado *MOPA*, está cuidadosamente diseñado para optimizar la potencia, consiguiendo pulsos de centenares de W con una potencia media de 5 W. A la vez debe trabajar con anchos de banda muy elevados, por encima de 1 GHz, para las que resulta difícil, si no imposible, aplicar técnicas de generación de pulsos intracavidad como el *Q-switch* mencionado en la sección 4.2.2.

En estas circunstancias, y para conseguir un funcionamiento más estable del láser, es aconsejable emplear ***modulación externa***, es decir, hacer que el láser trabaje en **continua** e introducir la codificación binaria en una unidad independiente a la salida.

El problema es que la modulación externa, en principio, es ***poco eficiente*** en términos de potencia, ya que se está "troceando" la salida del láser, redirigiéndola a una de dos salidas posibles. Con ello, una gran parte de la emisión se desperdicia, y los pulsos emitidos *tienen una potencia equivalente a la potencia media en continua*. La solución adoptada es generar un haz de **baja potencia** con un oscilador láser, **modular** la salida del mismo, y **amplificar** los pulsos ya modulados.

Las unidades elegidas –provisionalmente, la selección final no está hecha aún–, sigue la conservadora estrategia de emplear en el espacio elementos conocidos y probados. Así son todas ellas tecnologías cuyo funcionamiento está ampliamente respaldado por su uso en otras áreas (generalmente comunicaciones ópticas por fibra). El láser seleccionado como oscilador maestro es un láser de **realimentación distribuida** (DFB, *distributed feedback*) comercial, montado en una **fibra Bragg dopada** con yterbio. El oscilador va bombeado ópticamente por un láser de semiconductor a 980 nm. La salida del láser está a 1064 nm.

Los láseres DFB han sido desarrollados profusamente para CC.OO., al igual que las fibras Bragg y el dopaje del núcleo de la fibra con tierras raras (los amplificadores EDFA, de uso frecuente en CC.OO., están basados en fibras dopadas con erbio). En este producto comercial se aúnan estas tecnologías: se monta el láser en la propia fibra dopada, y se aprovecha la selección de longitud de onda que provee la difracción Bragg.

El modulador está también basado en una tecnología ampliamente utilizada en CC.OO. Se trata de un ***modulador Mach-Zehnder*** (MZ) de niobato de litio ($LiNbO_3$). Estos moduladores presentan bajas pérdidas de inserción y su ancho de banda a 3 dB está por encima de 10 GHz, con lo que puede dar cabida a los anchos de banda exigidos por la modulación PPM prevista.

El único inconveniente de este modulador es que su *relación de extinción* (que en último término determina la diferencia entre bits 1 y 0 en el canal) es relativamente baja, alrededor de 20 dB. Para evitar este problema, se propone unir dos moduladores MZ en **cascada**, consiguiendo una relación de extinción conjunta de 40 dB.

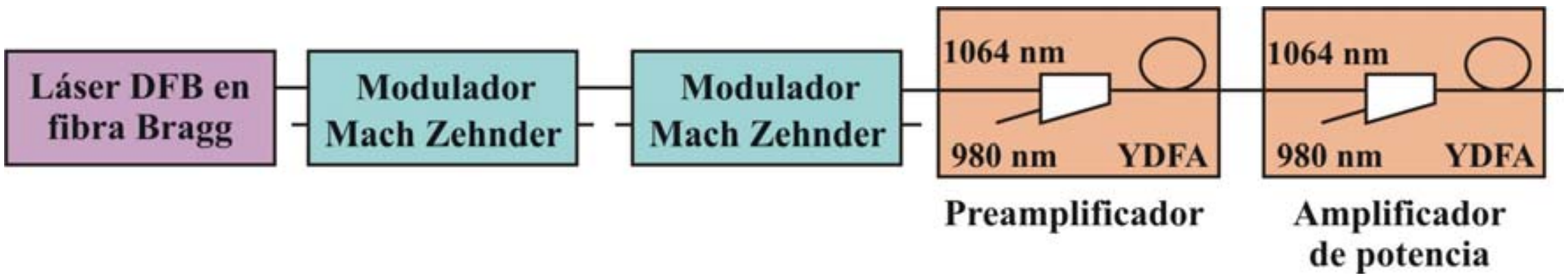

*Figura 6.35. Diagrama de bloques del MOPA del MTO*

El amplificador es también de fibra dopada, un ***YDFA*** (*ytterbium-doped fiber amplifier*), enteramente similar a los amplificadores EDFA de CC.OO., aunque sintonizado a 1064 nm. Al igual que sucedía con la etapa moduladora, en el esquema inicial se ha propuesto **duplicar** esta etapa, empleando un preamplificador que ataca un amplificador (ambos YDFA). De esta manera se consigue saturar el segundo amplificador, extrayendo eficientemente los 5 W de potencia media.

Es importante advertir que tanto el oscilador como los dos amplificadores están realizados con el **mismo material activo**, y se bombean ópticamente con diodos láser en la **misma banda** de 980 nm. Aunque la propuesta de diseño no lo comenta explícitamente, es de suponer que se estudien configuraciones de ***bombeo compartido***, empleando el mismo diodo para excitar los tres componentes, o bien usando **varios diodos** de modo que el bombeo pudiera hacerse indistintamente desde uno u otro, añadiendo redundancia a un elemento crítico como es el diodo láser de bombeo.

## 6.6.2. Telescopio emisor del MTO

Las especificaciones [53] para el proyecto *MLCD* establecen un espejo primario de 30,5 centímetros. Manteniendo este tamaño máximo de espejo primario es posible tener distintas posibilidades de diseño en cuanto al haz que emite el telescopio. A continuación se estudian dichas posibilidades.

### 6.6.2.a. Diseño del factor de truncamiento

Como ya se adelantó en el apartado 4.4.4, el factor de truncamiento determina la relación entre el radio del espejo primario y la anchura del haz gaussiano acoplado al telescopio cuando la potencia cae $1/e^2$. Este factor puede ajustarse cambiando la relación entre los diámetros y las distancias focales en el espejo primario o secundario.

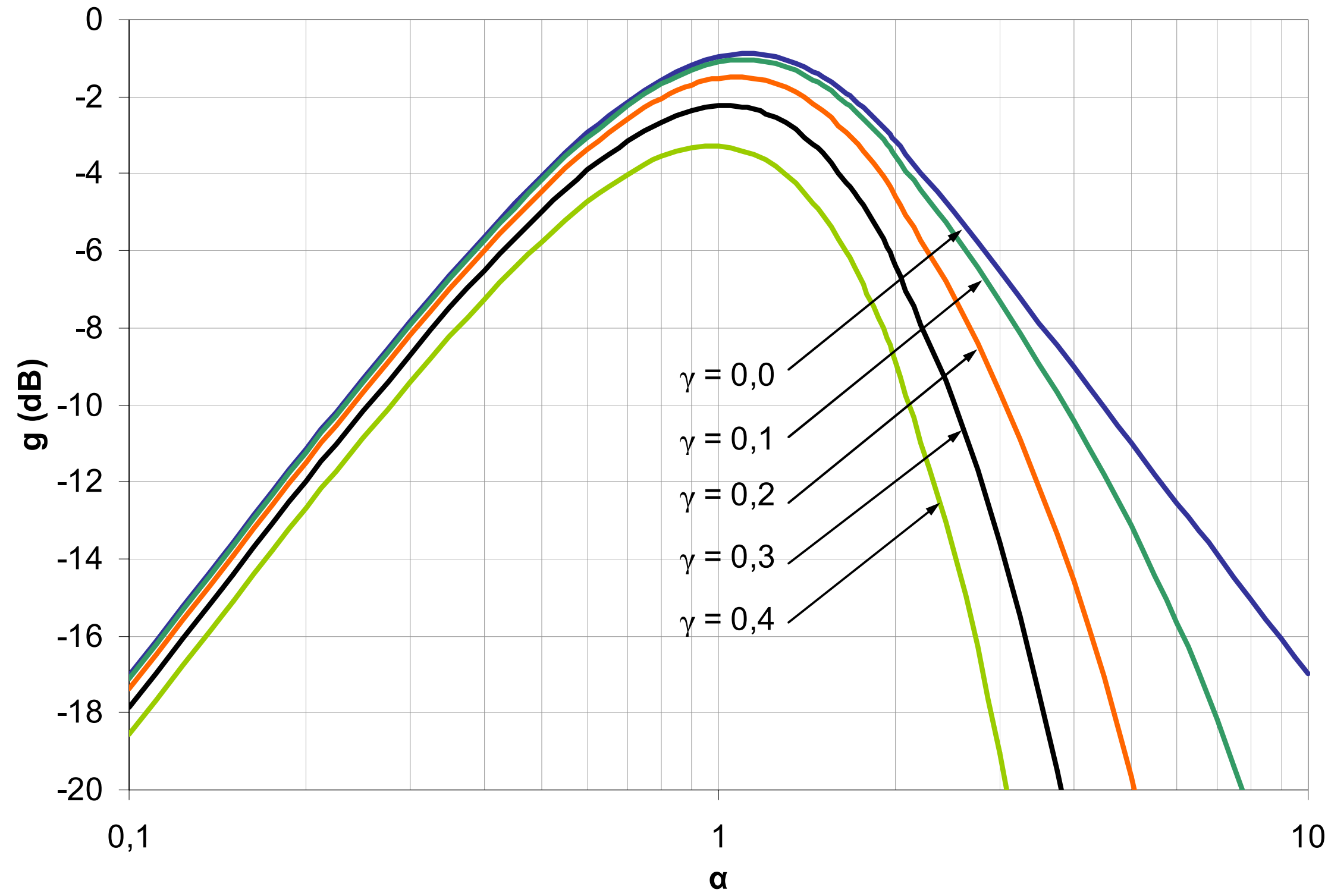

***Figura 6.36. Parámetro g de eficiencia de la ganancia para distintos valores del factor de bloqueo $\gamma$ y del factor de truncamiento $\alpha$.***

En la Figura 6.36 se muestra una representación gráfica calculada a partir de la ecuación (94). El parámetro de corrección g de la ganancia se ha expresado, como es habitual en las ganancias, en unidades logarítmicas, por lo que habría que sumárselo a la ganancia ideal (también previa conversión a dB).

Se puede observar en la Figura 6.36 que para un factor de bloqueo γ determinado (suponiendo fijo el radio del espejo primario y secundario) existe un factor de truncamiento α óptimo que hace que la ganancia real G sea máxima. De ello se deduce que el telescopio se diseñará para que $\alpha = \alpha_{óptimo}$.

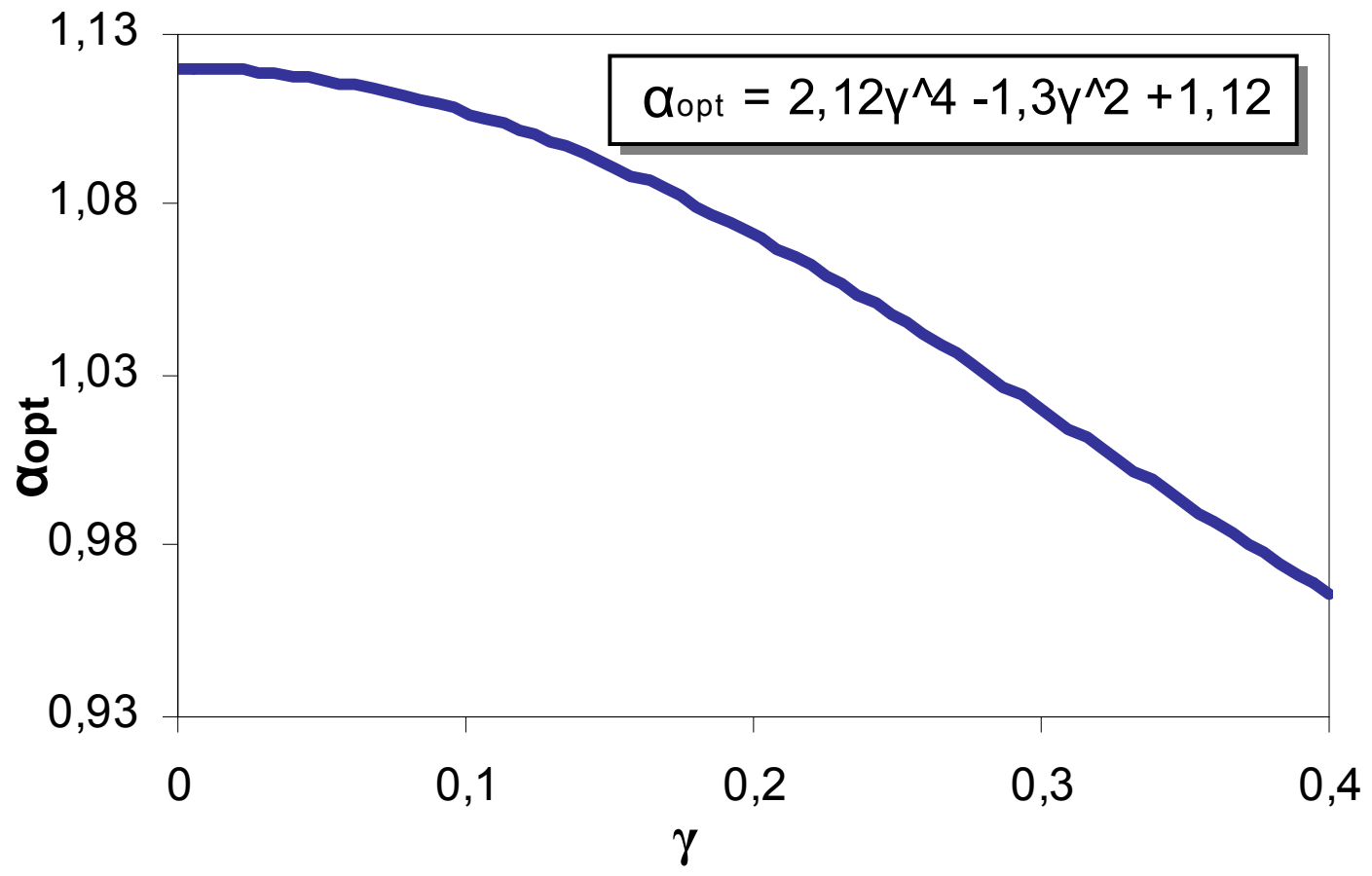

***Figura 6.37. Relación entre el factor de bloqueo γ y el factor de truncamiento óptimo $\alpha_{opt}$.***

La relación entre el factor óptimo de truncamiento y el factor de bloqueo, que servirá para conocer el valor de $\alpha_{óptimo}$ para cada factor de bloqueo, se puede obtener tomando los valores máximos de α de la Figura 6.36 para los distintos valores de γ. En la Figura 6.37 se presenta el resultado de la relación entre $\alpha_{óptimo}$ y γ obtenido mediante una interpolación polinomial bicuadrática de los valores de $\alpha_{óptimo}$ en función del factor de bloqueo γ.

### 6.6.2.b. Diseño del factor de bloqueo

El factor de bloqueo sirve para estimar la cantidad de la potencia del láser que no se emite a través de la apertura del telescopio debido al bloqueo del espejo secundario. Este factor se puede ajustar modificando el tamaño del espejo secundario del telescopio.

Si se toma de nuevo la ecuación (93) aproximando otra vez β a cero (campo lejano) y se representa la ganancia para una eficiencia g en función del ángulo θ se puede obtener el diagrama de radiación del telescopio, esto es, la variación de la ganancia con el ángulo de emisión

$$g = 2\alpha^2 \left[ \int_{\gamma^2}^{a} \exp\left(-\alpha^2 u\right) J_0 \left[ \frac{2\pi a}{\lambda} \operatorname{sen}(\theta) \sqrt{u} \right] du \right]^2 \qquad (150)$$

En la Figura 6.38 se presenta el cálculo realizado para la variación de la eficiencia g con el parámetro θ·(D/λ) (donde D representa el diámetro del espejo primario, por lo que D = 2a) para distintos valores del factor de bloqueo y tomando siempre un factor óptimo de truncamiento.

Para realizar una representación suficientemente precisa se utilizaron más de 200 pares de valores para cada una de las ganancias representadas en la gráfica. Dado que para la obtención de los valores de la eficiencia g era necesario realizar la integral del producto de una exponencial y una función de Bessel, se empleó un rutina implementada en el programa de cálculo matemático *Maplesoft Maple 8.0*, que proporcionaba el resultado de la integral para todos los valores de θ·(D/λ).

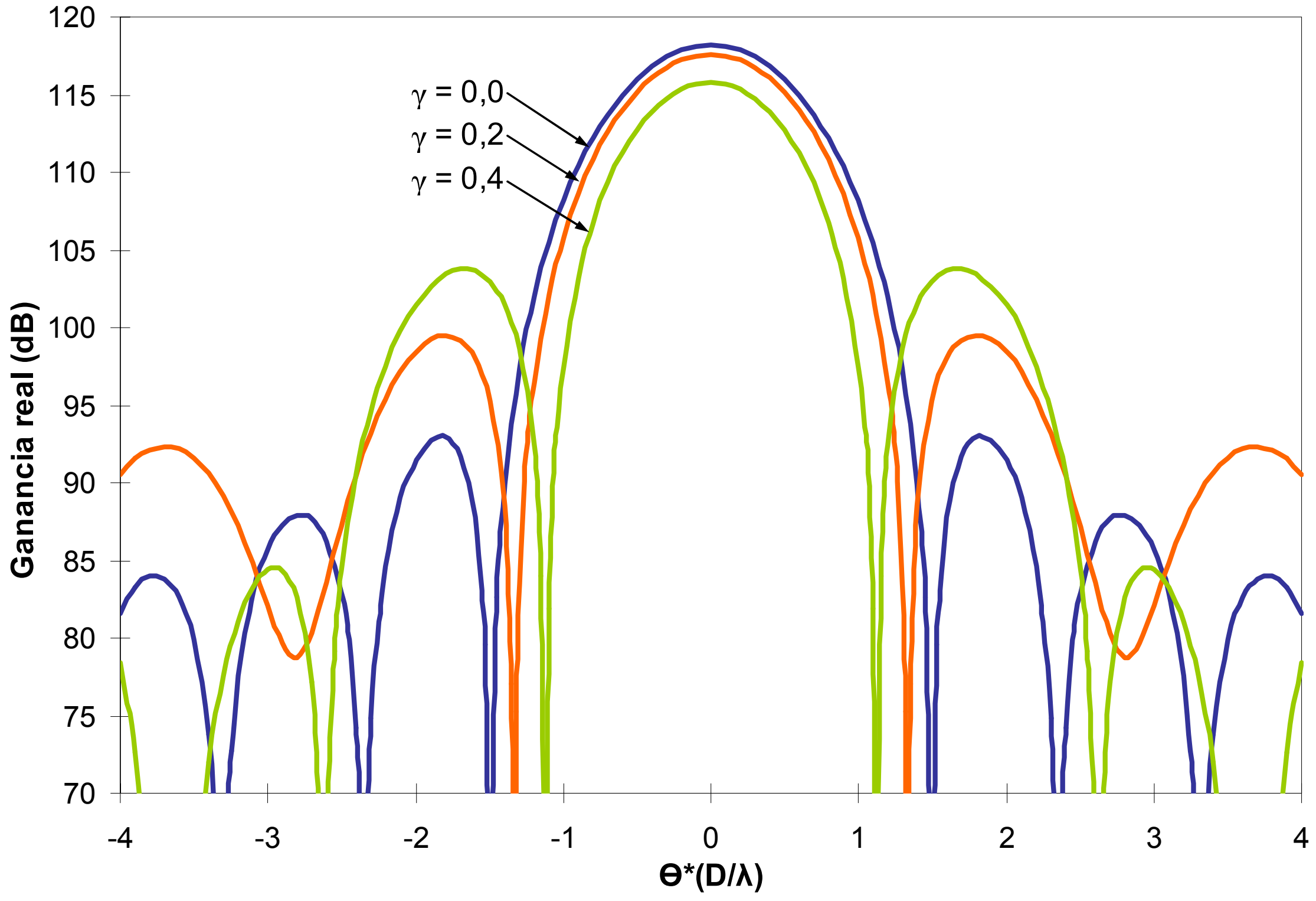

***Figura 6.38. Variación de la ganancia del telescopio con el parámetro θ·(D/λ) para distintos factores del bloqueo y un factor óptimo de truncamiento.***

Recuérdese que, tal como se vio en el apartado 4.4.1 este parámetro θ·(D/λ) era el que variaba según el criterio que se utilizara para estimar dónde se encontraba la mayor parte de la potencia (cuando la potencia cae tres dB, cuando cae 1/e, la potencia entre los primeros nulos, etc.).

Se comprueba en la Figura 6.38 que, para un factor de truncamiento constante y mínimo ($\alpha = \alpha_{opt}$), al aumentar el factor de bloqueo (para un tamaño fijo del espejo primario, aumentar el factor de bloqueo significa aumentar el tamaño del espejo secundario), disminuye la ganancia del telescopio.

Sin embargo, también puede observarse cómo al aumentar el factor de bloqueo el haz se hace más estrecho, por lo que surgen dos posibilidades de diseño: utilizar el mínimo factor de bloqueo posible consiguiendo así una ganancia mayor en la dirección de emisión o utilizar un factor mayor de bloqueo consiguiendo un haz más estrecho, lo que se traduce en que el efecto de la divergencia con la distancia sea menor. A continuación se presenta un cuadro en el que se evalúan estas dos posibilidades empleando valores del proyecto *MLCD*.

**EVALUACIÓN DE LAS POSIBILIDADES DE DISEÑO DEL FACTOR DE BLOQUEO**

Para evaluar ambas alternativas de diseño se considera un factor de bloqueo de $\gamma = 0{,}0$ y otro de $\gamma = 0{,}4$. La comparación se ha realizado de la siguiente manera: para cada factor de bloqueo (asumido un valor óptimo para el factor de truncamiento), se calcula la eficiencia de ganancia que le corresponde. Esta eficiencia (en unidades naturales) multiplica a una potencia de 320 W (resultado de utilizar una modulación 64-PPM con 5 W de potencia media que tiene el láser del *MTO*).

El efecto de la distancia (para la que se toma un valor medio entre la Tierra y Marte de 300 millones de kilómetros) se evalúa empleando el parámetro $\theta\cdot(D/\lambda)$ que corresponda en cada caso (será menor cuanto mayor sea el factor de bloqueo y también menor la divergencia). Midiendo en la Figura 6.38, se obtiene, para cada $\gamma$, un valor de $\theta_{3dB}\cdot(D/\lambda)$ (con el que, conocidos el diámetro de 30,5 cm del espejo primario del telescopio del *MTO* y la longitud de onda de 1,064 µm, se puede calcular el ángulo $\theta_{3dB}$). El área A del *spot* que llega a la Tierra se calcula tal como se vio en el apartado 3.1.1 (mediante el ángulo de difracción $\theta_{3dB}$ a la salida del telescopio y la distancia Marte-Tierra). Por último, la densidad de potencia S se obtiene como el cociente de la potencia previamente calculada y el área del *spot*.

Con los resultados de la Tabla 6.6, se comprueba que pese a que el área del *spot* que se recibe en la Tierra es menor para el caso de mayor factor de bloqueo, también la densidad de potencia es menor, por lo que se deduce que interesa utilizar factores de bloqueo lo más pequeños posibles para así lograr una máxima ganancia en la potencia transmitida, lo que resulta, pese a la mayor divergencia del haz, en una mayor densidad de potencia recibida. Además, si se tiene mayor densidad de potencia recibida y además una mayor área del *spot*, el sistema de apuntamiento no requiere tanta precisión como en el otro caso.

| $\gamma$ | g (veces) | g·P (W) | $\theta_{3dB}\cdot(D/\lambda)$ (rad) | $\theta_{3dB}$ (µrad) | A ($m^2$) | S ($pW/m^2$) |
|---|---|---|---|---|---|---|
| 0,0 | 0,81 | 259,2 | 0,58 | 4,05 | $1{,}16\cdot10^{12}$ | 223,45 |
| 0,4 | 0,47 | 150,4 | 0,5 | 3,49 | $0{,}86\cdot10^{12}$ | 174,88 |

***Tabla 6.6. Densidad de potencia recibida en función del factor de bloqueo.***

### 6.6.2.c. Ganancia real y pérdidas por apuntamiento

En la Tabla 6.7 se presentan los distintos valores que tomaría la ganancia del telescopio incorporado en el terminal láser del satélite *MTO*. Para la eficiencia g se utiliza un factor óptimo de truncamiento y una longitud de onda de 1,064 µm y para el cálculo de la ganancia ideal se toma el radio de espejo primario de 30,5 centímetros.

| Factor de bloqueo $\gamma$ | Factor óptimo de truncamiento $\alpha_{opt}$ | Eficiencia de ganancia g | Ganancia ideal $G_{ideal}$ (dB) | Ganancia real G (dB) |
|---|---|---|---|---|
| 0 | 1,1200 | -0,89 | 119,09 | 118,20 |
| 0,1 | 1,1072 | -1,04 | 119,09 | 118,05 |
| 0,2 | 1,0714 | -1,49 | 119,09 | 117,60 |
| 0,3 | 1,0202 | -2,24 | 119,09 | 116,85 |
| 0,4 | 0,9663 | -3,28 | 119,09 | 115,81 |

***Tabla 6.7. Ganancia del telescopio del MTO para distintos factores de bloqueo asumiendo factor óptimo de truncamiento y 30,5 cm para el espejo primario.***

Como se ha deducido, el factor de truncamiento que se empleará será el óptimo y el factor de bloqueo será el menor posible. De los valores estudiados se utilizará el de 0,1 (el menor de ellos). Por lo tanto, la ganancia real del telescopio transmisor que se utilizará en el balance de potencia será **118,05 dB**.

Generalmente el valor de ganancia de antena que se emplea en los cálculos de los balances de potencia corresponde a la ganancia máxima, es decir, en la dirección del eje paralelo a la emisión. En un enlace de comunicaciones ópticas en el que el haz es tan estrecho y además el terminal transmisor se halla en un satélite con la consiguiente vibración, el apuntamiento se vuelve una dificultad añadida. Por ello, más realista que emplear únicamente el valor máximo de la ganancia, es emplear un factor de pérdidas por desapuntamiento que corrija este valor máximo de la ganancia para aproximarlo al que se obtendría en un experimento real.

Conociendo la variación de la ganancia del telescopio con el ángulo de emisión, es posible calcular un parámetro que evalúe las pérdidas debidas a errores en el apuntamiento del haz. Este parámetro se calcula midiendo en la Figura 6.38 la ganancia en cada dirección de emisión (para lo que es necesario particularizar el diámetro del espejo primario del transmisor y la longitud de onda para obtener el ángulo $\theta_{3dB}\cdot(D/\lambda)$ a partir del $\theta_{3dB}\cdot(D/\lambda)$) y realizando la diferencia entre esta ganancia y la máxima.

En la Figura 6.39 se presenta el resultado del cálculo que se acaba de describir para distintos valores del factor de bloqueo y asumiendo de nuevo el valor óptimo del factor de truncamiento.

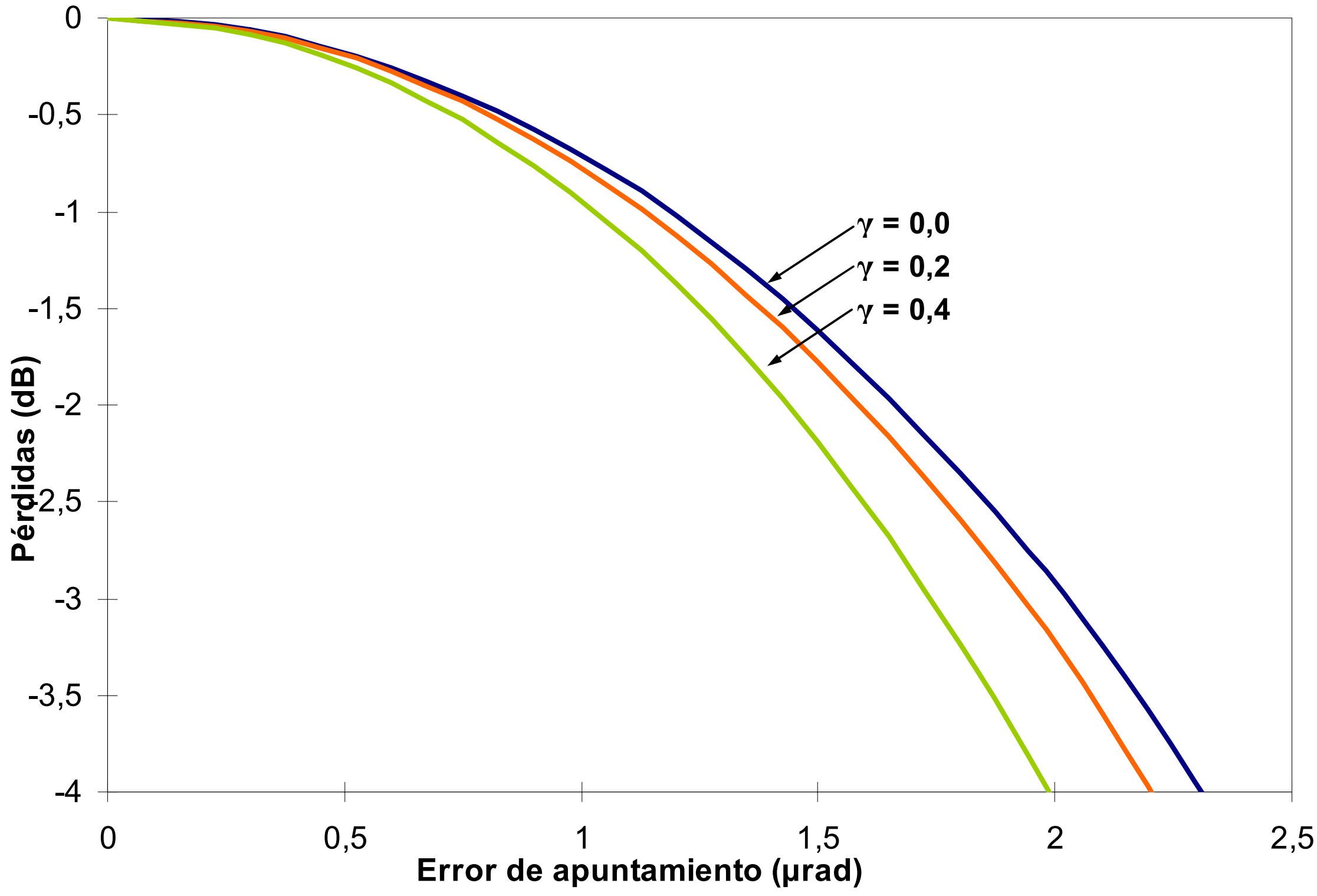

***Figura 6.39. Pérdidas por errores en el apuntamiento.***

Según un cálculo obtenido de [41] con una función densidad de probabilidad de Rice para evaluar qué porcentaje corresponde a qué valores de errores de apuntamiento en el terminal de comunicaciones del *MTO*, se concluye que el 99,88% del tiempo los errores de apuntamiento serán menores de 1,54 μrad. Este resultado se muestra en la Figura 6.40. Si se mide este error máximo en la Figura 6.39, para un factor de bloqueo de 0,1 (como el considerado en el cálculo de la ganancia), se obtienen las pérdidas correspondientes al mismo. Por lo tanto, se puede afirmar que en el 99,88 % del tiempo las perdidas debidas a errores en el apuntamiento del haz láser emitido serán inferiores a 2 dB. En el balance de potencia se empleará una corrección de la ganancia por pérdidas de apuntamiento de **2 dB**.

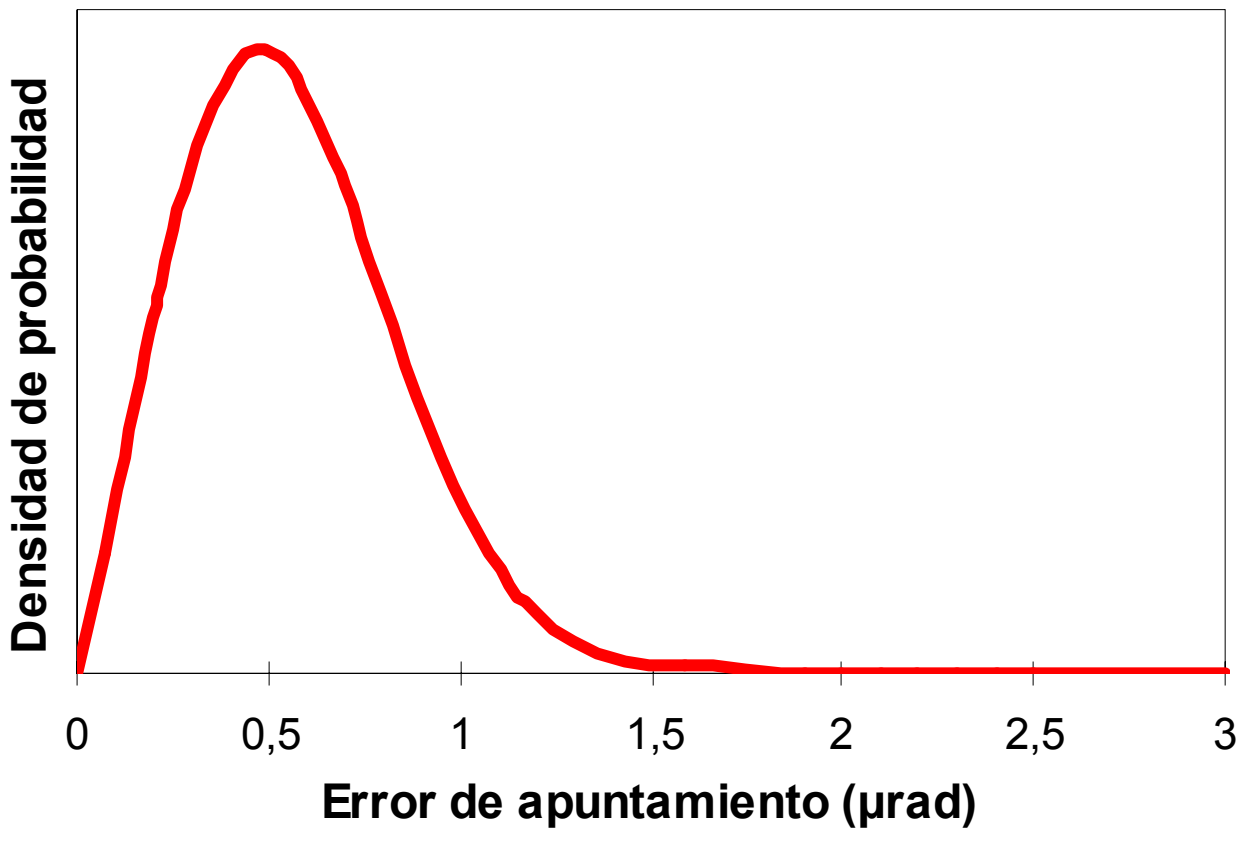

***Figura 6.40. Función densidad de probabilidad de Rice del error de apuntamiento [41].***

## 6.7. TERMINAL TERRESTRE

El terminal terrestre del *MLCD* estará formado por los siguientes elementos

- La ***antena receptora***, destinada a captar el haz y enfocarlo sobre el receptor terrestre. Estará un telescopio astronómico, incluyendo su propio sistema de apuntamiento y seguimiento.

- Uno o varios ***filtros ópticos***, destinados a **proteger** la óptica del telescopio y a **mejorar** la relación señal-ruido.

- El ***receptor*** propiamente dicho. Se estudiarán dos posibilidades, fotodiodo APD y fotomultiplicador.

### 6.7.1. Filtro óptico fijo

Su función es eliminar la mayor parte de la radiación luminosa no deseada procedente principalmente del Sol –y de los astros del cielo nocturno por la noche– con el fin de proteger la óptica del telescopio. La función del filtro es especialmente crítica durante el día, ya que tanto la luz del Sol como la radiación difusa de la atmósfera puede provocar ***calentamientos locales*** en la óptica del telescopio, provocando deformaciones de la misma. El uso del telescopio durante el día requerirá probablemente [45] un sistema de **control de temperatura** para evitar tales deformaciones.

El amplio espectro recibido puede reducirse significativamente interponiendo un filtro paso banda que estará centrado a la longitud de onda de emisión 1,064 μm. El área del filtro habrá de ser bastante grande, puesto que tiene que cubrir el espejo primario del telescopio. Por esta razón se optará por una solución simple y de poco peso, un ***filtro fijo interferencial***. En realidad no se tratará de un solo filtro, ya que no existen en el mercado elementos de ese tamaño. En su lugar se propone emplear un ***mosaico*** de filtros montado sobre una estructura de cuadrícula o de panal de abeja como el que se muestra en la Figura 6.41.

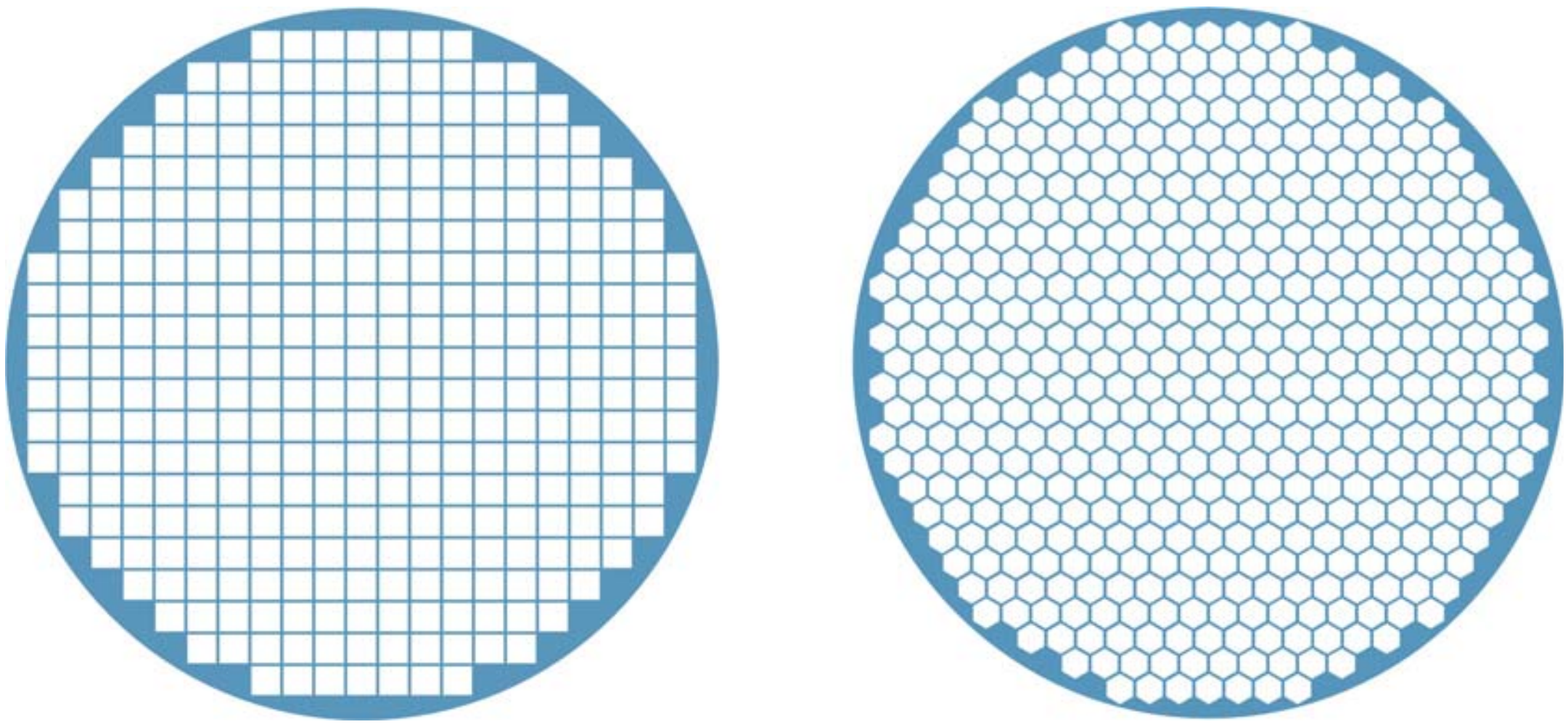

***Figura 6.41. Propuestas para el filtro óptico de protección solar de 349 elementos en cuadrícula y 484 elementos en panal.***

Este filtro óptico deberá tener un ancho de banda suficiente como para que no quede eliminada ni la frecuencia de la señal emitida ni sus correspondientes variaciones debidas al efecto Doppler. La propuesta que se hace en este proyecto es utilizar ***un segundo filtro*** a la entrada del detector para controlar el ruido de fondo. En estas circunstancias, la anchura del filtro fijo podrá hacerse significativamente mayor que las variaciones Doppler (véase apartado 6.3 y Tabla 6.1), de modo que éstas *no modifiquen de forma apreciable* la potencia de entrada de la señal. Puesto que los 35 GHz (Figura 6.42) de rango Doppler equivalen a 0,13 nm, se ha optado por un filtro en torno a **1 nm de ancho**, unas 8 veces mayor. Esta anchura espectral es suficiente para proteger la óptica sin comprometer la recepción de la señal. Posteriormente, justo antes del detector, se introducirá otro filtro más estrecho, posiblemente sintonizable, destinado a maximizar la relación señal-ruido.

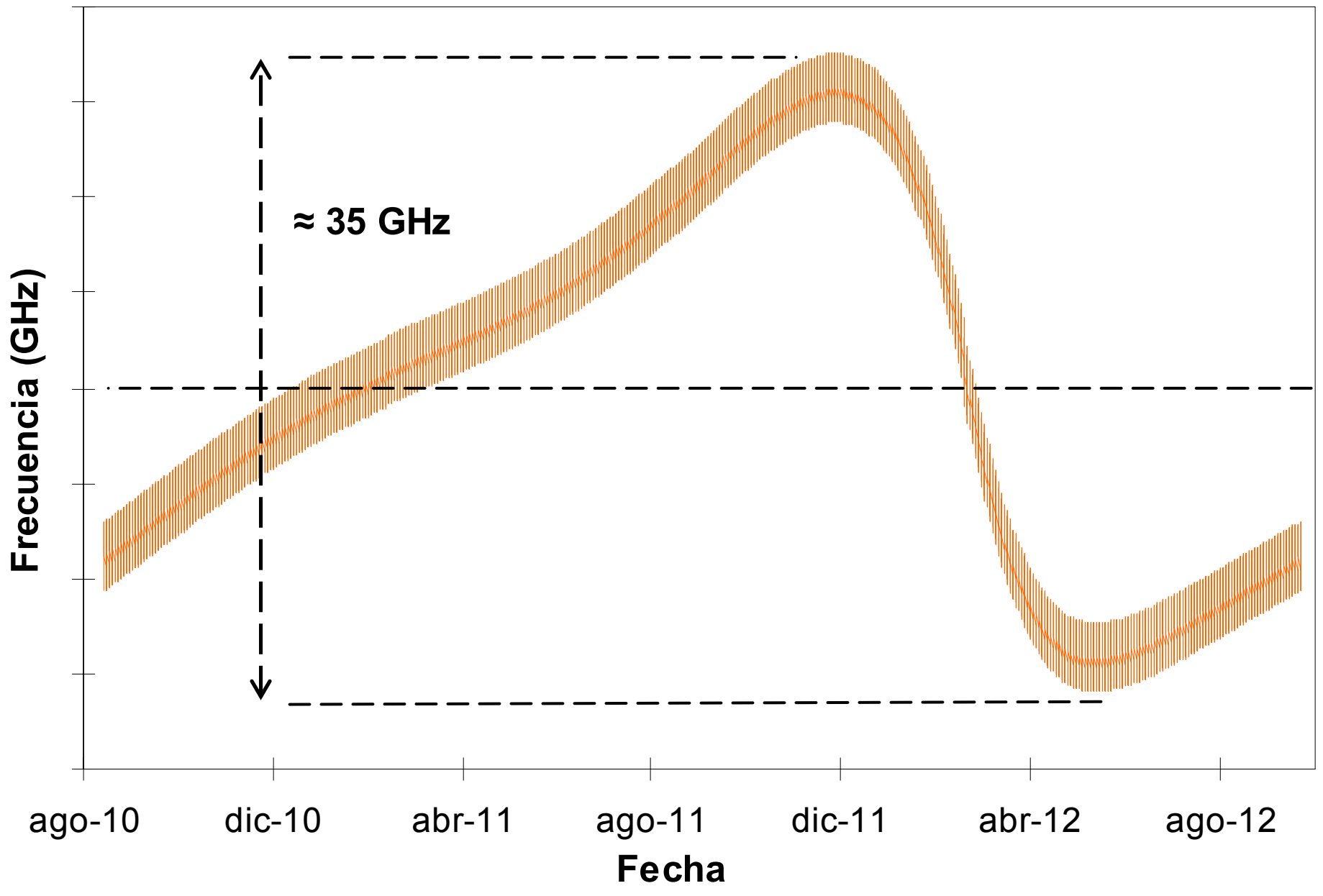

***Figura 6.42. Ancho de banda del filtro óptico fijo igual a la suma de los tres rangos de variación de la frecuencia por efecto Doppler.***

### 6.7.2. Filtro óptico sintonizable

El filtro solar deja pasar todas las variaciones Doppler de la frecuencia de la señal emitida. Para minimizar el ruido de fondo recogido por el fotodetector sería deseable recibir la señal con el ancho de banda más pequeño posible. Dado que las variaciones Doppler del satélite *MTO* y la rotación de la Tierra son muy rápidas, se adopta en este proyecto una solución eficiente, que consiste en emplear un filtro óptico sintonizable que vaya siguiendo la ***variación lenta*** de la frecuencia debida a la traslación de los planetas (que se consideraría constante durante cada periodo diario de observación) y deje pasar toda la ***variación rápida*** debida a los movimientos del satélite y la rotación de la Tierra (que tendría que estar ajustándose continuamente a lo largo de la jornada de observación).

Con este planteamiento, la variación lenta mencionada determinará la ***frecuencia central*** del filtro sintonizable y la variación rápida determinará el ***ancho de banda*** del mismo filtro. La variación de la frecuencia central del filtro a lo largo de los dos años de proyecto *MLCD* se muestra en la Figura 6.43.

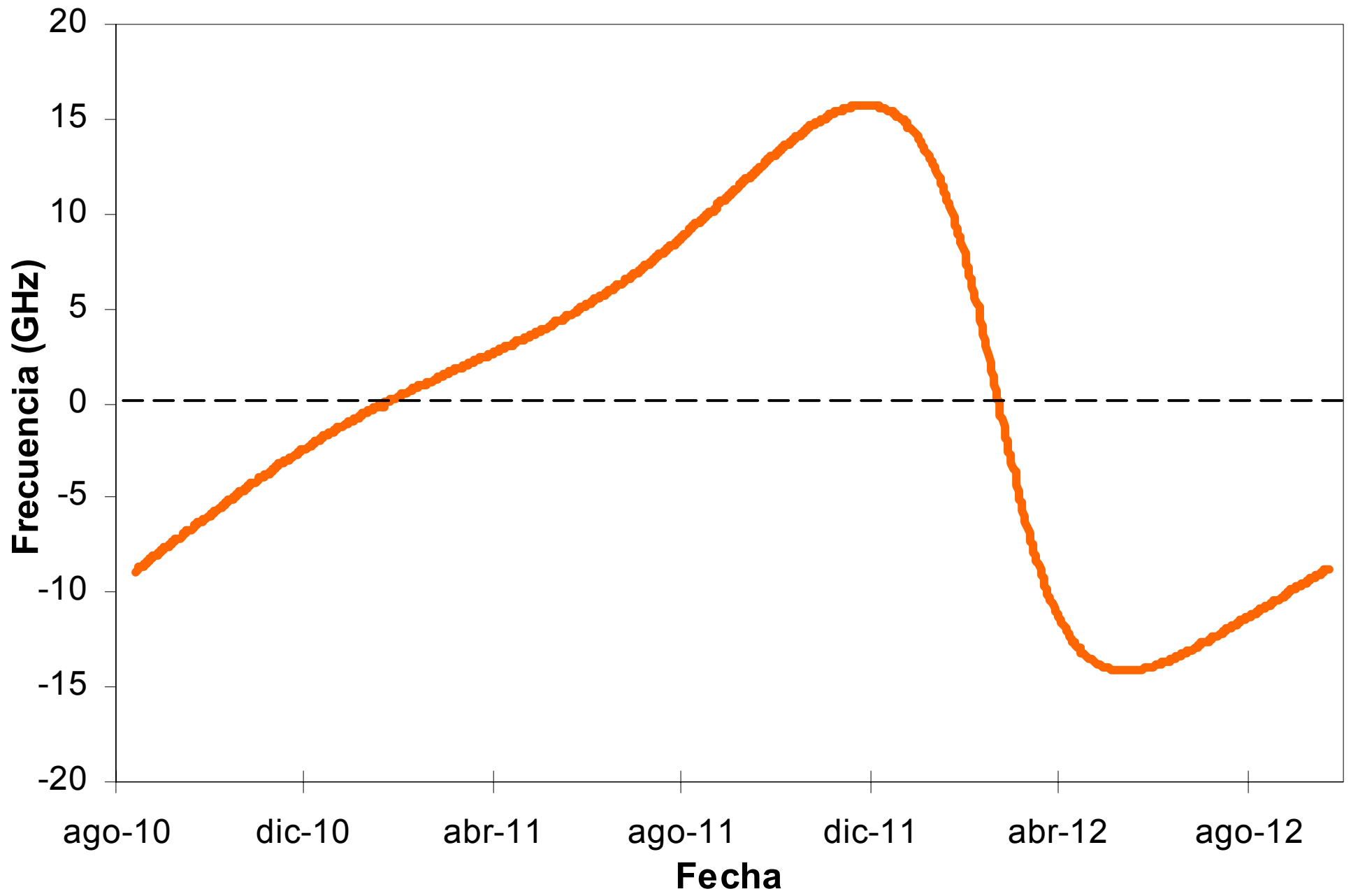

***Figura 6.43. Frecuencia central del filtro sintonizable durante los dos años de proyecto MLCD.***

El ancho de banda del filtro deberá ser superior a la suma de los dos rangos de variación de la frecuencia por efecto Doppler, correspondientes a los movimientos de rotación del *MTO* y la Tierra, mostrados en la Tabla 6.1, quedando

$$\Delta f = 0{,}7565 + 4{,}39315 = 5{,}150\,\text{GHz} \tag{151}$$

Esta variación traducida a longitud de onda queda

$$|\Delta\lambda| = \Delta f \frac{\lambda^2}{c} = 0{,}0194\,\text{nm} \tag{152}$$

El filtro deberá ser **mayor** que esta variación; si tuviera el mismo ancho de banda que la variación Doppler, en los momentos en que la longitud de onda estuviera en los extremos de variación, el filtro cortaría parte de la señal (la mitad exactamente, si se ha definido la anchura espectral del filtro a 3 dB). Por ello, se ha optado por un filtro de **anchura espectral de 0,1 nm**, equivalentes a esta λ a unos 26,5 GHz, que deja un margen suficiente para que la señal no sufra atenuación apreciable en el rango Doppler diario.

Si el ancho de banda que se acaba de calcular supusiera un problema en el sentido de que la cantidad de ruido que deja pasar degrada críticamente la relación señal-ruido, existe una solución óptima para emplear el mínimo ancho de banda posible a costa de aumentar la complejidad de los receptores. Si se controla exactamente cual es la frecuencia recibida en cada instante teniendo en cuenta (además de las variaciones lentas) las variaciones rápidas de frecuencia introducidas por la rotación de la Tierra y el movimiento orbital del *MTO* –para lo que habría que conocer exacta y simultáneamente la posición del *MTO* y del receptor terrestre–, se podría emplear el filtro óptico sintonizable más estrecho disponible tecnológicamente, siempre que su ancho de banda estuviera por encima de la anchura espectral de la fuente láser. Por ejemplo, los analizadores de espectros ópticos (OSA) emplean filtros sintonizables de hasta 25 o 50 pm, es decir, 20 veces menores que el elegido.

## 6.7.3. Ganancia

Para el proyecto *MLCD* se planea utilizar telescopios astronómicos. En este proyecto se aplican los tres principales telescopios de los observatorios considerados en el apartado 6.3 del cálculo del efecto Doppler.

De la misma manera que se hizo para el telescopio del terminal del *MTO*, los telescopios en la Tierra tendrán una ganancia debido a su directividad, que se calcula de forma similar. El parámetro que determina la ganancia del telescopio es el diámetro del espejo primario según la relación que se vio en la ecuación (89). En la Figura 6.44 puede observarse esta dependencia en el rango de tamaño de telescopios que se ha considerado.

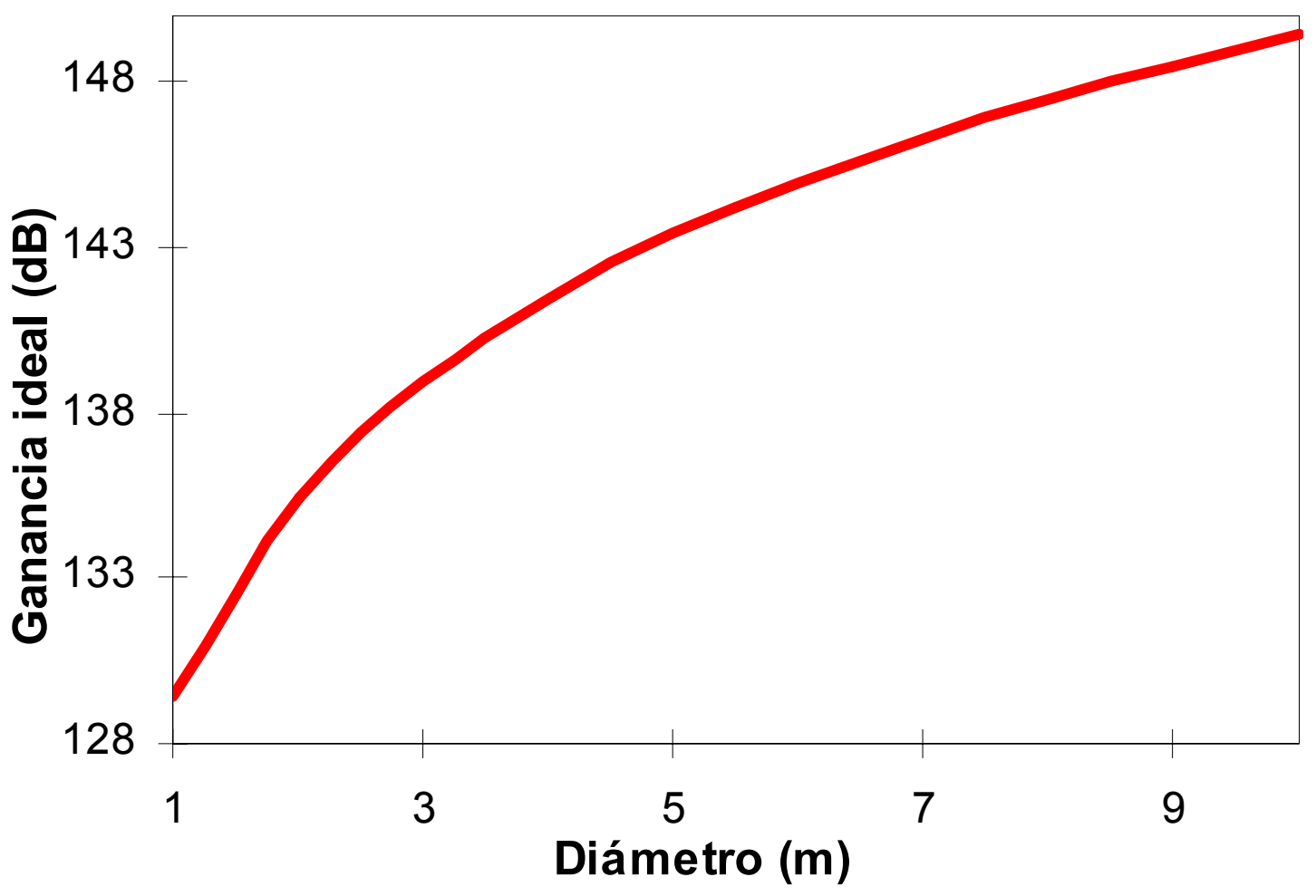

***Figura 6.44. Dependencia de la ganancia del telescopio con el diámetro del espejo primario.***

En el caso del telescopio receptor, la ganancia ideal se calcula de forma idéntica que en el emisor y el parámetro g sólo se considera dependiente del factor de bloqueo. Este parámetro de eficiencia de la ganancia se calcula de la siguiente manera [41]

$$g = 1 - \gamma^2 \tag{153}$$

En la Tabla 6.8 se muestran los valores de las ganancias para los tres telescopios considerados, incluyendo el factor de eficiencia g.

| Observatorio | ∅ espejo primario (m) | ∅ espejo secundario (m) | Ganancia ideal (dB) | Factor de bloqueo γ | Ganancia real (dB) |
|---|---|---|---|---|---|
| ***Roque de los Muchachos (La Palma, España)*** | 10,4 | 1,176 | 149,75 | 0,11 | 149,69 |
| ***Palomar (California, EEUU)*** | 5 | 1 | 143,38 | 0,2 | 143,21 |
| ***Stromlo (Canberra, Australia)*** | 2,3 | 0,3 | 136,64 | 0,13 | 136,56 |

***Tabla 6.8. Ganancias de los tres telescopios considerados en el proyecto.***

## 6.7.4. Diseño del campo de visión

El campo de visión de los telescopios receptores terrestres es un parámetro determinante para el enlace del proyecto. Tiene que ser tal que incluya en todo momento al satélite *MTO* para que el receptor pueda recibir la señal. Puesto que el satélite orbita alrededor de Marte, la solución más simple se consigue haciendo que en el campo de visión encaje siempre la órbita que describe dicho satélite.

La altura de la órbita del satélite *MTO* desde la superficie de Marte es de 4.450 km. Este valor equivale un radio orbital (desde el centro de Marte hasta la posición del satélite) de 7.840 km (recuérdese que el radio de Marte es de 3.390 km). El campo de visión que se considerará deberá incluir un rango de aproximadamente 18.000 km asumiendo un cierto margen de seguridad para compensar el efecto de turbulencias atmosféricas.

Esos 18.000 km que siempre debe incluir el campo de visión provocarán que, cuando la distancia Marte-Tierra sea menor, el campo de visión sea mayor y cuando dicha distancia aumente, el campo de visión disminuya. En una primera aproximación, se podría utilizar un **valor fijo** para el campo de visión que sería el calculado para la distancia mínima (el mayor de todos, para abarcar todos los casos). Dicha solución quedó descartada en la simulación porque un valor de campo de visión establecido al máximo valor, generaba una excesiva cantidad de ruido cuando Marte y la Tierra se encontraban a la máxima distancia.

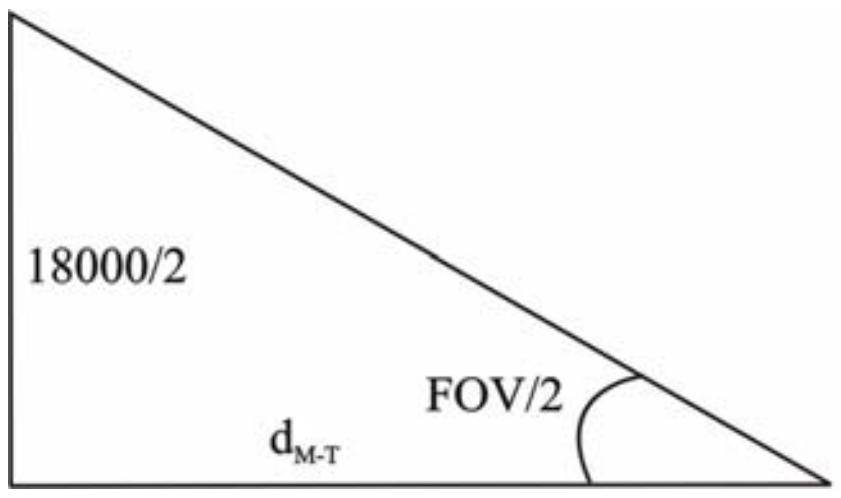

***Figura 6.45.Cálculo del campo de visión***

Finalmente, la opción que se ha adoptado en este proyecto es que el campo de visión se ajuste desde el receptor, conforme varía la distancia, de forma que siempre esté ajustado a la menor cantidad de ruido posible.

Por lo tanto, la expresión que calcula en la simulación el campo de visión es la siguiente

$$FOV = 2 \cdot \arctan\left(\frac{18000/2}{d_{M-T}}\right) \qquad (154)$$

siendo $d_{M-T}$ la distancia Marte-Tierra expresada en kilómetros. Para el cálculo del ruido de fondo deberá convertirse a estereoradianes para obtener $\Omega_R$.

Como ya se explicó en el apartado 5.1.2 dedicado al campo de visión, este ajuste se puede conseguir, para un tamaño de detector fijo, variando la focal de la lente de enfoque.

## 6.7.5. Ruido de fondo

El ruido de fondo que capta el receptor será distinto si es de día o de noche. Durante el día son dos las principales fuentes de ruido: por un lado, al estar el receptor terrestre permanentemente apuntando hacia Marte, la fracción de radiación solar que refleja el planeta –el ***albedo***– se recibe en el telescopio. Esta es una fuente constante de ruido causada por el hecho de tener siempre a Marte dentro del campo de visión. Por otro lado, el *scattering* que sufre la radiación solar al atravesar la atmósfera hace que, aunque no se apunte directamente al Sol, parte de su radiación entre al campo de visión. Esta aportación ruidosa se nombrará en adelante como ***ruido del cielo***.

Por la noche obviamente el ruido del cielo es inexistente. Se ha despreciado el efecto que tendría el ruido recibido por la inclusión esporádica de alguna estrella en el campo de visión así como la luz zodiacal [45]. Esta aportación ruidosa es despreciable frente al ruido de Marte que siempre es mayor.

La caracterización de la densidad espectral de este ruido con la distinción de si es de día o de noche, se ha simulado basándose en las ecuaciones (121) y (122), a las que se llegó en el apartado dedicado al ruido de fondo.

En cuanto al cálculo de la *densidad espectral de potencia de ruido de Marte*, que estará presente tanto de día como de noche, para poder aplicar la ecuación (122) (en el caso en que $\Omega_s \in \Omega_R$), es necesario (además de conocer el área del receptor $\pi(D/2)^2$ siendo D el diámetro del espejo primario) determinar la irradiancia de Marte. Este valor varía constantemente con la distancia Marte-Tierra y viene dado por la siguiente expresión [34]

$$H(\lambda)_M = alb \cdot \frac{H(\lambda)_S}{d^2_{M-S}} \left( \frac{R_M}{d_{M-T}} \right)^2 \tag{155}$$

siendo $H(\lambda)_S$ la irradiancia solar a una unidad astronómica ($H(\lambda)_S = 668$ $W \cdot m^{-2} \cdot \mu m^{-1}$ para 1,064 µm), $d_{M-S}$ la distancia entre el Sol y Marte en UA (1,52366 UA), $R_M$ el radio de Marte (3.390 km), $d_{M-T}$ la distancia entre la Tierra y Marte y alb el albedo de Marte, que es del 15 % (0,15) [21].

Como se puede deducir de la ecuación (155), cuanto más cerca estén los planetas, mayor será la irradiancia de Marte y consecuentemente mayor el ruido que introduce. La simulación de esta dependencia del ruido de Marte con la distancia durante los dos años de proyecto, para un telescopio de 10,4 metros de diámetro, se muestra en la Figura 6.46.

En lo que respecta al cálculo de la *densidad espectral de potencia del ruido del cielo*, el procedimiento adoptado ha sido acotar el ruido haciendo variar la radiancia del cielo ya que ésta es **distinta** según el ángulo de visión y el ángulo cenital del sol (Figura 6.47). Como se observa en la figura, el valor máximo de radiancia del cielo, que se utilizará como caso peor, es de unos 90 $W \cdot m^{-2} \cdot \mu m^{-1} \cdot sr^{-1}$. Para cada valor de radiancia, es inmediato obtener, mediante la ecuación (121) (en el caso en que $\Omega_r \in \Omega_s$), la densidad espectral de potencia de ruido debido al cielo, a través del campo de visión $\Omega_r$. La relación del ruido del cielo con el campo de visión implica que, cuando Marte y la Tierra están alejados, el campo de visión será menor y por tanto el ruido del cielo disminuirá. Sin embargo, para distancias menores, dado que el campo de visión se aumentará en el receptor para poder contemplar toda la órbita del satélite *MTO* alrededor de Marte, el ruido crecerá.

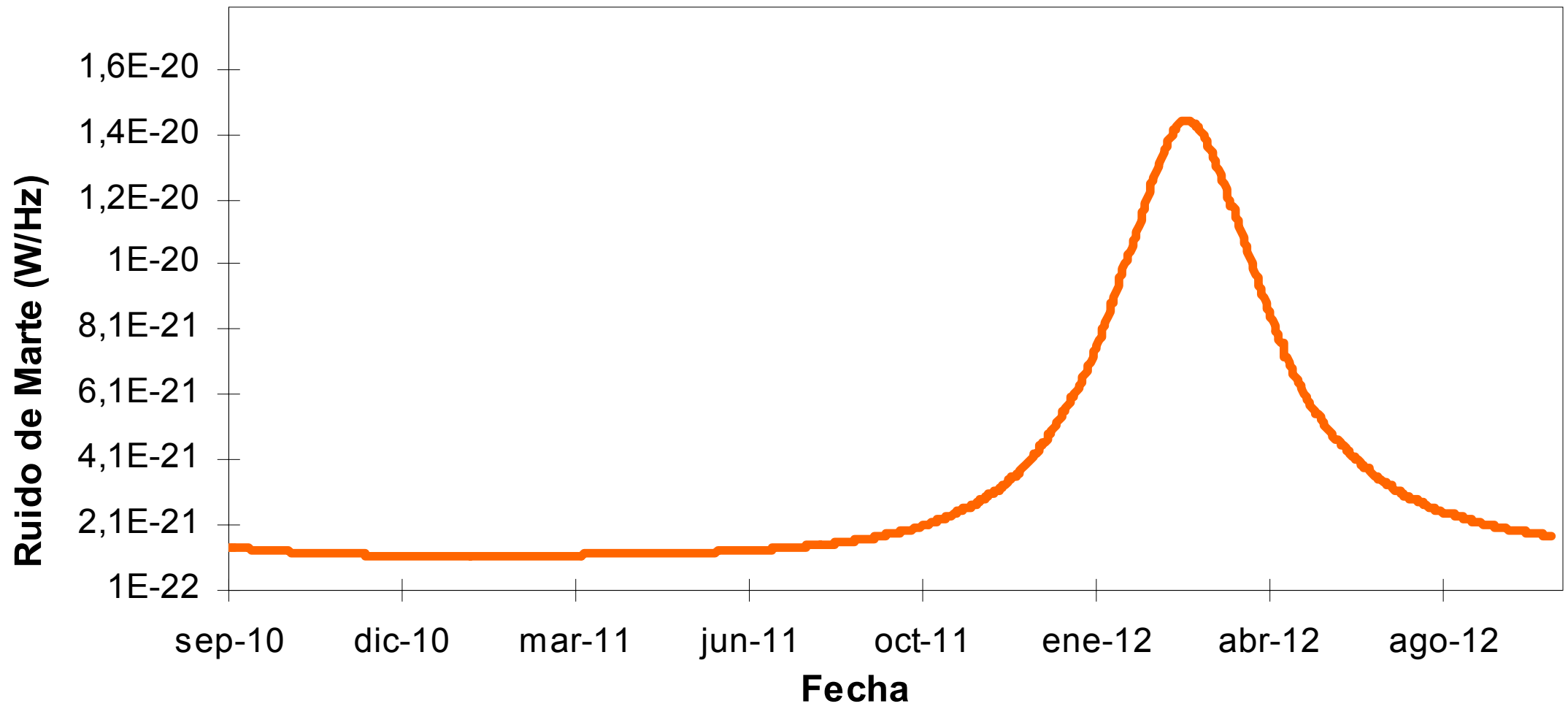

***Figura 6.46. Densidad de potencia de ruido de fondo debido al albedo de Marte recibido en un telescopio de 10,4 metros para los dos años de proyecto.***

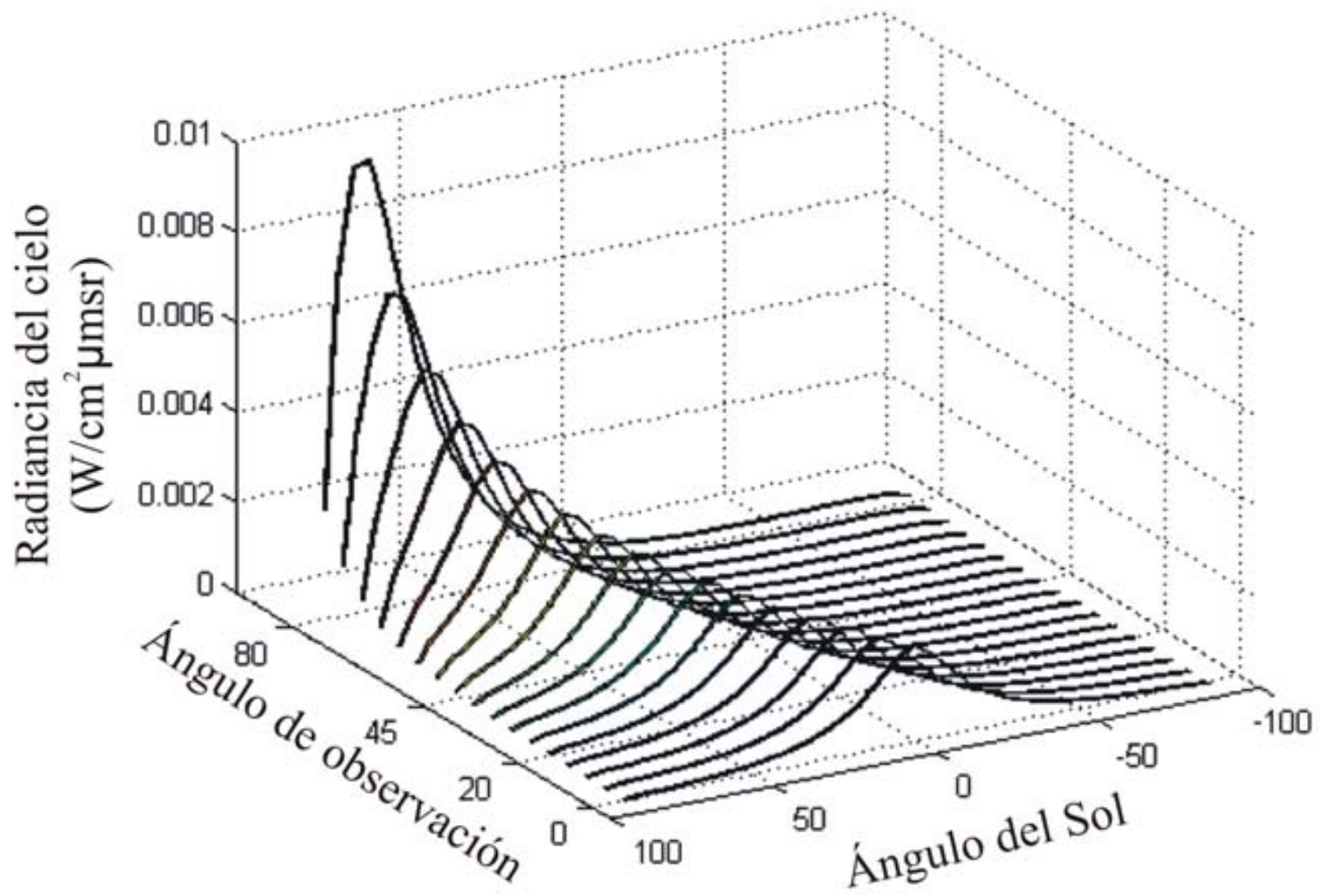

***Figura 6.47. Radiancia del cielo en función del ángulo del Sol y el ángulo de apuntamiento [47].***

### 6.7.6. Tasa binaria para una modulación m-PPM

Fijada una especificación de calidad para el enlace mediante el parámetro BER (del inglés *Bit Error Rate* o tasa de error de bit), que se desea mantener durante toda la comunicación, es posible determinar la relación señal-ruido que implica y consecuentemente la tasa binaria que será posible alcanzar.

En el proyecto *MLCD* se planea utilizar una modulación m-PPM. Esta técnica produce señales equienergéticas (recuérdese que la energía de cada pulso es la misma distinguiéndose unos símbolos de otros por la posición que ocupan dentro del intervalo de símbolo) y ortogonales (el producto escalar de todas las posibles señales es cero debido a ocupar posiciones diferentes en el intervalo de símbolo), por lo que el cálculo de la probabilidad de error de símbolo es idéntico a un esquema m-FSK, que viene dado por [59]

$$P_{em} \approx \frac{(m-1)}{2}\operatorname{erfc}\left[\sqrt{\frac{S}{2N}}\right] \tag{156}$$

donde erfc representa a la función de error complementaria cuya definición es

$$\operatorname{erfc}(x) = 1-\operatorname{erf}(x) = 1-\frac{2}{\sqrt{\pi}}\int_0^x e^{-t^2}dt = \frac{2}{\sqrt{\pi}}\int_x^{\infty} e^{-t^2}dt \tag{157}$$

Se puede suponer que todos los símbolos son igualmente probables y en este caso es posible establecer una relación entre la probabilidad de error de símbolo $P_{em}$ y un límite máximo para la probabilidad de error de bit BER

$$BER \leq \frac{m}{2(m-1)}P_{em} \tag{158}$$

por lo que la probabilidad de error de bit para m-PPM queda

$$BER \leq \frac{m}{4}\operatorname{erfc}\left[\sqrt{\frac{S}{2N}}\right] \tag{159}$$

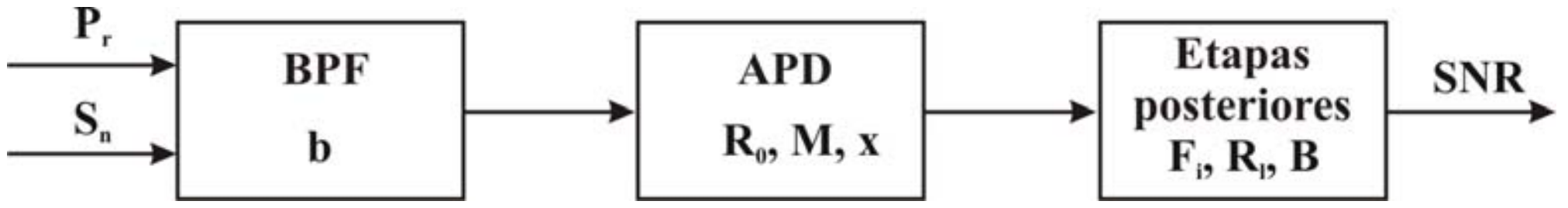

***Figura 6.48. Esquema de detección directa con un fotodiodo APD.***

Para un diagrama de bloques de un esquema de detección directa, como el de la Figura 6.48, y asumiendo como detector un fotodiodo APD, a continuación se evalúan las corrientes de señal y de ruido para el nivel "1" y "0" de una modulación m-PPM con el objeto de calcular la relación señal-ruido. Nótese en dicha figura que al receptor no sólo le llega una proporción de la señal emitida, es decir la $P_R$ resultante del balance de potencia, sino también una contribución procedente de la densidad espectral de potencia de ruido de fondo $S_n$. Se usará la notación con subíndice "1" para la presencia en el tiempo de *slot* tanto de la señal como del ruido de fondo y el subíndice "0" para indicar la presencia únicamente de ruido de fondo.

La corriente de señal, como ya se vio, es

$$I_s = R_0 P_r M \tag{160}$$

La corriente rms de ruido para los "1" viene dada por

$$\sigma_1^{\,2} = \underbrace{eM^2FR_0[P_r + S_n b]\cdot 2B}_{(a)} + \underbrace{M^2R_0^{\,2}S_n[2P_r + S_n b]\cdot 2B}_{(b)} + \underbrace{\frac{2kT}{R_L}\cdot 2B}_{(c)} \qquad (161)$$

y la corriente rms de ruido para los "0"

$$\sigma_0^{\,2} = \underbrace{eM^2FR_0S_n b\cdot 2B}_{(d)} + \underbrace{M^2R_0^{\,2}S_n^{\,2}b\cdot 2B}_{(e)} + \underbrace{\frac{2kT}{R_L}\cdot 2B}_{(c)} \qquad (162)$$

De la ecuación (161) la contribución por ruido *shot* del APD se corresponde con (a), el resultado del batido señal-ruido de fondo y ruido de fondo-ruido de fondo es (b) y la componente de ruido térmico es (c). De la ecuación (162) el ruido shot asociado al nivel "0" (resultante únicamente del ruido de fondo, no se envía señal en el nivel "0") se corresponde con (d) y el ruido correspondiente al batido ruido de fondo-ruido de fondo (en el nivel "0" no hay señal) es (e). Los términos (b) y (e) provienen del ruido que se genera de los distintos batidos entre el ruido de fondo y la señal deseada que pasa por el ancho de banda del filtro óptico. El ancho de banda b se corresponde con el del filtro óptico que debe ser mucho mayor que el ancho de banda B de la señal.

Las expresiones de señal y de ruido anteriores, dan como resultado la siguiente relación señal a ruido

$$\frac{S}{N} = \frac{R_0^{\,2}M^2P_r^{\,2}}{(\sigma_1^{\,2} + \sigma_0^{\,2})} \qquad (163)$$

Por lo tanto, la expresión que relaciona la tasa de error de bit con los parámetros del enlace queda

$$BER \le \frac{m}{4}\,\mathrm{erfc}\left[\sqrt{\frac{R_0^{\,2}M^2P_r^{\,2}}{2(\sigma_1^{\,2} + \sigma_0^{\,2})}}\right] \qquad (164)$$

El dato que interesa despejar de la ecuación (164) se encuentra en el denominador de la raíz, en especial dentro del factor común B (el ancho de banda de la señal modulada en m-PPM) que está contenido en el término ($\sigma_1^2 + \sigma_0^2$). El dato de interés mencionado es, por supuesto, el régimen binario $R_b$ que se relaciona, para una modulación m-PPM que codifica n bits mediante m símbolos, con el ancho de banda de la señal B de la siguiente manera

$$B = \frac{m}{n} R_b \qquad (165)$$

Cabe mencionar que este ancho de banda B es la principal limitación a la hora de usar una modulación m-PPM ya que es necesario el empleo de un fotodetector con este gran ancho de banda.

En la simulación, el resultado final del régimen binario $R_b$ se ha obtenido de la siguiente manera. Inicialmente se ha despejado la raíz cuadrada que está dentro de la función erfc de forma que se iguala a la función inversa de erfc de BER/(4m). Mediante el programa *MathWorks MatLab 6* se ha calculado dicho argumento de la función inversa de erfc que se muestra en la Tabla 6.9 para distintos valores de m (número de símbolos) y BER (tasa de error de bit).

| BER | m | $erfc^{-1}(BER/4m)$ |
|---|---|---|
| $10^{-6}$ | 2 | 3,361178563 |
| $10^{-6}$ | 4 | 3,458910737 |
| $10^{-6}$ | 8 | 3,554139891 |
| $10^{-6}$ | 16 | 3,647045573 |
| $10^{-6}$ | 32 | 3,737787041 |
| $10^{-6}$ | 64 | 3,82650631 |
| $10^{-6}$ | 128 | 3,913330647 |
| $10^{-6}$ | 256 | 3,99837462 |
| $10^{-6}$ | 512 | 4,081741789 |

| BER | m | $erfc^{-1}(BER/4m)$ |
|---|---|---|
| $10^{-9}$ | 2 | 4,241090013 |
| $10^{-9}$ | 4 | 4,320005385 |
| $10^{-9}$ | 8 | 4,397571017 |
| $10^{-9}$ | 16 | 4,47385297 |
| $10^{-9}$ | 32 | 4,548912124 |
| $10^{-9}$ | 64 | 4,622804727 |
| $10^{-9}$ | 128 | 4,69558287 |
| $10^{-9}$ | 256 | 4,767294908 |
| $10^{-9}$ | 512 | 4,837985817 |

***Tabla 6.9. Resultado de evaluar el término de la función inversa de error complementaria de la ecuación (164) para distintas BER y m.***

Inicialmente se evaluó la posibilidad de usar distintas combinaciones de BER y m. Tras varias simulaciones se llegó a la conclusión de que la BER de $10^{-6}$, exigida en el proyecto *MLCD*, dando calidad aceptable al enlace (un bit erróneo de cada millón), permite alcanzar mayores regímenes binarios. Por otro lado, el número de símbolos mejora en gran medida las prestaciones del enlace al aumentar la potencia transmitida en un factor m.

Por lo tanto, el criterio para la elección de m ha sido utilizar el mayor valor posible siempre que existan receptores que soporten los anchos de banda derivados de pulsos tan estrechos. Muy pocos fotodetectores adecuados para este proyecto llegan a algo más de los 2 GHz, por lo que la elección del número de símbolos ha tenido que hacerse en función de este ancho de banda y del régimen binario buscado. Los requerimientos de régimen binario para el proyecto *MLCD* contemplan demostrar un máximo de unos 50 megabits por segundo, lo que conduce, aplicando la ecuación (165) y asumiendo un cierto margen, a un ***número de símbolos óptimo*** de 256. Para una modulación 256-PPM y un régimen binario de 50 megabits por segundo, la señal será de 1,6 GHz. En comunicaciones ópticas es habitual emplear fotodetectores de ancho de banda igual a 0,7-0,8 veces el ancho de banda de la señal, por lo que, siguiendo este criterio, el dispositivo empleado para detectar la señal 256-PPM deberá contar con un ancho de banda superior a 1,1-1,2 GHz.

Una vez conocido el resultado de la función inversa de erfc, conocidas las características del fotodiodo empleado, la potencia recibida y el ruido de todas las fuentes se puede despejar el régimen binario ya que el ancho de banda B de la señal es factor común en el denominador. De esta manera se puede obtener un régimen binario variable con la distancia y el ruido de fondo $S_n$, que *variará al abrir y cerrar el campo de visión* en función de la distancia.

# 6.8. EVALUACIÓN DEL ENLACE

A continuación se realizará la evaluación teórica del enlace que se ha diseñado en este proyecto. Para ello se empleará el régimen binario que reflejará la capacidad del mismo. Además se discutirán las diferentes alternativas de diseño, la influencia de los elementos variables del sistema y se fundamentará el enlace diseñado.

## 6.8.1. Balance de potencia

A diferencia de los enlaces de comunicaciones ópticas guiadas, en los que es necesario hacer dos balances (de potencia y de *dispersión*), en este caso el enlace está claramente ***limitado por potencia***, por lo que la evaluación de la calidad del enlace se basa en este balance. La dispersión temporal de la señal es prácticamente nula, puesto que se produce únicamente, y en muy limitada cuantía, en el ***tramo atmosférico*** del enlace.

El balance de potencia permite determinar la potencia recibida a la entrada del receptor, conocida la potencia transmitida en función de las diversas pérdidas que sufre esta potencia durante el trayecto entre ambos terminales. En el balance se incluye además un margen de seguridad con el objetivo de contemplar cualquier imprevisto que pueda surgir durante el enlace. La ecuación que permite calcular el balance de potencia viene dada por [4]

$$\begin{aligned} P_r(\mathrm{dBm}) &= P_t(\mathrm{dBm}) + G_t(\mathrm{dB}) + G_r(\mathrm{dB}) + \\ &- L_{fs}(\mathrm{dB}) - L_{atm}(\mathrm{dB}) - L_{apunt}(\mathrm{dB}) - M(\mathrm{dB}) \end{aligned} \tag{166}$$

donde $G_t$ y $G_r$ son las ganancias de la antena transmisora y receptora respectivamente, $L_{fs}$ las pérdidas por espacio libre, $L_{atm}$ las pérdidas por atenuación atmosférica, $L_{apunt}$ las pérdidas por errores de apuntamiento y M el margen de seguridad. Todos estos valores han sido ya deducidos.

Según las especificaciones del proyecto *MLCD* [53], el *MTO* contará con un telescopio de 30,5 centímetros y una potencia media de láser de 5 W. La ganancia de su antena (apartado 6.6.2.c) será de 118,05 dB y la potencia transmitida por el terminal será de 1,28 kW (61,07 dBm) debido a que la potencia media es de 5 W y la modulación empleada es una 256-PPM, que (apartado 4.3.2) ofrece una relación entre potencia de pico y potencia media igual al número de símbolos (256 en este caso). En la Figura 6.49 se presenta la variación de la potencia recibida durante los dos años de proyecto considerados, en las siguientes condiciones

- utilizando como ***antena receptora*** el telescopio GTC de 10,4 metros de diámetro del observatorio de Roque de los Muchachos (La Palma) que ofrece una ganancia (apartado 6.7.3) de 149,69 dB.
- asumiendo una ***atenuación atmosférica*** de 2,89 dB –icluyendo *scattering* de gases (0,191 dB) y aerosoles (2,696 dB)– perteneciente a un modelo desértico. La ubicación receptora admitiría modelo marítimo o desértico, ya que se encuentra próxima al desierto del Sáhara. Se utiliza la segunda para asumir el caso peor.
- con un ***ángulo cenital*** de recepción de 70 grados (caso peor).
- con unas ***pérdidas por apuntamiento*** de 2 dB (también para caso peor).
- y un ***margen de seguridad*** de 5 dB.

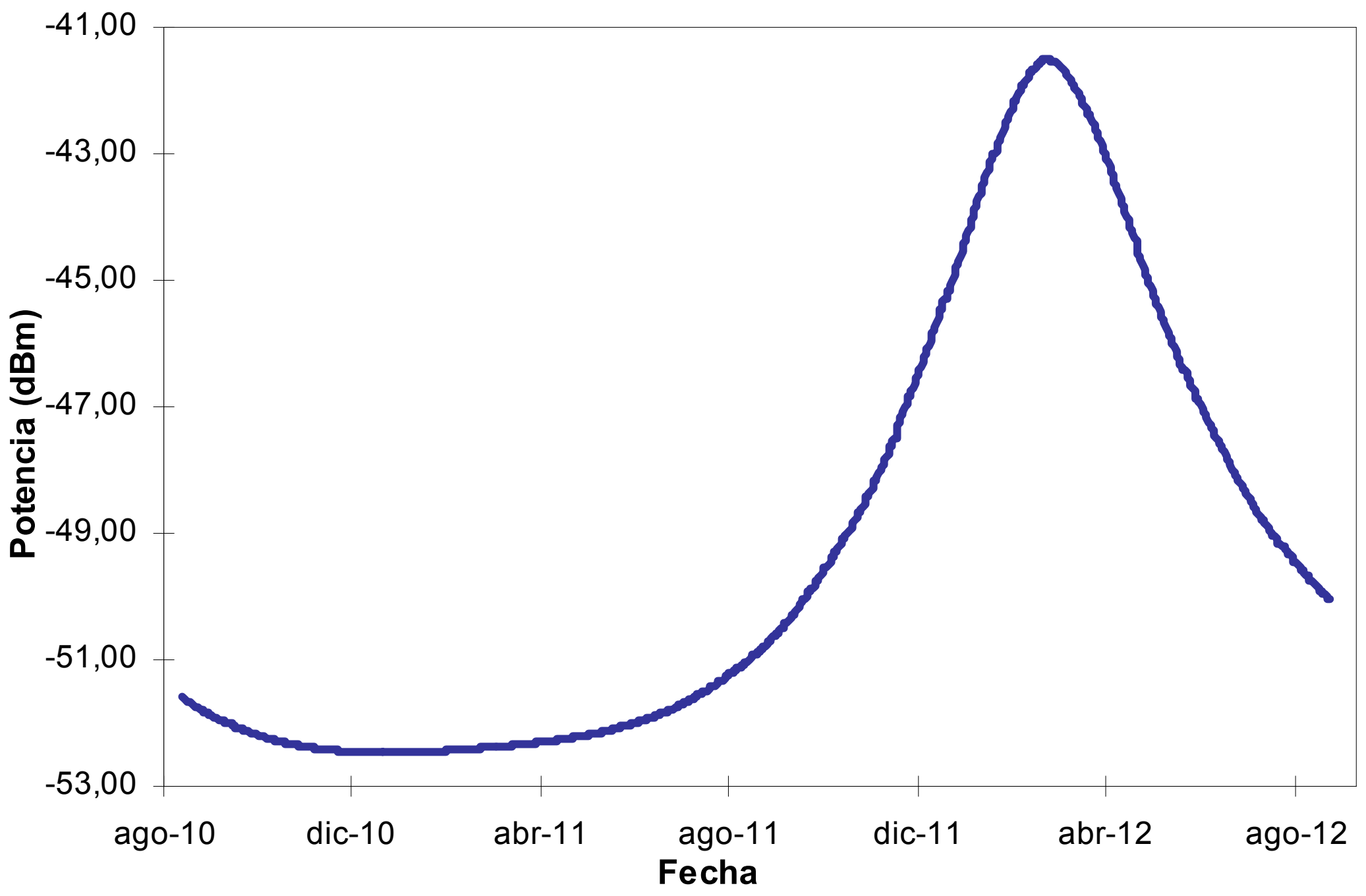

***Figura 6.49. Potencia recibida durante dos años de proyecto.***

## 6.8.2. Selección del detector

Con los resultados expuestos en la sección anterior, se puede hacer una preselección de detectores

- La potencia recibida es ***muy baja***, entre 10 y 20 dB inferior a las que usualmente se manejan en CC.OO. (la sensibilidad del receptor suele estar entre -30 y -35 dBm).

- En estas circunstancias, el uso de un fotodiodo PIN **queda descartado**. Cabría la posibilidad de emplear un PIN con un amplificador óptico (SOA o YDFA) acoplado como preamplificador. Sin embargo, no se han encontrado componentes comerciales que permitan el ***acoplo*** de la radiación incidente al sistema preamplificador-detector de un modo eficiente. Menos aún a la longitud de onda de trabajo (1064 nm).

- Por consiguiente, los candidatos serán en principio ***fotodiodos APD.***

- La ***longitud de onda*** de trabajo está ligeramente por encima del rango de funcionamiento de los ***fotodiodos APD de Si***. Si se escoge un APD de Si, su rendimiento cuántico no será óptimo.

- La ***responsividad***, a igualdad de rendimiento cuántico, es **la misma** para un fotodiodo de Si que para un APD III-V (siempre que la $\lambda$ de trabajo esté incluida en su rango, naturalmente). Así pues, se podría pensar en buscar fotodiodos InGaAs o InGaAsP, de los cuales existe una amplia variedad derivada de su uso en CC.OO. guiadas, tanto en segunda (1310 nm) como en tercera ventana (1550 nm).

- Por desgracia, el ***factor de multiplicación M*** de los fotodiodos III-V es sensiblemente inferior (alrededor de 30) que el de los fotodiodos de Si (hasta 300-400). por lo tanto, la selección inicial se focalizará sobre fotodiodos de Si, a pesar de su rendimiento cuántico relativamente pobre a la λ de trabajo.

- La baja potencia de señal hará que ***el ruido shot*** sea también bajo. Así pues, no resultará extraño que, a diferencia del modo habitual de funcionamiento de un APD, *el término dominante de ruido no sea el ruido shot*.

- Si el ruido predominante es ***térmico***, conviene maximizar la **resistencia de carga** $R_L$, con el fin de reducirlo.

- Por desgracia, si se incrementa $R_L$, también aumenta la ***constante RC*** del circuito, reduciendo su ancho de banda. La resistencia de carga quedará limitada a la *máxima que tolere el ancho de banda requerido*.

- Por último, si no se encuentra un fotodiodo APD que cumpla las especificaciones durante el periodo de la misión, se seleccionará como fotodetector un ***fotomultiplicador***.

La mayor dificultad en la selección del receptor basado en un fotodiodo APD ha sido la escasez de dispositivos que, cumpliendo los requisitos de ancho de banda de la señal 256-PPM, trabajen a la longitud de onda de 1,064 µm. Como se ha comentado, la búsqueda se reduce a APDs de silicio, los cuales, pese a no presentar la máxima eficiencia cuántica a 1,064 µm, posee una alta ganancia que los hace muy sensibles. Se han buscado componentes cuya aplicación original es la detección de pulsos rápidos de láseres Nd:YAG, cuya λ de trabajo coincide con la de este proyecto. El APD que presentaba mejores propiedades ha sido la versión mejorada (*enhanced*) del C5658 de Hamamatsu, cuyas principales características se especifican en la Tabla 6.10. Éste será el fotodiodo que se aplicará al cálculo del balance de potencia.

| **APD Hamamatsu C5658 *enhanced*** | | |
|---|---|---|
| Rango espectral | 400 – 1.090 | nm |
| Material | Si | - |
| Temperatura de trabajo | 0 a 50 | ºC |
| Diámetro área activa | 0,5 | mm |
| Eficiencia cuántica | 0,5 | - |
| Ganancia | 100 | - |
| Ancho de banda | 2,0 | GHz |
| Precio aproximado | 3.000 | € |

***Tabla 6.10. Características del fotodiodo C5658 empleado para la simulación.***

Con el fin de no complicar en exceso el receptor con un filtro que no incluya toda la variación Doppler (lo que obligaría a ser sintonizado constantemente en función de la posición instantánea de la Tierra y el *MTO*), se utiliza el filtro de 0,1 nm calculado en el apartado 6.7.2.

### 6.8.3. Resultados APD

Tras simular el enlace empleando las características del APD, se comprobó que, asumiendo

- los 5 dB de **margen de seguridad** mencionados,
- un modelo de **desierto**,
- y un **filtro óptico de 0,1 nm**,

el enlace no garantizaba los 1 Mbit/s durante los dos años considerados. Únicamente se logra el requerimiento desde el 4/1/2012 hasta el 10/5/2012, es decir algo más de cinco meses de un total de dos años.

Tras el pobre resultado obtenido, se procedió a ***relajar*** las condiciones para comprobar bajo cuales, si era posible, se podían lograr las especificaciones empleando el APD. Tras realizar varias simulaciones, la conclusión que se extrajo fue que únicamente se puede mantener un régimen binario mínimo de 1 Mbit/s durante los dos años *si se elimina el margen de seguridad y se emplea el modelo marítimo*. El resultado bajo estas condiciones se muestra en la Figura 6.50.

En la Figura 6.51 se muestra el periodo en el que el enlace supera 1 Mbit/s para distintas combinaciones de margen de seguridad y modelo atmosférico empleando el fotodiodo APD. Como se observa, incluso en el mejor de los casos el enlace queda cortado 21 días debido a que el ángulo Sol-Tierra-Marte se hace inferior a los tres grados mínimos establecidos. Esta circunstancia, obviamente, se producirá con cualquier detector.

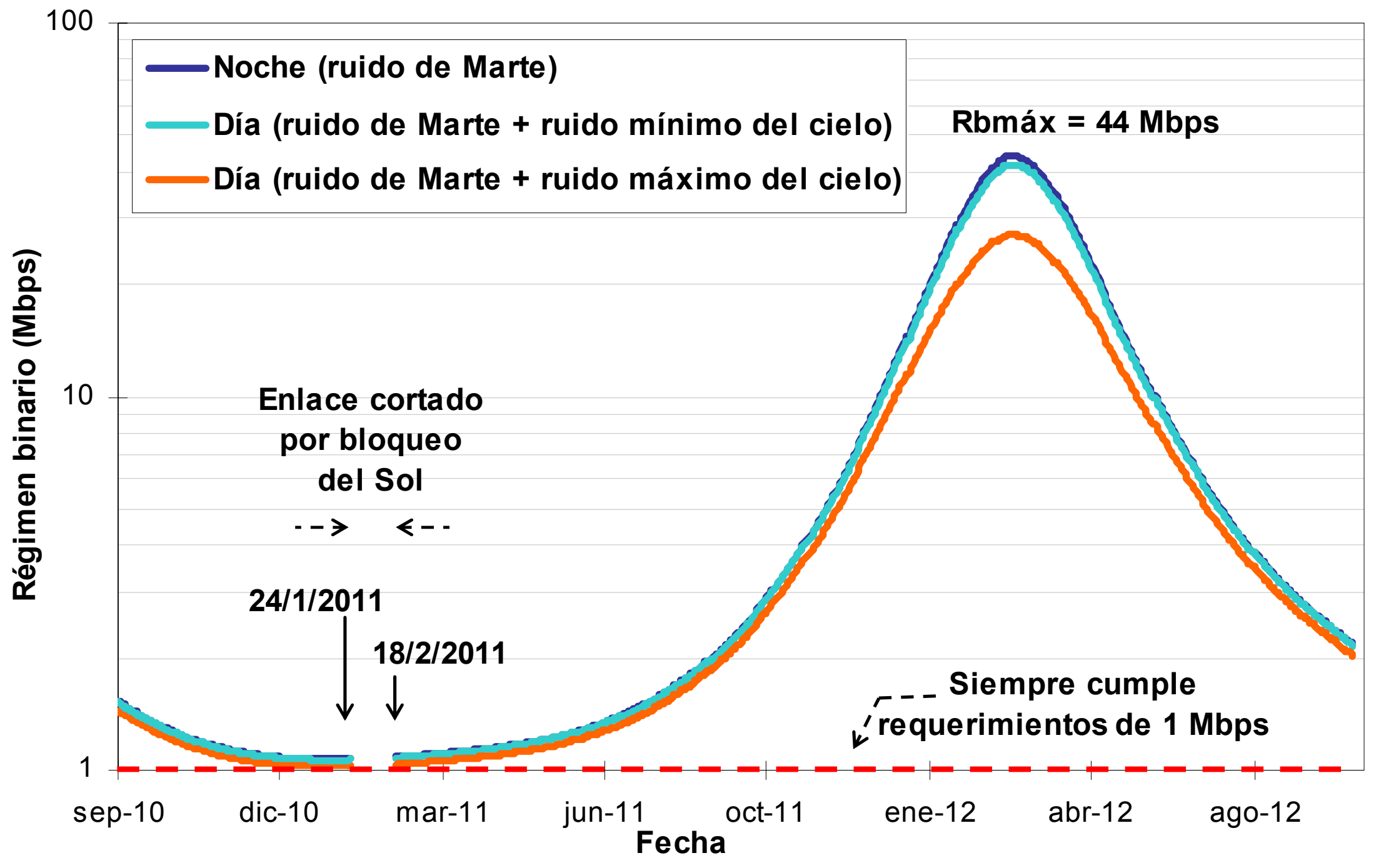

***Figura 6.50. Resultado del enlace para el receptor basado en APD bajo condiciones favorables.***

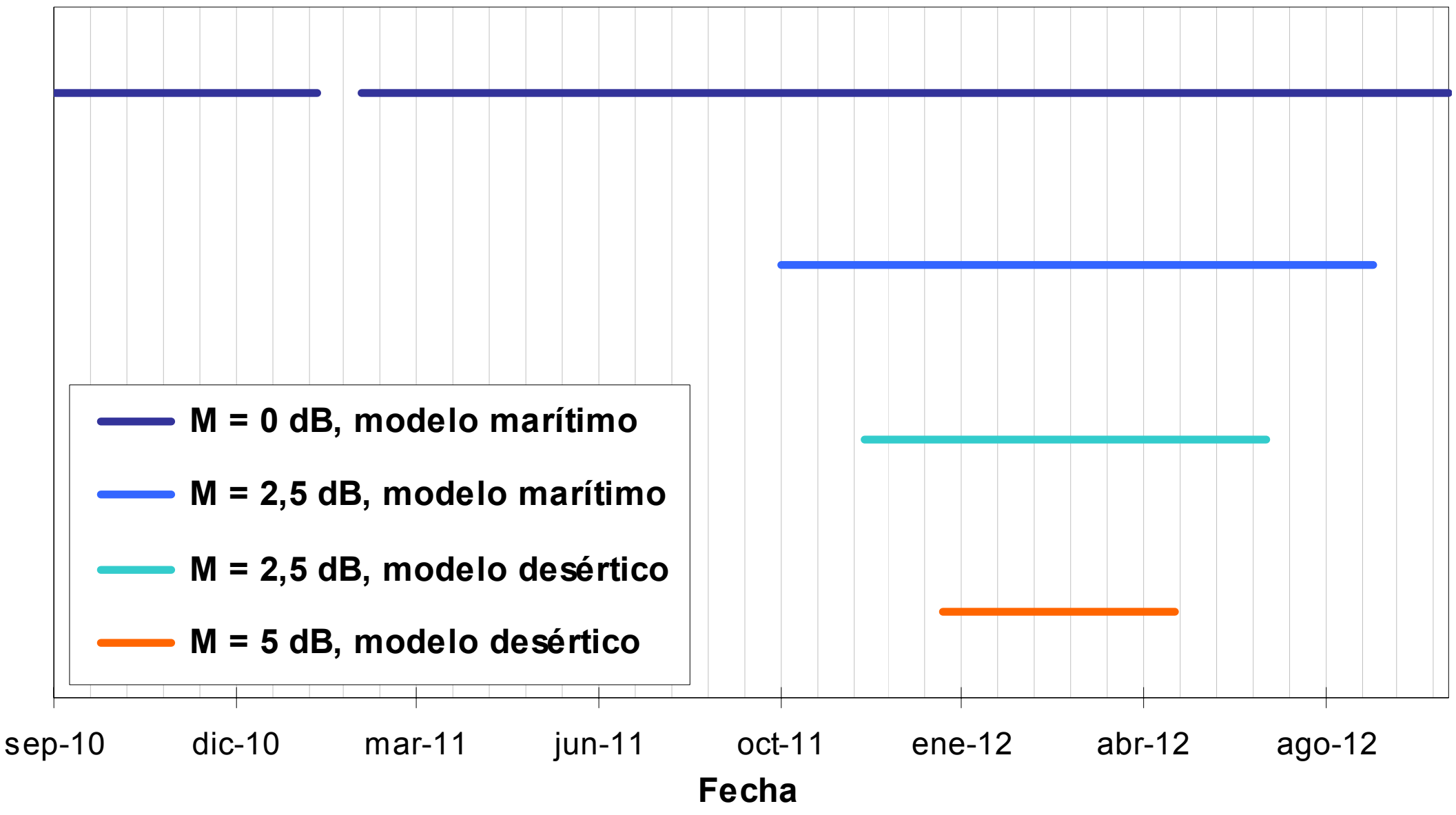

***Figura 6.51. Periodos en que Rb > 1Mbit/s para distintas combinaciones de margen de seguridad y modelo atmosférico usando el APD.***

Como puede verse en los resultados precedentes, el APD elegido ***está en el límite*** de cumplir especificaciones. El resultado puede juzgarse satisfactorio ya que es esperable que las especificaciones definidas en el proyecto *MLCD* estén en el **máximo rango alcanzable** con tecnologías estándar actuales.

Antes de rechazar *prematuramente* el APD como detector, conviene hacer algunas consideraciones. En efecto, el cálculo del balance de potencia, según se ha comprobado en apartados anteriores, es extremadamente complejo, interviniendo un gran número de variables. Pues bien, existe un conjunto de penalizaciones que se han tomado **en caso peor** y se han mantenido constantes en el cálculo. Entre ellas están

- el **ángulo cenital**, de 70º, que produce las mayores pérdidas de *scattering*;
- la recepción se hace **durante el día**, con el Sol muy próximo a Marte;
- el **factor de multiplicación** del APD (100) es constante, tomado de las características del fotodiodo. Este factor M ***no es óptimo***. Si pudiera modificarse, aunque fuese en un reducido rango, mejoraría la SNR;
- la **resistencia de carga** (50 Ω) es constante, tomada de las características del fotodiodo. Con ese valor, el ruido dominante es el ***ruido térmico***. Si pudiera incrementarse su valor, mejoraría sensiblemente la SNR. Es cierto que se reduciría paralelamente el **ancho de banda**, pero como el actual es 2 GHz, se dispone de un pequeño margen de mejora puesto que el ancho de banda necesario para el enlace es ligeramente superior a 1 GHz;
- las pérdidas de apuntamiento consideradas son las máximas (2 dB);
- Marte –y su ruido asociado– se ha considerado en ***fase llena***, sin tener en cuenta las posiciones relativas de la Tierra y el Sol respecto al mismo. En realidad Marte presenta fases, como la Luna, que vistas desde la Tierra pueden llegar a oscurecer un 40% del planeta [45];
- la función de distribución de partículas de aerosol es la que produce **mayor *scattering*** dentro del modelo considerado (2,76 dB en desierto y 0,78 dB en marítimo).

La mayor parte de factores apuntados suponen *pequeñas contribuciones* al cálculo del balance de potencia. No obstante, el conjunto de *casos peores* considerado permite afirmar que, en promedio, el enlace funcionaría con una tasa binaria superior a 1 Mbit/s durante una buena parte del periodo del proyecto *MLCD*.

En cualquier caso, no resulta aceptable un diseño de ingeniería sin márgenes de seguridad, o sujeto a factores ambientales externos que resultan imprevisibles. Tampoco hay que olvidar que la simulación se ha hecho para el *"caso mejor"* de elección de telescopio, el GTC o Gran Telescopio de Canarias, cuyo uso continuado en el enlace es ciertamente problemático. Para garantizar el mantenimiento del enlace en las condiciones exigidas, por lo tanto, se hace necesario elegir otro fotodetector.

### 6.8.4. Resultados PMT

En vista de los resultados obtenidos con el APD, se ha rediseñado el enlace empleando como fotodetector un fotomultiplicador *contador de fotones*. El dispositivo más idóneo, que se emplea para la simulación, ha sido el R3809U-68 de Hamamatsu, cuyas características se indican en la Tabla 6.11.

| **PMT Hamamatsu R3809U-68** | | |
|---|---|---|
| Tecnología | Placa microcanal | - |
| Rango espectral | 950-1.400 | nm |
| Material del fotocátodo | In/P-InGaAsP | - |
| Temperatura de trabajo | –90 a –70 | ºC |
| Refrigeración | $N_2$ líquido | - |
| Tensión de trabajo | 3.400 | V |
| Diámetro área activa | 2 | mm |
| Eficiencia cuántica | 0,5 | - |
| Ganancia | $2\cdot10^5$ | - |
| Tiempo de subida | 170 | ps |
| Tiempo de bajada | 450 | ps |
| Tiempo de tránsito | 100 | ps |
| Precio | 26.000 | € |

***Tabla 6.11. Características del fotomultiplicador empleado para la simulación.***

Según los valores relativos a la respuesta temporal del PMT se puede realizar una estimación aproximada de su ancho de banda a partir de sus tiempos de subida y bajada

$$B_{rec} \approx \frac{1}{170\ \text{ps} + 450\ \text{ps}} = 1{,}61\ \text{GHz} \tag{167}$$

Se comprueba que el ancho de banda del PMT cumple los requisitos de la señal 256-PPM, siendo superior a los 1,12 GHz calculados en el apartado 6.7.6. También cumple con el requisito de trabajar a la longitud de onda de 1,064 µm, en cuyo valor responde con una eficiencia cuántica de 0,5.

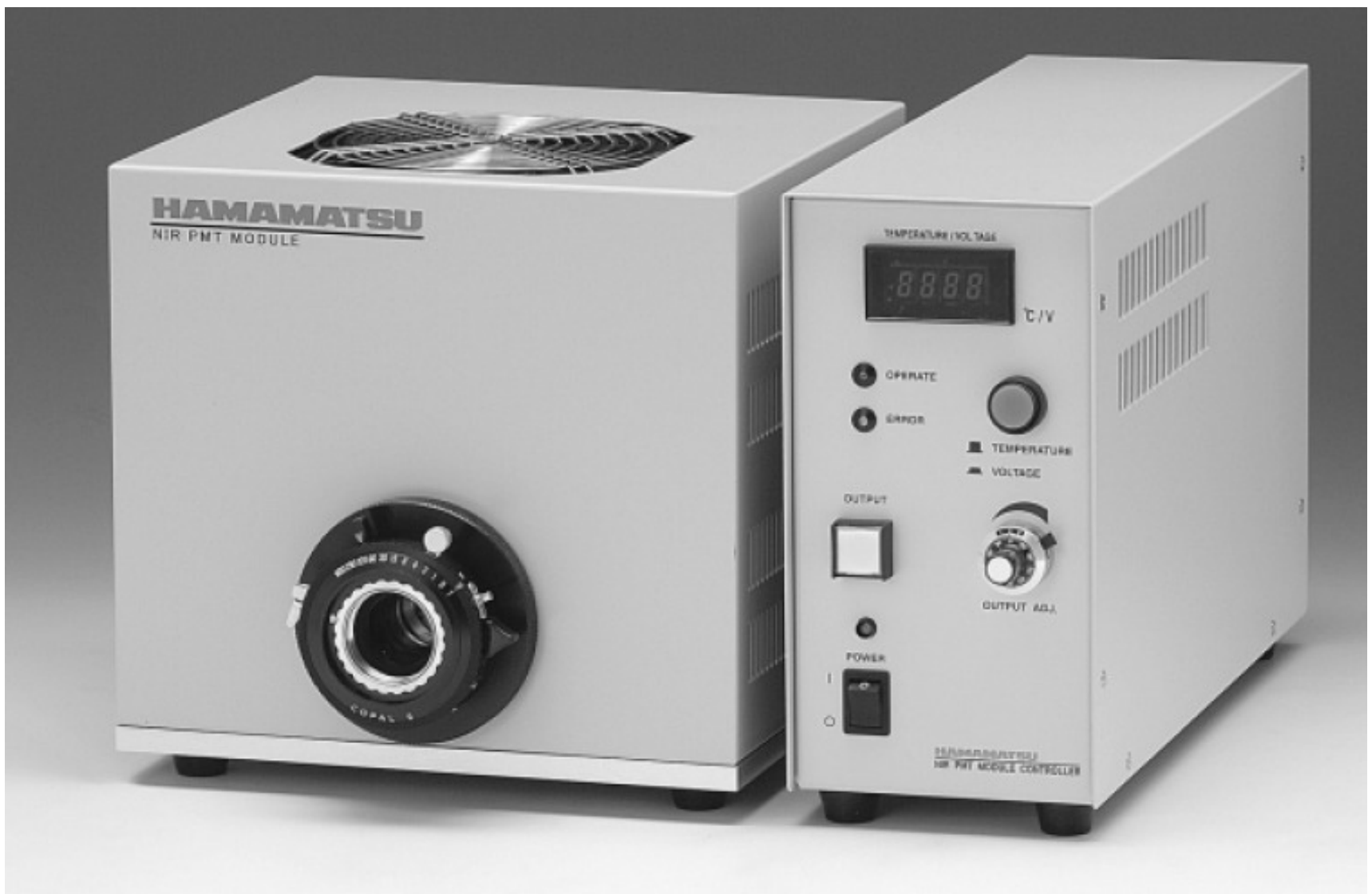

***Figura 6.52. Fotomultiplicador empleado para la simulación.***

El tratamiento para el cálculo del régimen binario con el PMT puede hacerse adaptando las fórmulas deducidas en el apartado 6.7.6, que ya se han utilizado para el receptor basado en APD [9]. Como factor de exceso de ruido se toma el valor típico que se mencionó en el apartado 5.4.4. El filtro óptico empleado será el mismo que en el APD.

En la Figura 6.53 se muestra el resultado del enlace empleando el PMT para las **condiciones originales** de margen de seguridad de 5 dB y modelo desértico. Como se observa el PMT cumple el requerimiento en cuanto al régimen binario mínimo de 1 Mbit/s durante los dos años de proyecto, a excepción de los inevitables 26 días en los que el ángulo Sol-Tierra-Marte es inferior a tres grados. El mínimo régimen binario que se alcanzaría es de **3,1 Mb/s** cuando la distancia Tierra-Marte es **máxima** y es **de día**. El máximo puede llegar hasta los casi **58 Mbit/s** en **mínima** distancia y de **noche**. En contraste con el APD, el PMT es *casi independiente de la potencia de señal* recibida, ya que su enorme ganancia (200.000) hace que la señal salga muy amplificada. Además esta misma ganancia hace que el ruido *shot* prevalezca siempre en comparación con el ruido térmico que resulta prácticamente despreciable en el PMT. Como desventaja está su elevado coste en comparación con el APD, aunque ciertamente este factor puede obviarse considerando las ingentes cantidades que se manejan en una misión espacial.

Hay que destacar que el enlace para el PMT se ha realizado asumiendo las ***peores condiciones*** para el modelo atmosférico y un amplio margen de seguridad de 5 dB. Unas suposiciones algo más optimistas podrían considerarse

- Si en lugar del modelo desértico se usa el modelo **marítimo** (para el caso en que todas las partículas fueran de agua –caso peor de modelo marítimo–) las pérdidas atmosféricas disminuirían de los 2,89 a los 0,97 dB,

- Además, reduciendo el **margen de seguridad** de 5 a 4 dB (que sigue siendo un valor aceptable), se concluye que podría emplearse la ganancia de recepción de 143,21 dB correspondiente al telescopio de 5 m del observatorio de Monte Palomar en California. Muy cerca de especificaciones quedaría la instalación de Calar Alto, en Almería, con su telescopio de 3,5 m

- La simulación realizada para estas condiciones asegura un régimen binario mínimo de **1,02 Mbit/s** en las **peores condiciones** de ruido de fondo de día + ruido de fondo de Marte. Probablemente el régimen binario sería algo superior debido a que el ***modelo continental*** (aplicable al observatorio mencionado) es más favorable que el modelo marítimo utilizado.

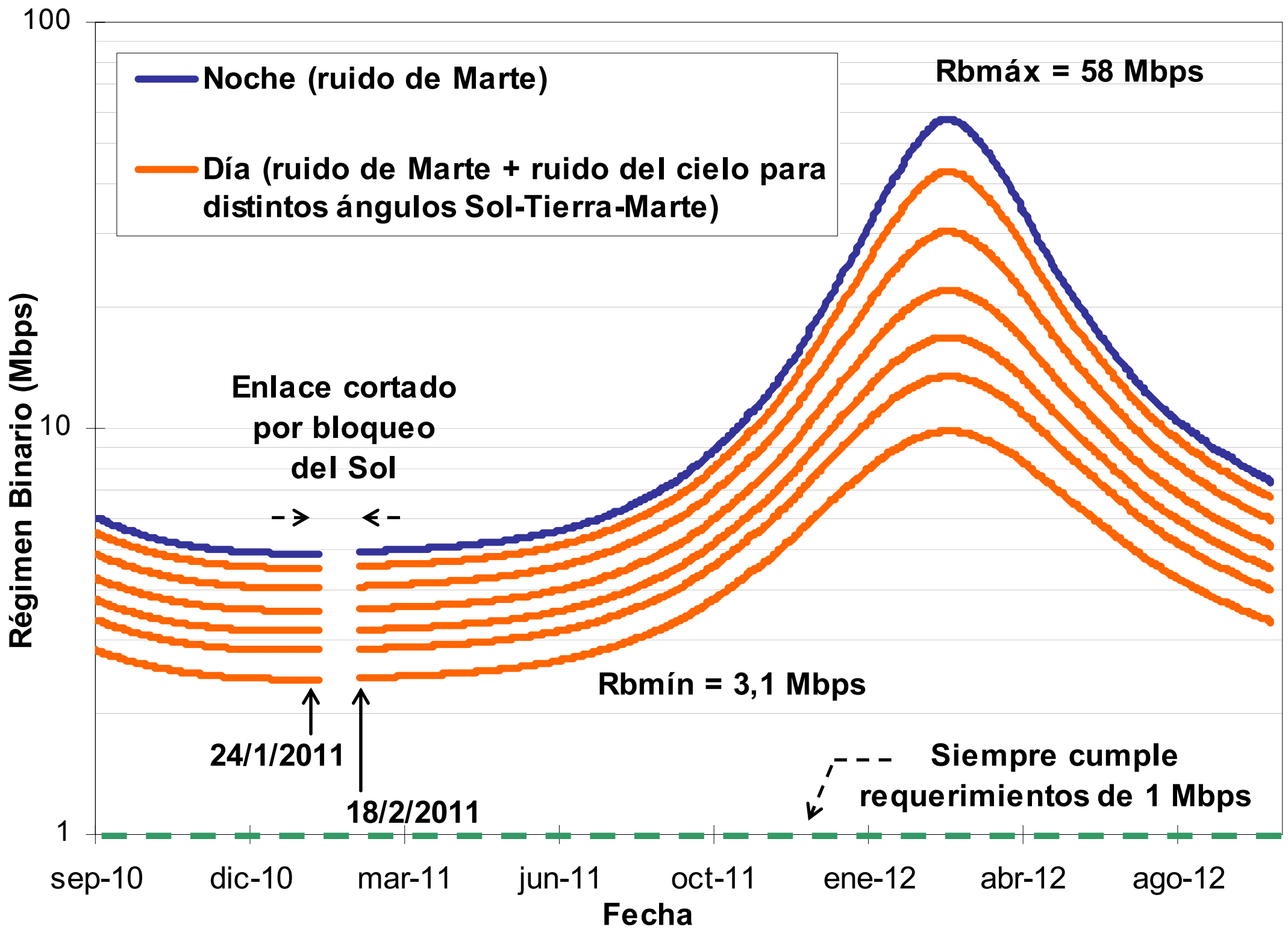

***Figura 6.53. Resultado del enlace para el receptor basado en PMT bajo condiciones desfavorables.***

# 7. Conclusiones y líneas futuras

## 7.1. CONCLUSIONES

- Se ha realizado un cálculo completo en que se demuestra la ***factibilidad*** de emplear comunicaciones ópticas en un enlace entre la Tierra y Marte.

- El enlace que se ha simulado ***cumple las especificaciones*** requeridas por el proyecto *MLCD* de la NASA actualmente en fase de diseño. En concreto, mantiene un régimen binario mínimo de 1 Mbit/s en las peores condiciones (a excepción de los inevitables periodos de bloqueo del Sol).

- Se ha implementado un ***simulador de dinámica orbital*** capaz de calcular, para un día determinado, la posición tridimensional heliocéntrica de cualquier objeto, conocidos sus parámetros orbitales característicos.

- Asimismo, el simulador permite realizar la ***animación de los movimientos orbitales*** con el fin de almacenar los resultados generados por dicha simulación.

- Se han evaluado las diferentes ***contribuciones al efecto Doppler*** que actúan en el viaje de la nave a Marte y en órbita marciana. Se ha demostrado que tales contribuciones condicionan el diseño de los filtros ópticos del receptor.

- Se han identificado las causas que generan ***atenuaciones*** en el canal de bajada: espacio libre –descontando las ganancias de las antenas–, errores de apuntamiento, limitaciones de la óptica, y perturbaciones atmosféricas.

- Se han descrito y evaluado diferentes ***modelos de atmósfera*** que permiten estimar pérdidas dependientes de la ubicación geográfica y determinadas condiciones ambientales.

- Se han identificado ***ubicaciones reales*** –observatorios astronómicos que pueden emplearse como estaciones receptoras en el proyecto–. Se han utilizado en cada caso los instrumentos ópticos de cada instalación.

- Se han seleccionado dos ***fotodetectores*** –un fotodiodo APD y un fotomultiplicador– como mejores opciones para el detector.

- Se ha demostrado que el ***fotodiodo APD*** se encuentra al límite de prestaciones, pudiendo dejar de cumplir especificaciones si se aúnan al mismo tiempo algunos factores adversos.

- Se ha demostrado que el ***fotomultiplicador microcanal*** cumple holgadamente las especificaciones en todas las condiciones, aunque en una sola de las ubicaciones consideradas. Si se relajan los márgenes de seguridad y/o se alivia el conjunto de factores adversos –todos ellos considerados en caso peor– el enlace sería factible en las tres ubicaciones elegidas.

## 7.2. LÍNEAS FUTURAS

La línea futura más obvia –y deseable– es abordar la ***construcción del sistema*** receptor del enlace, al menos a nivel de demostrador, para confrontar sus resultados con los cálculos que se han presentado aquí. Resulta ocioso indicar que el diseño de un enlace real de este tipo, debido a su carácter ampliamente multidisciplinar, corresponde a un buen número de grupos de investigación trabajando simultáneamente.

A lo largo de este proyecto se ha evolucionado desde lo general a lo particular, profundizando en los detalles críticos del enlace, para lo que necesariamente se ha tenido que acotar el estudio de unos elementos en detrimento de otros.

Un aspecto del enlace que podría mejorar sus prestaciones se basa en la posibilidad de utilizar diodos de avalancha como *contadores de fotones* mediante su funcionamiento en **modo Geiger**. Esta técnica podría proporcionar la posibilidad de emplear detectores extremadamente sensibles –del tipo del fotomultiplicador– a costes más reducidos.

Los APD, además, resultan más sencillos de montar en paralelo, formando un *array*, lo que permitiría **prescindir** de los grandes telescopios astronómicos y utilizar *arrays* de pequeños telescopios solidarios en su lugar. Al margen de una drástica reducción de costes, esta solución es ciertamente más viable, dada la enorme demanda de *tiempo de telescopio* que soportan las grandes instalaciones.

También relacionado con los detectores, convendría estudiar el uso de dispositivos con **mayor ancho de banda**, que permitirían emplear la modulación m-PPM con un *mayor número de símbolos*, lo que redundaría en que los pulsos transmitidos tendrían mayor potencia a igual consumo de potencia del láser. Asimismo, un aumento en la potencia media del láser emisor tendría el mismo efecto manteniendo el número de símbolos de la modulación m-PPM. Para ambos casos, cabe mencionar que potencias de pico varias veces mayores a las empleadas en este proyecto han sido ya demostradas, lo que hace factible esta posibilidad.

Otra mejora del sistema la podría constituir el uso de satélites como terminales receptores terrestres. Esto eliminaría la influencia perjudicial de la atmósfera permitiendo el empleo de telescopios de menor tamaño para iguales prestaciones y además ofrecería la posibilidad de usar longitudes de onda más pequeñas. Un cálculo previo realizado con el telescopio espacial Hubble resulta extraordinariamente revelador de las grandes posibilidades que se abrirían con el empleo de satélites terrestres.

# Bibliografía